\documentclass[11pt,fleqn]{book}
\usepackage{a4wide}
\usepackage{enumerate}
\usepackage{psfig,subfigure}
\usepackage{epsfig}
\usepackage[bookmarks]{hyperref}

\usepackage[nottoc,notlot,notlof]{tocbibind}

\begin{document}
%
%
%
%
%

\pagenumbering{gobble}
\section*{}
\setlength{\parindent}{0cm}
\begin{center}



\vspace*{1.cm}

{\Huge Modern Antennas and Microwave Circuits} \\

\vspace*{1.cm}

{\LARGE A complete master-level course} \\



\vspace*{4.cm}


{\large by} \\
{\large Prof. dr.ir. A.B. Smolders, Prof.dr.ir. H.J. Visser and dr. Ulf Johannsen} \\
\vspace*{1.cm}

{\large Electromagnetics Group} \\
{\large Center for Wireless Technology Eindhoven (CWTe)} \\
{\large Department of Electrical Engineering} \\
{\large Eindhoven University of Technology, The Netherlands} \\

\vspace{1cm}

{\large Version January 2022} \\

\vspace{3cm}

\includegraphics[width=3cm]{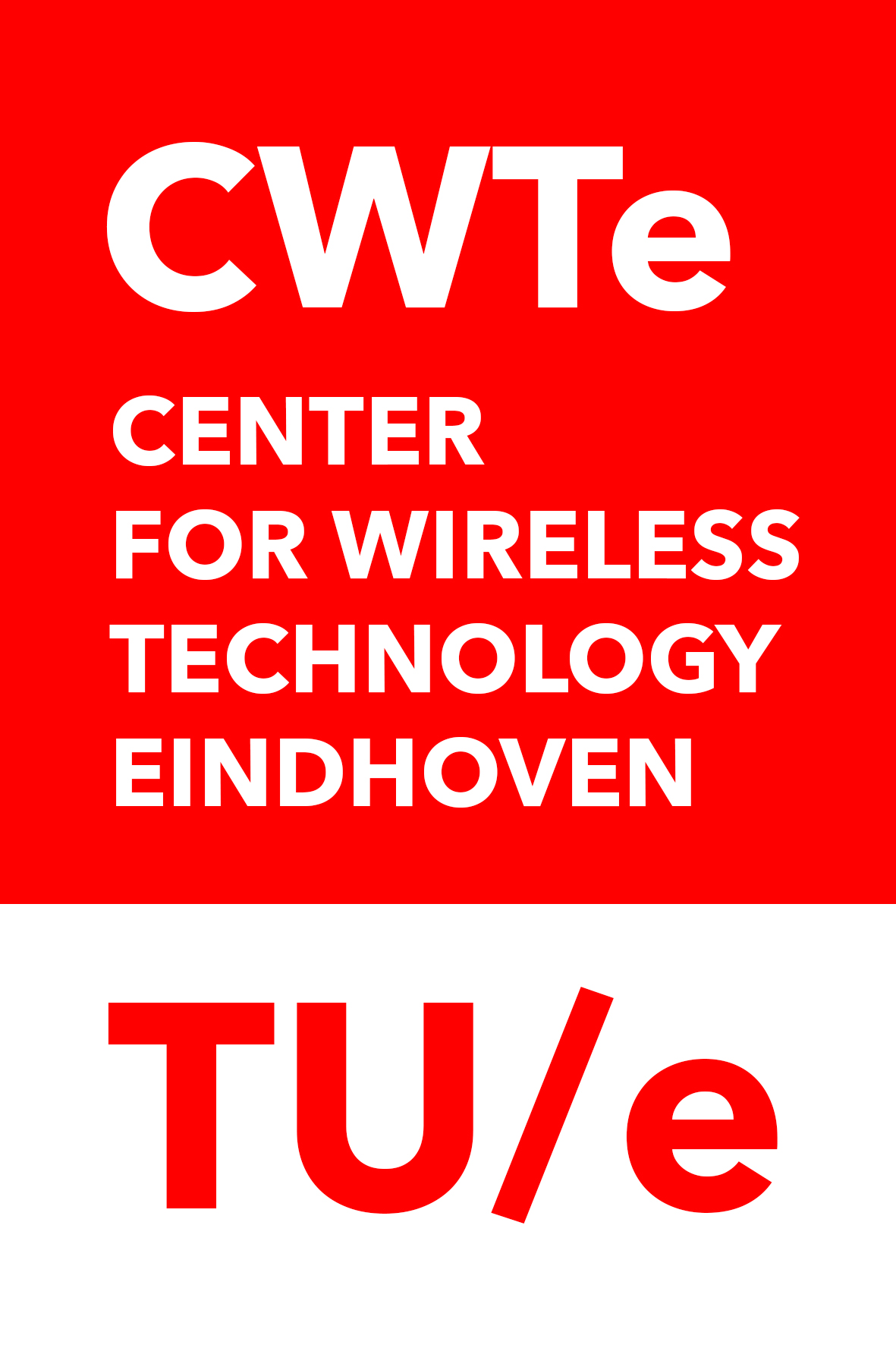} \\

\end{center}

\newcommand{\sinc}{{\mbox{sinc}}}
\newcommand{\sign}{{\mbox{sign}}}
\newcommand{\fG}{{\cal{G}}}
\newcommand{\eps}{{\epsilon_r}}
\newcommand{\dfG}{\bar{  \! \! \! \; \bar{\cal{G}}}}
\newcommand{\fF}{{\cal{F}}}
\newcommand{\dinttw}{\int \! \! \! \int}
\newcommand{\dintdr}{\int \! \! \! \int \! \! \! \int}
\newcommand{\dinti}{\int \! \! \! \int\limits_{-\infty}^{\infty}}
\newcommand{\dintia}{\int\limits_{-\infty}^{\infty} \int\limits_{0}^{\infty}}
\newcommand{\dintba}{\int\limits_{-\infty}^{0} \int\limits_{0}^{\infty}}
\newcommand{\dintab}{\int\limits_{0}^{\infty} \int\limits_{-\infty}^{0}}
\newcommand{\dintbb}{\int\limits_{-\infty}^{0} \int\limits_{-\infty}^{0}}
\newcommand{\dintaa}{\int\limits_{0}^{\infty} \int\limits_{0}^{\infty}}
\newcommand{\vE}{{\vec{E}}}
\newcommand{\vA}{{\vec{A}}}
\newcommand{\fA}{{\cal{A}}}
\newcommand{\dG}{\bar{\! \bar{G}}}
\newcommand{\vJ}{{\vec{J}}}
\newcommand{\vfJ}{{\vec{\cal{J}}}}
\newcommand{\dE}{\bar{\! \bar{E}}}
\newcommand{\dfQ}{\bar{\! \bar{\cal{Q}}}}
\newcommand{\dQ}{\bar{\! \bar{Q}}}
\newcommand{\dQa}{\bar{\bar{\widetilde{Q}}}}
\newcommand{\dsom}{{\sum_{m=-\infty}^{\infty}\!\sum_{n=-\infty}^{\infty}}}
\newcommand{\dsompq}{{\sum_{p=-\infty}^{\infty}\!\sum_{q=-\infty}^{\infty}}}
\newcommand{\dsomo}{{\sum_{m=1}^{\infty}\!\sum_{n=1}^{\infty}}}
\newcommand{\he}{{\hat{e}}}
\newcommand{\dints}{{\int \! \! \! \int\limits_{S_p}}}
\newcommand{\dintspr}{{\int \! \! \! \int\limits_{S_{probe}}}}
\newcommand{\fQ}{{\\cal{Q}}}
\newcommand{\som}{{\sum_{n=-\infty}^{\infty}}}
\newcommand{\somi}{{\sum_{m=-\infty}^{\infty}}}
\newcommand{\somam}{{\sum_{m=1}^{\infty}}}
\newcommand{\soman}{{\sum_{n=1}^{\infty}}}
\newcommand{\somie}{{\sum_{\stackrel{m=-\infty}{m \neq 0}} ^{\infty}}}
\newcommand{\inti}{{\int\limits_{-\infty}^{\infty}}}
\newcommand{\wo}{{\omega _o}}
\newcommand{\vr}{\vec{r}}
\newcommand{\jm}{\jmath}
\newcommand{\vfE}{\vec{\cal{E}}}
\newcommand{\fE}{{\cal{E}}}
\newcommand{\vfA}{\vec{\cal{A}}}
\newcommand{\w}{\omega}
\newcommand{\somN}{\sum _{j=1}^{N_{max}}}
\newcommand{\tinti}{\int \! \! \! \int \! \! \! \int\limits_{-\infty}^{\infty}}
\newcommand{\ve}{{\vec{e}}}
\newcommand{\tint}{\int \! \! \! \int \! \! \! \int}
\newcommand{\be}{\beta}
\newcommand{\al}{\alpha}
\newcommand{\ResM}{\begin{array}{c} Res \\ { \scriptstyle z=\frac{l2 \pi -k_oau}{k^x} \frac{W_x}{a}}  \end{array} }
\newcommand{\leZ}{l \in {\cal{Z}}}
\newcommand{\somil}{{\sum_{l=-\infty}^{\infty}}}
\newcommand{\di}{\displaystyle}
\newcommand{\intbo}[0]{    \int\limits_{-\infty}^{\infty}
\mbox{\makebox[-2.08em][l]{} \scriptsize
\begin{tabular}{c} $\;\cap$ \\ $\cup\;$ \end{tabular} } }
\newcommand{\intb}[0]{     \int\limits_{-\infty}^{\infty}
\hspace*{-0.65mm} \mbox{\makebox[-1.3em][l]{} \scriptsize $\cap$ }
}
\newcommand{\intbii}[0]{     \int\limits_{0}^{\infty}
\mbox{\makebox[-1.1em][l]{} \scriptsize $\cap$ }                                                }
\newcommand{\into}[0]{     \int\limits_{-\infty}^{\infty}
\mbox{\makebox[-1.43em][l]{} \scriptsize $\cup$ }                                                }
\newcommand{\res}[1]{      \!\!\mbox{\tiny
\begin{tabular}[t]{c} {\normalsize Res} \\ $ #1 $ \end{tabular} } \!\!    }

\section*{}
\setlength{\parindent}{1cm}

Modern Antennas and Microwave Circuits - A complete master-level course \\
A.B. Smolders, H.J. Visser, U. Johannsen.  \\
Eindhoven University of Technology, version January 2022 (First version 2019). \\
\\
A catalogue record is available from the Eindhoven University of Technology Library \\
ISBN: 978-90-386-4943-6 \\
\\
Subject headings: Antennas, microwave engineering, phased-arrays, wireless communications. \\
\\
\\
\copyright 2022 by A.B. Smolders, H.J. Visser, U. Johannsen, Eindhoven, The Netherlands.\\
All rights reserved. No part of this publication may be reproduced or transmitted
in any form or by any means, electronic, mechanical, including photocopy,
recording, or any information storage and retrieval system, without the prior
written permission of the copyright owner.

\pagenumbering{arabic}
\tableofcontents
\chapter{Introduction}
\label{chap:introductie}

\section{Purpose of this textbook}
This textbook provides all relevant material for Master-level courses in the domain of antenna systems. For example, at Eindhoven University of Technology, we use this book in two courses (total of 10 ECTS) of the master program in Electrical Engineering, distributed over a semester. The book includes comprehensive material on antennas and provides a solid introduction into microwave engineering, ranging from passive components to active circuits. We believe that this is the perfect mixture of know-how for a junior antenna expert. The theoretical material in this textbook can be supplemented by labs in which the students learn how to use state-of-the-art antenna and microwave design tools and test equipment, such as a vector network analyser or a near-field scanner.

Several chapters in this book can be used quite independent from each other. For example, chapter \ref{chap:Phased-Arrays} can be used in a dedicated phased-array antenna course without the need to use Maxwell-based antenna theory from chapter \ref{chap:AntennaTheory}. Similarly, chapter \ref{chap:Tlines} does not require deep back-ground knowledge in electromagnetics or antennas. The antenna theory presented in chapter \ref{chap:AntennaTheory} is inspired by the original lecture notes of dr. Martin Jeuken \cite{Jeuken}.

\section{Antenna functionality}

According to the Webster dictionary, an antenna is:
\begin{enumerate}[i.]
\item one of a pair of slender, movable, segmented sensory organs on the head of insects, myriapods, and crustaceans,
\item a usually metallic device (such as a rod or wire) for radiating or receiving radio waves.
\end{enumerate}
The second definition of an antenna is used within the domain of Electrical Engineering.
An antenna transforms an electromagnetic wave in free space in to a guided wave that propagates along a transmission line and vice-versa. In addition, antennas often provide spatial selectivity, which means that the antenna radiates more energy in a certain direction (transmit case) or is more sensitive in a certain direction (receive case).
Traditional antenna concepts, like wire antennas, are passive electromagnetic devices which implies that the reciprocity concept holds. As a result, the transmit and receive properties are identical.
In integrated antenna concepts, where the transmit and receive electronics are part of the antenna functionality, reciprocity cannot always be applied.

\section{History of antennas}

One of the principal characteristics of human beings is that they almost
continually send and receive signals to and from one another, \cite{Smolders}.
The exchange of meaningful signals is the heart of what is called communication. In its
simplest form, communication involves two people, namely the signal transmitter
and the signal receiver. These signals can take many forms. Words are the most
common form. They can be either written or spoken.
Before the invention of technical resources such as radio communication or telephone,
long-distance communication was very difficult and usually took a lot of time.
Proper long-distance communication was at that time
only possible by exchange of written words. Couriers were used to transport
the message from the sender to the receiver. Since the invention of
radio and telephone, far-distance communication or {\it tele}communication
is possible, not only of written words, but also of spoken words and
almost without any time delay between the transmission and
the reception of the signal.
Antennas have played an important part in the development of our present
telecommunication services. Antennas have made it possible  to
communicate at far distances, without the need of a physical connection
between the sender and receiver. Apart from the sender and receiver, there
is a third important element in a communication system, namely the propagation
channel. The transmitted signals may deteriorate when they propagate
through this channel.

\begin{figure}[hbt]
\centerline{\psfig{figure=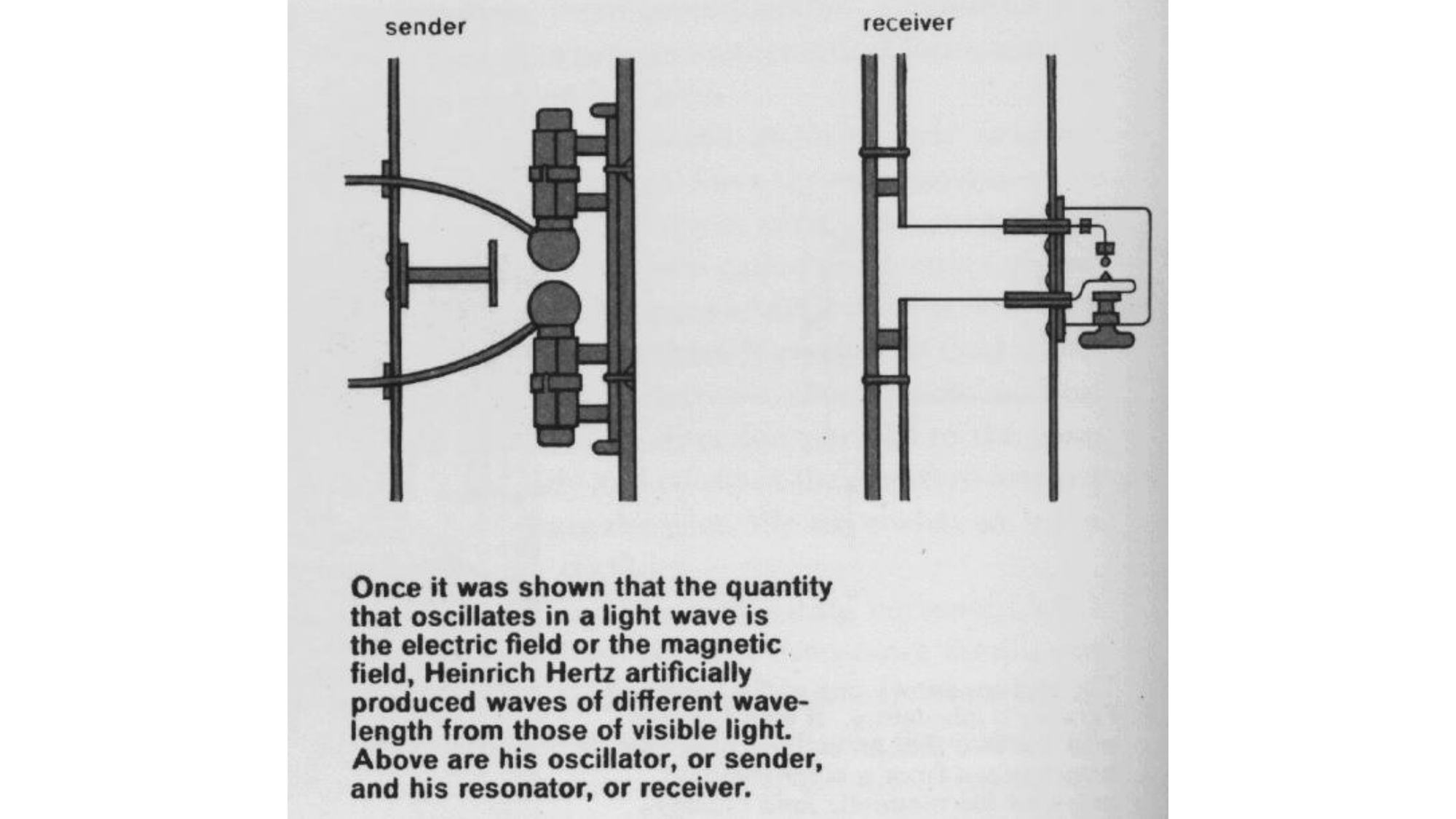,width=150mm}}
\caption{\it First radio system of Heinrich Hertz operating at a wavelength of about 8 meter \cite{Hertz}.}
\label{fig:Hertz}
\end{figure}

In 1886, Heinrich Hertz, who was a professor of physics at the Technical Institute
in Karlsruhe, was the first person who made a complete radio system
\cite{Hertz}.
When he produced sparks at a gap of the transmitting antenna, sparking also
occurred at the gap of the receiving antenna. Hertz in fact visualised the
theoretical postulations of James Clerk Maxwell. Hertz's first experiments
used wavelengths of about 8 meters, as illustrated in Fig. \ref{fig:Hertz}.
After Hertz the Italian Guglielmo Marconi became the motor behind the development of practical
radio systems \cite{Marconi}. He was not a famous scientist like Hertz, but he was obsessed
with the idea of sending messages with a wireless communication system.
He was the first who performed wireless communications across the Atlantic.
The antennas that Marconi used were very large wire antennas mounted onto
two 60-meter wooden poles. These antennas had a very poor efficiency, so
a lot of input power had to be used. Sometimes the antenna wires
even glowed at night.
In later years antennas were also used for other purposes such as radar systems and radio
astronomy. In 1953, Deschamps \cite{Deschamps} reported for the first time about
planar microstrip antennas, also known as patch antennas.
It was only in the early eighties that microstrip antennas became an interesting topic
among scientists and antenna manufacturers. Microstrip antennas are an example of so-called printed antennas which are metal structures printed on a substrate (e.g. printed-circuit-board (PCB)).
\begin{figure}[htb]
\centerline{\hbox{\psfig{figure=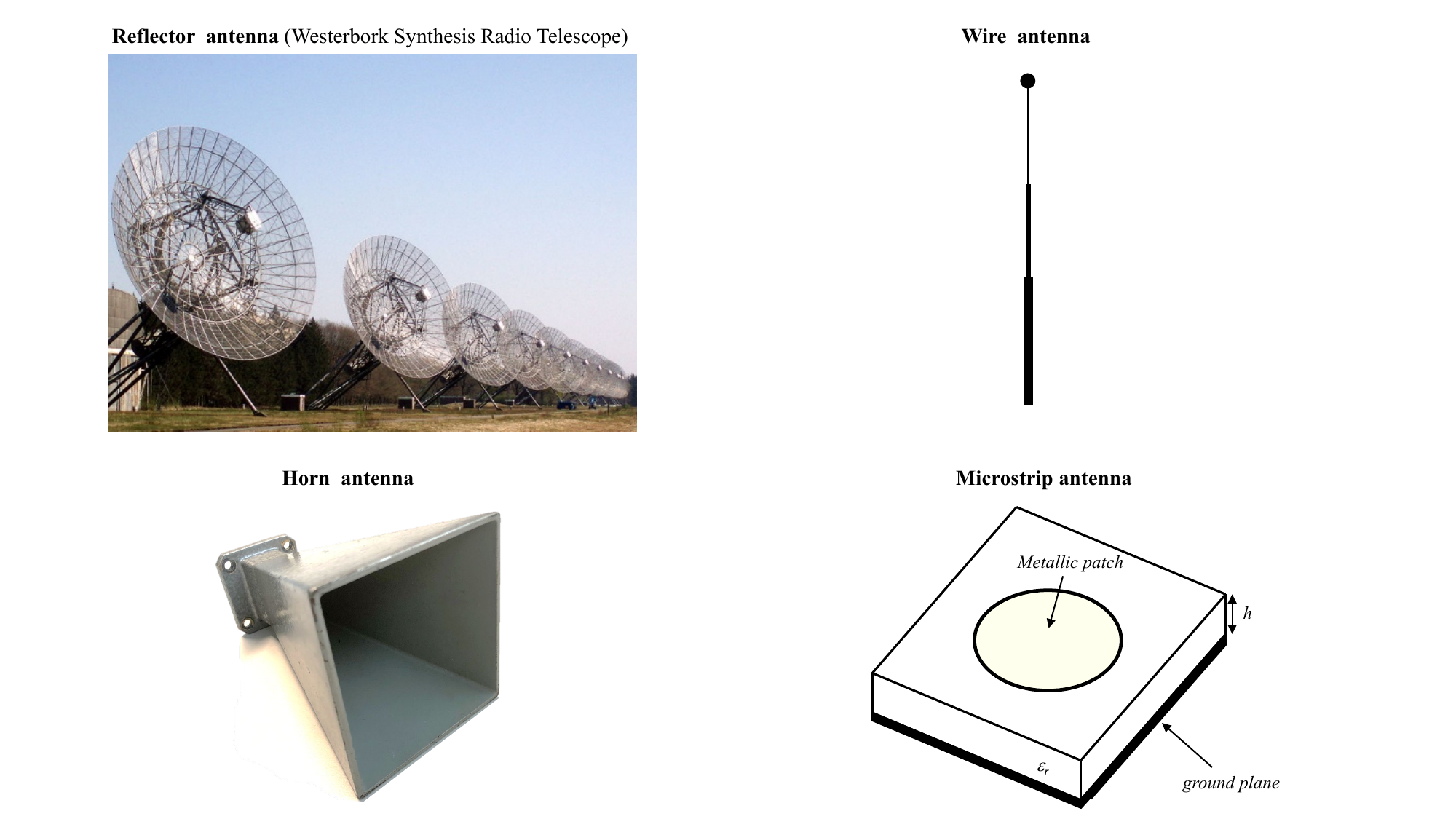,width=190mm}}}
\caption{{\it Some well-known antenna concepts.}}
\label{fig:antconc}
\end{figure}
Figure \ref{fig:antconc} shows some well-known antenna configurations which
are nowadays used in a variety of applications.
Reflector antennas, horn antennas, wire antennas and printed antennas all have become mass products.
Reflector antennas are often used for the reception of satellite television, whereas
wire antennas are commonly used to receive
radio signals with a car or portable radio receiver. Printed antennas are also widely used nowadays, for example in smart phones. Printed antennas can be realized as two-dimensional structures on PCBs or as three-dimensional structures on plastic substrates. More recently, printed antennas have even been realized on chips in integrated circuits (ICs)\cite{Adela_AoC}.

\begin{figure}[htb]
\centerline{\hbox{\psfig{figure=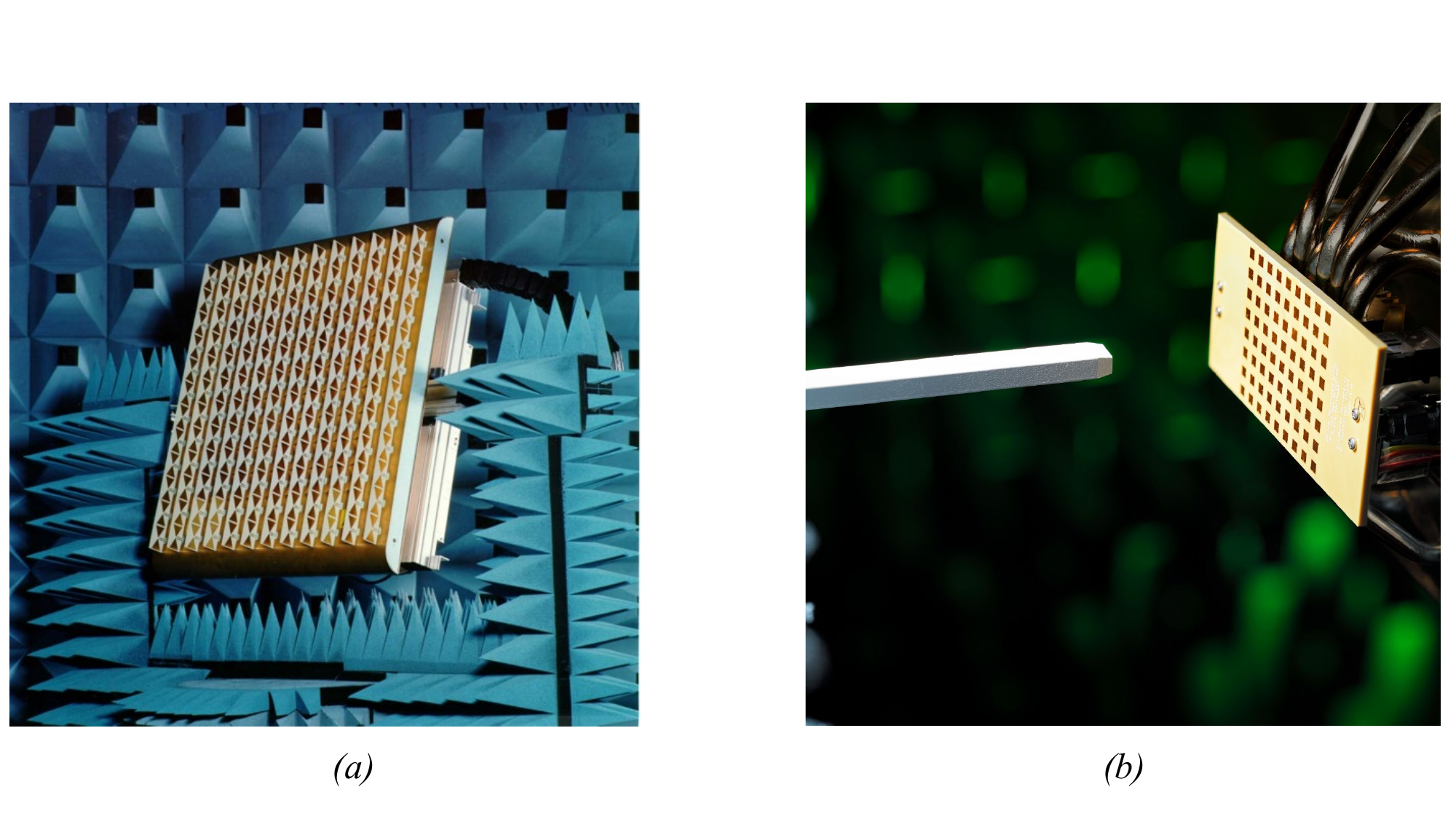,width=165mm}}}
\vspace{-0.5cm}
\caption{\it Examples of phased-array antennas. (a) The one-square-meter-array (OSMA) was a test-bed of the square-kilometer-array (SKA) and consists of an active center of 64 bow-tie antennas with electric beamforming. The active center is surrounded by two rows of passive bow-tie antennas \cite{Smolderscalibration}, \cite{OSMA1999}, \cite{SmoldersTHEA}.
(b) Dual-polarized active phased array prototype for mm-wave (26.5-29.5 GHz) 5G/beyond-5G wireless communications \cite{Biggelaar5Garray}, \cite{Bronckers5G}, \cite{Tessema}.}
\label{fig:arconcept}
\end{figure}

Reflector antennas provide high directivity (spatial selectivity), but have the disadvantage
that the main lobe of the antenna has to be steered in the desired direction by means
of a highly accurate mechanical steering mechanism. This means that
simultaneous communication with several points in space is not possible.
Wire and printed antennas are usually more omni-directional, resulting in a low directivity.
There are certain applications where these conventional antennas cannot
be used. These applications often require a phased-array antenna.
A phased-array antenna has the capability to communicate with several targets which may be
anywhere in space, simultaneously and continuously,
because the main beam of the antenna can be directed electronically into
a certain direction. Another advantage of phased-array antennas is the
fact that they are relatively flat. Figure \ref{fig:arconcept} shows two examples of phased arrays used in radio astronomy and in base stations for future mm-wave 5G and beyond-5G wireless communications, respectively.
In a phased-array system, three essential layers can be distinguished 1) an antenna layer, 2)
a layer with transmitter and receiver modules (T/R modules)
and 3) a signal-processing and control layer that controls
the direction of the main beam of the array. The antenna layer consists of
several individual antenna elements which are placed on a rectangular or on a triangular
grid.
Arrays of antennas can also be used to realize {\it multiple-input-multiple-output (MIMO)} communication systems. In such a system several non-correlated spatial communication channels can be realized between two devices (e.g. a base station and mobile user), resulting in a much higher channel capacity \cite{Hampton_MIMO}.

\section{Electromagnetic spectrum}

\begin{figure}[hbt]
\centerline{\psfig{figure=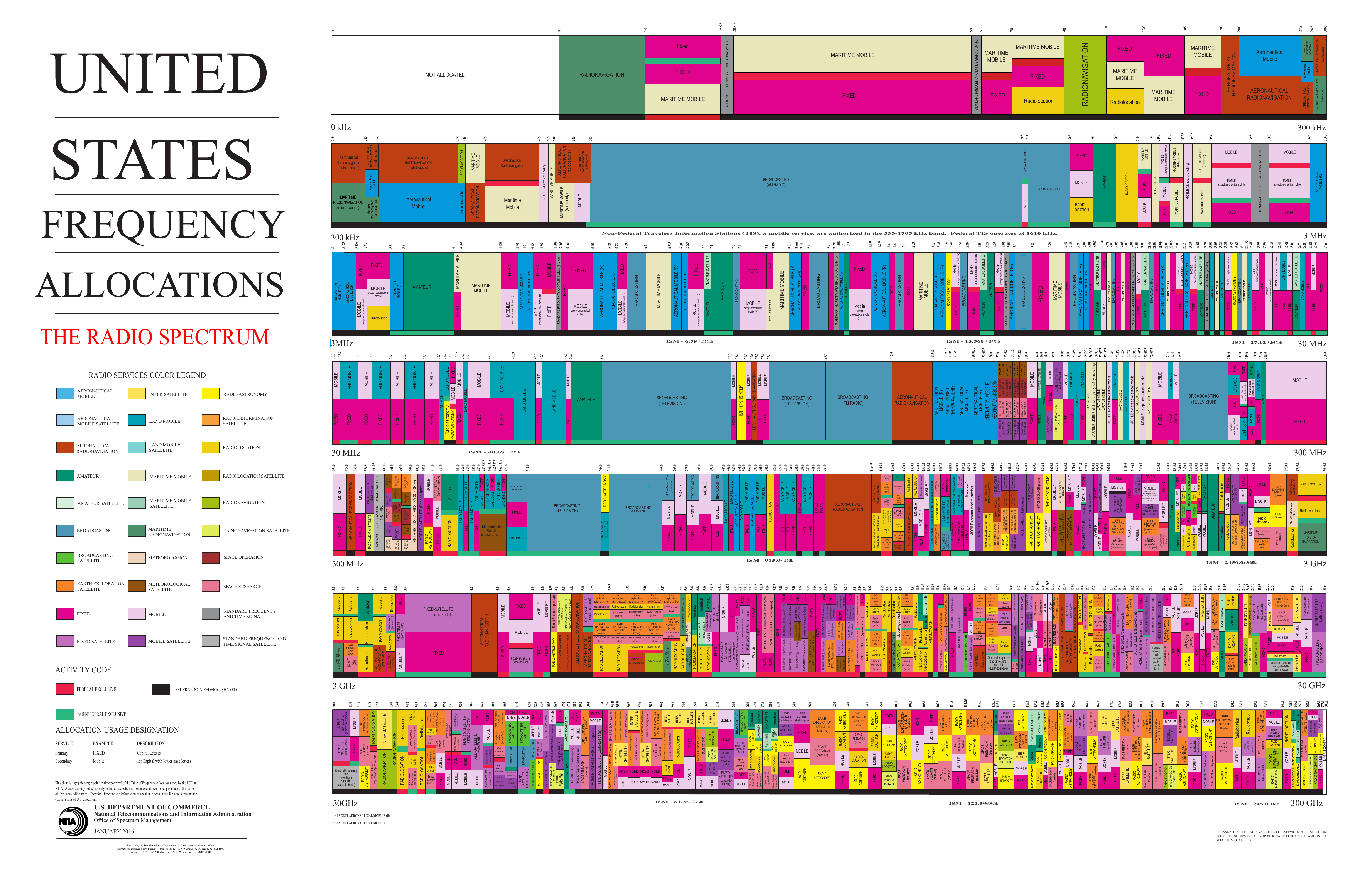,width=170mm}}
\caption{\it Example of spectrum allocation, frequency range 0-300 GHz, United States, 2016. Applications include wireless communication, satellite communications, military and commercial radar, radio astronomy and license-free Industrial-Scientific-Medical (ISM) bands. More information can be found at: www.ntia.doc.gov and www.itu.int}
\label{fig:spectrum}
\end{figure}

Antenna systems radiate electromagnetic energy at a particular frequency. The specific frequency range which can be used for a specific application is restricted and regulated worldwide by the International Telecommunication Union (ITU). The ITU organizes World Radiocommunication Conferences (WRCs) to determine the international Radio Regulations, which is the international treaty which defines the use of the world-wide radio-frequency spectrum. An example of such an allocation chart of the radio-frequency spectrum is shown in Fig. \ref{fig:spectrum}. This particular map is valid in the United States, but is very representative for the world-wide spectrum allocation.

\chapter{System-level antenna parameters}
\label{chap:fundpar}

\section{Coordinate system and time-harmonic fields}

\begin{figure}[hbt]
\centerline{\psfig{figure=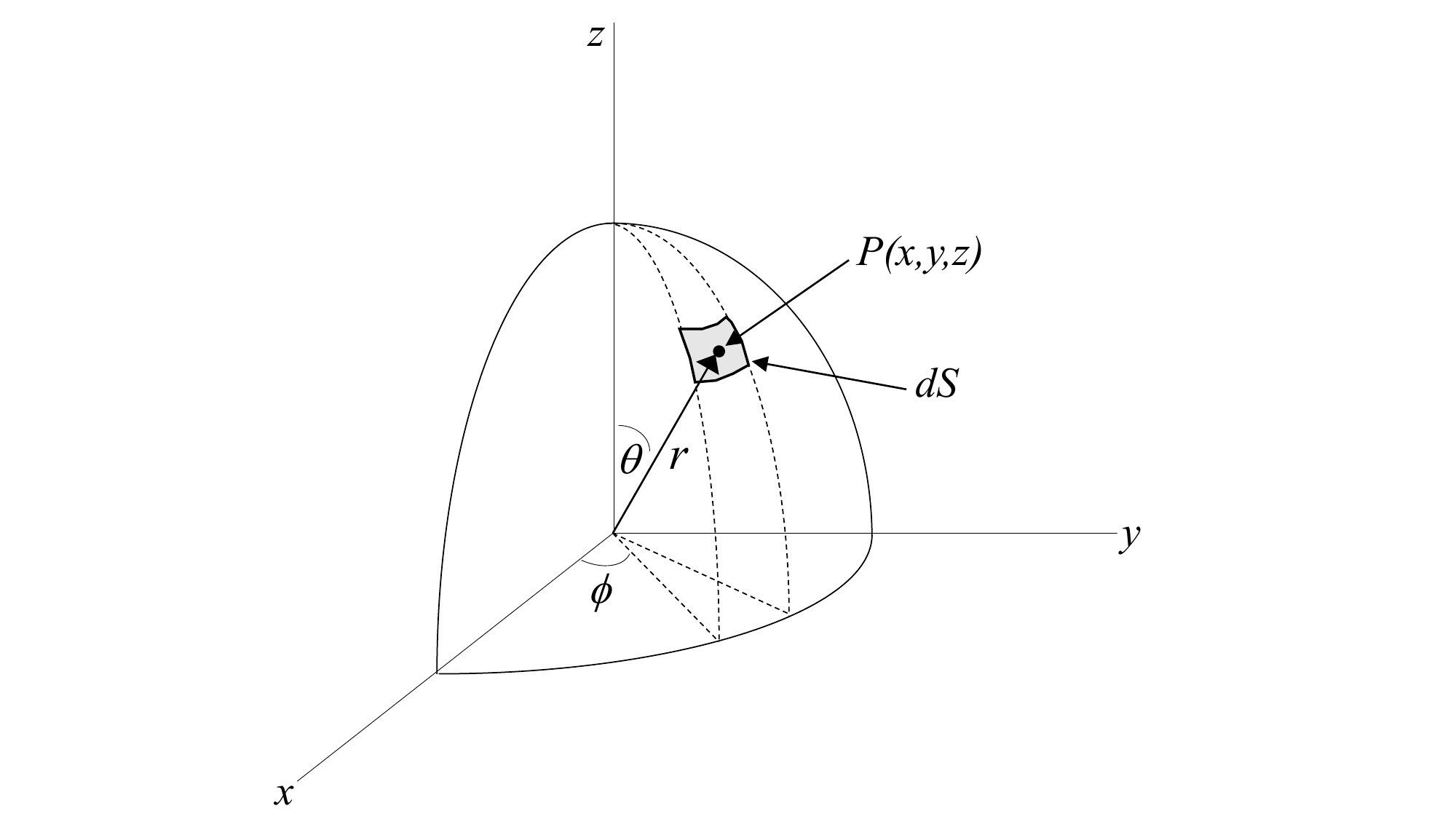,width=140mm}}
\caption{\it Spherical coordinate system with $dS=r^2\sin\theta d\theta d\phi=r^2 d\Omega$, where $d\Omega=\sin\theta d\theta d\phi$ and $\Omega$ is the solid angle.}
\label{fig:coordinaten}
\end{figure}

Radiation properties of antennas are generally expressed in terms of spherical coordinates. The spherical coordinate system is shown in Figure \ref{fig:coordinaten}. In this spherical coordinate system we will use unit vectors along the $r$, $\theta$ and $\phi$ directions which are denoted by $\vec{u}_r$, $\vec{u}_{\theta}$ and $\vec{u}_{\phi}$, respectively. A particular point $P(x,y,z)$ is indicated by the position vector $\vec{r}=x\vec{u}_x+y\vec{u}_y+z\vec{u}_z=\sqrt{x^2+y^2+z^2}\vec{u}_r=|\vec{r}|\vec{u}_r$.

In this book we will assume that all currents, voltages and electric and magnetic field components have a sinusoidal time variation. The instantaneous time-domain electric field $\vec{\cal E}$ is now related to a complex electric field $\vec{E}$ according to:
\begin{equation}
\displaystyle \vec{\cal E} (\vec{r},t)  =  \di Re[\vec{E}(\vec{r})e^{\jmath \omega t}],
 \label{eq:timeharmonic1}
\end{equation}
where $\omega=2\pi f$ is the angular frequency, $f$ is the frequency of operation and $Re[]$ denotes the real part of the complex variable.
The electric field $\vec{E}(\vec{r})$ in the space-frequency domain (phasor formulation) can now be expressed in terms of spherical coordinates:
\begin{equation}
\begin{array}{lcl}
\displaystyle \vec{E}(\vec{r}) &=&  E_r(\vec{r})\vec{u}_r +
E_{\theta}(\vec{r})\vec{u}_{\theta} +
E_{\phi}(\vec{r})\vec{u}_{\phi} \\
\displaystyle &=& |E_r(\vec{r})|e^{\jmath \varphi_{r}(\vec{r})}\vec{u}_r +
|E_{\theta}(\vec{r})|e^{\jmath \varphi_{\theta}(\vec{r})}\vec{u}_{\theta} +
|E_{\phi}(\vec{r})|e^{\jmath \varphi_{\phi}(\vec{r})}\vec{u}_{\phi}, 
\end{array}
 \label{eq:coordinatendefinitie}
\end{equation}
where $\varphi_{r}$, $\varphi_{\theta}$, $\varphi_{\phi}$ are the phases of the spherical components of the complex electric field.
By combining \ref{eq:timeharmonic1} and \ref{eq:coordinatendefinitie} we can write the time-domain electric field $\vec{\cal E} (\vec{r},t)$ in terms of the space-frequency domain components:
\begin{equation}
\displaystyle \vec{\cal E} (\vec{r},t)  = |E_r|\cos{(\omega t +\varphi_{r})}\vec{u}_r +
|E_{\theta}|\cos{(\omega t +\varphi_{\theta})} \vec{u}_{\theta} +
|E_{\phi}| \cos{(\omega t +\varphi_{\phi})} \vec{u}_{\phi},
 \label{eq:timeharmonic1b}
\end{equation}
where we omitted the $(\vec{r})$ dependence of the amplitudes and phases of the field components to simplify the notation.

\section{Field regions}

\begin{figure}[hbt]
\vspace{-1cm}
\centerline{\psfig{figure=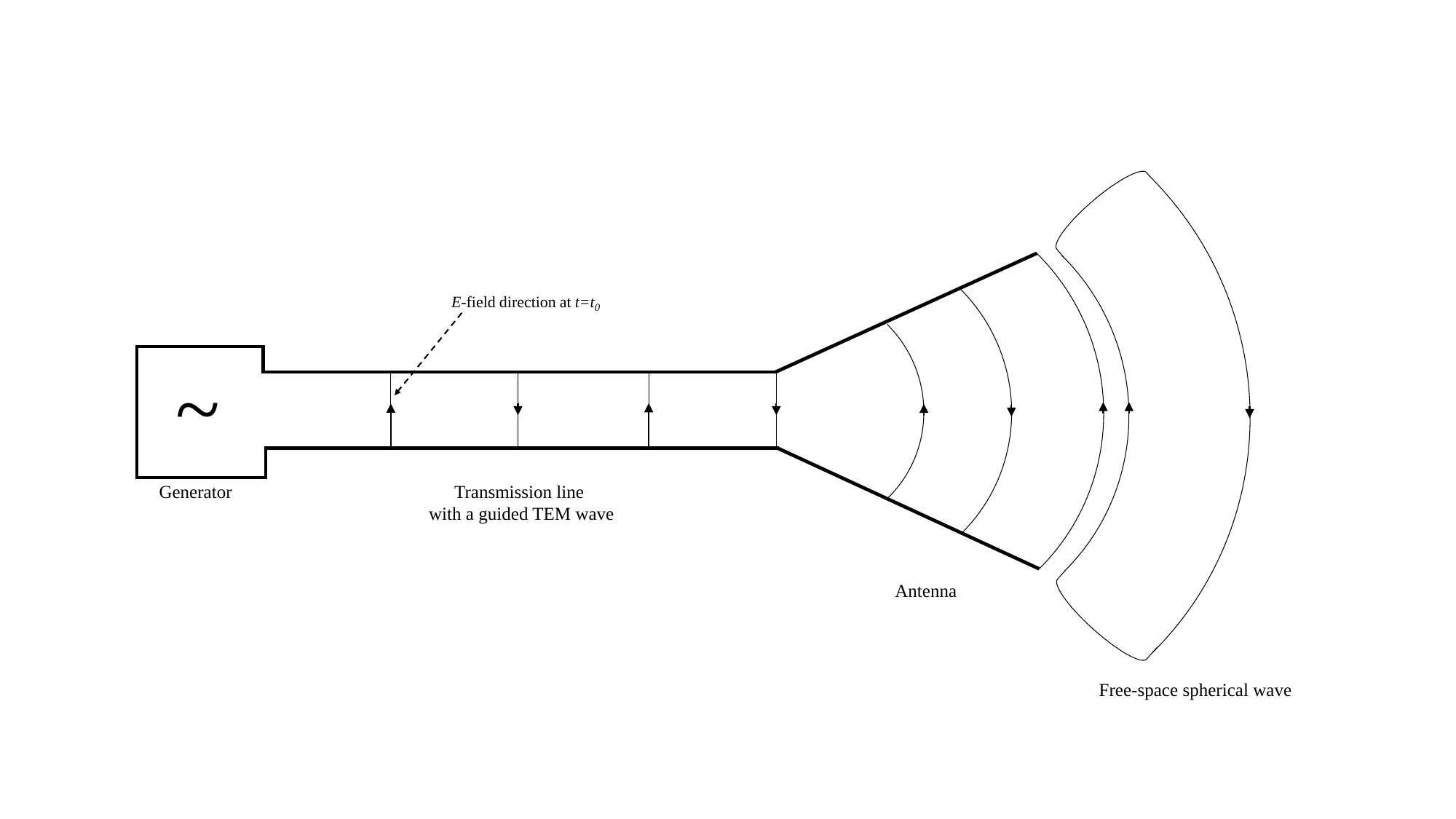,width=170mm}}
\vspace{-1cm}
\caption{\it Transition of a guided wave along a transmission line into radiated spherical waves in free space using an antenna.}
\label{fig:transantenne}
\end{figure}

An antenna transforms a guided electromagnetic (EM) wave that propagates along a transmission line into radiated waves that propagate in free space.
These radiated waves can, in turn, be received by another antenna. Due to reciprocity, the receiving antenna will transform the incident EM waves into a guided EM wave along a transmission line.
One could also interpret this as the transition of electrons in conductors to photons in free space.
The transition from a guided wave along a transmission line to radiated waves from the antenna is illustrated in Fig. \ref{fig:transantenne}.

\begin{figure}[hbt]
\centerline{\psfig{figure=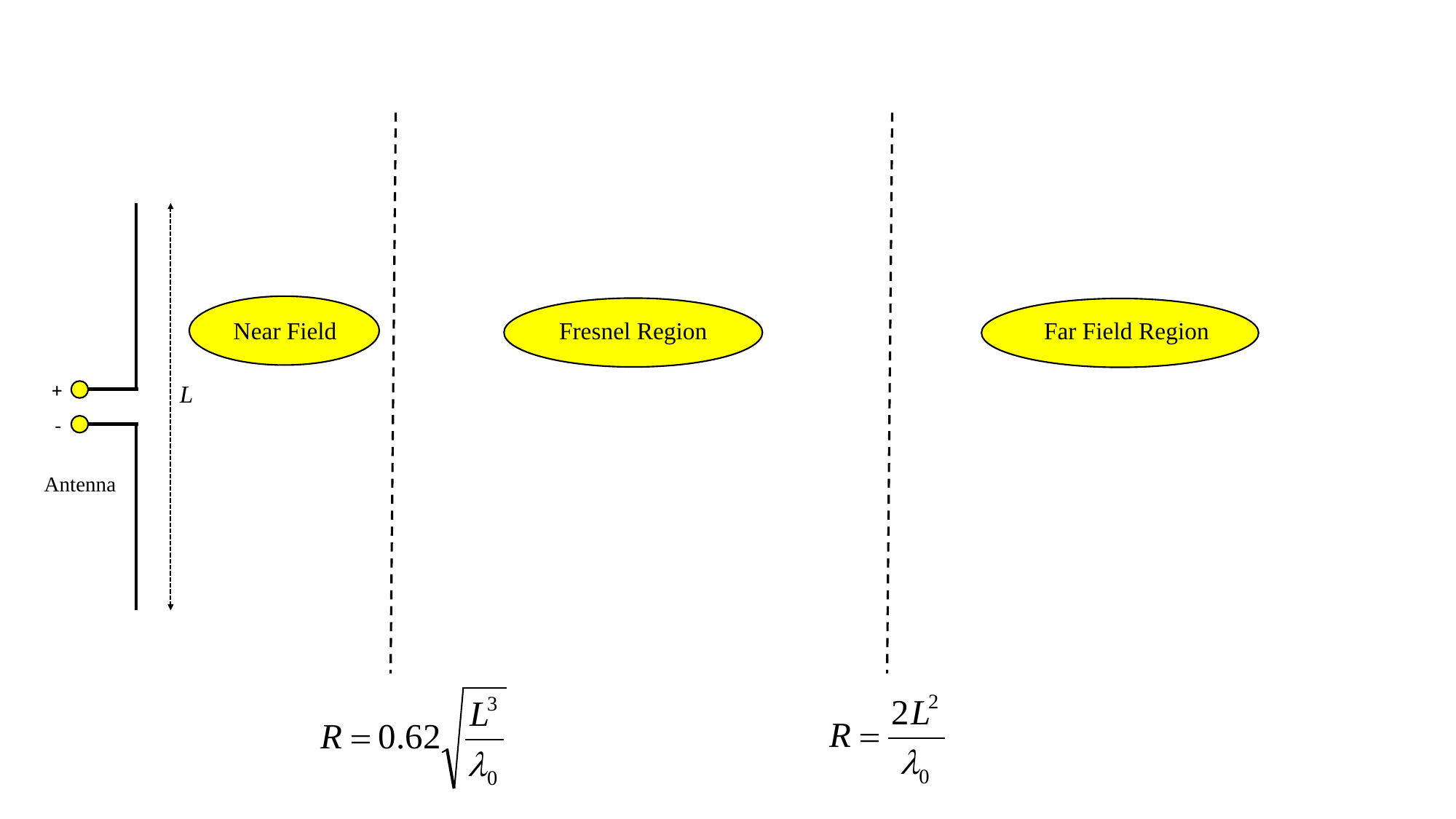,width=140mm}}
\caption{\it Definition of three field regions around an antenna.}
\label{fig:veldregio}
\end{figure}

The electromagnetic field radiated by an antenna can be divided into three main regions: 1) the {\it near-field} region, 2) transition or {\it Fresnel} region and 3) the {\it far-field} region.
Fig. \ref{fig:veldregio} shows an example of an horn antenna and the corresponding field regions.
In the far-field region, the electric and magnetic field components are orthogonal to each other and to the direction of propagation, resulting in transverse electromagnetic waves, also known as {\it plane waves}. For an antenna located in the origin of the coordinate system of Fig. \ref{fig:coordinaten}, the electromagnetic energy will flow in the $\vec{u}_r$ direction.
There is not a very clear transition between the three field regions. However, it is common to use the so-called {\it Fraunhofer distance} to define the start of the far-field region:
\begin{equation}
\displaystyle R = \frac{2L^2}{\lambda_0}, \label{eq:verreveld}
\end{equation}
where $R$ is the radial distance from the antenna, $L$ is the largest dimension of the antenna, $\di \lambda_0=\frac{c}{f}$ is the free-space wavelength of the radiated electromagnetic wave and $c=3 \cdot 10^{8}$ [m/s] is the speed of light. In chapter 4 we will show how criterium (\ref{eq:verreveld}) can be derived directly from the theoretical antenna framework that we will introduce.

\section{Far field properties}

Consider a transmit antenna located in the origin of the coordinate system of Fig. \ref{fig:coordinaten} and a receive antenna in the far-field region at a distance $\di R> \frac{2L^2}{\lambda_0}$. In this case, the incident field at the receiving antenna will be locally flat, similar to an incident plane wave with an equi-phase plane perpendicular to the direction of propagation. These plane waves in free space are transverse electromagnetic waves (TEM) with the following general form in the frequency domain:
\begin{equation}
\begin{array}{lcl}
\displaystyle E_y (z) &=& A e^{-\jmath k_0 z},
\end{array}
\label{eq:vlakkegolf}
\end{equation}
where $k_0=\frac{\omega}{c}$ is the free-space wavenumber and $A$ the amplitude.
Note that in (\ref{eq:vlakkegolf}), we have assumed that the plane wave propagates in the $+z$-direction.
For the sake of simplicity we only considered a component in the $\vec{u}_y$ direction. The corresponding magnetic field $\vec{H}$ is perpendicular both to the $\vec{E}$ field and the direction of propagation.

The transmit antenna will radiate spherical waves. In chapter 4 and 6 we will investigate the radiated fields in more detail. It will be shown that the far field in spherical coordinates $\vec{E}(\vec{r}) =E_{\theta}(\vec{r})
\vec{u}_{\theta}+E_{\phi}(\vec{r}) \vec{u}_{\phi}+E_r
(\vec{r})\vec{u}_r$ due to an antenna placed at the origin of our coordinate system at $\vec{r}=(0,0,0)$ can be expressed in the following form:
\begin{equation}
\begin{array}{lcl}
\displaystyle E_{\theta} (\vec{r}) &=& \displaystyle E_{\theta}
(\theta, \phi) \frac{e^{-\jmath k_0 r}}{r}, \\

\displaystyle E_{\phi} (\vec{r}) &=& \displaystyle E_{\phi}
(\theta, \phi) \frac{e^{-\jmath k_0 r}}{r}, \\

\displaystyle E_r (\vec{r}) &=& 0.

\end{array}
\label{eq:verreveldalgemeen}
\end{equation}
Since EM waves in the far-field locally behave as TEM waves, we can write the corresponding magnetic field $\vec{H}$ in terms of the electric field:
\begin{equation}
\displaystyle \vec{H}(\vec{r}) = \frac{1}{Z_0} \vec{u}_r \times
\vec{E} (\vec{r}),
 \label{eq:hverreveld}
\end{equation}
where $Z_0=\sqrt{\frac{\mu_0}{\epsilon_0}} = \ 377 \Omega$ is the intrinsic free-space impedance.
The radiated power density expressed in Watts per square meter [$W/m^2$], is known as the {\it Poynting vector} and describes the directional energy flux (the energy transfer per unit area per unit time) of an electromagnetic field.
The Pointing vector in the far-field region only has a component in the radial direction $\vec{u}_r$.
The time-average Poynting vector $\vec{S_p}(\vec{r})$ over a period $T=\frac{2\pi}{\omega}$  at a specific position $\vec{r}$ in the far-field is now given by:
\begin{equation}
\begin{array}{lcl}
\displaystyle \vec{S_p}(\vec{r}) &=& \displaystyle \frac{1}{T}
\int\limits_{0}^{T} \vec{S_p}(\vec{r}, t) dt  \\

\displaystyle &=& \displaystyle \frac{1}{T} \int\limits_{0}^{T}
\vec{{\cal E}}(\vec{r}, t) \times \vec{{\cal H}}(\vec{r},
t) dt \\

\displaystyle &=& \displaystyle \frac{1}{T} \int\limits_{0}^{T}
Re[\vec{E}(\vec{r}) e^{\jmath \omega t}] \times
Re[\vec{H}(\vec{r}) e^{\jmath \omega t}]dt \\

\displaystyle &=& \displaystyle \frac{1}{T} \int\limits_{0}^{T}
\frac{1}{2}(\vec{E}(\vec{r}) e^{\jmath \omega t} + {\vec{E}^*}
(\vec{r}) e^{-\jmath \omega t}) \times
\frac{1}{2}(\vec{H}(\vec{r}) e^{\jmath \omega t} +
{\vec{H}^*}(\vec{r}) e^{-\jmath \omega t})dt \\

\displaystyle &=& \displaystyle \frac{1}{4} [\vec{E}(\vec{r})
\times {\vec{H}^*}(\vec{r}) + \vec{E}^*(\vec{r}) \times
{\vec{H}}(\vec{r}) ] \\

\displaystyle &=& \displaystyle \frac{1}{2} Re[\vec{E}(\vec{r})
\times {\vec{H}^*}(\vec{r})].

\end{array}
\label{eq:vectorpointing}
\end{equation}
In the far-field region we can use relation (\ref{eq:hverreveld}). As a result, expression (\ref{eq:vectorpointing}) can be written in the following form:
\begin{equation}
\begin{array}{lcl}
\displaystyle \vec{S_p}(\vec{r}) &=&  \frac{1}{2}
Re[\vec{E}(\vec{r}) \times {\vec{H}^*}(\vec{r})] = \frac{1}{2}
Z_0^{-1} Re [\vec{E}(\vec{r}) \times (\vec{u}_r \times {\vec{E}^*}
(\vec{r}) )] \\

\displaystyle &=& \frac{1}{2} Z_0^{-1} Re [ (\vec{E}(\vec{r})
\cdot {\vec{E}^*} (\vec{r}))\vec{u}_r -(\vec{E}(\vec{r}) \cdot
\vec{u}_r)\vec{E}^* (\vec{r})] \\

\displaystyle &=& \frac{1}{2} Z_0^{-1} |\vec{E}(\vec{r})|^2
\vec{u}_r ,

\end{array}
\label{eq:pointingverreveld}
\end{equation}
since $\vec{E}(\vec{r}) \cdot \vec{u}_r =0$. We can conclude that the electromagnetic energy propagates in the radial direction $\vec{u}_r$.

\section{Radiation pattern}

\begin{figure}[hbt]
\centerline{\psfig{figure=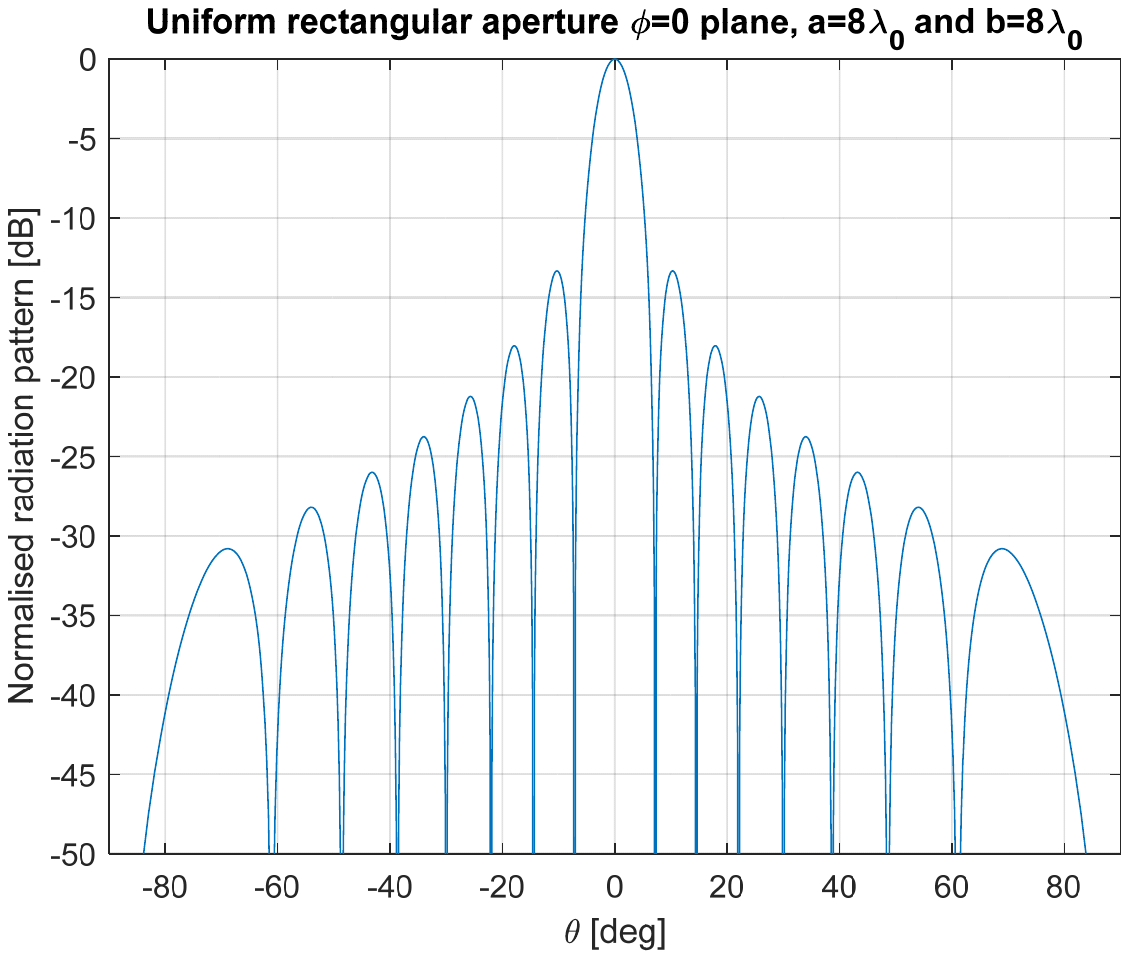,width=160mm}}
\vspace{-1cm}
\caption{\it Two-dimensional normalized radiation pattern. A cut in the $\phi=0^0$-plane is shown.
The antenna has a uniformly-illuminated square-shaped aperture of size $A=(8\lambda_0)^2$. The first sidelobe has a maximum level of -13.2 dB } \label{fig:tweedimradpat}
\end{figure}

\begin{figure}[hbt]
\vspace{-0.5cm}
\centerline{\psfig{figure=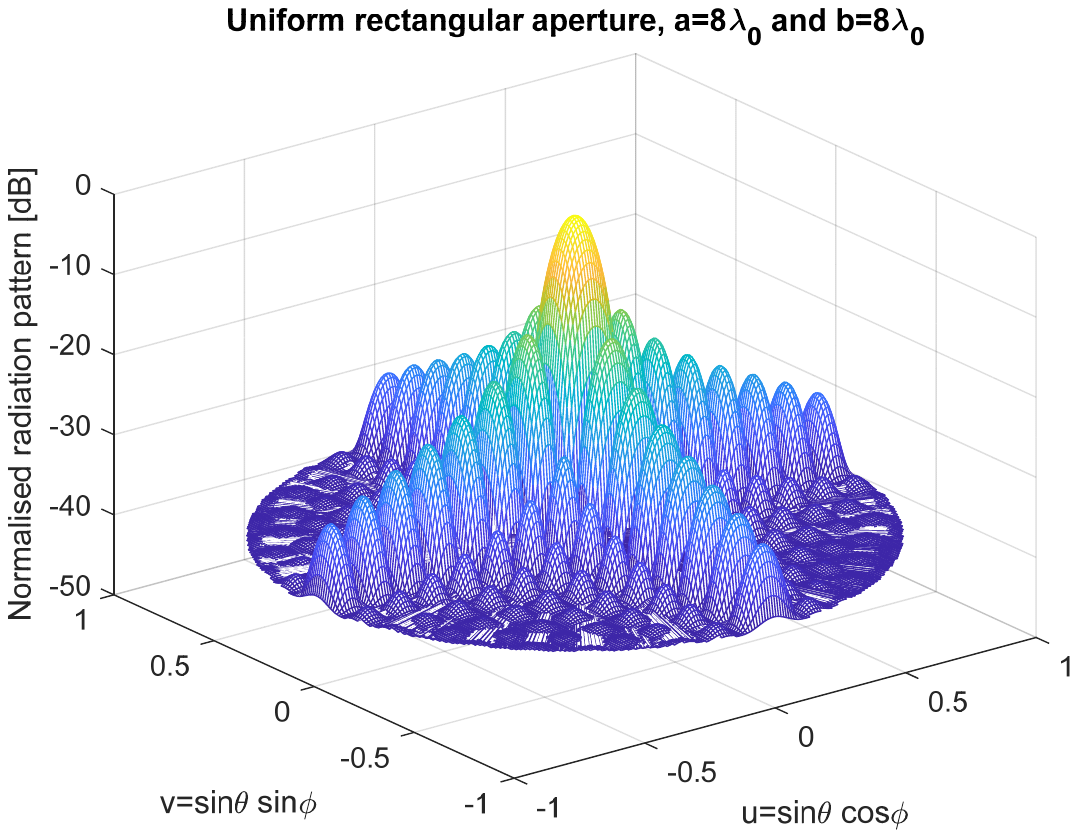,width=160mm}}
\vspace{-1.5cm}
 \caption{\it Three-dimensional normalized radiation pattern of the antenna described in Fig.\ref{fig:tweedimradpat}}
  \label{fig:driedimradpat}
\end{figure}

The radiation pattern is a graphical illustration of the radiated power in a certain direction in the far field region of the antenna.
From (\ref{eq:verreveldalgemeen}) we can observe that the electric field vector $\vec{E}$ has a $r^{-1}$ dependence in the far-field region. The radiated power per element of solid angle $d\Omega=sin{\theta}d\theta d\phi$ can be determined from (see also Fig. \ref{fig:coordinaten}):

\begin{equation}
\displaystyle P(\vec{r}) = P (\theta, \phi) =  | r^{2}
\vec{S_p}(\vec{r})|,
 \label{eq:radpower}
\end{equation}
where Poynting's vector is given by (\ref{eq:pointingverreveld}).
Expression (\ref{eq:radpower}) is only valid in the far-field region where the radiated power per element of solid angle is independent of $r$.
Let us assume that the maximum value of $P (\theta, \phi)$ occurs at the angle $(\theta,\phi)=(0,0)$. The {\it normalized radiation pattern} $F(\theta,\phi)$ is found from:
\begin{equation}
\displaystyle F(\theta, \phi) = \frac{P(\theta, \phi)}{P(0,0)}.
\label{eq:normradpat}
\end{equation}
Radiation patterns are always expressed in decibels [dB], that is using $10\log_{10}(F(\theta,\phi))$.
Some examples are shown in Fig. \ref{fig:tweedimradpat} and Fig. \ref{fig:driedimradpat}.
The {\it main beam}, also known as {\it main lobe}, of the antenna is directed towards the direction of maximum radiation, in this case towards $\theta=0$.
The other local maxima in the radiation pattern are called {\it sidelobes}.
The maximum sidelobe level is often one of the design parameters of an antenna, since the maximum sidelobe level determines the sensitivity of the antenna for (unwanted) interference from other directions.
The height of the sidelobes is determined by a number of factors, including the size, shape and type of antenna.
Some antennas, like small dipoles, do not have sidelobes: they only have a main lobe.
In chapter \ref{chap:Phased-Arrays} we will show in more detail how you can design array antennas with low sidelobes.
From the normalized radiation pattern we cannot determine all quality measures of an antenna. Therefore, we will
introduce some additional scalar antenna parameters, including {beam width, directivity, antenna gain, effective aperture, aperture efficiency and input impedance}. These parameters will be explained in more detail in the next sections.

\section{Beam width}

The beam width defines the beam area for which the radiated or received power is larger than half of the maximum power, the so-called {\it Half-Power Beam Width (HPBW)}. The beam width in the principle planes ($\theta=0$ and $\phi=0$ plane) is expressed by $\theta_{HP}$ or as $\phi_{HP}$.

\section{Directivity and antenna gain}

The directivity describes the beamforming capabilities of an antenna. It describes the concentration of radiated power in the main lobe w.r.t. all other directions.
Let us consider an antenna that generates a total radiated power $P_t$ [W].
When this antenna would be an {\it isotropic} radiator, the radiated power would be equally radiated over all directions with an uniform radiated power density of $\displaystyle \frac{P_t}{4\pi}$ [W per unit of solid angle].
Note that an isotropic radiator cannot exist in practise, it is only a theoretical reference that is used to define the directivity of a real antenna. The {directivity function} $D(\theta, \phi)$ of a real physical antenna is defined as the power density per unit of solid angle in the direction $(\theta,\phi)$ relative to an isotropic radiator:
\begin{equation}
\displaystyle D(\theta, \phi) = \frac{P(\theta, \phi)}{P_t /
4\pi}. \label{eq:richtfunctie}
\end{equation}
The maximum of the directivity-function is the directivity $D=\max[D(\theta,\phi)]$.
Antennas always have some losses, for example due to resistive losses in metal structures, dielectric losses in dielectric substrates or impedance-matching losses between antenna and the connected source.
Due to these losses not all the power that is provided by the source will be radiated.
When we substitute the total radiated power $P_t$ in (\ref{eq:richtfunctie}) by the total input power $P_{in}$ to the antenna, we get the well-known {\it antenna gain function} $G(\theta,\phi)$, which is given by:
\begin{equation}
\displaystyle G(\theta, \phi) = \frac{P(\theta,
\phi)}{P_{in}/4\pi}. \label{eq:gainfunctie}
\end{equation}
Similar to the directivity function, the antenna gain function will have a maximum, which is called the {\it antenna gain $G$}, with $G=\max[G(\theta,\phi)]$.
The antenna efficiency $\eta$ is a quantity that descibes how much of the input power towards the antenna is transformed into radiated power:
\begin{equation}
\displaystyle \eta = \frac{P_{t}}{P_{in}}. \label{eq:efficiency}
\end{equation}
and provides a relation between the directivity and antenna gain:
\begin{equation}
\displaystyle G = \eta D.
\label{eq:gaindir}
\end{equation}

\section{Circuit representation of antennas}

\begin{figure}[hbt]
\vspace{-1.5cm}
\centerline{\psfig{figure=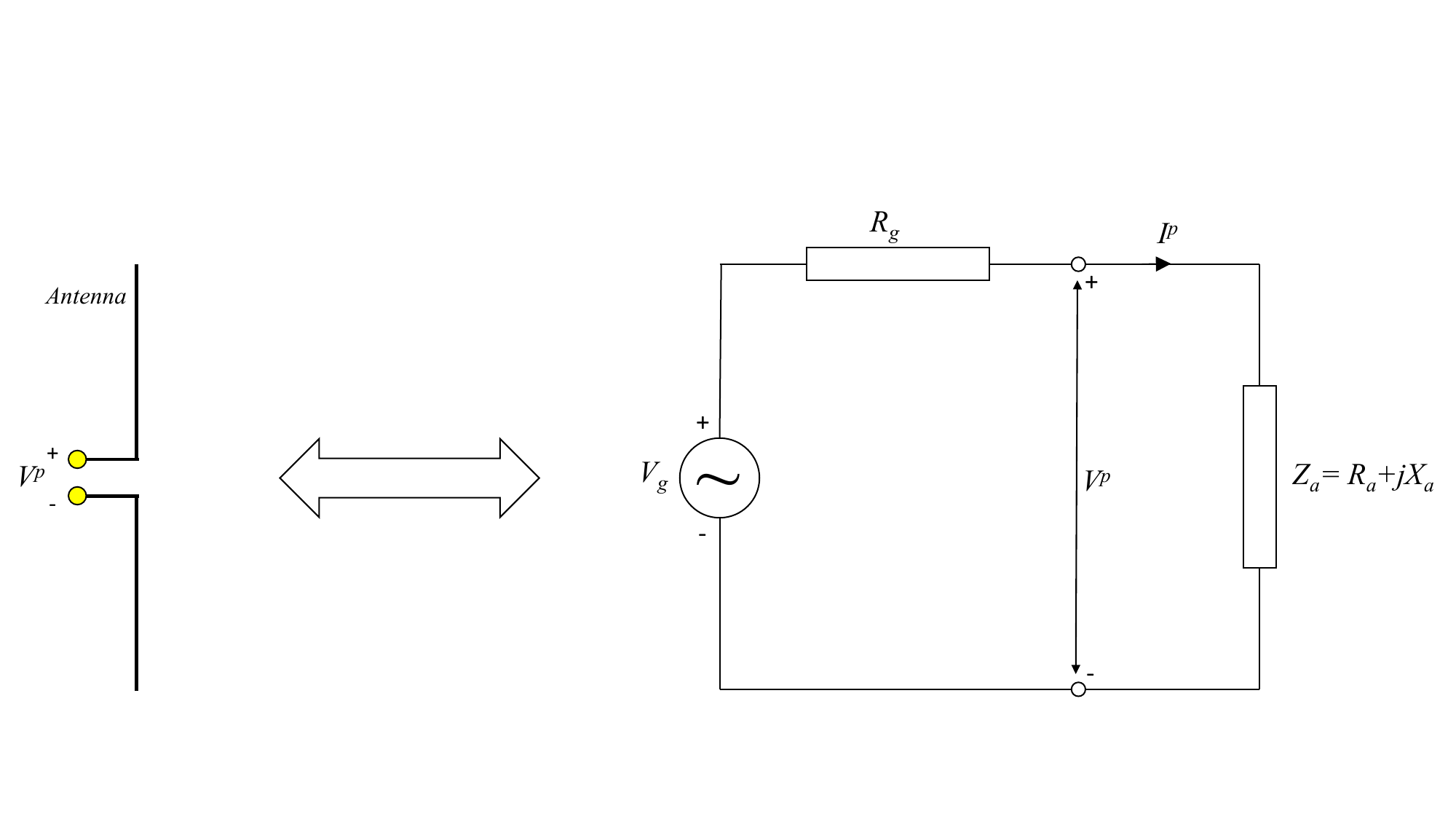,width=150mm}}
\vspace{-1cm}
\caption{\it Equivalent circuit model of an antenna. The antenna with impedance $Z_a$ is connected to a generator with internal impedance $R_g$. }
\label{fig:equivalentcircuit}
\end{figure}

Typically, antennas are connected to a generator (transmit mode) or detector (receive mode) by means of a transmission line.
More details on transmission line theory can be found in the next chapter.
Sometimes, antennas are directly connected to an amplifier, in which case power matching to a transmission line is not perse required.
The input impedance of the antenna is an important measure that determines how much of the power transported along the transmission line is actually radiated by the antenna.
The input impedance of an antenna will be complex in general:
\begin{equation}
\begin{array}{lcl}
\displaystyle Z_a & = & \di R_a + \jmath X_a \\
\displaystyle  & = & \di R_r+R_L + \jmath X_a,
\end{array}
\label{eq:Zindefinitie}
\end{equation}
where $R_a$ is the antenna resistance, $R_r$ represents the {\it radiated} losses and $R_L$ the real losses like Ohmic or dielectric losses. Furthermore, $X_a$ is the antenna reactance which is related to the stored reactive energy in the close vicinity of the antenna. Note that the {\it radiation resistance} $R_r$ is directly related to the total radiated power by the antenna and is in fact a measure that defines how well the antenna works.
Fig. \ref{fig:equivalentcircuit} shows the equivalent circuit of an antenna which is connected to a generator with an internal impedance $Z_g=R_g$.
We can define the radiation resistance $R_r$ by considering the antenna in transmit mode.
Let $I^p$ be the current flowing through the antenna and $P_t$ the total radiated power.
The radiation resistance $R_r$ is now defined as a resistance in which the power $P_t$ is dissipated when the current $I^p$ is flowing through the antenna.
In other words:
\begin{equation}
\displaystyle R_r = \frac{P_t}{\frac{1}{2}|I^p|^2}.
\label{eq:stralingsweerstand}
\end{equation}
The total radiated power $P_t$ can be found by integrating the radiated power density over a sphere.
Using relation (\ref{eq:pointingverreveld}) and
(\ref{eq:verreveldalgemeen}), we obtain the following expression for $P_t$:
\begin{equation}
\begin{array}{lcl}
\displaystyle P_t &=& \displaystyle \int\int \vec{S_p}(\vec{r})
\cdot \vec{u}_r r^2 d\Omega \\

\displaystyle &=& \displaystyle \frac{1}{2}Z_0^{-1}
\int\limits_{0}^{2\pi} \int\limits_{0}^{\pi}
[|E_{\theta}(\theta,\phi)|^2+|E_{\phi}(\theta,\phi)|^2] \sin\theta
d\theta d\phi ,

\end{array}
\label{eq:radiatedpower}
\end{equation}

At microwave frequencies it is, in general, not possible to measure voltages and currents directly. Therefore, we cannot directly measure the input impedance $Z_a$.
We will show in chapter 3 that it is possible to measure the (complex) amplitude of incident and reflected waves along transmission lines. In this way we can experimentally determine the input reflection coefficient of the antenna using a {\it vector network analyser} (VNA).
Now let us assume that the antenna is connected to a transmission line with a characteristic impedance equal to $Z_0^t$.
Usually $Z_0^t=50 \Omega$. Furthermore, we will assume that an incident TEM wave propagates along the transmission line with complex amplitude
$a$ and a corresponding reflected wave (in the opposite direction) with complexe amplitude $b$.
The reflection coefficient is now defined as the ratio between both complex amplitudes:
\begin{equation}
\displaystyle \Gamma = \frac{b}{a}. \label{eq:reflectiecoef}
\end{equation}
From transmission line theory (see chapter 3 for more details), it is well known that the relation between the reflection coefficient and the input impedance at the input of the antenna is given by:
\begin{equation}
\displaystyle \Gamma = \frac{Z_a-Z_0^t}{Z_a+Z_0^t}.
\label{eq:reflecversusZ}
\end{equation}
When the antenna is properly matched to a transmission line with $Z_a=Z_0^t$, we will obtain a reflection coefficient $\Gamma=0$.

\section{Effective antenna aperture}

Up to now, we have investigated the antenna parameters mainly by considering the antenna in transmit mode.
The receive properties of the antenna are, of course, also very important.
One of the key parameters that describes the antenna properties in receive mode is the {\it effective antenna aperture} $A_{e}$.
$A_{e}$ defines the equivalent surface in which the antenna absorbs the incident electromagnetic field that is dissipated by the
load impedance $Z_L$ which is connected to the terminals of the antenna.
Now let $S_p$ represent the power density $(W/m^2)$ of an incident plane wave.
The effective antenna aperture is now defined as the ratio between the power $P_r$ delivered to the load impedance and the power density $S_p$:
\begin{equation}
\displaystyle A_e = \frac{P_r}{S_p}. \label{eq:effectieveapertuur}
\end{equation}
Where we have assumed that the antenna is optimally matched to the load impedance, according to $Z_{a}=Z^*_L$.
In chapter \ref{chap:AntennaTheory}, we will introduce the reciprocity theorem. The reciprocity theorem defines the relation between the receive and transmit properties of an antenna.
We will show that there is an unique relation between the antenna gain $G$ and the effective antenna aperture $A_{e}$ according to:
\begin{equation}
\displaystyle G = \frac{4\pi A_e}{\lambda^2_0}.
\label{eq:gainversusAeff}
\end{equation}

\section{Polarization properties of antennas}
\label{sec:polarisatie}

The radiated electric field vector from an antenna will, in general, have two components, $E_{\theta}$ and $E_{\phi}$, both perpendicular to the direction of propagation $\vec{u}_r$ in the far-field region.
Both components (when described as phasors in the frequency domain) will usually have a relative phase difference with respect to each other.
With a phase difference equal to $0$, the resulting field will be {\it linearly polarized}. As a result, the direction of the electric field vector in the time domain will be constant.
When a phase difference between the $E_{\theta}$ and $E_{\phi}$ components exists, the direction of the realized total time-domain electric field $\vec{{\cal E}}(\vec{r},t)$ will vary over time with a period of $T=2\pi/\omega$.
The time-domain electric field vector will now describe an ellipse, corresponding to {\it elliptical polarization}.
Now let us take a closer look at the time-domain electric field components:
\begin{equation}
\begin{array}{lcl}
\displaystyle {\cal E}_{\theta}(\vec{r},t) & = & E_1 (\vec{r})
\cos{(\omega t -k_0 r)},
\\ \displaystyle {\cal E}_{\phi}(\vec{r},t) & = & E_2(\vec{r}) \cos{(\omega
t-k_0 r+\varphi)},
\\
\end{array}
\label{eq:poldeffield}
\end{equation}
where $\varphi$ is the phase difference between both electric-field components.
Let us consider the case that $r=0$.
We can find the elliptical trajectory of the electric field vector ${\cal E}=({\cal E}_{\theta},{\cal E}_{\phi})$ by eliminating the time-dependence.
We can rewrite (\ref{eq:poldeffield}) as:
\begin{equation}
\begin{array}{lcl}
\displaystyle {\cal E}_{\phi} & = & E_2( \cos{\omega t}
\cos{\varphi} - \sin{\omega t} \sin{\varphi}) \\

\displaystyle & = & \displaystyle E_2(\frac{{\cal
E}_{\theta}}{E_1} \cos{\varphi} - \sin{\omega t} \sin{\varphi}),
\\

\end{array}
\label{eq:Ephiexpand1}
\end{equation}
resulting in:
\begin{equation}
\displaystyle \left( \frac{{\cal E}_{\phi}}{E_2}- \frac{{\cal
E}_{\theta}}{E_1} \cos{\varphi} \right)^2  = \sin^2{\omega t}
\sin^2{\varphi} = \left( 1 - \cos^2{\omega t} \right)
\sin^2{\varphi} = \left( 1- \left[ \frac{{\cal E}_{\theta}}{E_1}
\right]^2 \right) \sin^2{\varphi}. \label{eq:Ephiexpand2}
\end{equation}
or:
\begin{equation}
\displaystyle \left( \frac{{\cal E}_{\theta}}{E_1} \right)^2+
\left(\frac{{\cal E}_{\phi}}{E_2} \right)^2  - \frac{2 {\cal
E}_{\theta} {\cal E}_{\phi}}{E_1 E_2} \cos{\varphi}=
\sin^2{\varphi}. \label{eq:Ephiexpand3}
\end{equation}
This equation describes an ellipse in the plane perpendicular to the direction of propagation $\vec{u}_r$. The ratio between the major and minor axis of the ellipse is defined as the {\it axial ratio} ($AR$), with $AR \geq 1$.
We can distinguish two special cases

\begin{enumerate}[i.]
 \item {\it Circular polarization} ($E_1=E_2=E$ and
 $\varphi= \pm \pi/2$).

 Equation (\ref{eq:Ephiexpand3}) now represents a circle. The field will be {\it circularly-polarized}. When $\varphi=\pi/2$, we obtain {\it Left-Hand Circular Polarization} (LHCP):
 \begin{equation}
 \displaystyle \vec{\cal E}_L = E \left( \vec{u}_{\theta} \cos{\omega t}
 - \vec{u}_{\phi} \sin{\omega t} \right),
 \label{eq:linkspol}
 \end{equation}
In case $\varphi=-\pi/2$, we obtain {\it Right-Hand Circular Polarization} (RHCP):
 \begin{equation}
 \displaystyle \vec{\cal E}_R = E \left( \vec{u}_{\theta} \cos{\omega t}
  + \vec{u}_{\phi} \sin{\omega t} \right).
 \label{eq:rechtspol}
 \end{equation}
  The axial ratio describes the quality of the circularly-polarized wave and is given by:
 \begin{equation}
 \displaystyle AR = \left| \frac{|E_L|+|E_R|}{|E_L|-|E_R|} \right|.
 \label{eq:AR}
 \end{equation}
 where $\vec{\cal E}_L=Re[\vec{E}_L e^{\jmath \omega t}]$.
 Perfect circular polarization is realized when $AR=1$.

 \item {\it Linear polarization} ($E_1 \neq E_2$ and $\varphi= \pm \pi$).
 Equation (\ref{eq:Ephiexpand3}) transforms into the equation of a straight line, with  corresponding axial ratio $AR=\infty$.

\end{enumerate}

\section{Link budget analysis: basic radio- and radar equation}

\begin{figure}[hbt]
\vspace{-1cm}
\centerline{\psfig{figure=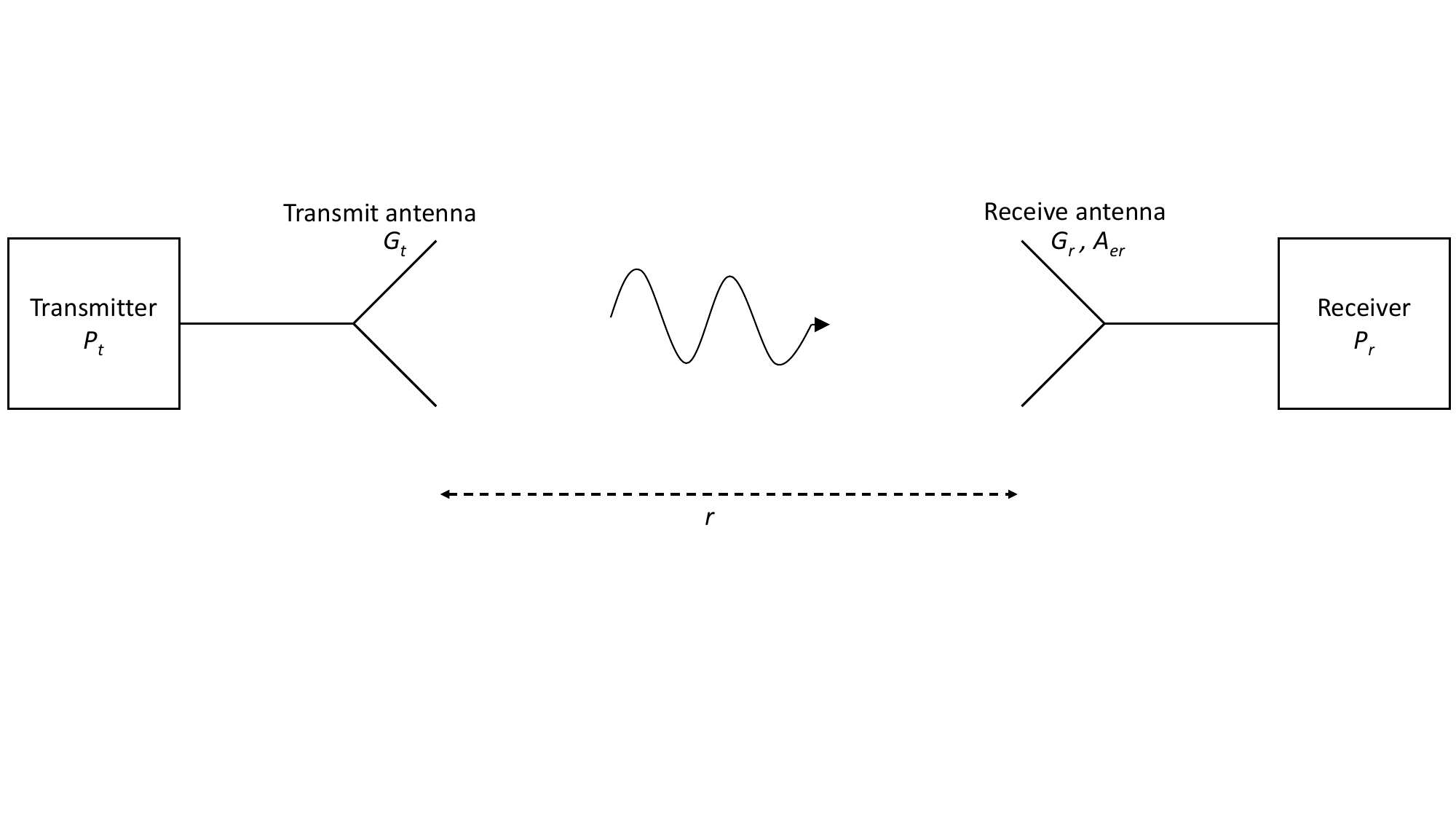,width=160mm}}
\vspace{-2.5cm}
\caption{\it Radio link between a transmit antenna and a receive antenna over a distance $r$.}
\label{fig:radiolink}
\end{figure}

The antenna is one of the components of a complete system. Examples of such systems include wireless communication systems (e.g. 4G/5G) and radar. The antenna parameters that we have introduced in the previous sections can be used to quantify the quality of a radio link or to determine the range of a radar.
Fig. \ref{fig:radiolink} shows a schematic illustration of a wireless link, consisting of a transmit and receive antenna. The transmit antenna is connected to a transmitter that can generate $P_t$ Watts of power.
We will assume that both antennas have the same polarization and orientation with respect to each other.
The power density in ($W/m^2$) at the receive antenna is now given by:
 \begin{equation}
 \displaystyle S_r  = \frac{P_t G_t}{4\pi r^2},
 \label{eq:Sontvangst}
 \end{equation}
where $G_t$ is the antenna gain of the transmit antenna as compared to an isotropic antenna.
The power received by the receive antenna with an effective aperture $A_{er}$ now becomes:
\begin{equation}
 \displaystyle P_r  = \frac{P_t G_t A_{er}}{4\pi r^2}.
 \label{eq:Pontvangst}
 \end{equation}
This equation is also known as the {\it radio equation} or {\it
Friis equation}. By using $\displaystyle A_{er}=\frac{\lambda_0^2
G_r}{4\pi}$ in (\ref{eq:Pontvangst}), we finally obtain:
\begin{equation}
 \displaystyle P_r  = \frac{P_t G_t G_r \lambda_0^2}{(4\pi)^2 r^2},
 \label{eq:Pontvangst2}
 \end{equation}
where $G_r$ is the antenna gain of the receiving antenna.
The term $\di \left(\frac{\lambda_0^2}{(4\pi)^2 r^2} \right)$ is also known as the {\it free-space path loss} and depends on frequency, since $\lambda_0=c/f$. However, this term is not really a loss, since we had assumed in our ideal radio link of Fig. \ref{fig:radiolink} that the entire system, including free-space, is lossless.  The frequency dependence comes from the relation between the antenna gain $G_r$ and the effective area $A_{er}$. It tells us that for a constant antenna gain, the effective area scales with $1/f^2$. This implies that at higher frequencies antennas with a much larger antenna gain need to be used in order to maintain the same range as compared to lower frequencies. Note that a large antenna gain also implies a more directive antenna with a narrow beam. This complicates the implementation of omni-directional wireless communication at millimeter-wave frequencies. Phased-arrays with electronic beam scanning offer a solution for this problem, see chapter \ref{chap:Phased-Arrays}.
\\

As an example of a link-budget analysis, consider the Bluetooth system operating in the $2.4-2.48$ GHz band. Assume omni-directional antennas ($G_t=G_r=1$), an output power of $P_t=1$ mW ($0$ dBm) and a receiver sensitivity $P_{r,min}=10^{-10}$ W ($=-70$ dBm). The maximum range $r_{max}$ of this system then becomes:
\begin{equation}
 \displaystyle r_{max}  = \sqrt{\frac{P_t G_t G_r \lambda_0^2}{(4\pi)^2 P_{r,min}}} \approx 30 \ [m].
 \label{eq:Rontvangst}
 \end{equation}

Let us now consider a radar system. In this case, we replace the receive antenna by an object with a radar cross section $\sigma [m^2]$. The radar cross section of an object depends on the size and shape (e.g. car or airplane) and determines how much of the incident power is reflected back to the radar.
We will assume that the transmit antenna can also be used as receive antenna, therefore $G_t=G_r=G$.
The latter is due to the reciprocity theorem which applies to passive microwave structures including antennas. The received power of the radar is now given by:
\begin{equation}
 \displaystyle P_r  = \frac{P_t G_t}{4\pi r^2} \frac{\sigma}{4\pi r^2} A_{er}
 = \frac{P_t G^2 \sigma \lambda_0^2}{(4\pi)^3 r^4}.
 \label{eq:Radareq1}
 \end{equation}
This equation is known as the {\it radar equation}. It is a first-order estimation of the received power and  shows that the probability of detection of an object decreases according to $r^4$.
In a real system, the radar equation will be somewhat more complicated. In addition, the maximum detection range can be increased by using advanced signal processing techniques. More background information can be found in Skolnik \cite{Skolnik}. Now assume that the receiver of the radar requires a minimum power of $P_{r,min}$ to detect an object. The maximum radar range $r_{max}$ now becomes:
\begin{equation}
 \displaystyle r_{max}  = \left[ \frac{P_t G^2 \sigma \lambda_0^2}{(4\pi)^3 P_{r,min}} \right]^{1/4}.
 \label{eq:Radareq2}
 \end{equation}
As an example, consider an X-band radar operating at $f=10$ GHz with $A_e=1 m^2$, $P_t=10$ kW, $\sigma=1 m^2$ and $P_{min}=10^{-13}$ W ($10\log_{10}P_{r,min}=-100$ dBm). This implies that $G=14000$
($G_{dB}=41$ dB). The maximum range now becomes $r_{max}=54$ km.

\chapter{Transmission line theory and microwave circuits}
\label{chap:Tlines}

\section{Introduction}

In this chapter we will extend the well-known circuit theory to transmission line theory. Circuit theory can be applied to electrical circuits in which the size of the individual (lumped) components, like resistors, capacitors and inductors, are much smaller than the electrical wavelength (size $\ll 0.1\lambda$). In case the physical dimension of a component or network becomes a significant fraction of the wavelength (size $> 0.1 \lambda$), we need to apply transmission line theory. We consider the transmission line to be a distributed network in which the voltages and currents vary along the location along the line. In this chapter, we will derive transmission line concepts by using circuit theory. Therefore, we do not need to solve Maxwell's equations explicitly to analyze networks based on transmission lines.

\section{Telegrapher equation}
\label{sec:Telegrapherequation}

\begin{figure}[hbt]
\centerline{\psfig{figure=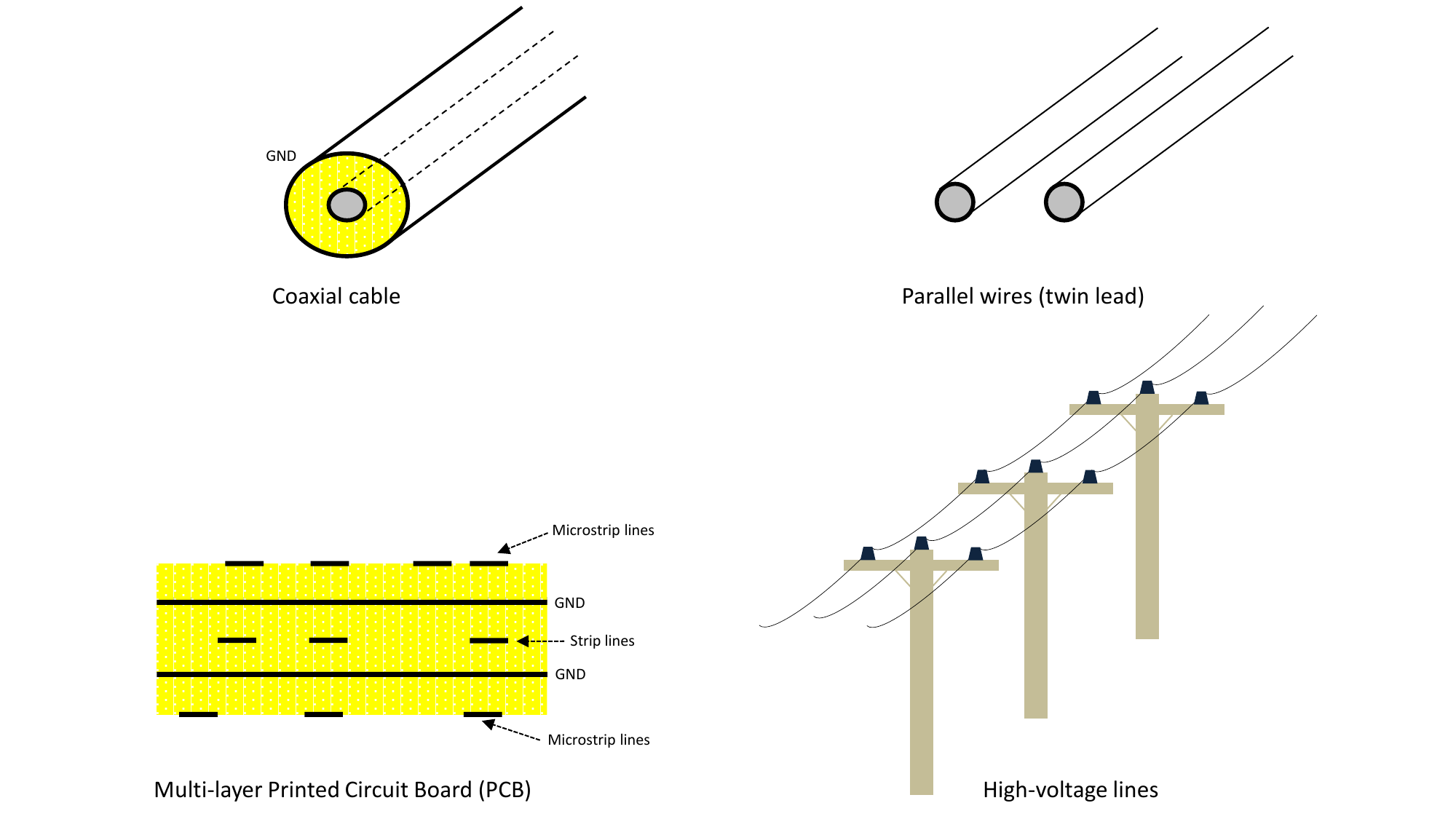,width=170mm}}
\caption{\it Examples of transmission lines used in various applications.}
\label{fig:Tlineexamples}
\end{figure}

Fig. \ref{fig:Tlineexamples} illustrates some examples of transmission lines. Well-known transmission lines used in many microwave applications are coaxial cables, microstrip lines and waveguides. In addition, long-range three-phase electrical power lines operating at 50 Hz should also be considered as transmission lines. From basic electromagnetics courses, it is known that a transverse electromagnetic wave (TEM) with zero cut-off frequency can propagate along transmission lines consisting of at least two metal conductors, e.g. a coaxial cable. These type of transmission lines also allow other modes, e.g. transverse electric (TE) or transverse magnetic (TM), to propagate above their cut-off frequencies. However, in this chapter we will assume that only a single TEM mode can propagate along the transmission line.
A plane wave is also an example of a TEM wave, see section \ref{chap:fundpar}.
Note that along a waveguide (see Fig. \ref{fig:Tlineexamples}) a TEM mode cannot exist, only TE and TM modes can propagate. However, the theory introduced in this chapter can also be applied to waveguides, as long as only a single mode can propagate.

\begin{figure}[hbt]
\vspace{-2cm}
\centerline{\psfig{figure=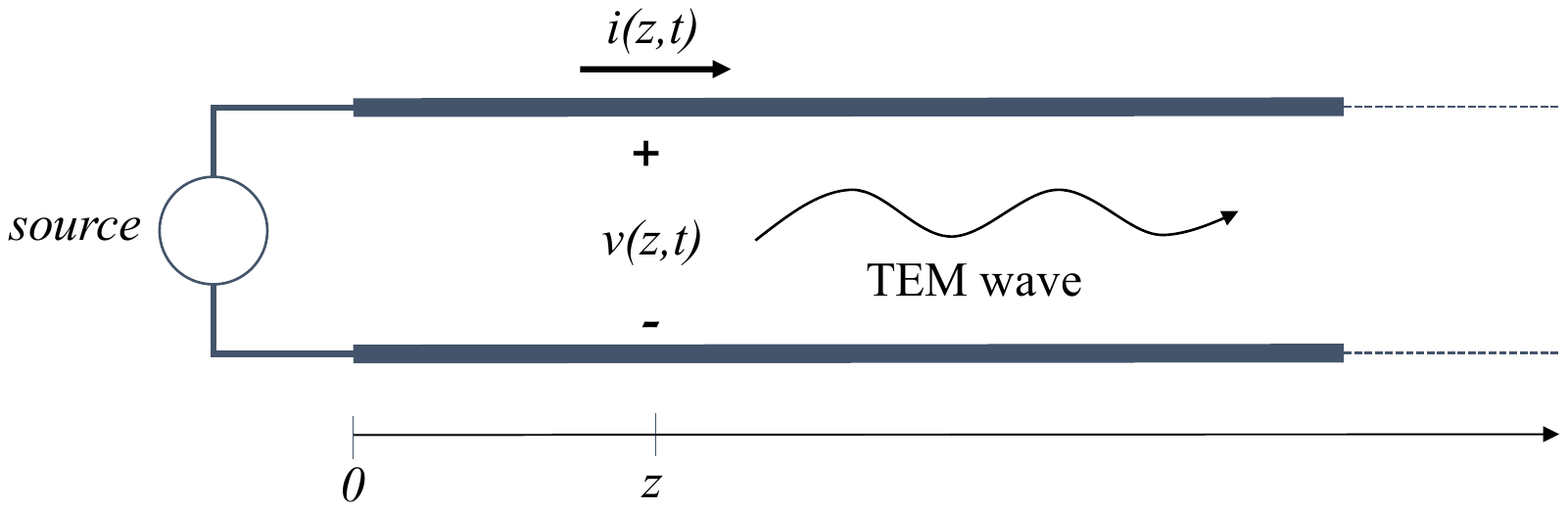,width=140mm}}
\vspace{-3cm}
\caption{\it Two-wire representation of a transmission line. A TEM wave propagates in the $+z$-direction.}
\label{fig:Tline1}
\end{figure}

Now let us consider a general transmission line that carries a TEM wave that propagates along the $+z$-direction, as illustrated in Fig. \ref{fig:Tline1}. The transmission line is connected to a source. As a result of this, a voltage $v(z,t)$ and current $i(z,t)$ will exist at the location $z$.

\begin{figure}[hbt]
\vspace{-1cm}
\centerline{\psfig{figure=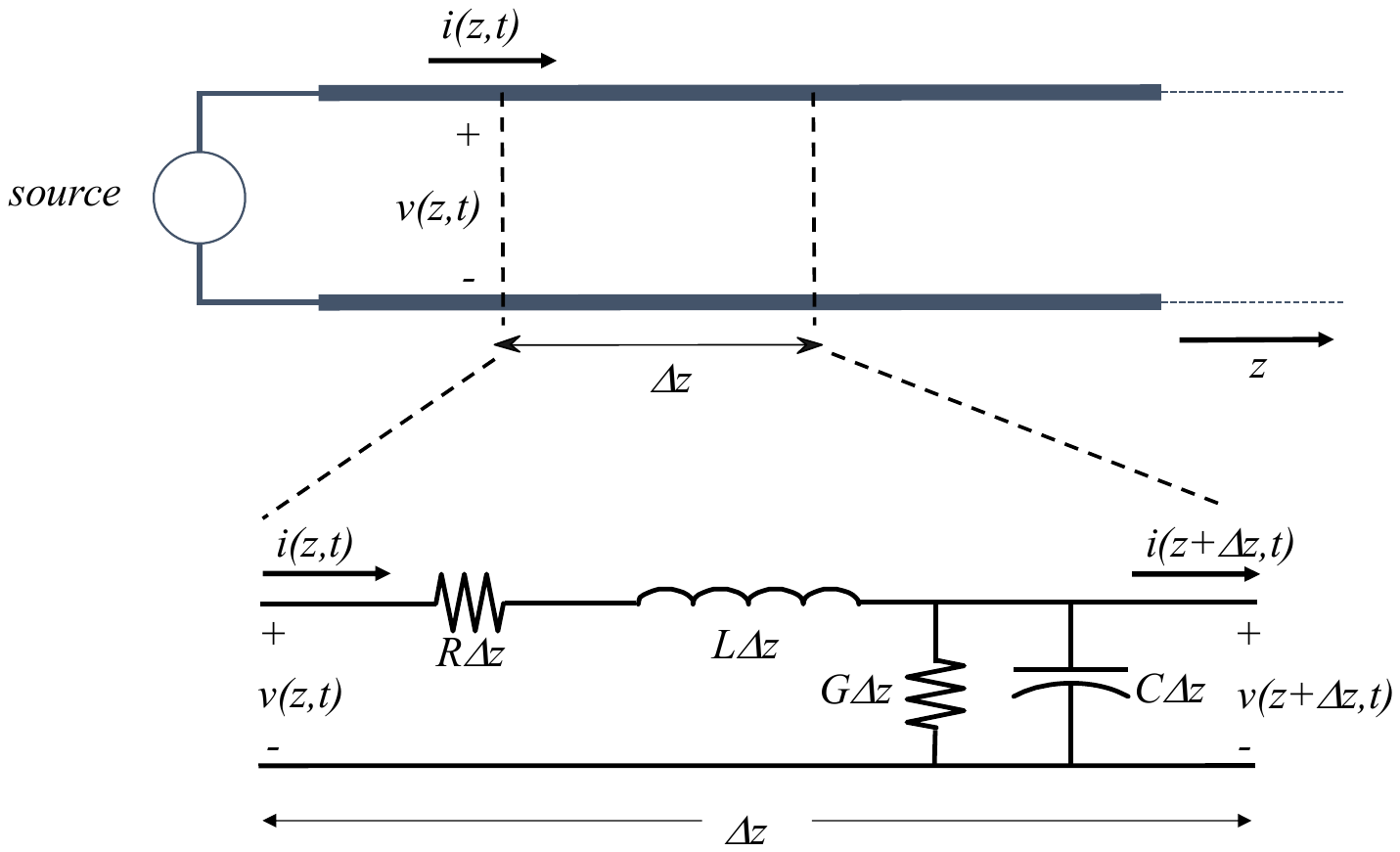,width=160mm}}
\vspace{-2cm}
\caption{\it Lumped element circuit representation of a small section $\Delta z$ of a transmission line.}
\label{fig:Tline2}
\end{figure}

We would like to apply Kirchhoff's voltage and current laws instead of Maxwells' equations to analyze the characteristics of the transmission line. This can be done by considering a small section between $z$ and $z+\Delta z$ of the transmission line, as illustrated in Fig. \ref{fig:Tline1}. Since this section is electrically very small ($\Delta z \ll \lambda$), we can model this small section using lumped-element circuit components as illustrated in Fig. \ref{fig:Tline2}, where the lumped elements are defined per unit length:
\begin{equation}
\begin{array}{lcl}
\displaystyle L &=& \mbox{ series inductance per unit length } [H/m], \\
\displaystyle C &=& \mbox{ shunt capacitance per unit length } [F/m], \\
\displaystyle R &=& \mbox{ series resistance per unit length } [\Omega/m], \\
\displaystyle G &=& \mbox{ shunt conductance per unit length } [S/m].
\end{array}
\label{eq:Lumpedelements}
\end{equation}

The series inductance $L\Delta z$ originates from the magnetic field that is induced by the current $i(z,t)$ flowing on the line and the series resistance $R\Delta z$ accounts for the related losses in the conductors. The shunt capacitor $C\Delta z$ is created by the voltage $v(z,t)$  which creates an electric field between the two wires. Dielectric losses related to dielectric material between the two wires are represented by the shunt conductance $G\Delta z$.
We can now apply Kirchhoff's voltage and current laws to the equivalent circuit of a small section $\Delta z$. Kirchhoff's voltage law provides:
\begin{equation}
\displaystyle v(z,t) - R \Delta z i(z,t) - L \Delta z \frac{\partial{i(z,t)}}{\partial{t}} -v(z+\Delta z,t) = 0.
\label{eq:KirchhoffV}
\end{equation}
Similarly, Kirchhoff's current law gives:
\begin{equation}
\displaystyle i(z,t) - G \Delta z v(z+ \Delta z,t) - C \Delta z \frac{\partial{v(z + \Delta z,t)}}{\partial{t}} -i(z+\Delta z,t) = 0.
\label{eq:KirchhoffI}
\end{equation}
Now divide (\ref{eq:KirchhoffV}) and (\ref{eq:KirchhoffI}) by $\Delta z$ and take the limit of $\Delta z \rightarrow 0$, resulting in  two differential equations with two unknowns ($v(z,t)$ and $i(z,t)$) which can be written in the following form:
\begin{equation}
\begin{array}{lcl}
\displaystyle \frac{\partial{v(z,t)}}{\partial{z}} &=& \di - Ri(z,t)-L \frac{\partial{i(z,t)}}{\partial{t}}, \\
\displaystyle \frac{\partial{i(z,t)}}{\partial{z}} &=& \di - Gv(z,t)-C \frac{\partial{v(z,t)}}{\partial{t}}.
\end{array}
\label{eq:TelegraphTime}
\end{equation}
Equation (\ref{eq:TelegraphTime}) is known as the {\it Telegrapher equation} in the time domain.
Since we assume a sinusoidal time variation of all field components in this book, the voltage and current can be written as:
\begin{equation}
\begin{array}{lcl}
\displaystyle v(z,t)  &=&  \di Re[ V(z) e^{\jmath \omega t}],\\
\displaystyle i(z,t)  &=&  \di Re[ I(z) e^{\jmath \omega t}],
\end{array}
 \label{eq:VItimeharmonic}
\end{equation}
in which $\omega=2\pi f$ and $f$ is the frequency of operation.
In this case the time derivative of $v(z,t)$ in (\ref{eq:TelegraphTime}) transforms into a simple multiplication in the frequency domain:
\begin{equation}
\displaystyle \frac{\partial{v(z,t)}}{\partial{t}}  \leftrightarrow  \jmath \omega V(z).\\
\end{equation}
By applying this in (\ref{eq:TelegraphTime}) we obtain the frequency-domain Telegrapher equation:
\begin{equation}
\begin{array}{lcl}
\displaystyle \frac{\partial{V(z)}}{\partial{z}} &=& \di - \left( R + \jmath \omega L \right) I(z), \\
\displaystyle \frac{\partial{I(z)}}{\partial{z}} &=& \di - \left( G + \jmath \omega C \right) V(z).
\end{array}
\label{eq:TelegraphFreq}
\end{equation}
The general solution of the frequency-domain Telegrapher equation is now a combination of a voltage and current wave propagating in the $+z$ direction and a wave propagating in the $-z$ direction:
\begin{equation}
\begin{array}{lcl}
\displaystyle V(z)&=& \di V^+_0 e^{-\gamma z} + V^-_0 e^{\gamma z}, \\
\displaystyle I(z)&=& \di I^+_0 e^{-\gamma z} + I^-_0 e^{\gamma z},
\end{array}
\label{eq:SolutionTelegraph1}
\end{equation}
where $\gamma$ is the {\it complex propagation constant}:
\begin{equation}
\displaystyle \gamma =  \alpha + \jmath \beta = \sqrt{(R+\jmath \omega L)(G + \jmath \omega C)},
 \label{eq:gammacomplex}
\end{equation}
in which $\alpha$ is the {\it attenuation} of the line and $\beta$ is the {\it phase constant}. Note that the wavenumber $k_0$ used in chapter \ref{chap:fundpar} and chapter \ref{chap:AntennaTheory} is strongly related to the phase constant $\beta$. However, we will use different symbols, since it is more common to use $k_0$ in the antenna community and $\beta$ in the microwave engineering community.
The term $e^{-\gamma z}$ corresponds to a wave propagating in the $+z$ direction and $e^{\gamma z}$ to a wave propagating in the $-z$ direction. Note that $V(z)$ and $I(z)$ in (\ref{eq:SolutionTelegraph1}) represent the {\it total} voltage and current, respectively, at the position $z$ along the line.
It can be easily verified that (\ref{eq:SolutionTelegraph1}) describes the general solution by substituting (\ref{eq:SolutionTelegraph1}) into the Telegrapher equation (\ref{eq:TelegraphFreq}).
By doing this, we can also obtain a relation between the complex voltage and current amplitudes:
\begin{equation}
\begin{array}{lcl}
\displaystyle I^+_0 &=& \di \frac{\gamma}{R+ \jmath \omega L} V^+_0, \\
\displaystyle I^-_0 &=& \di \frac{-\gamma}{R+ \jmath \omega L} V^-_0.
\end{array}
\label{eq:VIamplitudes}
\end{equation}
In addition, the {\it characteristic impedance} $Z_0$ of the transmission line is defined by:
\begin{equation}
\displaystyle Z_0 = \frac {V^+_0}{I^+_0} = \frac{R+\jmath \omega L}{\gamma}.
 \label{eq:Z0}
\end{equation}

The time-domain representation of the voltage along the transmission line can be obtained by using the frequency domain solution (\ref{eq:SolutionTelegraph1}) and relation (\ref{eq:VItimeharmonic}). So for a wave travelling along a transmission line in the positive $z$-direction we obtain the following time-domain representation:
\begin{equation}
\begin{array}{lcl}
\displaystyle v(z,t)&=& \di |V^+_0| cos\left( \omega t -\beta z + \phi^+ \right) e^{-\alpha z},
\end{array}
\label{eq:SolutionTelegraph1time}
\end{equation}
where it can be observed that $\alpha$ describes the exponential attenuation along the line and where the phase constant $\beta$ determines the {\it wavelength} on the line:
\begin{equation}
\displaystyle \lambda = \frac {2 \pi}{\beta}.
 \label{eq:Lambda}
\end{equation}

\vspace{0.5cm}

{ \underline{\it Question}}: Why is there a $-$ sign in the expression of $I^-_0$ in (\ref{eq:VIamplitudes})?

\section{The lossless transmission line}
\label{sec:losslessTline}

In the previous section we have derived the solution for the voltage and current along a transmission line with losses, resulting in a complex propagation constant $\gamma=\alpha + \jmath \beta$ and complex characteristic impedance $Z_0$. In a lot of practical applications, the losses along the transmission line are very small and can, therefore, be neglected for a first-order analysis. When the transmission line is lossless, the series resistance $R$ and the shunt conductance $G$ in Fig. \ref{fig:Tline2} can be neglected. Therefore, the propagation constant and characteristic impedance now become real:
\begin{equation}
\begin{array}{lcl}
\displaystyle \gamma & = & \di \jmath \beta = \jmath \omega \sqrt{LC}, \\
\di \beta & = & \di \omega \sqrt{LC}, \\
\di \alpha & = & 0, \\
\di Z_0 & = & \di \sqrt{\frac{L}{C}}.
\end{array}
 \label{eq:gammaZ0real}
\end{equation}
The corresponding wavelength and associated phase velocity of the wave along the line are given by:
\begin{equation}
\begin{array}{lcl}
\displaystyle \lambda & = & \di \frac{2\pi}{\beta} = \frac{2 \pi}{\omega \sqrt{LC}} \\
\di v_p & = & \di \frac{\omega}{\beta} = \frac{1}{\sqrt{LC}}.
\end{array}
 \label{eq:LambdaReal}
\end{equation}

{ \underline{\it Example}}

A transmission line operating at $f=500$ MHz has the following per unit length parameters:
\begin{equation}
\begin{array}{lcl}
\displaystyle L &=&  0.2 \mbox{ [$\mu$H/m]}, \\
\displaystyle C &=&  300 \mbox{ [pF/m]}, \\
\displaystyle R &=&  5 \mbox{ [$\Omega$/m]}, \\
\displaystyle G &=&  0.01 \mbox{ [S/m]}.
\end{array}
\end{equation}
Using (\ref{eq:gammacomplex}), we find that the complex propagation coefficient is calculated by:
\begin{equation}
\displaystyle \gamma =  \alpha + \jmath \beta = \sqrt{(R+\jmath \omega L)(G + \jmath \omega C)}=\sqrt{-590 +\jmath 10.99}=0.23+\jmath 24.3 \mbox{ [1/m]},
\end{equation}
in which the square root was evaluated according to Im$(\gamma)>0$.
Re-calculating this for a lossless line with $R=G=0$ gives:
\begin{equation}
\displaystyle \gamma =  \jmath \beta = \jmath \omega \sqrt{LC} = \jmath 24.3 \mbox{ [1/m]}.
\end{equation}

\section{The terminated lossless transmission line}
\label{sec:TerminatedLine}

\begin{figure}[hbt]
\vspace{-1cm}
\centerline{\psfig{figure=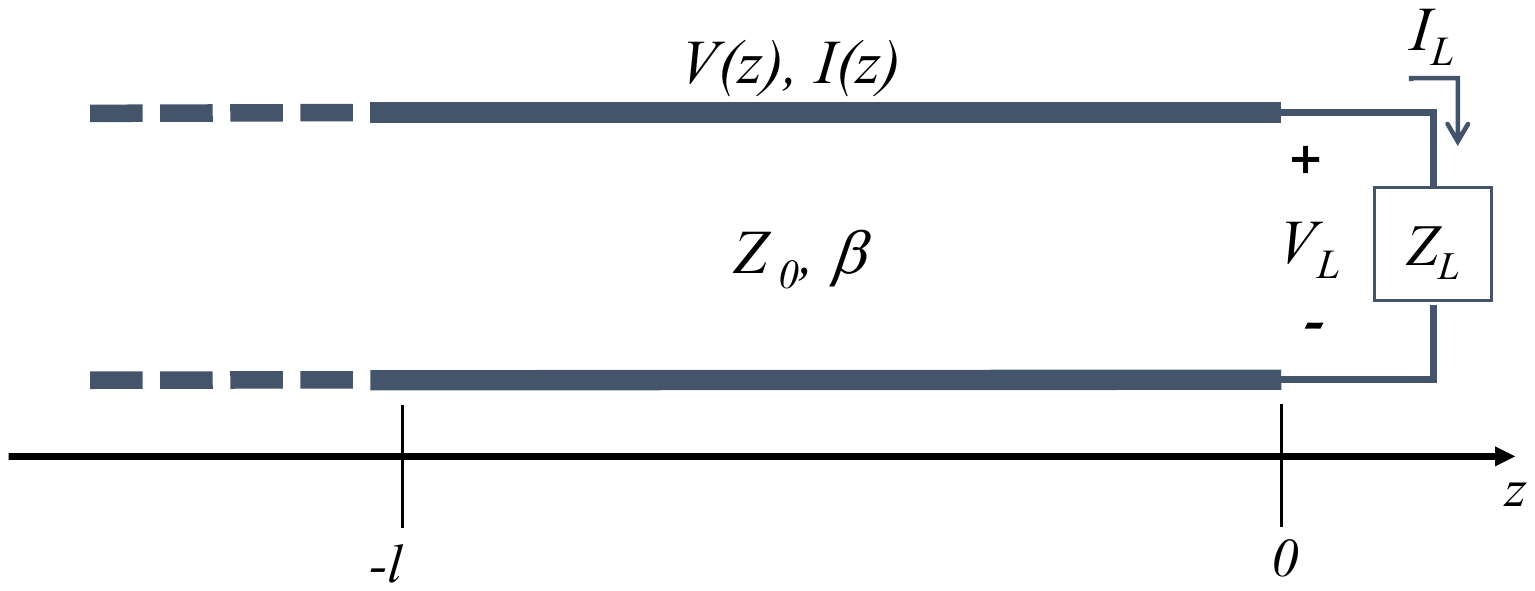,width=120mm}}
\vspace{-3cm}
\caption{\it The terminated lossless transmission line, terminated with a complex load impedance $Z_L$}
\label{fig:Tline3}
\end{figure}

Consider the terminated lossless transmission of Fig. \ref{fig:Tline3}. The line is terminated at $z=0$ with a complex load impedance $Z_L=R_L+\jmath X_L$. The total voltage and current at the position $z$ along the line are found by (\ref{eq:SolutionTelegraph1}) and (\ref{eq:VIamplitudes}) and using $\gamma=\jmath \beta$:
\begin{equation}
\begin{array}{lcl}
\displaystyle V(z)&=& \di V^+_0 e^{-\jmath \beta z} + V^-_0 e^{\jmath \beta z}, \\
\displaystyle I(z)&=& \di \frac{V^+_0}{Z_0} e^{-\jmath \beta z} - \frac{V^-_0}{Z_0} e^{\jmath \beta z}.
\end{array}
\label{eq:VITerminatedTline1}
\end{equation}
This equation still has two unknown coefficients $V^+_0$ and $V^-_0$. A relation between both coefficients can be found by using the relation between the voltage and current along the load impedance.
At $z=0$ this relation is given by:
\begin{equation}
\begin{array}{lcl}
\displaystyle Z_L &=& \di \frac{V_L}{I_L} = \frac{V(0)}{I(0)} =\frac{V^+_0 + V^-_0}{V^+_0 - V^-_0}Z_0.
\end{array}
\label{eq:ZL}
\end{equation}
Rewriting (\ref{eq:ZL}) results in:
\begin{equation}
\begin{array}{lcl}
\displaystyle V^-_0 &=& \di =\frac{Z_L-Z_0}{Z_L+Z_0}V^+_0.
\end{array}
\end{equation}
We can now introduce the {\it reflection coefficient} $\Gamma$ as:
\begin{equation}
\begin{array}{lcl}
\displaystyle \Gamma &=& \di =\frac{V^-_0}{V^+_0}=\frac{Z_L-Z_0}{Z_L+Z_0}.
\end{array}
\label{eq:Reflectioncoefficient}
\end{equation}
Some special terminations are:
\begin{equation}
\begin{array}{lcl}
\displaystyle Z_L &=& Z_0, \ \Gamma =0, \mbox{ matched load }, \\
\displaystyle Z_L &=& 0, \ \Gamma = -1, \mbox{ short circuit termination}, \\
\displaystyle Z_L &=& \infty , \ \Gamma = 1, \mbox{ open circuit termination}.
\end{array}
\end{equation}
In case of a matched load $Z_L=Z_0$, the voltage along the line is constant $|V(z)|=|V^+_0|$. This line is now also called to be "flat". In other cases when $\Gamma= |\Gamma|e^{\jmath \theta} \neq 0$, the magnitude of the voltage will not be constant over the line:
\begin{equation}
\begin{array}{lcl}
\displaystyle |V(z)| &=& \di =\left| V^+_0 \right|\left|1+\Gamma e^{2\jmath \beta z} \right|= \left| V^+_0 \right|\left|1+ \left| \Gamma \right| e^{\jmath (\theta-2 \beta l)} \right|, \mbox{ at } z=-l.
\end{array}
\end{equation}
The voltage along the line fluctuates according to:
\begin{equation}
\begin{array}{lcl}
\displaystyle V_{max} &=& \di \left| V^+_0 \right| \left|1+ \left| \Gamma \right|  \right|, \\
\displaystyle V_{min} &=& \di \left| V^+_0 \right| \left|1- \left| \Gamma \right|  \right|, \\
\displaystyle \mbox{VSWR} &=& \di \frac{V_{max}}{V_{min}} = \frac{1+ \left| \Gamma \right|}{1- \left| \Gamma \right|},
\end{array}
\label{eq:VSWR}
\end{equation}
where the VWSR is the voltage standing wave ratio which is a quantity that describes the fluctuation of the voltage along the line. The VSWR is an important quantity that is often used to specify the load conditions of active devices, like power amplifiers (PAs). PAs use semiconductor transistors which have a maximum break-down voltage. By specifying a maximum VWSR, we can avoid the maximum voltage along the line to exceed the break-down voltage of the transistor.
The reflection coefficient as defined in (\ref{eq:Reflectioncoefficient}) can be generalized to any point $z=-l$ on the transmission line according to:
\begin{equation}
\begin{array}{lcl}
\displaystyle \Gamma(z=-l) &=& \di \frac{V^-_0 e^{-\jmath \beta l}}{V^+_0 e^{\jmath \beta l}} = \Gamma e^{-2\jmath \beta l},
\end{array}
\label{eq:generalizedreflection}
\end{equation}
where $\Gamma=\Gamma (z=0)$.

\begin{figure}[hbt]
\vspace{-1cm}
\centerline{\psfig{figure=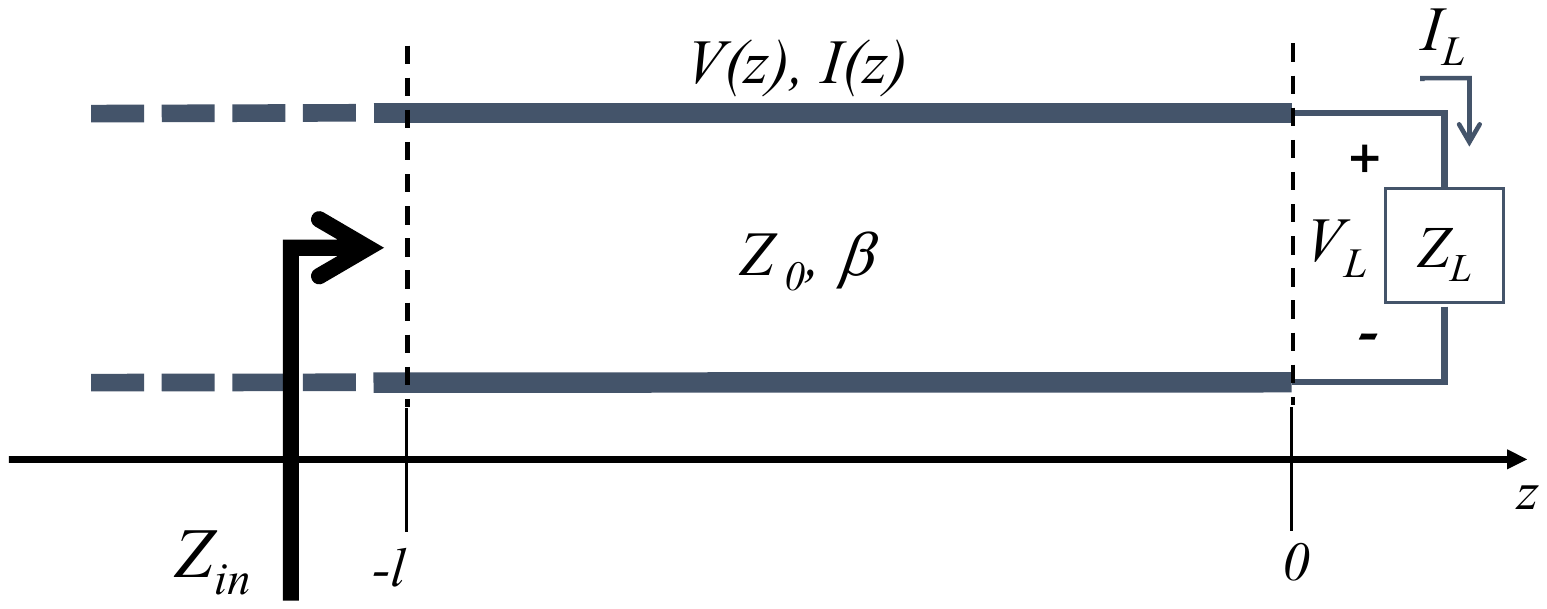,width=120mm}}
\vspace{-3cm}
\caption{\it The terminated lossless transmission line with characteristic impedance $Z_0$ and length $l$. We will assume that at the terminal $z=-l$ the terminated line is connected to another very long transmission line with characteristic impedance $Z_0$.}
\label{fig:Tline4}
\end{figure}

Now let us consider the configuration of Fig. \ref{fig:Tline4} and investigate the input impedance of the terminated line with length $l$. The input impedance $Z_{in}$ looking into the terminated line from the location $z=-l$ is now given by:
\begin{equation}
\begin{array}{lcl}
\displaystyle Z_{in} &=& \di \frac{V(z=-l)}{I(z=-l)}= \frac{V^+_0 \left[e^{\jmath \beta l}+\Gamma e^{-\jmath \beta l} \right]}{V^+_0 \left[e^{\jmath \beta l}-\Gamma e^{-\jmath \beta l} \right]} Z_0 \\
&=& \di Z_0 \frac{Z_L+\jmath Z_0 \tan \beta l}{Z_0+\jmath Z_L \tan \beta l}.
\end{array}
\label{eq:inputimpedance}
\end{equation}
Equation (\ref{eq:inputimpedance}) shows that the input impedance of a terminated line can be {\it tuned} to a specific value by changing the length $l$ of the transmission line. This can be very useful in matching circuits and filters, as we will see lateron.

\vspace{0.5cm}
{\underline{\it Example}}

Consider a lossless  transmission line of length $l$ which is short-circuited at the termination ($Z_L=0$). As expected in case of a short-circuit, we find from (\ref{eq:Reflectioncoefficient}) that the reflection at the termination $\Gamma=-1$.
The voltage and current along the line is determined from (\ref{eq:VITerminatedTline1}) and takes the form:
\begin{equation}
\begin{array}{lcl}
\displaystyle V(z) &=& \di -2 \jmath V^+_0 \sin \beta z ,\\
\di I(z) &=& \di  \frac{2V^+_0}{Z_0} \cos \beta z, \\
\displaystyle Z_{in}(z=-l) &=& \di \jmath Z_0  \tan \beta l.
\end{array}
\end{equation}
The corresponding $VSWR=\infty$ since the minimum voltage on the line $V_{min}=0$. When a power amplifier is connected to a short circuited transmission line, the output voltage at the output port of the amplifier may become much larger than the breakdown voltage. In this case, the amplifier will be destroyed.
\vspace{0.3cm}

When the terminated transmission line is not well matched to the connecting transmision line, that is $Z_{in} \neq Z_0$, part of the incident power will be reflected. Now let us assume that the load impedance is purely resistive $Z_L=R_L$. The incident power $P_{in}$ and reflected power $P_{r}$ can be related by using the magnitude of the voltage reflection coefficient:
\begin{equation}
\begin{array}{lcl}
\displaystyle \frac{P_{r}}{P_{in}} &=& \di \frac{\di \frac{\left|V_{r} \right|^2}{2 R_L}} {\di \frac{\left|V_{in} \right|^2}{2 R_L}} = |\Gamma|^2.
\end{array}
\label{eq:reflectedpower}
\end{equation}
Note that in (\ref{eq:reflectedpower}) we have used the time-average power which can be calculated from:
\begin{equation}
\begin{array}{lcl}
\displaystyle P_{av}(z) &=& \displaystyle \frac{1}{T} \int\limits_{0}^{T} {\cal P}(z,t) dt  \\
&=& \displaystyle \frac{1}{T} \int\limits_{0}^{T} {v}(z,t){i}(z,t) dt \\
&=& \di \frac{1}{2} Re \left[ V(z)I^*(z) \right],
\end{array}
\label{eq:Paverage}
\end{equation}
in which a similar derivation as used in (\ref{eq:vectorpointing}) was applied. Note that the time domain voltage and current can be expressed as:
\begin{equation}
\begin{array}{lcl}
\displaystyle {v}(z,t) &=&  \di Re \left[ V(z) e^{\jmath \omega t} \right] = Re \left[ |V(z)| e^{\jmath \omega t+\jmath \phi_v} \right] \\
\displaystyle {i}(z,t) &=&  \di Re \left[ I(z) e^{\jmath \omega t} \right] = Re \left[ |I(z)| e^{\jmath \omega t+\jmath \phi_i} \right].
\end{array}
\label{eq:VItimedomain}
\end{equation}

\vspace*{1cm} \hrulefill
 {\bf Exercise} \hrulefill \\
Derive the expression for time-average power, given by (\ref{eq:Paverage}), by using relation (\ref{eq:VItimedomain}).
\vspace{0.3cm}

\begin{figure}[hbt]
\vspace{-1cm}
\centerline{\psfig{figure=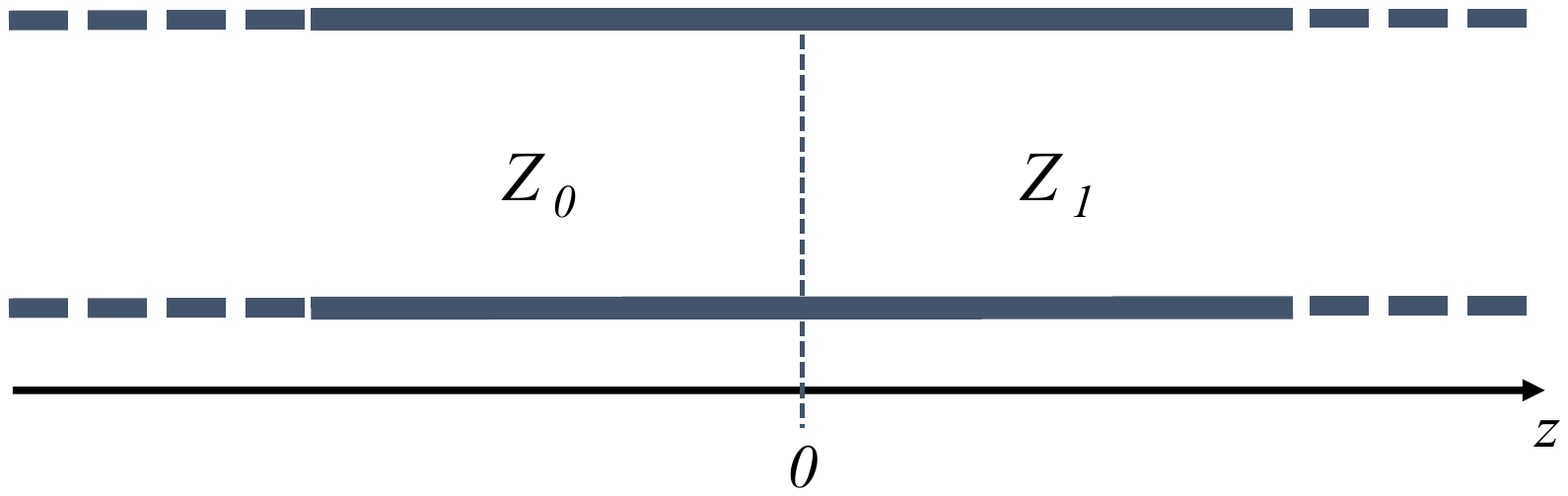,width=120mm}}
\vspace{-3cm}
\caption{\it Interface between two transmission lines causing insertion loss.}
\label{fig:Tline5}
\end{figure}

The {\it return loss} $RL$ is now defined as the amount of power reflected by the load:
 \begin{equation}
\begin{array}{lcl}
\displaystyle RL &=& \di -10 \log_{10} |\Gamma|^2 = \di -20 \log_{10} |\Gamma|.
\end{array}
\label{eq:returnloss}
\end{equation}
Related to return loss is the {\it insertion loss}. The insertion loss describes the loss of a transition between two transmission lines, as illustrated in Fig. \ref{fig:Tline5}. According to (\ref{eq:VITerminatedTline1}) the transmitted wave for $z>0$ takes the form:
\begin{equation}
\displaystyle V_1(z)= V^+_0 T e^{-\jmath \beta z},
\label{eq:Vtransmittedwave}
\end{equation}
where $T$ is the transmission coefficient:
\begin{equation}
\displaystyle T = 1+\Gamma=1+\frac{\di Z_1-Z_0}{\di Z_1+Z_0}= \frac{\di 2Z_1}{\di Z_1+Z_0}.
\label{eq:Tcoefficient}
\end{equation}
Let $P_{in}$ be the power of the incident wave and $P_1$ the power of the transmitted wave. The insertion loss $IL$ is now found as:
\begin{equation}
\begin{array}{lcl}
\displaystyle IL &=& \di -10 \log_{10} \left(\frac{\di P_1}{\di P_{in}} \right)=-10 \log_{10} \left(\frac{\frac{\di |V_1|^2}{\di2Z_1}}{\frac{\di |V^+_0|^2}{\di 2Z_0}} \right) \\
& =& -20 \log_{10}|T|+10\log_{10} \left(\di \frac{Z_1}{Z_0} \right).
\end{array}
\label{eq:insertionloss}
\end{equation}

\section{The quarter-wave transformer}
\label{sec:quarterwavetransformer}
\begin{figure}[hbt]
\vspace{-1cm}
\centerline{\psfig{figure=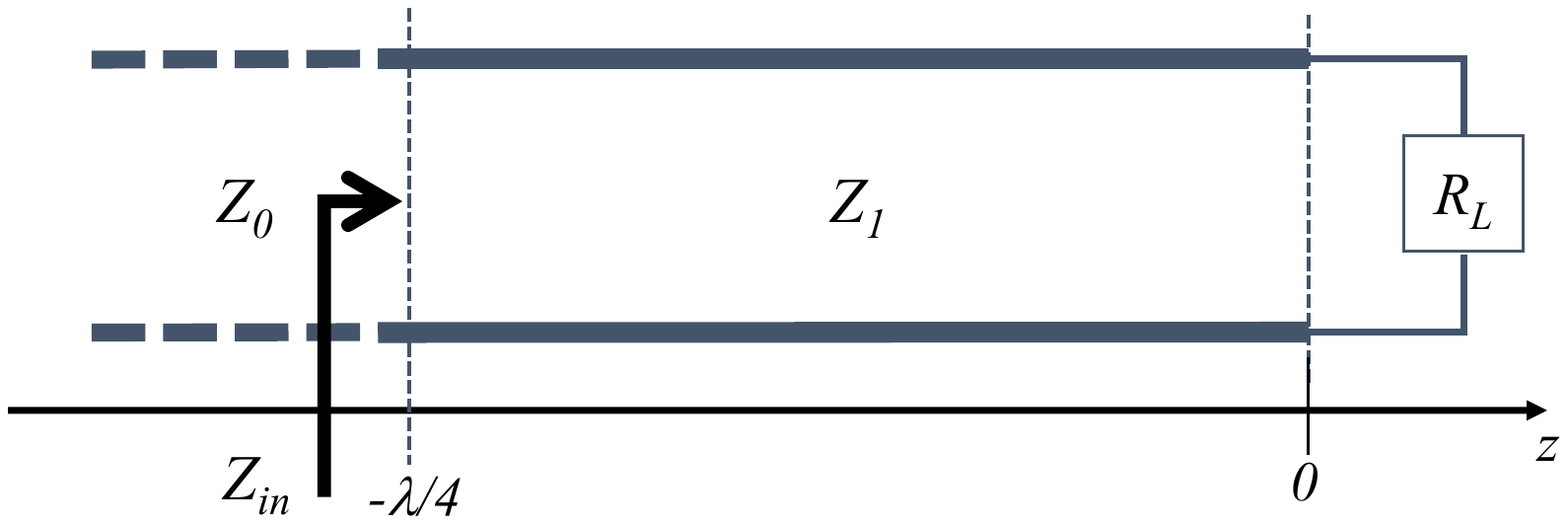,width=120mm}}
\vspace{-3cm}
\caption{\it The $\lambda/4$ transformer in which a $\lambda/4$-transmission line is connected to a terminated lossless transmission line, terminated with a load impedance $R_L$}
\label{fig:Lambda4transformer}
\end{figure}

A very nice application of the terminated transmission line is the so-called {\it $\lambda/4$ transformer} as illustrated in Fig. \ref{fig:Lambda4transformer}. A transmission line section of length $l=\lambda/4$ with a characteristic impedance $Z_1$ is terminated by a load impedance $R_L$. In this section we will only consider a resistive load. The transformer is connected to another transmission line with characteristic impedance $Z_0$ of which we will assume that the length is infinitely long or connected to a matched source with impedance $Z_s=Z_0$ (no reflections from the source).
The impedance looking into the transformer at $z=-\lambda/4$ can be calculated using (\ref{eq:inputimpedance}), where it should be noted that the quarterwave line has a characteristic impedance $Z_1$. Using $l=\lambda_0/4$ and $Z_L=R_L$ we get:
\begin{equation}
\begin{array}{lcl}
\displaystyle Z_{in} &=& \di Z_1 \frac{R_L+\jmath Z_1 \tan \beta l}{Z_1+\jmath R_L \tan \beta l}\\
&=& \di Z_1 \frac{R_L+\jmath Z_1 \tan \beta \lambda/4}{Z_1+\jmath R_L \tan \beta \lambda/4}\\
\displaystyle &=& \di \frac{Z^2_1}{R_L},
\end{array}
\label{eq:Lambda4transformerZin}
\end{equation}
where $\beta l=\left( \frac{2 \pi}{\lambda} \right) \left( \frac{\lambda}{4} \right)=\frac{\pi}{2}$.
From (\ref{eq:Lambda4transformerZin}) we find that the reflection coefficient at $z=-\lambda/4$ is zero when the following condition is satisfied:
\begin{equation}
\begin{array}{lcl}
\displaystyle \Gamma(z=-\lambda/4)=\frac{Z_{in}-Z_0}{Z_{in}+Z_0}=0  \longrightarrow Z_{in}=Z_0 \longrightarrow \di Z_1 = \sqrt{Z_0 R_L}.
\end{array}
\label{eq:Gammazerocondition}
\end{equation}

{\underline{\it Example}}

\begin{figure}[hbt]
\centerline{\psfig{figure=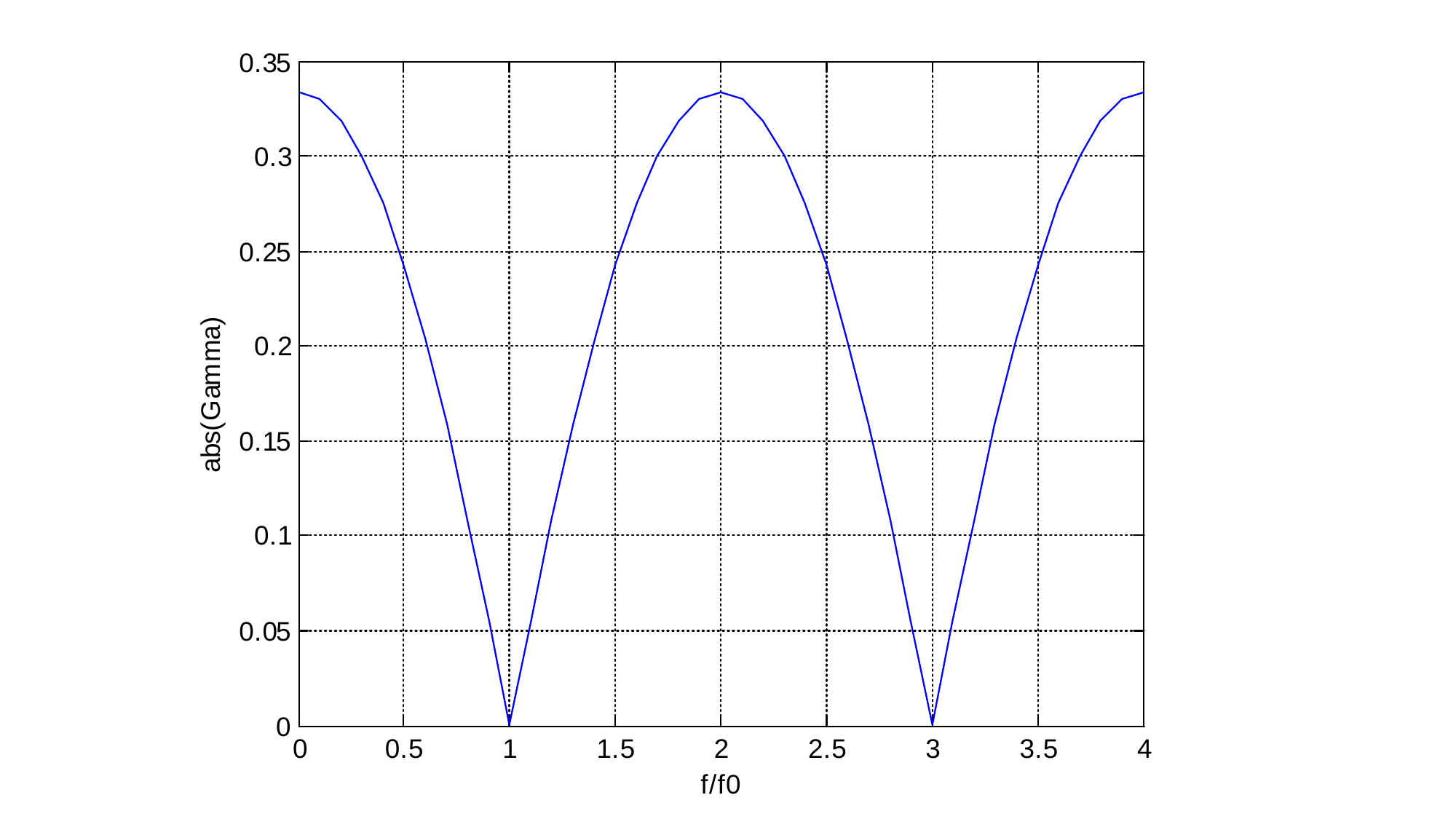,width=130mm}}
\caption{\it Magnitude of the input reflection coefficient versus the normalized frequency $f/f_0$ of a $\lambda/4$ transformer terminated with a load impedance $R_L=100 \ \Omega$}
\label{fig:GammaLambda4}
\end{figure}

Consider a load resistance $R_L=100 \ \Omega$ which needs to be matched to a long transmission line with characteristic impedance $Z_0=50 \ \Omega$. Design a quarter-wave transformer at the frequency $f_0$ and plot the magnitude of the input reflection coefficient versus the normalized frequency $f/f_0$.
From (\ref{eq:Gammazerocondition}) we find that $Z_1=\sqrt{Z_0 R_L}=70.71 \ \Omega$. The magnitude of the input reflection coefficient is found from:
\begin{equation}
\begin{array}{lcl}
\displaystyle |\Gamma| &=& \di \left| \frac{Z_{in}-Z_0}{Z_{in}+Z_0} \right|, \\
\displaystyle Z_{in} &=& \di Z_1 \frac{R_L+\jmath Z_1 \tan \beta l}{Z_1+\jmath R_L \tan \beta l} \\
\di \beta l $=$ \di \frac{\pi f}{2 f_0}.
\end{array}
\end{equation}
Fig. \ref{fig:GammaLambda4} shows a plot of the amplitude of the reflection coefficient of this quarter-wave transformer. We can observe that only at the design frequency an optimal matching is obtained. This is a clear drawback of a single-stage quarter-wave transformer and limits the frequency bandwidth of such a matching circuit. The bandwidth can be increased by using multiple stages. Note that the transformer is also matched at $l=n \lambda/4$ with $n=3,5,....$.

\section{The lossy terminated transmission line}
\label{sec:lossyTline}

\begin{figure}[hbt]
\vspace{-1cm}
\centerline{\psfig{figure=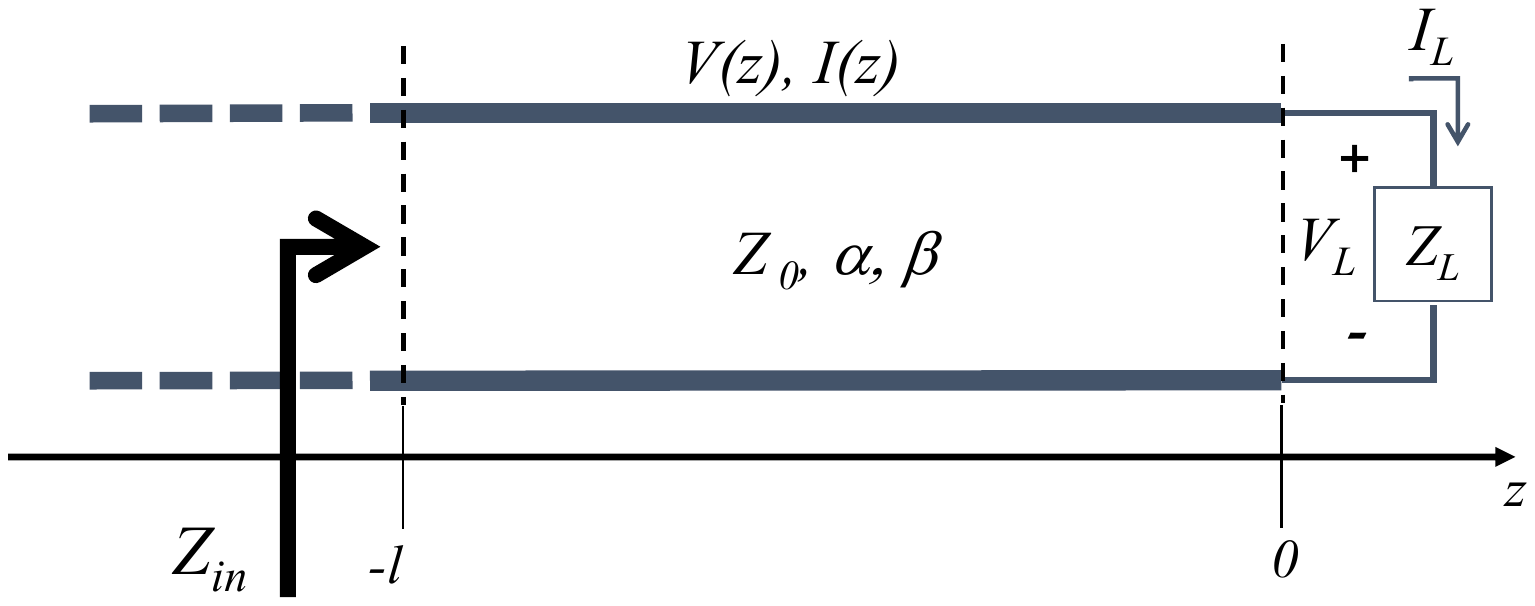,width=120mm}}
\vspace{-3cm}
\caption{\it The lossy terminated transmission line with $\gamma=\alpha+\jmath \beta$.}
\label{fig:Lossyterminatedline}
\end{figure}

In practise, transmission lines will always be lossy, although the theory of lossless lines is generally a very good first-order estimation. In case of a lossy terminated line, as illustrated in Fig. \ref{fig:Lossyterminatedline}, the voltage and current along the line are found using the general solution of the Telegrapher equation (\ref{eq:SolutionTelegraph1}) and can be written in the following form:
\begin{equation}
\begin{array}{lcl}
\displaystyle V(z)&=& \di V^+_0 \left( e^{-\gamma z} + \Gamma e^{\gamma z} \right), \\
\displaystyle I(z)&=& \di \frac{V^+_0}{Z_0} \left( e^{-\gamma z} - \Gamma e^{\gamma z} \right),
\end{array}
\label{eq:SolutionTelegraph2}
\end{equation}
where $\Gamma$ is the reflection coefficient at the termination $z=0$. The input impedance at $z=-l$ is now found by:
\begin{equation}
\begin{array}{lcl}
\displaystyle Z_{in} &=& \di \frac{V(z=-l)}{I(z=-l)}= \frac{V^+_0 \left[e^{\gamma l}+\Gamma e^{-\gamma l} \right]}{V^+_0 \left[e^{\gamma l}-\Gamma e^{-\gamma l} \right]} Z_0 \\
&=& \di Z_0 \frac{Z_L+\jmath Z_0 \tanh \gamma l}{Z_0+\jmath Z_L \tanh \gamma l}.
\end{array}
\label{eq:ZinLossyline}
\end{equation}
The power delivered at $z=-l$ is now:
\begin{equation}
\begin{array}{lcl}
\displaystyle P_{in} &=& \di \frac{1}{2} Re \left( V(-l) I^*(-l) \right) \\
\di &=& \di \frac{\left|V^+_0 \right|^2}{2Z_0} \left( e^{2 \alpha l} - |\Gamma|^2 e^{-2 \alpha l} \right).
\end{array}
\label{eq:PinLossyline}
\end{equation}
The corresponding power delivered to the load at $z=0$ is found by:
\begin{equation}
\begin{array}{lcl}
\displaystyle P_{L} &=& \di \frac{1}{2} Re \left( V(0) I^*(0) \right) \\
\di &=& \di \frac{\left|V^+_0 \right|^2}{2Z_0} \left( 1 - |\Gamma|^2 \right).
\end{array}
\label{eq:PLLossyline}
\end{equation}
The loss in the terminated line of length $l$ is found as the difference between $P_{in}$ and $P_{L}$:
\begin{equation}
\begin{array}{lcl}
\displaystyle P_{loss} &=& \di P_{in}-P_{L} = \frac{\left|V^+_0 \right|^2}{2Z_0} \left( \left[ e^{2\alpha l}-1 \right] +|\Gamma|^2 \left[1-e^{-2\alpha l} \right] \right).
\end{array}
\label{eq:PLossLossyline}
\end{equation}
The first term in (\ref{eq:PLossLossyline}) is the loss of the incident wave along the line with length $l$. The second term describes the loss of the reflected wave.

\section{Field analysis of transmission lines}
\label{sec:Fieldanalysis}
\begin{figure}[hbt]
\centerline{\psfig{figure=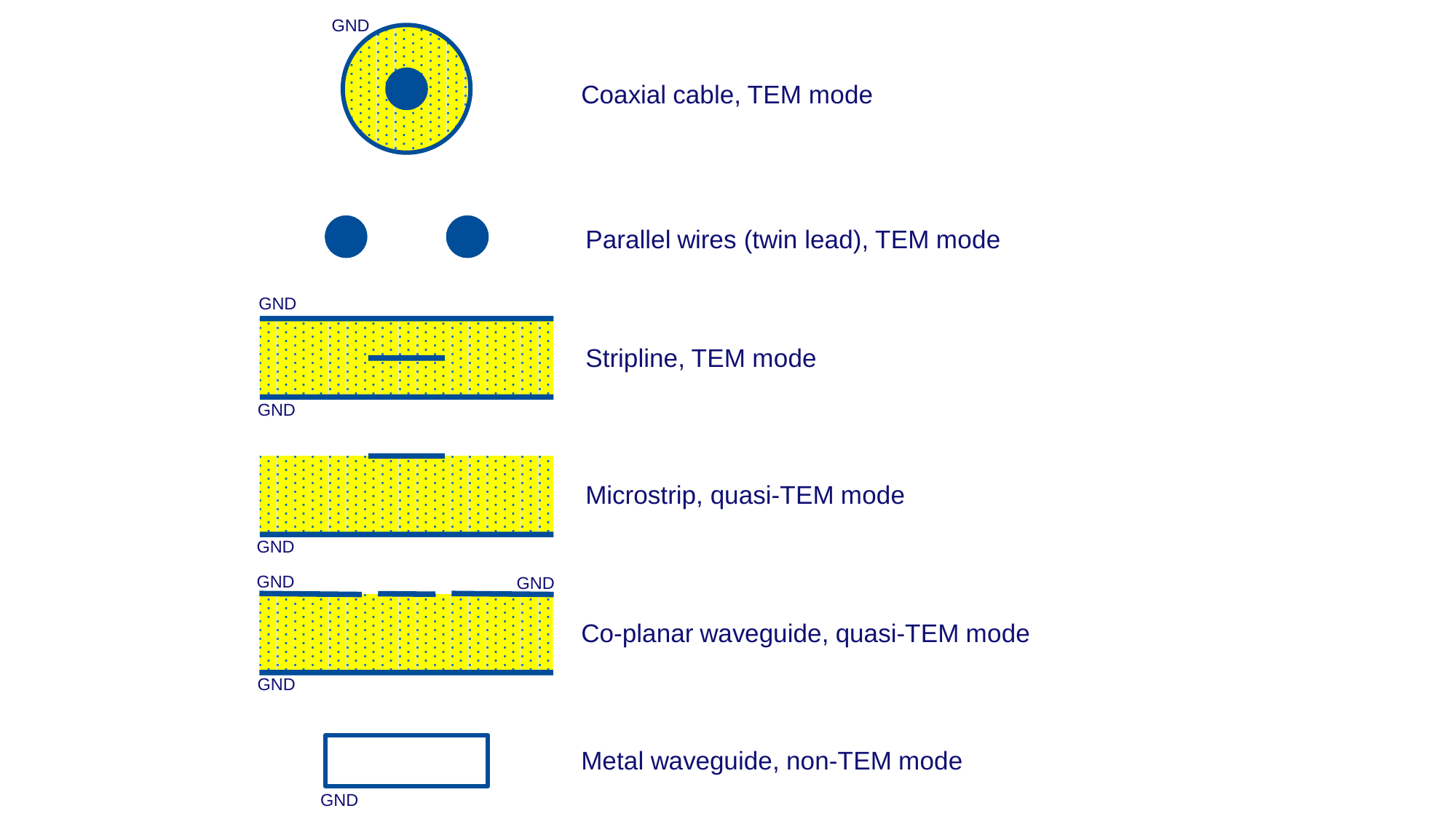,width=200mm}}
\caption{\it Cross sections of various transmission lines that support TEM, quasi-TEM or non-TEM wave propagation.}
\label{fig:Tlineoverview}
\end{figure}

\begin{figure}[hbt]
\centerline{\psfig{figure=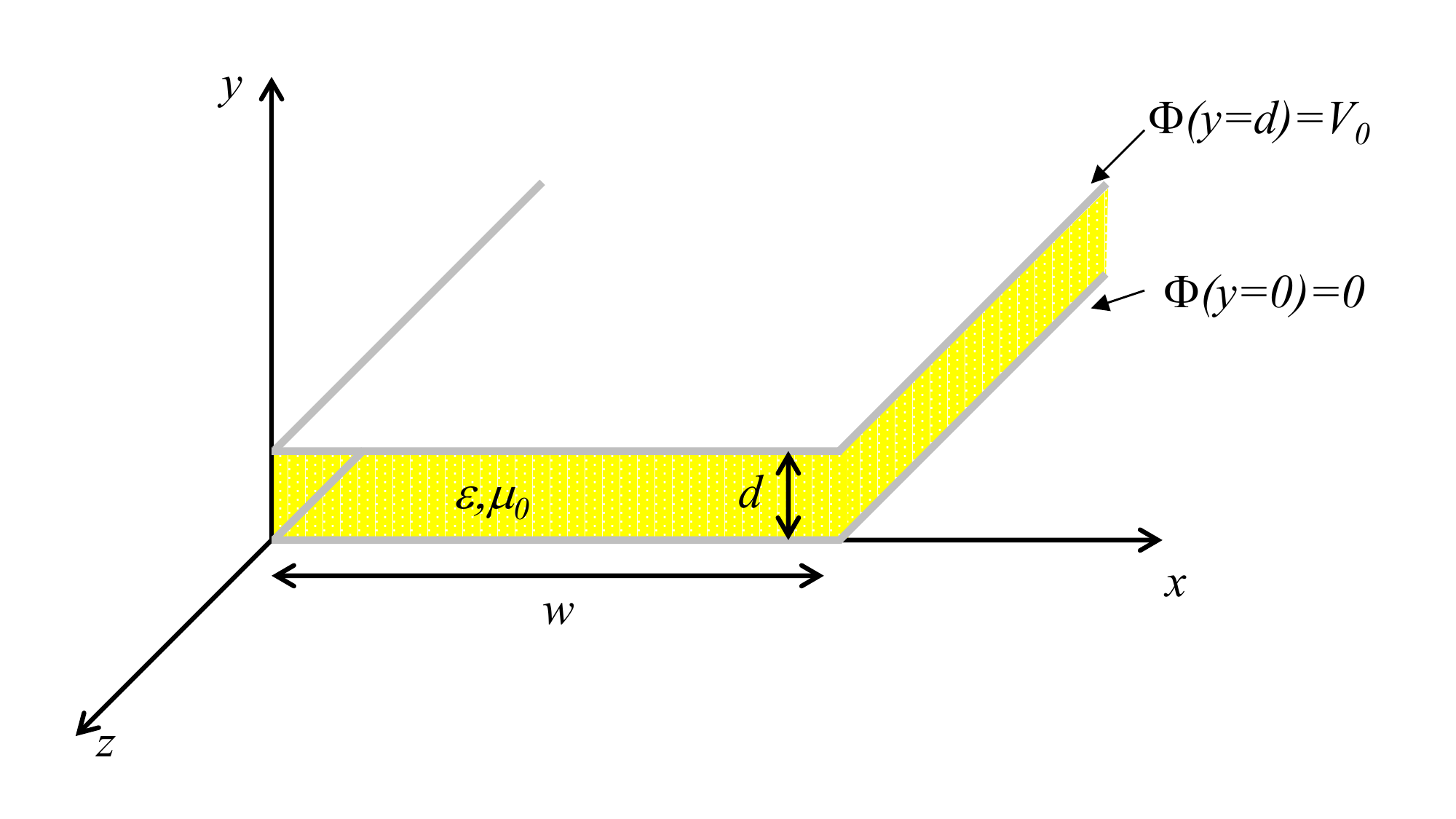,width=100mm}}
\caption{\it The parallel-plate transmission line that support a TEM wave.}
\label{fig:Parallelplate}
\end{figure}

Real transmission lines may take various shapes. Fig. \ref{fig:Tlineoverview} shows the cross section of various types of transmission lines that support TEM or quasi-TEM wave propagation. The physical parameters like width of a microstrip line and dielectric constant and loss tangent of the substrate can be translated into transmission parameters $Z_0, \alpha, \beta$ and/or in terms of $L, C, R, G$. However, for most transmission line types this cannot be done in an analytical way, since we need to solve Maxwells' equation using the boundary conditions on the metal structures. Approximate models can be applied, but numerical methods need to be used in order to obtain very accurate results. Nowadays, several open-source and commercial software packages, like QUCS and ADS are available with tools to determine the transmission line parameters for various types of transmission lines.
A type of transmission line for which we can easily solve Maxwells' equations is the parallel-plate waveguide as illustrated in Fig. \ref{fig:Parallelplate}. Although it is not a very practical transmission line, its characteristics are quite similar to more practical transmission lines, e.g. microstrip line. It consists of two large metallic plates separated by a dielectric substrate with permittivity $\epsilon=\eps \epsilon_0$ and permeability $\mu_0$. We will assume that the transmission line is lossless. In addition, it is assumed that the width $w \gg d$.
The TEM wave has zero cut-off frequency and propagates along the $z$ direction, therefore, $E_z=H_z=0$ and $\di \vec{E}=\vec{E}_t=(E_x \vec{u}_x +E_y \vec{u}_y ) e^{-\jmath \beta z}$. In this case, it can be shown that the transverse components of the electric and magnetic field can be found by solving Laplace's equation:
\begin{equation}
\begin{array}{lcl}
\displaystyle \nabla^2_t \vec{E_t}&=& 0, \\
\displaystyle \nabla^2_t \vec{H_t}&=& 0, \\
\end{array}
\label{eq:Laplace1}
\end{equation}
where the transverse electric field is $\vec{E}_t =E_x \vec{u}_x + E_y \vec{u}_y$ and $\nabla^2_t= \partial^2 /\partial x^2+ \partial^2 /\partial y^2$ resembles the Laplacian operator in the $(x,y)$ dimensions. From (\ref{eq:Laplace1}) we can observe that the transverse fields are the same as the static fields in a parallel plate configuration. From electrostatics, we know that the electric field in (\ref{eq:Laplace1}) can be expressed as a gradient of a scalar potential $\Phi(x,y)$:
\begin{equation}
\begin{array}{lcl}
\displaystyle \vec{E_t} &=& - \nabla_t \Phi(x,y),
\end{array}
\label{eq:Efieldscalarpot}
\end{equation}
This scalar potential $\Phi(x,y)$ satisfies Laplace's equation:
\begin{equation}
\begin{array}{lcl}
\displaystyle \nabla^2_t \Phi(x,y)&=& 0.
\end{array}
\label{eq:Laplace3}
\end{equation}
The boundary conditions of this differential equation are as indicated in Fig. \ref{fig:Parallelplate}:
\begin{equation}
\begin{array}{lcl}
\displaystyle \Phi(x,0)&=& 0, \\
\displaystyle \Phi(x,d)&=& V_0.
\end{array}
\label{eq:boundaryLaplace}
\end{equation}
A solution of $\Phi(x,y)$ cannot depend on $x$, since we have assumed that the width $w$ is very large (close to infinite). Therefore, $\Phi(x,y)$ takes the form:
\begin{equation}
\begin{array}{lcl}
\displaystyle \Phi(x,y)&=& \di \frac{V_0 y}{d},
\end{array}
\label{eq:scalarpot}
\end{equation}
which can be easily verified by substituting this in (\ref{eq:Laplace3}) and (\ref{eq:boundaryLaplace}).
The electric and magnetic fields between the plates are now obtained using (\ref{eq:Efieldscalarpot}):
\begin{equation}
\begin{array}{lcl}
\displaystyle \vec{E}(x,y,z) &=& \di - \nabla_t \Phi(x,y) e^{-\jmath \beta z} = -\vec{u}_y \frac{V_0}{d} e^{-\jmath \beta z}, \\
\di \vec{H}(x,y,z) &=& \di \frac{1}{\eta}  \vec{u}_z \times \vec{E}(x,y,z) = \vec{u}_x \frac{V_0}{\eta d} e^{-\jmath \beta z},
\end{array}
\label{eq:EHfield}
\end{equation}
where the propagation constant $\beta = \omega \sqrt{\epsilon \mu_0}$ and the intrinsic impedance $\eta=\sqrt{\mu_0/\epsilon}$. The corresponding voltage and current along the line are now given by:
\begin{equation}
\begin{array}{lcl}
\displaystyle V(z) &=& \di - \int_{0}^{d} E_y dy = V_0 e^{-\jmath \beta z} \\
\di I(z) &=& \di \int_{0}^w J_z dx = \int_{0}^w (-\vec{u}_y \times \vec{H} ) \cdot \vec{u}_z dx = \int_{0}^w H_x dx =  \frac{wV_0}{\eta d} e^{-\jmath \beta z}.
\end{array}
\label{eq:VIparallelplate}
\end{equation}
The transmission line parameters are then found by (with $R=G=0$):
\begin{equation}
\begin{array}{lcl}
\displaystyle Z_0 &=& \di \frac{V}{I} = \frac{\eta d}{w} \ [\Omega], \\
L & = & \di \frac{\mu_0}{|I|^2} \int \int |\vec{H}|^2 dS = \frac{\mu_0 d}{w} \ \left[\frac{H}{m} \right], \\
C & = & \di \frac{\epsilon}{|V_0|^2} \int \int |\vec{E}|^2 dS = \frac{\epsilon w}{d} \ \left[\frac{F}{m} \right], \\
\beta & = & \di \omega \sqrt{LC}=\omega \sqrt{\epsilon \mu_0} \left[\frac{1}{m} \right].
\end{array}
\label{eq:LCZparallelplat}
\end{equation}

\section{The Smith chart}
\label{sec:Smithchart}
\begin{figure}[hbt]
\centerline{\psfig{figure=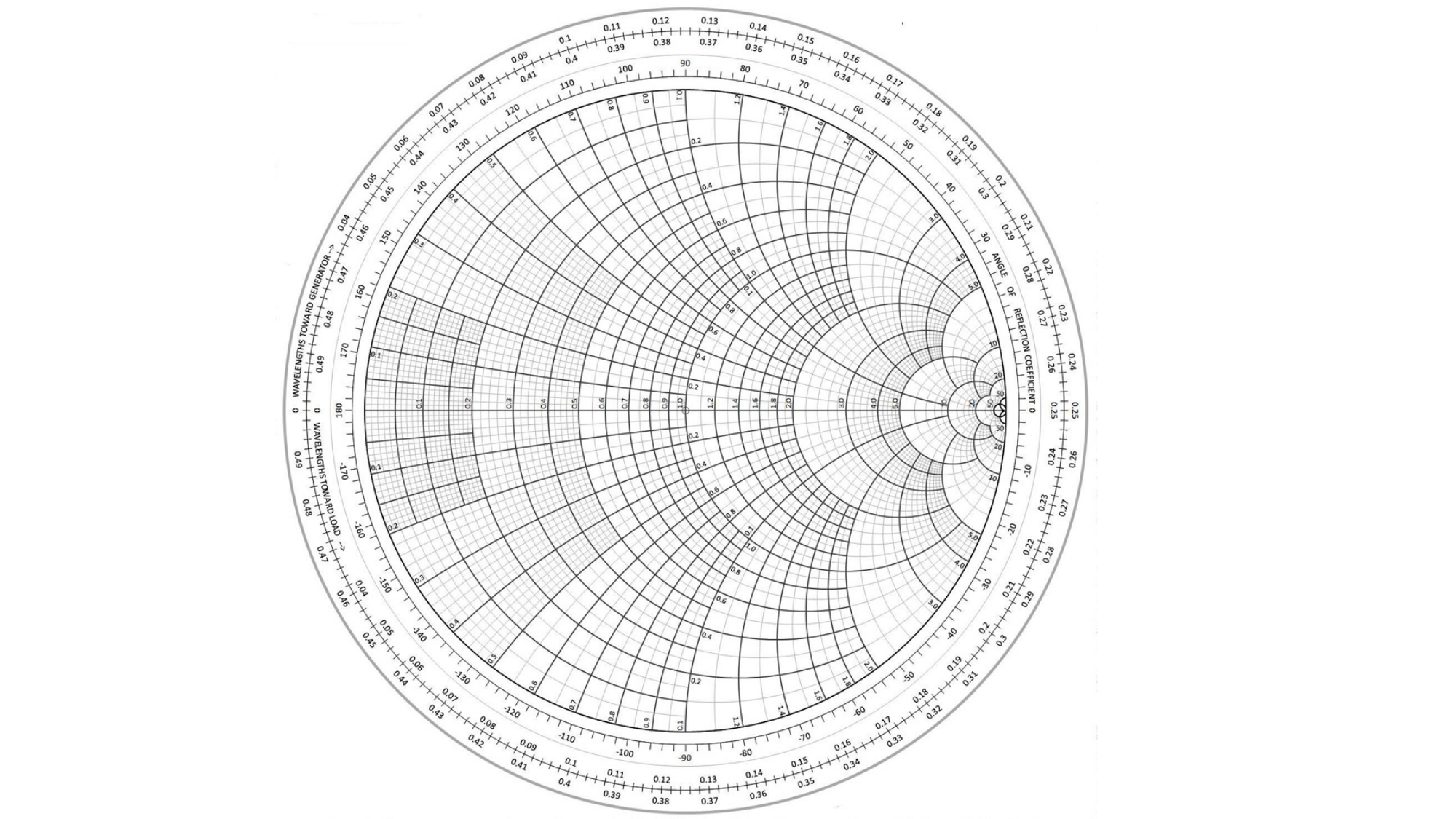,width=220mm}}
\caption{\it The Smith chart. The normalized impedance values $z_{in}=r_{in}+\jmath x_{in}=Z_{in}/Z_0$ are listed. Usually $Z_0=50 \ \Omega$.}
\label{fig:Smithchart}
\end{figure}

\begin{figure}[hbt]
\centerline{\psfig{figure=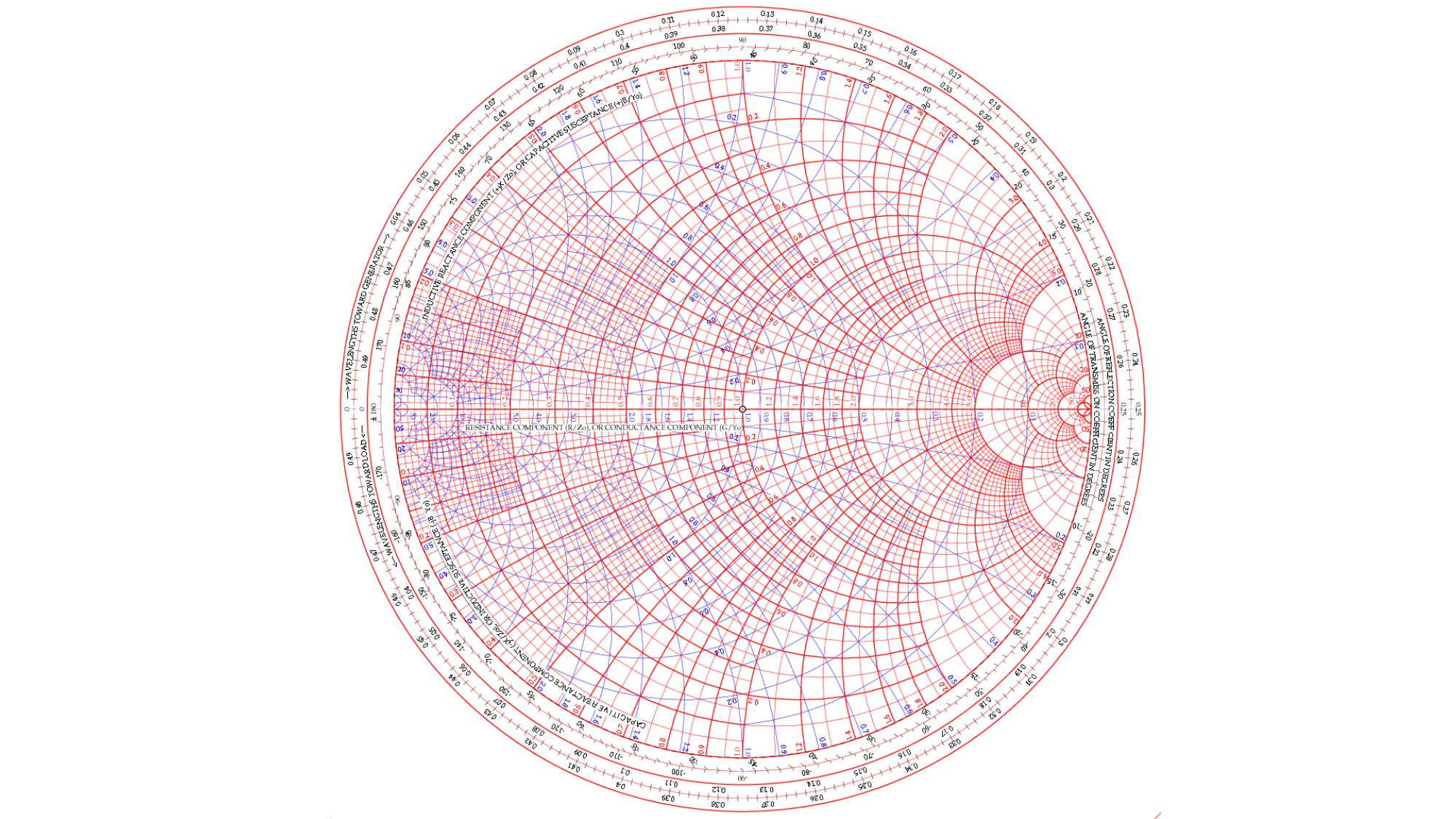,width=220mm}}
\caption{\it Smith chart with normalized impedance and admittance coordinates.}
\label{fig:Smithchartadm}
\end{figure}

The Smith chart is a very useful graphical tool to visualize the behaviour of transmission-line circuits. The Smith chart was developed by P. Smith in 1939, well before computer-based design tools had become available. The Smith chart is still de-facto de standard representation tool in microwave engineering. The Smith chart is basically built around the complex representation of the reflection coefficient of a transmission line circuit, for example as described in (\ref{eq:Gammazerocondition}) in case of a quarter-wave transformer. The reflection coefficient is complex and can be expressed in terms of amplitude and phase:
\begin{equation}
\begin{array}{lcl}
\displaystyle \Gamma=\frac{Z_{in}-Z_0}{Z_{in}+Z_0} = \Gamma_r + \jmath \Gamma_i =|\Gamma| e^{\jmath \phi_{\Gamma}}.
\end{array}
\label{eq:Gammapolar}
\end{equation}
When we would plot $\Gamma$ in the complex plane, we find that all values are within the unit circle, since $|\Gamma| \leq 1$. In case $Z_{in}=Z_0$, $\Gamma=0$ and we would end up in the center of the unit circle. The most common form of the Smith chart is shown in Fig. \ref{fig:Smithchart}.
Note that the normalized impedance values $z_{in}=r_{in}+\jmath x_{in}=Z_{in}/Z_0$ are listed in the Smith chart, where $r_{in}, x_{in}$ are the normalized resistance and reactance, respectively.
Let us now investigate some special cases in more detail. First assume that the transmission line circuit is purely resistive with $Z_{in}=R_{in}+\jmath X_{in}=R_{in}=r_{in} Z_0$. In this case we find that:
\begin{equation}
\begin{array}{lcl}
\displaystyle \Gamma=\frac{Z_{in}-Z_0}{Z_{in}+Z_0} =\frac{r_{in}-1}{r_{in}+1}.
\end{array}
\label{eq:GammaRin}
\end{equation}
Now $\Gamma$ is real and describes a line along the horizontal axis with $\Gamma=-1$ for $r_{in}=0$ (short), $\Gamma=0$ for $r_{in}=Z_0$ (matched load) and $\Gamma=1$ for $r_{in}=\infty$ (open). Usually, $Z_0=50 \ \Omega$ is used.
Another special case occurs when the load is purely reactive with $Z_{in}=\jmath X_{in}=\jmath x_{in} Z_0$. It can be shown that the reflection coefficient $\Gamma$ now describes a circle with amplitude $|\Gamma|=1$. This can be generalized to the case of a constant resistance and varying reactance. In this case we obtain circles as can be seen in Fig. \ref{fig:Smithchart}. The lines for a load with constant reactance and varying resistance can also be observed in the Smith chart and correspond to parts of large circles that extend the unit circle.
Smith charts with both impedance and admittance values are also available and are useful when designing serial or parallel matching circuits. Fig. \ref{fig:Smithchartadm} shows such a chart.

\vspace*{1cm} \hrulefill
 {\bf Exercise} \hrulefill \\
 Plot the following load conditions in the Smith chart:
\begin{enumerate}[i.]
 \item $Z_{in}=0$,
 \item $Z_{in}=\infty $,
 \item $Z_{in}=50 \ \Omega$,
 \item $Z_{in}=50 - \jmath 150 \ \Omega$.
\end{enumerate}

\section{Microwave networks}
\label{sec:microwavenetworks}
\begin{figure}[hbt]
\centerline{\psfig{figure=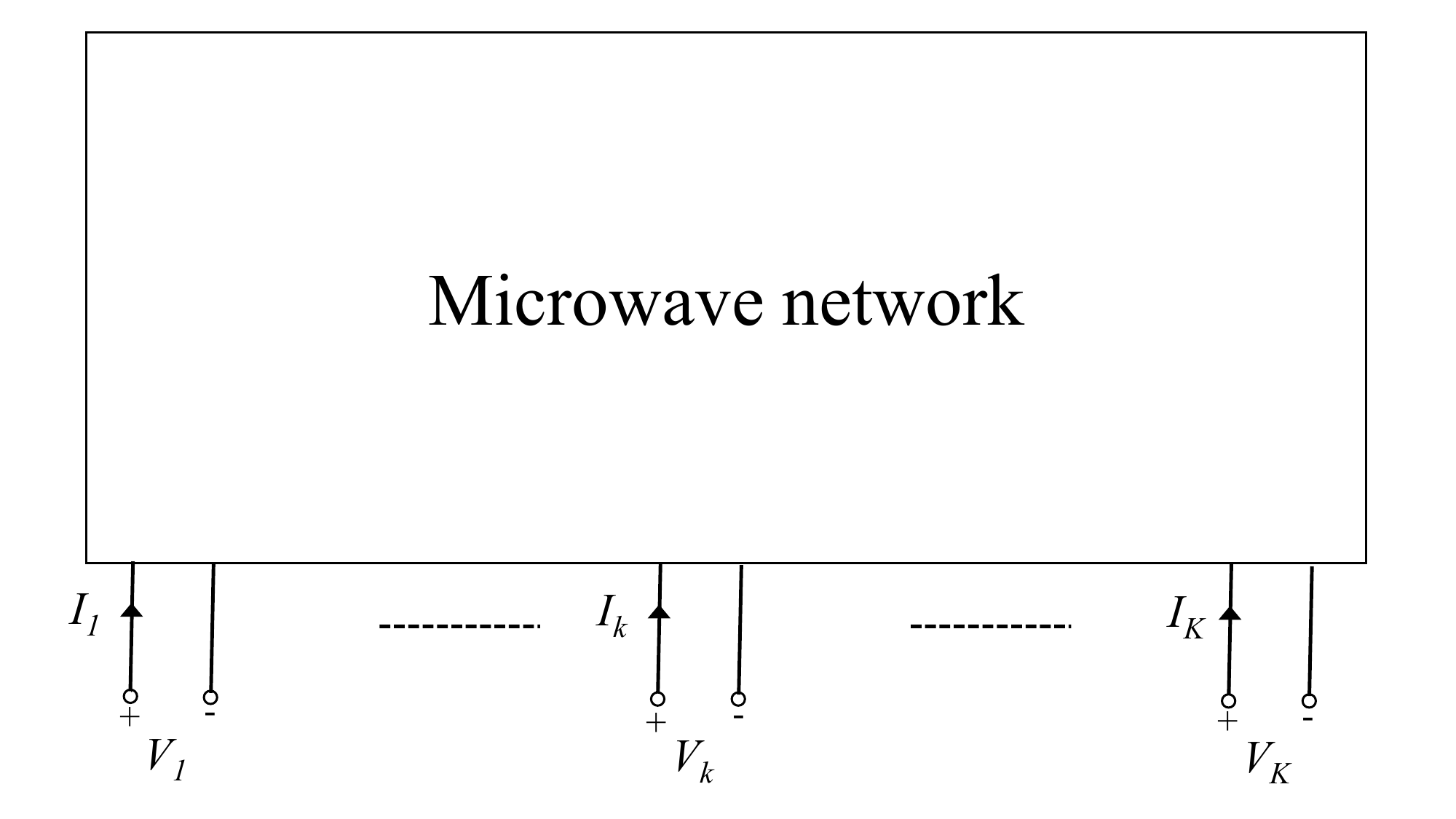,width=120mm}}
\caption{\it $K$-port microwave network with port voltages and currents.}
\label{fig:Kportmicrowavenetwork}
\end{figure}

Consider a microwave network with $K$-ports as illustrated in Fig. \ref{fig:Kportmicrowavenetwork}. Each port is connected to a transmission line on which TEM-modes can propagate. At port $k$, the total voltage and current is given by:
\begin{equation}
\begin{array}{lcl}
\displaystyle V_k &=& V^+_k+V^-_k,\\
\displaystyle I_k &=& I^+_k-I^-_k.
\end{array}
\label{eq:totalvoltagecurrentportk}
\end{equation}
The relation between the port voltages and port currents is defined by the {\it impedance matrix} $[Z]$:
\begin{equation}
\displaystyle \left[ \begin{array}{l} V_1 \\ V_2 \\ . \\ V_K \end{array}
\right] = \left[
\begin{array}{l}
Z_{11} \ \ Z_{12} \ \ . \ \ . \ \ Z_{1K} \\
Z_{21} \ \ Z_{22} \ \ . \ \ . \ \ Z_{2K} \\
\ \ . \ \ \ \  \ \ . \ \ \ \ . \ \ . \ \ \ \ . \\
Z_{K1} \ \ Z_{K2} \ \ . \ \ . \ \ Z_{KK}
 \end{array} \right]
 \left[ \begin{array}{l} I_1 \\ I_2 \\ . \\ I_K \end{array} \right].
 \label{eq:Impedancematrix}
\end{equation}
In most practical case we will investigate two-port microwave networks with $K=2$. The relation between port voltages and currents then reduces to:
\begin{equation}
\displaystyle \left[ \begin{array}{l} V_1 \\ V_2 \end{array}
\right] = \left[ \begin{array}{l} Z_{11} \ \ Z_{12} \\ Z_{21} \ \
Z_{22} \end{array} \right] \left[ \begin{array}{l} I_1 \\ I_2
\end{array} \right],
\end{equation}
where the individual components of the $Z$-matrix are found by:
\begin{equation}
\begin{array}{lcl}
\displaystyle Z_{11} = \left. \frac{V_1}{I_1} \right|_{I_2=0} & \ \ & \di Z_{12} = \left. \frac{V_1}{I_2} \right|_{I_1=0} \\
\displaystyle Z_{21} = \left. \frac{V_2}{I_1} \right|_{I_2=0} & \ \ & \di Z_{22} = \left. \frac{V_2}{I_2} \right|_{I_1=0}
\end{array}
\label{eq:Impedancematrix2}
\end{equation}

\begin{figure}[hbt]
\centerline{\psfig{figure=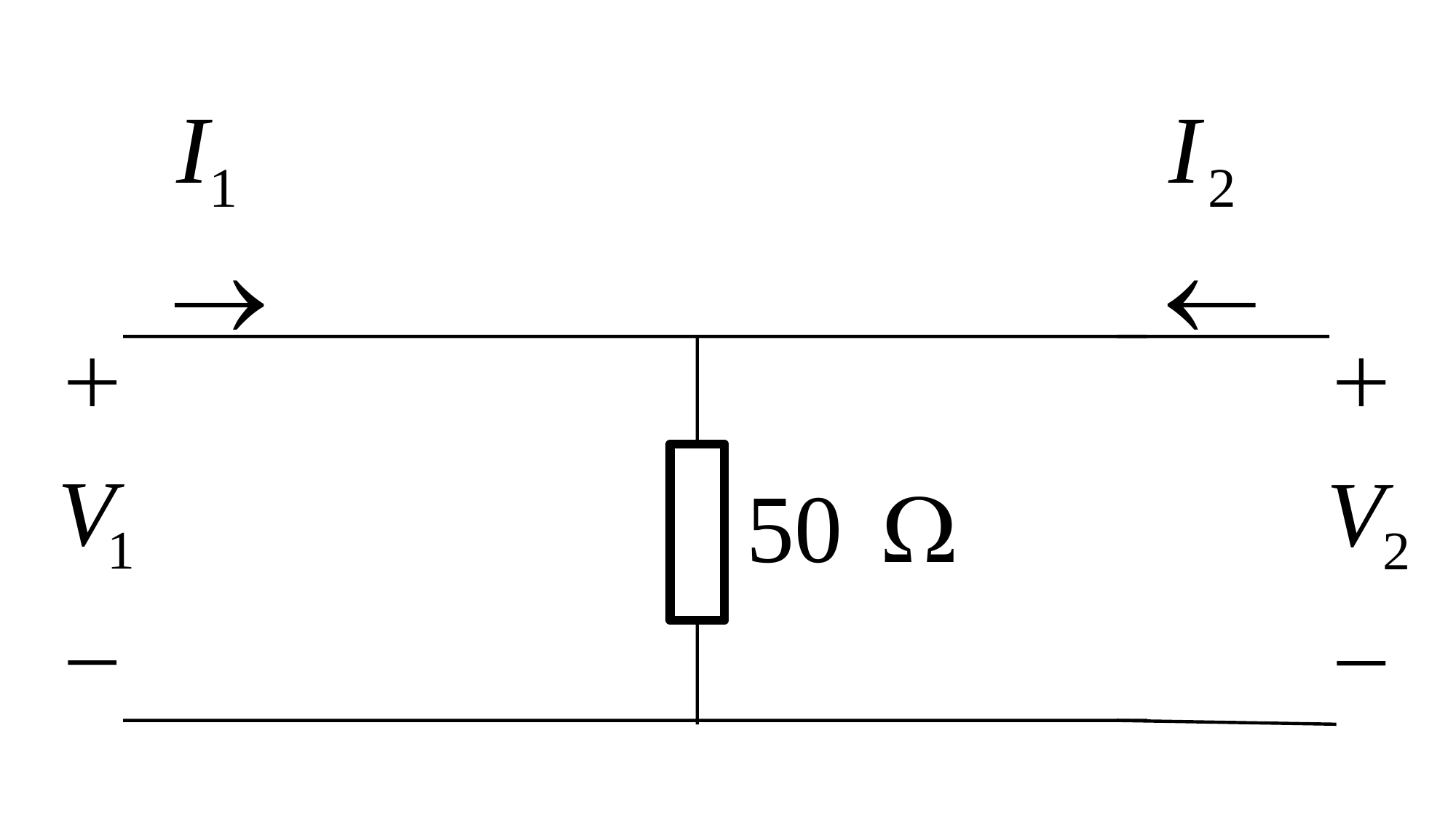,width=80mm}}
\caption{\it $2$-port microwave network with a parallel $50 \ \Omega$ resistor.}
\label{fig:Exampletwoport}
\end{figure}

\begin{figure}[hbt]
\centerline{\psfig{figure=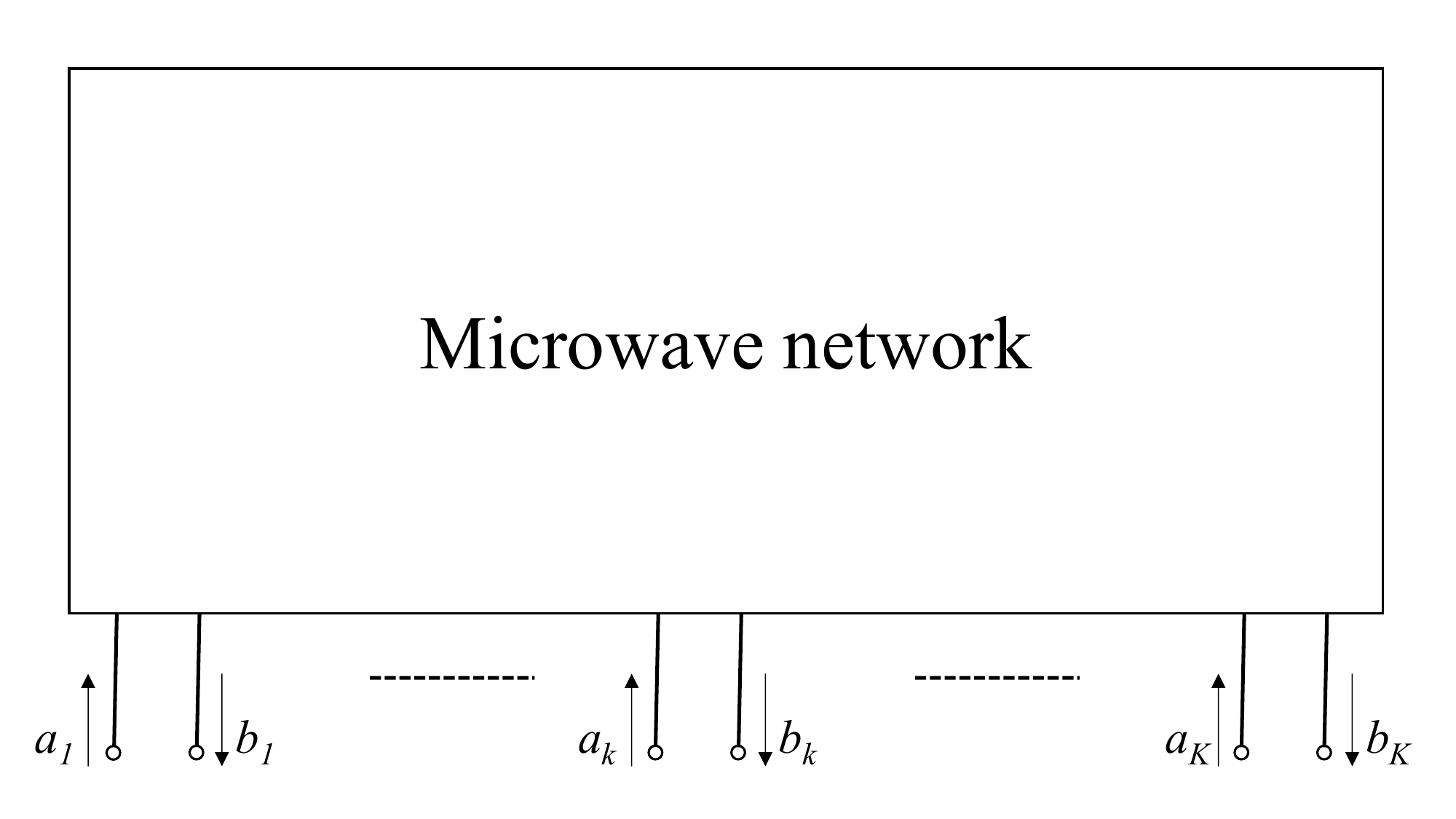,width=120mm}}
\caption{\it $K$-port microwave network described using  normalized complex amplitudes of the incident and reflected TEM waves}
\label{fig:KportnetworkSpar}
\end{figure}

\begin{figure}[hbt]
\centerline{\psfig{figure=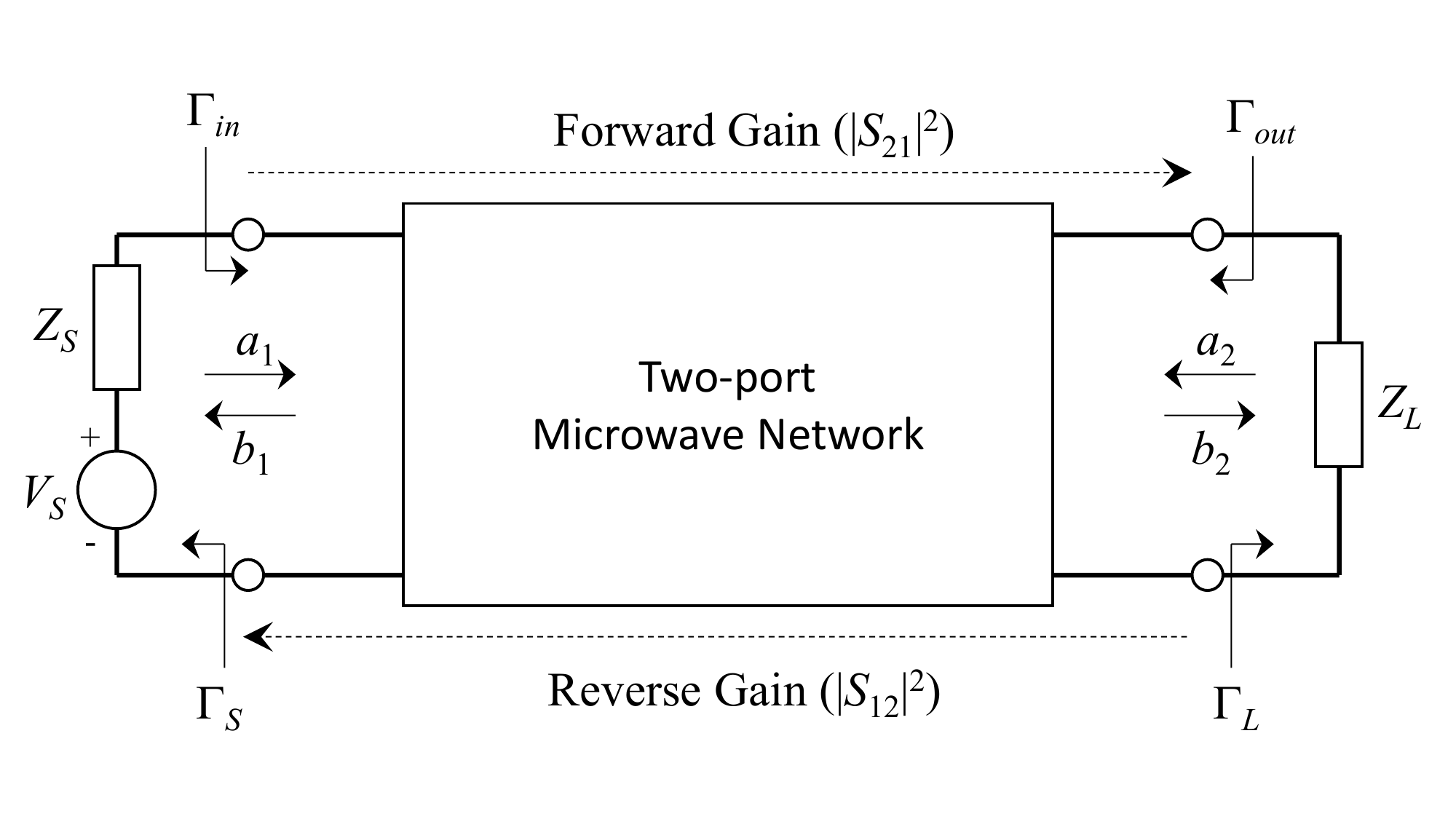,width=120mm}}
\caption{\it Two-port microwave network with input and output terminated with $Z_S$ and $Z_L$, respectively}
\label{fig:2portnetwork}
\end{figure}

\vspace{0.5cm}
\underline{{\it Example}}
Consider the network of Fig. \ref{fig:Exampletwoport}, where a $50 \ \Omega$ resistor is connected in parallel. The components of the impedance matrix are now found using (\ref{eq:Impedancematrix2}):
\begin{equation}
\begin{array}{lcl}
\displaystyle Z_{11} = \left. \frac{V_1}{I_1} \right|_{I_2=0} =50 \ \Omega & \ \ & \di Z_{12} = \left. \frac{V_1}{I_2} \right|_{I_1=0} =50 \ \Omega \\
\displaystyle Z_{21} = \left. \frac{V_2}{I_1} \right|_{I_2=0} =50 \ \Omega & \ \ & \di Z_{22} = \left. \frac{V_2}{I_2} \right|_{I_1=0}=50 \ \Omega
\end{array}
\end{equation}
\vspace{0.25cm}
The voltage-current relation of a microwave network can also be expressed by means of an {\it admittance matrix} $[Y]$. In case of a two-port network, we obtain:
\begin{equation}
\displaystyle \left[ \begin{array}{l} I_1 \\ I_2 \end{array}
\right] = \left[ \begin{array}{l} Y_{11} \ \ Y_{12} \\ Y_{21} \ \
Y_{22} \end{array} \right] \left[ \begin{array}{l} V_1 \\ V_2
\end{array} \right],
\end{equation}
where the individual components of the $Y$-matrix are found by:
\begin{equation}
\begin{array}{lcl}
\displaystyle Y_{11} = \left. \frac{I_1}{V_1} \right|_{V_2=0} & \ \ & \di Y_{12} = \left. \frac{I_1}{V_2} \right|_{V_1=0} \\
\displaystyle Y_{21} = \left. \frac{I_2}{V_1} \right|_{V_2=0} & \ \ & \di Y_{22} = \left. \frac{I_2}{V_2} \right|_{V_1=0}
\end{array}
\label{eq:Admittancematrix2}
\end{equation}
Although the impedance and admittance matrices describe the behaviour of a microwave network completely, it appears that this description is not very practical at microwave frequencies. This is due to the fact that we cannot measure the complex values of the total port voltages and currents in an accurate way at these frequencies. However, we can measure the amplitude and phase of the incident and reflected voltage waves in an accurate way by measuring the complex amplitude of the incident and reflected electric field of the TEM wave. This is done with a vector network analyzer (VNA). Let us consider again the $K$-port microwave network, but now described in terms of the complex amplitude of the incident and reflected waves at each port. This is illustrated in Fig. \ref{fig:KportnetworkSpar}. The normalized complex incident and reflected wave amplitudes are given by:
\begin{equation}
\begin{array}{lcl}
\displaystyle a_k &=& \di \frac{V^+_k}{\sqrt{Z_0}} = I^+_k \sqrt{Z_0} ,\\
\displaystyle b_k &=& \di \frac{V^-_k}{\sqrt{Z_0}} = I^-_k \sqrt{Z_0}.
\end{array}
\label{eq:abportk}
\end{equation}
Note that $Z_0=50 \ \Omega$ is usually used in measurements.
The relation between the incident waves $[a]$ and reflected waves $[b]$ is determined by the {\it scattering matrix} $[S]$, which for a two-port microwave network takes the following form:
\begin{equation}
\displaystyle \left[ \begin{array}{l} b_1 \\ b_2 \end{array}
\right] = \left[ \begin{array}{l} S_{11} \ \ S_{12} \\ S_{21} \ \
S_{22} \end{array} \right] \left[ \begin{array}{l} a_1 \\ a_2
\end{array} \right],
\label{eq:Scatteringmatrix}
\end{equation}
where the individual scattering parameters are found by:
\begin{equation}
\begin{array}{lcl}
\displaystyle S_{11} = \left. \frac{b_1}{a_1} \right|_{a_2=0} & \ \ & \di S_{12} = \left. \frac{b_1}{a_2} \right|_{a_1=0} \\
\displaystyle S_{21} = \left. \frac{b_2}{a_1} \right|_{a_2=0} & \ \ & \di S_{22} = \left. \frac{b_2}{a_2} \right|_{a_1=0}
\end{array}
\label{eq:Scatteringparameters}
\end{equation}
So the scattering parameter $S_{11}$ can be found by properly matching output port 2. The other parameters are found in a similar way. In case of a one-port network, we find that the input reflection coefficient $\Gamma_{in}=S_{11}$.
Now consider the two-port microwave network of Fig. \ref{fig:2portnetwork} with source impedance $Z_S$ at the input port and load impedance $Z_L$ at the output port. The parameter $|S_{21}|^2$ is known as the {\it forward gain} and $|S_{12}|^2$ is the {\it reverse gain}. Note that in case of passive microwave networks both the forward and reverse gain are smaller than 1 and $S_{21}=S_{12}$ due to the reciprocity principle.
The input and output reflection coefficients of this network are found by:
\begin{equation}
\begin{array}{lcl}
\displaystyle \Gamma_{in} & = & \di \frac{b_1}{a_1} = S_{11} + \frac{S_{12} S_{21}}{\di \frac{1}{\Gamma_L} - S_{22}} \\
\di \Gamma_{out} & = & \di \frac{b_2}{a_2} = S_{22} + \frac{S_{12} S_{21}}{\di \frac{1}{\Gamma_S} - S_{11}},
\end{array}
\label{eq:Inoutreflection}
\end{equation}
where $\Gamma_S$ and $\Gamma_L$ are the reflection coefficients when looking into the source and load, respectively:
\begin{equation}
\begin{array}{lcl}
\displaystyle \Gamma_{S} & = & \di \frac{Z_S-Z_0}{Z_S+Z_0}  \\
\displaystyle \Gamma_{L} & = & \di \frac{Z_L-Z_0}{Z_L+Z_0}.
\end{array}
\label{eq:Sourceloadreflection}
\end{equation}
Combining (\ref{eq:Inoutreflection}) and (\ref{eq:Sourceloadreflection}) we find that $\Gamma_{in}=S_{11}$ only if $Z_L=Z_0$. Similarly, we find that $\Gamma_{out}=S_{22}$ when $Z_S=Z_0$.

\begin{figure}[hbt]
\centerline{\psfig{figure=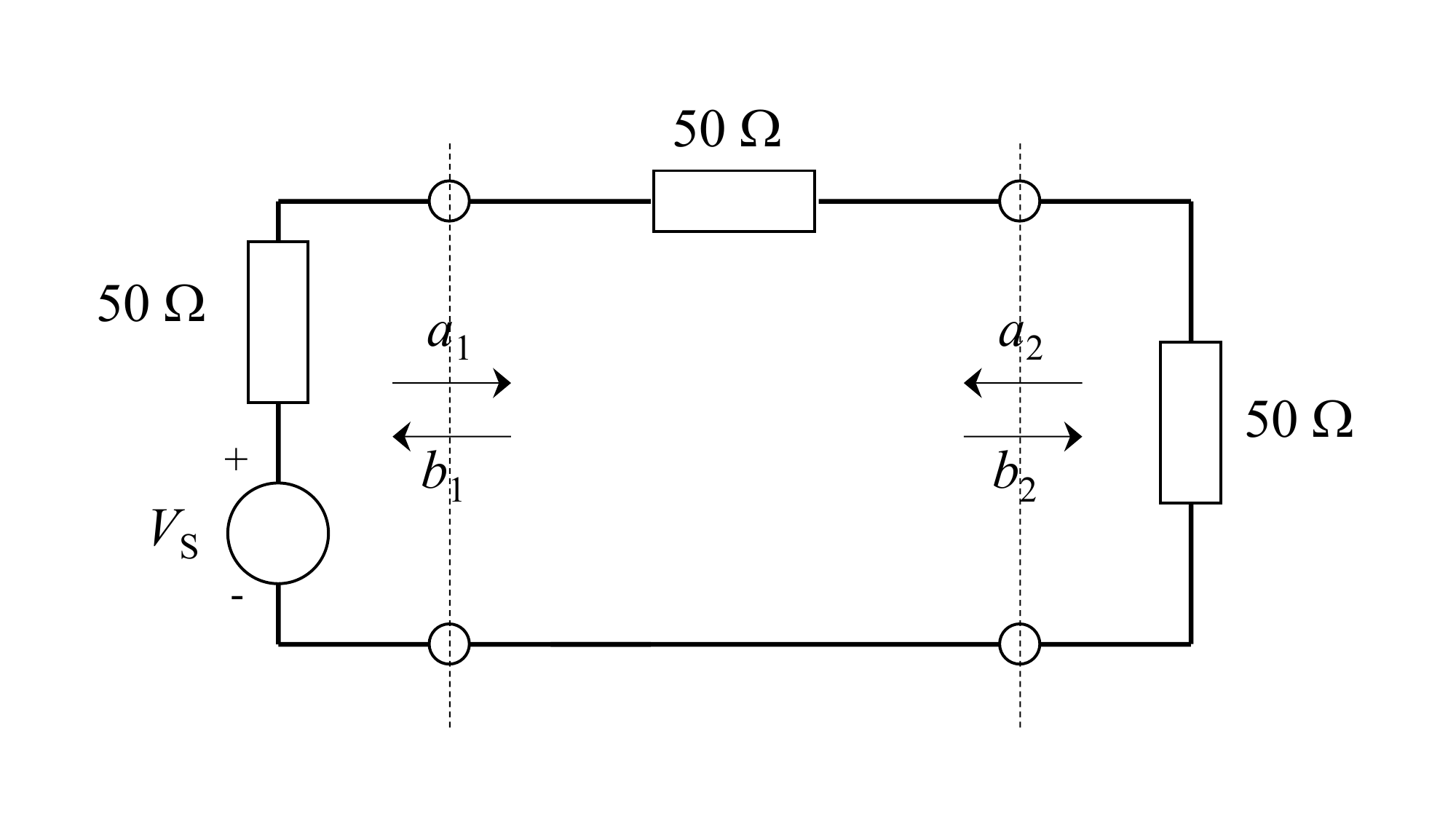,width=100mm}}
\vspace{-0.5cm}
\caption{\it Two-port microwave network consisting of a $50 \ \Omega$ lumped element connected in series between input and output ports}
\label{fig:2portexercise}
\end{figure}

\vspace*{1cm} \hrulefill
 {\bf Exercise} \hrulefill \\
 Consider the two-port network of Fig. \ref{fig:2portexercise}, where a $50 \ \Omega$ resistor is connected in series between both ports. The source and load impedance are $Z_S=Z_L=50 \ \Omega$.
Show that the scattering matrix takes the following form:
\begin{equation}
\displaystyle  \left[ \begin{array}{l} S_{11} \ \ S_{12} \\ S_{21} \ \
S_{22} \end{array} \right] = \left[ \begin{array}{l} \frac{1}{3} \ \ \frac{2}{3} \\ \frac{2}{3} \ \
\frac{1}{3} \end{array} \right].
\label{fig:Scatteringmatrix}
\end{equation}

\section{Power Combiners}
\label{sec:powercombiners}
\begin{figure}[hbt]
\centerline{\psfig{figure=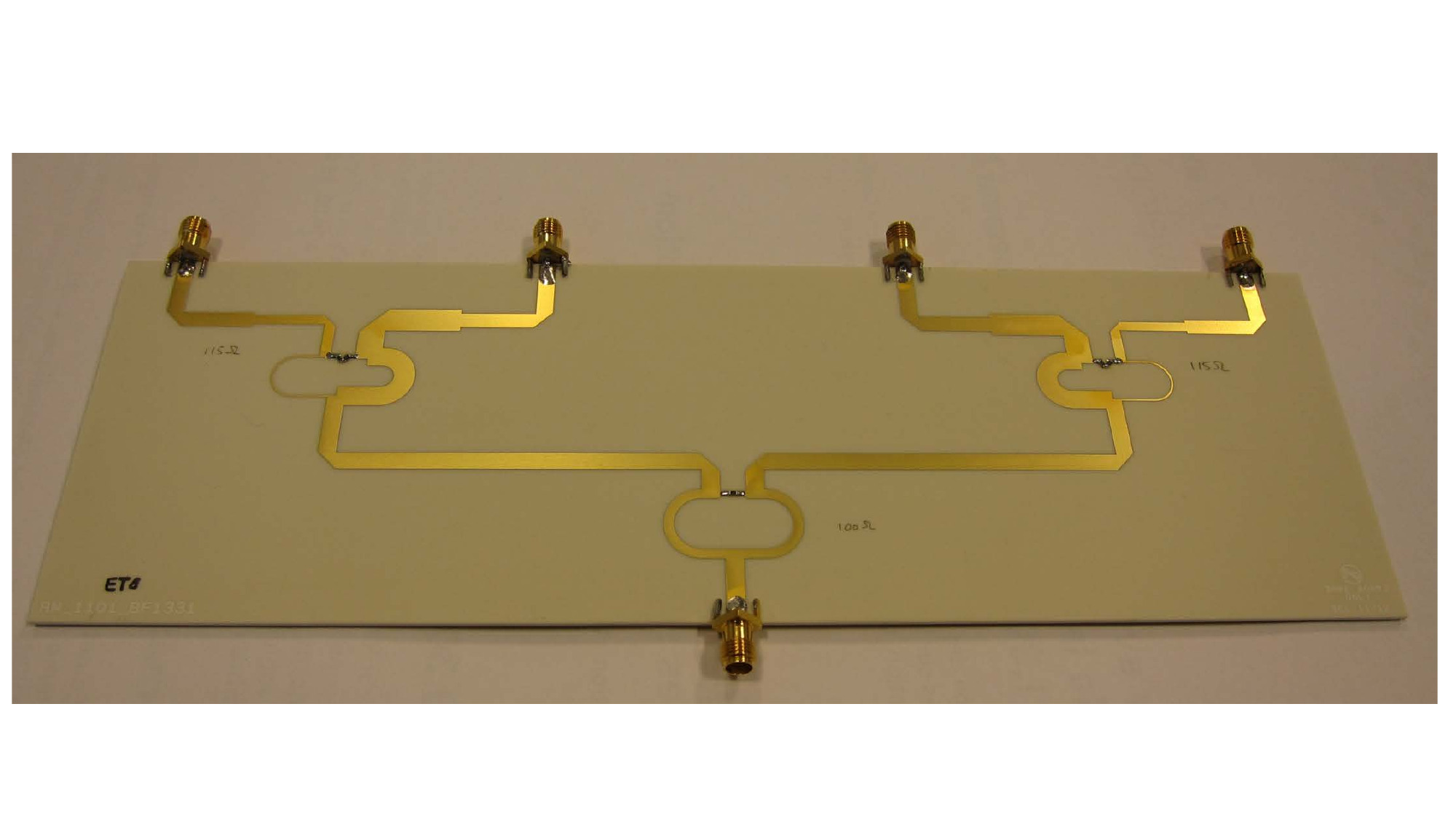,width=100mm}}
\caption{\it Example of a power combiner/divider with an unequal power division ratio, a 4:1 Wilkinson combiner used in base-station antennas, courtesy Rob Mestrom TU Eindhoven.}
\label{fig:Wilkinsonphoto}
\end{figure}

\begin{figure}[hbt]
\centerline{\psfig{figure=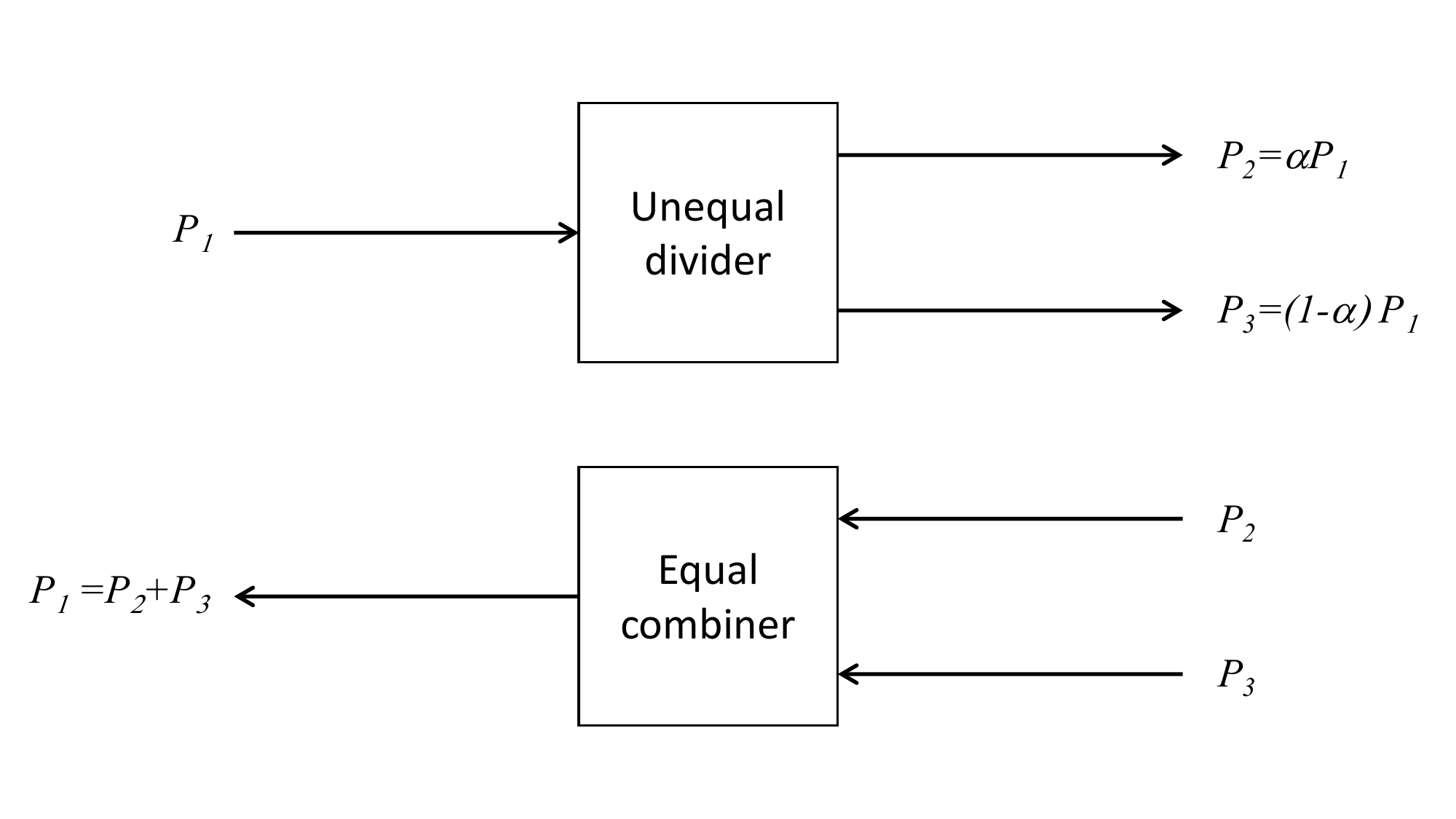,width=110mm}}
\caption{\it Power divider with unequal power division ratio $\alpha$ (top) and a power combiner with equal power division ratio (bottom).}
\label{fig:powercombiner}
\end{figure}

Power combiners and power splitters are one of the most commonly used microwave circuits. For example, in basestations for wireless communications, where a high {\it effective isotropic radiated power (EIRP)} is required to provide coverage of large macro-cells. A high EIRP can be achieved by combining the output power of several power amplifiers using a power combiner. Another application is in array antennas, where received signals from the individual antennas can be combined into a single output signal. An example of a 4-channel power combiner is shown in Fig. \ref{fig:Wilkinsonphoto}. Since power combiners are  passive transmission line circuits, reciprocity holds. As a result, power combiners can also be used as power dividers. The analysis of power combiners in this section is based on the description from David Pozar \cite{Pozar}.
Fig. \ref{fig:powercombiner} shows the basic power divider with unequal power division ratio $\alpha$ and a power combiner with equal power division ratio with $\alpha=0.5$.
It can be shown that the three-port network of Fig. \ref{fig:powercombiner} cannot be at the same time lossless, reciprocal and matched at all ports at the same time \cite{Pozar}. Therefore, any practical realization of a combiner or divider will be a compromise.
In the rest of this section we will investigate several well-known power combiner/divider types. Since the passive combiner/divider is a reciprocal device, we will only consider dividers in the remaining part of this section.

\begin{figure}[hbt]
\centerline{\psfig{figure=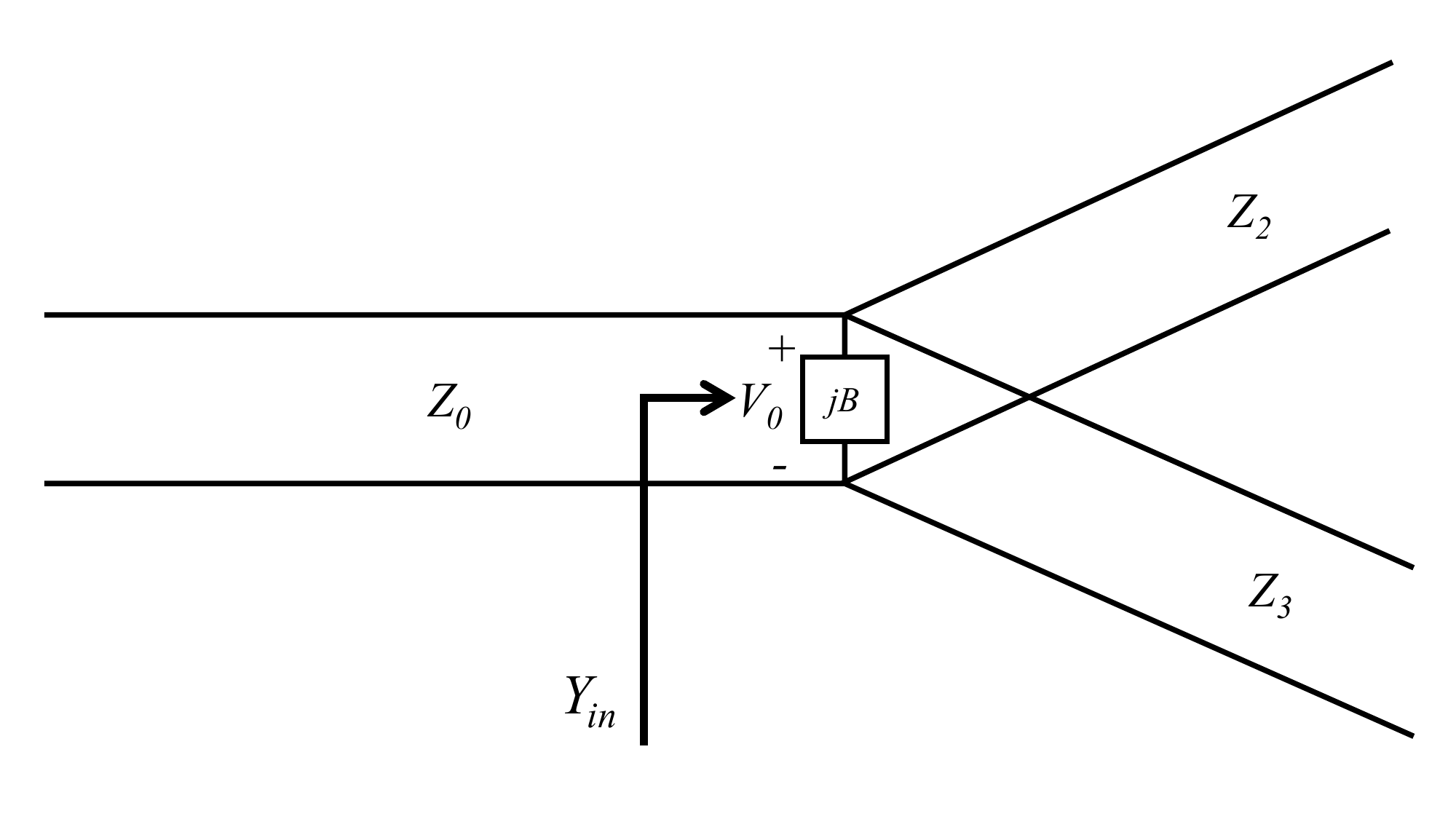,width=110mm}}
\caption{\it T-junction power divider with unequal power division ratio.}
\label{fig:Tjunction}
\end{figure}

Let us start by taking a closer look at the T-junction divider as illustrated in Fig. \ref{fig:Tjunction}. An input transmission line with characteristic impedance $Z_0$ is connected at the junction with two other transmission lines with characteristic impedance of $Z_2$ and $Z_3$, respectively. The additional susceptance $jB$ represents the fringing fields and higher-order modes that might exist at the T-junction. For the sake of simplicity, we will assume that $B=0$.
The input admittance at the input of the T-junction is now simply:
\begin{equation}
\begin{array}{lcl}
\displaystyle Y_{in} = \frac{1}{Z_2} + \frac{1}{Z_3}.
\end{array}
\end{equation}
For a matched input port it is required that $\di Y_{in}=\frac{1}{Z_0}$.
Now suppose that we would like to design an equal-split divider with $\alpha=0.5$ and $Z_0=50 \ \Omega$.
The input power delivered to the matched divider is:
\begin{equation}
\begin{array}{lcl}
\displaystyle P_{in} = \frac{V^2_0}{2 Z_0}.
\end{array}
\end{equation}
The corresponding output powers are given by:
\begin{equation}
\begin{array}{lcl}
\displaystyle P_{2} = \frac{V^2_0}{2 Z_2} = \alpha P_{in}=\frac{1}{2} P_{in}, \\
\displaystyle P_{3} = \frac{V^2_0}{2 Z_3} = (1-\alpha) P_{in}=\frac{1}{2} P_{in}.
\end{array}
\end{equation}
Both equations can be satisfied with the following choice for the characteristic impedances:
\begin{equation}
\begin{array}{lcl}
\displaystyle Z_{2} = 2 Z_0 = 100 \ \Omega, \\
\displaystyle Z_{3} = 2 Z_0 = 100 \ \Omega .
\end{array}
\end{equation}
The input impedance is now indeed $Z_{in}= 50 \ \Omega$ because of the parallel connection of two $100 \ \Omega$ impedances. However, the output ports are not matched and have a quite poor reflection coefficient:
\begin{equation}
\begin{array}{lcl}
\displaystyle \Gamma_{2} = \Gamma_{3}= \frac{Z_{out2} - Z_2}{Z_{out2} + Z_2} = \frac{33.3 - 100}{33.3 + 100} = -0.5,
\end{array}
\end{equation}
Note that poor output matching is not always a problem in an application. Another limitation of the simple T-junction divider is the poor isolation between the output ports, which could be a major issue in phased-array antennas.

\begin{figure}[hbt]
\centerline{\psfig{figure=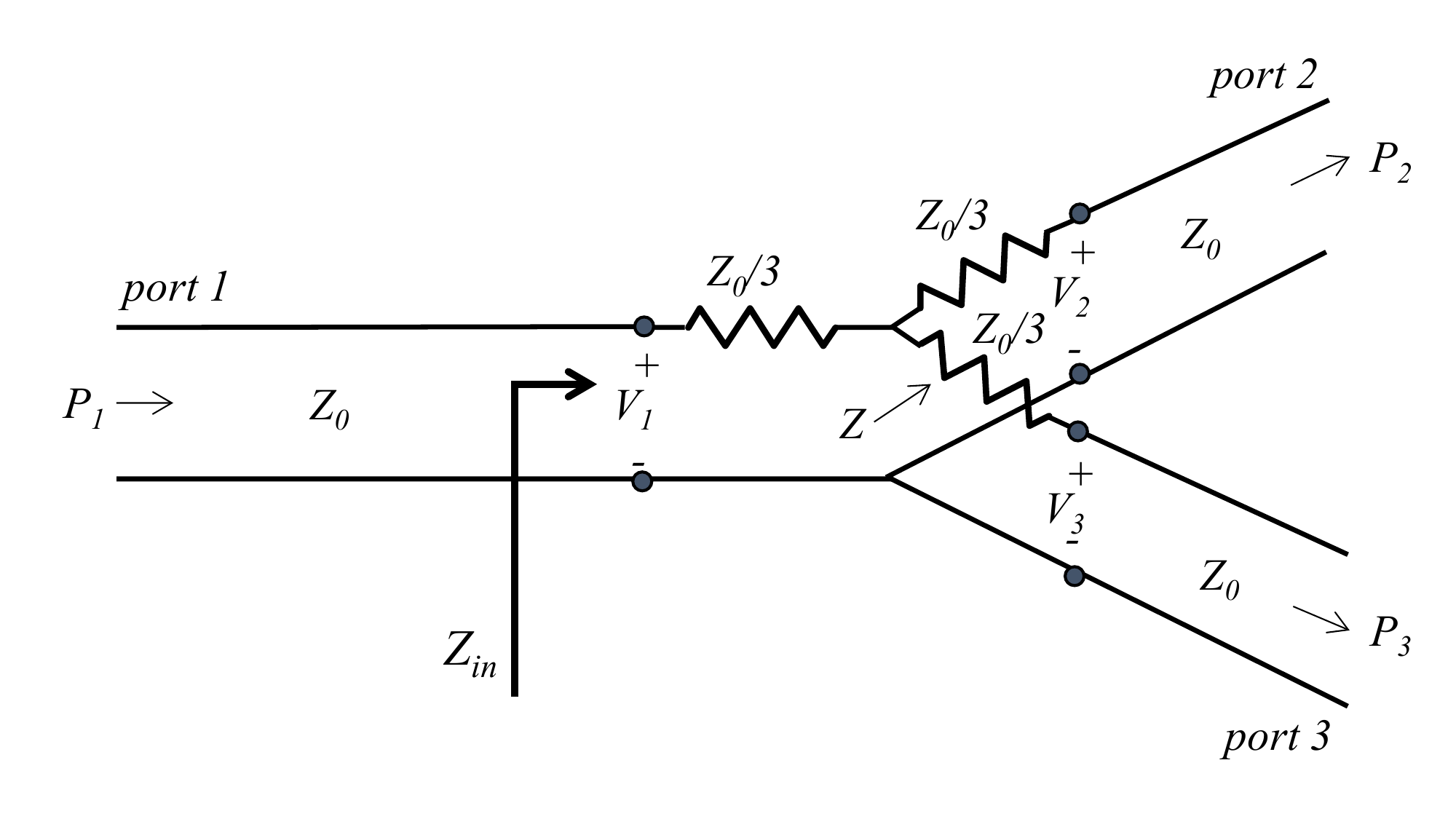,width=100mm}}
\caption{\it Resistive power divider with equal power division ratio $\alpha=0.5$}
\label{fig:Resistivedivider}
\end{figure}

A way to improve the poor output matching and poor isolation is by adding a star connection of lumped-element resistors at the junction, as illustrated in Fig. \ref{fig:Resistivedivider}.
In this case, an equal-split divider is created.
The impedance $Z$ looking via the $Z_0/3$ resistor into port 2 is given by:
\begin{equation}
\begin{array}{lcl}
\displaystyle Z = \frac{Z_0}{3}+Z_0 = \frac{4 Z_0}{3}.
\end{array}
\end{equation}
From the input port we then obtain a $Z_0/3$ resistor in series with a parallel connection of $Z$ at both output ports:
\begin{equation}
\begin{array}{lcl}
\displaystyle Z_{in} = \frac{Z_0}{3}+\frac{1}{2} \left( \frac{4 Z_0}{3} \right) = Z_0.
\end{array}
\end{equation}
As a result, the input is well matched. Due to symmetry, the two output ports are also matched. Therefore, all diagonal components of the scattering matrix $[S]$ are zero: $S_{11}=S_{22}=S_{33}=0$.
The other components of the scattering matrix can be found by determining the relation between the input voltage $V_1$ and the output voltages $V_2$ and $V_3$. Since all ports are matched, the total voltage is equal to the voltage amplitude of the TEM-wave travelling in the $+z$ direction, see also (\ref{eq:abportk}). The voltage at the central node is now easily found by using circuit theory:
\begin{equation}
\begin{array}{lcl}
\displaystyle V = V_1 \frac{2 Z_0/3}{Z_0/3+2 Z_0/3} = \frac{2 V_1}{3}.
\end{array}
\end{equation}
The corresponding output votages at port 2 and 3 then become:
\begin{equation}
\begin{array}{lcl}
\displaystyle V_2 = V_3= V \frac{Z_0}{Z_0+Z_0/3} = \frac{3 V}{4}= \frac{V_1}{2}.
\end{array}
\end{equation}
As a result, the three-port scattering matrix takes the following form:
\begin{equation}
\displaystyle [S] =  \left[ \begin{array}{l} S_{11} \ \ \ \ S_{21} \ \ \ \  S_{31} \\ S_{21} \ \ \ \ S_{22} \ \ \ \ S_{23} \\
S_{31} \ \ \ \ S_{32} \ \ \ \ S_{33} \end{array} \right]
= \frac{1}{2} \left[ \begin{array}{l} 0 \ \ \ \ 1 \ \ \ \  1 \\ 1 \ \ \ \ 0 \ \ \ \ 1 \\
1 \ \ \ \ 1 \ \ \ \ 0 \end{array} \right] .
\label{eq:Sresistivedivider}
\end{equation}
From (\ref{eq:Sresistivedivider}) we observe that $S_{21}= S_{31} = S_{23} = 1/2$, which corresponds to a power level of $-6$ dB with respect to the input power level. The output power at each of the output ports $\di P_2=P_3=\frac{1}{4} P_{in}$. Therefore, $50 \%$  of the input power is dissipated in the resistors. This is a major drawback in most microwave applications. However, in integrated circuits (ICs) resistive dividers could be useful, due to their small size. The power loss in such a circuit is not always a big issue since it can be compensated by adding additional amplification stages. Note that the isolation between the output ports of the resistive divider is also quite poor.

\begin{figure}[hbt]
\vspace{-1cm}
\centerline{\psfig{figure=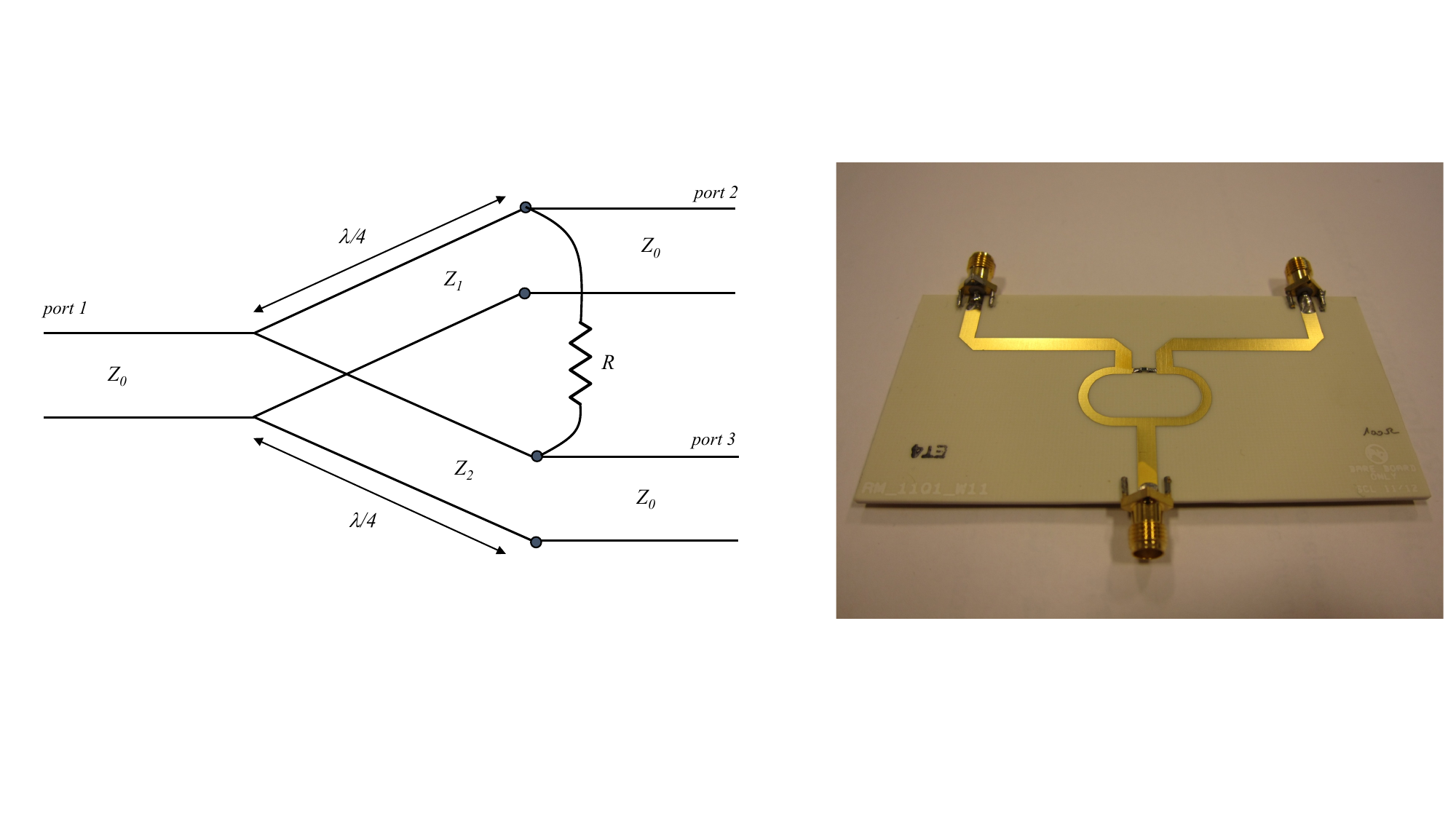,width=170mm}}
\vspace{-2cm}
\caption{\it Equal split Wilkinson power divider, $Z_1=Z_2=\sqrt{2} Z_0$ and $R=2 Z_0$}
\label{fig:Wilkinsonequalsplit}
\end{figure}

\begin{figure}[hbt]
\centerline{\psfig{figure=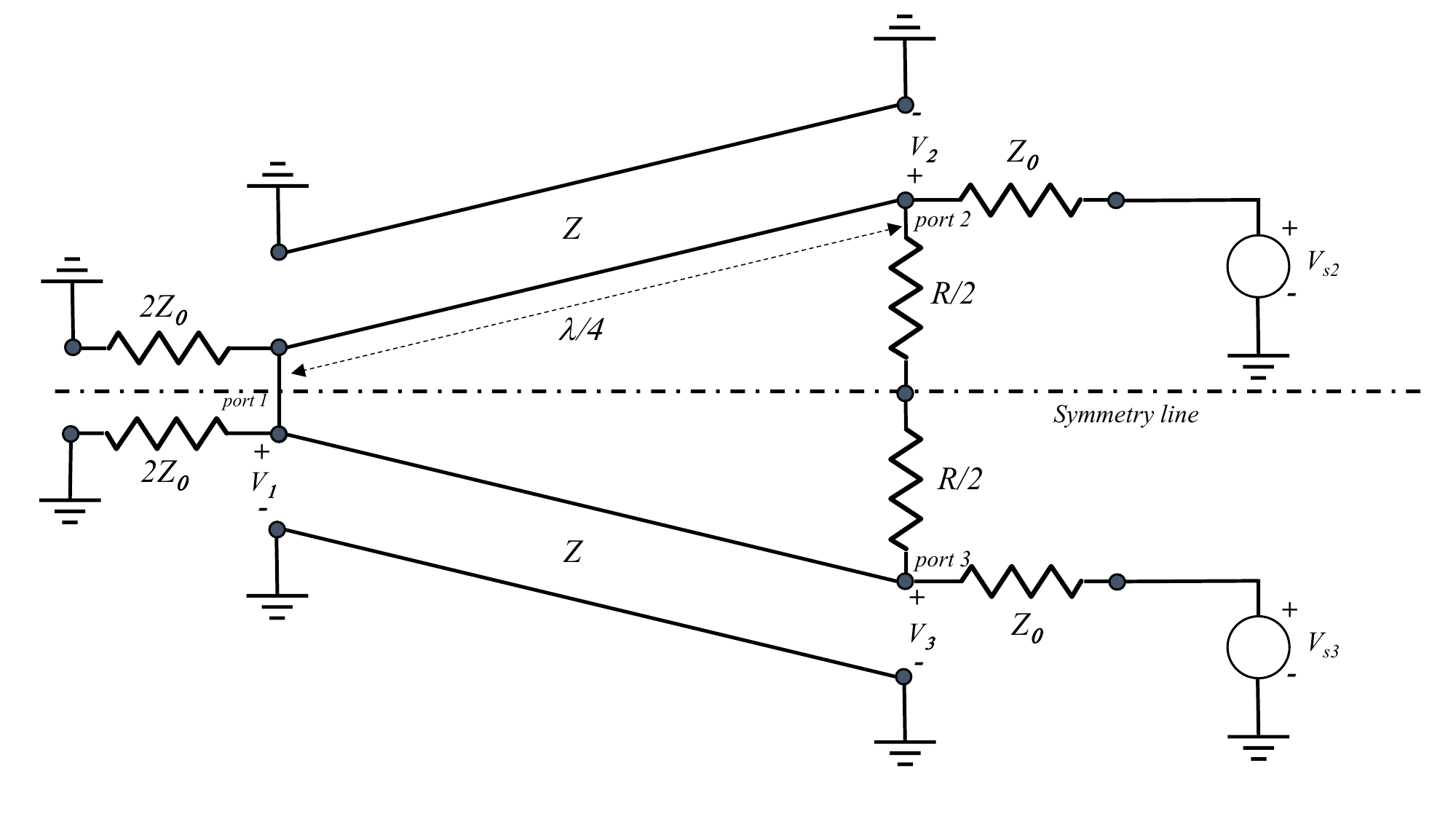,width=120mm}}
\caption{\it Symmetrical representation of the equal-split Wilkinson power divider.}
\label{fig:EvenOddmode1}
\end{figure}

So far we have investigated power dividers with major limitations in terms of matching, isolation or power dissipation.
A power divider which combines excellent matching properties, high isolation and low loss is the Wilkinson power divider.
Wilkinson dividers can be made for any division ratio, but we will only investigate the equal split case $\alpha=0.5$ in more detail in this section. Fig. \ref{fig:Wilkinsonequalsplit} shows the transmission line circuit of the equal-split Wilkinson power divider and a realization in microstrip technology. It consists of a T-junction with two quarter-wave transmission lines with $Z_1=Z_2=\sqrt{2} Z_0$ which are connected via a lumped-element resistor $R=2 Z_0$.
The Wilkinson divider can be analyzed by using the {\it even-odd mode} technique. In fact, we will excite the output ports with either the same voltage $2V_s$ (even mode) and with opposite voltage $2V_s$ and $-2V_s$ (odd mode). By using superposition, we then obtain the total voltage at the input port. From the voltages, we can determine the scattering matrix. The even-odd mode analysis starts by creating a new drawing that illustrates the symmetry of the Wilkinson power divider as shown in Fig. \ref{fig:EvenOddmode1}. All ports are terminated with a load impedance of $Z_0$. Furthermore, we will assume that $Z=\sqrt{2} Z_0$ and $R=2 Z_0$. Lateron, we will show that with these particular values the best performance is obtained.

\begin{figure}[hbt]
\centerline{\psfig{figure=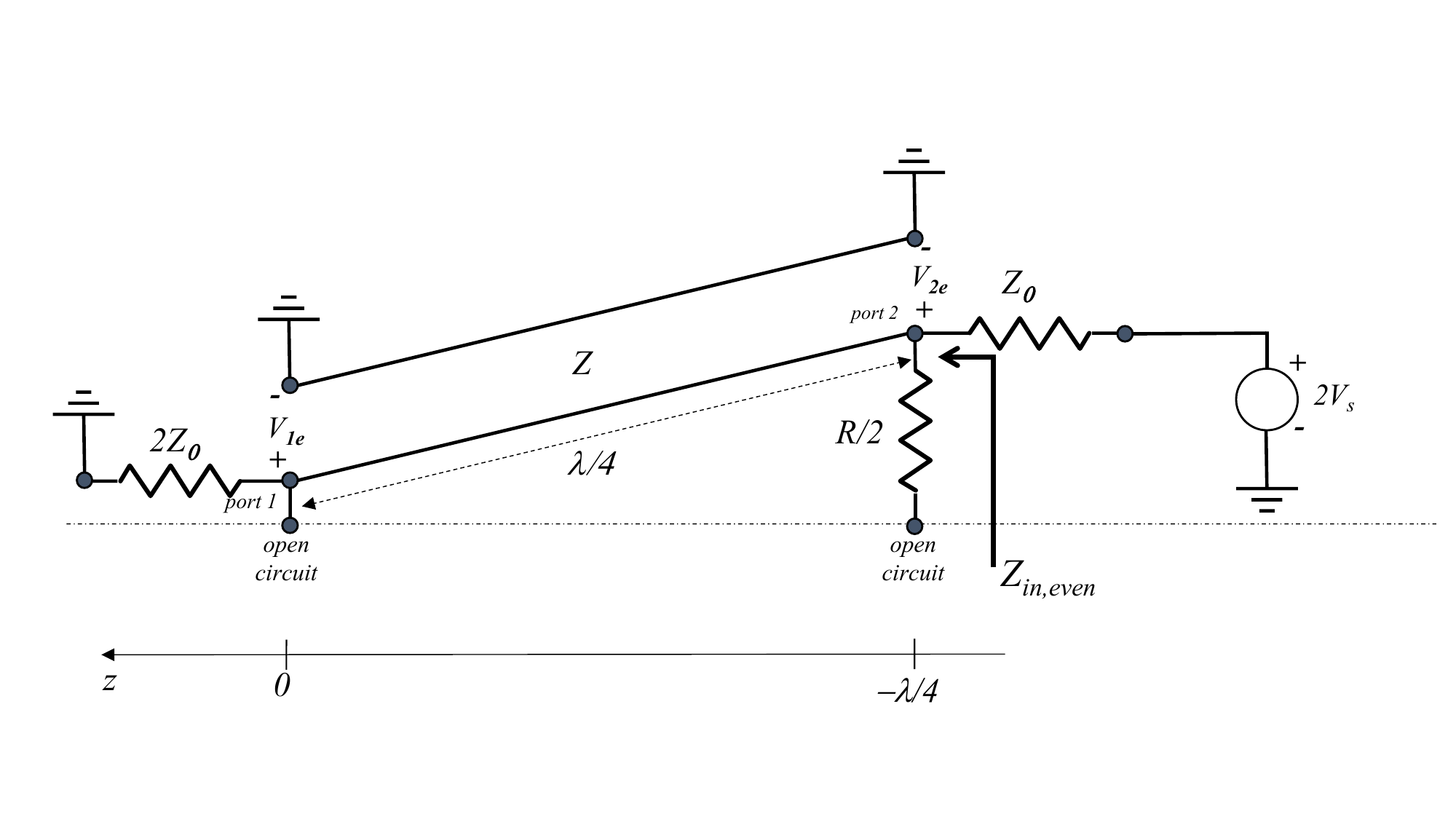,width=120mm}}
\vspace{-1cm}
\caption{\it Even-mode equivalent circuit of the equal-split Wilkinson power divider.}
\label{fig:EvenOddmode2}
\end{figure}

\vspace{0.5cm}
{\it Even-mode analysis} \\
In this case the source voltages at port 2 and port 3 are equal with $V_{s2}=V_{s3}=2V_s$.
Due to the symmetry in Fig. \ref{fig:EvenOddmode1}, no current will flow through the resistors $R/2$, which means that the symmetry line can be represented as an open-circuit. As a result, we can limit our analysis to the circuit shown in Fig. \ref{fig:EvenOddmode2}.
The resistor $R/2$ can now be neglected. The input impedance looking into port 2 is now easily found using the quarter-wave transformer equation (\ref{eq:Lambda4transformerZin}):
\begin{equation}
\begin{array}{lcl}
\displaystyle Z_{in,even}=\frac{Z^2}{2Z_0}=\frac{(\sqrt{2} Z_0)^2}{2Z_0}=Z_0.
\end{array}
\end{equation}
The voltage along the lossless $\lambda/4$ line with phase constant $\beta$ is found using:
\begin{equation}
\displaystyle V(z)=V_0^+ \left( e^{-\jmath \beta z} + \Gamma e^{\jmath \beta z} \right).
\end{equation}
As a result, the port voltages are given by:
\begin{equation}
\begin{array}{lcl}
\displaystyle V_{2e} &=& \di V(z=-\lambda/4) = \jmath V_0^+ \left( 1 - \Gamma \right) = V_s, \\
\di V_{1e} &= & \di V (0)= V_0^+ \left( 1 + \Gamma \right) = \jmath V_s \frac{\Gamma +1}{\Gamma -1},
\end{array}
\end{equation}
where $\Gamma$ is the reflection coefficient looking at port 1 into the load impedance $2Z_0$:
\begin{equation}
\begin{array}{lcl}
\displaystyle \Gamma &=& \di  \frac{2Z_0 - \sqrt{2} Z_0}{2Z_0 + \sqrt{2} Z_0}=\frac{2 - \sqrt{2} }{2 + \sqrt{2}}.
\end{array}
\end{equation}
The voltage at port 1 now becomes
\begin{equation}
\begin{array}{lcl}
\displaystyle V_{1e} &= & \di - \jmath V_s \sqrt{2}.
\end{array}
\end{equation}

\begin{figure}[hbt]
\vspace{-1cm}
\centerline{\psfig{figure=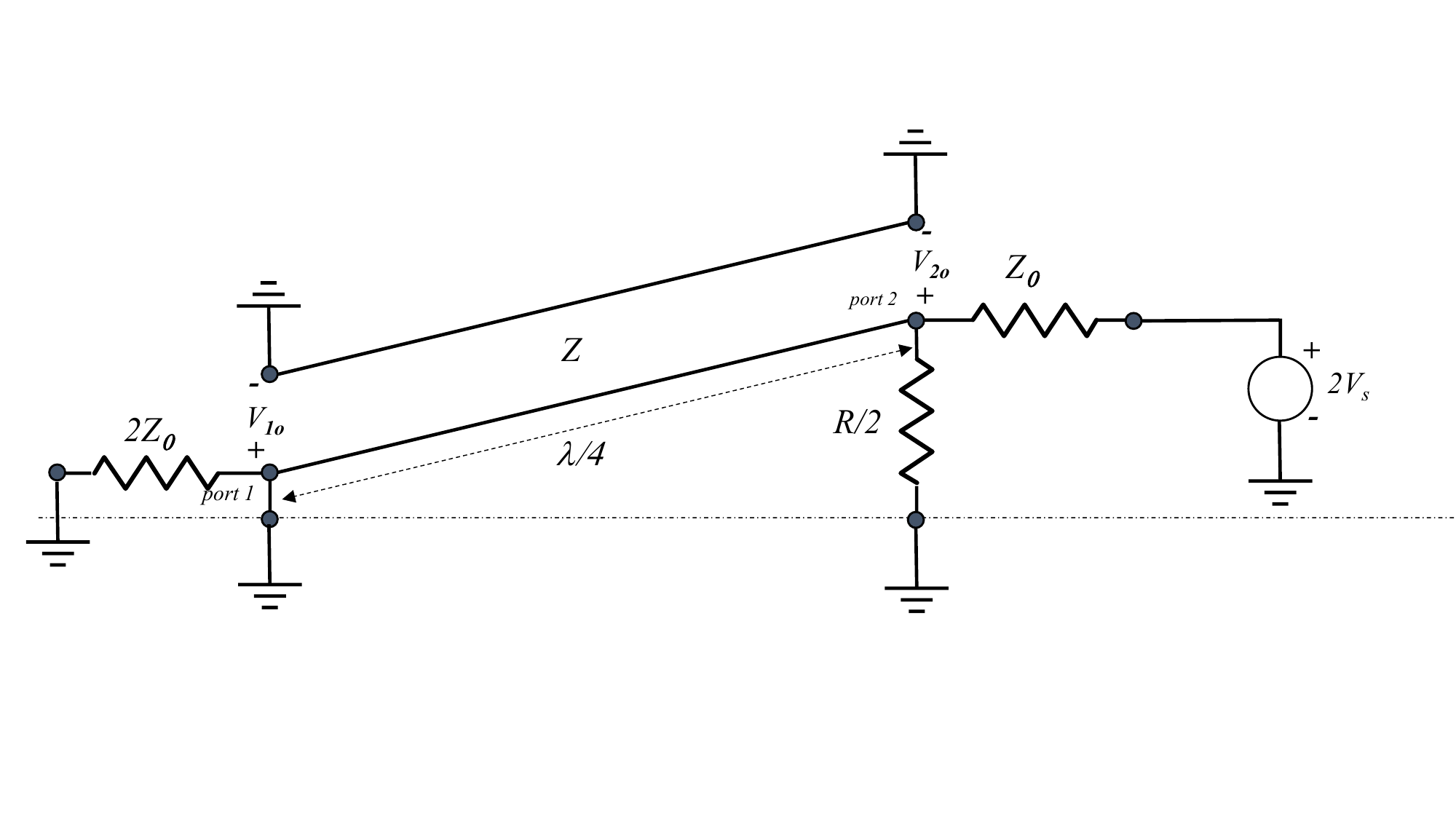,width=120mm}}
\vspace{-2cm}
\caption{\it Odd-mode equivalent circuit of the equal-split Wilkinson power divider.}
\label{fig:EvenOddmode3}
\end{figure}

\vspace{0.5cm}
{\it Odd-mode analysis} \\
The source voltages at port 2 and port 3 now have equal amplitudes but opposite sign, $V_{s2}=2V_s$ and $V_{s3}=-2V_s$. Due to the asymmetry, the voltage along the symmetry line is zero as illustrated in the equivalent circuit of Fig. \ref{fig:EvenOddmode3}.
The quarter-wave transformer now transforms the short at the end of the line to an open circuit at the start of the line at port 2. The remaining circuit is a voltage divider. Therefore, the voltages at port 1 and port 2 are given by:
\begin{equation}
\begin{array}{lcl}
\di V_{1o} &=& 0, \\
\displaystyle V_{2o} &= & \di \frac{1}{2} (2 V_s)=V_s.
\end{array}
\end{equation}

\vspace{0.5cm}
{\it Total solution: superposition of even-odd modes} \\
By applying superposition of the even- and odd-mode analysis we obtain the total solution when exciting port 2. In a similar way, the solution when exciting port 3 can be found.
In both the even and odd mode, port 2 is matched. Therefore $S_{22}=S_{33}=0$. In addition, it can be shown that with $Z=\sqrt{2} Z_0$ port 1 is also matched $S_{11}=0$.
The transfer functions when all ports are matched are now found by superimposing the port voltages:
\begin{equation}
\begin{array}{lcl}
\di S_{12} &=& \di S_{21} =  \frac{V_{1e}+V_{1o}}{V_{2e}+V_{2o}}= \frac{- \jmath}{\sqrt{2}}, \\
\di S_{13} &=& \di S_{31} =  \frac{- \jmath}{\sqrt{2}}, \\
\di S_{23} &=& \di S_{32} =0.
\end{array}
\end{equation}
The ideal Wilkinson divider has no losses since $|S_{21}|^2=|S_{12}|^2=1/2$, so when used as a divider, $50\%$ of the power is directed to each of the output ports. Furthermore, the isolation between the output ports is perfect since $S_{23} = \di S_{32} =0$. When we use it as a combiner to combine input signals from port 2 and port 3, the ideal Wilkinson combiner will also be lossless as long as both signals are identical, both in phase and amplitude. Any unbalance between the input signals will be dissipated in the resistor.
In a practical realization of the Wilkinson divider, there will be losses due to metal losses, dielectric losses and spurious radiation. In addition, the isolation ($20 \log_{10} |S_{32}|$) will be limited, typically to a value between -20 and -30 dB.

\section{Impedance matching and tuning}
\label{sec:Impedancematching}
\begin{figure}[hbt]
\vspace{-0.5cm}
\centerline{\psfig{figure=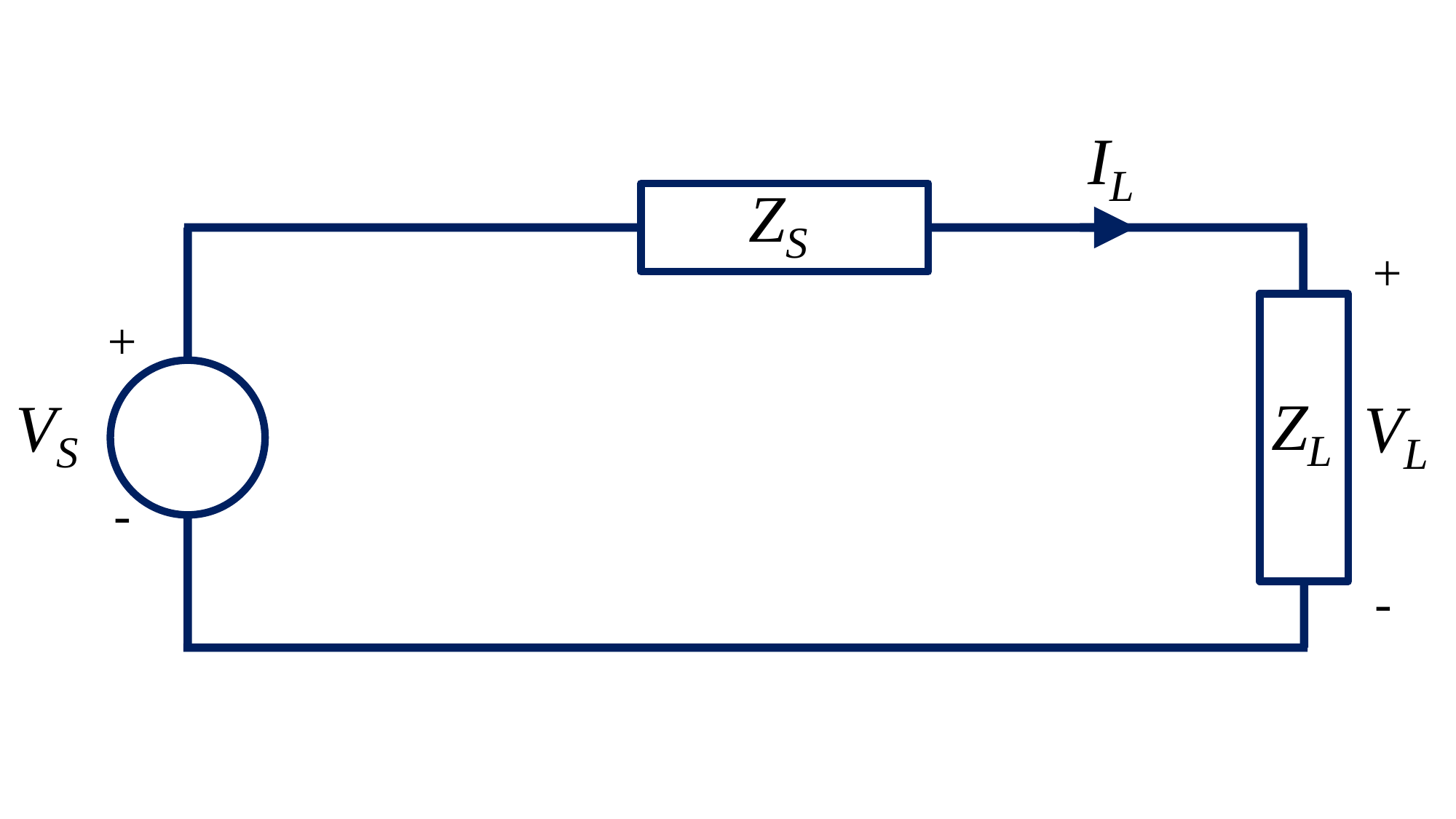,width=80mm}}
\vspace{-1cm}
\caption{\it Conjugate matching of a source impedance to a complex load impedance.}
\label{fig:conjugatematching1}
\end{figure}

\begin{figure}[hbt]
\centerline{\psfig{figure=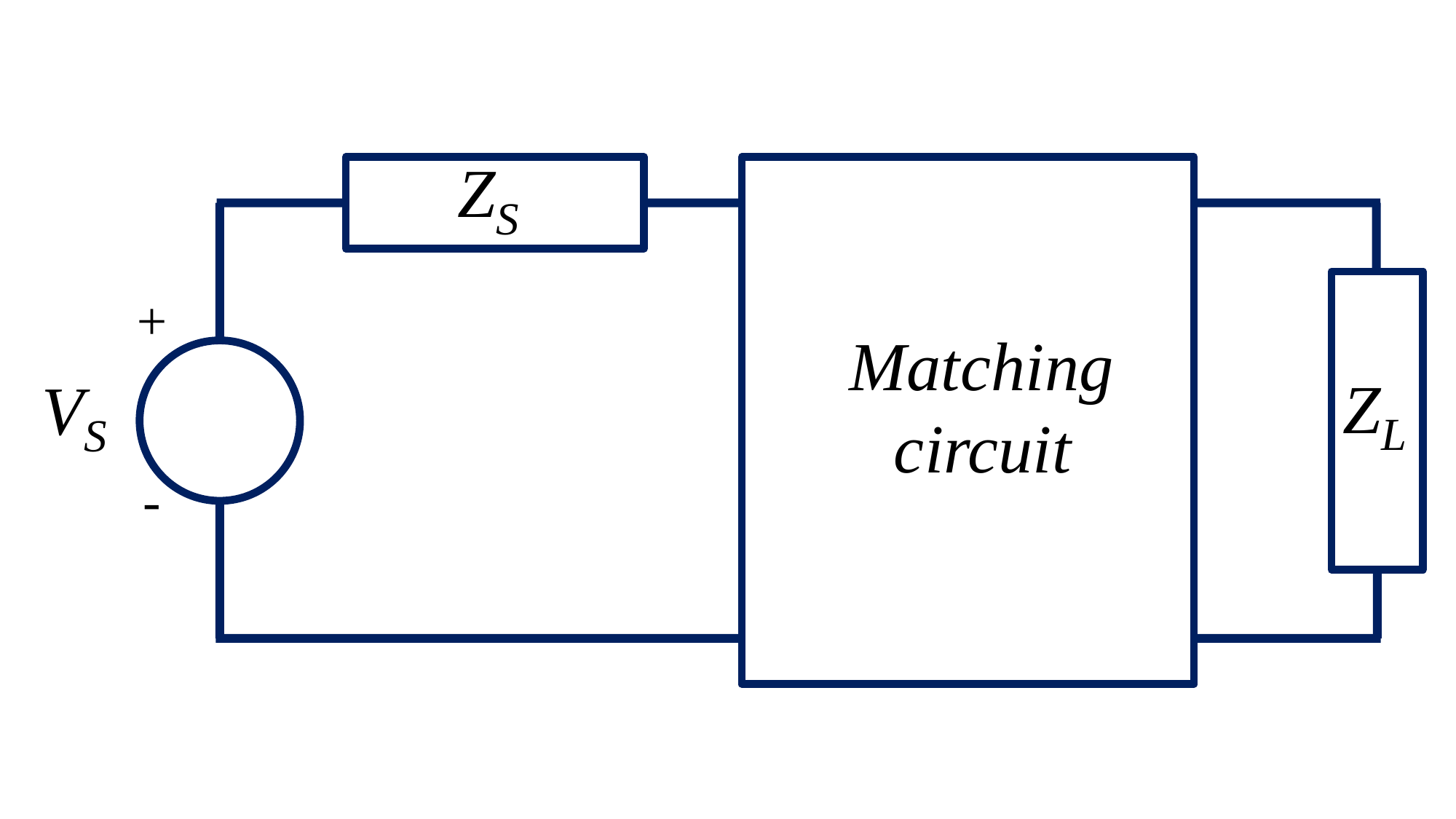,width=80mm}}
\vspace{-0.5cm}
\caption{\it Matching circuit to obtain a good matching between the source and load.}
\label{fig:Matchingcircuit}
\end{figure}

Consider the circuit of Fig. \ref{fig:conjugatematching1} where a source with complex source impedance $Z_S=R_S+\jmath X_S$ is connected to a complex load $Z_L=R_L+\jmath X_L$. The source could be a power amplifier and the load could be an antenna. The time-average power delivered to the load is found by:
\begin{equation}
\begin{array}{lcl}
\di P_{L} &=& \di \frac{1}{2} Re \left\{ V_L I^*_L \right\}.
\end{array}
\label{eq:PLoad}
\end{equation}
where
\begin{equation}
\begin{array}{lcl}
\di V_{L} &=& \di \frac{V_S Z_L}{Z_S+Z_L} , \\
\di I_{L} &=& \di \frac{V_S}{Z_S+Z_L}.
\end{array}
\label{eq:VLandIL}
\end{equation}
When substituting (\ref{eq:VLandIL}) in (\ref{eq:PLoad}) we obtain:
\begin{equation}
\begin{array}{lcl}
\di P_{L} &=& \di \frac{1}{2} Re \left\{ \frac{|V_S|^2 Z_L}{|Z_S+Z_L|^2} \right\}=
\frac{1}{2} \frac{|V_S|^2 R_L}{(R_S+R_L)^2+(X_S+X_L)^2}.
\end{array}
\label{eq:PLoad2}
\end{equation}
The maximum delivered power is now found by determining the value of $(R_L, X_L)$ for which the partial derivatives are equal to zero:
\begin{equation}
\begin{array}{lcl}
\di \frac{\partial P_{L}}{\partial R_L} &=& \di \frac{\partial P_{L}}{\partial X_L}=0.
\end{array}
\end{equation}
We find that this occurs when $R_L=R_S$ and $X_L=-X_S$, in other words when $Z_L=Z^*_S$. The maximum power transfer to the load occurs when the source and load impedances are conjugate matched to each other.
In a lot of situations direct conjugate matching of the source to load is not possible. In these cases, we can use a {\it matching circuit} as illustrated in Fig. \ref{fig:Matchingcircuit}
The matching circuit can be created using either lumped elements, like inductors (with inductance $L$ [H]) and capacitors (with capacitance $C$ [F]), or by using distributed transmission-line components. Note that lumped elements can only be used when the size of the component is much smaller as compared to the wavelength. In PCB technologies, surface-mount devices (SMD) are commonly used, with sizes down to SMD0201 (0.25 mm $\times$ 0.125 mm) and even SMD01005 (0.125 mm $\times$ 0.0675 mm). SMD components are typically used at frequencies up to 6 GHz. In semiconductor technologies, much smaller inductors and capacitors can be realized. As a result, lumped-element matching can be used up to much higher frequencies in integrated circuits.

\begin{figure}[hbt]
\centerline{\psfig{figure=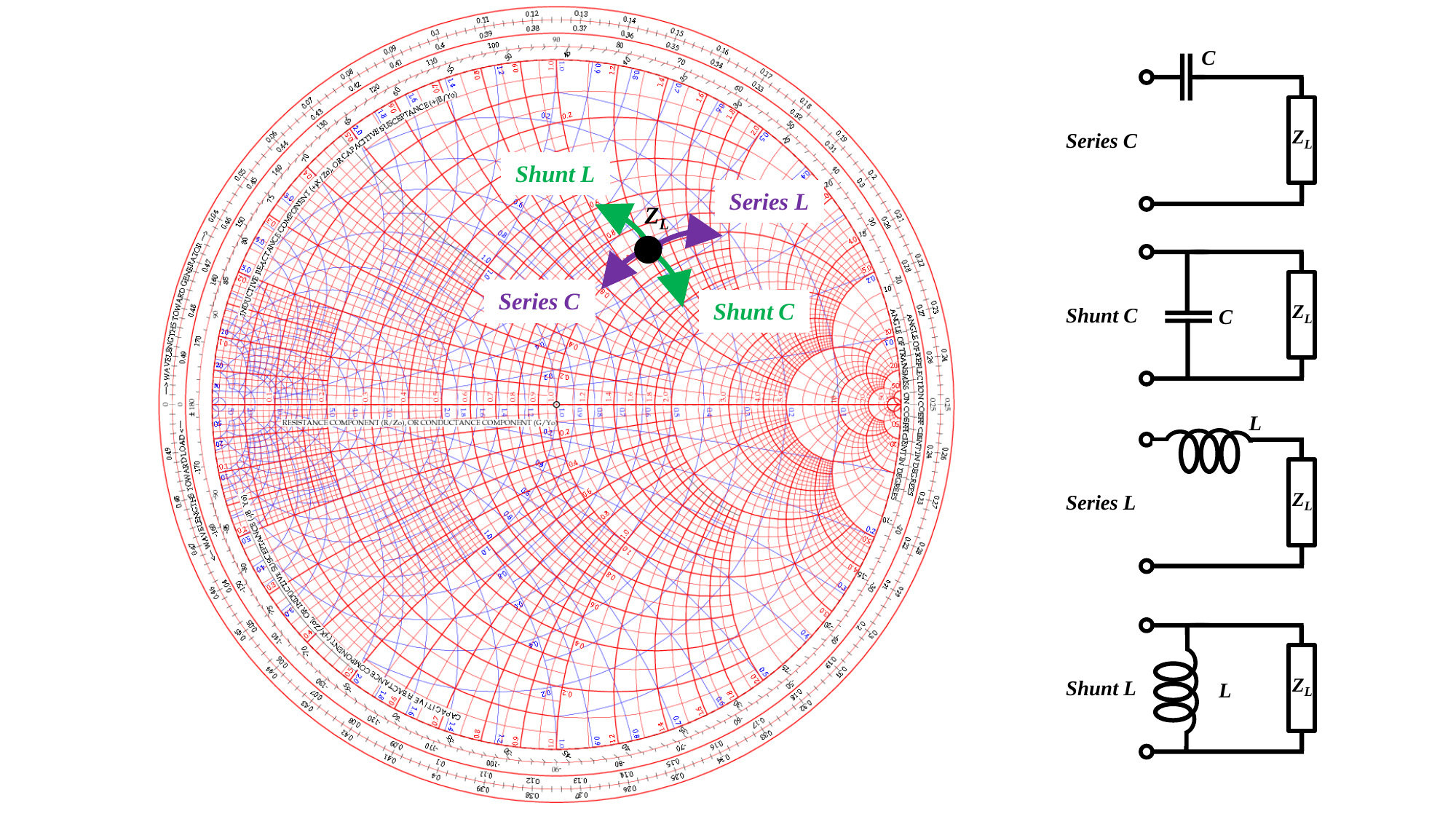,width=150mm}}
\caption{\it Lumped-element matching illustrated in a Smith chart.}
\label{fig:Smithmatching}
\end{figure}

Lumped-element matching can be realized by placing inductors and capacitors in series or parallel to the load impedance. The impedances of these components in the frequency domain are given by:
\begin{equation}
\begin{array}{lcl}
\di capacitor: \ \ Z & =& \di \frac{1}{\jmath \omega C}, \\
\di inductor: \ \ Z & =& \di \jmath \omega L,
\end{array}
\end{equation}
where $\omega=2 \pi f$.
The Smith chart is a useful tool to design such a network. Several commercial and open-source tools are available to design lumped-element matching circuits.
Suppose that we want to match a particular load impedance $Z_L$ to a source impedance $Z_S$. Fig. \ref{fig:Smithmatching} shows the degrees of freedom when placing an inductor or capacitor in series or in shunt (parallel) to the load impedance. By using several of these combinations (shunt-series) or (series-shunt), we can reach any point in the Smith chart. In addition, transmission line sections could be added. Note that ideal capacitors and inductors do not exist in practise at microwave frequencies. One has to take parasitics and losses into account. This will deteriorate the performance of a lumped-element matching circuit.

\vspace*{1cm} \hrulefill
 {\bf Exercise} \hrulefill \\
 Design a lumped-element matching circuit to match a complex load impedance $Z_L=200- \jmath 100 \ \Omega$ to a real load $Z_S=100 \ \Omega$. The frequency of operation is 1 GHz.
You can use the design tools ADS or QUCS (open-source, http://qucs.sourceforge.net/). Check you final result in a Smith chart using the strategy as shown in Fig. \ref{fig:Smithmatching}.

\begin{figure}[hbt]
\centerline{\psfig{figure=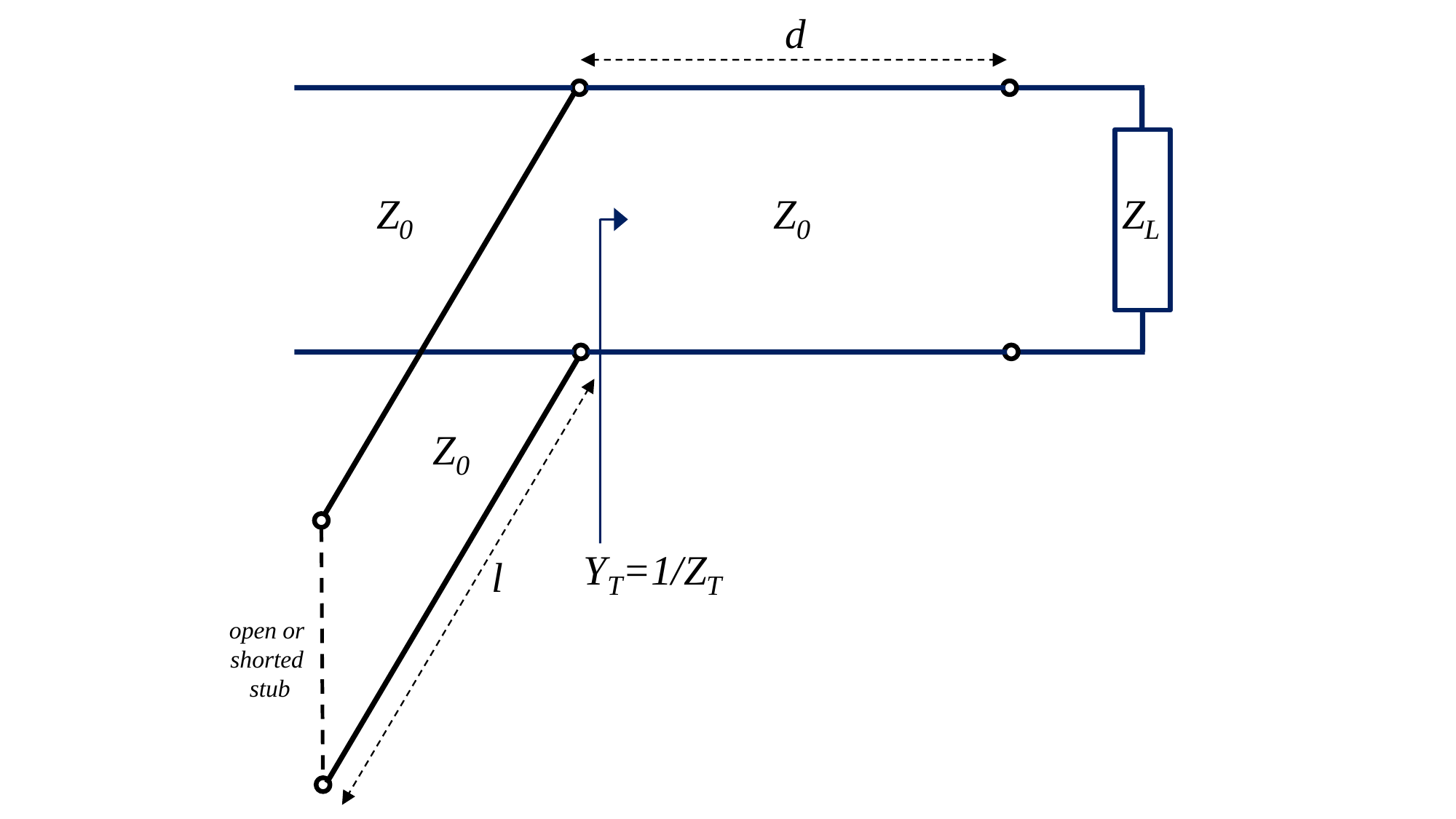,width=120mm}}
\caption{\it Single-stub matching circuit}
\label{fig:Singlestubmatching}
\end{figure}

\vspace{0.25cm}
Due to practical limitations of lumped elements at microwave frequencies, we generally prefer to use transmission-based matching circuits. When both the load and source impedance are real, we can use the quarter-wave transformer as discussed in section \ref{sec:quarterwavetransformer}.
Another matching strategy is to use stubs which are placed in parallel to a transmission-line transformer.
An example of such a circuit is shown in Fig. \ref{fig:Singlestubmatching}.
The stub is either open ended or shorted. The imput impedance is then obtained using (\ref{eq:inputimpedance}):
\begin{equation}
\begin{array}{lcl}
\di  shorted \ stub: \ \ & Z_{s}   =&  \di \jmath Z_0 \tan{\beta l}=\jmath \Omega Z_0= \jmath \Omega L, \\
\di open \ stub: \ \ & Z_{s}  =& \di \frac{Z_0}{\jmath \tan{\beta l}}=\frac{Z_0}{\jmath \Omega} = \frac{1}{\jmath \Omega C}.
\end{array}
\label{eq:shortedopenstub}
\end{equation}
Apparently, a shorted stub is the equivalent of an inductor and an open stub is the equivalent of a capacitor, with $l \leq \lambda/4$. This equivalence appears to be very useful when designing microwave filters, as we will discuss in more detail in section \ref{sec:microwavefilters}.

Impedance matching with the single-stub tuner of Fig. \ref{fig:Singlestubmatching} is now achieved as follows. Let $Y_T=1/Z_T$ be the admittance of the terminated transmission line with length $d$.
We can determine $Y_T$ using equation (\ref{eq:inputimpedance}):
\begin{equation}
\begin{array}{lcl}
\displaystyle Z_{T} &=& \di Z_0 \frac{Z_L+\jmath Z_0 \tan \beta d}{Z_0+\jmath Z_L \tan \beta d}, \\
\di Y_T & = & \di \frac{1}{Z_T} = G_T + \jmath B_T.
\end{array}
\label{eq:stubtuning}
\end{equation}
Now choose the length $d$ in such a way that $G_T=Y_0$. Furthermore, the shunt stub is designed to cancel the imaginary part of $Y_T$, so $Y_s=-\jmath B_T$.
Next to single-stub tuners, double stub tuners can be used to increase the degrees of freedom. One of the additional advantages of a double-stub tuner is that it can be designed in such a way that the line length between the two stubs can be fixed, which simplifies the construction of such a tuner.

\vspace*{1cm} \hrulefill
 {\bf Exercise} \hrulefill \\
 Determine the parameters $(l,d)$ of a single-stub tuner with a shorted stub to match a complex load impedance $Z_L=100-\jmath 60 \ \Omega$ to a $50 \ \Omega$ transmission line. Use transmission lines with $Z_0=50 \ \Omega$.
{\it Answer:} $d=0.125 \lambda$ and $l=0.118 \lambda$, where $\lambda$ is the wavelength in the transmission line.

\section{Microwave filters}
\label{sec:microwavefilters}
\begin{figure}[hbt]
\centerline{\psfig{figure=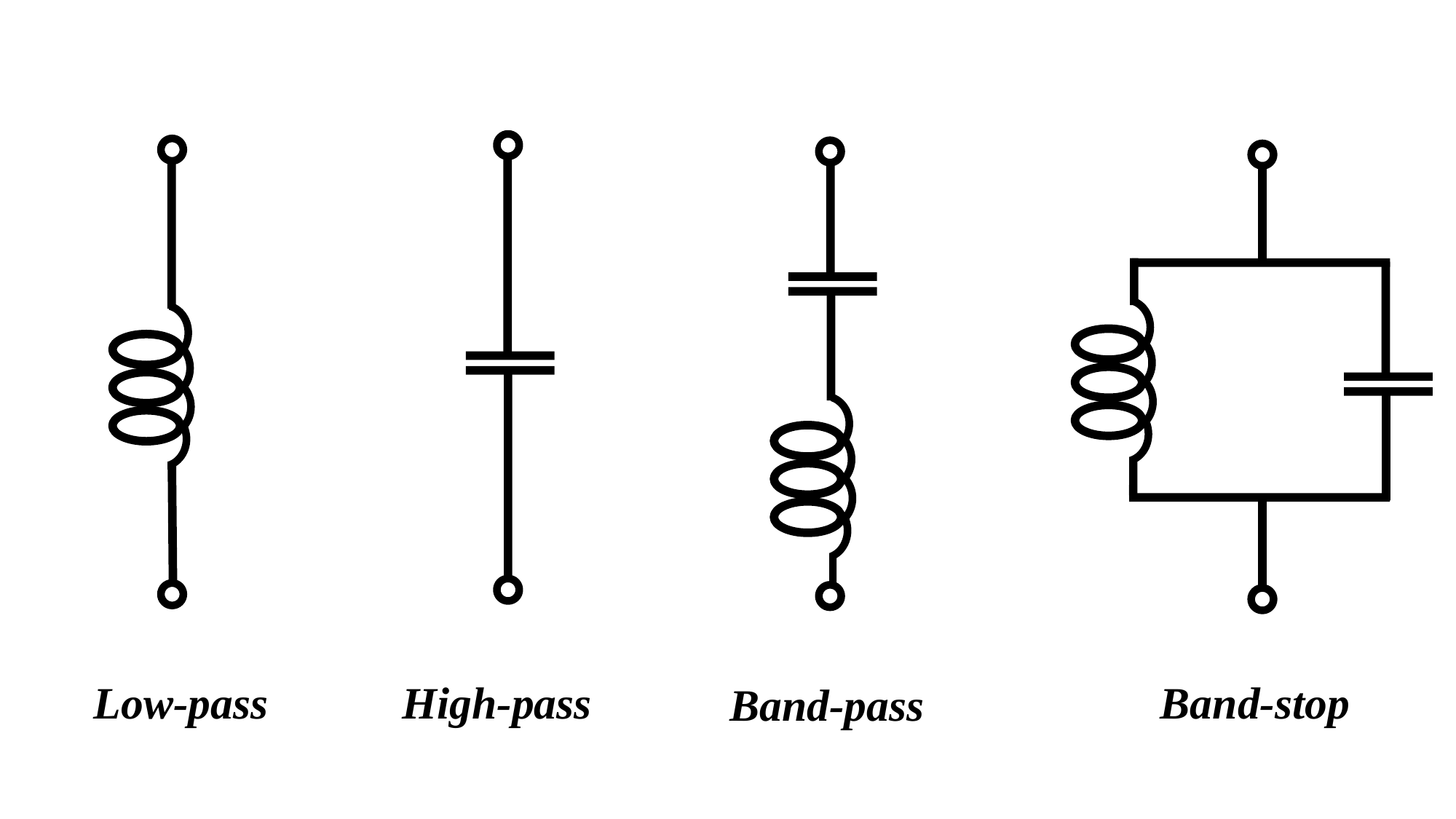,width=100mm}}
\caption{\it Lumped-element filters.}
\label{fig:Filtertypes}
\end{figure}

\begin{figure}[hbt]
\centerline{\psfig{figure=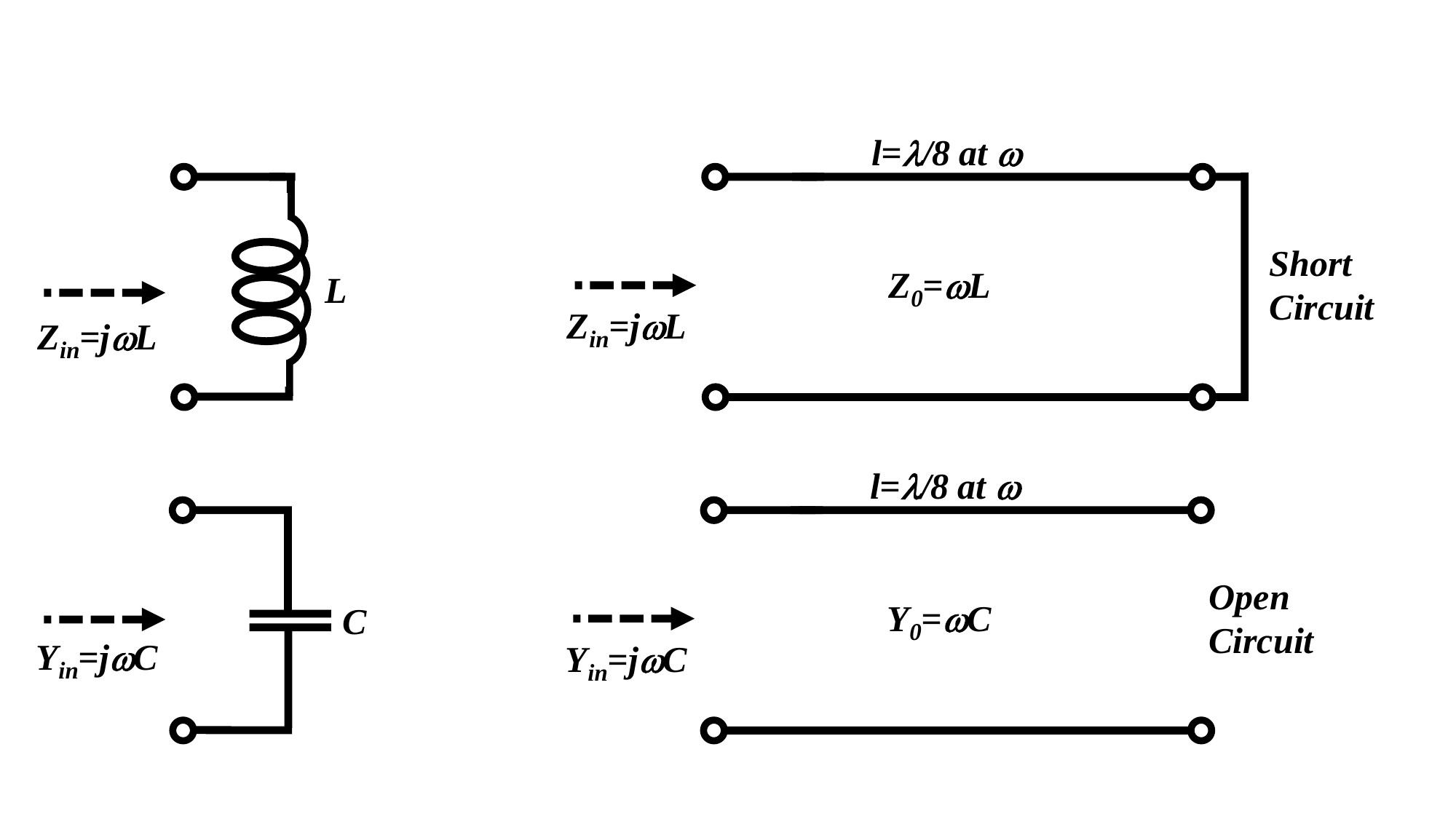,width=120mm}}
\caption{\it Equivalence of lumped-element L,C and shorted/open stubs with length of $\lambda/8$. This is also known as Richard's transformation.}
\label{fig:LCequivalence}
\end{figure}

\begin{figure}[hbt]
\centerline{\psfig{figure=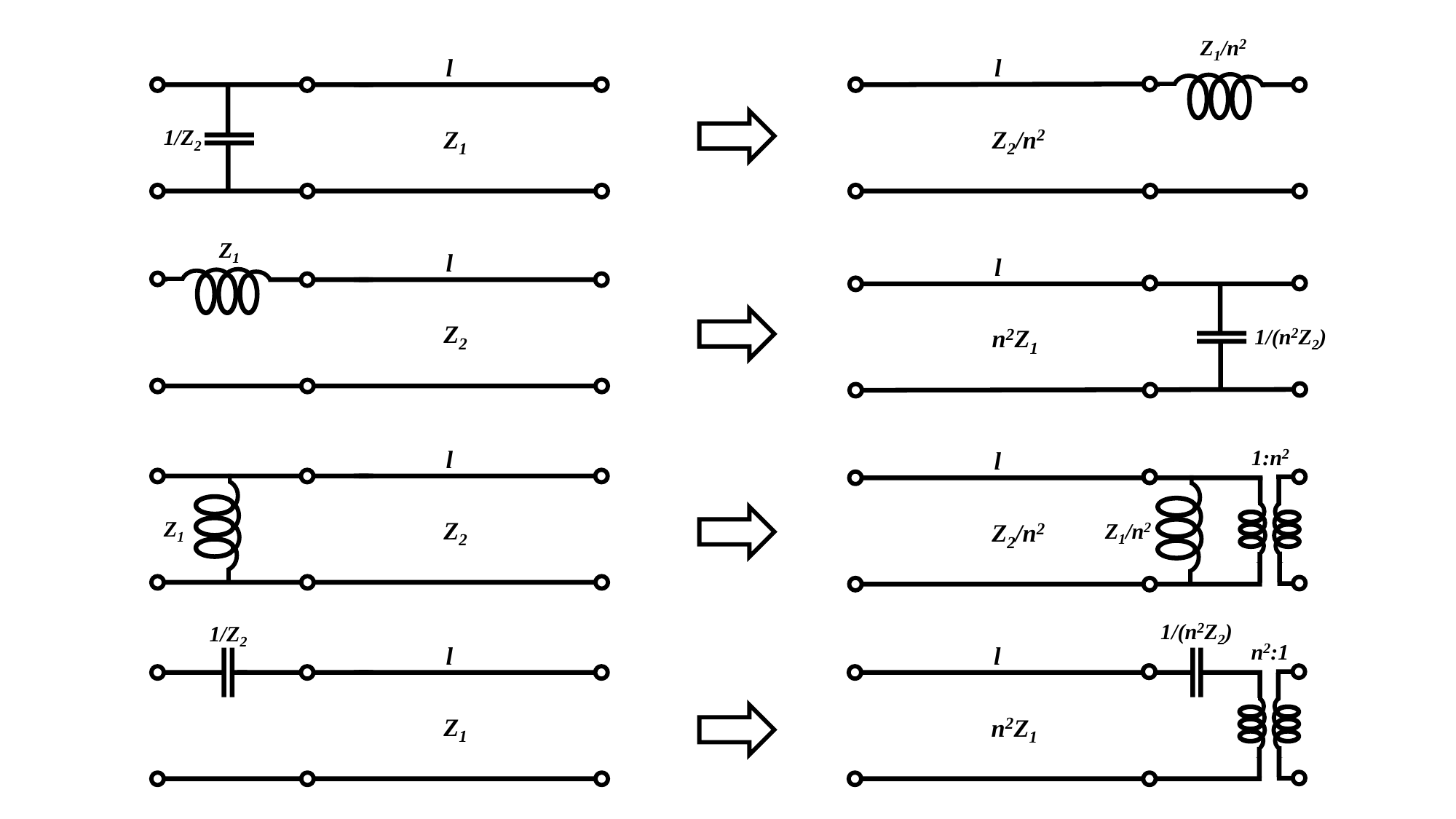,width=140mm}}
\caption{\it Kuroda's identities. $\di n^2=1+\frac{Z_2}{Z_1}$, $l=\lambda/8$.}
\label{fig:Kuroda}
\end{figure}

\begin{figure}[hbt]
\centerline{\psfig{figure=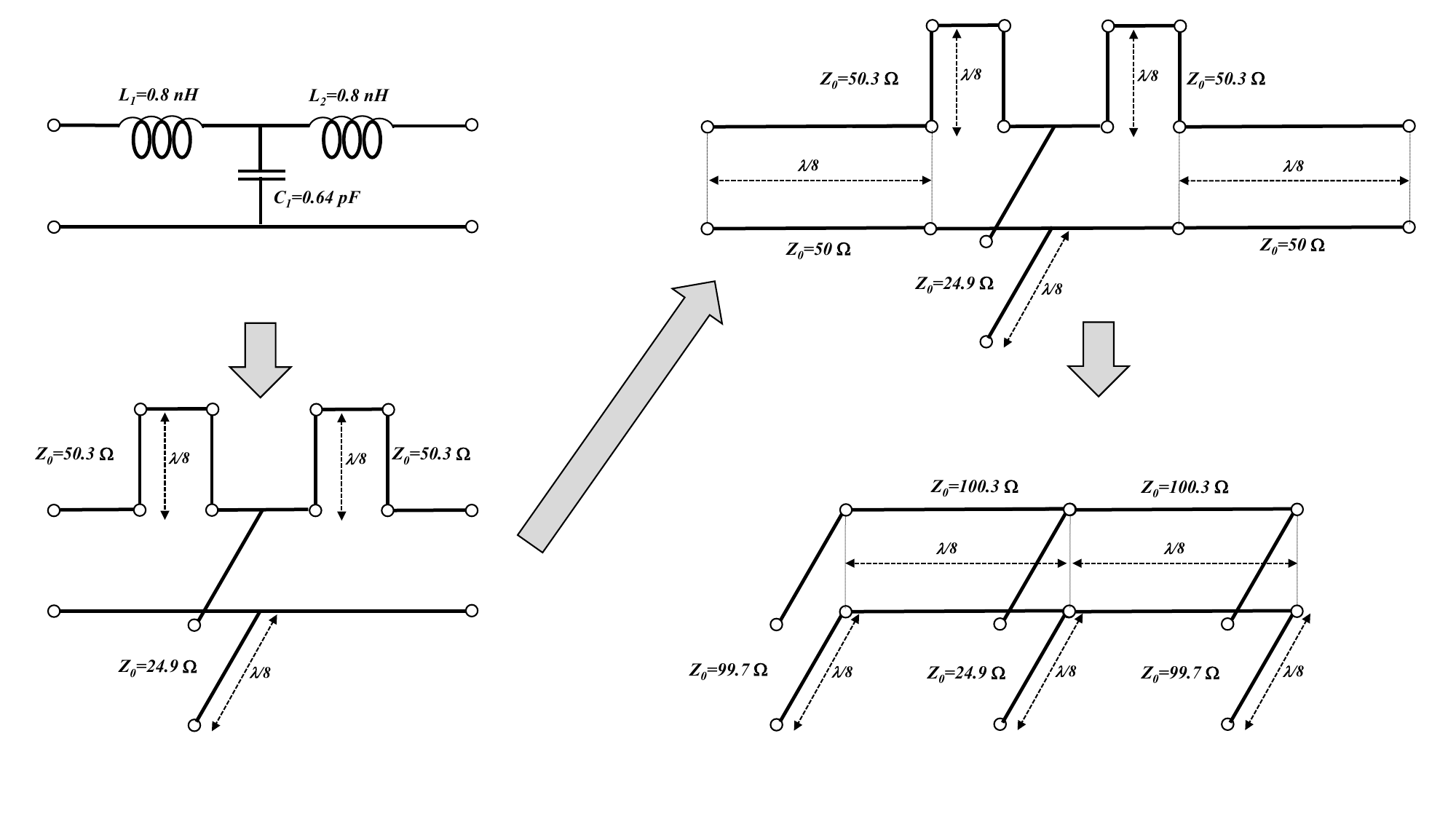,width=160mm}}
\vspace{-1cm}
\caption{\it Design of a transmission-line low-pass filter in 4 steps using Richard's transformation and Kuroda's identities with $n^2=1.995$. Maximally-flat filter with $N=3$, $R_0=50 \Omega$ and $f_{-3 dB}=10$ GHz.}
\label{fig:Lowpassdesign}
\end{figure}

\begin{figure}[hbt]
\centerline{\psfig{figure=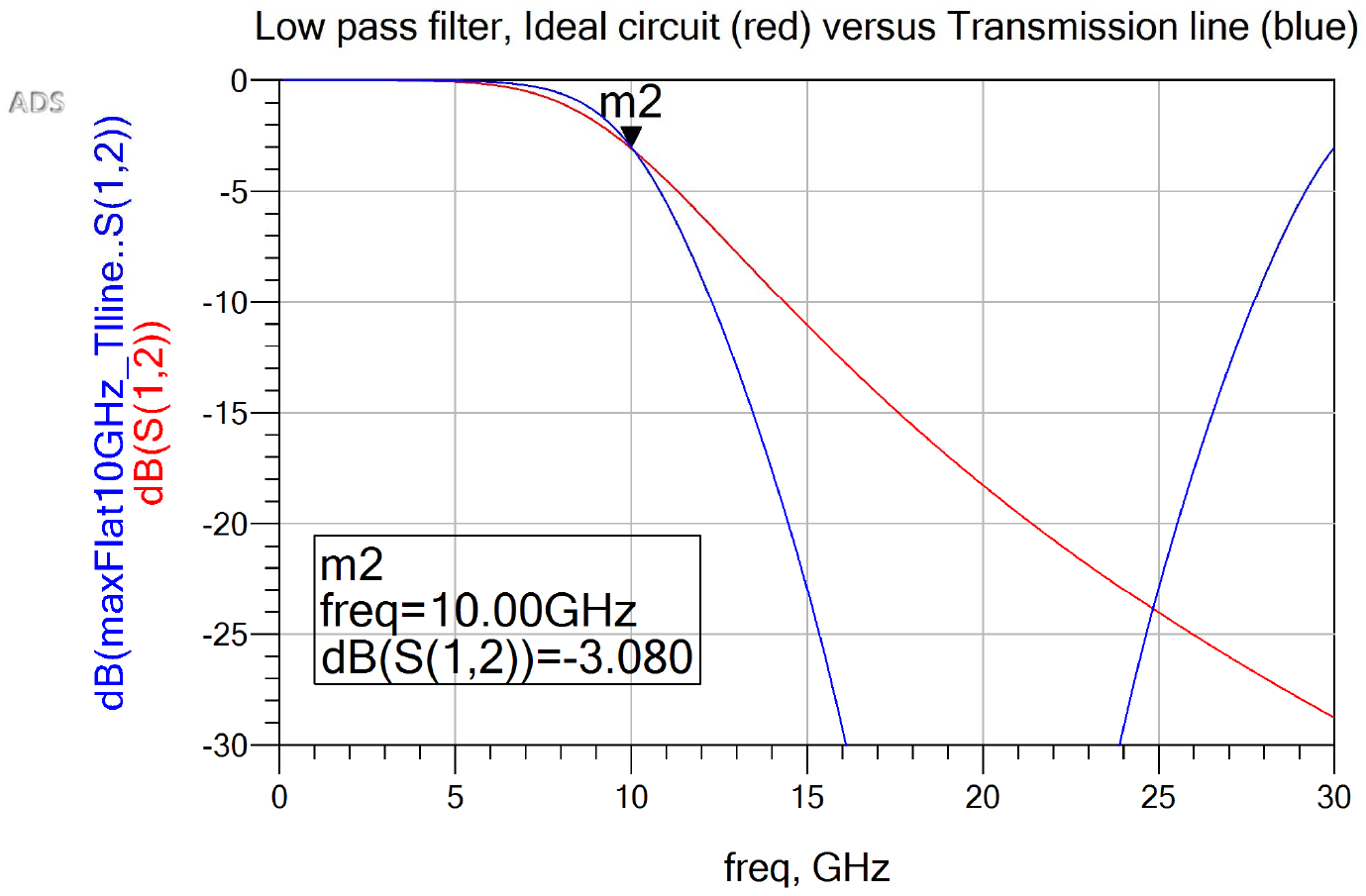,width=120mm}}
\vspace{-1cm}
\caption{\it Simulated performance of a low-pass filter, ideal lumped-element model (red) and transmission-line equivalent (blue).}
\label{fig:SimADSlowpass}
\end{figure}

Microwave filters can be used to manipulate the frequency response of a microwave system. Microwave filters are constructed using transmission line components. At lower frequencies ($<5$ GHz) lumped elements (L, C) can be used as well. Well-known filter characteristics are low-pass, high-pass, bandpass or bandstop. Fig. \ref{fig:Filtertypes} illustrates these four types using lumped elements. There is a lot of literature available for designing lumped-element filters \cite{Pozar}, \cite{CollinME}. In this section, we will assume that such an ideal lumped-element prototype filter is already known. We will show how such a lumped-element filter can be transformed into a microwave filter using transmission line components.
In the previous section in (\ref{eq:shortedopenstub}) we have already shown that shorted or open stubs are equivalent to inductors and capacitors. With the particular choice that the length of the stub $l=\lambda/8$ we find that $\di \Omega=\tan{\beta l} =\tan{\left( \frac{2\pi}{\lambda} \frac{\lambda}{8} \right)}=\tan{\frac{\pi}{4}}=1$ which results in:
\begin{equation}
\begin{array}{lcl}
\di  shorted \ stub: \ \ & Z_{s}  = \di \jmath Z_0 = &  \di \jmath L, \\
\di open \ stub: \ \ & Z_{s}  = \di  \frac{Z_0}{\jmath} = & \di \frac{1}{\jmath C},
\end{array}
\label{eq:shortedopenstublambda8}
\end{equation}
which resembles the equivalence of an inductor with inductance $L$ and capacitor with capacitance $C$ at $\omega =1$, respectively. Figure \ref{fig:LCequivalence} illustrates the equivalence between the open/shorted stub and L,C lumped elements for any value of the angular frequency $\omega=2 \pi f$.
Stubs can be used in parallel (see Fig. \ref{fig:Singlestubmatching}) or in series to a transmission line. A parallel stub can be easily realized in a practical PCB technology, for example as a shorted or open-ended microstrip line. A series stub cannot be easily realized. Therefore, it is most common to transform series stubs to parallel stubs using the so-called {\it Kuroda's identities}. Fig. \ref{fig:Kuroda} shows the four identities.
We will now show how Kuroda's identity can be used to design a low-pass filter in microstrip technology. We will start with a lumped-element prototype, corresponding to a third-order maximally flat low-pass filter prototype (see \cite{Pozar}, Table 8.3). The ideal circuit model and the corresponding steps towards the final transmission-line equivalent is provided in Fig. \ref{fig:Lowpassdesign}. The filter has a -3 dB roll-off at 10 GHz. We have assumed that both the input and output of the circuit are connected to $50 \Omega$ ports. The design starts with the ideal circuit model using two inductors and a parallel capacitor. Next the lumped elements are transformed to transmission line sections, in line with Fig. \ref{fig:LCequivalence}. Finally, by adding additional $\lambda/8$ line sections, we can apply Kuroda's identity (see Fig. \ref{fig:Kuroda}). The final design can be easily realized in microstrip technology on a PCB. The corresponding simulated performance of the original ideal circuit model and the final transmission-line low-pass filter is shown in Fig. \ref{fig:SimADSlowpass}.

\begin{figure}[hbt]
\centerline{\psfig{figure=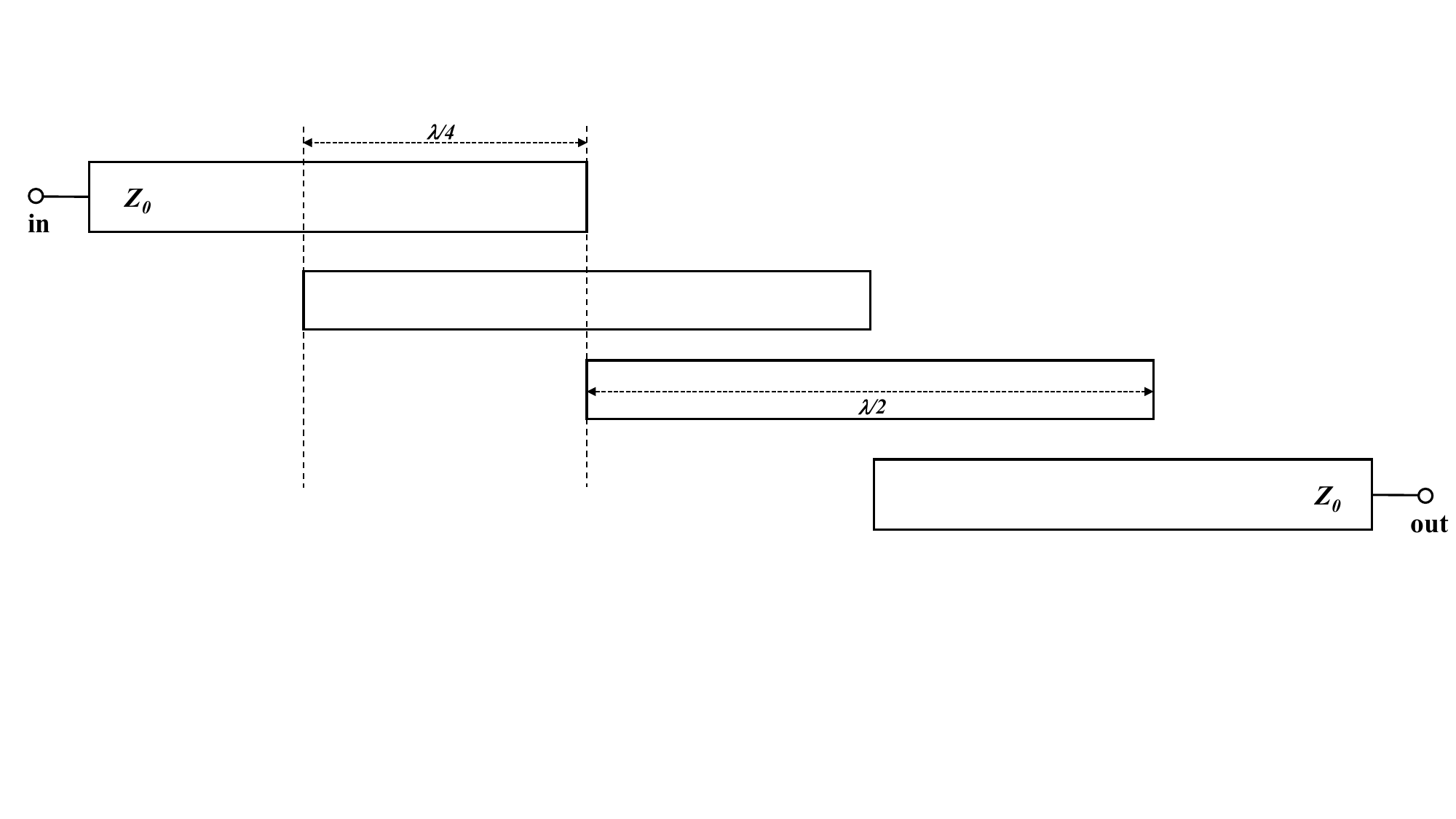,width=150mm}}
\vspace{-2cm}
\caption{\it A two-section ($N=2$) bandpass filter consisting of open-ended coupled lines.}
\label{fig:Coupledfilter}
\end{figure}

\begin{figure}[hbt]
\centerline{\psfig{figure=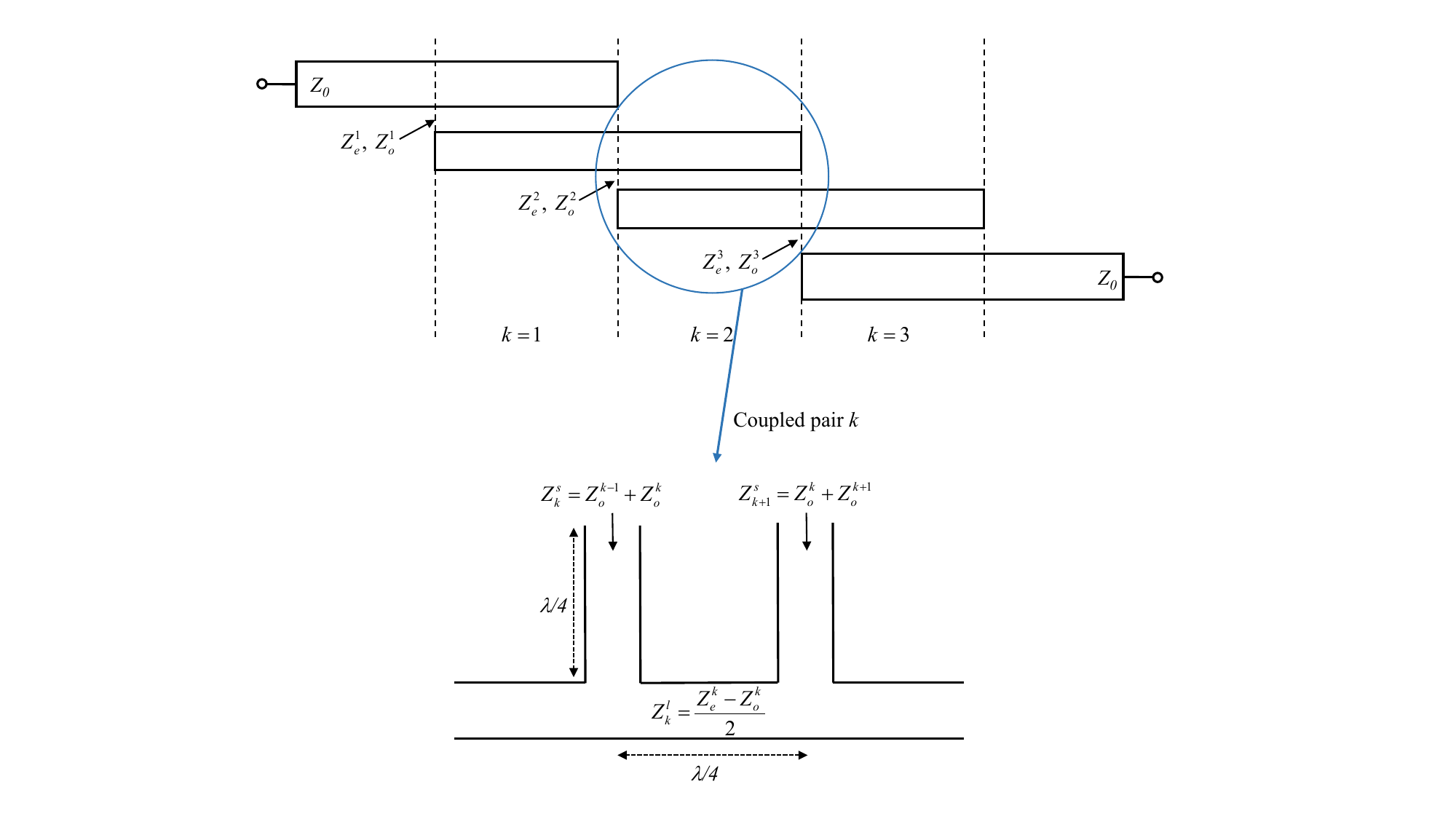,width=180mm}}
\caption{\it The equivalent transmission line circuit of the $k$-th coupled pair with $k=1..N+1$. Note that the characteristics impedances of the first and last stub are $Z_1^s=Z_o^1$ and $Z_{N+1}^s=Z_o^{N+1}$ .}
\label{fig:paircoupledlines}
\end{figure}

Bandpass filters are useful to suppress out-of-band interference. In this way, bandpass filters can help to reduce the linearity requirements of a receiver or reduce the out-of-band spurious emission generated by non-linearities of a transmitter. A very practical realization of a bandpass filter can be created by using parallel-coupled line sections, for example implemented in microstrip technology. An example of a two-section coupled-line filter is shown in Fig. \ref{fig:Coupledfilter}, where the top view of a microstrip implementation is shown. It uses two half-wavelength open-ended resonators which are coupled over a quarter-wave length to each other and to the connecting input and output transmission line. The characteristic impedance of the input and output transmission line is assumed to be equal to $Z_0$. The width of the resonators and the spacing between the resonators determine the characteristic impedances of the odd and even modes that can propagate along the coupled-line sections. By changing the resonators width and spacing, a specific bandpass characteristic can be realized. Note that short-circuited resonators can also be used.  One could argue that the half-wavelength resonators are in fact printed half-wavelength dipole antennas, as introduced in section \ref{sec-dunnedraadantene}. However, when implemented in microstrip technology using an electrically thin dielectric substrate, the electromagnetic field will be mainly concentrated between the metal strip and ground plane, and as a result, a strong coupling with the adjacent coupled lines will occur. Therefore, the (unwanted) spurious radiation will be very limited.
A single pair of coupled lines is shown in Fig. \ref{fig:paircoupledlines}. The two lines which are coupled over an electrical length $\lambda/4$ can be represented by the equivalent transmission line circuit as illustrated in Fig. \ref{fig:paircoupledlines} and consists of two open stubs which are placed in series and are separated by a quarter-wave line \cite{Matthaei}.
Now let us assume that Fig. \ref{fig:paircoupledlines} represents pair $k$ of a coupled-line filter, where we have a total of $N$ resonators and $N+1$ coupled pairs. Furthermore, let $Z_o^k$ and $Z_e^k$ represent the odd and even-mode characteristic impedance of the $k$-th coupled pair, with $k=1$..$N+1$. The corresponding characteristic impedances of the $\lambda/4$ stubs (indicated by $Z_k^s$) and of the $\lambda/4$ line (indicated by $Z_k^l$) of the equivalent circuit are provided in Fig. \ref{fig:paircoupledlines}.

In \cite{Matthaei} design equations have been derived for the design of Chebyshev coupled-line bandpass filters. The equations use a low-pass Chebyshev prototype as a starting point. Tables for such low-pass prototypes can be found in \cite{Pozar}, \cite{CollinME}. The design equations from \cite{Matthaei} are summarized below:
\begin{equation}
\begin{array}{lcl}
\di  \theta_1 & =& \di \frac{\pi \omega_1}{2\omega_0}, \ \ Q  = \di \frac{1}{\tan{\theta_1}}, \\
\di  K_1  & =&  \di K_{N+1} =\frac{Z_0}{\sqrt{g_{0} g_{1}}}, \\
\di  P & =& \di \sqrt{\frac{Q(Q^2+1)}{\di Q+\frac{1}{2(K_1/Z_0)^2}}}, \ \ \di  s  = \di Z_0 \left( \frac{P\sin\theta_1}{K_1/Z_0} \right)^2, \\
\di  Z_e^1 & =& \di Z_e^{N+1}= Z_0 (1+P\sin\theta_1), \\
\di  Z_o^1 & =& \di Z_o^{N+1}= Z_0 (1-P\sin\theta_1),
\end{array}
\label{eq:coupledfilterdesign1}
\end{equation}
and for the coupled pair with $k=2...N$ we have
\begin{equation}
\begin{array}{lcl}
\di  K_k & =& \di \frac{Z_0}{\sqrt{g_{k-1} g_{k}}}, \ \ \di  N_k  = \di \sqrt{\left(\frac{K_k}{Z_0}\right)^2+\left(\frac{\tan{\theta_1}}{2}\right)^2 },  \\
\di  Z_e^k & =& \di Z_e^{N+1-k} = s \left( N_k + \frac{K_k}{Z_0} \right),  \\
\di  Z_o^k & =& \di Z_o^{N+1-k} = s \left( N_k - \frac{K_k}{Z_0} \right),
\end{array}
\label{eq:coupledfilterdesign2}
\end{equation}
where $g_k$ are the component values of the low-pass prototype filter with cutoff radial frequency $\omega^`_1=1$. Furthermore, $\omega_0=2\pi f_0$ is the design radial frequency and $\omega_1=2\pi f_1$ is the radial frequency corresponding to the lower edge of the passband of the bandpass filter. The bandwidth of the filter is $2(f_0-f_1)/f_0$.

\begin{figure}[hbt]
\vspace{-1cm}
\centerline{\psfig{figure=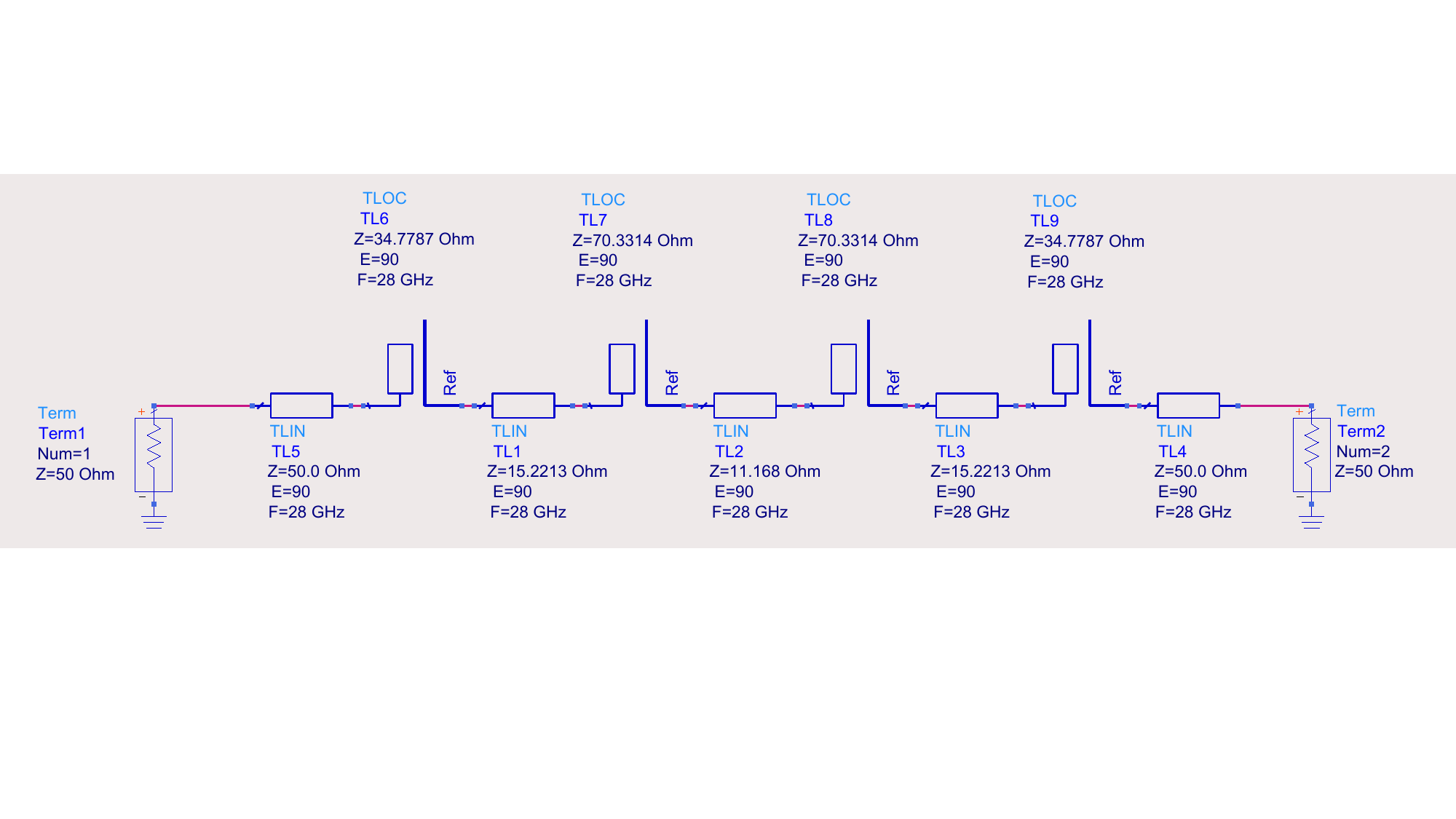,width=160mm}}
\vspace{-2.5cm}
\caption{\it Schematic ADS model of the two-section ($N=2$) Chebyshev bandpass filter consisting of open-ended coupled lines with a 3 dB passband ripple and $20\%$ bandwidth around the center frequency $f_0=28$ GHz. Note that ideal transmission line models have been used.}
\label{fig:schematicbandpass}
\end{figure}

\begin{figure}[hbt]
\centerline{\psfig{figure=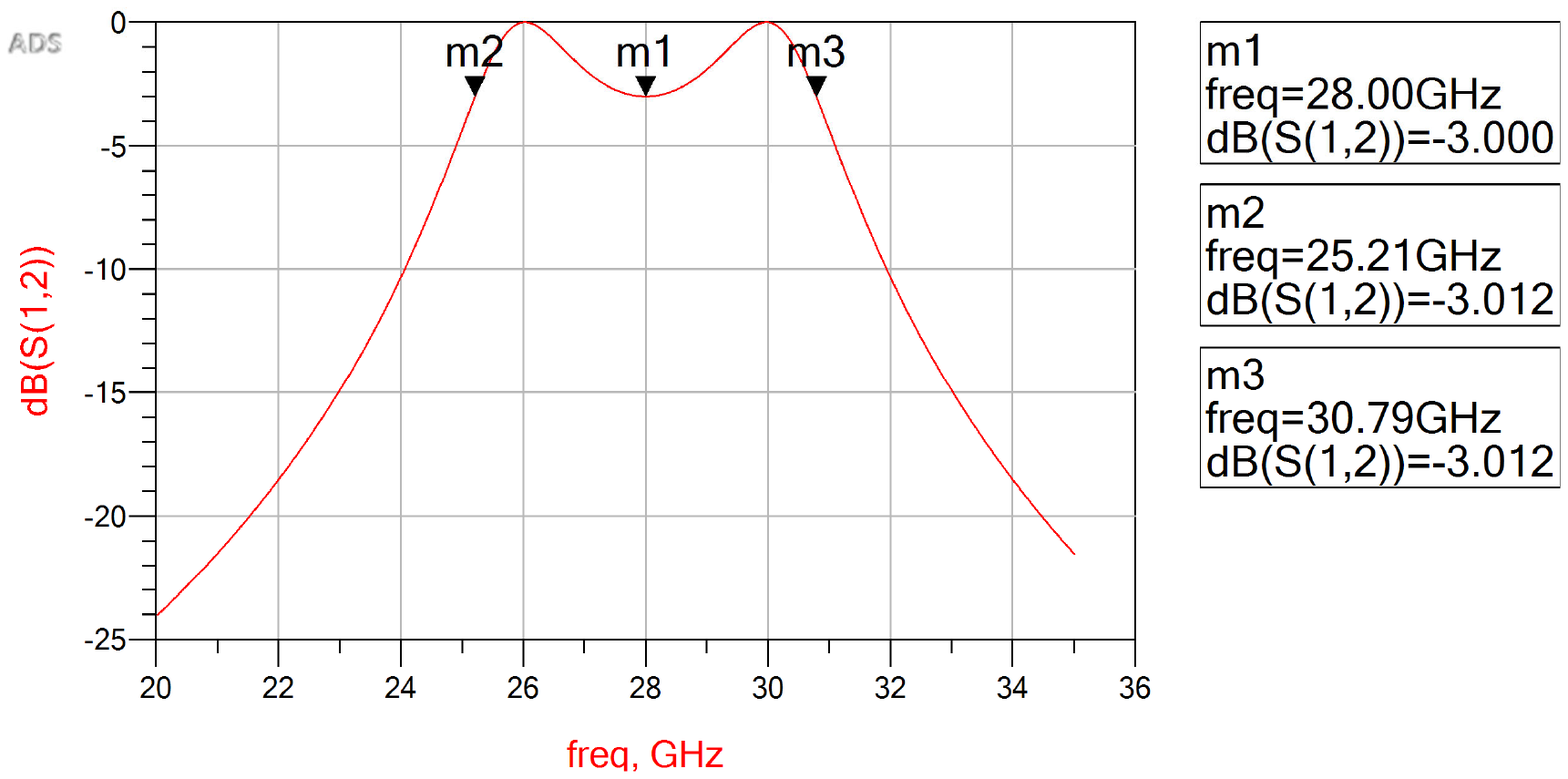,width=120mm}}
\vspace{-1cm}
\caption{\it Simulated performance of the two-section bandpass filter of Fig. \ref{fig:schematicbandpass}.}
\label{fig:S21bandpass}
\end{figure}

Let us now consider an example to illustrate the usage of design equations (\ref{eq:coupledfilterdesign1}) and (\ref{eq:coupledfilterdesign2}). We will design a two-section Chebyshev coupled-line filter with a $20\%$ bandwidth, 3 dB ripple in the passband and a center frequency at 28 GHz. The first step is to determine the low-pass prototype equivalent circuit. From \cite{Pozar} we find that the prototype values are: $g_0=1$, $g_1=3.1013$, $g_2=0.5339$, $g_3=5.8095$ with a cutoff radial frequency $\omega^`_1=1$. For the given bandwidth requirement we find that $\omega_1/\omega_0=0.9$. When substituting these values into the design equations (\ref{eq:coupledfilterdesign1}) and (\ref{eq:coupledfilterdesign2}) we find the following values for the coupled-line filter:
\begin{equation}
\begin{array}{lcl}
\di  Z_e^1 & =& \di 65.22 \ \Omega, \ \ \di  Z_o^1  = \di 34.78 \ \Omega,  \\
\di  Z_e^2 & =& \di 57.89 \ \Omega, \ \ \di  Z_o^2  = \di 35.55 \ \Omega.
\end{array}
\end{equation}
The characteristic impedances of the stubs and $\lambda/4$ sections (see also Fig. \ref{fig:paircoupledlines}) now become:
\begin{equation}
\begin{array}{lcl}
\di  Z_1^s & =& \di 34.78 \ \Omega, \ \ \di  Z_1^l  = \di 15.22 \ \Omega, \\
\di  Z_2^s & =& \di 70.33 \ \Omega, \ \ \di  Z_2^l  = \di 11.17 \ \Omega, \\
\di  Z_3^s & =& \di 70.33 \ \Omega, \ \ \di  Z_3^l  = \di 15.22 \ \Omega, \ \ \di  Z_4^s  = \di 34.78 \ \Omega.
\end{array}
\end{equation}
The final schematic of the bandpass filter is shown in Fig. \ref{fig:schematicbandpass} and the corresponding frequency response is illustrated in Fig. \ref{fig:S21bandpass}.

\vspace*{1cm} \hrulefill
 {\bf Exercise} \hrulefill \\
 Design a two-section coupled-line bandpass filter with a 0.5 dB equal-ripple (Chebyshev) and a $30\%$ bandwidth. The input and output are connected to $50 \Omega$ transmission lines. The low-pass prototype values are: $g_0=1$, $g_1=1.4029$, $g_2=0.7071$, $g_3=1.9841$.
{\it Answer:} $Z_1^s  = 24.75 \ \Omega, \  Z_1^l  = 25.25 \ \Omega,  Z_2^s  = 48.15 \ \Omega, \  Z_2^l  = 17.96 \ \Omega,  Z_3^s  = 48.15 \ \Omega, \ Z_3^l  = 25.25 \ \Omega, Z_4^s  = \di 24.75 \ \Omega.$

\section{Microwave amplifiers}
\label{sec:Microamp}

Antennas are almost always connected to amplifiers. In transmit mode, the antenna will be connected to a power amplifier (PA), whereas in receive mode the antenna will be connected to a low-noise amplifier (LNA). With the continuous increase of the operational frequency of new wireless applications, highly integrated active antenna concepts will need to be developed in order to combat the interconnect losses between antenna and amplifier. Therefore, it is crucial for the future antenna engineer to have a good understanding of amplifiers.

Microwave amplifiers can be realized in several semiconductor technologies, such as silicon-based technologies (CMOS and BiCMOS), and III-V technologies, such as gallium arsenide (GaAs) or gallium nitride (GaN). Discrete transistors can be used to realize an amplifier, but most commonly the amplifiers will be integrated in a more complex RF integrated circuit (RF-IC).
More background material on microwave amplifiers can be found in \cite{Gonzalez}.

\subsection{Power gain, available gain and transducer gain}
\begin{figure}[hbt]
\centerline{\psfig{figure=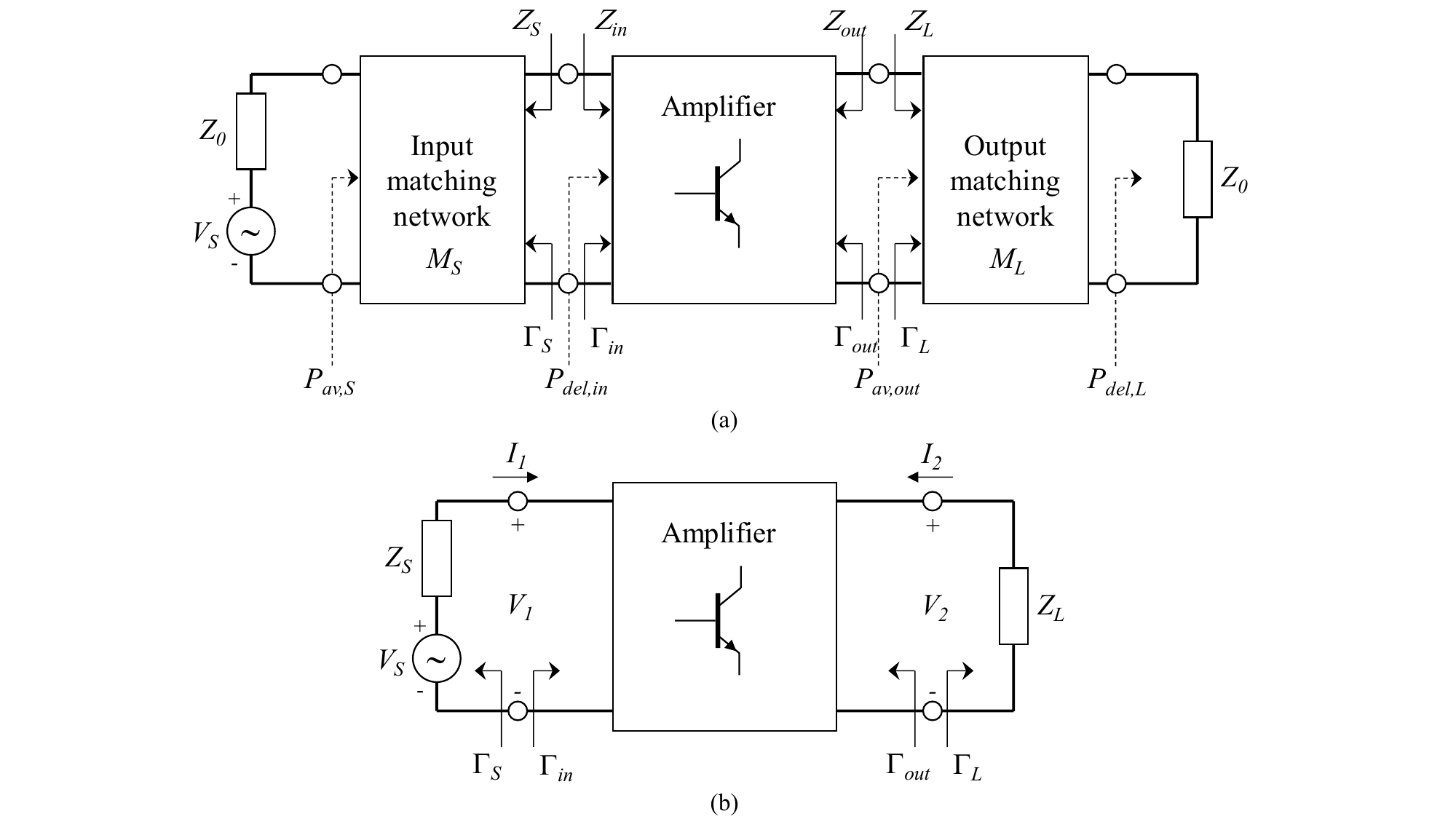,width=170mm}}
\caption{\it (a) General amplifier circuit including an input and output matching network. (b) Equivalent amplifier circuit.}
\label{fig:Amplifiercircuit1}
\end{figure}

For the analysis and design of amplifiers it is convenient to use three definitions for power gain. Consider the amplifier circuit of Fig. \ref{fig:Amplifiercircuit1}(a). In the center of this circuit we find the active device (transistor) which is described in terms of a two-port network.
The active device is connected to the source via an input matching network and at the output it is connected to a load impedance via an output matching network. Note that the combination of the output matching network and load impedance could be the antenna impedance in case of a power amplifier which is directly matched to an antenna. In a similar way, the source impedance and the input matching circuit could be the antenna impedance in case of a low-noise amplifier. Fig. \ref{fig:Amplifiercircuit1}(b) shows the equivalent circuit where the properties of the input and output matching networks are included in the source impedance $Z_S$ and load impedance $Z_L$, respectively. The three definitions that we will use are:
\begin{enumerate}[i.]
    \item {\it Delivered power gain} $\di G_{del}=\frac{P_{del,L}}{P_{del,in}}$, which is the ratio between the delivered power to the load $P_{del,L}$ and the delivered input power to the amplifier $P_{del,in}$.
    \item {\it Available power gain} $\di G_{av}=\frac{P_{av,out}}{P_{av,S}}$, which is the ratio between the available power at the output of the amplifier $P_{av,out}$ and the available power from the source $P_{av,S}$.
    \item {\it Transducer power gain} $\di G_T=\frac{P_{del,L}}{P_{av,S}}$, which is the ratio between the delivered power to the load $P_{del,L}$ and the available input power from the source $P_{av,S}$.
\end{enumerate}
In section \ref{sec:microwavenetworks} we already derived the expressions for the input and output reflection coefficients, given by (\ref{eq:Inoutreflection}) and (\ref{eq:Sourceloadreflection}). Now let $Z_{in}$ be the input impedance looking into the amplifier as indicated in Fig. \ref{fig:Amplifiercircuit1}. The voltage $V_1$ can be written as:
\begin{equation}
\begin{array}{lcl}
\displaystyle V_1 & = & \displaystyle  \frac{V_S Z_{in}}{Z_S+Z_{in}} = V^+_1 \left( 1+ \Gamma_{in} \right) .
 \end{array} \label{eq:V1amplifier}
\end{equation}
From (\ref{eq:V1amplifier}) we can determine $V^+_1$ and calculate the time-average input power to the amplifier, similar to the approach of section \ref{sec:lossyTline}:
\begin{equation}
\begin{array}{lcl}
\displaystyle P_{del,in} & = & \displaystyle  \frac{|V^+_1|^2}{2Z_0} \left(1 - |\Gamma_{in}|^2 \right) = \frac{|V_S|^2}{8Z_0}
  \frac{\left|1 - \Gamma_{S} \right|^2}{\left|1 - \Gamma_{S} \Gamma_{in} \right|^2} \left(1- |\Gamma_{in}|^2 \right),
 \end{array} \label{eq:Pinamplifier}
\end{equation}
where $\Gamma_{in}$, $\Gamma_S$ and $\Gamma_L$ are given by (\ref{eq:Inoutreflection}) and (\ref{eq:Sourceloadreflection}). In addition, we have expressed $Z_{in}$ in terms of $\Gamma_{in}$ by using
\begin{equation}
\begin{array}{lcl}
\displaystyle \Gamma_{in} & = & \displaystyle  \frac{Z_{in}-Z_0}{Z_{in}+Z_0}.
 \end{array} \label{eq:Zinamplifier}
\end{equation}
In a similar way we can determine the delivered power to the load:
\begin{equation}
\begin{array}{lcl}
\displaystyle P_{del,L} & = & \displaystyle  \frac{|V^-_2|^2}{2Z_0} \left(1 - |\Gamma_{L}|^2 \right) = \frac{|V_S|^2}{8Z_0}
  \frac{|S_{21}|^2 \left(1 - |\Gamma_{L}|^2 \right) \left|1-\Gamma_S \right|^2}{\left|1 - S_{22}\Gamma_{L} \right|^2 \left|1- \Gamma_{in} \Gamma_S  \right|^2},
 \end{array} \label{eq:PLamplifier}
\end{equation}
where we have written the incident and reflected voltage wave amplitudes in terms of the scattering coefficients using (\ref{eq:abportk}) and (\ref{eq:Scatteringmatrix}).
We then finally find the delivered power gain by using (\ref{eq:Pinamplifier}) and (\ref{eq:PLamplifier}):
\begin{equation}
\begin{array}{lcl}
\displaystyle G_{del} & = & \displaystyle \frac{P_{del,L}}{P_{del,in}}= \frac{|S_{21}|^2 \left(1 - |\Gamma_{L}|^2 \right)}
 {\left|1 - S_{22}\Gamma_{L} \right|^2 \left( 1- |\Gamma_{in}|^2 \right)}.
 \end{array} \label{eq:Gpamplifier}
\end{equation}
In a similar way we can determine the other two power gain definitions:
\begin{equation}
\begin{array}{lcl}
\displaystyle G_{av} & = & \displaystyle \frac{P_{av,out}}{P_{av,S}} = \frac{|S_{21}|^2 \left(1 - |\Gamma_{S}|^2 \right)}
 {\left|1 - S_{11}\Gamma_{S} \right|^2 \left( 1- |\Gamma_{out}|^2 \right)}, \\
 \displaystyle G_{T} & = & \displaystyle \frac{P_{del,L}}{P_{av,S}}= G_{av} M_L = G_{del} M_S ,
 \end{array} \label{eq:GavGtamplifier}
\end{equation}
where $M_S$ and $M_L$ represent the source and load mismatch factors, respectively, and are given by:
\begin{equation}
\begin{array}{lcl}
\displaystyle M_{S} & = & \displaystyle \frac{P_{del,in}}{P_{av,S}} = \frac{\left(1 - |\Gamma_{in}|^2 \right) \left(1 - |\Gamma_{S}|^2 \right)} {\left|1 - \Gamma_{in} \Gamma_{S} \right|^2}, \\
\displaystyle M_{L} & = & \displaystyle \frac{P_{del,L}}{P_{av,out}} = \frac{\left(1 - |\Gamma_{out}|^2 \right) \left(1 - |\Gamma_{L}|^2 \right)} {\left|1 - \Gamma_{out} \Gamma_{L} \right|^2}.
 \end{array} \label{eq:MSMLamplifier}
\end{equation}
Note that $\Gamma_{out}$ is given in (\ref{eq:Inoutreflection}).
Note that in case of a unilateral amplifier, e.g. $S_{12}=0$, the expressions for the input and output reflection coefficient (\ref{eq:Inoutreflection}) simplify to $\Gamma_{in}=S_{11}$ and $\Gamma_{out}=S_{22}$.

\begin{figure}[hbt]
\vspace{-5cm}
\centerline{\psfig{figure=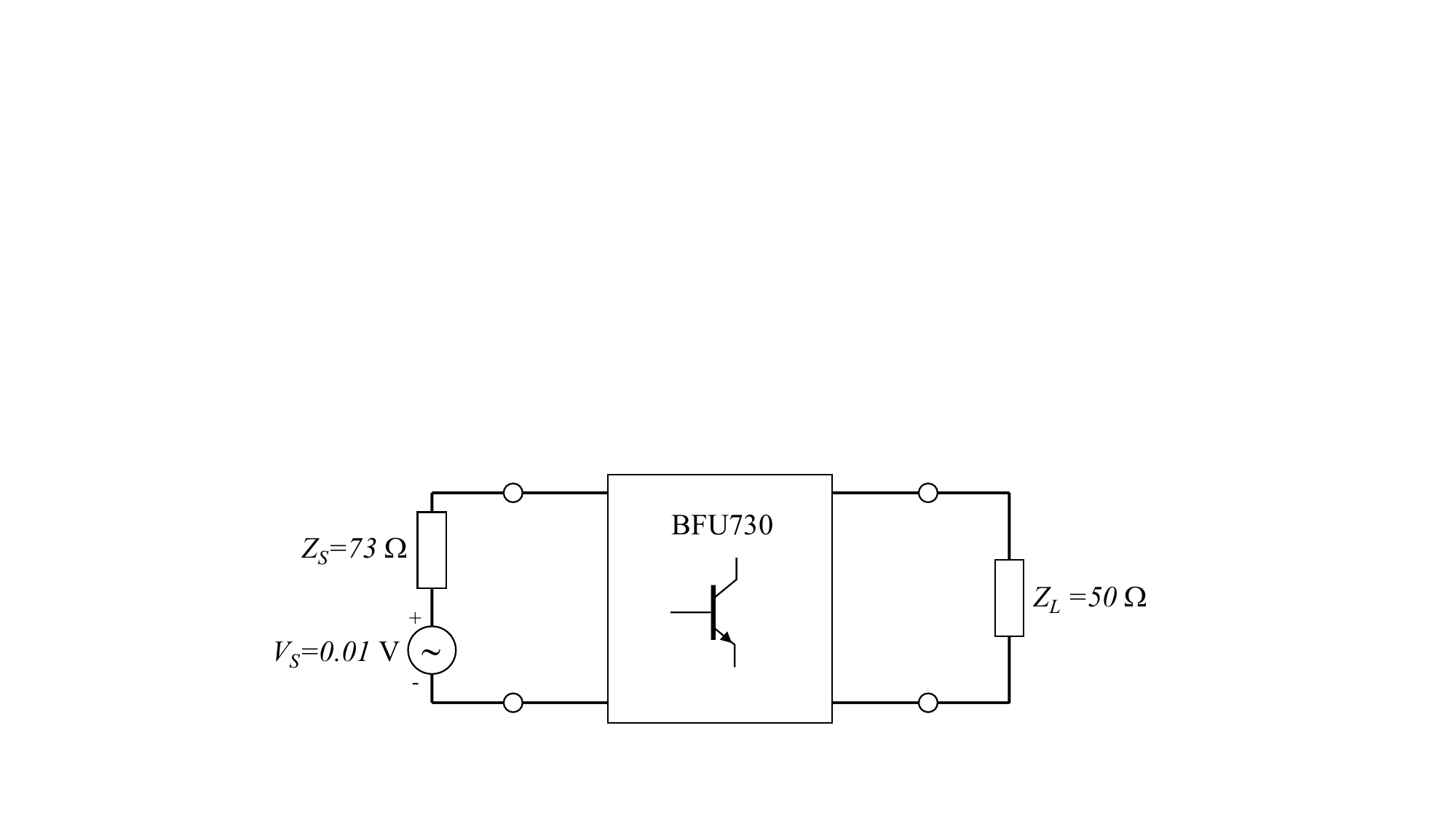,width=170mm}}
\caption{\it Amplifier based on the BFU730F transistor from NXP Semiconductors. The input is connected to an antenna with an input impedance of $73~\Omega$ which generates an equivalent time-harmonic voltage of $0.01$~V. The output is connected to a $50~\Omega$ load. }
\label{fig:ExampleAmplifier1}
\end{figure}

As an example, we will use the BFU730F transistor from NXP Semiconductors. A basic circuit (excluding biasing) using this amplifier is shown in Fig. \ref{fig:ExampleAmplifier1}. The amplifier is used as a low-noise amplifier which is connected at the input to a dipole antenna with an impedance of $73~\Omega$. The output of the amplifier is connected to a $50~\Omega$ load. The complex amplitude of the time-harmonic voltage source that represents the receiving antenna is equal to $V_S=0.01~$V.
The $S$-parameters of the BFU730F transistor in terms of magnitude and phase at 400 MHz are given by
\begin{equation}
\begin{array}{lcl}
\displaystyle S_{11} & = & \displaystyle 0.87, -28^{\circ} , \\
\displaystyle S_{12} & = & \displaystyle 0.01, 76^{\circ} , \\
\displaystyle S_{21} & = & \displaystyle 26.73, 159^{\circ} , \\
\displaystyle S_{22} & = & \displaystyle 0.96, -16^{\circ}.
 \end{array} \label{eq:SBFU730Famplifier}
\end{equation}
By using (\ref{eq:Inoutreflection}) we find the input and output reflection coefficients:
\begin{equation}
\begin{array}{lcl}
\displaystyle \Gamma_{in}  & = & 0.77 - \jmath 0.41 , \\
\displaystyle \Gamma_{out} & = & 0.89 - \jmath 0.31 , \\
 \end{array}
\end{equation}
The corresponding values for the three power gain definitions are found using (\ref{eq:Gpamplifier}) and (\ref{eq:GavGtamplifier}):
\begin{equation}
\begin{array}{lcl}
\displaystyle G_{del}  & = & 34.7~ \mbox{dB} , \\
\displaystyle G_{av} & = & 38.9~ \mbox{dB}, \\
\displaystyle G_T  & = & 29.7~ \mbox{dB} ,
 \end{array}
\end{equation}
and the corresponding source and load mismatch factors:
\begin{equation}
\begin{array}{lcl}
\displaystyle M_S  & = & 29.7 - 34.7 = -5.0~ \mbox{dB} , \\
\displaystyle M_L  & = & 29.7 - 38.9 = -9.2~ \mbox{dB}.
 \end{array}
\end{equation}
The power available from the source is
\begin{equation}
\begin{array}{lcl}
\displaystyle P_{av,S} & = & \di \frac{|V_S|^2}{8Z_S} = 0.17~ \mu \mbox{W}  ,
 \end{array}
\end{equation}
which corresponds to -37.7 dBm. The available power at the output of the amplifier is then $P_{av,out}=G_{av} P_{av,S}=1.3~\mbox{mW}$ which is equal to $+1.2~$dBm.

\subsection{Stability}
\begin{figure}[hbt]
\vspace{-1.5cm}
\centerline{\psfig{figure=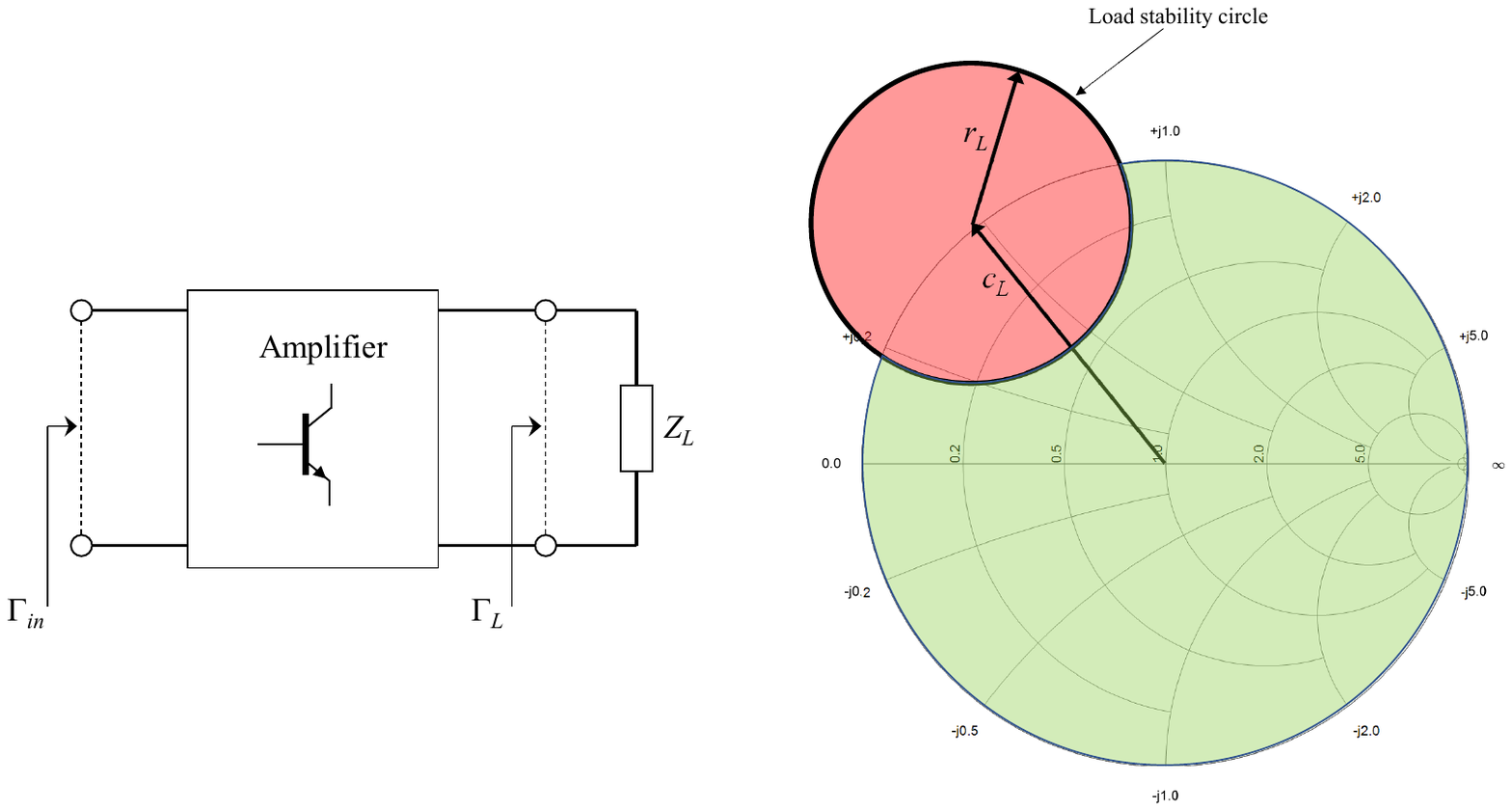,width=170mm}}
\vspace{-1.5cm}
\caption{\it Illustration of the output (load) stability circle in the Smith chart. It shows (green region) which loads (represented by $Z_L$ or $\Gamma_L$) result in load stability with $\Gamma_{in} < 1$. Note that we have assumed that $|S_{11}|<1$. }
\label{fig:ExampleAmplifier}
\end{figure}

One of the key aspects in the design of amplifiers is the so-called stability of the amplifier circuit. If $|\Gamma_{in}| > 1$ or $|\Gamma_{out}| > 1$ oscillation could be possible in the circuit of Fig. \ref{fig:Amplifiercircuit1}(a).
The amplifier is stable if the following two criteria are both satisfied:
\begin{equation}
\begin{array}{lcl}
\di |\Gamma_{in}| & = & \di \left| S_{11} + \frac{S_{12} S_{21}}{\di \frac{1}{\Gamma_L} - S_{22}} \right| < 1, \\
\di |\Gamma_{out}|& = & \di \left| S_{22} + \frac{S_{12} S_{21}}{\di \frac{1}{\Gamma_S} - S_{11}} \right| < 1.
\end{array}
\label{eq:StabilityGammainout}
\end{equation}
The amplifier is {\it unconditionally stable} if $|\Gamma_{in}| < 1$ and $|\Gamma_{out}| < 1$ for all passive source and load impedances, e.g. with $|\Gamma_{S}| < 1$ and $|\Gamma_{L}| < 1$. The amplifier is {\it conditionally stable} if (\ref{eq:StabilityGammainout}) is satisfied only for certain source and load impedances. Conditional stability can be easily visualized by using input and output stability circles in the Smith Chart. Let us consider the output stability circle in more detail. It defines the boundary in the Smith chart between the stable and unstable region for $\Gamma_L$ when $|\Gamma_{in}|=1$. The radius $r_L$ and center $c_L$ of this output stability circle are given by \cite{Gonzalez}:
\begin{equation}
\begin{array}{lcl}
\di r_{L} & = & \di \left| \frac{S_{12} S_{21}}{|S_{22}|^2-|\Delta|^2} \right| , \\
\di c_{L} & = & \di      \frac{(S_{22} - \Delta S^*_{11})^*}{|S_{22}|^2-|\Delta|^2} ,
\end{array}
\label{eq:OutputStability}
\end{equation}
where $\Delta$ is the determinant of the scattering matrix:
\begin{equation}
\di \Delta = S_{11} S_{22} - S_{12} S_{21}.
\label{eq:DeltaStability}
\end{equation}
Fig. \ref{fig:ExampleAmplifier} illustrates the use of the output stability circle in a Smith Chart.
In a similar way, source stability circles can be drawn in a Smith Chart to indicate the boundary between the stable and unstable region for $\Gamma_S$ when $|\Gamma_{out}|=1$. The corresponding radius $r_S$ and center $c_S$ of the input stability circle are given by:
\begin{equation}
\begin{array}{lcl}
\di r_{S} & = & \di \left| \frac{S_{12} S_{21}}{|S_{11}|^2-|\Delta|^2} \right| , \\
\di c_{S} & = & \di      \frac{(S_{11} - \Delta S^*_{22})^*}{|S_{11}|^2-|\Delta|^2}.
\end{array}
\label{eq:InputStability}
\end{equation}

Often we are only interested in knowing whether an amplifier is unconditionally stable within a certain frequency range for all source and load impedances. To check this we can use the {\it Rollett stability condition} \cite{Rollett}, expressed in terms of a condition for two parameters, $K$ and $\Delta$:
\begin{equation}
\begin{array}{lcl}
\di K        & = & \di  \frac{1 - |S_{11}|^2 - |S_{22}|^2 + |\Delta|^2}{2|S_{12} S_{21}|} > 1, \\
\di |\Delta| & = & \di  |S_{11} S_{22} - S_{12} S_{21}| < 1.
\end{array}
\label{eq:UnconditionalStability1}
\end{equation}
As an alternative for the $K$-$\Delta$ condition, the single parameter ($\mu$-test) introduced in \cite{Edwards1992} can also be used to check if the transistor is unconditionally stable at a certain frequency:
\begin{equation}
\begin{array}{lcl}
\di \mu        & = & \di  \frac{1 - |S_{11}|^2}{|S_{22} - S^*_{11} \Delta | + |S_{12} S_{21}|} > 1,
\end{array}
\label{eq:UnconditionalStability2}
\end{equation}
As an example, we can check the stability of the BFU730F transistor by substituting the scattering parameters of (\ref{eq:SBFU730Famplifier}) in to (\ref{eq:UnconditionalStability1}) or (\ref{eq:UnconditionalStability2}). We then find that $K=0.039$ and $|\Delta|=0.836$. The corresponding $\mu$-test value is $\mu=0.38$.  This means that this device is {\underline{not}} unconditionally stable at this frequency.

\subsection{Amplifier design using constant gain circles}
\begin{figure}[hbt]
\vspace{-1.5cm}
\centerline{\psfig{figure=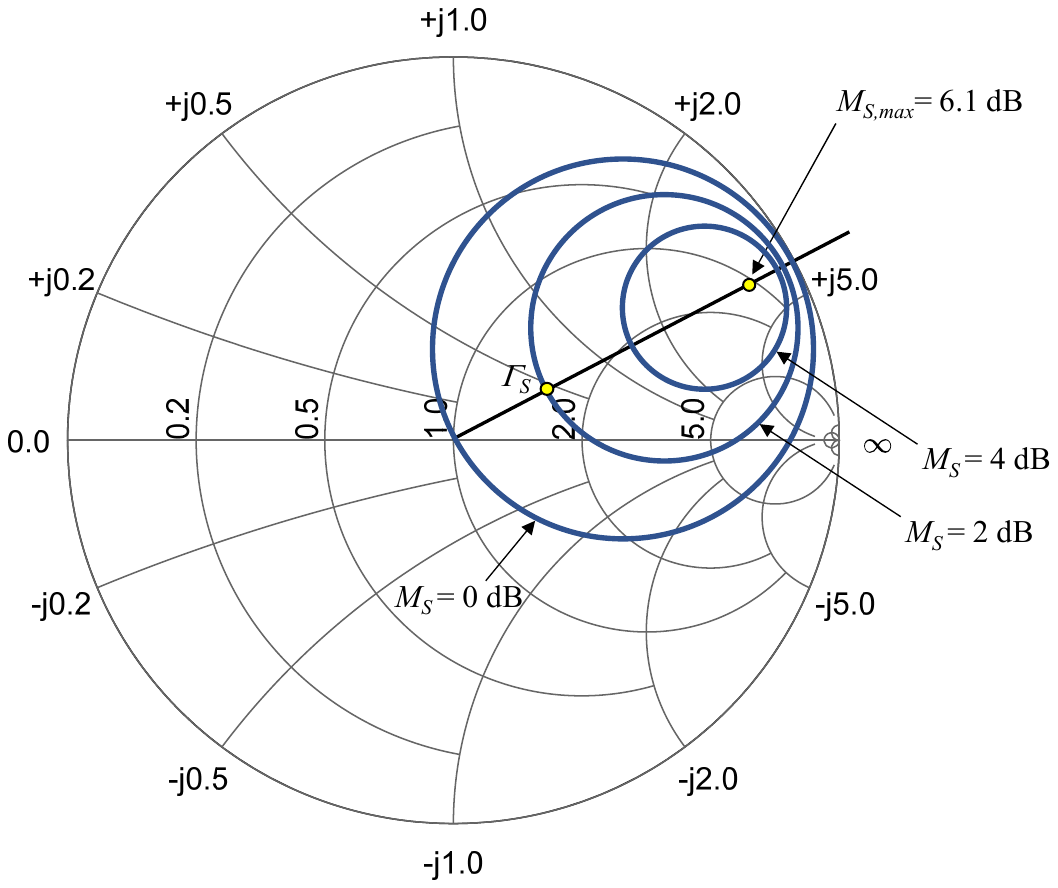,width=170mm}}
\vspace{-1.5cm}
\caption{\it Constant-gain circle of the BFU730F transistor at 400 MHz plotted in the source $(\Gamma_S)$ plane. For $M_S=0~$dB, we find $c_S=0.495, 28^{\circ}$, $r_S=0.495$; for $M_S=2~$dB, we find $c_S=0.627, 28^{\circ}$, $r_S=0.356$; for $M_S=4~$dB, we find $c_S=0.753, 28^{\circ}$, $r_S=0.215$. The transistor is assumed to be unilateral.}
\label{fig:ConstantGainCircleBFU730}
\end{figure}

\begin{figure}[hbt]
\vspace{-1.5cm}
\centerline{\psfig{figure=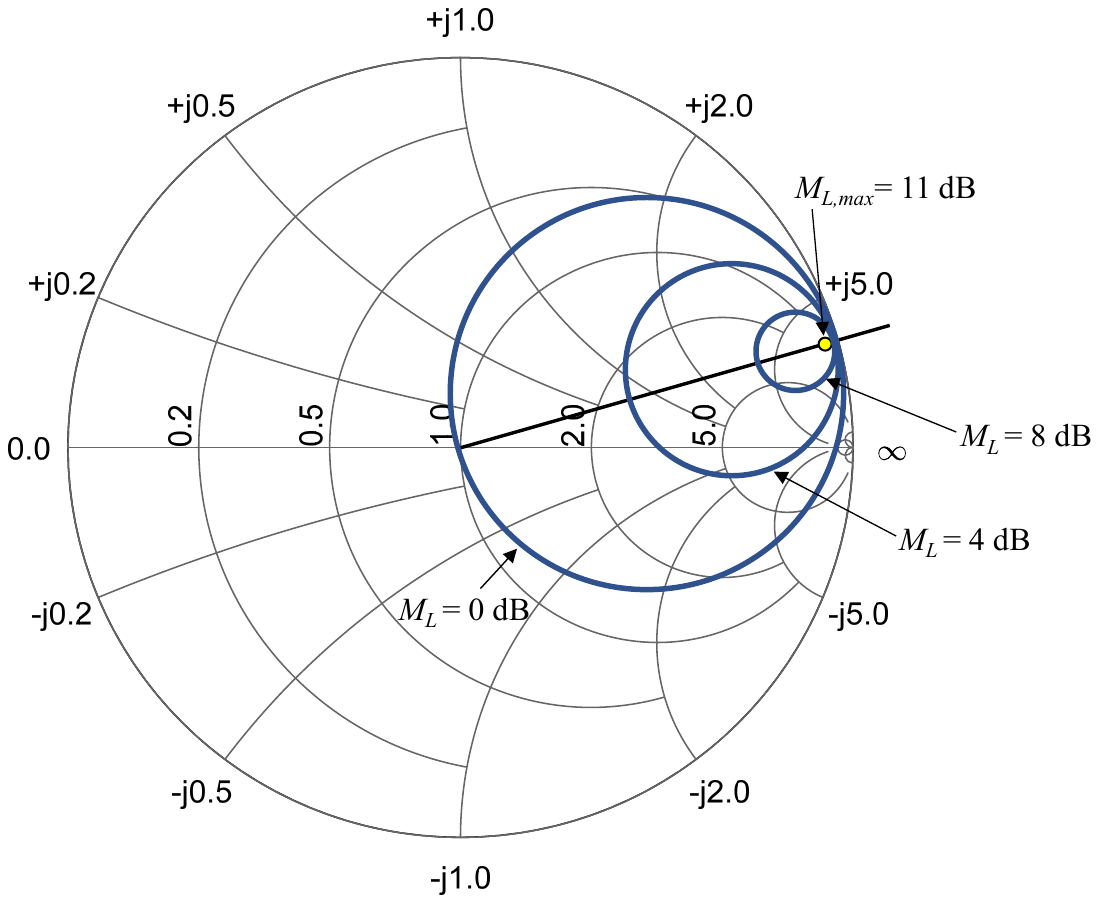,width=170mm}}
\vspace{-1.5cm}
\caption{\it Constant-gain circle of the BFU730F transistor at 400 MHz plotted in the load $(\Gamma_L)$ plane. For $M_L=0~$dB, we find $c_L=0.5, 16^{\circ}$, $r_L=0.5$; for $M_L=4~$dB, we find $c_L=0.73, 16^{\circ}$, $r_L=0.27$; for $M_L=8~$dB, we find $c_L=0.89, 16^{\circ}$, $r_L=0.1$. The transistor is assumed to be unilateral.}
\label{fig:ConstantGainCircleBFU730Load}
\end{figure}

The overall gain of the amplifier of Fig. \ref{fig:Amplifiercircuit1}(a) can be maximized by carefully choosing the input and output matching networks, specified by $M_S$ and $M_L$. The main drawback of this strategy is that the overall circuit will become quite narrow band in general. Therefore, it is better to compromise on the achievable gain in order to realize a more broadband amplifier. By plotting so-called {\it constant gain circles} in the Smith chart a trade-off can be made between realized gain and input and output matching.
We will consider a unilateral amplifier with $S_{12}=0$. The unilateral transducer gain can now be written in the following form:
\begin{equation}
\begin{array}{lcl}
 \displaystyle G_{Tu} & = & \displaystyle M_S |S_{21}|^2 M_L ,
 \end{array} \label{eq:GTuniamplifier}
\end{equation}
where the source and load mismatch factors are now given by:
\begin{equation}
\begin{array}{lcl}
\displaystyle M_{S} & = & \displaystyle  \frac{\left(1 - |\Gamma_{S}|^2 \right)} {\left|1 - S_{11} \Gamma_{S} \right|^2}, \\
\displaystyle M_{L} & = & \displaystyle  \frac{\left(1 - |\Gamma_{L}|^2 \right)} {\left|1 - S_{22} \Gamma_{L} \right|^2}.
 \end{array} \label{eq:MSMLuniamplifier1}
\end{equation}
The factor $M_S$ is maximised when $\Gamma_S=S^*_{11}$ with maximum value:
\begin{equation}
\begin{array}{lcl}
\displaystyle M_{S,max} & = & \displaystyle  \frac{1} {1-|S_{11}|^2}. \\
 \end{array} \label{eq:MSMLuniamplifier2}
\end{equation}
We can now introduce a normalized value $g_S$ with
\begin{equation}
\begin{array}{lcl}
\displaystyle g_{S} & = & \displaystyle  \frac{M_S} {M_{S,max}} = M_S \left( 1-|S_{11}|^2 \right).
 \end{array} \label{eq:gSuniamplifier}
\end{equation}
For a constant value of $g_S$, expression (\ref{eq:gSuniamplifier}) will be a circle in the $\Gamma_S$ plane when plotted in a Smith chart with radius and center given by:
\begin{equation}
\begin{array}{lcl}
\di r_{S} & = & \di  \frac{\left( 1-|S_{11}|^2 \right) \sqrt{1-g_S}}{1 - |S_{11}|^2 (1-g_S)}, \\
\di c_{S} & = & \di  \frac{g_S S^*_{11}}{1 - |S_{11}|^2 (1-g_S)}.
\end{array}
\label{eq:ConstantgaincircleSource}
\end{equation}
If we again consider the BFU730F transistor and assume that this transistor is unilateral with $S_{12}=0$ and where the other scattering parameters are given by (\ref{eq:SBFU730Famplifier}) at 400 MHz, we obtain the constant-gain circles as illustrated in Fig. \ref{fig:ConstantGainCircleBFU730}. Observe that the circle for $M_S=0~$dB passes the centre of the Smith chart at $\Gamma_S=0$. It can be shown that this is always the case for any device.
The maximum  value for this device is $M_S=6.1~$dB. In order to improve/simplify the matching towards the source, we could compromise a few dB, for example by choosing the $M_S=2~$dB circle in Fig. \ref{fig:ConstantGainCircleBFU730}. In this case we would choose the point $\Gamma_S$ on the circle closest to the center of the Smith chart as illustrated in Fig. \ref{fig:ConstantGainCircleBFU730}.
In a similar way, we can find the constant gain circles in the load $(\Gamma_L)$ plane. The radius and center in a Smith chart are given by:
\begin{equation}
\begin{array}{lcl}
\di r_{L} & = & \di  \frac{\left( 1-|S_{22}|^2 \right) \sqrt{1-g_L}}{1 - |S_{22}|^2 (1-g_L)}, \\
\di c_{L} & = & \di  \frac{g_L S^*_{22}}{1 - |S_{22}|^2 (1-g_L)}.
\end{array}
\label{eq:ConstantgaincircleLoad}
\end{equation}
where $g_L$ is the normalized value given by:
\begin{equation}
\begin{array}{lcl}
\displaystyle g_{L} & = & \displaystyle  \frac{M_L} {M_{L,max}} = M_L \left( 1-|S_{22}|^2 \right).
 \end{array} \label{eq:gLuniamplifier}
\end{equation}
The constant gain circles in the load plane for the BFU730F transistor at 400 MHz are shown in Fig. \ref{fig:ConstantGainCircleBFU730Load}.

\section{Low-noise amplifiers}
\label{sec:LNA}

\subsection{Noise in microwave circuits}
\begin{figure}[hbt]
\centerline{\psfig{figure=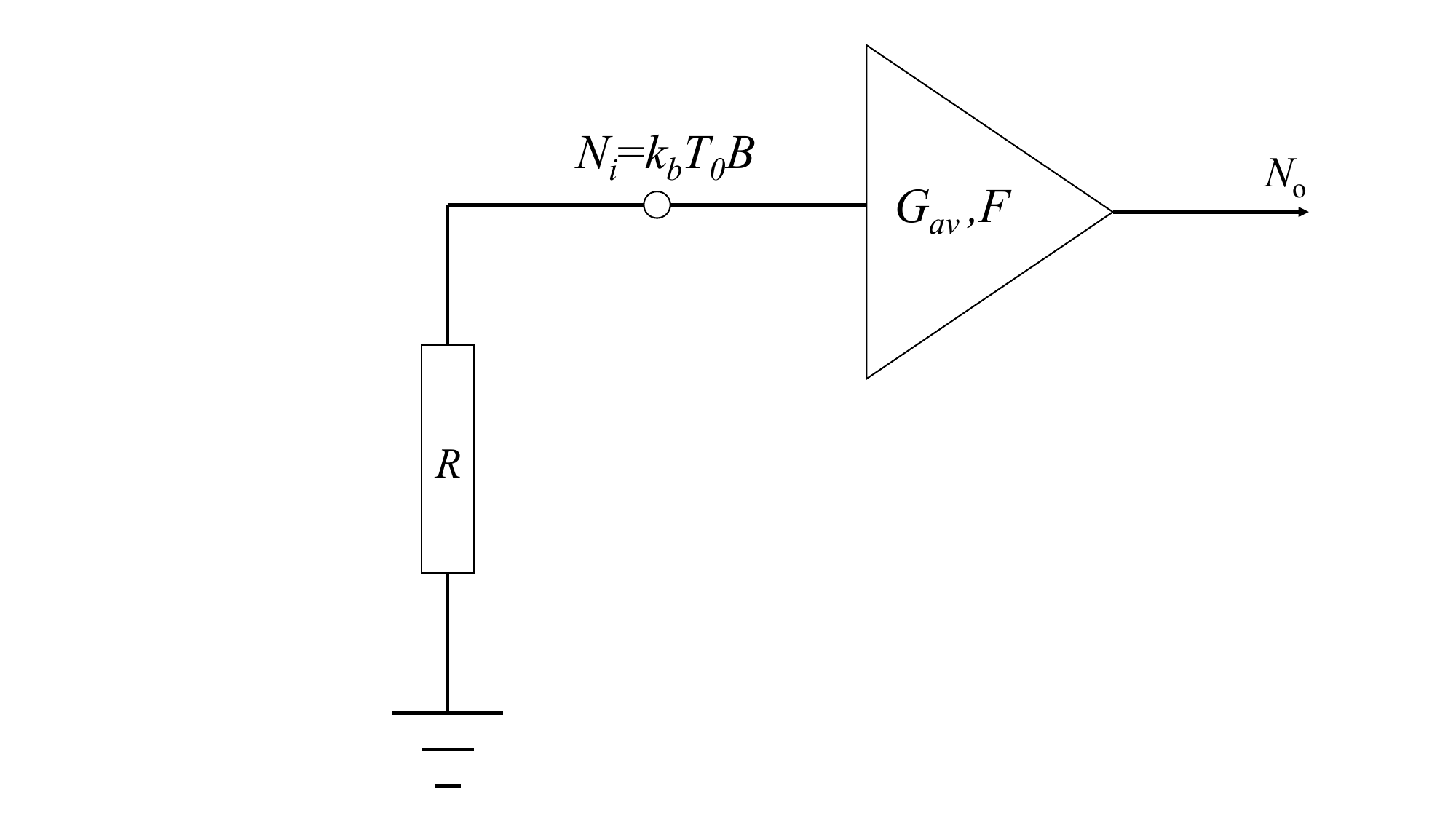,width=100mm}}
\caption{\it Thermal noise generated by a resistor at the input of a microwave amplifier.}
\label{fig:noiseamplifier1}
\end{figure}

In 1926 John Johnson at Bell Labs observed for the first time that a resistor generates noise in the absence of external current biasing. This so-called thermal noise or Johnson noise is due to the random motion of charges in the resistor material.
Consider the resistor connected at the input of a microwave amplifier, as illustrated in Fig. \ref{fig:noiseamplifier1}. The available noise power generated by this resistor at the input is given by:
\begin{equation}
\begin{array}{lcl}
\displaystyle N & = & \displaystyle  k_b T_0 B ,
 \end{array} \label{eq:NoisekTB}
\end{equation}
where $k_b=1.38\times 10^{-23}~$J/K is Boltmann's constant and $B$ [Hz] is the noise bandwidth of the system. Note that by adding additional resistors in series or in parallel at the input would not affect the available noise power, since the available noise power is independent of the resistor value. As a result, the available noise power at the input of any receiver will always be equal to -174 dBm/Hz at room temperature ($T_0=300~$K). Note that dBm refers to decibel-milliwatts, where 0 dBm corresponds to 1 mW of power.
The noise figure $F$ of a two-port microwave network is defined as the ratio between the input signal-to-noise ratio (SNR) and output SNR:
\begin{equation}
\begin{array}{lcl}
\displaystyle F & = & \displaystyle  \frac{\di \left( \frac{S_i}{N_i} \right)}{\di \left( \frac{S_o}{N_o} \right) }.
 \end{array} \label{eq:Noisefigure1}
\end{equation}
Since any microwave network will add noise to the available input noise power, we find that $F>1$. Usually the noise figure is expressed in [dB] according to:
\begin{equation}
\begin{array}{lcl}
\displaystyle NF & = & \displaystyle  10 \log_{10}(F) ~~[\mbox{dB}].
 \end{array} \label{eq:Noisefigure1}
\end{equation}

\vspace*{0.5cm} \hrulefill
 {\bf Exercise} \hrulefill \\
 Determine the available noise power at the input of a Bluetooth receiver operating in the 2.4 GHz band at room temperature. The channel bandwidth is 1 MHz.
{\it Answer:} $N_i=-114~$dBm. \\
\hrulefill \\
\vspace*{0.25cm}

Let us go back to the amplifier of Fig. \ref{fig:noiseamplifier1}. The amplifier is characterized by the available power gain $G_{av}$ and noise figure $F$. By using relation (\ref{eq:Noisefigure1}) and (\ref{eq:NoisekTB}), we can write the noise figure in the following form:
\begin{equation}
\begin{array}{lcl}
\displaystyle F & = & \displaystyle  \frac{\di \left( \frac{S_i}{N_i} \right)}{\di \left( \frac{S_o}{N_o} \right) }  =  \displaystyle  \frac{\di \left( \frac{S_i}{k_b T_0 B}\right) }{\di \left( \frac{G_{av} S_i}{N_o}\right) }
 = \frac{N_o}{G_{av} k_b T_0 B} .
 \end{array} \label{eq:Noise1}
\end{equation}
The noise which is added by the amplifier $\Delta N^{amp}_i$ seen {\it at the input} of a noise-free amplifier with available power gain $G_{av}$ is given by:
\begin{equation}
\begin{array}{lcl}
\displaystyle \Delta N^{amp}_i & = & \displaystyle  \frac{N_o-G_{av}N_i}{G_{av}}=(F-1)k_b T_0 B .
 \end{array} \label{eq:Noiseinputamplifier}
\end{equation}
Apart from the noise figure, it is often also convenient to express the noise performance of a microwave network in terms of the equivalent noise temperature $T_e$ with
\begin{equation}
\begin{array}{lcl}
\displaystyle T_e & = & \displaystyle  (F-1) T_0.
 \end{array} \label{eq:Noisetemperature1}
\end{equation}

\begin{figure}[hbt]
\centerline{\psfig{figure=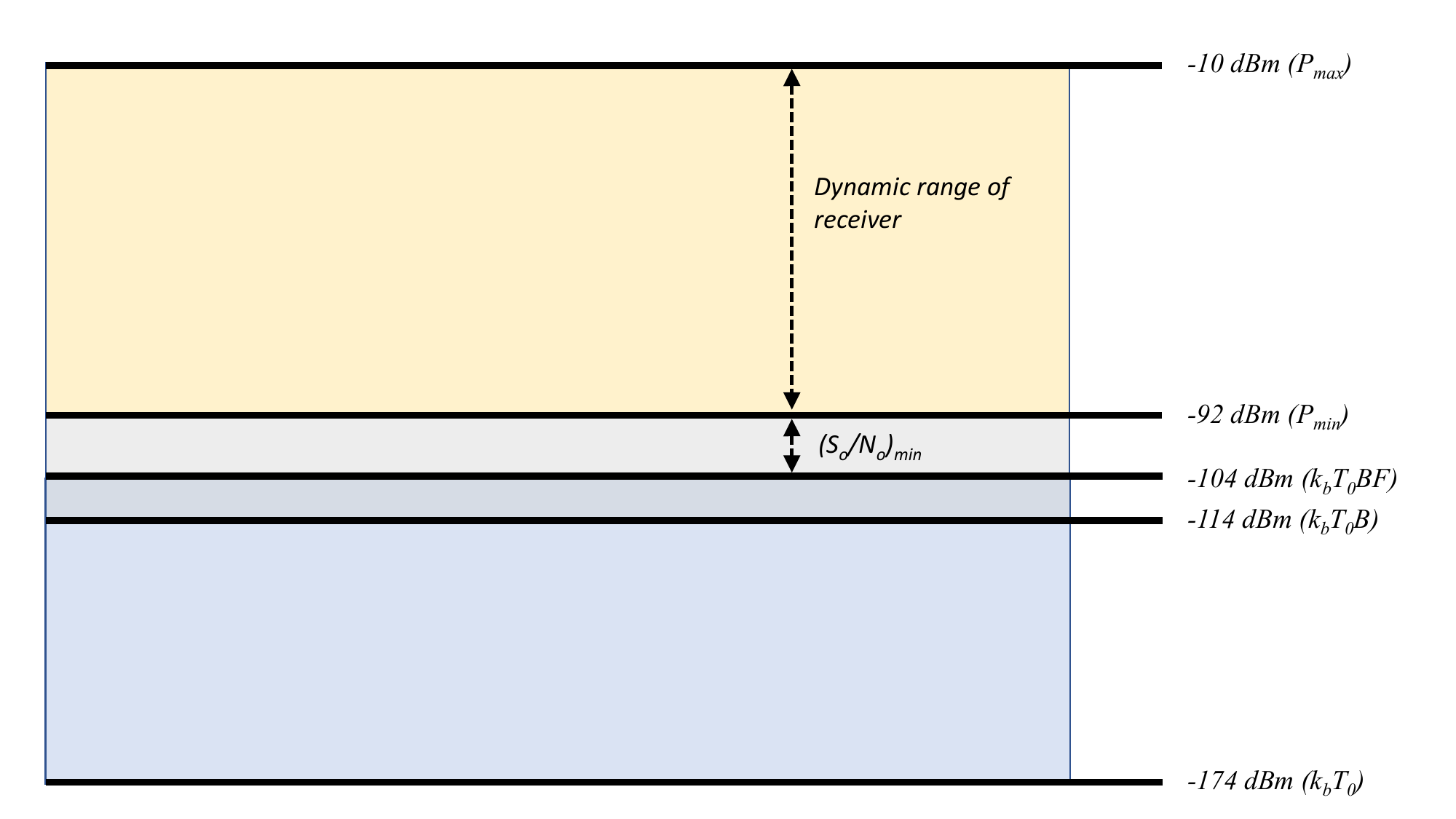,width=160mm}}
\caption{\it Typical power level diagram of a Bluetooth receiver. All power levels are defined at the input of the receiver at room temperature ($T_0=300~$K). The channel bandwidth $B=1~$MHz, the minimum required SNR for demodulation is $S_o/N_o=12~$dB, the receiver noise figure $NF=10~$dB and the maximum tolerated input power $P_{max}=-20~$dBm.}
\label{fig:leveldiagram}
\end{figure}

In order to visualize the noise performance and dynamic range of microwave receivers, a power level diagram can be a very useful tool. Fig. \ref{fig:leveldiagram} illustrates an example of such a level diagram. In this case a Bluetooth receiver is considered operating in the 2.4-2.48 GHz band with an instantaneous channel bandwidth of 1 MHz. All power levels are determined at the input of the receiver. The equivalent input noise power is $k_b T_0 B F$. The minimum required SNR required to demodulate the modulated received signal is assumed to be $S_o/N_o=12~$dB. From the level diagram we can determine the dynamic range in which the receiver has to operate.

\begin{figure}[hbt]
\centerline{\psfig{figure=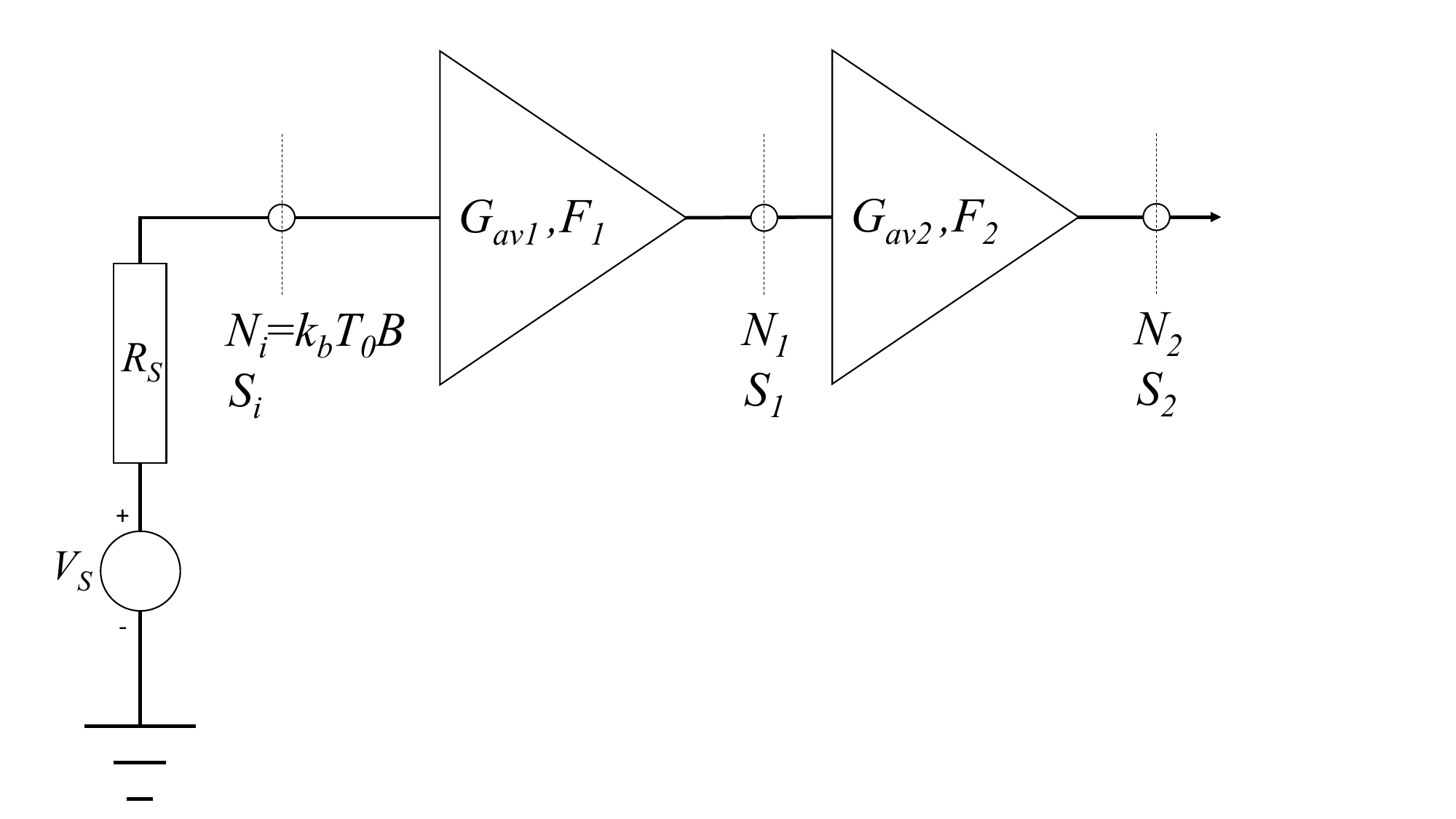,width=120mm}}
\caption{\it Cascaded two-stage amplifier.}
\label{fig:cascadednoise}
\end{figure}

Microwave receivers usually consist of several cascaded stages, each stage being characterized by an available power gain and noise figure. In Fig. \ref{fig:cascadednoise} a two-stage network is shown, consisting of a source and two amplifier stages.
At the output of the first stage we find:
\begin{equation}
\begin{array}{lcl}
\displaystyle S_1 & = & \displaystyle  G_{av1} S_{i}, \\
\displaystyle N_1 & = & \displaystyle  G_{av1} k_b T_0 B F_1.
 \end{array} \label{eq:Friis1}
\end{equation}
As a consequence, the input signal power and noise power for the second stage are given by $S_1$ and $N_1$ and provide at the output of the second stage:
\begin{equation}
\begin{array}{lcl}
\displaystyle S_2 & = & \displaystyle  G_{av2} S_{1} = G_{av1} G_{av2} S_{i}, \\
\displaystyle N_2 & = & \displaystyle  G_{av2} N_1 + G_{av2} (F_2-1) k_b T_0 B = G_{av2} k_b T_0 B [G_{av1} F_1+(F_2-1)],
 \end{array} \label{eq:Friis2}
\end{equation}
where relation (\ref{eq:Noiseinputamplifier}) was used. The total noise figure of the cascaded two-stage network is then given by
\begin{equation}
\begin{array}{lcl}
\displaystyle F_{tot} & = & \displaystyle  \frac{\di \left( \frac{S_i}{N_i} \right)}{\di \left( \frac{S_2}{N_2} \right) }  =  \displaystyle  \frac{\di \left( \frac{S_i}{k_b T_0 B}\right) }{\di \left( \frac{G_{av1} G_{av2} S_i}{G_{av2}k_b T_0 B[G_{av1} F_1+(F_2-1)] }\right) }
 = F_1+ \frac{F_2-1}{G_{av1}} .
 \end{array} \label{eq:Friis3}
\end{equation}
Equation (\ref{eq:Friis3}) is also known as the {\it Friis formula} named according to the Danish electrical engineer Harald Friis. As a consequence of this relation, the noise contribution of the second stage is significantly reduced thanks to the available power gain of the first stage. In receiver systems the first amplifier determines the noise performance and should have a low noise figure and reasonable power gain. Often special (and more expensive) semiconductor technologies are used to realize the first stage, for example by using III-V technologies such as Gallium Arsenide (GaAs).
Similar to (\ref{eq:Friis3}) we can also determine the equivalent noise temperature of a cascaded two-stage network, resulting in:
\begin{equation}
\begin{array}{lcl}
\displaystyle T_e^{tot} & = & \displaystyle T_{e1}+\frac{T_{e2}}{G_{av1}},
 \end{array} \label{eq:Friis4}
\end{equation}
where $T_{e1}$ and $T_{e2}$ are the equivalent noise temperatures of the first and second stage, respectively, which can be found using (\ref{eq:Noisetemperature1}).
Friis formula can be extended to a cascaded network with $N$ stages, providing:
\begin{equation}
\begin{array}{lcl}
\displaystyle F_{tot} & = & \displaystyle F_1+ \frac{F_2-1}{G_{av1}}+\frac{F_3-1}{G_{av1}G_{av2}}+~\cdot \cdot \cdot~+ \frac{F_N-1}{G_{av1}G_{av2}~\cdot \cdot \cdot~G_{avN-1}},\\
\displaystyle T_e^{tot} & = & \displaystyle T_{e1}+ \frac{T_{e2}}{G_{av1}}+\frac{T_{e3}}{G_{av1}G_{av2}}+~\cdot \cdot \cdot~+ \frac{T_{eN}}{G_{av1}G_{av2}~\cdot \cdot \cdot~G_{avN-1}}.\\
 \end{array} \label{eq:Friis5}
\end{equation}

\vspace*{0.5cm} \hrulefill
 {\bf Exercise} \hrulefill \\
 Determine the noise figure of a two-stage amplifier at room temperature. The first stage is a high-performance GaAs low-noise amplifier with a 15 dB power gain and 1.6 dB noise figure. The second stage is realized in low-cost Silicon with a power gain of 25 dB and 4 dB noise figure. Also determine the overall equivalent noise temperature.
{\it Answer:} $NF_{tot}=1.65~$dB and $T_e^{tot}= 138.1~$K. \\
\hrulefill \\
\vspace*{0.25cm}

\subsection{Noise Matching}
\begin{figure}[hbt]
\vspace{-1.5cm}
\centerline{\psfig{figure=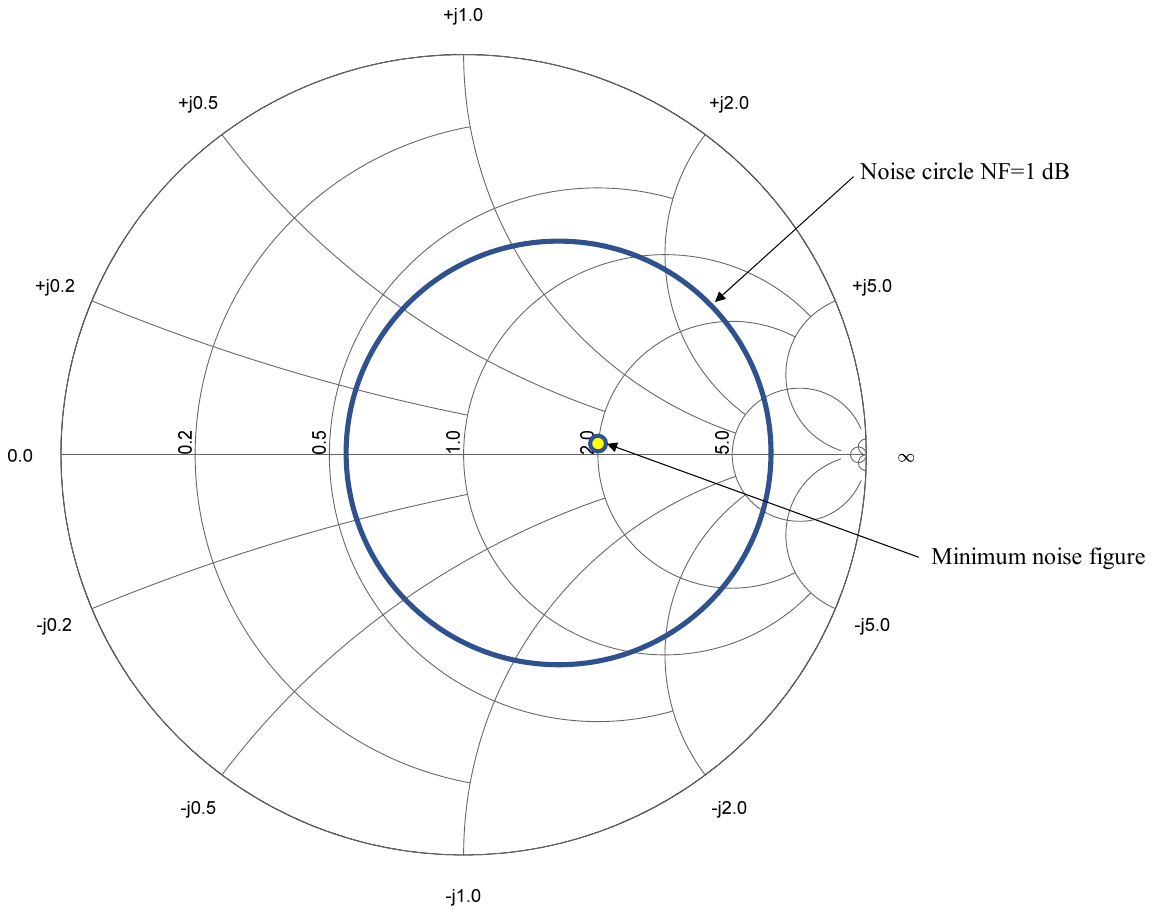,width=170mm}}
\vspace{-1.5cm}
\caption{\it Noise circle for $NF=1~$dB corresponding to an amplifier based on the BFU730F transistor at 400 MHz biased at $I_{c}=10~$mA and $V_{CE}=2~$V. When the source impedance falls within the noise circle, the resulting noise figure will be below or equal to 1~dB. The minimum noise figure of 0.57~dB is obtained at $Z_{opt}=100+\jmath 5.2~\Omega$ or $\Gamma_{opt}= 0.334+\jmath 0.023$}
\label{fig:NoisecircleBFU730F}
\end{figure}

Consider a low-noise amplifier which is connected to a source with complex source impedance $Z_S=R_S+\jmath X_S$. The noise figure of this amplifier can be determined from the following relation \cite{Gonzalez}, \cite{Ludwig}:
\begin{equation}
\begin{array}{lcl}
\displaystyle F & = & \displaystyle F_{min} + \frac{R_n}{G_S} \left|Y_S-Y_{opt} \right|^2,
 \end{array} \label{eq:noisematching1}
\end{equation}
where the source admittance is $Y_S=1/Z_S=G_S+\jmath B_S$. The other three parameters in (\ref{eq:noisematching1}) quantify the noise performance of a particular single-stage low-noise amplifier:
\begin{itemize}
\item{$F_{min}$, the minimum noise figure at particular DC-biasing of the transistor,}
\item{$R_n=1/G_n$, the equivalent noise resistance of the transistor,}
\item{$Y_{opt}$, the optimum source admittance that provides the minimum noise figure.}
\end{itemize}
Relation (\ref{eq:noisematching1}) can also be rewritten in terms of impedance values:
\begin{equation}
\begin{array}{lcl}
\displaystyle F & = & \displaystyle F_{min} + \frac{G_n}{R_S} \left|Z_S-Z_{opt} \right|^2,
 \end{array} \label{eq:noisematching2}
\end{equation}
Sometimes (e.g. in case of ultra low-noise amplifiers for radio astronomy) it is also convenient to convert (\ref{eq:noisematching1}) to the equivalent noise temperature:
\begin{equation}
\begin{array}{lcl}
\displaystyle T_e & = & \displaystyle T_{min} + \frac{R_n}{G_S} \left|Y_S-Y_{opt} \right|^2,
 \end{array} \label{eq:noisematching3}
\end{equation}
where the minimum noise temperature is $T_{min}=(F_{min}-1)T_0$.
It appears to be very convenient for the design of a particular low-noise amplifier to use the Smith chart. In the Smith chart we can indicate the region for the source impedance $Z_S$ for which the amplifier has a noise figure below a certain value. In the same Smith chart other regions can be indicated for $Z_S$ in which other relevant performance indicators, such as the available power gain $G_{av}$, is higher than a certain minimum required value. If the regions overlap, a suitable choice for $Z_S$ can be obtained that satisfies all requirements.
In order to be able to use (\ref{eq:noisematching1}) in a Smith chart, we need to convert the admittance values to reflection parameters. By using
\begin{equation}
\begin{array}{lcl}
\displaystyle Y_{opt} & = & \displaystyle Y_0 \frac{1-\Gamma_{opt}}{1+\Gamma_{opt}}, \\
\displaystyle Y_{S} & = & \displaystyle Y_0 \frac{1-\Gamma_{S}}{1+\Gamma_{S}},
 \end{array} \label{eq:noisematching4}
\end{equation}
we can transform (\ref{eq:noisematching1}) in the following form:
\begin{equation}
\begin{array}{lcl}
\displaystyle F & = & \displaystyle F_{min} + \frac{4R_n}{Z_0} \frac{\left|\Gamma_S-\Gamma_{opt} \right|^2}
{(1- \left|\Gamma_S \right|^2)\left|1+ \Gamma_{opt} \right|^2},
\end{array} \label{eq:noisematching5}
\end{equation}
where $\Gamma_{opt}$ is the optimum source reflection coefficient that would result in a minimum noise figure and $\Gamma_S$ is the realized source reflection coefficient. It can be shown that (\ref{eq:noisematching5}) represents a circle for $\Gamma_S$ in the Smith Chart with center $O_N$ and radius $R_N$ given by \cite{Ludwig}:
\begin{equation}
\begin{array}{lcl}
\displaystyle O_N & = & \displaystyle \frac{\Gamma_{opt}}{1+N}, \\
\displaystyle R_N & = & \displaystyle \frac{1}{1+N} \sqrt{N^2+N\left(1- \left|\Gamma_{opt} \right|^2 \right)}, \\
\displaystyle N & = & \displaystyle \frac{F-F_{min}}{4R_n/Z_0} \left| 1+\Gamma_{opt} \right|^2, \\
\end{array} \label{eq:noisematching6}
\end{equation}
where $N$ is also known as the noise figure parameter.
Let us now consider an example to illustrate how noise circles can be used. Consider the SiGe bipolar transistor BFU730F from NXP Semiconductors.
From the datasheet we find that at 400 MHz, with a DC collector current of $I_{c}=10~$mA and collector-emitter voltage $V_{CE}=2~$V, we get $NF_{min}=0.57~$dB, $R_n=6~\Omega$ and $Z_{opt}=100+\jmath 5.2~\Omega$. Suppose we want to design an amplifier using this transistor with a noise figure below 1~dB. The noise circle that defines the area in which $\Gamma_S$ can be selected to obtain a maximum noise figure of 1~dB is given by:
\begin{equation}
\begin{array}{lcl}
\displaystyle N & = & \displaystyle 0.443, \\
\displaystyle O_N & = & \displaystyle 0.2315+\jmath 0.016, \\
\displaystyle R_N & = & \displaystyle 0.53. \\
\end{array} \label{eq:noisematching7}
\end{equation}
The resulting noise circle is shown in Fig. \ref{fig:NoisecircleBFU730F}. We can observe that with a $50~\Omega$ source connected at the input of the transistor, a noise figure below $1~$dB will be obtained.

\vspace*{0.5cm} \hrulefill
 {\bf Exercise} \hrulefill \\
 Determine expression (\ref{eq:noisematching5}) from (\ref{eq:noisematching1}) and (\ref{eq:noisematching4}). \\
\hrulefill \\
\vspace*{0.25cm}

\chapter{Antenna Theory}
\label{chap:AntennaTheory}

In this chapter we will introduce the general properties of radiated fields from antenna sources.
We will use Maxwell's equations in the space-frequency domain as a starting point in which the antenna is represented by a current or charge distribution.
Next, we will determine the electromagnetic field in the far field of an arbitrarily electric current or charge distribution. Based on this, several elementary antenna structures will be analyzed, such as wire antennas and loop antennas.
Finally, we will introduce magnetic sources and the Lorentz-Larmor theorem in order to extend the theoretical framework towards the analysis of aperture antennas, such as horns, reflector antennas and microstrip antennas.

\section{Maxwell's equations}
\label{sec:antennatheory}

Maxwell's equations describe the propagation of electromagnetic waves generated by a current or charge distribution in any medium.
In the time domain Maxwell's equations are given by
\begin{equation}
\begin{array}{lcl}
\displaystyle \nabla \times \vec{\cal{E}}(\vec{r},t) &=&
\displaystyle - \frac{\partial \vec{\cal B}(\vec{r},t)}{\partial
t},
\\

\displaystyle \nabla \times \vec{\cal{H}}(\vec{r},t) &=&
\displaystyle \frac{\partial \vec{\cal{D}}(\vec{r},t)}{\partial t}
+ \vec{{\cal J}_e}(\vec{r},t), \\

\displaystyle \nabla \cdot \vec{\cal B} (\vec{r},t) & = & 0, \\
\displaystyle \nabla \cdot \vec{\cal D} (\vec{r},t) & = &
\displaystyle  {\cal \rho}_e (\vec{r},t),

\end{array}
\label{eq:maxwelltijd}
\end{equation}
where $\vec{\cal E}$ is the electric field in Volt per meter [V/m],  $\vec{\cal D}$ the displacement current in Coulomb per $\mbox{m}^2$ [$\mbox{C/m}^2$], $\vec{\cal B}$ the magnetic flux density in Weber per
$\mbox{m}^2$ [$\mbox{W/m}^2$], $\vec{\cal H}$ the magnetic field in Ampere per
meter [A/m], ${\cal \rho}_e$ the electric charge distribution in Coulomb
per $\mbox{m}^3$ [$\mbox{C/m}^3$] and $\vec{{\cal J}_e}$ the electric current distribution in Ampere
per $\mbox{m}^2$ [$\mbox{A/m}^2$]. In addition, the continuity equation provides a relation between the electric current and charge distribution:
\begin{equation}
\begin{array}{lcl}

\displaystyle \nabla \cdot \vec{\cal J}_e (\vec{r},t) +
\frac{\partial {\cal \rho}_e (\vec{r},t)}{\partial t} & = & 0.

\end{array}
\label{eq:continuiteitsvgl}
\end{equation}
Note that the continuity equation is not an independent equation, but can be derived from Maxwell's equations (\ref{eq:maxwelltijd}).

Since the propagation of electromagnetic waves from a transmit antenna to a receive antenna usually takes place in free-space, we can state that:
\begin{equation}
\begin{array}{lcl}

\displaystyle \vec{\cal D} (\vec{r},t)  & = & \displaystyle
\epsilon_0 \vec{\cal E}(\vec{r},t) , \\

\displaystyle \vec{\cal B} (\vec{r},t)  & = & \displaystyle  \mu_0
\vec{\cal H}(\vec{r},t).

\end{array}
\label{eq:freespacevoorwaarden}
\end{equation}
Furthermore, we will again assume a time-harmonic dependence of all field components, currents and charges, with angular frequency $\omega=2\pi f$, where $f$ is the frequency expressed in Hertz [Hz]. In this way, the electric field is expressed as:
\begin{equation}
\begin{array}{lcl}

\displaystyle \vec{\cal E} (\vec{r},t)  & = & \displaystyle Re
\left[ \vec{E} (\vec{r}) e^{\jmath \omega t} \right].

\end{array}
\label{eq:harmonischveld}
\end{equation}
A similar relation holds for the other components of (\ref{eq:maxwelltijd}).
Maxwell's equations now reduce to the following set of equations:
\begin{equation}
\begin{array}{lcl}
\displaystyle \nabla \times \vec{E}(\vec{r}) &=& \displaystyle -
\jmath \omega \mu_0 \vec{H}(\vec{r}), \\

\displaystyle \nabla \times \vec{H}(\vec{r}) &=& \displaystyle
\jmath \omega \epsilon_0 \vec{E}(\vec{r}) + \vec{J_e}(\vec{r}),
\\

\displaystyle \nabla \cdot \vec{H} (\vec{r}) & = & 0, \\
\displaystyle \nabla \cdot \vec{E} (\vec{r}) & = & \displaystyle
\frac{\rho_e (\vec{r})}{\epsilon_0}.

\end{array}
\label{eq:maxwellfreq}
\end{equation}

Next step is to determine the electric and magnetic fields,
$\vec{E}(\vec{r})$ and $\vec{H}(\vec{r})$,  due to the source distributions, given by
$\vec{J}_e(\vec{r})$ and  $\rho_e(\vec{r})$. From these frequency-domain fields, we can then determine the physical (time-domain) fields $\vec{\cal E}(\vec{r},t)$ and $\vec{\cal
H}(\vec{r},t)$ by using relation (\ref{eq:harmonischveld}).
Note that in (\ref{eq:maxwellfreq}), we have used the linear properties of Maxwell's equations, which allows to interchange the vector operators $\nabla \times$ and $\nabla \cdot$ with the operator $Re[]$. In the rest of the derivation, we will omit the use of  $(\vec{r})$ in our notation.
From $\nabla \cdot \vec{H} = 0$ we can conclude that $\vec{H}$
can be expressed in terms of a vector potential $\vec{A}_e$ according to
\begin{equation}
\displaystyle \vec{H} = \frac{1}{\mu_0} \nabla \times \vec{A}_e.
\label{eq:vectorpot1}
\end{equation}
Note that $\vec{A}_e$ does not have a unique solution, since the vector $\vec{A}^1_e=\vec{A}_e + \nabla \phi_e$, with $\phi_e$
an arbitrarily scalar potential, provides exactly the same solution of  $\vec{H}$.
Substitution of (\ref{eq:vectorpot1}) in the first equation of  (\ref{eq:maxwellfreq}) provides
\begin{equation}
\displaystyle \nabla \times \vec{E} = -\jmath \omega \nabla \times
\vec{A}_e, \label{eq:rotEenA}
\end{equation}
which leads to
\begin{equation}
\displaystyle \vec{E} = -\jmath \omega \vec{A}_e - \nabla \phi_e,
\label{eq:EenA}
\end{equation}
where $\phi_e$ is an unknown scalar function of
$\vec{r}$. Substituting  (\ref{eq:EenA}) in the second equation of  (\ref{eq:maxwellfreq}) gives
\begin{equation}
\displaystyle \nabla \times \nabla \times \vec{A}_e = \jmath
\omega \epsilon_0 \mu_0 \vec{E} + \mu_0 \vec{J}_e = k_0^2
\vec{A}_e - \jmath \omega \epsilon_0 \mu_0 \nabla \phi_e + \mu_0
\vec{J}_e. \label{eq:EenA2}
\end{equation}
Where $k_0$ is the wavenumber in free space with
$k_0^2=\omega^2 \epsilon_0 \mu_0 = \omega^2/c^2$, with $c$ the speed of light.
By using the vector property $\nabla
\times \nabla \times \vec{A}_e = \nabla \nabla \cdot \vec{A}_e -
\nabla^2 \vec{A}_e$, (\ref{eq:EenA2}) becomes
\begin{equation}
\displaystyle \nabla^2 \vec{A}_e + k_0^2 \vec{A}_e = \nabla \nabla
\cdot \vec{A}_e + \jmath \omega \epsilon_0 \mu_0 \nabla \phi_e
-\mu_0 \vec{J}_e. \label{eq:Helmholtz1}
\end{equation}
Up to now, we did not specify the scalar potential $\phi_e$. We will choose $\phi_e$ according to
\begin{equation}
\displaystyle \nabla \cdot \vec{A}_e = - \jmath \omega \epsilon_0
\mu_0 \phi_e . \label{eq:Lorentzijk}
\end{equation}
This particular choice is known as the {\it Lorenz-gauge}. Equation
(\ref{eq:Helmholtz1}) can now be re-written in the well-known  {\it Helmholtz
equation}
\begin{equation}
\displaystyle \nabla^2 \vec{A}_e + k_0^2 \vec{A}_e = -\mu_0
\vec{J}_e. \label{eq:Helmholtz2}
\end{equation}
Note that we also need to satisfy the last equation of
(\ref{eq:maxwellfreq}). This results in
\begin{equation}
\displaystyle \nabla \cdot (\epsilon_0 \vec{E}) = - \jmath \omega
\epsilon_0 \nabla \cdot \vec{A}_e -\epsilon_0 \nabla^2 \phi_e =
\rho_e. \label{eq:EenRho}
\end{equation}
Again, we apply the Lorenz-gauge, which provides us
\begin{equation}
\displaystyle \nabla^2 \phi_e + k_0^2 \phi_e = -
\frac{\rho_e}{\epsilon_0}. \label{eq:Helmholtzrho}
\end{equation}

\begin{figure}[hbt]
\centerline{\psfig{figure=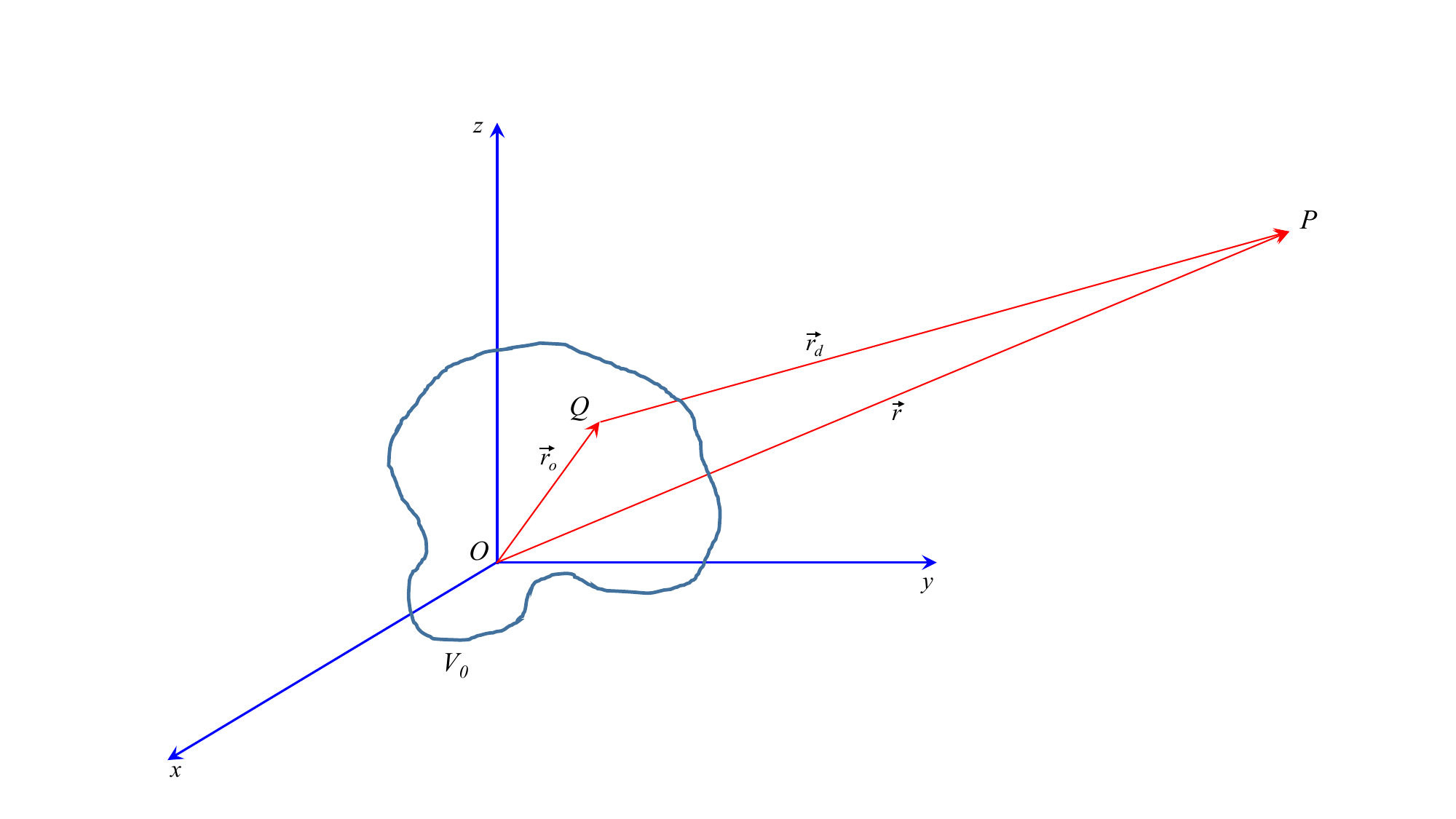,width=170mm}}
\caption{\it The antenna current and charge distribution are located within the volume $V_0$}
\label{fig:JinVolume}
\end{figure}

\begin{figure}[hbt]
\centerline{\psfig{figure=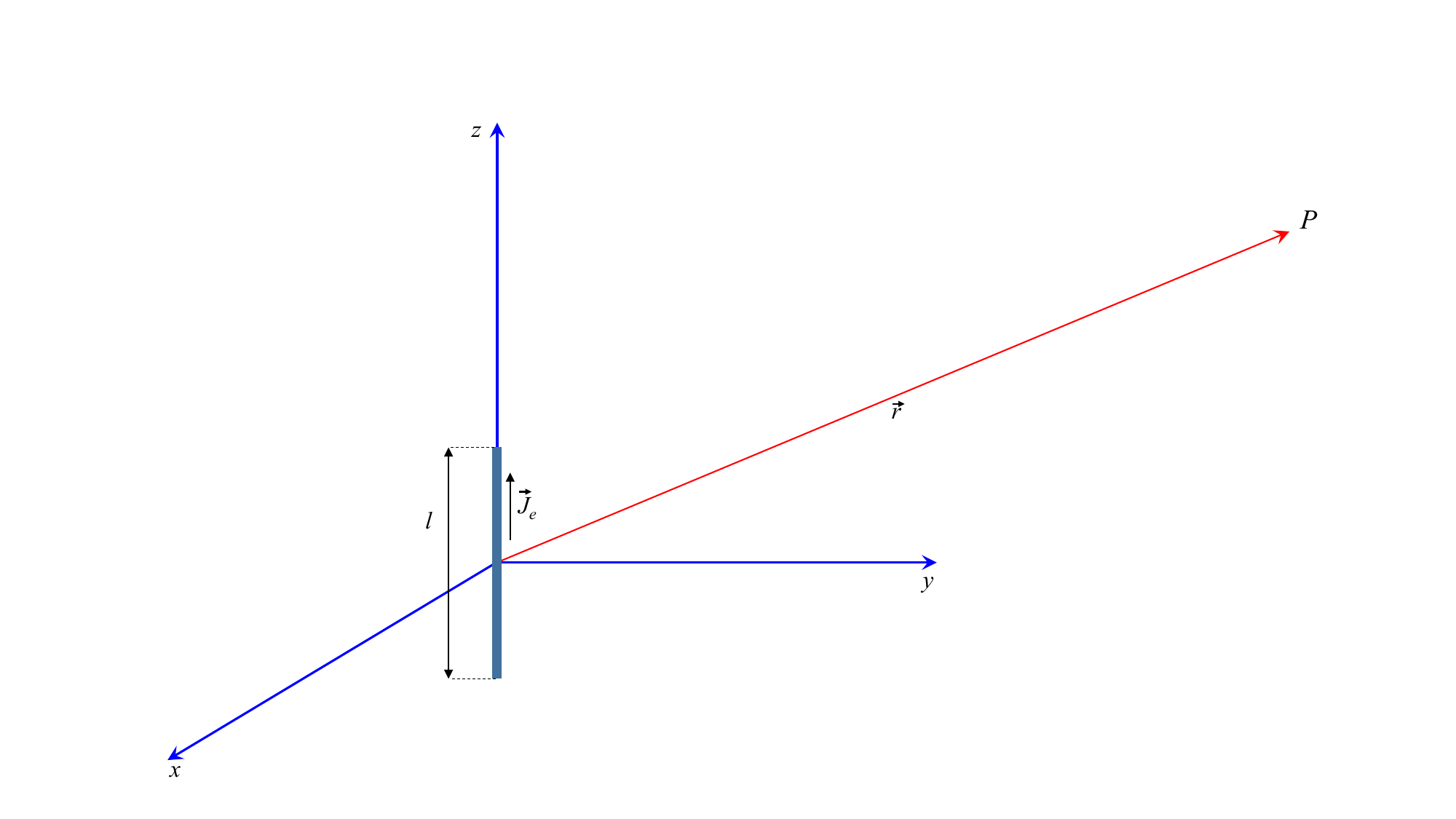,width=170mm}}
\caption{\it Electric dipole located in the origin and oriented along the
$z$-axis with $\vec{J}_e=I_0 l \vec{u}_z$.}
\label{fig:eldipole}
\end{figure}

When considering antennas, the sources which are represented by the electric current density $\vec{J}_e$ and charge distribution $\rho_e$, will be located within a finite-size volume $V_0$. This is illustrated in Fig. \ref{fig:JinVolume}.
As a first step, we will determine the solution of the Helmholtz equation for an electric dipole (point source) oriented along the $z$-axis located at the origin of the coordinate system
$(x,y,z)=(0,0,0)$ , where $\vec{u}_z$ represents the unit vector along the $z$-direction. The configuration is illustrated in Fig. \ref{fig:eldipole}.
Now let $I_0$ be the total current flowing through the dipole and $l$ the length of the dipole. Characteristic for the electric dipole is the fact that the length is much smaller than the free-space wavelength, so $l \ll \lambda_0$. The product $I_0 l=p$ is also called the "dipole-moment".  The current density along the electric dipole can now te expressed as:
\begin{equation}
\displaystyle \vec{J}_e = I_0l \delta(x) \delta(y) \delta(z)
\vec{u}_z = I_0l \delta(\vec{r}) \vec{u}_z. \label{eq:Edipool}
\end{equation}
Here $\delta(r)$ represents the three-dimensional delta function. Since the current has only a component in the $z$ direction, the vector potential will also only have a component in the $z$-direction, i.e.
$\vec{A}_e=A_{ez}\vec{u}_z$. The vector potential now satisfies the following equation
\begin{equation}
\displaystyle \nabla^2 A_{ez} + k_0^2 A_{ez} = -\mu_0 I_0l
\delta(\vec{r}). \label{eq:HelmholtzEdipol1}
\end{equation}
The scalar quantity $A_{ez}$ needs to represent a spherical behavior, since we have assumed a point source. As a result
$A_{ez}(\vec{r})=A_{ez}(r)$. Therefore, we can rewrite (\ref{eq:HelmholtzEdipol1}) in the following form
\begin{equation}
\displaystyle \frac{1}{r^2}\frac{d}{dr}\left( r^2
\frac{dA_{ez}}{dr} \right) + k_0^2 \vec{A}_{ez} = -\mu_0 I_0l
\delta(\vec{r}). \label{eq:HelmholtzEdipol2}
\end{equation}
Outside the origin, we now find two possible solutions of this differential equation:
\begin{equation}
\begin{array}{lcl}
 \displaystyle A^1_{ez} & =& \displaystyle \frac{C}{r} e^{\jmath k_0
 r} ,\\
\displaystyle A^2_{ez} & =& \displaystyle \frac{C}{r} e^{-\jmath
k_0 r} .
\end{array}
\end{equation}
where $C$ is a constant. The first solution $A^1_{ez}$ represents a spherical wave propagating towards the source, whereas the second solution
$A^2_{ez}$ represents a wave which propagates away from the source. This is due to the fact that we have assumed a
$\displaystyle e^{\jmath \omega t}$ time dependence of the fields.
We will use the second solution, where we still need to determine the constant $C$.
We can determine $C$ by substituting $k_0 \downarrow 0$
in (\ref{eq:HelmholtzEdipol2}), which transforms into the Poisson equation known from electrostatics with solution
\begin{equation}
\begin{array}{lcl}
 \displaystyle A_{ez} & =& \displaystyle \frac{\mu_0 I_0 l}{4\pi r}.
\end{array}
\end{equation}
As a result, we find that $\displaystyle C= \frac{\mu_0 I_0 l}{4 \pi}$.
The general solution of the vector potential $\vec{A}_e$ of an electric dipole along the $z$-axis now takes the following form:
\begin{equation}
 \displaystyle \vec{A}_e = \vec{u}_z A_{ez}  =  \vec{u}_z \frac{\mu_0 I_0 l}{4 \pi r} e^{-\jmath k_0
 r} .
 \label{eq:vecpoteldipole}
\end{equation}
The vector potential in (\ref{eq:vecpoteldipole}) has the same direction as the dipole. This can be generalized to all directions.
Therefore, the vector potential of an arbitrarily oriented dipole located at $\vec{r}_0=\vec{u}_x x_0 + \vec{u}_y y_0 +
\vec{u}_z z_0$ can be found by substituting  $r$ in expression
(\ref{eq:vecpoteldipole}) by  $|\vec{r}-\vec{r}_0|=
\sqrt{(x-x_0)^2+(y-y_0)^2+(z-z_0)^2}$. We now obtain
\begin{equation}
\begin{array}{lcl}
\displaystyle \vec{A}_e (\vec{r}) & = & \displaystyle I_0 \vec{l}
(\vec{r}_0) \frac{\mu_0}{4\pi} \frac{e^{-\jmath k_0
|\vec{r}-\vec{r}_0|}}{|\vec{r}-\vec{r}_0|} \\

& = & \displaystyle I_0 \vec{l} (\vec{r}_0) G(\vec{r},\vec{r}_0).
\end{array}
\label{eq:Greensfunctie}
\end{equation}
The function $G(\vec{r},\vec{r}_0)$ is known as the
{\it Greens function} and represents the response of a point source.
The Greens function plays an important role in numerical electromagnetic methods (e.g. Method-of-Moments) which are used to analyze complex antenna structures.
More details can be found in chapter \ref{chap:NumEM}.

The vector potential of a general electric current distribution $\vec{J}_e$ within the volume $V_0$ can be obtained by applying the superposition principle.
The solutions of expression (\ref{eq:Helmholtz2}) and
(\ref{eq:Helmholtzrho}) for an observation point $P$ outside the source volume $V_0$ can now be easily found by:
\begin{equation}
\displaystyle \vec{A}_e (\vec{r}) = \frac{\mu_0}{4\pi} \int_{V_0}
\vec{J}_e(\vec{r}_0) \frac{e^{-\jmath k_0
|\vec{r}-\vec{r}_0|}}{|\vec{r}-\vec{r}_0|} dV_0,
\label{eq:vectorpot}
\end{equation}
\begin{equation}
\displaystyle \phi_e (\vec{r}) = \frac{1}{4\pi \epsilon_0}
\int_{V_0} \rho_e(\vec{r}_0) \frac{e^{-\jmath k_0
|\vec{r}-\vec{r}_0|}}{|\vec{r}-\vec{r}_0|} dV_0,
\label{eq:scalarpot}
\end{equation}
where $\vec{r}_0$ indicates a source coordinate in the volume $V_0$ and $\vec{r}$ indicates an observation point at which the fields are being determined.
It should be noted that the solutions of (\ref{eq:Helmholtz2}) and
(\ref{eq:Helmholtzrho}) should meet the Lorenz-gauge. By using the notation
\begin{equation}
\displaystyle \varphi (\vec{r},\vec{r}_0) = \frac{1}{4\pi}
\frac{e^{-\jmath k_0 |\vec{r}-\vec{r}_0|}}{|\vec{r}-\vec{r}_0|},
\label{eq:varphidef}
\end{equation}
we find that
\begin{equation}
\begin{array}{lcl}
\displaystyle \nabla_r \cdot \vec{J}_e (\vec{r}_0)
\varphi(\vec{r},\vec{r}_0)& = & \varphi(\vec{r},\vec{r}_0)
\nabla_r \cdot \vec{J}_e (\vec{r}_0) + ( \vec{J}_e (\vec{r}_0)
\cdot \nabla_r \varphi(\vec{r},\vec{r}_0)) \\

& = & \displaystyle  \vec{J}_e (\vec{r}_0) \cdot \nabla_r
\varphi(\vec{r},\vec{r}_0) \\

& = & \displaystyle  - \vec{J}_e (\vec{r}_0) \cdot \nabla_{r_0}
\varphi(\vec{r},\vec{r}_0).
\end{array}
\end{equation}
Note that the operator $\nabla_r$ is related to $\vec{r}$ and not to $\vec{r}_0$. Furthermore, we have used the fact that
$\nabla_r \varphi(\vec{r},\vec{r}_0)= - \nabla_{r_0}
\varphi(\vec{r},\vec{r}_0)$. As a next step, we will determine $\nabla_r \cdot
\vec{A}_e$.
\begin{equation}
\begin{array}{lcl}
\displaystyle \nabla_r \cdot \vec{A}_e (\vec{r}) & = &
\displaystyle \mu_0 \int_{V_0} \nabla_r \cdot \vec{J}_e
(\vec{r}_0) \varphi(\vec{r},\vec{r}_0) dV_0 \\

& = & \displaystyle \mu_0 \int_{V_0} \varphi(\vec{r},\vec{r}_0)
\nabla_{r_0} \cdot \vec{J}_e (\vec{r}_0) dV_0 - \mu_0 \int_{V_0}
\nabla_{r_0} \cdot \vec{J}_e (\vec{r}_0)
\varphi(\vec{r},\vec{r}_0)  dV_0

\end{array}
\label{eq:nablaA}
\end{equation}
By applying Gauss theorem and by selecting a volume $V_0$ much larger than the source region, the second integral on the right-hand side of equation (\ref{eq:nablaA}) will vanish. In the first integral we can replace $\vec{J}_e(\vec{r}_0)$ by means of the continuity equation
(\ref{eq:continuiteitsvgl}) by $\rho_e$. This results in the Lorenz-gauge:
\begin{equation}
\begin{array}{lcl}
\displaystyle \nabla_r \cdot \vec{A}_e (\vec{r}) & = &
\displaystyle - \jmath \omega \mu_0 \int_{V_0} \vec{\rho}_e
(\vec{r}_0) \varphi(\vec{r},\vec{r}_0) dV_0 \\

& = & \displaystyle -  \jmath \omega \epsilon_0 \mu_0 \phi_e
(\vec{r}).

\end{array}
\label{eq:lorentzijkbewijs}
\end{equation}
By using the Lorenz-gauge we are able to express the fields $\vec{E}$ and $\vec{H}$ directly in terms of the vector potential $\vec{A}_e$.
\begin{equation}
\displaystyle \vec{H} = \frac{1}{\mu_0} \nabla \times \vec{A}_e,
\label{eq:HinA}
\end{equation}
\begin{equation}
\begin{array}{lcl}

\displaystyle \vec{E} & = & \displaystyle  -\jmath \omega
\vec{A}_e - \nabla \phi_e \\

& = & \displaystyle  -\jmath \omega \vec{A}_e + \frac{\nabla
\nabla \cdot \vec{A}_e}{\jmath \omega \epsilon_0 \mu_0} \\

& = & \displaystyle  \frac{1}{\jmath \omega \epsilon_0 \mu_0}
 \left[ \nabla \times \nabla \times \vec{A}_e + \nabla^2 \vec{A}_e
 +k_0^2 \vec{A}_e \right] \\

 & = & \displaystyle  \frac{1}{\jmath \omega \epsilon_0 \mu_0}
 \left[ \nabla \times \nabla \times \vec{A}_e - \mu_0 \vec{J}_e
 \right].

\end{array}
\label{eq:EinA}
\end{equation}
Outside the source region, we now find that
\begin{equation}
\begin{array}{lcl}
\displaystyle \vec{H} & = & \displaystyle  \frac{1}{\mu_0} \nabla
\times \vec{A}_e, \\

\displaystyle \vec{E} & = & \displaystyle  \frac{1}{\jmath \omega
\epsilon_0 \mu_0} \nabla \times \nabla \times \vec{A}_e.

\end{array}
\label{eq:HenEinA}
\end{equation}
Since the vector potential $\vec{A}_e$ only depends on $\vec{J}_e$ (and does not depend on the charge distribution $\rho_e$), we only need to know the electric current distribution $\vec{J}_e$ on a particular antenna structure in order to determine the radiated electromagnetic fields.
This is a consequence of the applied Lorenz-gauge.

\section{Radiated fields and far-field approximation}
\label{sec:radiatedfields}

In chapter \ref{chap:fundpar} we have discussed the various field regions of antennas.
Since most antennas are used to transfer information over long distances, we are primarily interested in the fields far away from the antenna.
In this section we will determine a general analytic expression of the electromagnetic field in the so-called far-field region of the antenna.
The derivation starts with the expression of the vector potential (\ref{eq:vectorpot}), which we derived in the previous section.
Combining this with (\ref{eq:HenEinA}) results in
\begin{equation}
\begin{array}{lcl}
\displaystyle \vec{H} & = & \displaystyle  \frac{1}{\mu_0} \nabla
\times \vec{A}_e \\
 \displaystyle & = &  \displaystyle \int_{V_0} \nabla_r \times
 \vec{J}_e (\vec{r}_0) \varphi(\vec{r},\vec{r}_0) dV_0,

\end{array}
\end{equation}
where $\varphi$ has been defined in \ref{eq:varphidef}. Furthermore, operator $\nabla_r \times$ is performed on the field point $\vec{r}$.
We now know that
\begin{equation}
\displaystyle \nabla_r \times
 \vec{J}_e (\vec{r}_0) \varphi(\vec{r},\vec{r}_0) =
 \varphi(\vec{r},\vec{r}_0)\nabla_r \times \vec{J}_e (\vec{r}_0) +
 \nabla_r \varphi(\vec{r},\vec{r}_0) \times \vec{J}_e (\vec{r}_0).
\end{equation}
Note that $\nabla_r \times \vec{J}_e (\vec{r}_0)=0$ and
\begin{equation}
\displaystyle \nabla_r \varphi(\vec{r},\vec{r}_0) = \frac{1}{4\pi}
e^{-\jmath k_0 |\vec{r}-\vec{r}_0|} \nabla_r\left(
\frac{1}{|\vec{r}-\vec{r}_0|} \right) + \frac{1}{4\pi
|\vec{r}-\vec{r}_0|} \nabla_r \left( e^{-\jmath k_0
|\vec{r}-\vec{r}_0|} \right).
\end{equation}
By using the following relations
\begin{equation}
\displaystyle  \nabla_r\left( \frac{1}{|\vec{r}-\vec{r}_0|}
\right) = - \frac{\vec{r}-\vec{r}_0}{|\vec{r}-\vec{r}_0|^3},
\end{equation}
and
\begin{equation}
\displaystyle \nabla_r \left( e^{-\jmath k_0 |\vec{r}-\vec{r}_0|}
\right) = -\jmath k_0 e^{-\jmath k_0 |\vec{r}-\vec{r}_0|}
\frac{(\vec{r}-\vec{r}_0)}{|\vec{r}-\vec{r}_0|},
\end{equation}
we find that
\begin{equation}
\begin{array}{lcl}
\displaystyle \vec{H}  & = &  \displaystyle \frac{1}{4\pi}
\int_{V_0} \left(-\jmath k_0 - \frac{1}{r_d} \right)
\frac{e^{-\jmath k_0 r_d}}{r_d} \vec{u}_{r_d} \times \vec{J}_e
(\vec{r}_0) dV_0,

\end{array}
\label{eq:HinrPQcoord1}
\end{equation}
where we have introduced the coordinate $\vec{r}_d$ according to
$\vec{r}_d=\vec{r}-\vec{r}_0$ with $r_d=|\vec{r}_d|$ and where the unit-vector in the direction $\vec{r}_d$ is given by
$\displaystyle \vec{u}_{r_d}=\frac{\vec{r}_d}{r_d}$.

We will now derive an approximation of the magnetic field $\vec{H}$, which is only valid at a large distance from the antenna system.  The distance
$r_d$ between a field point $P$ with coordinates $(x,y,z)$ and the source point with coordinates $(x_0,y_0,z_0)$ is given by
\begin{equation}
\displaystyle r_d = \sqrt{ \left( (x-x_0)^2 + (y-y_0)^2 +
(z-z_0)^2 \right)}.
\end{equation}
This can also be written in the following form
\begin{equation}
\displaystyle r_d = \sqrt{ \left( r^2 -2(xx_0 + yy_0 + zz_0) +
(x_0^2+y_0^2 + z_0^2) \right)}, \label{eq:rPQ}
\end{equation}
with $r^2=x^2+y^2+z^2$. In the far-field region $x_0, y_0$ and
$z_0$ will be very small as compared to $r$. As a result, we can express $r_d$ in terms of a binomial distribution according to
\begin{equation}
\begin{array}{lcl}
\displaystyle r_d  & = &  \displaystyle r \left[ 1+\frac{1}{2}
\left(-\frac{2}{r^2}\left( xx_0+yy_0+zz_0
\right)+\frac{x_0^2+y_0^2+z_0^2}{r^2} \right) \right. \\ & &
\displaystyle \left. - \frac{1}{8} \left( -\frac{2}{r^2}\left(
xx_0+yy_0+zz_0 \right)+\frac{x_0^2+y_0^2+z_0^2}{r^2} \right)^2 +
\cdot \cdot \cdot \cdot \right] \\

& = &  \displaystyle r \left[ 1 -
\frac{xx_0+yy_0+zz_0}{r^2}+\frac{x_0^2+y_0^2+z_0^2}{2r^2}-\frac{(xx_0+yy_0+zz_0)^2}{2r^4}+
\cdot  \cdot  \cdot \right] \\

& \approx &  \displaystyle r  - \frac{xx_0+yy_0+zz_0}{r} \\

& = &  \displaystyle r  - \vec{u}_r \cdot \vec{r}_0,

\end{array}
\end{equation}
with $\displaystyle \vec{u}_{r}= \frac{\vec{r}}{r}$. This approximation is only valid if the term  $\displaystyle
\frac{x_0^2+y_0^2+z_0^2}{2r}$ is small enough. The corresponding exponential factor will be close to 1.
The criterion which is often used in literature is
\begin{equation}
\displaystyle \frac{x_0^2+y_0^2+z_0^2}{2r} < \frac{\lambda_0}{16}.
 \label{eq:farfield2}
\end{equation}
Consider an antenna with maximum dimension $L$ and assume that the center of the antenna is located in the origin of our coordinate system.
Equation(\ref{eq:farfield2})now corresponds to the well-known far-field criterion introduced in chapter \ref{chap:fundpar}
\begin{equation}
\displaystyle R > \frac{2L^2}{\lambda_0}. \label{eq:verreveld2}
\end{equation}
We can now also replace the variable $r_d$ in the denominator of $\displaystyle \frac{e^{-\jmath k_0 r_d}}{r_d}$
by $r$ in expression (\ref{eq:HinrPQcoord1}) when the following conditions are fulfilled
\begin{equation}
\displaystyle \max(xx_0+yy_0+zz_0) \ll r^2,
\end{equation}
and
\begin{equation}
\displaystyle \max(x_0^2+y_0^2+z_0^2) \ll r^2.
\end{equation}
The second condition is fulfilled when $r \gg L$. In addition, the first condition is then also fulfilled since
$x/r$, $y/r$ and $z/r$ are both of order 1. If $r \gg \lambda_0$, we can replace $\displaystyle \left( -\jmath k_0 +\frac{1}{r_d} \right)$
by $-\jmath k_0$. By applying these approximations, we now obtain the following expression for the magnetic field $\vec{H}$
\begin{equation}
\begin{array}{lcl}
\displaystyle \vec{H}  & = &  \displaystyle \frac{-\jmath k_0
e^{-\jmath k_0 r} }{4\pi r} \int_{V_0}  \vec{u}_{r_d} \times
\vec{J}_e (\vec{r}_0) e^{\jmath k_0  \vec{u}_r \cdot \vec{r}_0} dV_0,
\end{array}
\label{eq:HinrPQcoord2}
\end{equation}
If our observation point $P$ is located far away from the antenna, the vector $\vec{u}_{r_d}$ and vector $\vec{u}_r$ will be parallel to eachother. As a consequence, we can replace $\vec{u}_{r_d}$ by $\vec{u}_r$. The unit vector $\vec{u}_r$ does not depend on the source point $Q$ and can, therefore, by moved outside the integral. We then finally obtain
\begin{equation}
\begin{array}{lcl}
\displaystyle \vec{H}  & = &  \displaystyle \frac{-\jmath k_0
e^{-\jmath k_0 r} }{4\pi r} \vec{u}_{r} \times  \int_{V_0}
\vec{J}_e (\vec{r}_0) e^{\jmath k_0  \vec{u}_r \cdot \vec{r}_0} dV_0,
\end{array}
\label{eq:Hverreveld}
\end{equation}
Expression (\ref{eq:Hverreveld}) leads to several important conclusions:
\begin{enumerate}[i.]
    \item $\vec{H}$ has an $\frac{1}{r}$ dependency.
    \item From $e^{-\jmath k_0 r}$ and the time-harmonic behavior $e^{\jmath \omega t}$,
    we can conclude that the waves propagate away from the source.
    \item $\vec{H} \cdot \vec{u}_r =0$. This implies that the radial component vanishes in the far-field region.
    \item The magnetic field can now be expressed as
    \begin{equation}
    \displaystyle
    \vec{H}=\frac{e^{-\jmath k_0 r}}{r}
    (\vec{u}_{\theta}H_{\theta} (\theta,\phi) + \vec{u}_{\phi}H_{\phi}
    (\theta,\phi)).
    \end{equation}
    The components $H_{\theta} (\theta,\phi)$ and $H_{\phi}
    (\theta,\phi))$ can be determined in the following way. We can express the electric current density in the following form
     \begin{equation}
    \displaystyle
    \vec{J}_e= J_{er} \vec{u}_r+J_{e\theta} \vec{u}_{\theta} + J_{e \phi} \vec{u}_{\phi}.
    \end{equation}
    We then obtain
    \begin{equation}
    \displaystyle
    \vec{u}_r \times \vec{J}_e= J_{e\theta} \vec{u}_{\phi} - J_{e \phi}
    \vec{u}_{\theta}.
    \end{equation}
    The field $\vec{H}$ can now also be written as
     \begin{equation}
    \displaystyle
    \vec{H}=\displaystyle \frac{-\jmath k_0
    e^{-\jmath k_0 r}}{4\pi r} \int_{V_0}
    (J_{e\theta} \vec{u}_{\phi} - J_{e \phi}
    \vec{u}_{\theta} ) e^{\jmath k_0  \vec{u}_r \cdot \vec{r}_0} dV_0,
    \end{equation}
    This specific formulation of the far field was already used in
    (\ref{eq:verreveldalgemeen}) in order to determine the radiation pattern of an antenna.
    \item The field $\vec{H}$ in the observation point $P$ can be interpreted as the superposition of the contributions from all source points.
    Note that the term $\displaystyle e^{\jmath k_0
     \vec{u}_r \cdot \vec{r}_0}$ corresponds to a phase difference between source point $Q$ $(x_0,y_0,z_0)$ and the origin $(0,0,0)$
    \end{enumerate}

The final step in the derivation of the radiated fields from a general antenna is the determination of the electric field $\vec{E}(\vec{r})$
using expresssion (\ref{eq:HenEinA})
\begin{equation}
\begin{array}{lcl}

\displaystyle \vec{E} & = & \displaystyle  \frac{1}{\jmath \omega
\epsilon_0 \mu_0}
 \nabla \times \nabla \times \vec{A}_e.

\end{array}
\end{equation}
The vector operator $\nabla \times \vec{A}_e$ was already applied previously resulting in (\ref{eq:HinrPQcoord1}). The electric field $\vec{E}$ now takes the form
\begin{equation}
\begin{array}{lcl}
\displaystyle \vec{E}  & = &  \displaystyle \frac{1}{4\pi \jmath
\omega \epsilon_0} \int_{V_0} \nabla_r \times \left[ \left(-\jmath
k_0 - \frac{1}{r_d} \right) \frac{e^{-\jmath k_0 r_d}}{r_d^2}
\vec{r}_d \times \vec{J}_e (\vec{r}_0) \right] dV_0,

\end{array}
\label{eq:EinrPQcoord1}
\end{equation}
Note that the operator $\nabla_r$ is not only a function of the vector
$\vec{r}_0$ but also of $\vec{r}$ due to the factor $\vec{r}_d=\vec{r}-\vec{r}_0$.
With $\displaystyle
\psi(\vec{r},\vec{r}_0)= \left(-\jmath k_0 - \frac{1}{r_d} \right)
\frac{e^{-\jmath k_0 r_d}}{r_d^2}$ and $\vec{a}(\vec{r},\vec{r}_0)
= \vec{r}_d \times \vec{J}_e (\vec{r}_0)$, we obtain
\begin{equation}
\begin{array}{lcl}
\displaystyle \nabla_r \times \psi \vec{a}  & = &  \displaystyle
\nabla_r \psi \times \vec{a} + \psi \nabla_r \times \vec{a}, \\

\nabla_r \times \vec{a}  & = &  \displaystyle \nabla_r \times
\left[ (\vec{r}-\vec{r}_0) \times \vec{J}_e(\vec{r}_0) \right] \\

& = & -2  \vec{J}_e(\vec{r}_0), \\

\displaystyle \nabla_r \psi & = &  \displaystyle \left(-\jmath k_0
- \frac{1}{r_d} \right) \nabla_r \left( \frac{e^{-\jmath k_0
r_d}}{r_d^2} \right) +  \frac{e^{-\jmath k_0 r_d}}{r_d^2} \nabla_r
\left( -\jmath k_0 - \frac{1}{r_d} \right) \\

& = &  \displaystyle \left(-\jmath k_0 - \frac{1}{r_d} \right)
\nabla_r \left( \frac{e^{-\jmath k_0 r_d}}{r_d^2} \right) +
\frac{e^{-\jmath k_0 r_d}}{r_d^2} \frac{1}{r_d^2} \vec{u}_{r_d} .

\end{array}
\label{eq:nablapsi}
\end{equation}
Furthermore,
\begin{equation}
\begin{array}{lcl}
 \displaystyle \nabla_r \left( \frac{e^{-\jmath k_0
r_d}}{r_d^2} \right) & = & \displaystyle -2 e^{-\jmath k_0 r_d}
\frac{1}{r_d^3} \vec{u}_{r_d} + \frac{1}{r_d^2} (-\jmath k_0)
e^{-\jmath k_0 r_d} \vec{u}_{r_d}.
\end{array}
\end{equation}
By substituting this expression in (\ref{eq:nablapsi}) we obtain
\begin{equation}
\begin{array}{lcl}
 \displaystyle \nabla_r \psi & = & \displaystyle \frac{e^{-\jmath k_0
 r_d}}{r_d^2} k_0^2 \left[ -1 -\frac{3}{\jmath k r_d} +
 \frac{3}{(k_0 r_d)^2} \right] \vec{u}_{r_d}.
\end{array}
\end{equation}
When we apply the same kind of far-field approximation as used for $\vec{H}$ we find that
\begin{equation}
\begin{array}{lcl}
\displaystyle \vec{E}  & = &  \displaystyle \frac{-k_0^2}{\jmath
\omega \epsilon_0} \frac{e^{-\jmath k_0 r}}{4\pi r}  \int_{V_0}
\vec{u}_{r_d} \times \left[ \vec{u}_{r_d} \times \vec{J}_e
(\vec{r}_0) \right]  e^{\jmath k_0  \vec{u}_r \cdot \vec{r}_0} dV_0.
\end{array}
\label{eq:Everreveld1}
\end{equation}
By substituting $\vec{u}_{r_d}$ by $\vec{u}_r$, we finally arrive at the expression for $\vec{E}$ in the far-field region of the antenna:
\begin{equation}
\begin{array}{lcl}
\displaystyle \vec{E}  & = &  \displaystyle \frac{-k_0^2}{\jmath
\omega \epsilon_0} \frac{e^{-\jmath k_0 r}}{4\pi r} \vec{u}_r
\times \vec{u}_r \times \int_{V_0} \vec{J}_e (\vec{r}_0) e^{\jmath
k_0  \vec{u}_r \cdot \vec{r}_0} dV_0.
\end{array}
\label{eq:Everreveld2}
\end{equation}
Note that the expressions of $\vec{E}$ and $\vec{H}$ given by (\ref{eq:Everreveld2}) and (\ref{eq:Hverreveld}) can also be obtained by substituting
the operator $\nabla \times$ in equation (\ref{eq:HenEinA}) by $(-\jmath k_0) \vec{u}_r \times$.
Applying the vector identity $\vec{A} \times [\vec{B} \times
\vec{C}]=(\vec{A} \cdot \vec{C})\vec{B} -  (\vec{A} \cdot
\vec{B})\vec{C}$ to expression (\ref{eq:Everreveld2}) we find
\begin{equation}
\begin{array}{lcl}
\displaystyle \vec{E}  & = &  \displaystyle \frac{-k_0^2}{\jmath
\omega \epsilon_0} \frac{e^{-\jmath k_0 r}}{4\pi r}  \int_{V_0}
\left[ (\vec{u}_r \cdot \vec{J}_e (\vec{r}_0)) \vec{u}_r -
\vec{J}_e (\vec{r}_0) \right] e^{\jmath k_0  \vec{u}_r \cdot \vec{r}_0}
dV_0 \\

& = &  \displaystyle \frac{k_0^2}{\jmath \omega \epsilon_0}
\frac{e^{-\jmath k_0 r}}{4\pi r}  \int_{V_0}
\left(\vec{u}_{\theta} J_{e \theta} (\vec{r}_0)) + \vec{u}_{\phi}
J_{e \phi} (\vec{r}_0) \right) e^{\jmath k_0
 \vec{u}_r \cdot \vec{r}_0} dV_0.
\end{array}
\label{eq:Everreveldthetaphi}
\end{equation}
From (\ref{eq:Everreveldthetaphi}) we can conclude that
$\vec{E} \cdot \vec{u}_r=0$, which implies that the radial component of $\vec{E}$ vanishes in the far-field region of the antenna.
As a result, the electric field $\vec{E}$ is perpendicular to $\vec{u}_r$.

By combining expression (\ref{eq:Hverreveld}) and
(\ref{eq:Everreveld2}) we can observe that
\begin{equation}
\begin{array}{lcl}
\displaystyle \vec{E}  & = &  \displaystyle Z_0 \vec{H} \times
\vec{u}_r, \\
 \displaystyle \vec{H}  & = &  \displaystyle \frac{1}{Z_0}
\vec{u}_r \times  \vec{E},
\end{array}
\label{eq:EHrelatie}
\end{equation}
where $Z_0=\sqrt{\frac{\mu_0}{\epsilon_0}} \approx 377 \Omega$ is the free-space wave impedance. From (\ref{eq:EHrelatie})
we find that the electromagnetic waves in the far-field region behave locally as TEM waves.
The Pointing vector is now readily found from
(\ref{eq:EHrelatie})
\begin{equation}
\begin{array}{lcl}
\displaystyle \vec{S}_p(\vec{r})  & = &  \displaystyle \frac{1}{2}
Re \left[ \vec{E} \times \vec{H}^* \right] \\

& =& \displaystyle \frac{1}{2 Z_0} Re \left[ \vec{E} \times
(\vec{u}_r \times \vec{E}^*) \right] \\

& =& \displaystyle \frac{1}{2 Z_0} Re \left[ (\vec{E} \cdot
\vec{E}^*)\vec{u}_r - (\vec{E} \cdot \vec{u}_r)\vec{E}^* \right]
\\

& =& \displaystyle \frac{1}{2 Z_0} |\vec{E}|^2 \vec{u}_r,

\end{array}
\label{eq:PointingS}
\end{equation}
since $\vec{E} \cdot \vec{u}_r$. As expected, the electromagnetic energy indeed flows in the radial direction.
\\
\\

{\Large \bf Summary} \\ For an antenna described by an electric current distribution
$\vec{J}_e (\vec{r}_0)$ within a finite volume
$V_0$, we can express the corresponding electromagnetic fields in the far-field region by

\begin{equation}
\begin{array}{lcl}
\displaystyle \vec{H}  & = &  \displaystyle \frac{-\jmath k_0
e^{-\jmath k_0 r} }{4\pi r} \vec{u}_{r} \times  \int_{V_0}
\vec{J}_e (\vec{r}_0) e^{\jmath k_0  \vec{u}_r \cdot \vec{r}_0} dV_0,
\\
\displaystyle \vec{E}  & = &  \displaystyle \frac{-k_0^2}{\jmath
\omega \epsilon_0} \frac{e^{-\jmath k_0 r}}{4\pi r} \vec{u}_r
\times \vec{u}_r \times \int_{V_0} \vec{J}_e (\vec{r}_0) e^{\jmath
k_0  \vec{u}_r \cdot \vec{r}_0} dV_0, \\

\displaystyle \vec{E}  & = &  \displaystyle Z_0 \vec{H} \times
\vec{u}_r, \\

\displaystyle Z_0  & = &  \displaystyle
\sqrt{\frac{\mu_0}{\epsilon_0}} \approx 377 \Omega.

\end{array}
\label{eq:Samenvatting}
\end{equation}

\section{Electric dipole}
The most elementary antenna is the electric dipole. The electric dipole is an infinitely small current element, which can be approximated in practice by a small linear antenna with a length much smaller than the free-space wavelength $\lambda_0$.
The electric dipole is located in the origin of our coordinate system and directed along the
$z$-axis, as indicated in Fig. \ref{fig:eldipole}. The current distribution $\vec{J}_e$ along the electric dipole can now be expressed in terms of the delta function:
\begin{equation}
\displaystyle \vec{J}_e = I_0l \delta(x) \delta(y) \delta(z)
\vec{u}_z = I_0l \delta(\vec{r}) \vec{u}_z. \label{eq:Edipool2}
\end{equation}
Where $I_0$ is the total current which we assume to be constant along the electric dipole.
The length of the antenna is $l$. The delta-function $\delta(z)$ belongs to the class of generalized functions with the property:
\begin{equation}
\displaystyle \gamma(z_0) = \int\limits_{-\infty}^{\infty}
\delta(z-z_0) \gamma(z) dz. \label{eq:deltafunctie}
\end{equation}
The delta-function can be visualized as follows
\begin{equation}
\displaystyle \delta(z) = \left\{
\begin{array}{lr}
\displaystyle \frac{1}{w} & -\frac{w}{2} < z < \frac{w}{2} \\
\displaystyle 0 &  |z| > \frac{w}{2}
\end{array}
\right. ,
 \label{eq:deltafunctie2}
\end{equation}
where $w$ needs to be very small. In order to determine whether (\ref{eq:Edipool2}) indeed represents an electric dipole, we need to investigate the corresponding charge distribution $\rho_e$ which can be found by applying the continuity equation
(\ref{eq:continuiteitsvgl}):
\begin{equation}
\displaystyle \rho_e (\vec{r}) = -\frac{1}{\jmath \omega} \nabla
\cdot \vec{J}_e (\vec{r})=\frac{-I_0l}{\jmath \omega} \delta(x)
\delta(y) \delta^{\prime}(z) .
 \label{eq:ladingdipool}
\end{equation}
By considering the derivative of (\ref{eq:deltafunctie2}),
$\delta^{\prime}(z)$, we can represent(\ref{eq:ladingdipool}) by a set of two point charges with opposite sign and located on the $z$-axis at $z=w/2$ and
$-w/2$, respectively, since
\begin{equation}
\displaystyle \delta^{\prime}(z) = \frac{1}{w} \left[ \delta
\left(z+\frac{w}{2}\right)- \delta \left(z-\frac{w}{2}\right)
\right]. \label{eq:afgeleidedeltafunctie}
\end{equation}

Let us now determine the electromagnetic fields in the far-field region of the electric dipole by using the general expressions derived in chapter
\ref{sec:radiatedfields}:
\begin{equation}
\begin{array}{lcl}
\displaystyle \vec{H}  & = &  \displaystyle \frac{-\jmath k_0
e^{-\jmath k_0 r} }{4\pi r} \vec{u}_{r} \times  \int_{V_0}
\vec{J}_e (\vec{r}_0) e^{\jmath k_0  \vec{u}_r \cdot \vec{r}_0} dV_0,
\\

\displaystyle \vec{E}  & = &  \displaystyle Z_0 \vec{H} \times
\vec{u}_r, \\
\end{array}
\label{eq:Samenvatting2}
\end{equation}
Substituting the current distribution of the electric dipole (\ref{eq:Edipool2})
in the above equation and using the fact that
\begin{equation}
\displaystyle \vec{u}_r \times \vec{u}_z = -\vec{u}_{\phi}
\sin{\theta}
\end{equation}
leads us to the magnetic field far away from the dipole
\begin{equation}
\displaystyle \vec{H} = H_{\phi} \vec{u}_{\phi} = \frac{\jmath k_0
I_0 l e^{-\jmath k_0 r} }{4\pi r} \sin{\theta} \vec{u}_{\phi}.
\label{eq:Hvelddipool}
\end{equation}
The corresponding electric field in the far-field region is now given by
\begin{equation}
\displaystyle \vec{E} = E_{\theta} \vec{u}_{\theta} = \frac{\jmath
k_0 Z_0 I_0 l e^{-\jmath k_0 r} }{4\pi r} \sin{\theta}
\vec{u}_{\theta}, \label{eq:Evelddipool}
\end{equation}
where we have used the following relation
\begin{equation}
\displaystyle \vec{u}_{\phi} \times \vec{u}_r = \vec{u}_{\theta}.
\end{equation}
The time-average power flux density (Pointing vector) expressed in [$\mbox{W/m}^2$] is found by using equation (\ref{eq:PointingS}), which provides
\begin{equation}
\begin{array}{lcl}
\displaystyle \vec{S}_p(\vec{r}) & =& \displaystyle \frac{1}{2 Z_0}
|\vec{E}|^2 \vec{u}_r \\

 & =& \displaystyle \frac{1}{2} \frac{k_0^2 Z_0 (I_0 l)^2}{(4 \pi
 r)^2} \sin^2{\theta} \vec{u}_r.

\end{array}
\label{eq:PointingSdipool}
\end{equation}
From (\ref{eq:PointingSdipool}) we can conclude that the electric dipole does not radiate uniformly in all directions.
In the direction along the $z$-axis $(\theta=0)$ the dipole does not radiate al all.
The power density $\vec{S}_p(\vec{r})$ decreases as $r^{-2}$ with increasing distance from the dipole. However, since the total power radiates through a sphere with radius $r$ and surface $4\pi r^2$, the total radiated power is independent of $r$.
The total radiated power by the electric dipole is then found by
\begin{equation}
\begin{array}{lcl}
\displaystyle P_t &=& \displaystyle \int\int \vec{S}_p(\vec{r})
\cdot \vec{u}_r r^2 d\Omega \\

\displaystyle &=& \displaystyle \int\limits_{0}^{2\pi}
\int\limits_{0}^{\pi} \frac{1}{2} \frac{k_0^2 Z_0 (I_0 l)^2}{(4
\pi)^2} \sin^3{\theta} d\theta d\phi \\

\displaystyle &=& \displaystyle \frac{k_0^2 Z_0 (I_0 l)^2}{12
\pi},

\end{array}
\label{eq:Ptotaldipool}
\end{equation}
where we have used the fact that
\begin{equation}
 \int\limits_{0}^{\pi} \sin^3{\theta} d\theta = \frac{4}{3}.
\end{equation}
As expected, the total power $P_t$ does not depend on $r$.

\begin{figure}[hbt]

\centerline{\psfig{figure=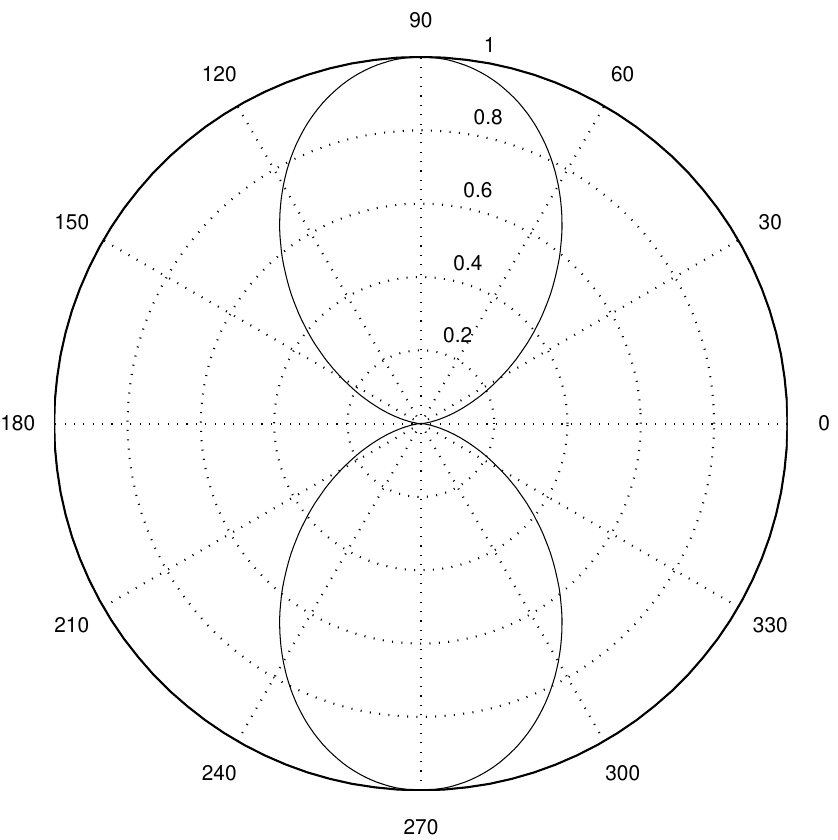,width=90mm}}

 \caption{\it Radiation pattern versus $\theta$ of an electric dipole of length $l$.
 Note that $\theta=0^0$ coincides with the positive $z$-axis}
  \label{fig:stralingdipool}
\end{figure}

Let us now explore some fundamental antenna parameters of the electric dipole.
According to (\ref{eq:normradpat}), the normalized radiation pattern is given by
\begin{equation}
\displaystyle F(\theta, \phi) = \frac{P(\theta, \phi)}{P(\pi/2,0)}
= \frac{P(\theta)}{P(\pi/2)} = \sin^2{\theta} .
\label{eq:normradpatdipool}
\end{equation}
This pattern is plotted in Fig. \ref{fig:stralingdipool}. The 3-dB beam width of the electric dipole is $90^0$.
The directivity is the maximum of the directivity function, given in (\ref{eq:richtfunctie}) by:
\begin{equation}
\displaystyle D(\theta, \phi) = \frac{P(\theta, \phi)}{P_t / 4\pi}
= \frac{3}{2} \sin^2{\theta} , \label{eq:richtfunctiedipool}
\end{equation}
which implies that the directivity is $\displaystyle D= \frac{3}{2}$.
Finally, let us determine the input impedance of the electric dipole.
The input impedance of the ideal electric dipole is purely resistive, so $Z_a=R_a$. $R_a$,  also known as the radiation resistance, can be found by using relation (\ref{eq:stralingsweerstand}).
Combining this with (\ref{eq:Ptotaldipool}) provides:
\begin{equation}
\displaystyle R_a = \frac{P_t}{\frac{1}{2}|I_0|^2} = 2 \frac{k_0^2
Z_0 l^2}{12 \pi} = 80 \pi^2 \left( \frac{l}{\lambda_0} \right)^2.
\end{equation}
The length $l$ of the dipole is small as compared to the wavelength.
For example, let $l=0.01 \lambda_0$, which results in an input impedance of $R_a = 0.08 \Omega$.
This imples that the electric dipole has a very low radiation resistance which will makes it difficult to match to a transmission line since most commonly used transmission lines have a much higher characteristic impedance (often $Z^t_0=50 \Omega$). This will result in a poorly matched antenna and a high corresponding reflection coefficient
(\ref{eq:reflecversusZ}) at the antenna port.
Electric dipoles are only used in very specific applications in which power-matching is not important.
A nice example is low-frequency radio-astronomy \cite{SKA}.

The half-wavelength dipole has much better radiation and matching properties. This type of antenna will be explored in the next section.

\section{Thin-wire antennas}
\label{sec-dunnedraadantene}

\begin{figure}[hbt]
\centerline{\psfig{figure=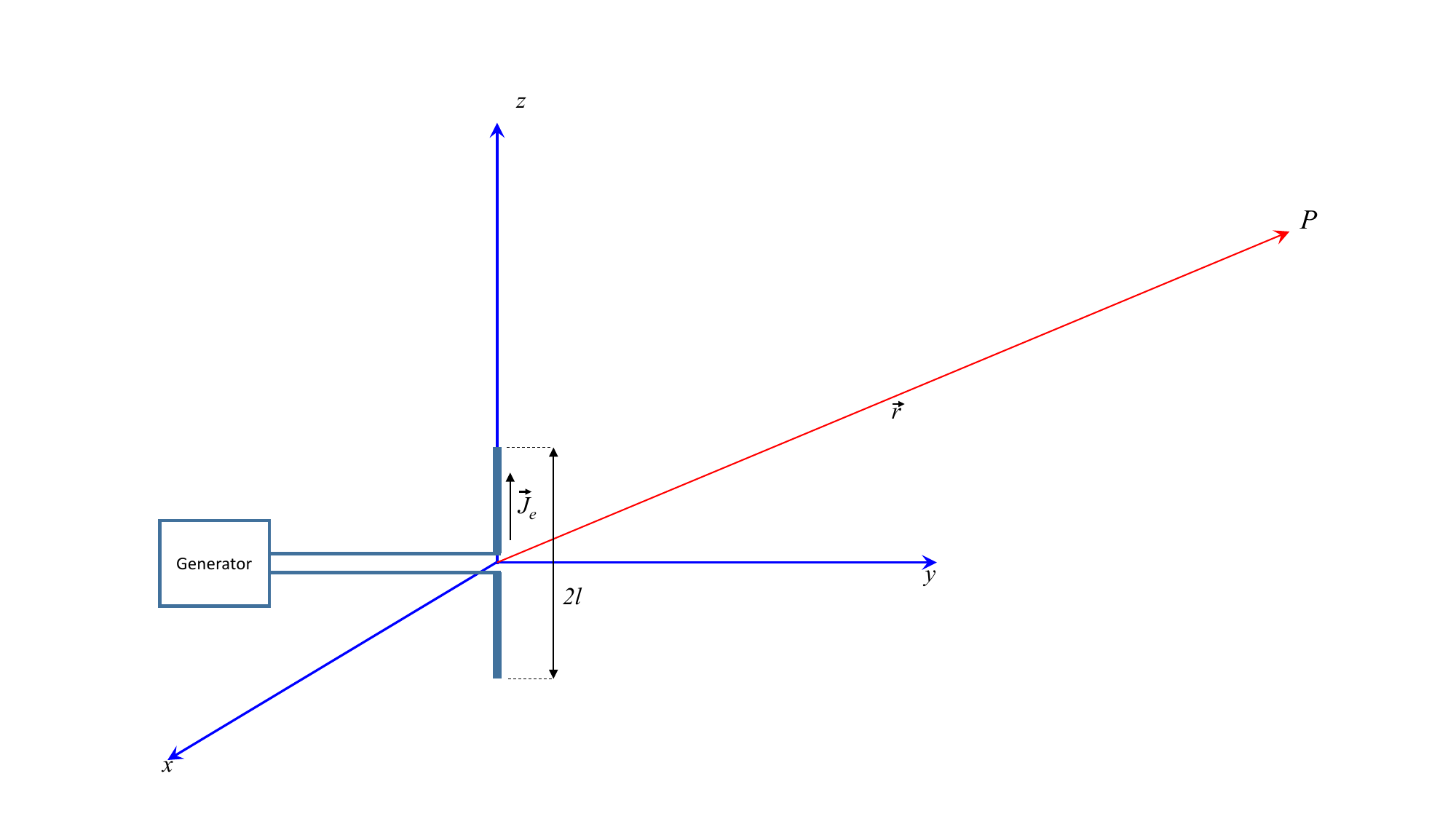,width=185mm}}
\caption{\it Thin (lineair) wire antenna of length $2l$ connected to a two-wire transmission line. The center of the wire is located in the origin of the coordinate system. The wire antenna is directed along the $z$-axis.} \label{fig:draadantenne}
\end{figure}

In the previous section we found that the radiation resistance of an electric dipole is very small, which makes it difficult to match to a transmission line.
The matching properties can be improved by increasing the length of the dipole, resulting in a so-called linear wire antenna.
This antenna consists of a cylindrical metal wire with diameter $d_0$ and length $2l$, which has a small gap in the center of the wire to accommodate a balanced feed using a so-called two-wire transmission line, also known as {\it twin lead} or {\it Lecher line}.
Furthermore, we will assume that the diameter $d_0$ of the wire very small compared to the wavelength $d_0 \ll \lambda_0$.
Fig. \ref{fig:draadantenne} illustrates the linear wire antenna connected to a transmission line.

\begin{figure}[hbt]
\centerline{\psfig{figure=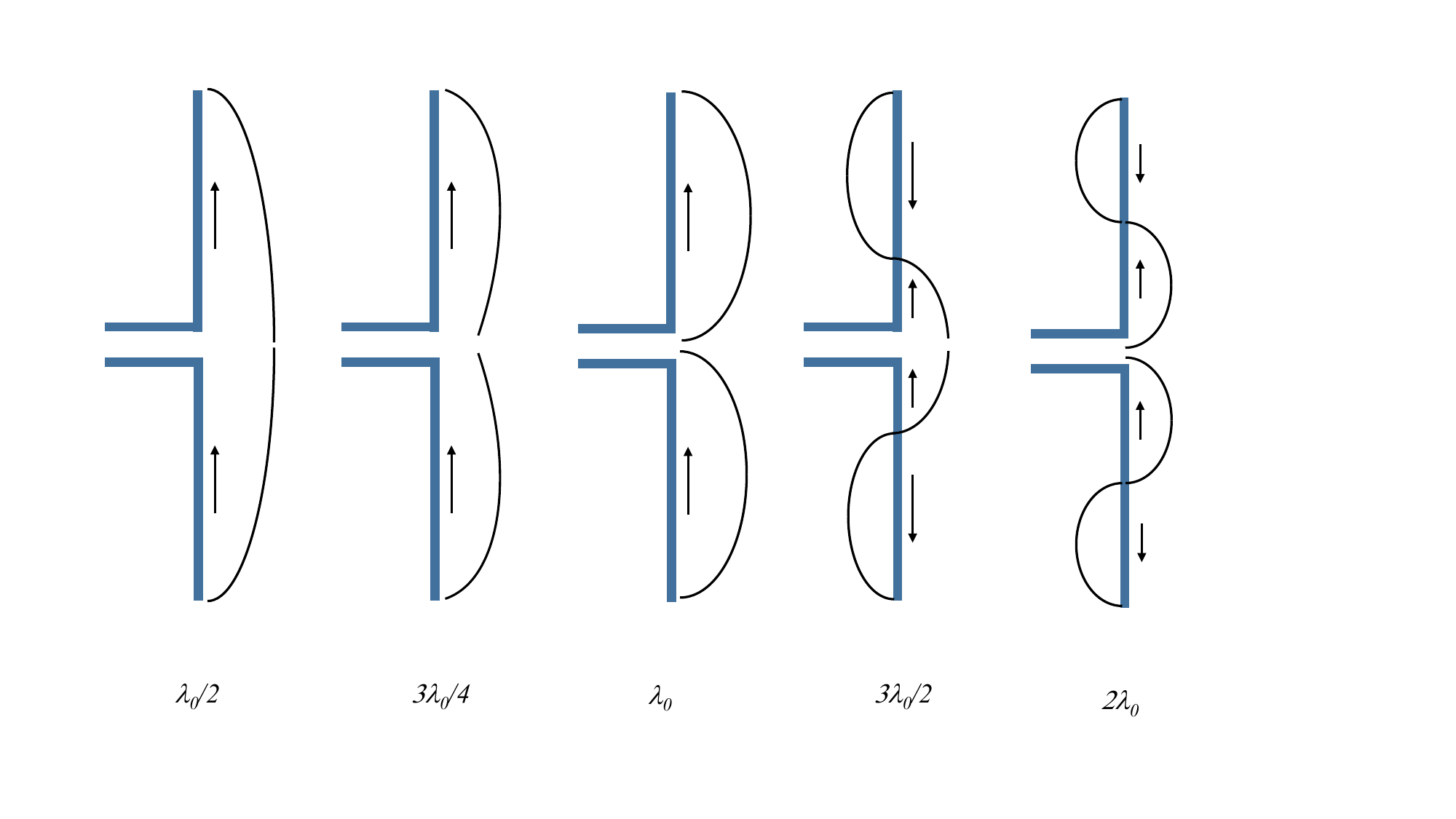,width=160mm}}
\caption{\it Current distribution along a thin wire antenna for various lengths according to (\ref{eq:Jdraadantenne}).}
\label{fig:stroomdraadantenne}
\end{figure}

The current distribution along the two-wire transmission line will be sinusoidal, when we neglect radiation and other losses.
The wire antenna can now be interpreted as the last part of an open-ended two-wire transmission line.
The radiation from this folded part of the transmission line is much larger as compared to the spurious radiation from the transmission line itself.
This is, of course, our aim, since we want to construct an efficient antenna.
Nevertheless, the current distribution along the wire will remain the same (first-order approximation) as the un-folded open-ended transmission line, so with the sinusoidal current distribution. The current distribution at the end point of the antenna must be equal to zero.
Furthermore, we will assume that the gap at the center of the wire antenna is negligibly small.
With all these assumptions, the current distribution $\vec{J}_e$ along the wire takes the following form:
\begin{equation}
\displaystyle \vec{J}_e = I_0 \delta(x) \delta(y) \sin\left[k_0
(l-|z|) \right] \vec{u}_z, \label{eq:Jdraadantenne}
\end{equation}
for $-l < z < l$. Lateron in this book we will use numerical methods to determine the exact current distribution along the wire.
It will be shown that (\ref{eq:Jdraadantenne}) is an accurate approximation of the current along the wire antenna when the diameter of the wire is smaller than $\lambda_0/100$.
Some examples of the current distribution for various lengths of the wire antenna are illustrated in Fig. \ref{fig:stroomdraadantenne}.
The electric field in the far-field region of the wire antenna can be determined using expression
(\ref{eq:Samenvatting}):
\begin{equation}
\begin{array}{lcl}
\displaystyle \vec{E}  & = &  \displaystyle \frac{-k_0^2}{\jmath
\omega \epsilon_0} \frac{e^{-\jmath k_0 r}}{4\pi r} \vec{u}_r
\times \vec{u}_r \times \int_{V_0} \vec{J}_e (\vec{r}_0) e^{\jmath
k_0  \vec{u}_r \cdot \vec{r}_0} dV_0,
\end{array}
\end{equation}
Since the current is directed along the $z$-axis we can rewrite
$e^{\jmath k_0  \vec{u}_r \cdot \vec{r}_0}$ into the expression $e^{\jmath
k_0 z_0 \cos{\theta}}$. Furthermore,
\begin{equation}
\displaystyle \vec{u}_r \times \vec{u}_r \times \vec{u}_z =
\vec{u}_{\theta} \sin{\theta}.
\end{equation}
As a result, we can conclude that the electric field $\vec{E}$ only has one component, being $E_{\theta}$. The radiated field is linearly polarized and takes the following form
\begin{equation}
\displaystyle E_{\theta}  = \frac{-k_0^2 I_0}{\jmath \omega
\epsilon_0} \frac{e^{-\jmath k_0 r}}{4\pi r} \sin{\theta}
\int\limits_{-l}^{l} \sin\left[ k_0 (l-|z_0|) \right] e^{\jmath
k_0 z_0 \cos{\theta}} dz_0,
 \label{eq:Edraadantenne1}
\end{equation}

As a first step towards a closed-form solution of  (\ref{eq:Edraadantenne1}), we will determine the integral by dividing it into two parts:
\begin{equation}
\begin{array}{lcl}
\displaystyle I_{z_0} & = & \displaystyle  \int\limits_{-l}^{l}
\sin\left[ k_0 (l-|z_0|) \right] e^{\jmath k_0 z_0 \cos{\theta}}
dz_0 \\

\displaystyle  & = & \displaystyle  \int\limits_{0}^{l} \sin\left[
k_0 (l-z_0) \right] e^{\jmath k_0 z_0 \cos{\theta}} dz_0 +
\int\limits_{-l}^{0} \sin\left[ k_0 (l+z_0) \right] e^{\jmath k_0
z_0 \cos{\theta}} dz_0 \\

\displaystyle  & = & \displaystyle \int\limits_{0}^{l} \sin\left[
k_0 (l-z_0) \right] e^{\jmath k_0 z_0 \cos{\theta}} dz_0 +
\int\limits_{0}^{l} \sin\left[ k_0 (l-z_0) \right] e^{-\jmath k_0
z_0 \cos{\theta}} dz_0 \\

\displaystyle  & = & \displaystyle 2 \int\limits_{0}^{l}
\sin\left[ k_0 (l-z_0) \right] \cos(k_0 z_0 \cos{\theta}) dz_0 \\

\displaystyle  & = & \displaystyle \int\limits_{0}^{l} \sin( k_0
l- k_0 z_0 + k_0 z_0 \cos{\theta}) dz_0 + \int\limits_{0}^{l}
\sin( k_0 l- k_0 z_0 - k_0 z_0 \cos{\theta}) dz_0 \\

& = & \displaystyle \frac{2 \left[ \cos(k_0 l \cos{\theta})
-\cos(k_0 l) \right]}{k_0 \sin^2{\theta}}.

\end{array}
 \label{eq:Intdraadantenne}
\end{equation}
Substitution of (\ref{eq:Intdraadantenne}) in
(\ref{eq:Edraadantenne1}) provides:
\begin{equation}
\displaystyle E_{\theta} = \jmath Z_0 I_0 \frac{e^{-\jmath k_0
r}}{2\pi r} \left[ \frac{ \cos(k_0 l \cos{\theta}) -\cos(k_0
l)}{\sin{\theta}} \right] .
 \label{eq:Ethetadraadantenne}
\end{equation}
Note that $E_{\theta}$ does not depend on the $\phi$ coordinate, which could have been expected as a result of the symmetry.
The corresponding magnetic field is now easily found, since in the far-field region we have
\begin{equation}
\begin{array}{lcl}
 \displaystyle \vec{H}  & = &  \displaystyle \frac{1}{Z_0}
\vec{u}_r \times  \vec{E} \\
 & = &  \displaystyle \frac{1}{Z_0} \vec{u}_r \times E_{\theta}
 \vec{u}_{\theta} \\
& = &  \displaystyle \frac{1}{Z_0} \vec{u}_{\phi} E_{\theta}.
\end{array}
\end{equation}
As a result, only the $H_{\phi}$ component is non zero and is given by
\begin{equation}
\begin{array}{lcl}
 \displaystyle H_{\phi}  & = &  \displaystyle \frac{1}{Z_0}
 E_{\theta} \\
& = &  \displaystyle \frac{\jmath I_0 e^{-\jmath k_0 r}}{2\pi r}
\left[ \frac{ \cos(k_0 l \cos{\theta}) -\cos(k_0 l)}{\sin{\theta}}
\right] .
\end{array}
\end{equation}
Note that also $H_{\phi}$ does not depend on the $\phi$ coordinate.
The Pointing vector (power density in [$\mbox{W/m}^2$]) is found by using the expression (\ref{eq:PointingS}):
\begin{equation}
\begin{array}{lcl}
\displaystyle \vec{S}_p(\vec{r})  & = &  \displaystyle \frac{1}{2
Z_0} |\vec{E}|^2 \vec{u}_r \\
 & = &  \displaystyle \frac{Z_0 |I_0|^2}{8\pi^2 r^2} \left[ \frac{
\cos(k_0 l \cos{\theta}) -\cos(k_0 l)}{\sin{\theta}} \right]^2
\vec{u}_r.
\end{array}
\label{eq:PointingSdraadant}
\end{equation}

We will now determine the normalized radiation pattern for several wire antennas with varying length.
The radiation pattern is found by using expression (\ref{eq:normradpat}).
\begin{enumerate}[i.]
    \item {\bf $\lambda_0/2$ wire antenna.}

    When $2l=\lambda_0/2$ we find that $E_{\theta}$ is equal to
    \begin{equation}
    \displaystyle E_{\theta} = \jmath Z_0 I_0 \frac{e^{-\jmath k_0 r}}{2\pi r} \left[ \frac{
    \cos((\pi/2) \cos{\theta})}{\sin{\theta}} \right] .
    \end{equation}
    The radiation pattern of the $\lambda_0/2$ dipole antenna is shown in Fig. \ref{fig:radpatlambda2}. Note that the radiation pattern is quite similar to the radiation pattern of the electric dipole, which is also shown in the figure.
    The 3 dB beam width of the $\lambda_0/2$ dipole is $78^0$, while the beam width of an electric dipole is $90^0$.
    \begin{figure}[hbt]
    \centerline{\psfig{figure=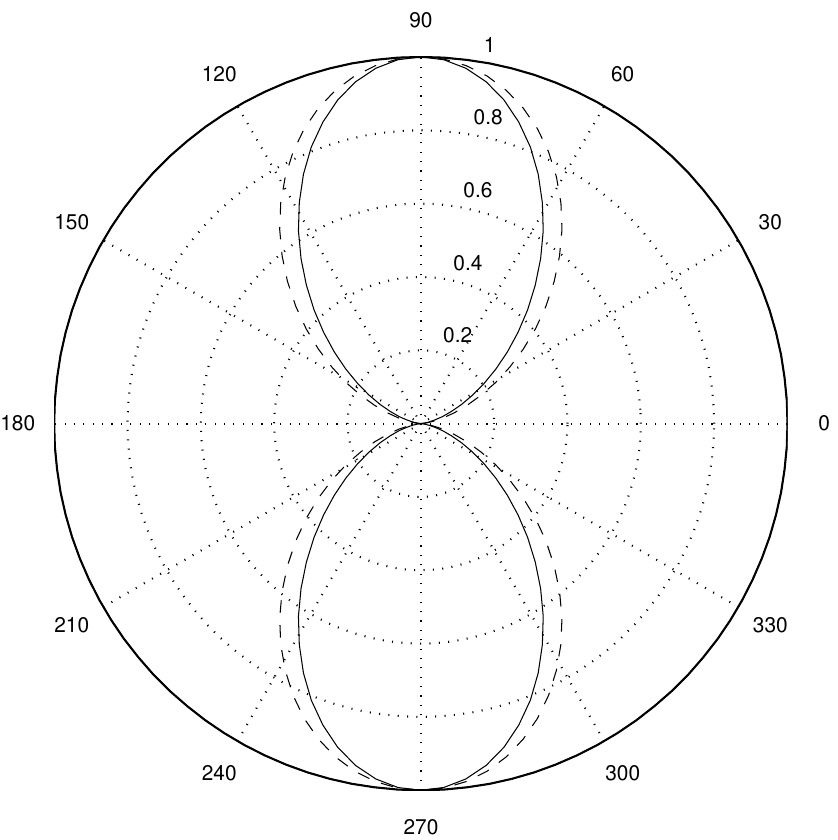,width=90mm}}

    \caption{\it Radiation pattern versus $\theta$ of an $\lambda_0/2$ wire antenna (full line)
    and of an electric dipole (dotted line).}
    \label{fig:radpatlambda2}
    \end{figure}

    \item {\bf $\lambda_0$ wire antenna.}

    Now $2l=\lambda_0$ and $E_{\theta}$ is given by
    \begin{equation}
    \displaystyle E_{\theta} = \jmath Z_0 I_0 \frac{e^{-\jmath k_0 r}}{2\pi r} \left[ \frac{
    \cos(\pi \cos{\theta})+1}{\sin{\theta}} \right] .
    \end{equation}
    The corresponding radiation pattern of the $\lambda_0$ antenna is shown in Fig. \ref{fig:radpatlambda}.  The 3 dB beam width is $47^0$.
    \begin{figure}[hbt]
    \centerline{\psfig{figure=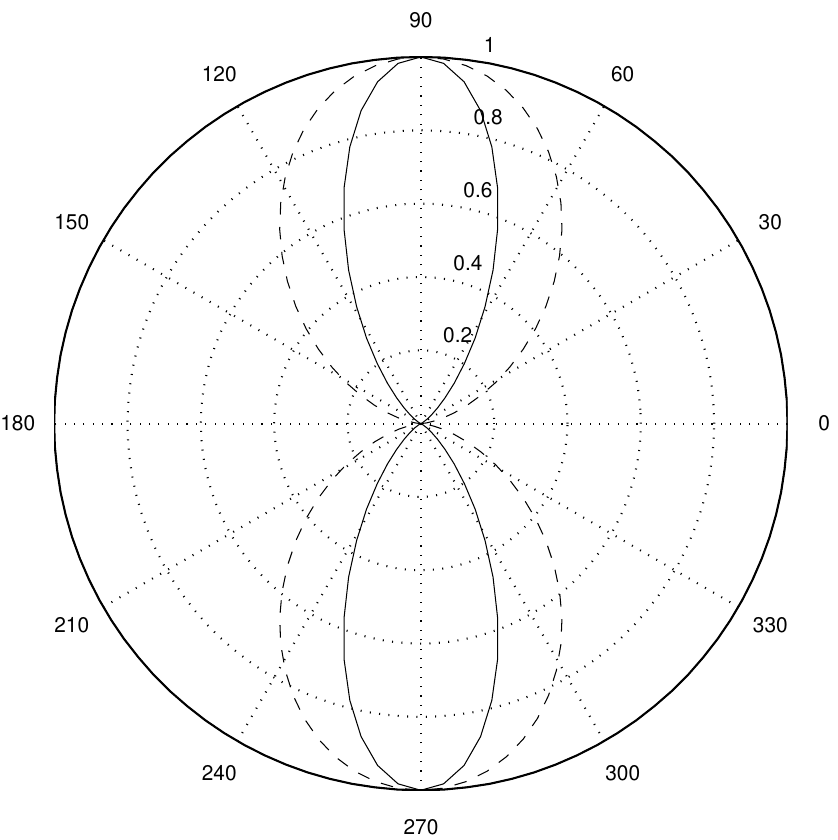,width=90mm}}

    \caption{\it Radiation pattern versus $\theta$ of an $\lambda_0$ wire antenna (full line)
    and of an electric dipole (dotted line).}
    \label{fig:radpatlambda}
    \end{figure}

    \item {\bf $3\lambda_0/2$ wire antenna.}

    With $2l=3\lambda_0/2$ we find that $E_{\theta}$ takes the form:
    \begin{equation}
    \displaystyle E_{\theta} = \jmath Z_0 I_0 \frac{e^{-\jmath k_0 r}}{2\pi r} \left[ \frac{
    \cos((3\pi/2) \cos{\theta})}{\sin{\theta}} \right] .
    \end{equation}
    The corresponding radiation pattern of the $3\lambda_0/2$ antenna is shown in Fig. \ref{fig:radpatlambda32}. We observe multiple beams in various directions.
    In addition, the antenna does not radiate energy in several other directions (pattern {\it nulls}). Although the beam width is much smaller now, the antenna generates multiple beams, which might not be very useful in most applications.
    \begin{figure}[hbt]
    \centerline{\psfig{figure=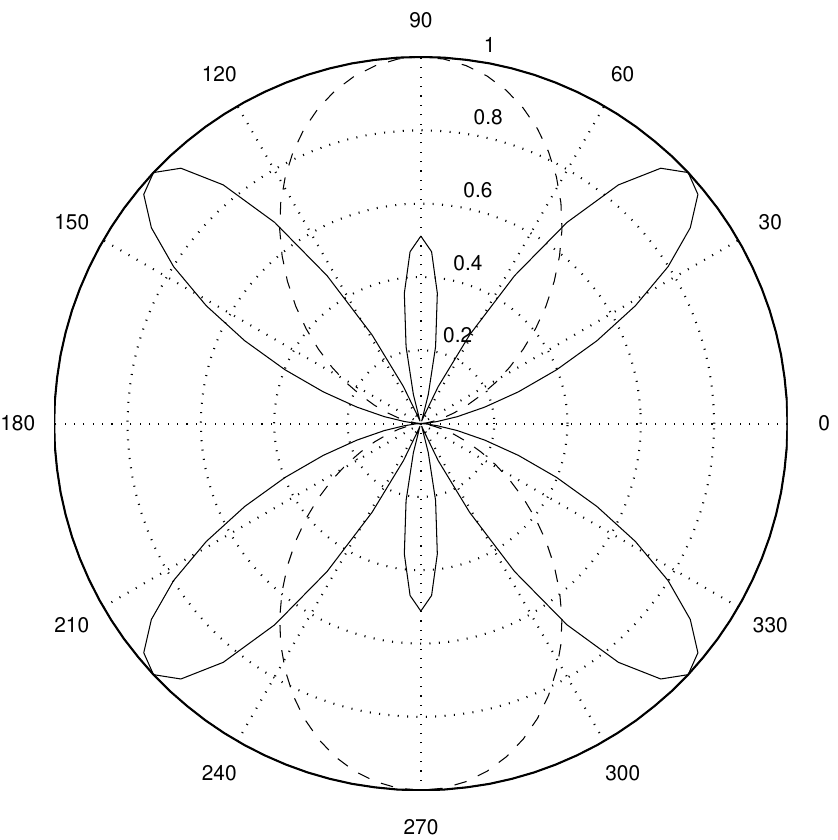,width=90mm}}

    \caption{\it Radiation pattern versus $\theta$ of an $3\lambda_0$ wire antenna (full line)
    and of an electric dipole (dotted line).}
    \label{fig:radpatlambda32}
    \end{figure}

\end{enumerate}
In order to determine the other antenna parameters of the wire antenna, such as the directivity and radiation resistance,
we first need to calculate the total radiated power
$P_t$ by integrating the power density over a sphere according to (\ref{eq:radiatedpower})
\begin{equation}
\begin{array}{lcl}
\displaystyle P_t &=& \displaystyle \int\int \vec{S}_p(\vec{r})
\cdot \vec{u}_r r^2 d\Omega \\

\displaystyle &=& \displaystyle \frac{Z_0 |I_0|^2}{8\pi^2} 2\pi
\int\limits_{0}^{\pi} \left[ \frac{ \cos(k_0 l \cos{\theta})
-\cos(k_0 l)}{\sin{\theta}} \right]^2 \sin{\theta} d\theta \\
\displaystyle &=& \displaystyle \frac{Z_0 |I_0|^2}{4\pi} I_p.

\end{array}
\label{eq:Ptdipool}
\end{equation}
A general analytic solution of the integral $I_p$ is not easily found.
However, in case of a $\lambda_0/2$-wire antenna we can express the integral $I_p$ in terms of the cosine integral.
By applying the coordinate transformation $u=\cos\theta$ and $d\theta=-du/\sin{\theta}$ we can write integral $I_p$
in the following form
\begin{equation}
\begin{array}{lcl}
\displaystyle I_p &=& \displaystyle \int\limits_{0}^{\pi} \left[
\frac{ \cos(k_0 l \cos{\theta}) -\cos(k_0 l)}{\sin{\theta}}
\right]^2 \sin{\theta} d\theta \\

&=& \displaystyle \int\limits_{-1}^{+1} \left[ \frac{\cos^2(\pi
u/2)}{1-u^2} \right] du \\

&=& \displaystyle \frac{1}{2} \int\limits_{-1}^{+1} \left[
\frac{\cos^2(\pi u/2)}{1+u} \right] du +  \frac{1}{2}
\int\limits_{-1}^{+1} \left[ \frac{\cos^2(\pi u/2)}{1-u} \right]
du \\

&=& \displaystyle \frac{1}{2} \int\limits_{0}^{2\pi} \left[
\frac{\cos^2((t-\pi)/2)}{t} \right] dt +  \frac{1}{2}
\int\limits_{0}^{2\pi} \left[ \frac{\cos^2((\pi-s)/2)}{s} \right]
ds \\

&=& \displaystyle \int\limits_{0}^{2\pi} \left[
\frac{\cos^2((t-\pi)/2)}{t} \right] dt \\

&=& \displaystyle \frac{1}{2} \int\limits_{0}^{2\pi} \left[
\frac{1+\cos(t-\pi)}{t} \right] dt \\

&=& \displaystyle \frac{1}{2} \int\limits_{0}^{2\pi} \left[
\frac{1-\cos{t}}{t} \right] dt,

\end{array}
\label{eq:Ipint1}
\end{equation}
where we have used the substitutions $1+u=t/\pi$ and
$1-u=s/\pi$. Furthermore, we have used the fact that $\cos^2(x/2)=(1+\cos{x})/2$.
The integral can now be expressed in terms of the well-known cosine integral $ci(x)$
\begin{equation}
\displaystyle ci(x) = - \int\limits_{x}^{\infty} \frac{\cos{t}}{t}
dt.
\end{equation}
The cosine integral is well tabulated (for example in the famous book of I.~S. Gradsteyn and I.~M. Ryzhik, ``{Table of
integrals, series and products}'', tabel (8.230),
\cite{Gradsteyn}) and is a standard function in Matlab (function {\it cosint}).
Expression (\ref{eq:Ipint1}) now takes the form
\begin{equation}
\begin{array}{lcl}
\displaystyle I_p &=& \displaystyle \frac{1}{2}
\int\limits_{0}^{2\pi} \left[ \frac{1-\cos{t}}{t} \right] dt \\

&=& \displaystyle \frac{1}{2} \left( {\cal C} + \ln(2\pi) -
ci(2\pi) \right) \\

&=& \displaystyle \frac{2.437}{2},

\end{array}
\label{eq:Ipint2}
\end{equation}
where ${\cal C}=0.577$ is Euler's constant. The total radiation power is now found from:
\begin{equation}
\begin{array}{lcl}
\displaystyle P_t &=& \displaystyle \frac{Z_0 |I_0|^2}{4\pi} I_p
\\
&=& \displaystyle \frac{1}{2} 73.1 |I_0|^2.

\end{array}
\label{eq:Ptdipool2}
\end{equation}
By using definition (\ref{eq:stralingsweerstand}) of the radiation resistance, we easily find that
\begin{equation}
\displaystyle R_a = 73.1 \Omega.
\end{equation}
This suggest that $R_a$ is not frequency dependent. However, this is not true, since we have assumed that the antenna length $2l=\lambda_0/2$.
Therefore, this antenna is a resonating structure, which implies that the input impedance is strongly frequency dependent.
The resonant wire antenna provides a purely resistive load $R_a$ seen at the center of the wire antenna (see Fig. \ref{fig:draadantenne}).
In general, the load impedance at the termination of the two-wire transmission line will take a complex form according to $Z_a=R_a+\jmath X_a$.
Only in case of a resonating wire antennas with $2l=\lambda_0/2$ and a small diameter $d_0 \ll \lambda_0$, we find a purely resistive load impedance with $X_a=0$ .
The calculation of the complex input impedance $Z_a=R_a+\jmath X_a$ of a wire antenna outside resonance is more complex and can only be done by using numerical techniques such as the {\it Method of Moments}. In chapter \ref{chap:NumEM} we will investigate this numerical method in more detail.

Finally, we can determine the directivity of the $\lambda_0/2$ wire antenna.
By using the definition of the directivity function (\ref{eq:richtfunctie}), we obtain
\begin{equation}
\displaystyle D = \frac{\max(P(\theta, \phi))}{P_t / 4\pi} =
\frac{\displaystyle \frac{Z_0 |I_0|^2}{8 \pi^2}}{\displaystyle
\frac{73.1|I_0|^2/2}{4\pi}} = 1.64 .
\label{eq:richtfunctiedraadantenne}
\end{equation}
The corresponding directivity expressed in Decibels ($10log_{10}D$) is $D_{dB} = 2.15$ dB, which is slightly higher than the directivity of the electric dipole $\displaystyle D=1.5$ or $D_{dB} = 1.76$ dB.

\section{Wire antenna above a ground plane}
\label{sec-draadbovengrond}

\begin{figure}[hbt]
\centerline{\psfig{figure=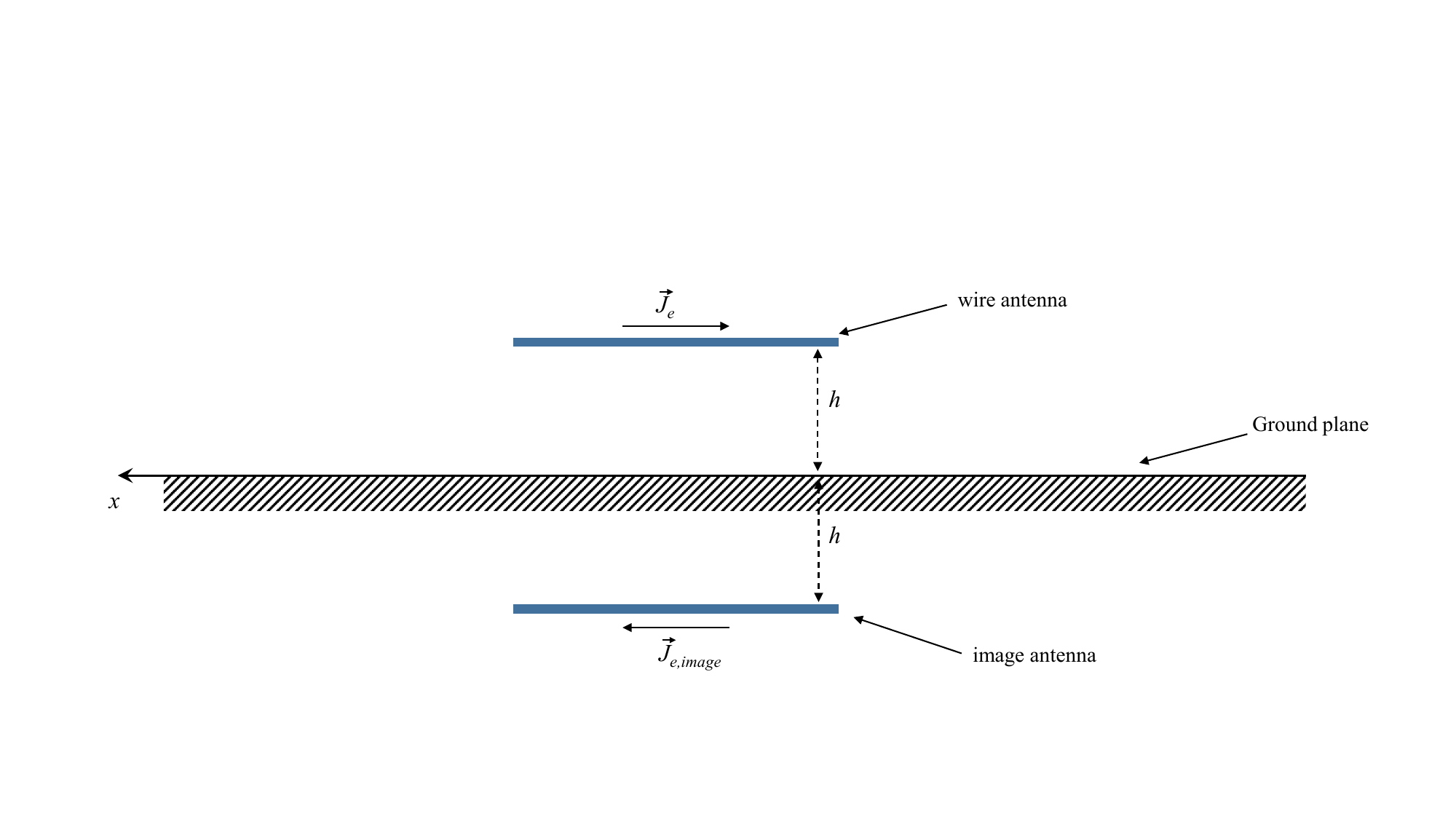,width=180mm}}
\caption{\it Thin horizontal wire antenna of length
$2l=\lambda_0/2$ above a perfectly conducting infinite ground plane.
The antenna is situated along the $x$-axis ($y=0$).}
\label{fig:hordipgrondvlak}
\end{figure}

Up to now we have assumed that the antenna is located in free space. However, in practise the antenna will often be placed close to objects, e.g. earth surface or in case of printed antennas close to a ground plane of the printed-circuit board (PCB).
As a result, the radiation properties and input impedance will change.
In this section we will investigate the effect of a perfectly conducting ground plane on the radiation properties of wire antennas.
We will assume that the horizontally-oriented resonating wire antenna of length $2l=\lambda_0/2$ is located above the conducting ground plane at a height $h$.
The configuration is shown in Fig. \ref{fig:hordipgrondvlak}.
In order to solve the antenna problem of Fig. \ref{fig:hordipgrondvlak}, we can use the so-called {\it image method}.
This method is based on the fact that the tangential component of the electric field $\vec{E}$ vanishes at the perfectly-conducting ground plane.
Exactly the same effect can be obtained by removing the conducting plate and placing a second horizontal wire antenna ({\it image antenna}) at a distance $2h$ from the original wire antenna, as illustrated in Fig. \ref{fig:hordipgrondvlak}.
The current distribution on the image antenna is $180^0$ out of phase with respect to the current distribution on the original horizontal wire antenna.
In this way, we have defined an equivalent model of the original wire antenna above a ground plane.
The equivalent problem can be easily solved by applying the superposition principle and by using the expressions found in section \ref{sec-dunnedraadantene}.
Note that we have assumed that the current distribution of the resonant wire antenna is still purely sinusoidal and not affected by the presence of the ground plane.

\begin{figure}[hbt]
\centerline{\psfig{figure=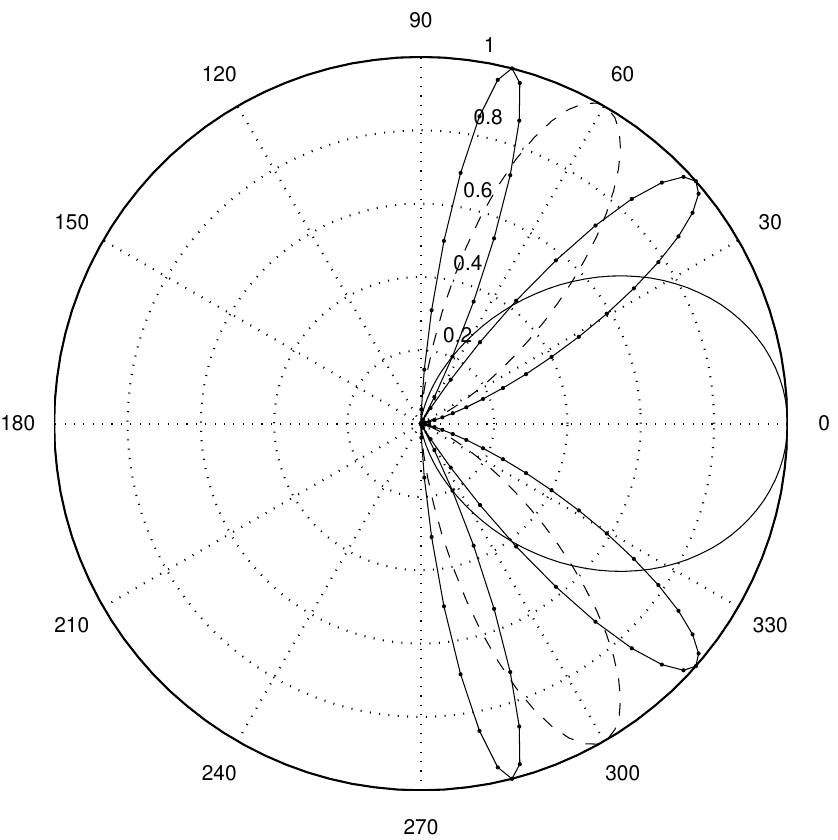,width=90mm}}

\caption{\it Normalized radiation pattern versus $\theta$ in
the $\phi=90^0$ plane of a horizontal wire antenna above a perfectly-conducting infinite metal ground plane. A linear scale is used.
With $h=0.1\lambda_0$ full line, $h=0.5\lambda_0$ dotted line and
$h=\lambda_0$ line with dotts.} \label{fig:Praddraadgrond}
\end{figure}

It is now interesting to investigate the radiation pattern of the horizontal wire antenna above a ground plane in more detail in the
$\phi=90^0$-plane. Note that we had assumed that the antenna is located along the $x$-axis.
Based on the symmetry in the $y-z$ plane, the radiation pattern of a wire antenna in free-space would have been independent of the angle $\theta$.
The ground plane makes the antenna $\theta$-dependent.
Now let $\vec{E}^o$ be the radiated electric field from a wire antenna in free-space and $\vec{E}^i$ the corresponding radiated electric field of the image antenna.
By applying the superposition principle we obtain the total field in the far-field region as:
\begin{equation}
\begin{array}{lcl}
 \displaystyle \vec{E}^{tot} & = & \displaystyle \vec{E}^o +
 \vec{E}^i \\
& = & \displaystyle \vec{E}_{free} \left[ e^{\jmath k_0 h
\cos{\theta}} - e^{-\jmath k_0 h \cos{\theta}} \right] \\

& = & \displaystyle \vec{E}_{free} 2 \jmath \sin(k_0 h
\cos{\theta}),

\end{array}
\label{eq:Edraadgrondvlak}
\end{equation}
where the vector $\vec{E}_{free}$ represents the electric field of a horizontal wire antenna in free space and located at the origin of the coordinate system.
The factor $2 \jmath \sin(k_0 h
\cos{\theta})$ describes the effect of the ground plane.
The corresponding radiation pattern, normalized to an horizontal wire antenna in free-space (so ignoring the factor $\vec{E}_{free}$ in (\ref{eq:Edraadgrondvlak})) is now given for the $\phi=90^0$ plane by:
\begin{equation}
\begin{array}{lcl}
\displaystyle F_r(\theta) & = & \displaystyle
\frac{P(\theta,\phi=90^0)}{P_{free}(\theta,\phi=90^0)} \\

& = & \displaystyle 4 \sin^2(k_0 h \cos{\theta}).
\end{array}
\end{equation}
The antenna does not radiate energy in the horizontal plane with $\theta=90^0$. Fig. \ref{fig:Praddraadgrond}
shows the radiation pattern versus the height $h$ above the ground plane.
The directivity of a horizontal wire antenna above a ground plane is larger as compared to a wire antenna in free space.
This is directly related to the fact that the antenna does not radiate energy in the half-space below the ground plane.
The directivity depends on the height $h$ and can be up to a factor 4 (6 dB) higher as compared to a free-space wire antenna.
The beam width is also strongly dependent on the height $h$ (See Fig.\ref{fig:Praddraadgrond}). In case $h=0.1\lambda_0$,
the beam width in the $\phi=90^0$-plane is $\theta_{HP}=94^0$.

\section{Folded dipole}

\begin{figure}[hbt]
\vspace{-0.5cm}
\centerline{\psfig{figure=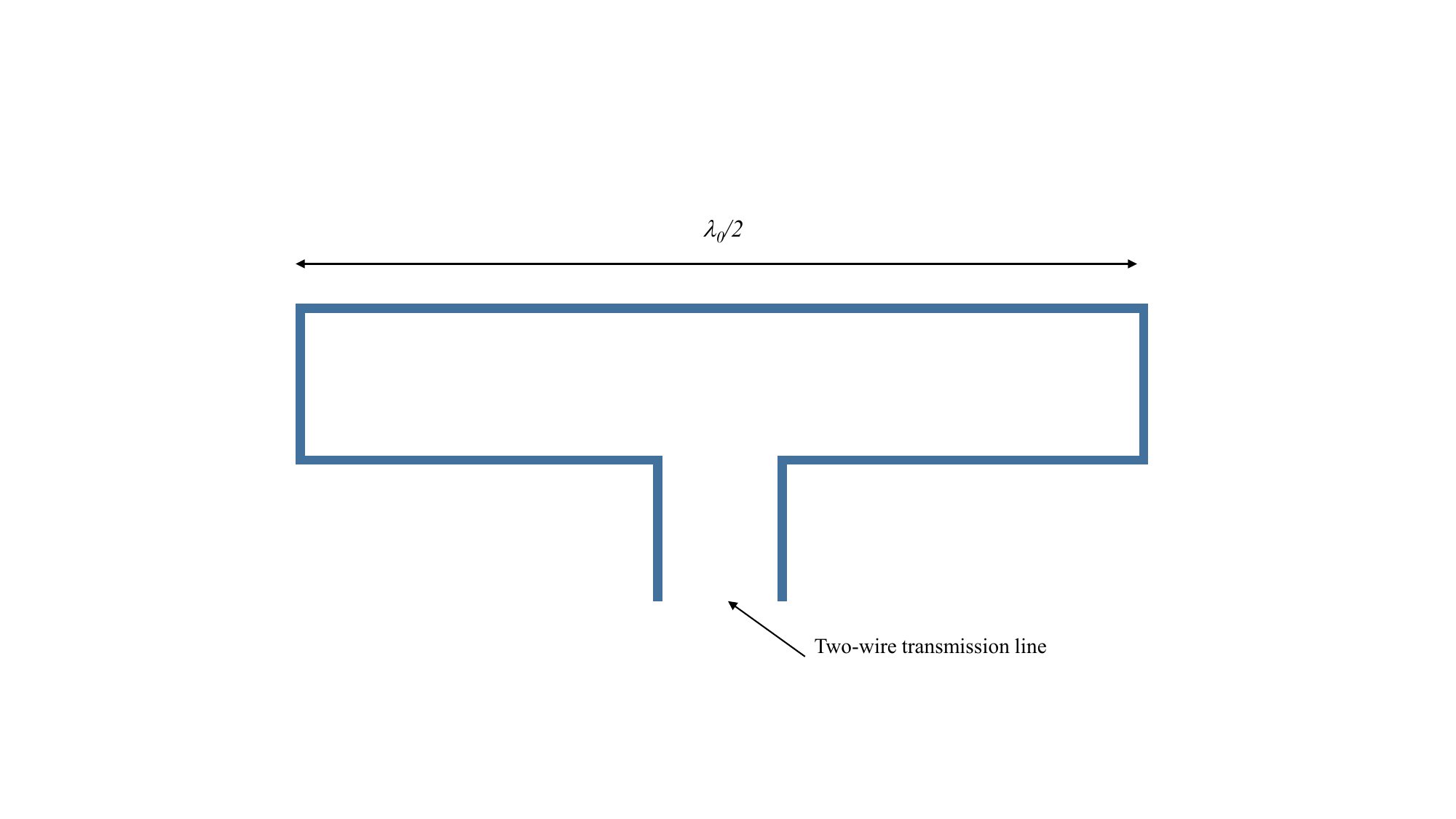,width=180mm}}
\vspace{-2cm}
\caption{\it Folded $\lambda_0/2$ dipole.}
\label{fig:foldeddipole}
\end{figure}

We have previously shown that a basic  $\lambda_0/2$ wire antenna has an input impedance at resonance of approx. $73 \Omega$. A two-wire transmission line (Lecherline or twin-lead) has a somewhat higher characteristic impedance, for example $Z_0^t=300 \Omega$. As a result a matching circuit would be required to match the antenna to the transmission line, resulting in additional losses and bandwidth limitations.
By modifying the basic $\lambda_0/2$ wire antenna to the structure of Fig.  \ref{fig:foldeddipole} we can realize a much higher input impedance. This antenna structure is called {folded dipole}. It can be shown that the input impedance of the $\lambda_0/2$ folded dipole is about 4 times higher as compared to the normal $\lambda_0/2$ wire antenna, so  $Z_{in} \approx
4 \cdot 73 =292 \Omega$. The radiation properties of the folded $\lambda_0/2$ dipole are quite similar to the characteristics of a normal $\lambda_0/2$ dipole.

\section{Loop antennas}

\begin{figure}[hbt]
\centerline{\psfig{figure=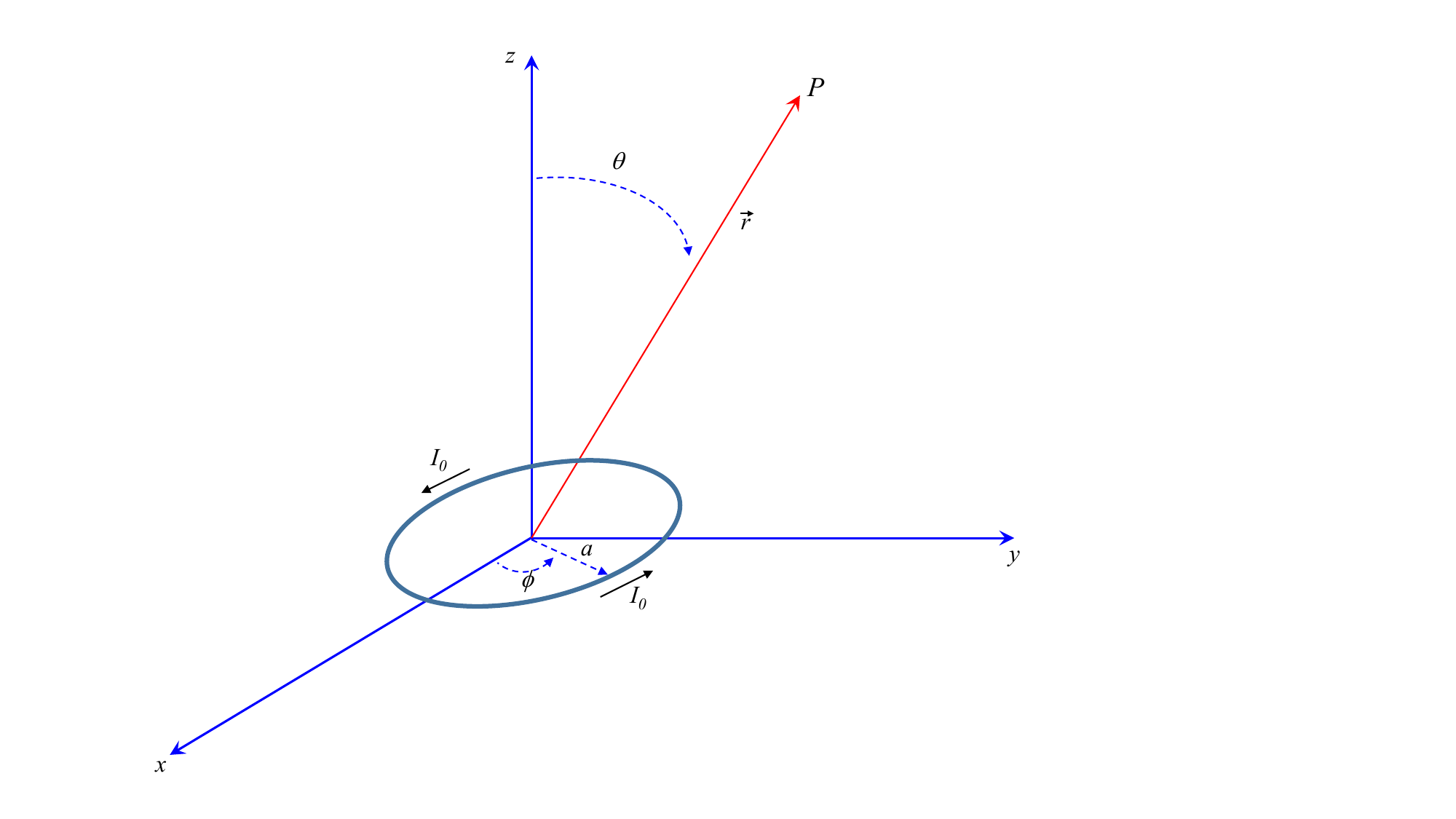,width=180mm}}
\caption{\it Loop antenna consisting of a circular thin metal wire with radius $a$ carrying a constant current $I_0$. The antenna is situated in the $x-y$-plane.} \label{fig:loopantenne}
\end{figure}

The {\it loop antenna} consists of a circular thin metal wire with radius $a$. In this section we will assume that the current $I_0$ along the loop is constant, which is only a valid approximation when the radius is very small as compared to the wavelength, i.e $a \ll \lambda_0$. For larger loop antennas, this can only be realized by using additional phase-shifters in the loop. Figure \ref{fig:loopantenne} shows the geometry of the loop antenna.
The electric current distribution $\vec{J}_e$ can now be represented by:
\begin{equation}
\displaystyle \vec{J}_e (\vec{r}_0) = I_0 \delta\left(
\sqrt{x_0^2+y_0^2}-a \right) \delta(z_0) \vec{u}_{\phi_0}.
\end{equation}
In chapter \ref{sec:radiatedfields} we found that the far field can be determined with:
\begin{equation}
\begin{array}{lcl}
\displaystyle \vec{H}  & = &  \displaystyle \frac{-\jmath k_0
e^{-\jmath k_0 r} }{4\pi r} \vec{u}_{r} \times  \int_{V_0}
\vec{J}_e (\vec{r}_0) e^{\jmath k_0  \vec{u}_r \cdot \vec{r}_0} dV_0,
\\

\displaystyle \vec{E}  & = &  \displaystyle Z_0 \vec{H} \times
\vec{u}_r.
\end{array}
\label{eq:Hveldloop1}
\end{equation}
The integration for the loop antenna is carried out over a circle with radius $r_0=a$. We will use polar coordinates to facilitate the integration. Furthermore
\begin{equation}
\begin{array}{lcl}
\displaystyle \vec{u}_r \cdot \vec{r}_0  & = & \displaystyle
[\cos{\phi}\sin{\theta}\vec{u}_x+\sin{\phi}\sin{\theta}\vec{u}_y
+\cos{\theta}\vec{u}_z] \cdot [a\cos{\phi_0}\vec{u}_x+a\sin{\phi_0}\vec{u}_y]
\\
 & = & \displaystyle a\cos(\phi-\phi_0)\sin{\theta}.

\end{array}
\end{equation}
Substituting this in (\ref{eq:Hveldloop1}) provides
\begin{equation}
\begin{array}{lcl}
\displaystyle \vec{H}  & = &  \displaystyle \frac{-\jmath k_0
e^{-\jmath k_0 r} }{4\pi r} (I_0a) \vec{u}_{r} \times
\int\limits_{0}^{2\pi} \vec{u}_{\phi_0} e^{\jmath k_0 a
\cos(\phi-\phi_0)\sin{\theta}} d\phi_0 .

\end{array}
\label{eq:Hveldloop2}
\end{equation}
Due to the symmetry of the geometry, we can assume that the fields will not depend on the $\phi$ coordinate. Therefore we can use $\phi=0$ in
(\ref{eq:Hveldloop2}). In addition,
$\displaystyle \vec{u}_{\phi_0}=-\vec{u}_x\sin{\phi_0} +
\vec{u}_y\cos{\phi_0}$. In the $\phi=0$ plane it is found that
$\vec{u}_{\phi}=\vec{u}_y$. We can now divide the integral in
(\ref{eq:Hveldloop2}) into two parts according to
\begin{equation}
\begin{array}{lcl}
\displaystyle \vec{I}_{\phi_0}  & = &  \displaystyle
\int\limits_{0}^{2\pi} \vec{u}_{\phi_0} e^{\jmath k_0 a
\cos{\phi_0}\sin{\theta}} d\phi_0 \\
  & = &  \displaystyle - \vec{u}_x \int\limits_{0}^{2\pi} \sin{\phi_0} e^{\jmath k_0 a
\cos{\phi_0}\sin{\theta}} d\phi_0 +  \vec{u}_y
\int\limits_{0}^{2\pi} \cos{\phi_0} e^{\jmath k_0 a
\cos{\phi_0}\sin{\theta}} d\phi_0 \\ & = &  \displaystyle
\vec{u}_y \int\limits_{0}^{2\pi} \cos{\phi_0} e^{\jmath k_0 a
\cos{\phi_0}\sin{\theta}} d\phi_0.

\end{array}
\label{eq:Iloop1}
\end{equation}
The first integral is equal to zero, since the integrand is an odd function with respect to $\phi_0$. The remaining integral can be analytically solved:
\begin{equation}
\begin{array}{lcl}
\displaystyle \vec{I}_{\phi_0}  & = &  \displaystyle \vec{u}_y
\int\limits_{0}^{2\pi} \cos{\phi_0} e^{\jmath k_0 a
\cos{\phi_0}\sin{\theta}} d\phi_0 \\

& = &  \displaystyle 2 \vec{u}_y \int\limits_{0}^{\pi}
\cos{\phi_0} e^{\jmath k_0 a \cos{\phi_0}\sin{\theta}} d\phi_0 \\

& = &  \displaystyle -2 \vec{u}_y \int\limits_{-\pi/2}^{\pi/2}
\sin{\phi_0} e^{-\jmath k_0 a \sin{\phi_0}\sin{\theta}} d\phi_0 \\

& = &  \displaystyle 2 \vec{u}_y \int\limits_{0}^{\pi/2}
\sin{\phi_0} \left( e^{\jmath k_0 a \sin{\phi_0}\sin{\theta}} -
e^{-\jmath k_0 a \sin{\phi_0}\sin{\theta}} \right) d\phi_0 \\

& = &  \displaystyle 4\jmath \vec{u}_y \int\limits_{0}^{\pi/2}
\sin{\phi_0} \sin(k_0 a \sin{\phi_0}\sin{\theta}) d\phi_0 \\

& = &  \displaystyle 2\pi \jmath J_1(k_0 a \sin{\theta}) \vec{u}_y
\\

& = &  \displaystyle 2\pi \jmath J_1(k_0 a \sin{\theta})
\vec{u}_{\phi}

\end{array}
\label{eq:Iloop2}
\end{equation}
where $J_1(z)$ is the Bessel function of the first kind with order 1 and argument $z$. Bessel functions are well tabulated (e.g. \cite{Gradsteyn}) and are available in Matlab and other mathematical tools.  Finally, we obtain the following expression for the magnetic field in the far-field region using $\vec{u}_r \times
\vec{u}_{\phi}= - \vec{u}_{\theta}$:
\begin{equation}
\displaystyle \vec{H}  = H_{\theta} \vec{u}_{\theta} = - \frac{k_0
a I_0 e^{-\jmath k_0 r}}{2 r} J_1(k_0 a \sin{\theta}).
\label{eq:Hveldloop3}
\end{equation}
Since $\vec{u}_{\theta} \times \vec{u}_r = -
\vec{u}_{\phi}$, we can conclude that the corresponding electric field will only have a $\phi$ component:
\begin{equation}
\displaystyle \vec{E}  = E_{\phi} \vec{u}_{\phi} = \frac{k_0 a Z_0
I_0 e^{-\jmath k_0 r}}{2 r} J_1(k_0 a \sin{\theta}).
\label{eq:Eveldloop}
\end{equation}

\subsection{Small loop antennas: magnetic dipole}
\label{sec:smallloop}

A small loop antenna, also known as {\it magnetic dipole}, consists of a circular metal wire with radius $a \ll \lambda_0$ and with a constant current $I_0$ along the loop.
We can use expression (\ref{eq:Hveldloop3}) and (\ref{eq:Eveldloop}) to determine the far fields, where we can approximate the Bessel function $J_1(z)$ for small argument $z$ according to
\begin{equation}
\displaystyle J_1(z) \approx \frac{z}{2}.
\end{equation}
By substituting this in (\ref{eq:Hveldloop3}) and (\ref{eq:Eveldloop}), we find that the far fields are given by:
\begin{equation}
\begin{array}{lcl}
 \displaystyle H_{\theta} & = & \displaystyle
- \frac{k_0^2 (\pi a^2 I_0) e^{-\jmath k_0 r}}{4 \pi r}
\sin{\theta}
\\ \displaystyle E_{\phi} & = &  \displaystyle \frac{k_0^2 Z_0 (\pi a^2 I_0) e^{-\jmath
k_0 r}}{4 \pi r} \sin{\theta}.
\end{array}
\label{eq:EHveldmagdipool}
\end{equation}
The factor $\pi a^2 I_0$ is also known as the "transmitter-moment" and is the product of the current
$I_0$ and the surface of the loop. Usually it is denoted by $m$ with
\begin{equation}
\displaystyle m=\pi a^2 I_0.
\end{equation}
There is a strong analogy between the fields radiated by a magnetic dipole according to (\ref{eq:EHveldmagdipool}) and the fields from the electric dipole, given by (\ref{eq:Hvelddipool}) and (\ref{eq:Evelddipool}). The radiation patterns of the magnetic and electric dipole are exactly the same.
Now let us assume that both an electric as well as an magnetic dipole are located in the origin of our coordinate system.
Furthermore, assume that the electric dipole is oriented along the $z$-axis and the magnetic dipole is situated in the $x$-$y$-plane.
When we choose the dimensions of both dipoles and current to fulfill the relation
\begin{equation}
\displaystyle p=-m k_0, \label{eq:penmrelatie}
\end{equation}
the corresponding electromagnetic field generated by these two sources will fullfill the following relation
\begin{equation}
\displaystyle \vec{E}= \jmath Z_0 \vec{H}.
\label{eq:EHmageldipool1}
\end{equation}
The minus sign in (\ref{eq:penmrelatie}) indicates that the currents in the electric and magnetic dipoles are $180^0$ out of phase. When the currents are in phase, i.e. $p=mk_0$, we find the following form of the radiated electromagnetic field:
\begin{equation}
\displaystyle \vec{E}= - \jmath Z_0 \vec{H}.
\label{eq:EHmageldipool2}
\end{equation}
Electromagnetic fields satisfying (\ref{eq:EHmageldipool1})
or (\ref{eq:EHmageldipool2}) have the unique property that the radiated far fields are circularly polarized in all directions.
Using (\ref{eq:EHmageldipool1}) we find that
\begin{equation}
\displaystyle E_{\phi}= \jmath E_{\theta},
\end{equation}
for all directions $(\theta,\phi)$. Therefore, we satisfy the relation between $E_{\theta}$ and $E_{\phi}$ as described in section \ref{sec:polarisatie} for a perfectly left-handed circularly-polarized field. Fields that satisfy
(\ref{eq:EHmageldipool2}) are right-handed circularly polarized for all directions $(\theta,\phi)$.

\vspace{1cm} \hspace{-0.5cm} \hrulefill
 {\bf Exercise} \hrulefill \\
 Calculate the directivity of a magnetic dipole.

\subsection{Large loop antennes}

\begin{figure}[hbt]
\centerline{\psfig{figure=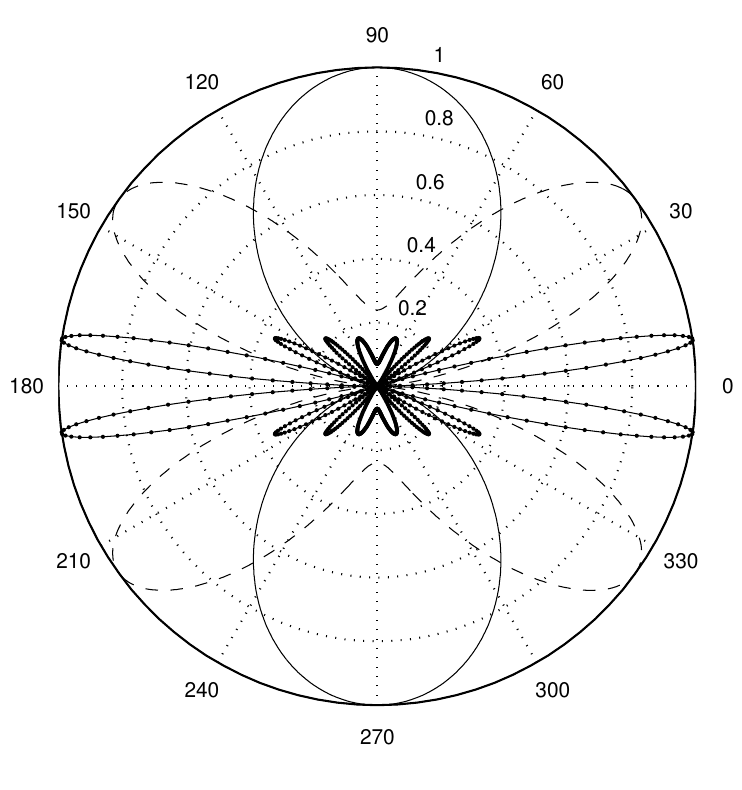,width=90mm}}

\caption{\it Normalized radiation pattern versus $\theta$ of a loop antenna with uniform current distribution for various loop dimensions. With $a=\lambda_0/20$ full line, $a=\lambda_0/2$ dotted line and $a=2 \lambda_0$ line with dots.} \label{fig:Stralingloop}
\end{figure}

The electromagnetic fields in the far-field region of a loop antenna with uniform current distribution along the loop was already determined in (\ref{eq:Hveldloop3}) and
(\ref{eq:Eveldloop}). We can use these expressions to determine the radiation pattern, resulting in
\begin{equation}
\begin{array}{lcl}
\displaystyle F(\theta) &  = & \displaystyle
\frac{P(\theta)}{\max(P(\theta))} \\ &  = & \displaystyle
\frac{|J_1(k_0 a \sin{\theta})|^2}{\max(|J_1(k_0 a
\sin{\theta})|^2)} .
\end{array}
\label{eq:Fradloop}
\end{equation}
The shape of the pattern will be similar to the behavior of the Bessel function $J_1 (z)$. Fig.
\ref{fig:Stralingloop} shows the normalized radiation pattern of a loop antenna with radius $a=\lambda_0/20$,
$a=\lambda_0/2$, $a=2 \lambda_0$, respectively. Note that the direction of maximum radiation moves towards $\theta=0$ with increasing loop radius (check Fig. \ref{fig:loopantenne}) .

\section{Magnetic sources}
\label{sec:magbron}

In order to be able to introduce aperture antennas, it is useful to extend Maxwell's equations, given by (\ref{eq:maxwelltijd}).
We will do this by introducing magnetic currents and associated magnetic charge distributions.
Note that both magnetic currents as well as magnetic charges do not exist in the real-world: both are non-physical quantities..
However, it will help us to introduce several new electromagnetic concepts in an elegant way.
For that purpose, we will introduce the magnetic current distribution $\vec{J}_m (\vec{r})$ and the magnetic charge density $\rho_m (\vec{r})$.
When we assume that there are no electric currents $\vec{J}_e (\vec{r})=\vec{0}$ and electric charges $\rho_e
(\vec{r})=0$, Maxwell's equations take the following form in the space-frequency domain:
\begin{equation}
\begin{array}{lcl}
\displaystyle \nabla \times \vec{E}(\vec{r}) &=& \displaystyle -
\jmath \omega \mu_0 \vec{H}(\vec{r})- \vec{J}_m (\vec{r}), \\

\displaystyle \nabla \times \vec{H}(\vec{r}) &=& \displaystyle
\jmath \omega \epsilon_0 \vec{E}(\vec{r}) ,
\\

\displaystyle \nabla \cdot \vec{H} (\vec{r}) & = & \displaystyle
\frac{\rho_m (\vec{r})}{\mu_0}, \\
 \displaystyle \nabla \cdot
\vec{E} (\vec{r}) & = & 0.

\end{array}
\label{eq:maxwellfreqJm}
\end{equation}
By combining the first and third equation from (\ref{eq:maxwellfreqJm}) we observe that the following continuity equation yields:
\begin{equation}
\begin{array}{lcl}

\displaystyle \nabla \cdot \vec{J}_m (\vec{r}) + \jmath \omega
{\rho}_m (\vec{r}) & = & 0.

\end{array}
\label{eq:continuiteitsvglJm}
\end{equation}
The solution of these equations related to magnetic sources can be found by using a similar approach as used in section \ref{sec:antennatheory} for the case of electric sources.
As a first step, we will introduce a vector potential, in this case the {\it electric
vector potential} $\vec{A}_m$, using the fourth equation in (\ref{eq:maxwellfreqJm}). As a result, we find that
\begin{equation}
\displaystyle \vec{E} = -\frac{1}{\epsilon_0} \nabla \times
\vec{A}_m. \label{eq:vectorpot1Jm}
\end{equation}
This gives $\nabla \times \vec{H}=-\jmath \omega \nabla \times
\vec{A}_m$. We can now write $\vec{H}$ in the form:
\begin{equation}
\displaystyle \vec{H} = -\jmath \omega \vec{A}_m - \nabla \phi_m,
\label{eq:HenAJm}
\end{equation}
where $\phi_m$ is (yet) an arbitrarily scalar function.
By substituting this relation into the first equation of (\ref{eq:maxwellfreqJm}) and using the Lorenz-gauge according to
\begin{equation}
\displaystyle \nabla \cdot \vec{A}_m = - \jmath \omega \epsilon_0
\mu_0 \phi_m , \label{eq:LorentzijkJm}
\end{equation}
leads towards the well-known Helmholtz equation for the vector potential $\vec{A}_m$, given by
\begin{equation}
\displaystyle \nabla^2 \vec{A}_m + k_0^2 \vec{A}_m = -\epsilon_0
\vec{J}_m. \label{eq:Helmholtz2Jm}
\end{equation}
A similar expression can be found for the scalar potental $\phi_m$ directly from the Lorenz-gauge that we used.
In section \ref{sec:antennatheory} we solved Helmholtz equation for an arbitrarily electric current distribution
$\vec{J}_e$ by first looking for the solution for an elementary point source.
The general solution was then found by applying the superposition concept, since any electric current distribution
$\vec{J}_e$ can be written as a continuous superposition of point sources.
By applying the same strategy here, we find that the electric vector potential $\vec{A}_m$ due to an arbitrarily magnetic current distribution $\vec{J}_m$ takes the following form:
\begin{equation}
\displaystyle \vec{A}_m (\vec{r}) = \frac{\epsilon_0}{4\pi}
\int_{V_0} \vec{J}_m(\vec{r}_0) \frac{e^{-\jmath k_0
|\vec{r}-\vec{r}_0|}}{|\vec{r}-\vec{r}_0|} dV_0,
\label{eq:vectorpotJm}
\end{equation}
where we have assumed that the observation point $P$, indicated by the vector $\vec{r}$, is located outside the source region $V_0$.
The electromagnetic fields outside the source region $V_0$ are now found by
\begin{equation}
\begin{array}{lcl}
\displaystyle \vec{E} & = & \displaystyle  - \frac{1}{\epsilon_0}
\nabla \times \vec{A}_m, \\

\displaystyle \vec{H} & = & \displaystyle  \frac{1}{\jmath \omega
\epsilon_0 \mu_0} \nabla \times \nabla \times \vec{A}_m.

\end{array}
\label{eq:HenEinAJm}
\end{equation}
Similar to section \ref{sec:radiatedfields}, we can determine the electromagnetic fields far away from the sources (far-field region) by using several approximations.
The fields in the far-field region than take the following form:
\begin{equation}
\begin{array}{lcl}
\displaystyle \vec{E}  & = &  \displaystyle \frac{\jmath k_0
e^{-\jmath k_0 r} }{4\pi r} \vec{u}_{r} \times  \int_{V_0}
\vec{J}_m (\vec{r}_0) e^{\jmath k_0 \vec{u}_r \cdot \vec{r}_0}
dV_0,
\\
\displaystyle \vec{H}  & = &  \displaystyle \frac{-k_0^2}{\jmath
\omega \mu_0} \frac{e^{-\jmath k_0 r}}{4\pi r} \vec{u}_r \times
\vec{u}_r \times \int_{V_0} \vec{J}_m (\vec{r}_0) e^{\jmath k_0
\vec{u}_r \cdot \vec{r}_0} dV_0, \\

\displaystyle \vec{E}  & = &  \displaystyle Z_0 \vec{H} \times
\vec{u}_r, \\
\end{array}
\label{eq:verreveldJm}
\end{equation}
We again can conclude that the fields in the far-field region locally behave as TEM-waves ($E$-field perpendicular to $H$ and perpendicular to the propagation direction $\vec{u}_r$).  The strong analogy between magnetic and electric sources is known in literature as the {\it duality concept}. Always keep in mind that magnetic currents and charges are mathematical quantities that cannot exist in nature.

We have now introduced all tools to determine the radiated electromagnetic fields from a general source, consisting of both electric and (fictive) magnetic currents and charges. We can do this by using expressions (\ref{eq:Samenvatting})
and (\ref{eq:verreveldJm}) and by applying the superposition principle.

\vspace*{1cm}  \hrulefill
 {\bf Exercise} \hrulefill \\ In the origin of our coordinate system a magnetic current distribution exist with the following properties:
\begin{equation}
\displaystyle \vec{J}_m(\vec{r}_0) = I_{0m} l \delta(x) \delta(y)
\delta(z) \vec{u}_z ,
\end{equation}
where $I_{0m}$ is the total magnetic current. Show that the radiated electromagnetic fields due to this source are identical to the fields due to the magnetic dipole (small loop antenna) as discussed in section \ref{sec:smallloop}.

\section{Lorentz-Larmor theorem}
\begin{figure}[hbt]
  \centerline{\psfig{figure=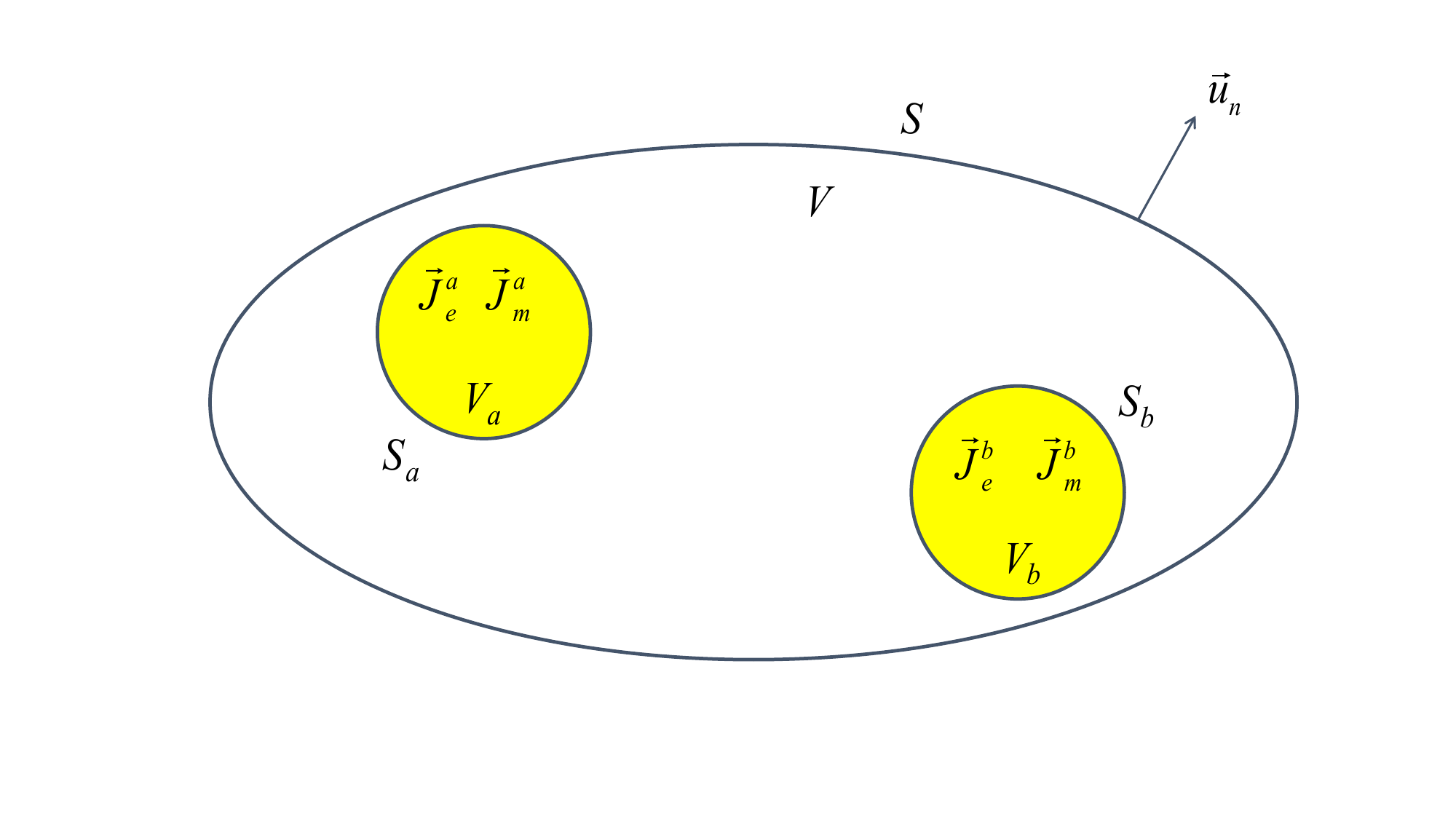,width=140mm}}
  \caption{\it Source region $V^a$ enclosed by the surface $S^a$ and source region $V^b$ enclosed by surface $S^b$. Both regions are located within volume $V$ with corresponding surface $S$.
  Within source region $V^a$ both electric and magnetic current distributions $\vec{J}_e^a$ and $\vec{J}_m^a$ exist and the corresponding charge distributions $\rho_e^a$ and $\rho_m^a$. Similar for source region $V^b$.}
  \label{fig:brongebieda}
\end{figure}

In order to analyze aperture antenna, it is convenient to introduce
the {\it Lorentz-Larmor theorem}. This concept defines the electromagnetic relation between two situations A and B. Later in this section we will show that we can apply this theorem to replace a particular antenna problem, represented by a source region $V^a$ with electric and magnetic (volume-) current distributions
$\vec{J}_e^a$ and $\vec{J}_m^a$, by an equivalent and simpler antenna problem.
In literature this concept is also known as the {\it Equivalence principle} \cite{Harrington1}.
Figure \ref{fig:brongebieda} illustrates source region $V^a$ bounded by the surface $S^a$.
The sources $\vec{J}_e^a$ and $\vec{J}_m^a$ generate an electromagnetic field $\vec{E}^a$ and $\vec{H}^a$ outside source region $V^a$.
It will be shown that the same fields {\it outside} $V^a$ can also be generated by an equivalent electric surface-current distribution $\vec{J}_{es}^a$ on $S^a$
and an equivalent magnetic surface-current distribution $\vec{J}_{ms}^a$ on
$S^a$ where
\begin{equation}
\begin{array}{lcl}
\displaystyle \vec{J}_{es}^a & = & \displaystyle \vec{u}_n \times
\vec{H}^a, \\ \displaystyle \vec{J}_{ms}^a & = & \displaystyle
\vec{E}^a \times \vec{u}_n,
\end{array}
\label{eq:equivalentebronnen1}
\end{equation}
where $\vec{u}_n$ is the normal vector on surface $S^a$. The equivalent sources (\ref{eq:equivalentebronnen1}) produce within $V^a$ an zero electromagnetic field and outside $V^a$ an electromagnetic field equal to the original problem.

\vspace*{1cm}  \hrulefill
 {\bf Intermezzo: derivation of Lorentz-Larmor theorem} \hrulefill \\
The derivation of the Lorentz-Larmor theorem provides interesting insight into the relation and electromagnetic interaction between two source regions.
As a first step, we will formulate Maxwell's equations in the space-frequency domain for situation A where the fields
$\vec{E}^a$ and $\vec{H}^a$ are generated only by the sources in volume $V^a$ (see also Fig. \ref{fig:brongebieda}):
\begin{equation}
\begin{array}{lcl}
\displaystyle \nabla \times \vec{E}^a  &=& \displaystyle - \jmath
\omega \mu_0 \vec{H}^a - \vec{J}_m^a,
\\

\displaystyle \nabla \times \vec{H}^a &=& \displaystyle \jmath
\omega \epsilon_0 \vec{E}^a +\vec{J}_e^a ,
\\

\displaystyle \nabla \cdot \vec{H}^a & = & \displaystyle
\frac{\rho_m^a}{\mu_0}, \\
 \displaystyle \nabla \cdot
\vec{E}^a & = &  \displaystyle  \frac{\rho_e^a}{\epsilon_0}.

\end{array}
\label{eq:maxwellfreqJmJeA}
\end{equation}
In situation B, a source distribution $\vec{J}_e^b$ and $\vec{J}_m^b$ exist within the volume $V^b$ together with the corresponding charge distributions $\rho_e^b$ and $\rho_m^b$. The fields generated by these sources (so without the situation A sources) comply to the following Maxwell equations :
\begin{equation}
\begin{array}{lcl}
\displaystyle \nabla \times \vec{E}^b  &=& \displaystyle - \jmath
\omega \mu_0 \vec{H}^b - \vec{J}_m^b,
\\

\displaystyle \nabla \times \vec{H}^b &=& \displaystyle \jmath
\omega \epsilon_0 \vec{E}^b +\vec{J}_e^b ,
\\

\displaystyle \nabla \cdot \vec{H}^b & = & \displaystyle
\frac{\rho_m^b}{\mu_0}, \\
 \displaystyle \nabla \cdot
\vec{E}^b & = &  \displaystyle \frac{\rho_e^b}{\epsilon_0}.

\end{array}
\label{eq:maxwellfreqJmJeB}
\end{equation}
The divergence of the vector function $\vec{E}^a \times \vec{H}^b -
\vec{E}^b \times \vec{H}^a$ is:
\begin{equation}
\displaystyle \nabla \cdot \left[ \vec{E}^a \times \vec{H}^b -
\vec{E}^b \times \vec{H}^a \right] = \nabla \cdot \left[ \vec{E}^a
\times \vec{H}^b \right] - \nabla \cdot \left[ \vec{E}^b \times
\vec{H}^a \right]. \label{eq:divEH}
\end{equation}
The first term of the right-hand side of equation (\ref{eq:divEH}) can be written as
\begin{equation}
\begin{array}{lcl}
 \displaystyle \nabla \cdot \left[
\vec{E}^a \times \vec{H}^b \right] & = & \displaystyle \vec{H}^b
\cdot \nabla \times \vec{E}^a - \vec{E}^a \cdot \nabla \times
\vec{H}^b \\
 & = & \displaystyle \vec{H}^b
\cdot \left( -\jmath \omega \mu_0 \vec{H}^a - \vec{J}_m^a \right)
- \vec{E}^a \cdot \left(\jmath \omega \epsilon_0 \vec{E}^b +
\vec{J}_e^b \right) \\

& = & \displaystyle  -\jmath \omega \mu_0 \vec{H}^b \cdot
\vec{H}^a - \vec{H}^b \cdot \vec{J}_m^a  - \jmath \omega
\epsilon_0 \vec{E}^a \cdot \vec{E}^b - \vec{E}^a \cdot
\vec{J}_e^b,
\end{array}
\label{eq:divEH1a}
\end{equation}
where we have used the vector identity $\nabla \cdot
(\vec{A} \times \vec{B}) = \vec{B} \cdot \nabla \times \vec{A} -
\vec{A} \cdot \nabla \times \vec{B}$. In a similar way, we can determine the second term in the right-hand side of equation (\ref{eq:divEH}) by exchange of  $a$ and $b$ in (\ref{eq:divEH1a}):
\begin{equation}
\begin{array}{lcl}
 \displaystyle \nabla \cdot \left[
\vec{E}^b \times \vec{H}^a \right] & = & \displaystyle -\jmath
\omega \mu_0 \vec{H}^a \cdot \vec{H}^b - \vec{H}^a \cdot
\vec{J}_m^b  - \jmath \omega \epsilon_0 \vec{E}^b \cdot \vec{E}^a
- \vec{E}^b \cdot \vec{J}_e^a .
\end{array}
\label{eq:divEH1b}
\end{equation}
Combining (\ref{eq:divEH1a}) and (\ref{eq:divEH1b}) provides the following result
\begin{equation}
\displaystyle \nabla \cdot \left[ \vec{E}^a \times \vec{H}^b -
\vec{E}^b \times \vec{H}^a \right] = \left[\vec{E}^b \cdot
\vec{J}_e^a - \vec{H}^b \cdot \vec{J}_m^a \right] -
\left[\vec{E}^a \cdot \vec{J}_e^b - \vec{H}^a \cdot \vec{J}_m^b
\right] . \label{eq:divEH2}
\end{equation}
Applying Gauss theorem according to
\begin{equation}
\int\limits_{V} \nabla \cdot \vec{A} dV = \int\limits_{S} \vec{A}
\cdot \vec{u}_n dS
\end{equation}
on the volume $V$ enclosed by surface $S$ (see Fig. \ref{fig:brongebieda}) provides
\begin{equation}
\displaystyle \int\limits_S \left[ \vec{E}^a \times \vec{H}^b -
\vec{E}^b \times \vec{H}^a \right] \cdot \vec{u}_n dS =
\int\limits_V \left( \left[\vec{E}^b \cdot \vec{J}_e^a - \vec{H}^b
\cdot \vec{J}_m^a \right] - \left[\vec{E}^a \cdot \vec{J}_e^b -
\vec{H}^a \cdot \vec{J}_m^b \right] \right) dV .
\label{eq:lorentztheorema}
\end{equation}
This relation is known as {\it Lorentz' theorem}. In order to derive the Lorentz-Larmor theorem,
we will apply Lorentz' theorem on several special cases.

\begin{enumerate}[i.]
\item {\bf Volume $V$ encloses the entire free-space $V_{\infty}$.} \\
The corresponding surface $S$ is now a sphere $S_{\infty}$ with infinite large radius.
The sources $(\vec{E}^a,\vec{H}^a)$ and $(\vec{E}^b,\vec{H}^b)$ are located within a finite region. At large distances from these sources we have, according to (\ref{eq:Samenvatting}) and
(\ref{eq:verreveldJm}), the following relations:
\begin{equation}
\begin{array}{lcl}
\displaystyle \vec{H}^a & = & \displaystyle \frac{1}{Z_0}
\vec{u}_r \times \vec{E}^a, \\ \displaystyle \vec{H}^b & = &
\displaystyle \frac{1}{Z_0} \vec{u}_r \times \vec{E}^b.
\end{array}
\end{equation}
Furthermore, for a point with location vector $\vec{r}$ on the sphere $S_{\infty}$ we know that
$\vec{u}_r=\vec{u}_n$. Therefore, the integrand of the left-hand side of expression (\ref{eq:lorentztheorema})
can be written in the following form
\begin{equation}
\begin{array}{lcl}
\displaystyle \left[ \vec{E}^a \times \vec{H}^b - \vec{E}^b \times
\vec{H}^a \right] \cdot \vec{u}_n & = & \displaystyle \left[
\vec{u}_n \times \vec{E}^a \right] \cdot \vec{H}^b - \left[
\vec{u}_n \times \vec{E}^b \right] \cdot \vec{H}^a \\
 & = & \displaystyle \left[
\vec{u}_r \times \vec{E}^a \right] \cdot \vec{H}^b - \left[
\vec{u}_r \times \vec{E}^b \right] \cdot \vec{H}^a \\

& = & \displaystyle Z_0 \vec{H}^a \cdot \vec{H}^b - Z_0 \vec{H}^b
\cdot \vec{H}^a \\ & = & 0
\end{array}
\end{equation}
The surface integral in (\ref{eq:lorentztheorema}) will vanish when $S=S_{\infty}$.
Therefore, expression (\ref{eq:lorentztheorema}) can be reduced to
\begin{equation}
\displaystyle \int\limits_{V_{\infty}} \left[\vec{E}^b \cdot
\vec{J}_e^a - \vec{H}^b \cdot \vec{J}_m^a \right] dV =
\int\limits_{V_{\infty}} \left[\vec{E}^a \cdot \vec{J}_e^b -
\vec{H}^a \cdot \vec{J}_m^b \right] dV .
\label{eq:lorentztheorema1}
\end{equation}
Although the integration in (\ref{eq:lorentztheorema1}) is performed over the entire volume $V_{\infty}$, it is evident that the integration can be limited to the (finite) source volume.

\item {\bf Source distribution B is an electric dipole.} \\
Assume that our source distribution is an electric dipole with dipole moment $p=1$ and direction $\vec{u}_b$ located at $\vec{r}$ outside the source regions. This is illustrated in Fig. \ref{fig:Bisdipool}.
\begin{figure}[hbt]
  \centerline{\psfig{figure=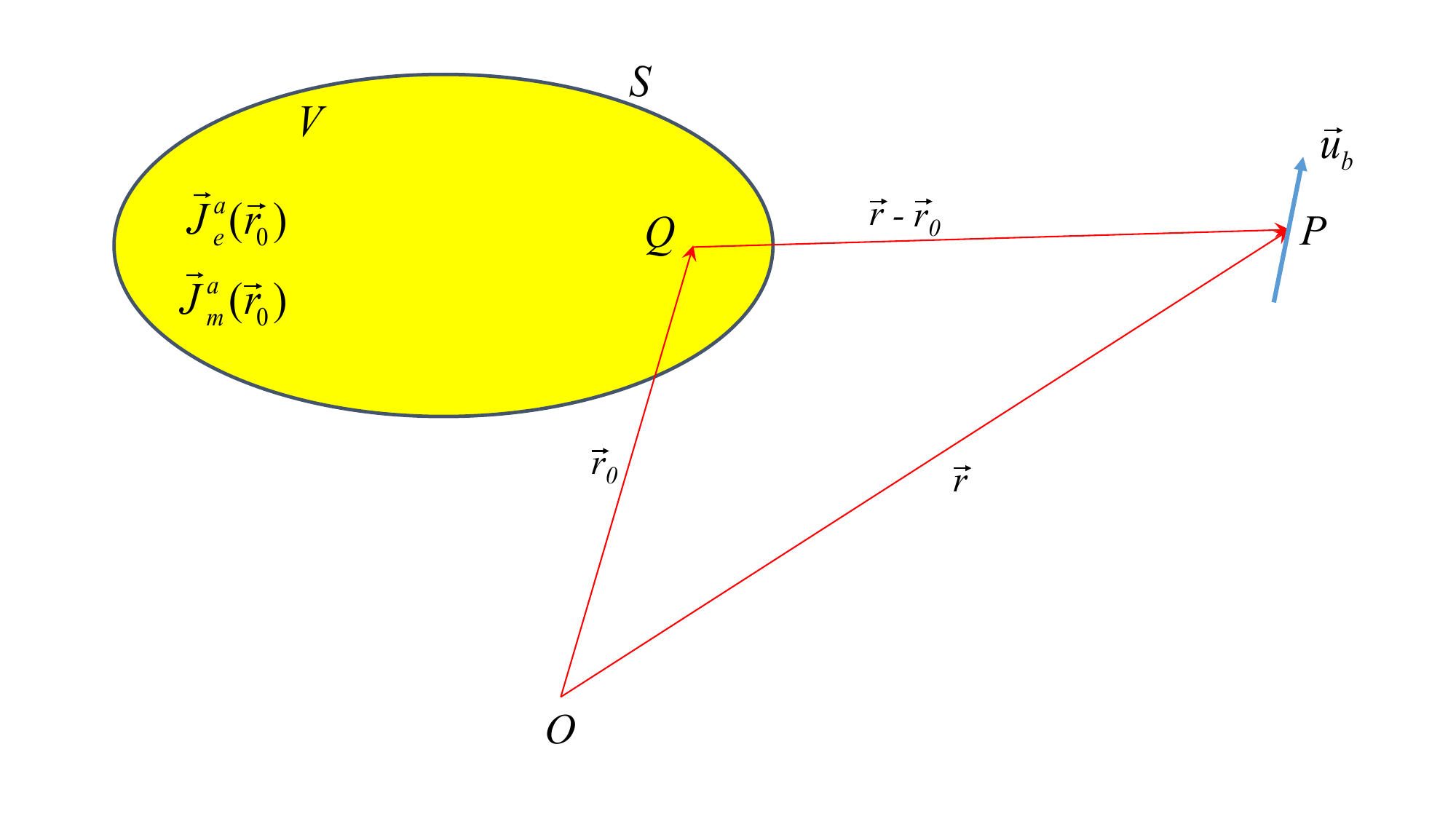,width=140mm}}
  \caption{\it Source distribution B is an electric dipole directed towards $\vec{u}_b$
  with dipole moment $p=1$.}
  \label{fig:Bisdipool}
\end{figure}
We now obtain
\begin{equation}
\begin{array}{lcl}
\displaystyle \vec{J}_m^b & = & 0, \\ \displaystyle \vec{J}_e^b &
= &  \displaystyle  \vec{u}_b \delta(\vec{r}-\vec{r}_0),
\end{array}
\end{equation}
where the vector $\vec{r}$ is fixed and $\vec{r}_0$ variable.
When using this relation in (\ref{eq:lorentztheorema1}) we find that
\begin{equation}
\displaystyle  \int\limits_{V_{\infty}} \vec{u}_b \cdot \vec{E}^a
(\vec{r}_0) \delta(\vec{r}-\vec{r}_0) dV_0 = \int\limits_{V^a}
\left[\vec{E}^b (\vec{r}_0) \cdot \vec{J}_e^a (\vec{r}_0) -
\vec{H}^b (\vec{r}_0) \cdot \vec{J}_m^a (\vec{r}_0) \right] dV_0,
\label{eq:lorentztheorema2a}
\end{equation}
or
\begin{equation}
\displaystyle  \vec{u}_b \cdot \vec{E}^a (\vec{r}) =
\int\limits_{V^a} \left[\vec{E}^b (\vec{r}_0) \cdot \vec{J}_e^a
(\vec{r}_0) - \vec{H}^b (\vec{r}_0) \cdot \vec{J}_m^a (\vec{r}_0)
\right] dV_0. \label{eq:lorentztheorema2b}
\end{equation}
In the special case that we only have electric sources within $V^a$,  expression (\ref{eq:lorentztheorema2b})
takes the form:
\begin{equation}
\displaystyle  \vec{u}_b \cdot \vec{E}^a (\vec{r}) =
\int\limits_{V^a} \vec{E}^b (\vec{r}_0) \cdot \vec{J}_e^a
(\vec{r}_0)  dV_0. \label{eq:lorentztheorema2c}
\end{equation}
It appears that the electric field $\vec{E}^a(\vec{r})$ due to source distribution A can be found by determining the electromagnetic fields of an electric dipole with $p=1$ located at $\vec{r}$ with direction $\vec{u}_b$ and substituting it in the volume integral.
By choosing the direction of the dipole $\vec{u}_b$ to be
$\vec{u}_x$, $\vec{u}_y$ and $\vec{u}_z$, respectively, we can determine the Cartesian components of  $\vec{E}^a(\vec{r})$. In section \ref{sec:antennatheory} we have determined the field generated by an electric dipole.  Using relation (\ref{eq:vectorpot}) and (\ref{eq:HenEinA}), we can rewrite (\ref{eq:lorentztheorema2c}) in the following form
\begin{equation}
\displaystyle  \vec{u}_b \cdot \vec{E}^a (\vec{r}) =
\frac{1}{\jmath \omega \epsilon_0 \mu_0} \int\limits_{V^a}
\vec{J}_e^a (\vec{r}_0) \cdot \nabla_{r_0} \times \nabla_{r_0}
\times \vec{A}_e(\vec{r},\vec{r}_0) dV_0,
\label{eq:lorentztheorema2d}
\end{equation}
with
\begin{equation}
\displaystyle \vec{A}_e (\vec{r}) = \frac{\mu_0}{4\pi} \vec{u}_b
\frac{e^{-\jmath k_0 |\vec{r}-\vec{r}_0|}}{|\vec{r}-\vec{r}_0|}.
\label{eq:vectorpotdipool}
\end{equation}
Now introduce the variable $\varphi$ according to
\begin{equation}
\displaystyle \varphi(\vec{r},\vec{r}_0) = \frac{1}{4\pi}
\frac{e^{-\jmath k_0 |\vec{r}-\vec{r}_0|}}{|\vec{r}-\vec{r}_0|}.
\label{eq:defvarphi}
\end{equation}
We can rewrite (\ref{eq:lorentztheorema2d}) into:
\begin{equation}
\begin{array}{lcl}
\displaystyle  \vec{u}_b \cdot \vec{E}^a (\vec{r}) & = &
\displaystyle \frac{1}{\jmath \omega \epsilon_0} \int\limits_{V^a}
\vec{J}_e^a (\vec{r}_0) \cdot \nabla_{r_0} \times \nabla_{r_0}
\times \vec{u}_b \varphi dV_0 \\
 & = & \displaystyle  \frac{1}{\jmath \omega \epsilon_0} \int\limits_{V^a}
\vec{J}_e^a (\vec{r}_0) \cdot \nabla_{r} \times \nabla_{r} \times
\vec{u}_b \varphi dV_0 \\ & = & \displaystyle  \frac{1}{\jmath
\omega \epsilon_0} \int\limits_{V^a} \vec{u}_b \cdot \nabla_{r}
\times \nabla_{r} \times \vec{J}_e^a (\vec{r}_0)  \varphi dV_0 \\
 & = & \displaystyle \vec{u}_b \cdot \frac{1}{\jmath
\omega \epsilon_0}  \nabla_{r} \times \nabla_{r} \times
\int\limits_{V^a} \vec{J}_e^a (\vec{r}_0)  \varphi dV_0,
\end{array}
\label{eq:lorentztheorema2e}
\end{equation}
where we have used the relation
\begin{equation}
\begin{array}{l}
\displaystyle  \nabla_{r_0} \times \nabla_{r_0} \times \vec{u}_b
\varphi = \nabla_{r} \times \nabla_{r} \times \vec{u}_b \varphi,
\\
\displaystyle \vec{c} \cdot \nabla \times \nabla \times \vec{u}_b
\varphi = \vec{u}_b \cdot \nabla \times \nabla \times \vec{c}
\varphi.
\end{array}
\label{eq:hulpvectorrelaties}
\end{equation}
Note that expression (\ref{eq:lorentztheorema2e}) is identical to
(\ref{eq:vectorpot}) in combination with (\ref{eq:HenEinA}). We
can determine $\vec{H}^a$ in a similar way by using an unit magnetic dipole directed along $\vec{u}_b$ and located at $\vec{r}$, while $\vec{J}_e^a=\vec{0}$.

\item {\bf All sources in volume $V$.} \\
Consider a volume in which the sources $\vec{J}_e^a$,
$\vec{J}_m^a$,  $\vec{J}_e^b$ and  $\vec{J}_m^b$ and the corresponding charge distributions exist, see Fig.
\ref{fig:lorentzgevaliii}. Outside $V$, which is the volume
$V_2=V_{\infty}-V$, no sources exist.
\begin{figure}[hbt]
  \centerline{\psfig{figure=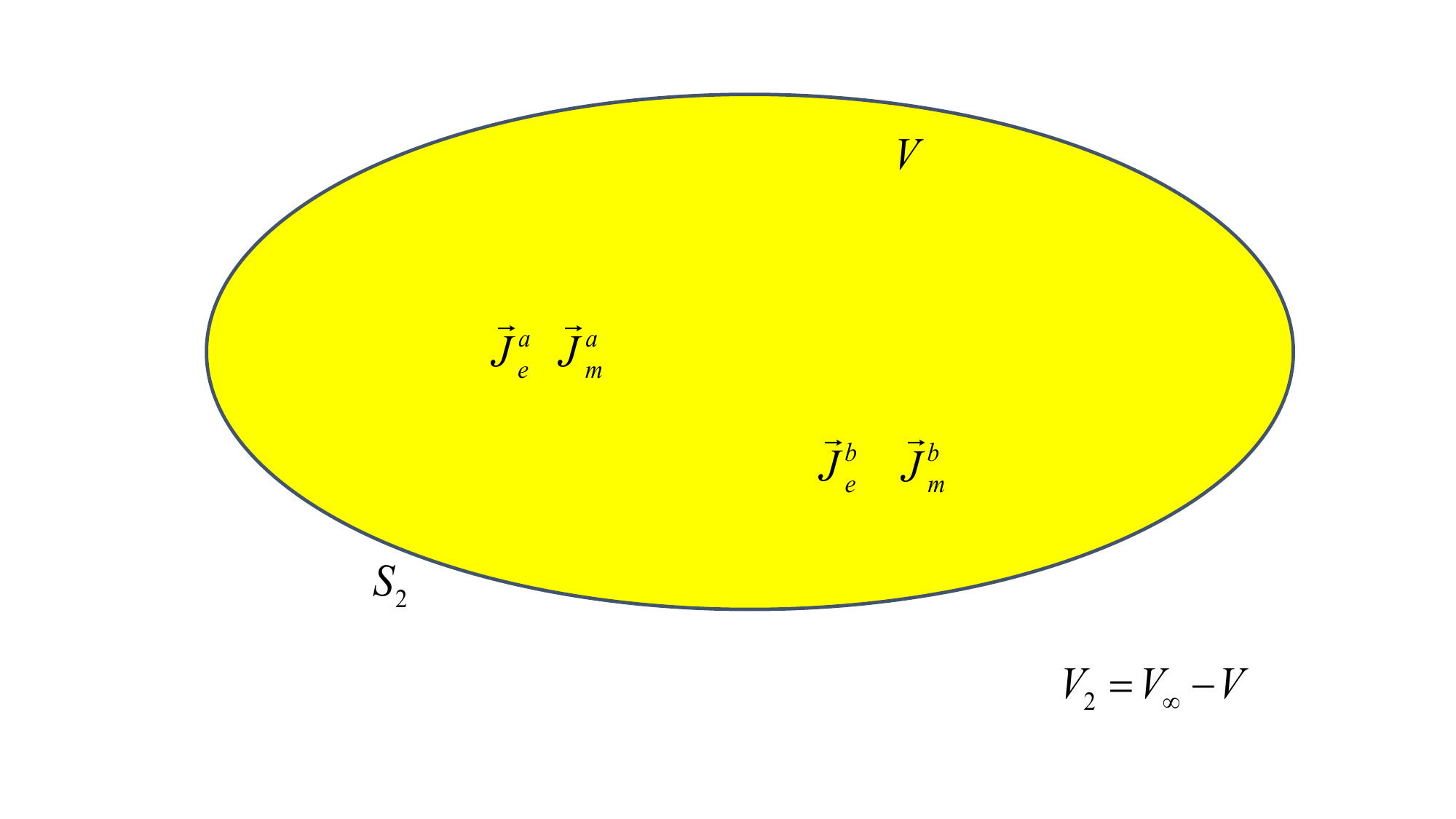,width=140mm}}
  \caption{\it All sources are located within volume $V$.}
  \label{fig:lorentzgevaliii}
\end{figure}
We will again apply Lorentz' theorem (\ref{eq:lorentztheorema}) on volume $V_2=V_{\infty}-V$.
This volume is bounded by the surface $S_2$ and by $S_{\infty}$. Since there are no sources within $V_2$, the volume integral in (\ref{eq:lorentztheorema}) is equal zero. Therefore, we find that
\begin{equation}
\displaystyle \int\limits_{S_2+S_{\infty}}  \left[ \vec{E}^a
\times \vec{H}^b - \vec{E}^b \times \vec{H}^a \right] \cdot
\vec{u}_n dS = 0,
\label{eq:Lorentzgevaliii1}
\end{equation}
in which $\vec{u}_n$ is the unit vector which is directed outwards from $V_2$.
Based on the our findings in case i), the contribution over $S_{\infty}$ will be zero. As a result
(\ref{eq:Lorentzgevaliii1}) will take the form:
\begin{equation}
\displaystyle \int\limits_{S_2}  \left[ \vec{E}^a \times \vec{H}^b
- \vec{E}^b \times \vec{H}^a \right] \cdot \vec{u}_n dS = 0.
\label{eq:Lorentzgevaliii2}
\end{equation}

\item {\bf Source distribution A in volume $V$ and source distribution B in volume $V_1$.} \\
Consider the volume $V$ in which the sources $\vec{J}_e^a$,
$\vec{J}_m^a$ and corresponding charge distributions exist.
Outside $V$, in $V_1$, we find the sources
$\vec{J}_e^b$ and $\vec{J}_m^b$ with corresponding charge distributions
(see Fig. \ref{fig:lorentzgevaliv}). Outside $V+V_1$ no sources exist.
\begin{figure}[hbt]
  \centerline{\psfig{figure=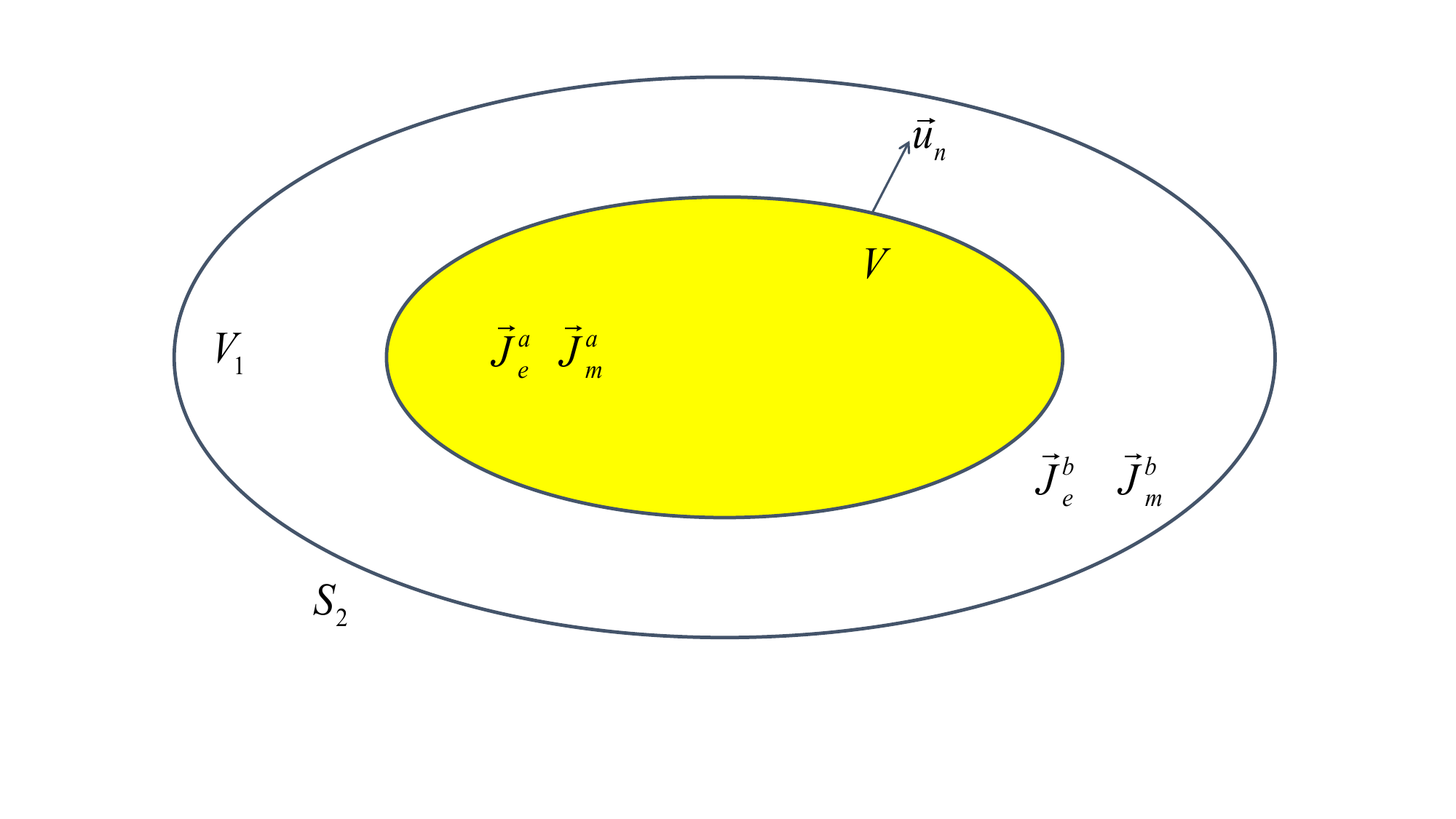,width=140mm}}
  \vspace{-1.5cm}
  \caption{\it Source distribution A is located in volume $V$
  and source distribution B is located in volume $V_1$.}
  \label{fig:lorentzgevaliv}
\end{figure}
By again applying Lorentz' theorem (\ref{eq:lorentztheorema}) on volume $V_1$ provides:
\begin{equation}
\displaystyle - \int\limits_S \left[ \vec{E}^a \times \vec{H}^b -
\vec{E}^b \times \vec{H}^a \right] \cdot \vec{u}_n dS = -
\int\limits_{V_1}  \left[\vec{E}^a \cdot \vec{J}_e^b - \vec{H}^a
\cdot \vec{J}_m^b \right]  dV ,
\label{eq:lorentzgevaliv1}
\end{equation}
where we have used relation (\ref{eq:Lorentzgevaliii2}), while the minus-sign in front of the surface integral accounts for the fact that $\vec{u}_n$ is directed inwards $V_1$ instead of outwards
$V_1$. Assume that source distribution B consists of an electric dipole with direction $\vec{u}_b$ and dipole moment $p=1$. Furthermore assume that $\vec{J}_m^b=0$. This is the same situation as used in case ii).
From (\ref{eq:lorentzgevaliv1}) we arrive at
\begin{equation}
\displaystyle \vec{u}_b \cdot \vec{E}^a (\vec{r}) = \int\limits_S
\left[ \vec{E}^a \times \vec{H}^b \right] \cdot \vec{u}_n - \left[
\vec{E}^b \times \vec{H}^a \right] \cdot \vec{u}_n dS.
\label{eq:lorentzgevaliv2}
\end{equation}
Using vector relation $(\vec{A} \times \vec{B}) \cdot
\vec{C} = - (\vec{A} \times \vec{C}) \cdot \vec{B}$ finally provides
\begin{equation}
\displaystyle \vec{u}_b \cdot \vec{E}^a (\vec{r}) = \int\limits_S
\left( \left[\vec{u}_n \times \vec{H}^a \right] \cdot \vec{E}^b -
\left[ \vec{E}^a \times \vec{u}_n \right] \cdot \vec{H}^b \right)
dS. \label{eq:lorentzgevaliv3}
\end{equation}

\end{enumerate}
\vspace*{0.5cm} With these four special cases, we can make our final last step in the derivation of the Lorentz-Larmor theorem.  For that purpose, we should compare expression (\ref{eq:lorentzgevaliv3}) and
(\ref{eq:lorentztheorema2b}). Both expressions are derived for the same situation as shown in Fig.
\ref{fig:Bisdipool}. Therefore, we can formulate the following important conclusion:

\vspace*{1cm} {\bf \it The sources $\vec{J}_e^a$ and $\vec{J}_m^a$
and the corresponding charge distributions within $V$ generate outside $V$ an electromagnetic field $\vec{E}^a$, $\vec{H}^a$.
The volume $V$ is enclosed by surface $S$. The fields $\vec{E}^a$, $\vec{H}^a$ outside $V$ can also be assumed to be generated by an electric {\underline surface} current distribution
$\vec{J}_{es}^a$ on $S$ and an magnetic {\underline surface} current density $\vec{J}_{ms}^a$ on $S$, where
\begin{equation}
\begin{array}{lcl}
\displaystyle \vec{J}_{es}^a & = & \displaystyle \vec{u}_n \times
\vec{H}^a, \\ \displaystyle \vec{J}_{ms}^a & = & \displaystyle
\vec{E}^a \times \vec{u}_n.
\end{array}
\label{eq:equivalentebronnen2}
\end{equation}
These equivalent sources generate inside volume $V$ an electromagnetic field equal to zero.}
\\

  \hrulefill
 {\bf End intermezzo: derivation of Lorentz-Larmor theorem} \hrulefill \\
\vspace*{1cm}

We can further derive the expression for the electric field at location
$\vec{r}$ outside the source region $V$ (\ref{eq:lorentzgevaliv3})
by applying the derived expressions for the dipole fields $\vec{E}^b$ and $\vec{H}^b$.
From (\ref{eq:HenEinA}) we find that
\begin{equation}
\begin{array}{lcl}
\displaystyle \vec{E}^b (\vec{r}_0) & = & \displaystyle
\frac{1}{\jmath \omega \epsilon_0} \nabla_{r_0} \times
\nabla_{r_0} \times \vec{u}_b \varphi, \\ \displaystyle \vec{H}^b
(\vec{r}_0) & = & \displaystyle \nabla_{r_0} \times \vec{u}_b
\varphi,
\end{array}
\label{eq:HenEinVarphi}
\end{equation}
where $\varphi$ is defined in (\ref{eq:defvarphi}).
Substitution of(\ref{eq:HenEinVarphi}) in
(\ref{eq:lorentzgevaliv3}) provides
\begin{equation}
\displaystyle \vec{u}_b \cdot \vec{E}^a (\vec{r}) =
\frac{1}{\jmath \omega \epsilon_0} \int\limits_S \left[\vec{u}_n
\times \vec{H}^a (\vec{r_0}) \right] \cdot  \nabla_{r_0} \times
\nabla_{r_0} \times \vec{u}_b \varphi dS - \int\limits_S \left[
\vec{E}^a (\vec{r_0}) \times \vec{u}_n \right] \cdot \nabla_{r_0}
\times \vec{u}_b \varphi dS. \label{eq:lorentzlamorE1}
\end{equation}
By using the vector relations provided in (\ref{eq:hulpvectorrelaties}) and relation  $\nabla_{r_0}
\times \vec{u}_b \varphi = -\nabla_r \varphi \times \vec{u}_b$
we obtain
\begin{equation}
\displaystyle \vec{E}^a (\vec{r}) =  \nabla_{r} \times
\int\limits_S \left[ \vec{u}_n \times \vec{E}^a (\vec{r_0})
\right] \varphi dS + \frac{1}{\jmath \omega \epsilon_0} \nabla_{r}
\times \nabla_{r} \times \int\limits_S \left[\vec{u}_n \times
\vec{H}^a (\vec{r_0}) \right] \varphi dS.
\label{eq:lorentzlamorE2}
\end{equation}
with
\begin{equation}
\displaystyle \varphi(\vec{r},\vec{r}_0) = \frac{1}{4\pi}
\frac{e^{-\jmath k_0 |\vec{r}-\vec{r}_0|}}{|\vec{r}-\vec{r}_0|}.
\label{eq:defvarphi2}
\end{equation}
The corresponding magnetic field $\vec{H}^a
(\vec{r})$ outside source region $V$ can be found by placing a magnetic dipole of unit strength and direction $\vec{u}_b$ at the location indicated by the position vector $\vec{r}$. From (\ref{eq:lorentzgevaliv1})
we than obtain
\begin{equation}
\begin{array}{lcl}
\displaystyle \vec{u}_b \cdot \vec{H}^a (\vec{r}) & = &
\displaystyle - \int\limits_S \left( \left[ \vec{E}^a \times
\vec{H}^b \right] \cdot \vec{u}_n - \left[ \vec{E}^b \times
\vec{H}^a \right] \cdot \vec{u}_n \right) dS \\ & = &
\displaystyle - \int\limits_S \left( \left[ \vec{u}_n \times
\vec{E}^a \right] \cdot \vec{H}^b + \left[ \vec{u}_n \times
\vec{H}^a \right] \cdot \vec{E}^b \right) dS .
\label{eq:lorentzHveld1}
\end{array}
\end{equation}
The (magnetic) dipole fields $\vec{E}^b$ and $\vec{H}^b$ have already been derived in section (\ref{sec:magbron}) and are given by (\ref{eq:HenEinAJm2}). This results in
\begin{equation}
\begin{array}{lcl}
\displaystyle \vec{E}^b (\vec{r}_0) & = & \displaystyle  -
\nabla_{r_0} \times \vec{u}_b \varphi, \\

\displaystyle \vec{H}^b (\vec{r}_0) & = & \displaystyle
\frac{1}{\jmath \omega \mu_0} \nabla_{r_0} \times \nabla_{r_0}
\times \vec{u}_b \varphi.
\end{array}
\label{eq:HenEinAJm2}
\end{equation}
Substitution of (\ref{eq:HenEinAJm2}) in (\ref{eq:lorentzHveld1})
results in
\begin{equation}
\displaystyle \vec{u}_b \cdot \vec{H}^a (\vec{r}) = \int\limits_S
\left[\vec{u}_n \times \vec{H}^a (\vec{r_0}) \right] \cdot
\nabla_{r_0} \times \vec{u}_b \varphi dS - \frac{1}{\jmath \omega
\mu_0} \int\limits_S \left[\vec{u}_n \times \vec{E}^a (\vec{r_0})
\right] \cdot  \nabla_{r_0} \times \nabla_{r_0} \times \vec{u}_b
\varphi dS . \label{eq:lorentzHveld2}
\end{equation}
By again applying vector relations we can rewrite
(\ref{eq:lorentzHveld2}) in the following form
\begin{equation}
\displaystyle \vec{H}^a (\vec{r}) =  \nabla_{r} \times
\int\limits_S \left[\vec{u}_n \times \vec{H}^a (\vec{r_0}) \right]
\varphi dS - \frac{1}{\jmath \omega \mu_0} \nabla_{r} \times
\nabla_{r} \times \int\limits_S \left[\vec{u}_n \times \vec{E}^a
(\vec{r_0}) \right] \varphi dS . \label{eq:lorentzHveld3}
\end{equation}

\vspace*{1cm}  \hrulefill
 {\bf Summary} \hrulefill \\ {\it Assume that the sources $\vec{J}_e$, $\vec{J}_m$ and corresponding charge distributions exist within volume $V$, enclosed by the surface $S$,
The generated electromagnetic field outside $V$ is then found by:
\begin{equation}
\begin{array}{lcl}
\displaystyle \vec{E} (\vec{r})&  = & \displaystyle \nabla_{r}
\times \int\limits_S \left[ \vec{u}_n \times \vec{E} (\vec{r_0})
\right] \varphi dS + \frac{1}{\jmath \omega \epsilon_0} \nabla_{r}
\times \nabla_{r} \times \int\limits_S \left[\vec{u}_n \times
\vec{H} (\vec{r_0}) \right] \varphi dS, \\ \displaystyle \vec{H}
(\vec{r}) & = & \displaystyle  \nabla_{r} \times \int\limits_S
\left[\vec{u}_n \times \vec{H} (\vec{r_0}) \right] \varphi dS -
\frac{1}{\jmath \omega \mu_0} \nabla_{r} \times \nabla_{r} \times
\int\limits_S \left[\vec{u}_n \times \vec{E} (\vec{r_0}) \right]
\varphi dS ,
\end{array}
 \label{eq:lorentzLamortheorema}
\end{equation}
where $\vec{J}_{es} = \vec{u}_n \times \vec{H}$ and $\vec{J}_{ms}
= \vec{E} \times \vec{u}_n$ are the equivalent electric and magnetic surface current distribution on the surface $S$, respectively.
When $\vec{r}$ is within volume $V$ we find that
\begin{equation}
\begin{array}{lcl}
\displaystyle \vec{E}(\vec{r}) & = & \vec{0}, \\ \displaystyle
\vec{H}(\vec{r}) & = & \vec{0}.
\end{array}
\end{equation}
This is the {\underline{Lorentz-Larmor theorem}} also referred to as {\it equivalence principle} }  \vspace*{1cm}

It is now fairly straightforward to derive expressions for the electromagnetic fields far away from the sources, in the so-called far-field region.
The integrand in (\ref{eq:lorentzLamortheorema}) consists of products of vector functions and scalar functions, where the vector function does not depend on $\vec{r}$, while the operator $\nabla_r \times$ operates only on $\vec{r}$. Formulas with such a structure have been discussed before in section \ref{sec:radiatedfields}. We concluded that for far-field points $\vec{r}$ the following relations can be used:
\begin{equation}
\begin{array}{lcl}
\displaystyle \nabla_r \times & \rightarrow & \displaystyle
-\jmath k_0 \vec{u}_r \times, \\ \displaystyle \nabla_r \times
\nabla_r \times & \rightarrow & \displaystyle (-\jmath k_0)^2
\vec{u}_r \times \vec{u}_r \times.
\end{array}
\label{eq:verreveldtransformatie}
\end{equation}
Furthermore, in the far-field region the following approximation can be used:
\begin{equation}
\displaystyle \frac{e^{-\jmath k_0
|\vec{r}-\vec{r}_0|}}{|\vec{r}-\vec{r}_0|} \approx
\frac{e^{-\jmath k_0 r}}{r} e^{\jmath k_0 \vec{u}_r \cdot
\vec{r}_0}. \label{eq:verreveldrelatie}
\end{equation}
Substitution of (\ref{eq:verreveldtransformatie}) and
(\ref{eq:verreveldrelatie}) in (\ref{eq:lorentzLamortheorema})
provides the following result
\begin{equation}
\begin{array}{lcl}
\displaystyle \vec{E} (\vec{r})&  = & \displaystyle \frac{-\jmath
k_0 e^{-\jmath k_0 r}}{4 \pi r} \vec{u}_r \times \int\limits_S
\left( \left[ \vec{u}_n \times \vec{E} (\vec{r_0}) \right] -
\vec{u}_r \times \left[\vec{u}_n \times Z_0 \vec{H} (\vec{r_0})
\right] \right) e^{\jmath k_0 \vec{u}_r \cdot \vec{r}_0} dS,
\\ \displaystyle Z_0 \vec{H} (\vec{r}) & = & \displaystyle
\frac{-\jmath k_0 e^{-\jmath k_0 r}}{4 \pi r} \vec{u}_r \times
\int\limits_S \left( \left[\vec{u}_n \times Z_0 \vec{H}
(\vec{r_0}) \right]+  \vec{u}_r \times \left[\vec{u}_n \times
\vec{E} (\vec{r_0}) \right] \right)  e^{\jmath k_0 \vec{u}_r \cdot
\vec{r}_0}  dS .
\end{array}
 \label{eq:lorentzLamortheoremaverreveld}
\end{equation}
This can also be written in a more compact form:
\begin{equation}
\begin{array}{lcl}
\displaystyle \vec{E} (\vec{r})&  = & \displaystyle \frac{-\jmath
k_0 e^{-\jmath k_0 r}}{4 \pi r} \vec{F}_a(\vec{u}_r),\\
\displaystyle Z_0 \vec{H} (\vec{r}) & = & \displaystyle
\frac{-\jmath k_0 e^{-\jmath k_0 r}}{4 \pi r} \vec{u}_r \times
\vec{F}_a(\vec{u}_r) .
\end{array}
 \label{eq:lorentzLamortheoremaverreveld2}
\end{equation}
with
\begin{equation}
\begin{array}{lcl}
\displaystyle \vec{F}_a (\vec{u}_r)&  = & \displaystyle \vec{u}_r
\times \int\limits_S \left( \left[ \vec{u}_n \times \vec{E}
(\vec{r_0}) \right] - \vec{u}_r \times \left[\vec{u}_n \times Z_0
\vec{H} (\vec{r_0}) \right] \right) e^{\jmath k_0 \vec{u}_r \cdot
\vec{r}_0} dS.
\end{array}
\end{equation}
The subscript $a$ is used to indicate that the function $F_a$ is found by integrating the fields on the antenna aperture (= surface $S$). Note that the following relation also holds in the far-field:
\begin{equation}
\displaystyle \vec{E} = Z_0 \vec{H} \times \vec{u}_r.
\end{equation}
The formulas (\ref{eq:lorentzLamortheoremaverreveld2}) can now be used to determine the radiation properties of arbitrarily-shaped aperture antennas.
A proper choice of the surface $S$ is very important. In the next sections we will show how $S$ can be chosen in a convenient way for some well-known aperture antenna concept. In this way, we are able to find an analytic expression of the radiated fields.

\section{Horn antennas}
\begin{figure}[hbt]
 \vspace{-0.5cm}
  \centerline{\psfig{figure=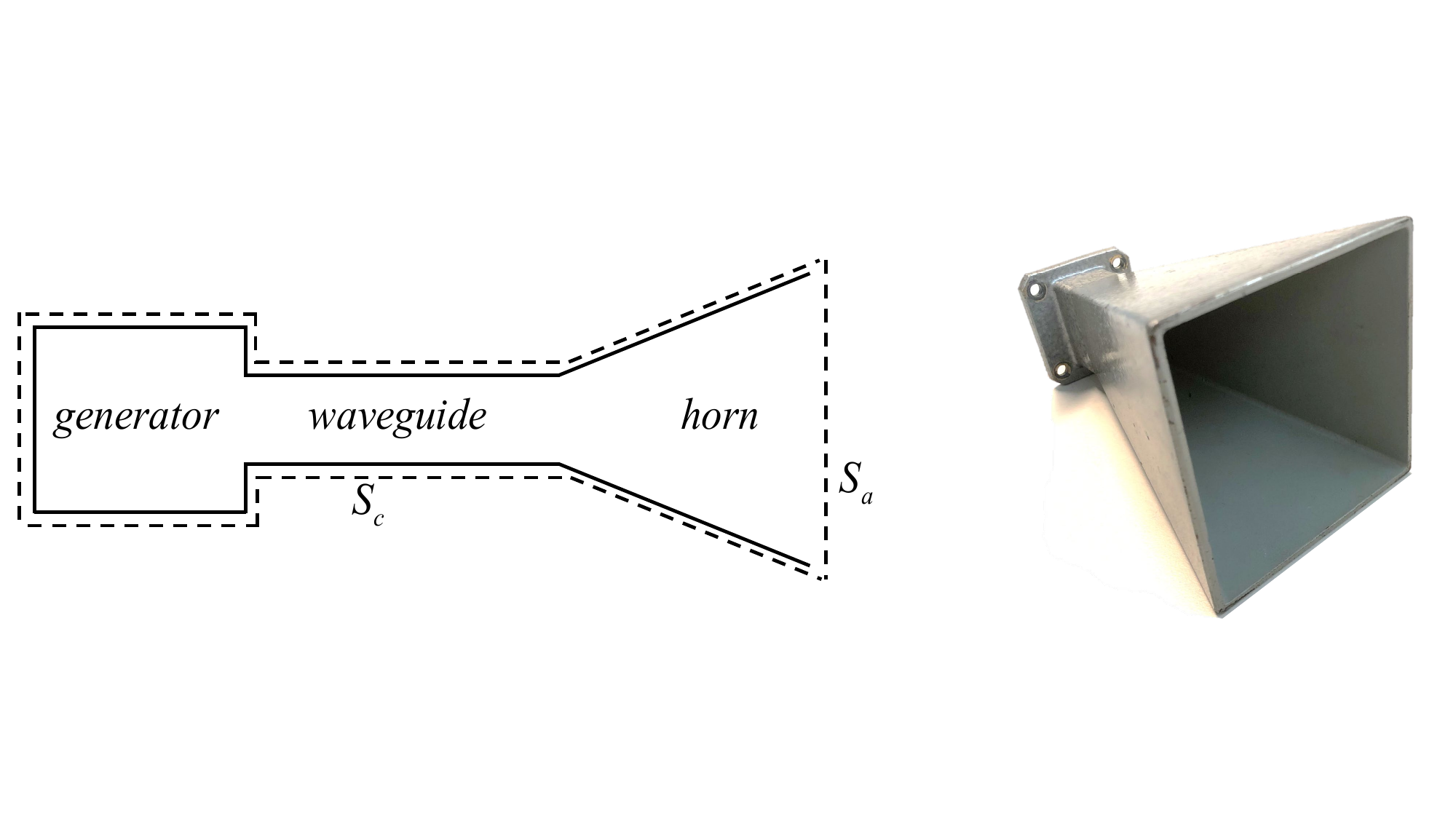,width=140mm}}
  \vspace{-2cm}
  \caption{\it Example of horn antennas and choice of the enclosed surface $S$ as $S_a+S_c$.}
  \label{fig:hoornantenne2}
\end{figure}

\begin{figure}[hbt]
\centerline{\psfig{figure=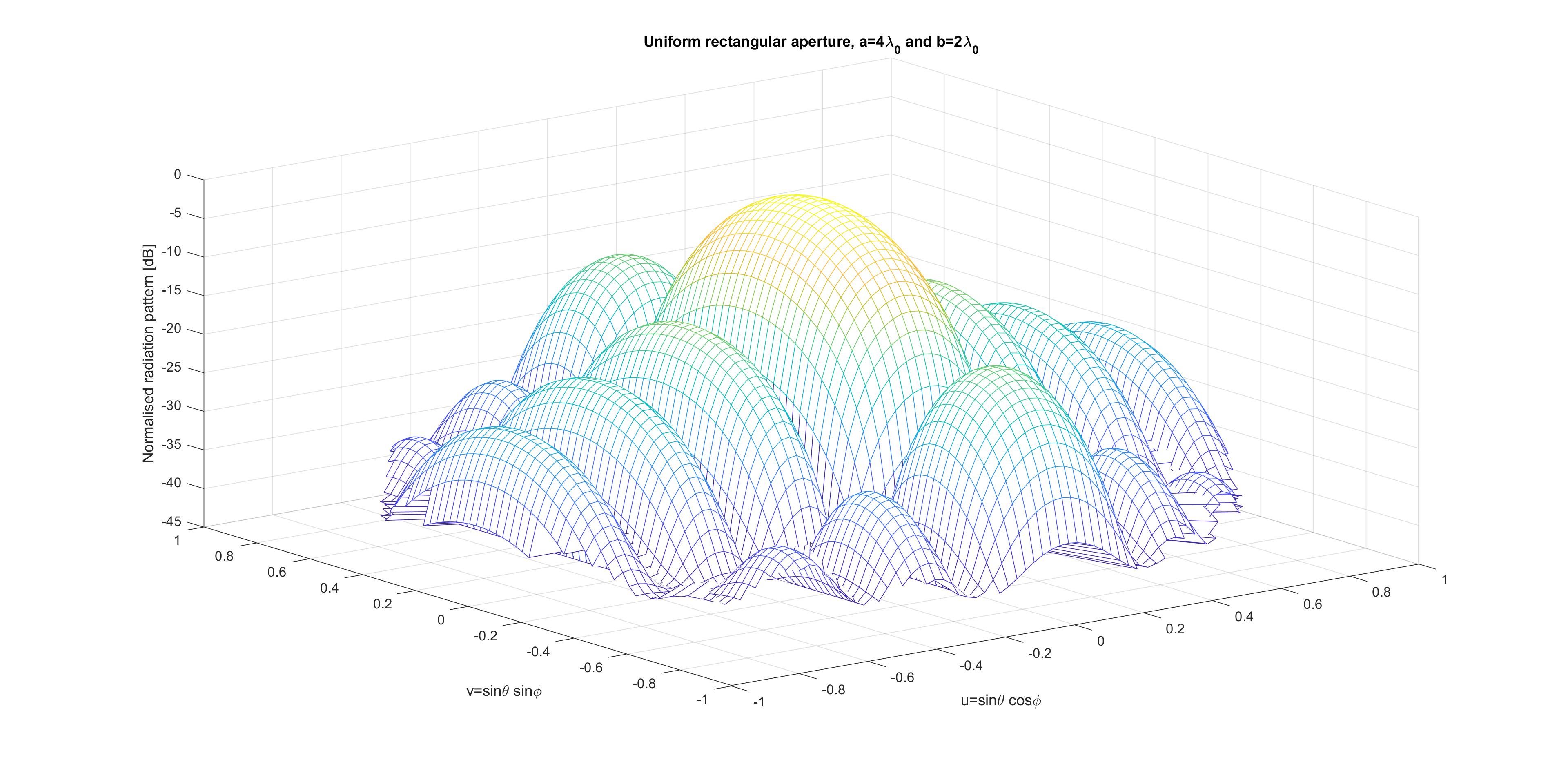,width=140mm}}

\caption{\it Three-dimensional normalized radiation pattern versus $(\theta,\phi)$ of a uniform
aperture with dimensions $a=4\lambda_0$ and $b=2\lambda_0$. All
values are in dB.} \label{fig:uniapertuurradpat3d}
\end{figure}

\begin{figure}[hbt]
\centerline{\psfig{figure=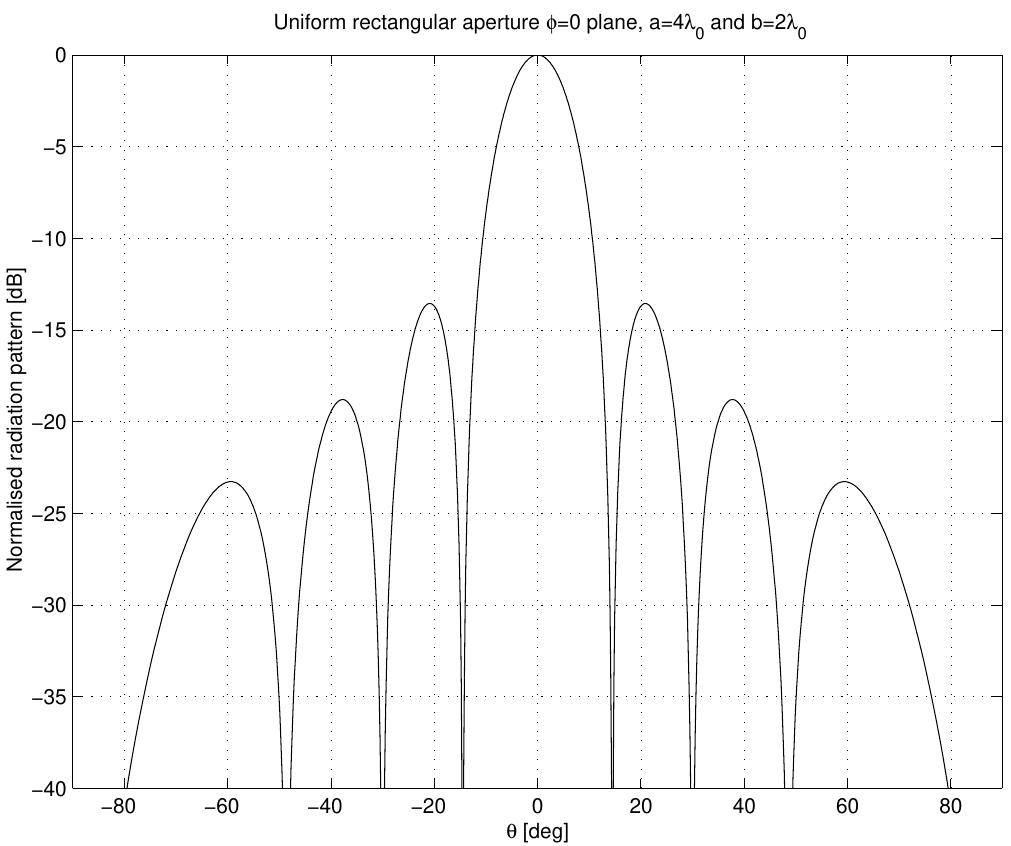,width=90mm}}

\caption{\it Cross section in the $\phi=0$ plane (H-plane) of the normalized radiation pattern versus $\theta$ of a uniform
aperture with dimensions $a=4\lambda_0$ and $b=2\lambda_0$. All
values are in dB. The 3 dB beam width in this plane is
$\theta_{HP}=2 \arcsin(0.433\lambda_0/a)=12.4^0$ and directivity $D=20$ dBi.}
\label{fig:uniapertuurradpat2d}
\end{figure}

A well known example of an aperture antenna is the horn antenna as illustrated in Fig. \ref{fig:hoornantenne2}.
Horn antennas are available with rectangular, square and circular apertures and are usually fed from a waveguide.
Fig. \ref{fig:hoornantenne2} also shows how we could select the closed surface $S$ with the equivalent sources in order to apply the Lorentz-Larmor concept as discussed in the previous section.
$S$ consists of two part, namely $S_a$ and $S_c$. $S_a$ is the actual radiating aperture of the horn antenna, while $S_c$ contains the outer (metallic) surface of the horn antenna and generator. Therefore, all sources are enclosed by the surface $S$.
We can now make the following assumptions:
\begin{itemize}
\item{The outer surface of the antenna and the generator are perfectly electric conductors (PEC). This implies that the tangential electric field along this surface is zero: $\vec{u}_n \times \vec{E} = \vec{0}$
on $S_c$.}
\item{The (secondary) currents along the outer surface of the antenna and generator are very small and can be ignored in our analysis, i.e. $\vec{u}_n \times \vec{H} =\vec{0}$
on $S_c$.}
\end{itemize}
As a result, the integration in (\ref{eq:lorentzLamortheoremaverreveld}) is limited to $S_a$.
\begin{figure}[hbt]
    \centerline{\psfig{figure=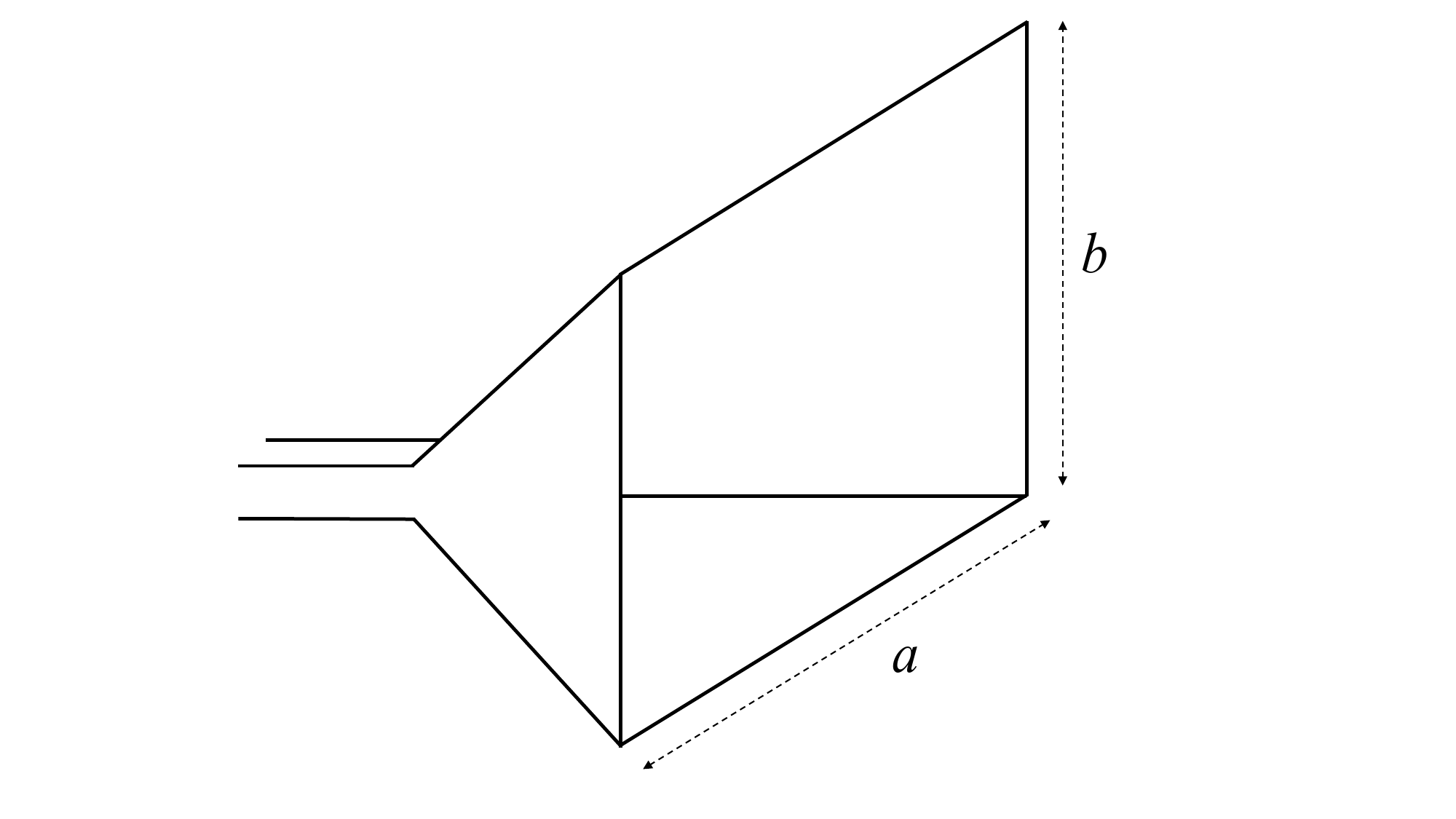,width=140mm}}
  \caption{\it Rectangular uniform aperture with $\vec{E}_a = E_0 \vec{u}_y$ and
  $\displaystyle \vec{H}_a = -\frac{E_0}{Z_0} \vec{u}_x $ for $-\frac{a}{2} < x < \frac{a}{2}$ and
  $-\frac{b}{2} < y < \frac{b}{2}$.}
  \label{fig:uniapertuur}
\end{figure}
As a first-order approximation we will assume that the aperture fields are uniformly distributed in the $x$-$y$
plane, as indicated in Fig. \ref{fig:uniapertuur}.
Note that the derivation for the case of a more realistic aperture distribution is quite similar, although the mathematics are more complex or even require a numerical integration.  The fields of the uniformly illuminated aperture can be described as:
\begin{equation}
\begin{array}{lcl}
\displaystyle \vec{E}_a & = & \displaystyle E_0 \vec{u}_y, \\
\displaystyle \vec{H}_a & = & \displaystyle - \frac{E_0}{Z_0}
\vec{u}_x,
\end{array}
\end{equation}
valid for the rectangular region $-\frac{a}{2} < x < \frac{a}{2}$ and $-\frac{b}{2} < y
< \frac{b}{2}$. Furthermore, $E_0$ is a constant and we have assumed that the characteristic impedance is equal to
$Z_0$. The equivalent electric and magnetic surface currents on $S_a$ are now given by
\begin{equation}
\begin{array}{lcl}
\displaystyle \vec{J}_{ms} & = & \displaystyle \vec{E}_a \times
\vec{u}_n = E_0 \vec{u}_y \times \vec{u}_z = E_0 \vec{u}_x,
\\ \displaystyle \vec{J}_{es} & = & \displaystyle \vec{u}_n \times \vec{H}_a
= - \frac{E_0}{Z_0} \vec{u}_z \times \vec{u}_x =  -
\frac{E_0}{Z_0} \vec{u}_y.
\end{array}
\label{eq:JmJeuniapertuur}
\end{equation}
We can calculate the electric field in the far-field region by applying expression (\ref{eq:lorentzLamortheoremaverreveld}) on aperture $S=S_a$. We now find
\begin{equation}
\begin{array}{lcl}
\displaystyle \vec{E} (\vec{r})&  = & \displaystyle \frac{-\jmath
k_0 e^{-\jmath k_0 r}}{4 \pi r} \vec{u}_r \times
\int\limits_{-a/2}^{a/2} \int\limits_{-b/2}^{b/2} \left(
-\vec{J}_{ms} -Z_0 \vec{u}_r \times \vec{J}_{es} \right) e^{\jmath
k_0 \vec{u}_r \cdot \vec{r}_0} dx_0 dy_0 \\ 
& = & \displaystyle
\frac{\jmath k_0 e^{-\jmath k_0 r}}{4 \pi r} \left( \vec{L}_m+\vec{L}_e
\right),
\end{array}
 \label{eq:Euniapertuur1}
\end{equation}
where the integrals $\vec{L}_m$ and $\vec{L}_e$ are given by
\begin{equation}
\begin{array}{lcl}
\displaystyle \vec{L}_m &  = & \displaystyle \int\limits_{-a/2}^{a/2}
\int\limits_{-b/2}^{b/2} \vec{u}_r \times \vec{J}_{ms} (\vec{r}_0)
e^{\jmath k_0 \vec{u}_r \cdot \vec{r}_0} dx_0 dy_0, \\
\displaystyle \vec{L}_e & = & \displaystyle Z_0 \int\limits_{-a/2}^{a/2}
\int\limits_{-b/2}^{b/2} \vec{u}_r \times
 \vec{u}_r \times
\vec{J}_{es} (\vec{r}_0)  e^{\jmath k_0 \vec{u}_r \cdot
\vec{r}_0} dx_0 dy_0.
\end{array}
 \label{eq:IntegralenLmLe}
\end{equation}
We can determine the integrals easily by writing the in- and outer-products in (\ref{eq:IntegralenLmLe}) in terms of spherical coordinates  $(r,\theta,\phi)$:
\begin{equation}
\begin{array}{lcl}
\displaystyle \vec{u}_r \times \vec{u}_x & = & \displaystyle
\vec{u}_r \times \left(\vec{u}_{\theta} \cos{\theta} \cos{\phi} -
\vec{u}_{\phi} \sin{\phi} \right) = \vec{u}_{\phi}\cos{\theta}
\cos{\phi} + \vec{u}_{\theta} \sin{\phi}, \\
 \displaystyle
\vec{u}_r \times \vec{u}_r \times \vec{u}_y & = & \displaystyle
-\vec{u}_{\theta}\cos{\theta} \sin{\phi} - \vec{u}_{\phi}
\cos{\phi}, \\
 \displaystyle
\vec{u}_r \cdot \vec{r}_0 & = & (\vec{u}_x x_0 + \vec{u}_y y_0)
\cdot (\vec{u}_x
\sin{\theta}\cos{\phi}+\vec{u}_y\sin{\theta}\sin{\phi}+\vec{u}_z\cos{\theta})\\
 \displaystyle & = & x_0 \sin{\theta}\cos{\phi} +
 y_0\sin{\theta}\sin{\phi} = x_0 u + y_0 v,
\end{array}
 \label{eq:Inenuitproducten}
\end{equation}
with $u=\sin{\theta}\cos{\phi}$ and $v=\sin{\theta}\sin{\phi}$.
Substitution of (\ref{eq:Inenuitproducten}) and
(\ref{eq:JmJeuniapertuur}) in (\ref{eq:IntegralenLmLe}) provides for the first integral $\vec{L}_m$
\begin{equation}
\begin{array}{lcl}
\displaystyle \vec{L}_m &  = & \displaystyle E_0 \left(
\vec{u}_{\phi}\cos{\theta} \cos{\phi} + \vec{u}_{\theta}
\sin{\phi} \right) \int\limits_{-a/2}^{a/2}
\int\limits_{-b/2}^{b/2} e^{\jmath k_0 (x_0 u + y_0 v)} dx_0 dy_0,
\\ \displaystyle  & = & \displaystyle abE_0 \left(
\vec{u}_{\phi}\cos{\theta} \cos{\phi} + \vec{u}_{\theta}
\sin{\phi} \right) \left[\frac{\sin(k_0 au/2)}{k_0 au/2}\right]
\left[\frac{\sin(k_0 bv/2)}{k_0 bv/2}\right],
\end{array}
 \label{eq:IntegraalLm}
\end{equation}
where we have used the well-known integral
\begin{equation}
\displaystyle \int\limits_{-a/2}^{a/2} e^{\jmath \beta x} dx = a
\left[\frac{\sin(\beta a/2)}{\beta a/2} \right].
\end{equation}
Similarly, we can determine the integral $\vec{L}_e$ as
\begin{equation}
\begin{array}{lcl}
\displaystyle \vec{L}_e &  = & \displaystyle abE_0 \left(
\vec{u}_{\theta}\cos{\theta} \sin{\phi} + \vec{u}_{\phi}
\cos{\phi} \right) \left[\frac{\sin(k_0 au/2)}{k_0 au/2}\right]
\left[\frac{\sin(k_0 bv/2)}{k_0 bv/2}\right].
\end{array}
 \label{eq:IntegraalLe}
\end{equation}
Finally, by substituting (\ref{eq:IntegraalLe}) and (\ref{eq:IntegraalLm}) in
(\ref{eq:Euniapertuur1}) we obtain the following expression for the electric field in the far-field region of the horn antenna
\begin{equation}
\begin{array}{lcl}
\displaystyle \vec{E} (\vec{r})&  = &  \displaystyle \frac{\jmath
k_0 e^{-\jmath k_0 r}}{4 \pi r} \left( \vec{L}_m+\vec{L}_e \right) \\ &  = &
\displaystyle \frac{\jmath k_0 ab E_0 e^{-\jmath k_0 r}}{4 \pi r}
\left[ \vec{u}_{\phi} \cos{\phi} (\cos{\theta}+1) +
\vec{u}_{\theta} \sin{\phi} (\cos{\theta} +1) \right]
\left[\frac{\sin(k_0 au/2)}{k_0 au/2}\right] \left[\frac{\sin(k_0
bv/2)}{k_0 bv/2}\right] \\
 &  = &
\displaystyle E_{\theta} \vec{u}_{\theta} + E_{\phi}
\vec{u}_{\phi}.
\end{array}
 \label{eq:Euniapertuur2}
\end{equation}
We can observe that $\vec{E}$ has two components,
$E_{\theta} \vec{u}_{\theta}$ and $E_{\phi} \vec{u}_{\phi}$. In the principle planes of the antenna, i.e. the $\phi=0$ plane (H-plane) and $\phi=\pi/2$ plane (E-plane), one of both components is equal to zero. The magnetic field $\vec{H}$ in the far-field region is easily found by using the far-field relation $\vec{E} =
Z_0 \vec{H} \times \vec{u}_r$. The radiation pattern is determined according to the approach in chapter \ref{chap:fundpar} by using relation (\ref{eq:pointingverreveld}) up to
(\ref{eq:normradpat}). The resulting normalized radiation pattern then takes the following form:
\begin{equation}
\displaystyle F(\theta,\phi) =
\frac{|E_{\theta}(\theta,\phi)|^2+|E_{\phi}(\theta,\phi)|^2}{|E_{\theta}(0,0)|^2+|E_{\phi}(0,0)|^2}.
\end{equation}
The corresponding half-power beam width is given by $\theta_{HP}=2 \arcsin(0.433\lambda_0/a)$ in the H-plane and $\theta_{HP}=2 \arcsin(0.433\lambda_0/b)$ in the E-plane, respectively. When the dimension of the horn increases, the beam width decreases, so the antenna will become more directive. The directivity is found from $D=4\pi \frac{ab}{\lambda_0}$.
Fig. \ref{fig:uniapertuurradpat3d} shows the three-dimensional radiation pattern in [dB] for a uniform aperture with dimensions $a=4\lambda_0$ and $b=2\lambda_0$. The corresponding cross section in the $\phi=0$ plane ($v=0$) is illustrated in Fig.
\ref{fig:uniapertuurradpat2d}. The first sidelobe has a peak at  $-13.26$
dB w.r.t. the main lobe. The 3-dB beam width in the $\phi=0$ plane of this aperture antenna is $\theta_{HP}=12.4^0$ and the directivity $D=20$ dBi.


\section{Reflector antennas}
\begin{figure}[hbt]
\centerline{\psfig{figure=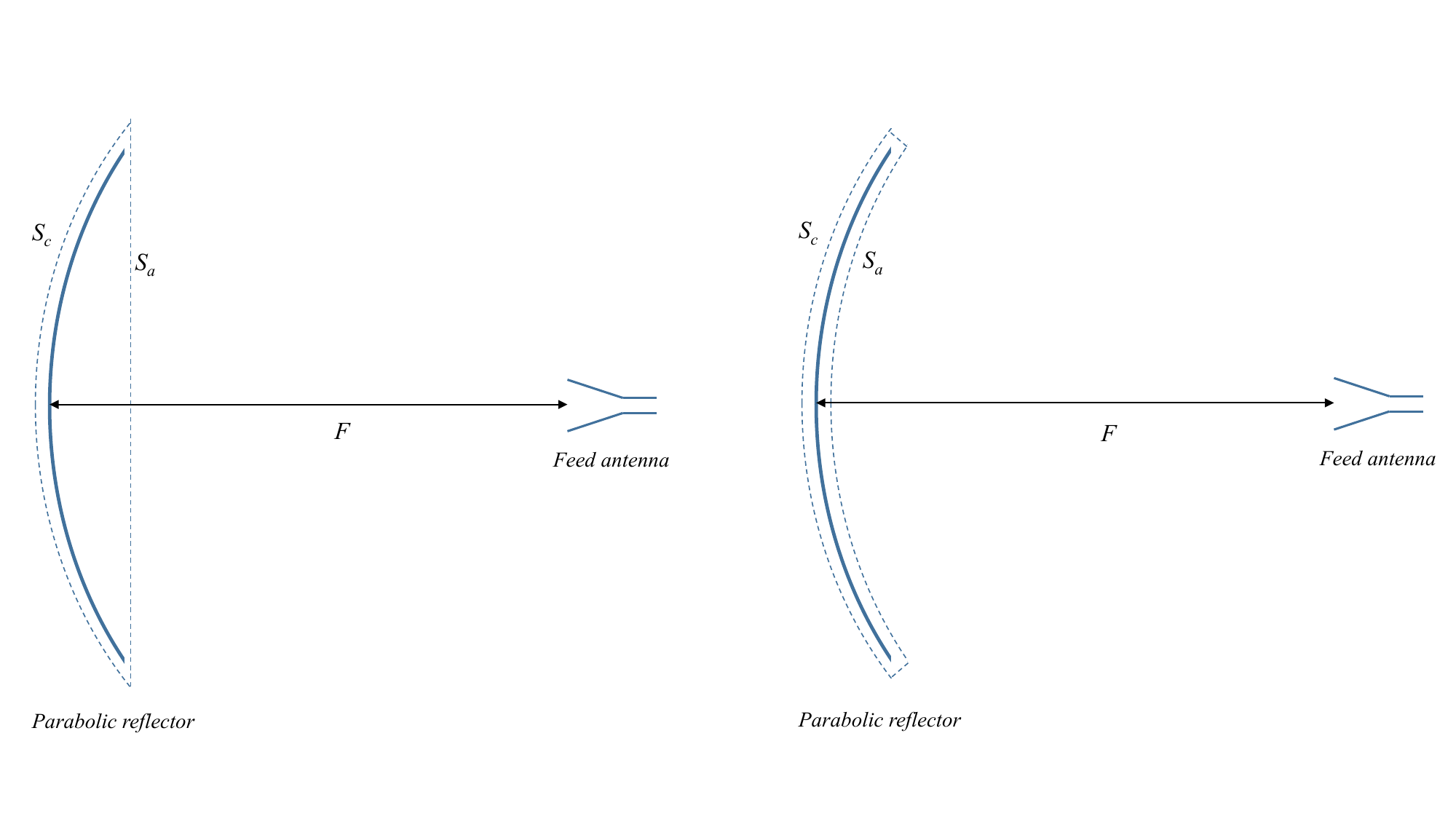,width=170mm}}
  \caption{\it Parabolic reflector (cross section) with feed placed in the focal point.
  (a) The choice of the surface $S_a$ according to the aperture method, (b) surface $S_a$ is selected according to the current-integration method..}
  \label{fig:reflector3}
\end{figure}

Reflector antennas can take many shapes, e.g. parabolic, spherical or cylindrical. The parabolic reflector antennas is the most commonly used type. Reflector antennas provide a large antenna gain and associated effective antenna aperture, which makes them very suited for long-range wireless applications, such as satellite communication, point-to-point (PtP) backhauling and long-range radar. The parabolic reflector antenna is fed by a small antenna element({\it feed}), such as a dipole antenna or horn antenna. The feed is placed in the focal point of the reflector and illuminates the reflector. As a result of this, electric currents are generated on the metal reflector surface.  These currents generate the scattered field of the reflector.  Fig. \ref{fig:reflector3} shows an illustration of the parabolic reflector and feed antenna.
The electromagnetic field in the far-field region can be determined using equation  (\ref{eq:lorentzLamortheoremaverreveld}). In this case, we will choose $S=S_a+S_c$ as indicated in Fig. \ref{fig:reflector3}a. The sources of the scattered field are now all located within the surface $S$. We will assume that the reflector is made of a perfectly-conducting metallic material. Furthermore, we will ignore the currents flowing on the back side of the reflector.  As a result
\begin{equation}
\begin{array}{lc}
\displaystyle \vec{u}_n \times \vec{E} = \vec{0} & on \ \ S_c, \\
\displaystyle \vec{u}_n \times \vec{H} = \vec{0} & on \ \ S_c.
\end{array}
\end{equation}
Therefore, we can limit the integration in (\ref{eq:lorentzLamortheoremaverreveld}) to the aperture $S_a$. The radiation pattern of the feed antenna located in the focal point $F$ determines the field distribution over the aperture plane $S_a$. This method is known as the  {\it aperture method}. However, it is also possible to select the surface $S$ as indicated in Fig. \ref{fig:reflector3}b.
In this case we find that $\vec{u}_n \times \vec{E} = \vec{0}$ on
$S_a$, and as a result (\ref{eq:lorentzLamortheoremaverreveld}) reduces to
\begin{equation}
\begin{array}{lcl}
\displaystyle \vec{E} (\vec{r})&  = & \displaystyle \frac{\jmath
k_0 e^{-\jmath k_0 r}}{4 \pi r} \vec{u}_r \times \vec{u}_r \times
\int\limits_S \left[\vec{u}_n \times Z_0 \vec{H}_a (\vec{r_0})
\right] e^{\jmath k_0 \vec{u}_r \cdot \vec{r}_0} dS,
\end{array}
 \label{eq:Estroomintmethode}
\end{equation}
The magnetic field in the aperture $\vec{H}_a$ on $S_a$ can be easily determined from the radiation characteristics of the feed antenna located in the focal point $F$. This method is known as the {\it current integration method}.

\begin{figure}[hbt]
\centerline{\psfig{figure=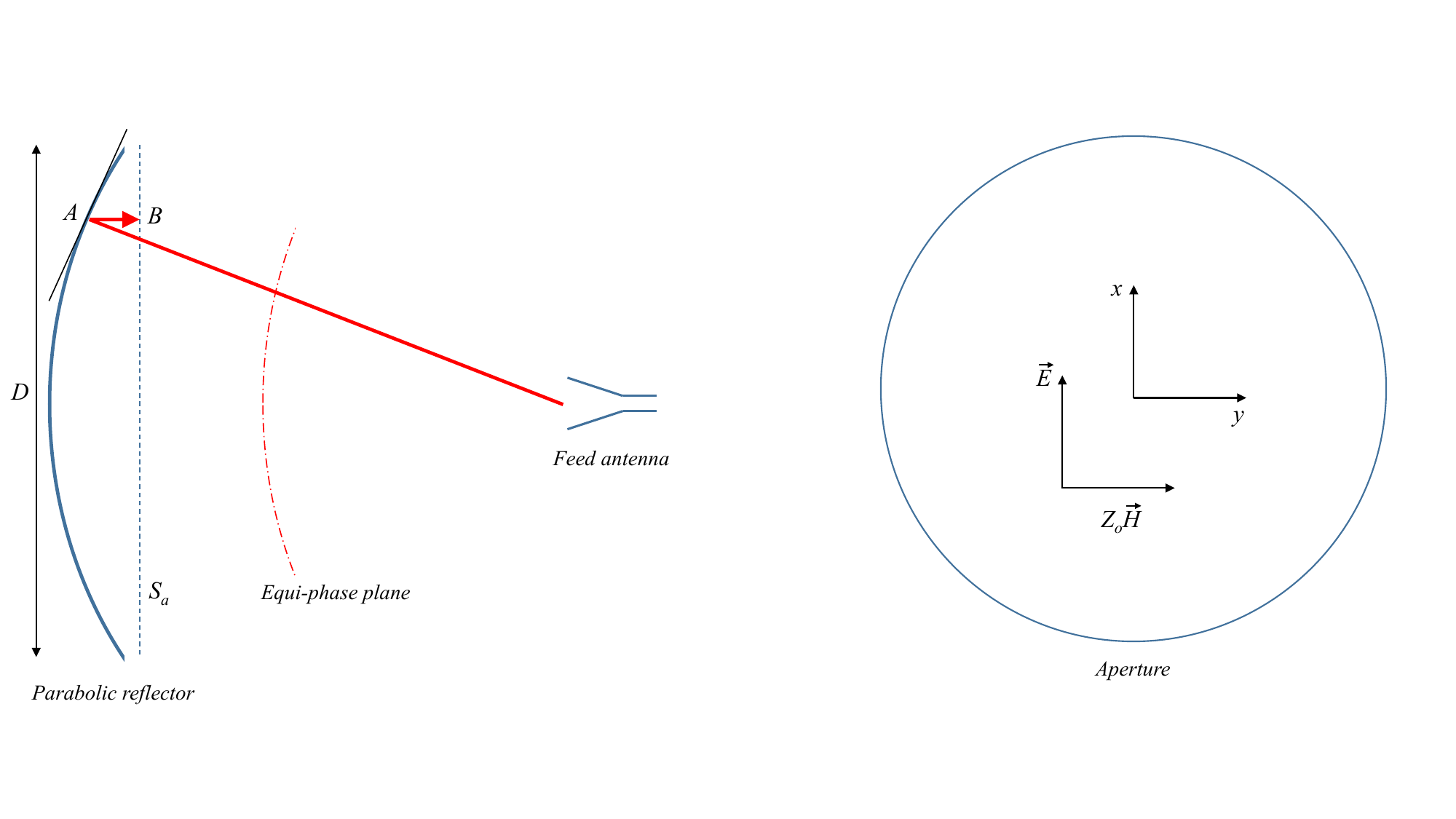,width=170mm}}
  \caption{\it Circularly-symmetric parabolic reflector with feed located in the focal point. The radius of the circularly symmetric aperture is $a=D/2$.}
  \label{fig:reflector5}
\end{figure}

In this section we will use the apeture method to determine the radiation characteristics of a parabolic reflector antenna. Now consider the configuration of Fig. \ref{fig:reflector5}. The reflector has a diameter $D=2a$.
A feed antenna is located in the focal point $F$.
Since the feed antenna is relative small (typically $0.5\lambda_0 - 4 \lambda_0$), a point $A$ on the reflector will be located in the far-field region of the feed antenna, that is the distance between the feed and the reflector is larger than  $\displaystyle
\frac{2L^2}{\lambda_0}$, with $L$ the largest dimension of the feed antenna.
In the far-field region of the feed antenna, the equi-phase phase front is spherical. Perpendicular to these planes, we can define so-called {\it rays} directed along the Pointing vector. Now assume that a ray that originates from the feed hits the reflector at point $A$ on the reflector surface, as indicated in Fig. \ref{fig:reflector5}.
We will assume that the reflector at point $A$ is locally flat.
Furthermore, the incident wave at point $A$ is locally flat as well, represented as a plane wave.
As a result, we can use the well-known reflection rules which yield for a plane wave incident on a large and perfectly-conduction flat plate. This provides us the field in point $B$, see Fig. \ref{fig:reflector5}. In this way, we can determine the fields in the aperture $S_a$.
The radiation characteristics of the feed antenna have a very large impact on the final radiation characteristics of the reflector antenna. In many applications the feed is designed in such a way that the edges of the reflector are less illuminated as compared to the center region. This is called {\it amplitude tapering}.

\begin{figure}[hbt]
\centerline{\psfig{figure=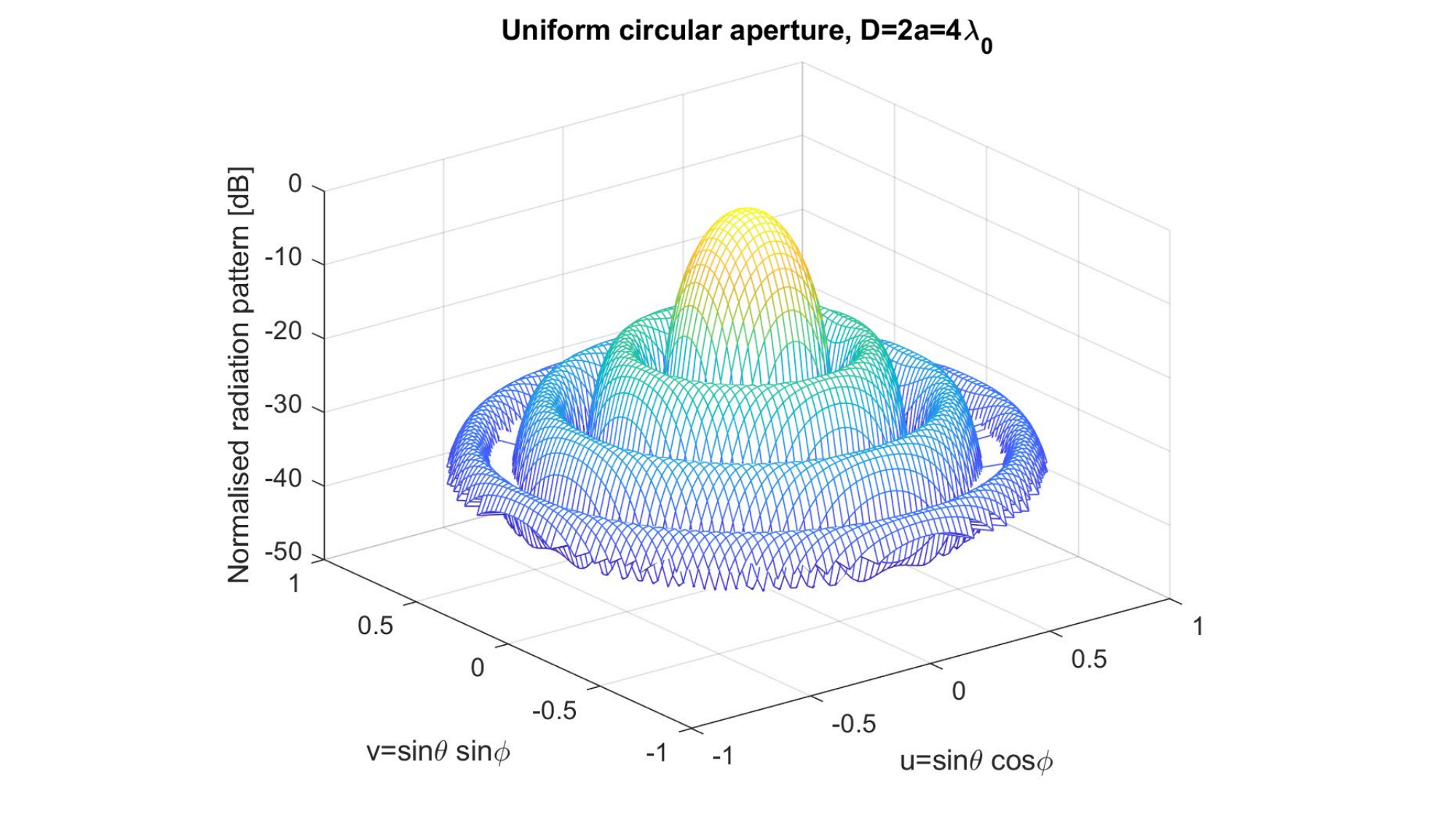,width=140mm}}

\caption{\it Three-dimensional radiation pattern versus
$(\theta,\phi)$ of a parabolic reflector with uniform aperture illumination with diameter $D=2a=4\lambda_0$. All values are in dB.}
\label{fig:reflectorradpat3d}
\end{figure}

\begin{figure}[hbt]
\centerline{\psfig{figure=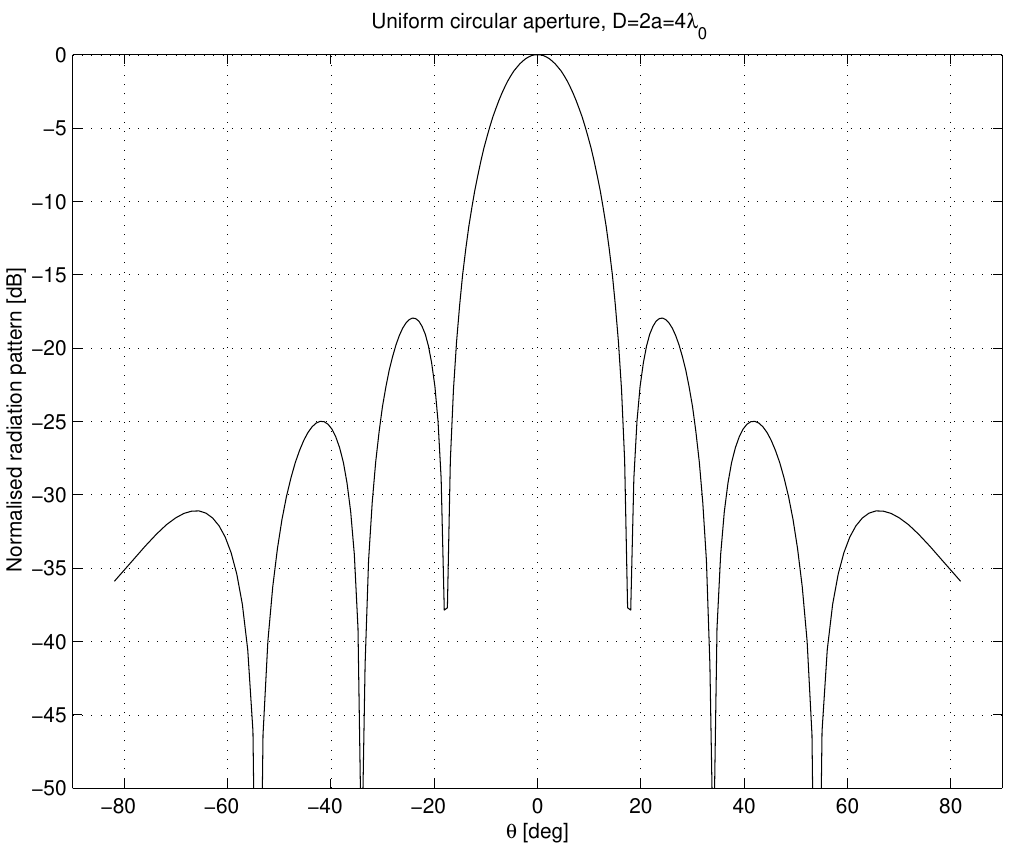,width=90mm}}

\caption{\it Radiation pattern in the $\phi=0$-plane $\theta$ of a parabolic reflector with uniform aperture illumination with diameter $D=2a=4\lambda_0$. All values are in dB.
The 3 dB beam width in this plane is $\theta_{HP} = 1.02 \lambda_0/(2a) =
14.6^0$ }
\label{fig:reflectorradpat2d}
\end{figure}

For the sake of simplicity, we will first assume that the feed antenna provides a uniform illumination of the aperture $S_a$. Now let the incident electric field $\vec{E}_i$ coincide with the $x$-direction and the magnetic field $\vec{H}_i$ with the $y$-direction, as illustrated in Fig. \ref{fig:reflector5}. In a circular-symmetric aperture, we then have the following field distribution:
\begin{equation}
\begin{array}{lcl}
\displaystyle  \vec{E}_a & = & E_0 \vec{u}_x, \\ \displaystyle Z_0
\vec{H}_a & = & E_0 \vec{u}_y,
\end{array}
\end{equation}
where $E_0$ is a constant. The electric field in a point $P$ far away from the aperture, can be calculated by using expresion (\ref{eq:lorentzLamortheoremaverreveld})
\begin{equation}
\begin{array}{lcl}
\displaystyle \vec{E} (\vec{r})&  = & \displaystyle \frac{-\jmath
k_0 e^{-\jmath k_0 r}}{4 \pi r} \vec{u}_r \times \int\limits_{S_a}
\left( \left[ \vec{u}_n \times \vec{E}_a (\vec{r_0}) \right] -
\vec{u}_r \times \left[\vec{u}_n \times Z_0 \vec{H}_a (\vec{r_0})
\right] \right) e^{\jmath k_0 \vec{u}_r \cdot \vec{r}_0} dS.
\end{array}
 \label{eq:Eparareflector1}
\end{equation}
We have already shown in section \ref{sec:radiatedfields} that the electric field $\vec{E}$ in the far-field region only has two components: $E_{\theta}$ and $E_{\phi}$.
Therefore, we can express $\vec{u}_r
\times \left[ \vec{u}_n \times \vec{E}_a (\vec{r_0})  -
\vec{u}_r \times \vec{u}_n \times Z_0 \vec{H}_a (\vec{r_0})
\right]$ in the following form:
\begin{equation}
\begin{array}{lcl}
\displaystyle \vec{M} & = & \displaystyle \vec{u}_r \times \left[
\vec{u}_n \times \vec{E}_a  - \vec{u}_r \times
\vec{u}_n \times Z_0 \vec{H}_a \right] \\
 & = & \displaystyle E_0 \vec{u}_r \times \left[ \vec{u}_z \times
\vec{u}_x  - \vec{u}_r \times  \vec{u}_z \times
\vec{u}_y \right]
\\
& = & \displaystyle E_0 (1+\cos{\theta}) \left(\vec{u}_{\phi}
\sin{\phi} - \vec{u}_{\theta} \cos{\phi} \right).
\end{array}
\end{equation}
Furthermore, the inner product $(\vec{r}_0 \cdot \vec{u}_r)$
can be written in polar coordinates according to
\begin{equation}
\vec{u}_r \cdot \vec{r}_0 = r_0 \sin{\theta}\cos(\phi-\phi_0).
\end{equation}
The electric field in the far-field region then takes the following form
\begin{equation}
\begin{array}{lcl}
\displaystyle \vec{E} (\vec{r})&  = & \displaystyle \frac{-\jmath
k_0 e^{-\jmath k_0 r}}{4 \pi r} (1+\cos{\theta})
\left(\vec{u}_{\phi} \sin{\phi} - \vec{u}_{\theta} \cos{\phi}
\right) \int\limits_0^{2\pi} \int\limits_0^a E_0 e^{\jmath k_0 r_0
\sin{\theta}\cos(\phi-\phi_0)} r_0 dr_0 d\phi_0.
\end{array}
 \label{eq:Eparareflector2}
\end{equation}
Since we have assumed a uniform aperture illumination, $E_0$ is constant and can be put outside the integral. The $\phi_0$ integral can be expressed in terms of a Bessel function according to
\begin{equation}
\displaystyle  \int\limits_0^{2\pi}  e^{\jmath k_0 r_0
\sin{\theta}\cos(\phi-\phi_0)} d\phi_0 = 2\pi J_0(k_0 r_0
\sin{\theta}),
\label{eq:phiintegratiereflector}
\end{equation}
where $J_0$ is the Bessel function of the first kind with order zero.
The remaining integration w.r.t. $r_0$ can now be expressen in closed-form as:
\begin{equation}
\begin{array}{lcl}
\displaystyle \int\limits_0^a J_0(k_0 r_0 \sin{\theta}) r_0 dr_0 &
= & \displaystyle a^2 \int\limits_0^1 J_0(k_0 a \rho \sin{\theta})
\rho d\rho \\ & = & \displaystyle a^2 \frac{J_1(k_0 a
\sin{\theta})}{k_0 a \sin{\theta}},
\end{array}
\end{equation}
where $J_1$ is the Bessel function of the first kind with order one.
Inserting this in (\ref{eq:Eparareflector2}) provides
\begin{equation}
\begin{array}{lcl}
\displaystyle E_{\theta} &  = & \displaystyle \frac{\jmath a^2 k_0
E_0 e^{-\jmath k_0 r}}{2 r} (1+\cos{\theta})\cos{\phi}
\frac{J_1(k_0 a \sin{\theta})}{k_0 a \sin{\theta}} , \\
\displaystyle E_{\phi} &  = & \displaystyle \frac{-\jmath a^2 k_0
E_0 e^{-\jmath k_0 r}}{2 r} (1+\cos{\theta})\sin{\phi}
\frac{J_1(k_0 a \sin{\theta})}{k_0 a \sin{\theta}}.
\end{array}
 \label{eq:Eparareflector3}
\end{equation}
The radiation pattern is found by using relations (\ref{eq:pointingverreveld})-
(\ref{eq:normradpat}):
\begin{equation}
\begin{array}{lcl}
\displaystyle F(\theta) & = & \displaystyle
\frac{|E_{\theta}(\theta,\phi)|^2+|E_{\phi}(\theta,\phi)|^2}
{|E_{\theta}(0,0)|^2+|E_{\phi}(0,0)|^2} \\ & = & \displaystyle
(1+\cos{\theta})^2 \left[ \frac{J_1(k_0 a \sin{\theta})}{k_0 a
\sin{\theta}} \right]^2.
\end{array}
\label{eq:reflectorradpat}
\end{equation}
Note that we used the relation $\displaystyle \lim_{x
\rightarrow 0} \frac{J_1 (z)}{z} = \frac{1}{2}$.
The radiation pattern (in [dB]) is shown in Fig. \ref{fig:reflectorradpat3d} in case of a uniformly-illuminated aperture with a
diameter $D=2a=4\lambda_0$. The cross section in the $\phi=0$-plane is
shown in Fig. \ref{fig:reflectorradpat2d}. The first sidelobe is found at a level of $-17.6$ dB w.r.t. the main beam and the 3-dB beam width is $\theta_{HP} = 1.02 \lambda_0/(2a) = 14.6^0$.
The beam width scales directly with the size of the reflector: the larger $D$, the smaller the beam width. Note that the first sidelobe of a circular aperture is lower than the first sidelobe of a rectangular aperture (-17.6 dB versus -13.3 dB).

Up to now, we have assumed a uniform illumination of the aperture: we have assumed $E_0$ to be constant over the aperture surface $S_a$. The drawback of a uniform distribution is the relative high first sidelobe level of -17.6 dB. We will now extend our analysis to a more general aperture distribution. For that purpose, we will assume that the aperture distribution on $S_a$ can be expressed in the following form:
\begin{equation}
\begin{array}{lcl}
\displaystyle  \vec{E}_a(r_0) & = & W(r_0) \vec{u}_x, \\
\displaystyle Z_0 \vec{H}_a (r_0) & = & W(r_0) \vec{u}_y,
\end{array}
\end{equation}
with aperture distribution (or weighting) function given by
\begin{equation}
\displaystyle W(r_0) = E_0 \left[ 1-\left(\frac{r_0}{a}\right)^2
\right]^p.
\end{equation}
The specific case $p=0$, which provides a uniform illumination, is already discussed in this section. For other valus of $p$ we have to return to expression
(\ref{eq:Eparareflector2}) and replace $E_0$ by the aperture distribution function $W(r_0)$.
This provides us with the following integral
\begin{equation}
\displaystyle  \int\limits_{S_a} W(r_0) e^{\jmath k_0 (r_0 \cdot
\vec{u}_r)} dS =  E_0 \int\limits_{0}^{a}  \int\limits_{0}^{2\pi}
\left[ 1-\left(\frac{r_0}{a}\right)^2 \right]^p e^{\jmath k_0 r_0
\sin{\theta}\cos(\phi-\phi_0)} d\phi_0 r_0 dr_0.
\end{equation}
The $\phi_0$ integration is done in a similar way as in
(\ref{eq:phiintegratiereflector}). As a result we obtain the following integral along $r_0$:
\begin{equation}
\displaystyle 2 \pi E_0 \int\limits_{0}^{a} \left[
1-\left(\frac{r_0}{a}\right)^2 \right]^p J_0(k_0 r_0 \sin{\theta})
r_0 dr_0.
\end{equation}
Substituting $\rho=r_0/a$ and $u_a=k_0 a \sin{\theta}$ results in
\begin{equation}
\displaystyle 2 \pi a^2 E_0 \int\limits_{0}^{1} \left[ 1-\rho^2
\right]^p  J_0(u_a \rho) \rho d\rho.
\end{equation}
This integral can be expressed in terms of a Bessel function by applying the integral of Sonine (see  \cite{Watson}, p.373)
\begin{equation}
\displaystyle  J_{\mu+\nu+1} (z) =
\frac{z^{\nu+1}}{2^{\nu}\Gamma_f(\nu+1)} \int\limits_{0}^{\pi/2}
J_{\mu} (z \sin{\theta}) \sin^{\mu+1}{\theta}
\cos^{2\nu+1}{\theta}d\theta,
\end{equation}
Here $\Gamma_f$ is the Gamma function \cite{Gradsteyn} for which $\Gamma(p+1)=p!$. Now let $\mu=0$, $\nu=p$, $z=u_a$ and
$\sin{\theta}=\rho$. We then find
\begin{equation}
\displaystyle  \frac{J_{p+1}(u_a)}{u_a^{p+1}} =
\frac{1}{2^p\Gamma_f(p+1)} \int\limits_{0}^{1} J_0 (u_a \rho) \left[
1-\rho^2 \right]^p \rho d\rho.
\end{equation}
As a result, the integral takes the form
\begin{equation}
\displaystyle  \int\limits_{S_a} W(r_0) e^{\jmath k_0 (r_0 \cdot
\vec{u}_r)} dS =  2\pi a^2 E_0 2^p p! \frac{J_{p+1}(u_a)}{u_a^{p+1}}.
\label{eq:intapertuurreflector}
\end{equation}
The electric field in the far-field region is now easily found by inserting  (\ref{eq:intapertuurreflector}) in (\ref{eq:Eparareflector2}).
Some calculated results are summarized in table \ref{tab:resultatenreflector}.
From this table we can conclude that the uniform illumination ($p=0$) provides the highest directivity and antenna gain. However, the first sidelobe level is relative high.
By using a weighted aperture distribution (tapering), we can reduce the sidelobe level at the expense of a reduced directivity and corresponding efficiency, which is as low as $44 \%$ for $p=2$. The requirements for a particular application will determine the optimal shape of the aperture distribution function $W(r_0)$.
\begin{table}[htb]
 \begin{center}
  \begin{tabular}{|l|c|c|c|}
   \hline
   \multicolumn{1}{|l|}{\it \small  Antenna parameter} &
   \multicolumn{1}{|c|}{\it \small $p=0$} &
   \multicolumn{1}{|c|}{\it \small $p=1$} &
   \multicolumn{1}{|c|}{\it \small $p=2$} \\
   \hline
    3-dB beam width in degrees  & $\displaystyle 29.2 \frac{\lambda_0}{a}$ & $\displaystyle 36.4 \frac{\lambda_0}{a}$
      & $\displaystyle 42.1 \frac{\lambda_0}{a}$ \\
    Beam width between zeros in degrees  & $\displaystyle 69.9 \frac{\lambda_0}{a}$ & $\displaystyle 93.4 \frac{\lambda_0}{a}$
      & $\displaystyle 116.3 \frac{\lambda_0}{a}$ \\
    First sidelobe level [dB]  & -17.6 & -24.6 & -30.6 \\
    Directivity  & $\displaystyle \left( \frac{2\pi a}{\lambda_0} \right)^2$
    & $\displaystyle 0.75 \left( \frac{2\pi a}{\lambda_0} \right)^2$
    & $\displaystyle 0.56 \left( \frac{2\pi a}{\lambda_0} \right)^2$ \\
   \hline
  \end{tabular}
 \end{center}
  \caption{{\it Overview of the most relevant antenna parameters as a function of the order $p$ of the aperture distribution function $W=E_0 \left[ 1-(r_0/a)^2 \right]^p$ in case of a parabolic reflector antenna with diameter $D=2a$.}}
 \label{tab:resultatenreflector}
\end{table}

\section{Microstrip antennas}
\begin{figure}[hbt]
\centerline{\psfig{figure=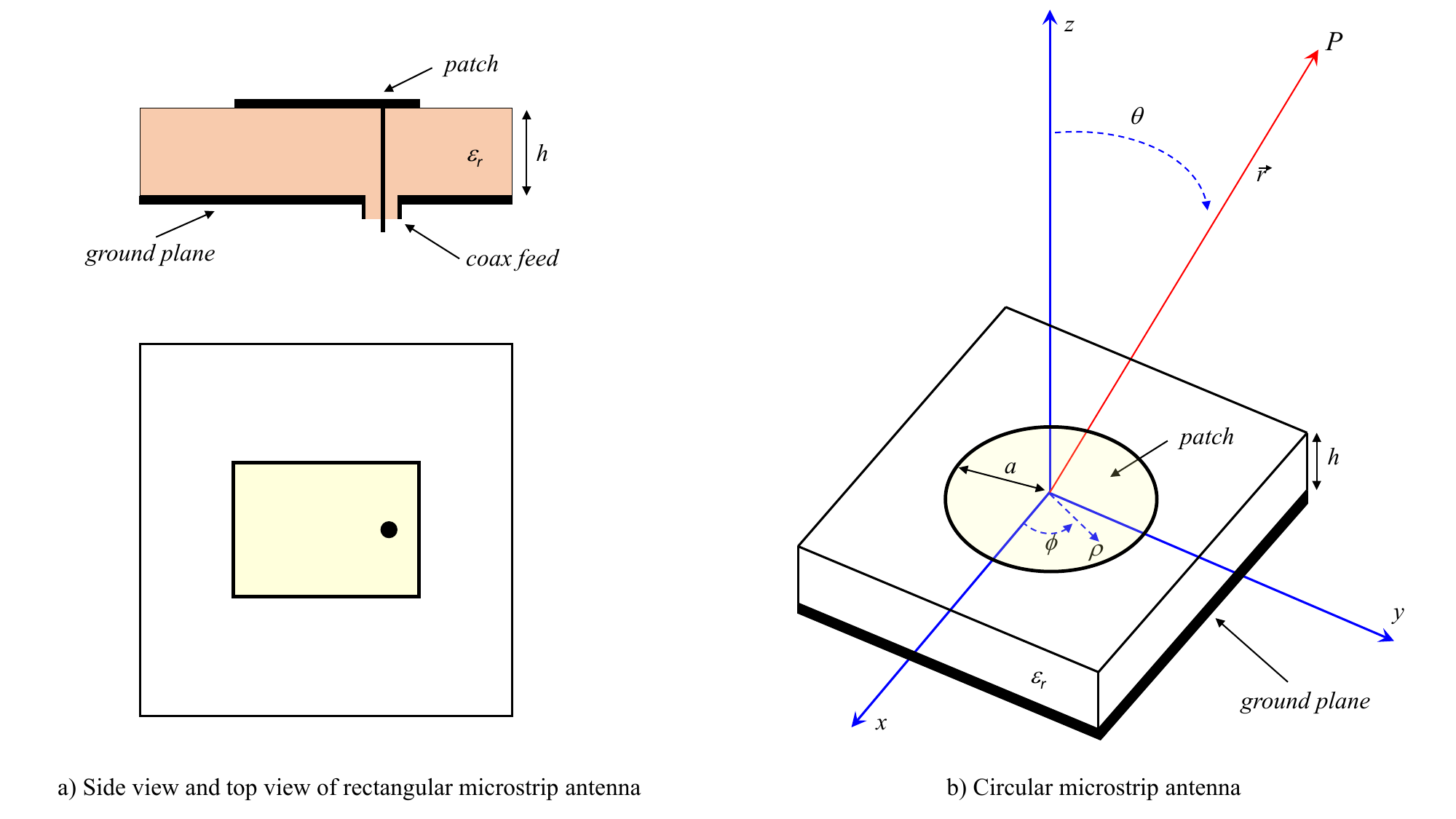,width=170mm}}
\caption{\it Rectangular and circular microstrip antenna fed by a coaxial cable. The metallic patch is printed on a substrate with permittivity $\epsilon_r$.}
\label{fig:micantenne2}
\end{figure}

Microstrip antennas, also known as {\it patch antennas} or {\it printed antennas} are resonant antennas which can be realized on printed-circuit boards (PCBs).
In order to determine the radiation properties of microstrip antennas, we can use the same method as used for aperture antennas in the previous sections.
Microstrip antennas can take many forms and shapes. We will limit ourselves to some basic structures. Fig. \ref{fig:micantenne2} shows the side- and top-view of a rectangular and circular microstrip antenna fed by a coaxial cable. The inner conductor of the coaxial feed is connected to the metal patch and the outer conductor is connected to the ground plane.
The dielectric substrate has a thickness $h$ and dielectric constant $\epsilon_r$. The dielectric losses are indicated by the loss tangent $\tan{\delta}$ and will be neglected here.
Microstrip antennas can be manufactured with standard PCB etching techniques to realize printed copper structures with accuracies well below $100 \mu m$. The patch can also be fed by a microstrip line or by using a resonant slot in the ground plane.

In literature, various models have been introduced to determine the radiation properties of microstrip antennas, including analytical models such as the cavity model and transmission line model, and numerical models such as the method of moments (MoM)\cite{Smolders}.
In this section we will explore the cavity model in more detail for the case of a circular microstrip antenna as illustrated in Fig. \ref{fig:micantenne2}. We will assume that all metal surfaces (patch and ground plane) are perfect electric conductors (PEC).  Furthermore, we will assume that the thickness of the substrate $h$ is very thin as compared to the wavelength, i.e. $h \ll \lambda_0$. We will use cylindrical coordinates  with $(\rho,\phi,z)$ to describe the antenna configuration. With these assumptions, we can introduce the following simplifications:
\begin{enumerate}[i.]
\item The electric field will be concentrated between the circular patch and the ground plane within a cylinder with radius $a$ and height $h$. This cylinder is called {\it cavity}. The cavity is filled with a homogeneous diectric material with permittivity $\epsilon_r$.

\item Due to the small thickness of the substrate, the fields within the cavity will not depend on the $z$-coordinate.

\item The electric field has only a component in the $z$ direction, so $\vec{E}=E_z \vec{u}_z$.

\item The current distribution $\vec{J}_e$ along the edge $\rho=a$ of the circular patch will be directed parallel to the edge of the patch. This implies that the magnetic field given by $\vec{H}=\vec{u}_z \times \vec{J}_e$, will be perpendicular to the side walls of the cavity, so $\vec{u}_n \times
\vec{H}=\vec{0}$. Note that $\vec{u}_n$ is the normal on the side walls of the cavity.

\end{enumerate}
With these assumptions, our problem is reduced to a cavity structure with the following boundary conditions:
\begin{equation}
\begin{array}{ll}
\displaystyle \vec{u}_n \times \vec{H} = \vec{0} & \ \ \ \ \ \ \ \
\  on \ the \ side walls  \\ \displaystyle \vec{u}_n \times \vec{E} =
\vec{0} & \ \ \ \ \ \ \ \ \  on \ the \ top \ and \ bottom \ side \ of \ the \ cavity,
\end{array}
\label{eq:cavitybound1}
\end{equation}
where $\vec{u}_n$ is the normal vector on the surface under consideration. Since $\vec{E}$ will only have a $z$-component, we can express the condition that $ \vec{u}_n \times \vec{H} = \vec{0}$ when
$\rho=a$ in terms of a condition for $E_z$, since
\begin{equation}
\displaystyle \vec{H}=\frac{-1}{\jmath \omega \mu_0} \nabla \times
\vec{E}.
\end{equation}
As a result, we can express the components of the magnetic field in the following form:
\begin{equation}
\begin{array}{lcl}
\displaystyle H_{\rho} & = & \displaystyle \frac{-1}{\jmath \omega
\mu_0} \frac{1}{\rho} \frac{\partial E_z}{\partial \phi}, \\
\displaystyle H_{\phi} & = & \displaystyle \frac{1}{\jmath \omega
\mu_0} \frac{\partial E_z}{\partial \rho}.
\end{array}
\end{equation}
Apparantly, on the side walls the following condition holds
\begin{equation}
\begin{array}{ll}
\displaystyle \frac{\partial E_z}{\partial \rho} = 0, & \ \ \ \ \
\ \ \ when \ \rho=a.
\end{array}
\end{equation}
Inside the cavity, we can derive the Helmholtz equation for the electric field  $\vec{E}=E_z \vec{u}_z$ by using Maxwell's equations for a source-free region (\ref{eq:maxwellfreq}). We now obtain
\begin{equation}
\nabla^2 \vec{E}+ k^2\vec{E} = \vec{0},
\end{equation}
where $k=k_0 \sqrt{\epsilon_r}$ represents the wave number in the substrate.
The Helmholtz equation in cylindrical coordinates then takes the following form (with $\displaystyle
\frac{\partial E_z}{\partial z}=0$)
\begin{equation}
\displaystyle \frac{\partial^2 E_z}{\partial \rho^2} +
\frac{1}{\rho} \frac{\partial E_z}{\partial \rho} +
\frac{1}{\rho^2} \frac{\partial^2 E_z}{\partial \phi^2} + k^2 E_z
= 0.
\end{equation}
The solution of this second-order differential equation is
\begin{equation}
\displaystyle E_z = E_0 J_n(k \rho) \cos(n \phi),
\end{equation}
where $J_n(k \rho)$ is the Bessel function of the first kind with order $n$ and $E_0$ is a constant which is determined by the applied voltage at the feed point. The other field components are given by
\begin{equation}
\begin{array}{lcl}
\displaystyle H_{\rho} & = & \displaystyle \frac{n}{\jmath \omega
\mu_0 \rho}  E_0 J_n(k \rho) \sin(n \phi), \\ \displaystyle
H_{\phi} & = & \displaystyle \frac{k}{\jmath \omega \mu_0} E_0
\frac{\partial J_n(k \rho)}{\partial \rho} \cos(n \phi).
\end{array}
\end{equation}
All other field components are zero, so $E_{\rho}=E_{\phi}=H_{z}=0$.
The side wall of the cavity acts as a magnetic wall with $\vec{u}_n \times \vec{H} = \vec{0}$ or
$H_{\phi}=0$. This provides us the resonance condition:
\begin{equation}
\displaystyle \frac{\partial J_n(k a)}{\partial \rho} = 0.
\end{equation}
Since $\displaystyle \frac{\partial J_n(k a)}{\partial \rho}$ has multiple zeros for a particular values of $n$, we will find multiple resonance frequencies for a given radius $a$ and $n$. Now let $K_{nm}$ be the $m$-th zero of $\displaystyle \frac{\partial
J_n(k a)}{\partial \rho}$. Table \ref{tab:nulpuntenJn} summarizes a number of zeros with corresponding values of $k a$.
\begin{table}[htb]
 \begin{center}
  \begin{tabular}{|c|c|}
   \hline
   \multicolumn{1}{|c|}{\it \small Mode $(n,m)$} &
   \multicolumn{1}{|c|}{\it \small zero $ka$} \\
   \hline
       $0,1$ & $0$ \\
       $1,1$ & $1.841$ \\
       $2,1$ & $3.054$ \\
       $0,2$ & $3.832$ \\
       $3,1$ & $4.201$ \\
       $4,1$ & $5.317$ \\
       $1,2$ & $5.331$ \\
       $5,1$ & $6.416$ \\
   \hline
  \end{tabular}
 \end{center}
  \caption{{\it Zeros of $\displaystyle \frac{\partial
J_n(ka)}{\partial \rho}$.}}
 \label{tab:nulpuntenJn}
\end{table}
The resonance frequency corresponding to a particular mode can be determined from the following equation
\begin{equation}
\displaystyle ka = \frac{2 \pi f}{c} \sqrt{\epsilon_r}a=K_{nm},
\end{equation}
resulting in
\begin{equation}
\displaystyle f_{nm} = \frac{K_{nm} c}{2 \pi a \sqrt{\epsilon_r}}.
\label{eq:resfreq}
\end{equation}
The mode with $(n,m)=(1,1)$ provides the lowest resonance frequency for a particular radius $a$ of the circular patch. Note that the cavity model does not account for fringing fields at the edges of the cavity. These fringing fields will slightly change the resonance frequency. This effect can be included in formula (\ref{eq:resfreq}) by introducing an effective
radius $a_e$. Closed-form expressions for $a_e$ can be found in literature \cite{Bahl}-\cite{Balanis}:
\begin{equation}
\displaystyle a_e = a \sqrt{1+\frac{2h}{\pi a \epsilon_r} \left[\ln\left(\frac{\pi a}{2h}\right) +1.7726 \right]}.
\label{eq:aeffective}
\end{equation}

\begin{figure}[hbt]
\centerline{\psfig{figure=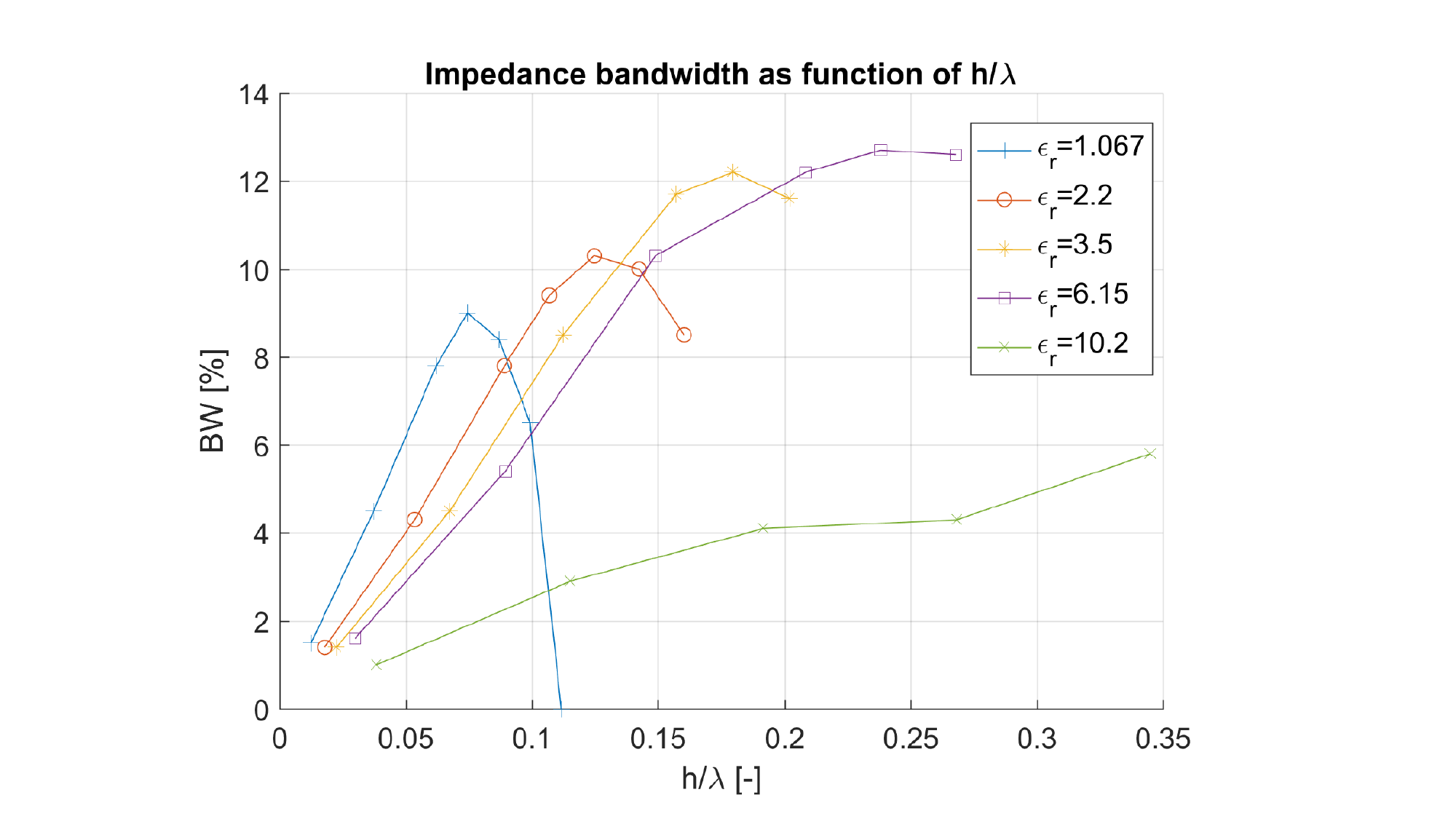,width=140mm}}

\caption{\it Bandwidth of a single-layer microstrip antennas versus electrical height of the dielectric substrate.}
\label{fig:Bandwidthmicrostrip}
\end{figure}

Near resonance it is quite easy to match the input impedance of a microstrip antennas to a transmission line, such as a $50 \Omega$ coax cable. The bandwidth of microstrip antennas (defined as the frequency band over which the antenna can be matched to a transmission line with a characteristic impedance  $Z_0$) strongly depends on the electrical thickness of the substrate.
An accurate estimation of the input impedance of microstrip antennas can be best obtained using numerical methods, such as the Method-of-Moments (MoM), see also chapter \ref{chap:NumEM}.
Fig. \ref{fig:Bandwidthmicrostrip} shows the realized bandwidth of a coax-fed microstrip antenna versus electrical thickness, according to MoM simulation from \cite{Smolders}.
A larger bandwidth can be obtained by using stacked patches \cite{Smolders} or by using a resonant slot feed in the ground plane.

\begin{figure}[hbt]
\centerline{\psfig{figure=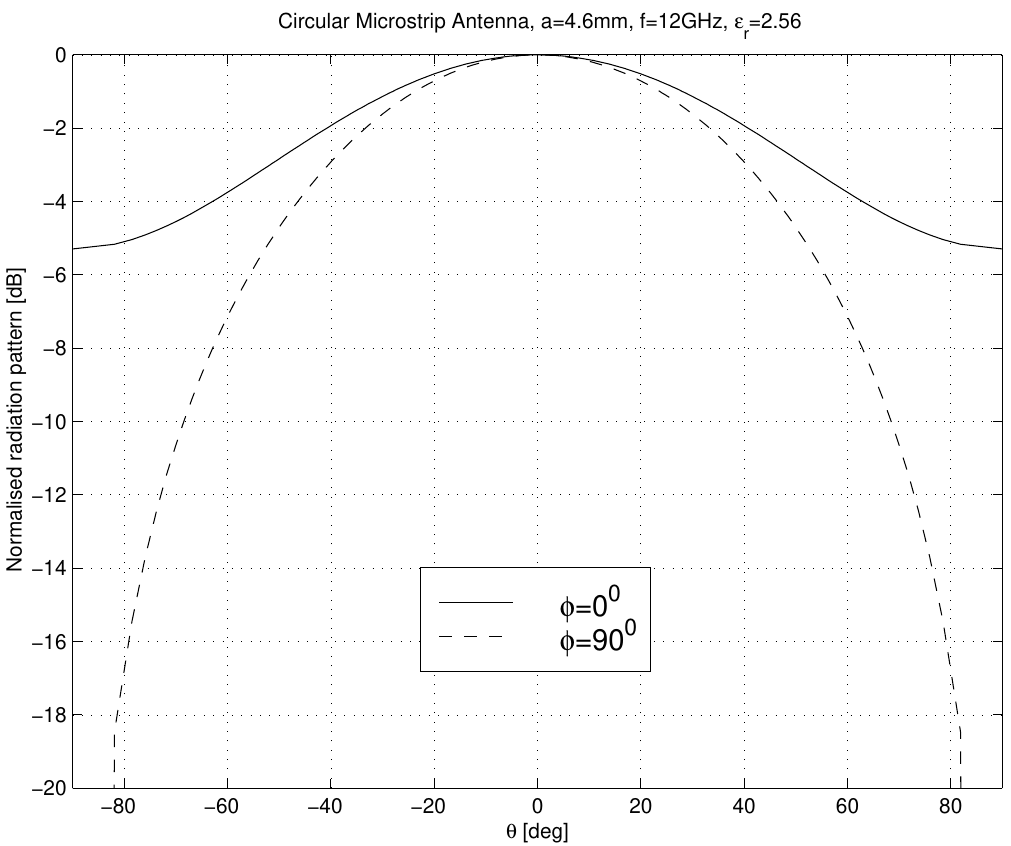,width=90mm}}

\caption{\it Radiation pattern of a circular microstrip antenna printed on an electrically thin substrate in the $\phi=0$ plane and $\phi=90^0$-plane. The antenna operates in the fundamental $(n,m)=(1,1)$-mode at $f=12$ GHz. The radius of the patch $a=4.6$ mm and
$\epsilon_r=2.56$. All values are plotted in dB.}
\label{fig:radpatmicant}
\end{figure}

The radiation properties of microstrip antennas can be easily determined by using the aperture method according to the Lorentz-Larmor theorem. We will select the surface $S$ as $S_p+S_a+S_g$, where $S_p$ is the patch, $S_a$ the side wall of the cavity and $S_g$ the ground plane.
We can also apply the image principle (see section
\ref{sec-draadbovengrond}) in order to account for the ground plane.
At the side wall of the cavity we then obtain the following magnetic current distribution
\begin{equation}
\begin{array}{ll}
\displaystyle \vec{J}_{ms} = 2\vec{E} \times \vec{u}_n = 2E_z
\vec{u}_{\phi} & \ \ \ \ \ \ \ \ when \ \rho=a \ \ and \ 0 < z < h.
\end{array}
\end{equation}
As a result, we find that the field components $E_{\theta}$ and
$E_{\phi}$ in the far-field region are given by
\begin{equation}
\begin{array}{lcl}
\displaystyle E_{\theta} & = & \displaystyle \frac{\jmath^n ha k_0
E_0 J_n(ka)e^{-\jmath k_0 r}}{2r} \cos{n\phi} \left[ J_{n+1}(k_0 a
\sin{\theta}) - J_{n-1}(k_0 a \sin{\theta}) \right], \\
\displaystyle E_{\phi} & = & \displaystyle \frac{\jmath^n ha k_0
E_0 J_n(ka)e^{-\jmath k_0 r}}{2r} \cos{\theta} \sin{n\phi} \left[
J_{n+1}(k_0 a \sin{\theta}) + J_{n-1}(k_0 a \sin{\theta}) \right],
\end{array}
\end{equation}
where we have assumed that the patch is fed by a transmission line in the $\phi=0$ plane.
For the fundamental mode with
$(n,m)=(1,1)$ we obtain:
\begin{equation}
\begin{array}{lcl}
\displaystyle E_{\theta} & = & \displaystyle \frac{\jmath ha k_0
E_0 J_1(ka)e^{-\jmath k_0 r}}{2r} \cos{\phi} \left[ J_2(k_0 a
\sin{\theta}) - J_0(k_0 a \sin{\theta}) \right], \\ \displaystyle
E_{\phi} & = & \displaystyle \frac{\jmath ha k_0 E_0
J_1(ka)e^{-\jmath k_0 r}}{2r} \cos{\theta} \sin{\phi} \left[
J_2(k_0 a \sin{\theta}) + J_0(k_0 a \sin{\theta}) \right].
\end{array}
\end{equation}
Fig. \ref{fig:radpatmicant} shows the radiation pattern in the $\phi=0^0$ plane and the $\phi=90^0$ plane of a circular microstrip antenna operating in the $(n,m)=(1,1)$-mode and resonating at $f=12$ GHz. The radius of the patch is $a=4.6$mm and substrate permittivity  $\epsilon_r=2.56$. The 3-dB beam width is $104^0$ in the $\phi=0^0$ plane
and $80^0$ in the $\phi=90^0$ plane. Due to its omni-directional character, microstrip antennas are very usefull in phased-array antennas that require wide-scan electronic beamforming.

\chapter{Numerical electromagnetic analysis}
\label{chap:NumEM}

In chapter \ref{chap:AntennaTheory} we have determined the radiation characteristics of several antenna types with a given electric and/or magnetic current distribution.
In section \ref{sec-dunnedraadantene} we have investigated wire antennas, where we assumed a sinusoidal current distribution along the wire.  When the antenna is operated close to its resonance, this will be a very good approximation of the real current distribution.
However, at other frequencies the current distribution will be different.
In general, it is not easy (or not possible) to derive an accurate analytic description of the currents on antenna structures, even not for a fairly simple $\lambda_0/2$ dipole (wire) antenna.
For that purpose, antenna designers use numerical methods to determine the current distribution on the antenna structure. Several numerical approaches are described in literature, including the Finite-Element-Method (FEM), Finite-Difference-Time-Domain (FDTD) and the Method-of-Moments (MoM), see for example in \cite{Sadiku}.
One of the more common methods for antenna analysis is MoM, which is implemented in several commercial microwave and antenna design tools, for example in {\it Momentum} which is part of the EDA tool {\it Advanced Design System (ADS)} \cite{ADS}. In this section we will show in a step-by-step approach how MoM can be used to determine the current distribution on a cylindrical wire antenna.

\section{Method of moments (MoM)}
\label{sec:MoM}
\begin{figure}[hbt]
\centerline{\psfig{figure=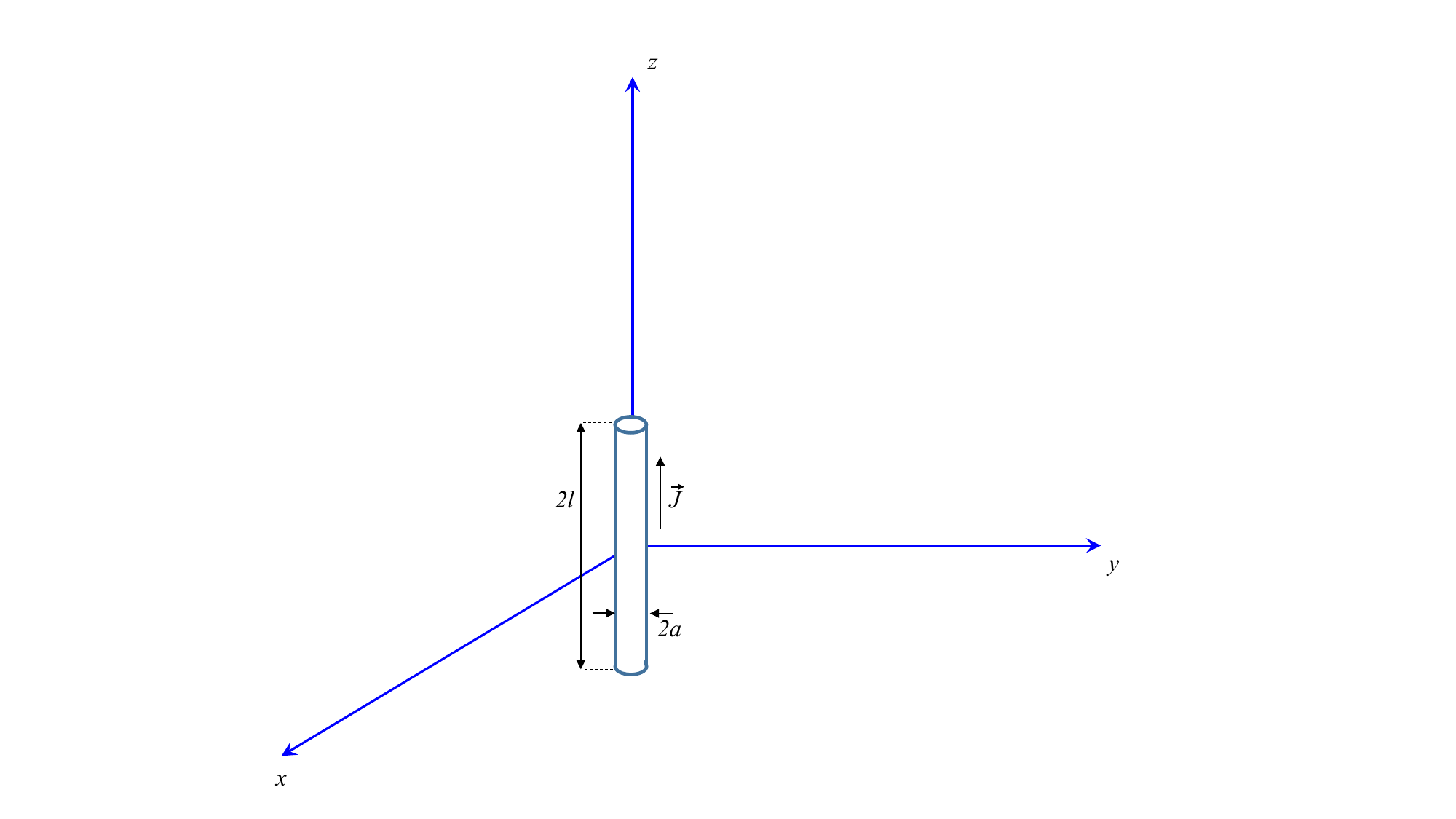,width=140mm}}
\caption{\it Electrically-thin cylindrical wire antenna of length $2l$ and
diameter $2a$. The center of the wire is located at the origin of the coordinate system and is oriented along the $z$-axis.} \label{fig:cildraadant}
\end{figure}

\begin{figure}[hbt]
\vspace{-1cm}
\centerline{\psfig{figure=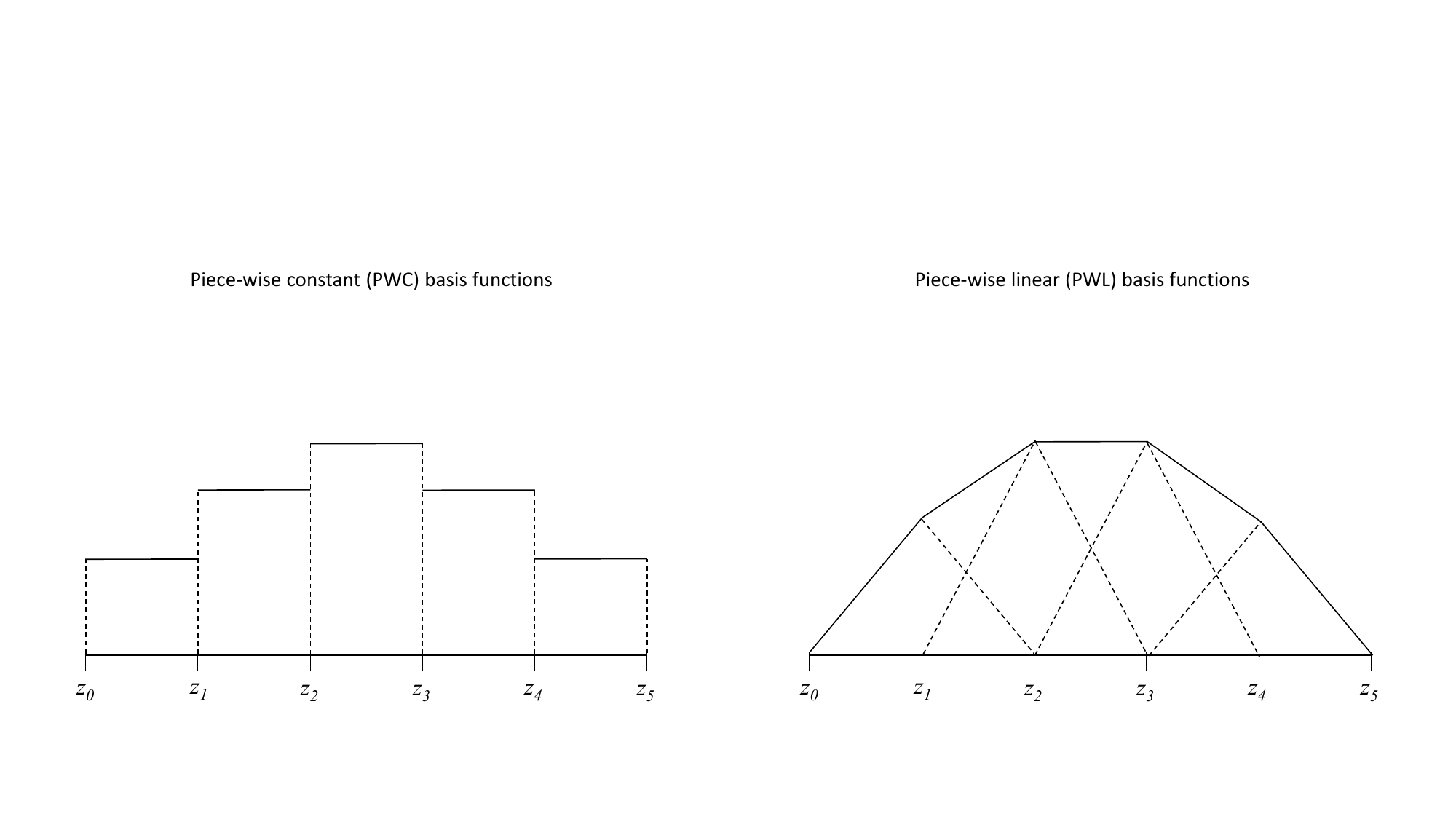,width=170mm}}
\vspace{-1.5cm}
\caption{\it Piece-wise constant (PWC) and Piece wise linear (PWL) approximation of a function.} \label{fig:PWLmodes}
\end{figure}

The method of moments is a general numerical procedure for solving integral equations. One of the first books which describes MoM for electromagnetic problems is the book of Harrington
\cite{HarringtonMoM}. In this section we will illustrate how MoM can be used to determine the current distribution along a simple cylindrical wire antenna. Note that MoM can also be used to determine equivalent magnetic currents.
The following step-wise approach is used:
\begin{itemize}
\item{determine the integral equation of the antenna problem at hand,}
\item{discretize the antenna structure in small segments (also known as sub-domains),}
\item{derive the matrix equation for the (still) unknown current distribution along the antenna, }
\item{calculate the elements of the matrix and solve the matrix equation.}
\end{itemize}
Now consider the cylindrical wire antenna in free space as illustrated in Fig. \ref{fig:cildraadant}. The antenna is symmetrically oriented along the $z$-axis with length $2l$ and diameter $2a$. The antenna is fed in the center at $z=0$. This is exactly the same structure as discussed in section
\ref{sec-dunnedraadantene}.
We will assume that the diameter of the wire is much smaller than the wavelength, so $2a \ll \lambda_0$. Therefore, we can ignore electric currents at the end surfaces of the cylinder, and as a result, the electric current distribution will only have a component along the $z$-axis, i.e. $\vec{J}=J_{z} \vec{u}_z$. Note that we have removed the subscript $e$ in $\vec{J}$ in order to simplify the notation in the rest of this section.
The antenna configuration of Fig. \ref{fig:cildraadant} imposes two boundary conditions:
\begin{enumerate}[i]
\item{The tangential field at the perfectly-conducting cylinder with diameter $2a$ has to be equal to zero.}
\item{The current at both ends of the cylinder is zero.}
\end{enumerate}
In our MoM approach we will only apply the first boundary condition directly. This boundary condition can be describes as
\begin{equation}
\begin{array}{l}
\displaystyle \vec{u}_{n} \times {\vec{\cal{E}}}^{tot} (\vec{r}) =
\vec{u}_{n} \times ({\vec{\cal{E}}}^{ex} (\vec{r}) +
 {\vec{\cal{E}}}^s (\vec{r})) = \vec{0}, \ \ \ \vec{r} \epsilon S_0,
\end{array}
\label{eq:boundcondition}
\end{equation}
where the surface $S_0$ is the outer surface of the cylinder and where ${\vec{\cal{E}}}^{ex} (\vec{r})$ and ${\vec{\cal{E}}}^s
(\vec{r})$ are the excitation field and the scattered field, respectively.  Furthermore, $\vec{u}_{n}$ is the normal vector on the metal cylinder.
The scattered field is the field due to the induced currents on the wire antenna. The excitation field is the field induced by the voltage source which is connected to the differential port at the center of the antenna.
The scattered field can be expressed in terms of the (yet) unknown electric current distribution  ${\vec{J}}$ and the free-space Greens' function $G(\vec{r},\vec{r}_0)$ by using expressions (\ref{eq:HenEinA}), (\ref{eq:Greensfunctie}) and
(\ref{eq:vectorpot}) from chapter \ref{chap:AntennaTheory}.
In case the antenna is not located in free-space, i.e. in case of a microstrip antenna, we can still use the proposed approach. However, the Greens function will be more complicated, see for example in  \cite{Smolders} for the case of multi-layer microstrip antennas.  For our free-space configuration of Fig. \ref{fig:cildraadant} we can express equation (\ref{eq:boundcondition}) in the following form:
\begin{equation}
\begin{array}{ll}
\displaystyle \vec{u}_{n} \times {\vec{{E}}}^{ex} (\vec{r}) &
=
 \di \vec{u}_{n} \times \jmath \omega \mu_0
 \dinttw G(\vec{r}, \vec{r}_0) \vec{J} (\vec{r}_0) dS_0 \\
 & \di + \vec{u}_{n} \times \nabla \left( \frac{\jmath \omega \mu_0}{k_0^2} \nabla \cdot
 \dinttw G (\vec{r}, \vec{r}_0) \vec{J} (\vec{r}_0) dS_0 \right),
 \vec{r} \epsilon S_0.
\end{array}
\label{eq:integralequation}
\end{equation}
Integral equation (\ref{eq:integralequation}) can be solved numerically by using the method of moments.
The first step is the expansion of the yet unknown current distribution $\vec{J}$ on the wire antenna in {\it expansion functions} according to
\begin{equation}
\begin{array}{ll}
\displaystyle \vec{J} (x,y,z) & = \di \sum_{n} I_n \vec{J}_{n}
(x,y,z),
\end{array}
\label{eq:expansion1}
\end{equation}
where $I_n$ are the mode coefficients which we need to determine.  The expansion functions $\vec{J}_{n} (x,y,z)$ are also called {\it basis functies} and can take many forms. When we want to have an exact solution of the surface current distribution $\vec{J}$, we need to use an infinite summation in
(\ref{eq:expansion1}). In addition, the set of basis functions need to be complete in the sense that all physical effects are included.  In practice we will limit the summation in
(\ref{eq:expansion1}) to a maximum number of basis functions of $n=N_{max}$.
In this way, we can approximate the exact solution with a limited computational effort.
There are many types of basis functions. We can subdivide them into two main categories, that is in  {\it global} basis functions and {\it local} basis functions.
Global basis functions are non-zero over the entire domain, in our example over the entire cylindrical wire, whereas local basis functions are only non-zero over a very small part of the entire domain.
For that reason, local basis functions are also known as {\it sub-domain} basis functies.
Two examples of sub-domain basis functions are shown in Fig. \ref{fig:PWLmodes},
where piece-wise constant (PWC) and overlapping piece-wise linear (PWL) basis functions are shown that approximate a particular function.

When using $N_{max}$ basis functions to describe the yet unknown current distribution on the antenna, we obtain
\begin{equation}
\begin{array}{lcl}
\displaystyle \vec{J} (x,y,z) & = &
 \di \sum_{n}^{N_{max}} I_n \vec{J}_{n} (x,y,z).
\end{array}
\label{eq:expansion2}
\end{equation}
The scattered electric field $\vec{E}^s (x,y,z)$ can be written in terms of the current distribution $\vec{J} (x,y,z)$
\begin{equation}
 \di \vec{E}^s (x,y,z) = L \{ \vec{J} (x,y,z) \} ,
\label{eq:linearoper}
\end{equation}
where $L$ is a linear operator. By combining
(\ref{eq:linearoper}) and (\ref{eq:expansion2}) we obtain
\begin{equation}
\displaystyle \vec{E}^s (x,y,z) =
 \sum_{n=1}^{N_{max}} I_{n} L \{ \vec{J}_{n} (x,y,z) \} =
\sum_{n=1}^{N_{max}} I_{n} \vec{E}^s_n (x,y,z) . \label{eq:eexpan}
\end{equation}
Substitution of the expansion (\ref{eq:eexpan}) in equation
(\ref{eq:boundcondition}) provides
\begin{equation}
\begin{array}{lc}
\displaystyle \vec{u}_{n} \times \left( \sum_{n=1}^{N_{max}} I_{n}
\vec{E}^s_n(x,y,z) +
 \vec{E}^{ex}(x,y,z) \right) = \vec{0} & ,
\end{array}
\label{eq:boundeq2}
\end{equation}
on the outer surface $S_0$ of the wire antenna. Let us now introduce the residue $\vec{R}$ according to
\begin{equation}
\begin{array}{lc}
\displaystyle \vec{R}(x,y,z) = \vec{u}_{n} \times \left(
\sum_{n=1}^{N_{max}} I_{n} \vec{E}^s_n (x,y,z) +
 \vec{E}^{ex}(x,y,z) \right).
\end{array}
\label{eq:defres}
\end{equation}
This residue has to be zero over the entire outer surface of the wire antenna.
We will relax this requirement by {\it weighting} the residue to zero with respect to some weighting functions $\vec{J}_m (x,y,z)$, such that
\begin{equation}
\begin{array}{lcc}
\displaystyle {\langle \vec{R} \ ; \ \vec{J}_{m} \rangle }
 = \dinttw_{S_m} \vec{R}(x,y,z) \cdot \vec{J}_{m} (x,y,z) \ dS = 0 &,
\end{array}
\label{eq:resweight}
\end{equation}
for $m=1,2...,N_{max}$, where $S_m$ represents the surface on which the weighting function  $\vec{J}_{m}$ is nonzero. The set of weighting functions $\vec{J}_{m}$, is also known as a set of {\it test functions}. Often the set of test functions is the same as the set of expansion functions in  (\ref{eq:expansion1}). This particular choice is called the Galerkin's method \cite{Harrington1}.
Inserting (\ref{eq:defres}) in (\ref{eq:resweight}) provides the following set of $N_{max}$ linear equations
\begin{equation}
\begin{array}{l}
\displaystyle
  \sum_{n=1}^{N_{max}} I_n \dinttw_{S_m}
 \vec{E}^s_n (x,y,z) \cdot \vec{J}_m (x,y,z)  \ dS
+ \dinttw_{S_m} \vec{E}^{ex} (x,y,z) \cdot \vec{J}_m (x,y,z) \ dS
= 0 ,
\end{array}
\end{equation}
for $m=1,2...,N_{max}$. We can write this set of linear equations in a more compact form:
\begin{equation}
\begin{array}{l}
\displaystyle
  \sum_{n=1}^{N_{max}} I_n Z_{mn} + V^{ex}_m = 0 ,
\end{array}
\end{equation}
for $m=1,2...,N_{max}$. In matrix notation we now obtain:
\begin{equation}
\displaystyle
 [Z] [I] + [V^{ex}] = [0],
\label{eq:mateq}
\end{equation}
where the elements of the matrix $[Z]$ and the elements of the excitation vector
$[V^{ex}]$ are given by
\begin{equation}
\begin{array}{ll}
\displaystyle Z_{mn} & \di = \dinttw_{S_m} \vec{E}^s_n (x,y,z)
 \cdot \vec{J}_m (x,y,z) \ dS, \\
\displaystyle V^{ex}_m  & \di =  \dinttw_{S_m} \vec{E}^{ex}
(x,y,z)
 \cdot \vec{J}_m (x,y,z) \ dS,
\end{array}
\label{eq:zenv}
\end{equation}
The matrix $[Z]$ includes $N_{max} \times N_{max}$ elements, $[I]$ is
a vector with $N_{max}$ unknown mode coefficients and
$[V^{ex}]$ is the excitation vector with $N_{max}$ elements. The matrix equation (\ref{eq:mateq}) can be solved rather easily by using standard numerical linear-algebra routines such as those available in MATLAB. In practise it will turn out that the largest computational effort is claimed for the calculation of the elements of the matrix $[Z]$. After matrix equation (\ref{eq:mateq}) has been solved, we obtain a solution for the mode coefficients and we can find an approximation of the current distribution $\vec{J}$ on the antenna surface by inserting the mode coefficients in expression  (\ref{eq:expansion2}). Once the current distribution is known, we can calculated the far-field pattern and other antenna characteristics by using the concepts from chapters \ref{chap:fundpar} and \ref{chap:AntennaTheory}.

The elements of the matrix $[Z]$ can be found by using the expression for $\vec{E}^s$ as applied in (\ref{eq:integralequation}), which results in:
\begin{equation}
\begin{array}{ll}
\displaystyle Z_{mn} & \di = \dinttw_{S_m} \left[ -\jmath \omega
\mu_0 \dinttw_{S_n} G(\vec{r}, \vec{r}_0) \vec{J}_n (x_0,y_0,z_0)
dS_0 \right. \\ & \left. \ \ \ \ - \di \nabla \left( \frac{\jmath
\omega \mu_0}{k_0^2} \nabla \cdot \dinttw_{S_n} G (\vec{r},
\vec{r}_0) \vec{J}_n (x_0,y_0,z_0) dS_0 \right) \right] \cdot
\vec{J}_m (x,y,z) \ dS,
\end{array}
\label{eq:ZinG}
\end{equation}
In general it will not be possible to find an analytic expression for the elements of $[Z]$.
We will need to use numerical methods. Note that, depending on the particular antenna structure under investigation, several numerical problems (e.g. singularities) might occur while evaluating (\ref{eq:ZinG}). More on this can be found in \cite{Sadiku} and other literature.

We will  now return to our example, the cylindrical wire antenna as illustrated in Fig.
\ref{fig:cildraadant}. Since we have assumed that the diameter $2a \ll \lambda_0$, the electric current distribution will only have a component along the $z$-axis, i.e.
$\vec{J}=J_{z} \vec{u}_z$. Therefore, the expansion and test functions will only have a component in the $z$ direction, so $\vec{J}_m=J_{zm} \vec{u}_z$. We can now write an element of the matrix $[Z]$ in the following form
\begin{equation}
\begin{array}{ll}
\displaystyle Z_{mn} & \di = -\jmath \omega \mu_0 \int\limits_{m}
\left(1+ \frac{1}{k_0^2} \frac{\partial^2}{\partial z^2} \right) \int\limits_{n}
 G(z,z_0) J_{zn} (z_0) dz_0 J_{zm} (z) \ dz \\
& \di = -\jmath \omega \mu_0 \int\limits_{m}  \int\limits_{n}
\left(1+ \frac{1}{k_0^2} \frac{ \partial^2}{ \partial z^2} \right)
 G(z,z_0) J_{zn} (z_0) dz_0 J_{zm} (z) \ dz
\end{array}
\label{eq:Zdraadant1}
\end{equation}
where the Green's function $G(z,z_0)$ is given by
\begin{equation}
\di G(z,z_0)=\frac{\di e^{-\jmath k_0 R_a}}{4 \pi R_a},
\end{equation}
with
\begin{equation}
\di R_a = \sqrt{a^2+(z-z_0)^2}.
\end{equation}
Note that $R_a$ is nonzero for $z=z_0$. This is the result of our choice (in fact approximation) to define the expansion function on the outer cylindrical surface $(\rho=a)$ of the wire antenna, whereas the test functions are defined along the axis of the wire antenna ($\rho=0$).
In this way, we avoid a singularity of the Green's function $G(z,z_0)$ at
$z=z_0$. The integral equation for the wire antenna that leads towards expression  (\ref{eq:Zdraadant1}) for $Z_{mn}$ is known as the {\it Pocklington integral equation}.
Next to this equation,  we can also derive an alternative integral equation, known as the {\it Hall\'{e}n integral equation}, see also \cite{HarringtonMoM},\cite{Sadiku}.

In our example, we will choose piece-wise constant (PWC) expansion- and test functions. The
$m$-th basis function of this set is defined as
\begin{equation}
\di J_{zm}(z)= \left\{
\begin{array}{lc}
1, & \di z_{m-1} < z < z_{m+1}, \\ 0, & \mbox{elsewhere,}
\end{array} \right.
\label{eq:PWCfunctions}
\end{equation}
where the index $m=1,2,..,N_{max}$. For the wire antenna of Fig. \ref{fig:cildraadant}, we get $z_0=-l$ and $\di z_{N_{max}+1}=l$. Note that the PWC basis functions do not explicitly satisfy the physical requirement that the current at both ends of the wire should be equal to zero.
This requirement can be satisfied when using PWL basis functions (see Fig. \ref{fig:PWLmodes}).
For that reason, PWL basis functions will provide a more accurate result for a given maximum of sub-domains.
By substituting (\ref{eq:PWCfunctions}) for PWC functions in the expression for $Z_{mn}$ (\ref{eq:Zdraadant1}), we obtain \begin{equation}
\begin{array}{ll}
\displaystyle Z_{mn} & \di = -\jmath \omega \mu_0
\int\limits_{z_{m-1}}^{z_{m+1}} \int\limits_{z_{n-1}}^{z_{n+1}}
\left(1+ \frac{1}{k_0^2} \frac{ \partial^2}{\partial z^2} \right)
 \frac{\di e^{-\jmath k_0 R_a}}{4 \pi R_a}  dz_0 \ dz
\end{array}
\label{eq:Zdraadant2}
\end{equation}
The double derivative w.r.t. $z$ in (\ref{eq:Zdraadant2}) can easily be obtained analytically, see  \cite{Kraus}, \cite{Balanis}. The resulting double integral then needs to be determined numerically, for example by using standard integration routines available in MATLAB.
The last step in the calculation of the mode coefficients with (\ref{eq:mateq}) is the calculation of the excitation vector $[V^{ex}]$. The excitation field in expression (\ref{eq:zenv})
can be an incident plane wave or a locally generated field. In this example, we will assume that a voltage generator is connected to the input port of the wire antenna, located at the center of the wire.  The voltage generator can be approximated by a so-called  {\it delta-gap generator}, which generates locally a very strong electric field directed along the $z$-axis.
This excitation field can be expressed as follows:
\begin{equation}
\di \vec{E}^{ex} = V_g \delta (z-z_g) \vec{u}_z,
\end{equation}
where $V_g$ represent the generator voltage and where $z_g$ is the position of the delta-gap generator. When we connect the delta-gap
generator at the center of mode $m=i$, with
$z_g=z_i$, the excitation vector $[V^{ex}]$ takes the following form:
\begin{equation}
 [V^{ex}] =
\left(
\begin{array}{l}
 \ V_1^{ex} \\
 \  V_2^{ex} \\
   .......  \\
 \  V_i^{ex} \\
   .......  \\
 \  V_{N_{max}}^{ex}
\end{array} \right)
 =
\left(
\begin{array}{l}
 \ 0 \\
 \  0 \\
   .......  \\
 \  V_g \\
   .......  \\
 \  0
\end{array} \right).
\label{eq:Vexdeltagap}
\end{equation}
As a result, only element $i$ of the vector $[V^{ex}]$ is nonzero.
We have now determined all elements of $[Z]$ and $[V^{ex}]$ in
matrix equation (\ref{eq:mateq}). By solving this equation (numerically), we can determine the unknown mode coefficients and related current distribution along the wire antenna.
The far-field pattern is then determined by using the expressions introduced in section \ref{sec:radiatedfields}.
The input impedance is easily found as the relation between the input voltage and input current:
\begin{equation}
\di Z_{in} = \frac{V_g}{I_i},
\end{equation}
where $I_i$ is the mode coefficient of mode $i$.


\chapter{Phased Arrays and Smart Antennas}
\label{chap:Phased-Arrays}

\begin{figure}[hbt]
  \centerline{\epsfig{figure=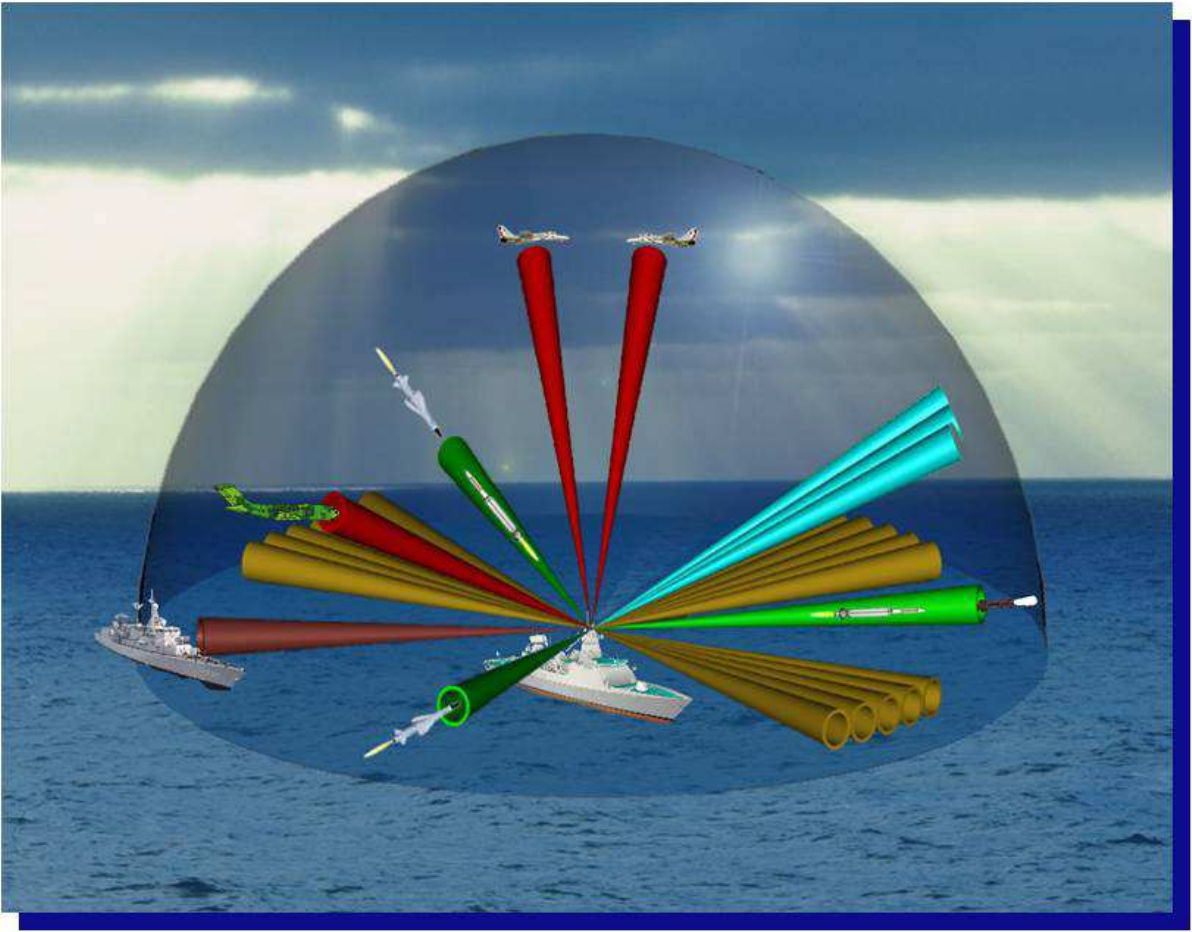,height=7.0cm,clip=}}
  \caption{\it Example of a phased-array radar operating at X-band frequencies, courtesy: Thales Netherlands.}
  \label{fig:apar}
  \end{figure}

\begin{figure}[hbt]
  \centerline{\epsfig{figure=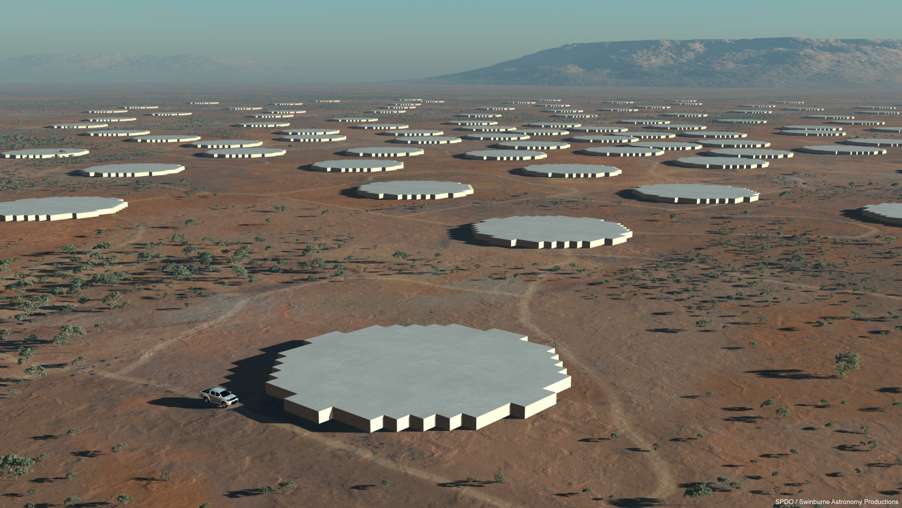,height=7.0cm,clip=}}
  \caption{\it Example of a large phased-array antenna used in radio astronomy, \cite{SKA}.}
  \label{fig:ska}
  \end{figure}

Phased-array antennas with electronic beamsteering capabilities have been developed in the period between 1970-1990 mainly for military radar applications to replace mechanically rotating antenna systems.
Reflector antennas have a high antenna gain, but have the disadvantage
that the main lobe of the antenna has to be steered in the desired direction by means
of a highly accurate mechanical steering mechanism. This means that
simultaneous communication with several points in space is not possible.
Small antennas, like a dipole or microstrip antenna have a close to omni-directional radiation pattern, but have a very low antenna gain. Therefore, small antennas cannot be used for large-distance communication or sensing.
By combining a large number of small antennas, we can create an array of antennas. When each of these individual antenna elements is provided with an adjustable amplitude and phase, we obtain a phased-array antenna.
In this way, a phased-array antenna can communicate with several targets which may be
anywhere in space, simultaneously and continuously,
because the main beam of the antenna can be directed electronically into
a certain direction. Another advantage of phased-array antennas is the
fact that they are usually relatively flat.
Examples of modern phased array systems are shown in Fig. \ref{fig:apar} and Fig. \ref{fig:ska}  used in radar and radio astronomy, respectively.
Phased arrays are also very suited for future millimeter-wave 5G and beyond 5G wireless communication systems, where a high antenna gain is required in combination with electronic beamsteering.

In this chapter, we will first investigate the general properties of phased arrays, starting with linear arrays of isotropic radiators.
Then, we will extend the theory to two-dimensional arrays.
As a next step, we will introduce real antenna elements, like dipoles and microstrip antennas \cite{Adela2018}. We will show that mutual coupling between the elements can limit the performance of the array. Finally, we will investigate the effect of errors, noise and present a method to calibrate the array to compensate for errors.
In this chapter we will again assume to have time-harmonic signals and waves.

\section{Linear phased arrays of isotropic antennas}

\subsection{Array factor}
\begin{figure}[hbt]
  \centerline{\epsfig{figure=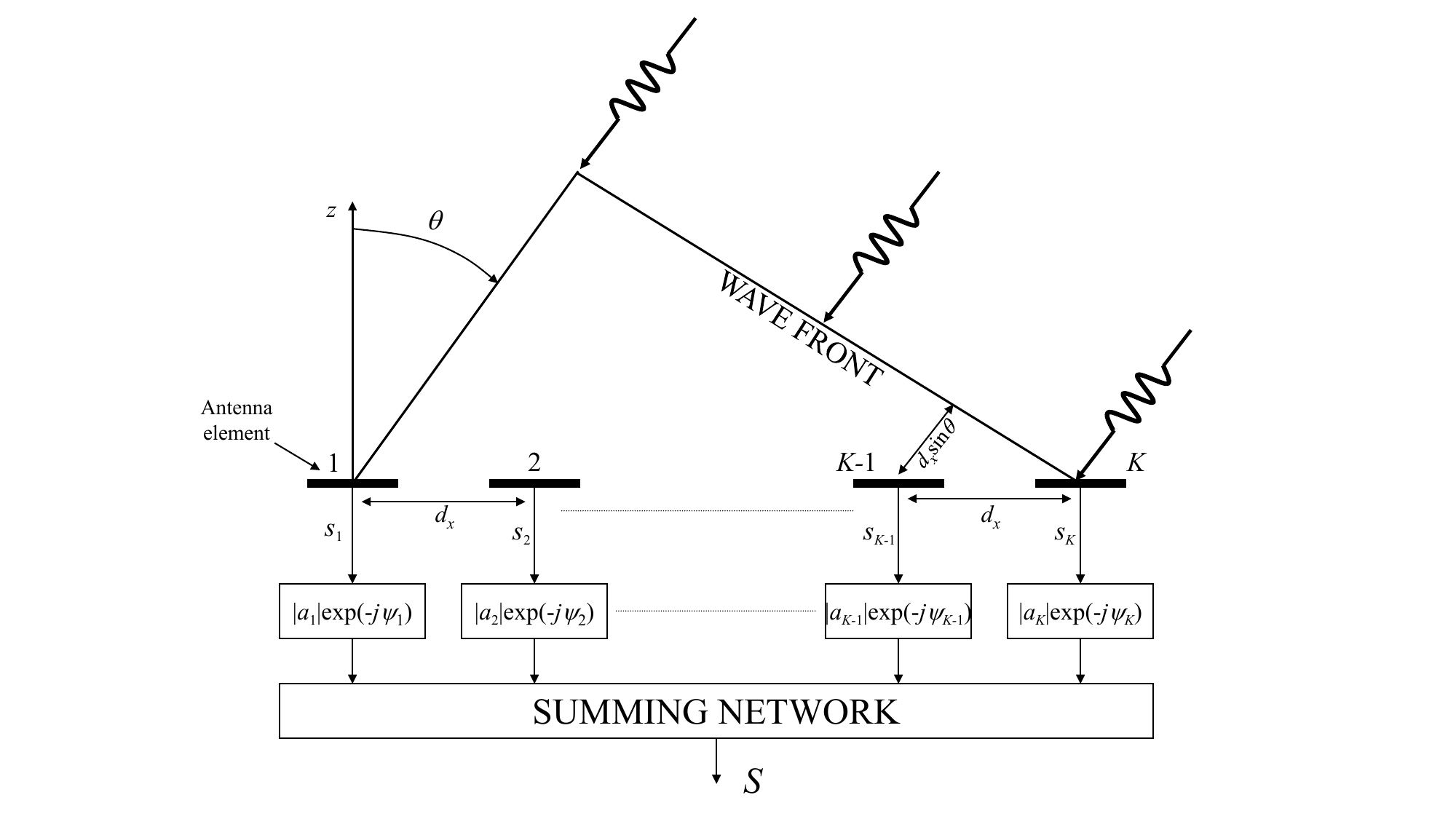,width=160mm}}
  \caption{\it Linear phased-array antenna consisting of $K$ identical antenna elements, separated by a distance $d_x$.
  The signals received by each of the antenna elements are multiplied by a complex coefficient with amplitude $|a_{k}|$ and phase $-\psi_{k}$.}
  \label{fig:lineairarray}
  \end{figure}

We will investigate the basic principle of operation of linear arrays by considering the array in receive mode, as illustrated in Fig. \ref{fig:lineairarray}.
Due to reciprocity, the behaviour of the array antenna in transmit mode is exactly the same. However, in practical active phased-array systems, the functionality of the transmit electronics may differ from the receive electronics.
The linear array of Fig. \ref{fig:lineairarray} consists of $K$ identical antenna elements with
$k=1....K$, separated by an element spacing of $d_x$. Element
$k=1$ is located in the origin of the coordinate system at $(x,y,z)=(0,0,0)$.
In this section, we will assume that all antenna elements are isotropic (omni-directional) radiators. Although isotropic radiators cannot exist in reality, they allow us to investigate the basic array properties. Each antenna element $k$ includes a circuit which can perform a complex weighting of the received signal $s_k$, by multiplication with the complex coefficient $\displaystyle
a_k=|a_k|e^{-\jmath \psi_k}$. This complex coefficient consists of an amplitude $|a_k|$ ({\it taper}) and a phase shift $-\psi_k$.
The set of complex coefficients is also often called the {\it array illumination}.
In general the complex weighting will be realized by means of an electronic circuit and can be done both in the analog and digital domain.
Finally, all signals are summed in the summing network.

Now let us investigate the receive properties of a linear array of isotropic antennas in more detail.
We will assume that a time-harmonic electromagnetic plane wave is incident on the array from an angle $\theta_0$ with respect to the $z$-axis.
This plane wave can be assumed to be generated by another (transmit) antenna located far away from the array.
The received signal of array element $k$ can now be written as:
\begin{equation}
\displaystyle s_k = e^{\jmath k_0 (k-1) d_x \sin{\theta_0}},
\label{eq:signaalelementk}
\end{equation}
where we have normalized the received signal $s_1$ of element 1 to be equal to 1.
In addition, $\displaystyle k_0=\frac{2\pi}{\lambda_0}$ is the free-space wavenumber.
The received signals at each array element can be physically interpreted as the voltage along the output port of each antenna element or can represent the (complex) amplitude of the
guided wave along the transmission line which is connected to the antenna element, for example in case of a waveguide or a microstrip transmission line.
Since the incident plane wave illuminates the array with an angle $\theta_0$, a phase difference will occur between the signals received by each of the array elements.
This phase difference could also be interpreted as a time delay, since element $k$ will receive the signal a fraction $\tau$ earlier as compared to array element ($k-1$), where
$\tau$ is given by:
\begin{equation}
\displaystyle \tau = \frac{d_x \sin{\theta_0}}{c},
\label{eq:timedelay}
\end{equation}
where $c=3.10^8$ is the speed of light in free-space. By using the superposition concept of electromagnetic waves we can write the total received signal at the output of the array of Fig. \ref{fig:lineairarray} in the following form:
\begin{equation}
\displaystyle S(\theta_0) = \sum_{k=1}^K |a_k| e^{\jmath [k_0
(k-1) d_x \sin{\theta_0}-\psi_k]}. \label{eq:Arfactorlin1}
\end{equation}
The function $S(\theta_0)$ is called the {\it array factor}. The array
factor is the response of the array of Fig. \ref{fig:lineairarray} on an incident plane wave that hits the array surface under the angle $\theta_0$, expressed in the coefficients
$a_k=|a_k|e^{-\jmath \psi_k}$ with $k=1...K-1$. The array factor is a periodic function of $\sin{\theta_0}$ with a period of $\lambda_0/d_x$.

A similar analysis could be done for an array in transmit mode. Besides some constants and the factor $e^{-\jmath k_0r}/r$, we would obtain a similar expression as (\ref{eq:Arfactorlin1}) for the electromagnetic field far away from the array (far field).

The maximum of the array factor occurs when the applied phase shift $-\psi_k$ by the electronics connected to each of the receiving antenna elements exactly compensates the phase of each of the received signals. This occurs when:
\begin{equation}
\displaystyle \psi_k = k_0 (k-1) d_x \sin{\theta_0}.
\label{eq:faselineairarray}
\end{equation}
By changing the set of applied phases $\psi_k$, we can manipulate the direction $\theta_0$ for which the array is maximized.
In other words, by changing the set of phases $\psi_k$, we can electronically scan the beam of the array. Because of this, this type of antenna is called a {\it phased array}.

Let us now assume that we use a set of excitation coefficients $a_k$ of which the phase $\psi_k$ is given by (\ref{eq:faselineairarray}).
The resulting array will have a maximum sensitivity when incident plane waves hit the array with an angle of $\theta=\theta_0$ w.r.t. to the $z$-axis. Let us now introduce the variables
$u$ and $u_0$ according to:
\begin{equation}
\begin{array}{l}
\displaystyle u= \sin{\theta} \\ \displaystyle u_0 =
\sin{\theta_0}
\\
\end{array}
\end{equation}
If we now assume that a plane wave is incident on the array under an angle
$\theta$, we can re-write the array factor (\ref{eq:Arfactorlin1}) by means of the following expression
\begin{equation}
\begin{array}{lcl}
\displaystyle S(u) & = & \displaystyle \sum_{k=1}^K |a_k|
e^{\jmath [k_0 (k-1) d_x (\sin{\theta}-\sin{\theta_0})]} \\ & = &
\displaystyle \sum_{k=1}^K |a_k| e^{\jmath [k_0 (k-1) d_x
(u-u_0)]}.
 \end{array} \label{eq:Arfactorlin2}
\end{equation}

\subsection{Uniform amplitude tapering}
\label{subsec:unitaper}
In case of a uniform amplitude weighting ({\it tapering}), where $|a_k|=1$
for $k=1...K$, we can express the summation of (\ref{eq:Arfactorlin2}) in terms of an analytical expression.
By introducing $\beta=k_0 d_x (u-u_0)$, we can re-write the array factor:
\begin{equation}
\displaystyle S(\beta)  =  \displaystyle \sum_{k=1}^K e^{\jmath
(k-1) \beta}. \label{eq:Arfactorlin3}
\end{equation}
By multiplying both left- and right-side of (\ref{eq:Arfactorlin3}) with $\displaystyle e^{\jmath \beta}$ we obtain:
\begin{equation}
\displaystyle S(\beta)e^{\jmath \beta}  =  e^{\jmath \beta}
\sum_{k=1}^K  e^{\jmath (k-1) \beta} = \sum_{k=1}^K e^{\jmath k
\beta}. \label{eq:Arfactorlin4}
\end{equation}
By subtracting the relation (\ref{eq:Arfactorlin3}) from (\ref{eq:Arfactorlin4}), we get:
\begin{equation}
\displaystyle S(\beta)\left( e^{\jmath \beta}-1 \right)  =
 e^{\jmath K \beta} -1. \label{eq:Arfactorlin5}
\end{equation}
In other words:
\begin{equation}
\begin{array}{lcl}
\displaystyle S(\beta) & = & \displaystyle  \left[ \frac{e^{\jmath
K \beta}-1}{ e^{\jmath \beta}-1} \right] \\ & = & \displaystyle
e^{\jmath (K-1) \beta/2} \left[ \frac{e^{\jmath K
\beta/2}-e^{-\jmath K \beta/2}}{e^{\jmath \beta/2}-e^{-\jmath
\beta/2}} \right] \\ & = & \displaystyle e^{\jmath (K-1) \beta/2}
\left[ \frac{\sin(K \beta/2)}{\sin(\beta/2)} \right].
\end{array}
\label{eq:Arfactorlin6}
\end{equation}
Expressed in $u$ coordinates, we then finally find the following expression for the array factor of an array with uniform tapering:
\begin{equation}
\begin{array}{lcl}
\displaystyle S(u) & = & \displaystyle e^{\jmath (K-1) k_0 d_x
(u-u_0)/2} \left[ \frac{\sin(K k_0 d_x (u-u_0)/2)}{\sin(k_0 d_x
(u-u_0)/2)} \right].
\end{array}
\label{eq:Arfactorlin7}
\end{equation}
The antenna radiation pattern can by found by normalizing the received power w.r.t. its maximum value, according to the definition introduced in chapter \ref{chap:fundpar}.
When we apply this to our array factor, we obtain the following expression for the normalized (power) radiation pattern of the linear array:
\begin{equation}
\begin{array}{lcl}
\displaystyle \frac{|S(u)|^2}{|S(u)|^2_{max}} & = & \displaystyle
\left[ \frac{\sin(K k_0 d_x (u-u_0)/2)}{K \sin(k_0 d_x (u-u_0)/2)}
\right]^2.
\end{array}
\label{eq:Arfactorlin8}
\end{equation}
For small values of the term $k_0 d_x (u-u_0)/2$, this reduces to
\begin{equation}
\begin{array}{lcl}
\displaystyle \frac{|S(u)|^2}{|S(u)|^2_{max}} & \approx &
\displaystyle  \left[ \frac{\sin(K k_0 d_x (u-u_0)/2)}{K k_0 d_x
(u-u_0)/2} \right]^2.
\end{array}
\label{eq:Arfactorlin9}
\end{equation}
Expression (\ref{eq:Arfactorlin9}) is exactly equal to the radiation pattern of a uniformly illuminated continuous line source of length
$L=Kd_x$. This can easily be verified by comparing (\ref{eq:Arfactorlin9}) with expression (\ref{eq:Euniapertuur2}), which resembles the radiation
pattern of a uniform aperture.

\begin{figure}[hbt]
\centerline{\psfig{figure=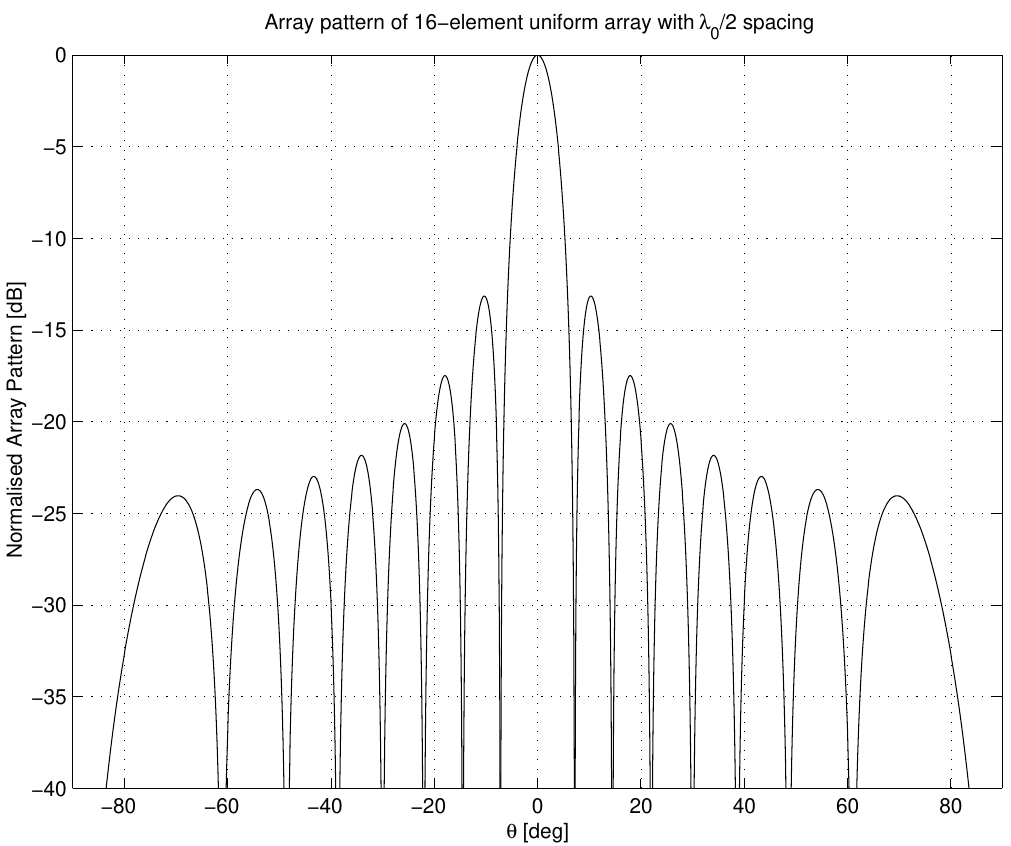,width=90mm}}

\caption{\it Normalized radiation pattern according to (\ref{eq:Arfactorlin8}) of a uniformly illuminated linear array of 16 isotropic antennas.
The main beam is located at broadside ($\theta=0^0$). The element spacing is $d_x=\lambda_0/2$. }
\label{fig:linarunitheta0}
\end{figure}

\begin{figure}[hbt]
\centerline{\psfig{figure=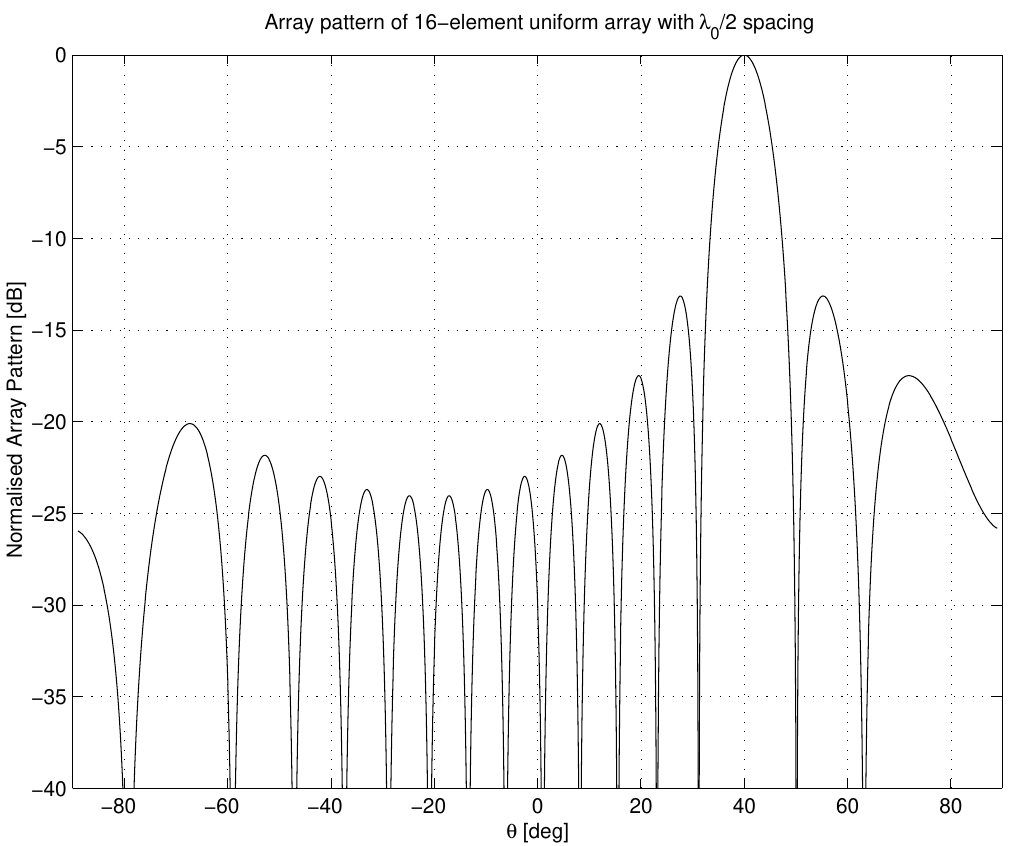,width=90mm}}

\caption{\it Normalized radiation pattern according to (\ref{eq:Arfactorlin8}) of a uniformly illuminated linear array of 16 isotropic antennas.
The main beam is located at an angle of $\theta=40^0$. The element spacing is $d_x=\lambda_0/2$. }
\label{fig:linarunitheta40}
\end{figure}

As an example, let us investigate a linear array with uniform tapering consisting of
$K=16$ isotropic antenna elements, separated by a distance $d_x=\lambda_0/2$.
The normalized radiation pattern can be found by using equation (\ref{eq:Arfactorlin8}).
Fig. \ref{fig:linarunitheta0} shows the radiation pattern in case the main beam of the array is directed towards the angle $\theta_0=0^0$.
Fig. \ref{fig:linarunitheta40} shows the radiation pattern when the main beam is scanned towards
$\theta_0=40^0$.
The first sidelobe of the radiation pattern is located at a level of -13.2 dB below the main lobe level.
The height of the first sidelobe is directly related to our choice of a uniform tapering.
It can be shown that the 3 dB beam width of a linear array with uniform tapering for large values of $K$
equals:
$\displaystyle \theta_{HP} \approx \frac{0.8858 \lambda_0}{K d_x
cos{\theta_0}}$ \cite{Hansen}. With $K=16$, $d_x=\lambda_0/2$ and
$\theta_0=0^0$, we obtain $\theta_{HP} \approx 6.3^0$. The exact 3 dB
beam width in Fig. \ref{fig:linarunitheta0} is
$\theta_{HP} = 6.2^0$.


\subsection{DFT/FFT relation}
When we investigate expression (\ref{eq:Arfactorlin2}) more carefully, we can recognize a Discrete Fourier
Transformatie (DFT). The DFT can be written in the following general form:
\begin{equation}
\begin{array}{lcll}
\displaystyle F(n) & = & \displaystyle \sum_{k=1}^K f(k)
e^{\frac{\jmath 2 \pi (k-1)(n-1)}{K}} &  \displaystyle 1 \leq n
\leq K.
 \end{array} \label{eq:DFTdef}
\end{equation}
We can rewrite the array factor (\ref{eq:Arfactorlin2}) in a similar form as (\ref{eq:DFTdef}) by choosing the sample points in the $u$
space according to:
\begin{equation}
\displaystyle u-u_0 = \frac{\lambda_0 (n-1)}{d_x K}.
\end{equation}
For the specific case that the number of array elements $K$ is a power of 2, the array factor simplifies in to a Fast Fourier Transform (FFT).
By using FFT processing, we can efficiently compute the radiation pattern of very large phased arrays.
FFT processing is also often used in the digital back-end of phased array antennas with digital beamforming.

\subsection{Grating lobes}
Previously, we have already shown that the main beam of a linear array can be scanned to any angle $\theta_0$ by applying a linear phase distribution along the array
according to (\ref{eq:faselineairarray}).
The array factor as defined in (\ref{eq:Arfactorlin2}) is a periodical function of the variable $u-u_0$.
In other words: besides the main beam, the array factor can have more maxima in the visible space, the so-called {\it grating lobes}.
The existence and exact location of the grating lobes w.r.t. the main beam depends on the element spacing $d_x$.
In most practical systems, $d_x$ will be chosen in such as way as to avoid the existence of grating lobes in the visible space,
where the visible space is defined by  $-1 \leq u=\sin\theta \leq 1$.
A grating lobe will appear in the direction $u_m$ when the argument of the exponent in expression
(\ref{eq:Arfactorlin2}) can be written as an integer number of $2\pi$
\begin{equation}
\displaystyle k_0 (k-1) d_x (u_m-u_0) = 2\pi m,
\end{equation}
where $m$ is an integer number and where $\displaystyle
k_0=\frac{2\pi}{\lambda_0}$. For a given scan angle of the main beam $u_0$, a grating lobe will appear in the visible space when $|u_m=1|$ or $\theta_m=\pm 90^0$.
The relation between the element spacing and the maximum scan angle $\theta_0^{max}$ of the main beam of the linear array without the appearance of a grating lobe is given by:
\begin{equation}
\displaystyle \frac{d_x}{\lambda_0} \leq
\frac{1}{1+|\sin{\theta_0^{max}}|}. \label{eq:gratinglinar}
\end{equation}
A lot of applications require scanning over the entire visible space.
By substituting $\theta_0^{max}=90^0$ in expression(\ref{eq:gratinglinar}), we can observe that grating lobes are avoided when the
element spacing $d_x \leq
\lambda_0/2$. This corresponds to the well-known Nyquist sampling
criterium used in analog-to-digital convertors (ADC).

\subsection{Amplitude tapering and pattern synthesis}
\begin{figure}[hbt]
\centerline{\psfig{figure=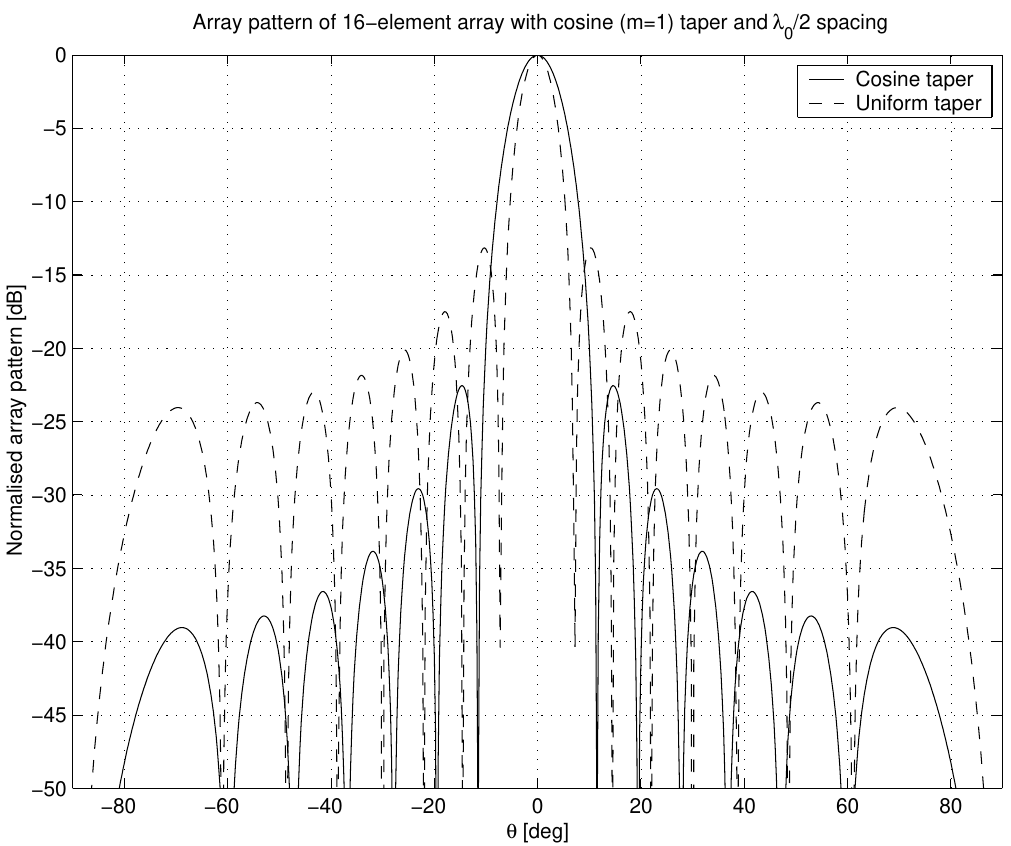,width=90mm}}

\caption{\it Normalized radiation pattern of a linear array with a cosine taper $(m=1)$ with $h=0$ and $K=16$ array elements.
The highest sidelobe is found at a level of -23 dB. The main beam is directed towards broadside $\theta=0^0$.
The element distance $d_x=\lambda_0/2$. The corresponding radiation pattern of an array with uniform tapering is also shown. }
\label{fig:linarcostaper1}
\end{figure}

\begin{figure}[hbt]
\centerline{\psfig{figure=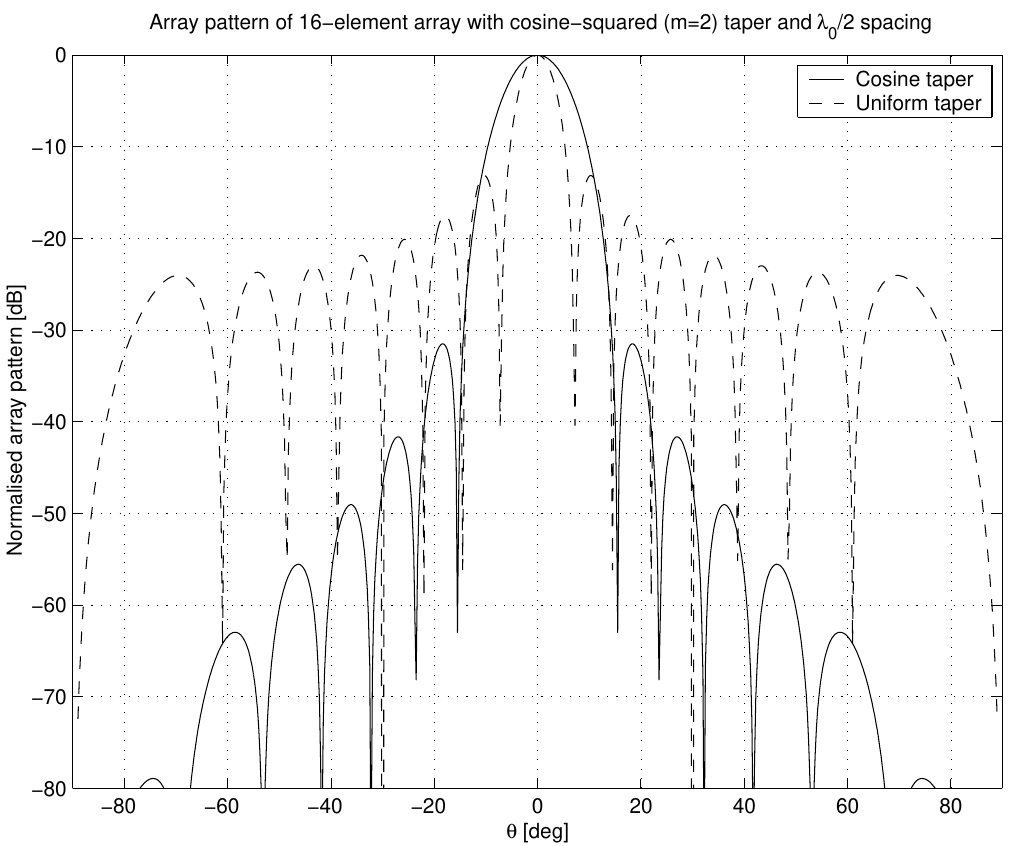,width=90mm}}

\caption{\it Normalized radiation pattern of a linear array with a cosine-square taper $(m=2)$ with $h=0$ and $K=16$ array elements.
The highest sidelobe is found at a level of -32 dB. The main beam is directed towards broadside $\theta=0^0$.
The element distance $d_x=\lambda_0/2$. The corresponding radiation pattern of an array with uniform tapering is also shown. }
\label{fig:linarcostaper2}
\end{figure}

\begin{figure}[hbt]
\centerline{\psfig{figure=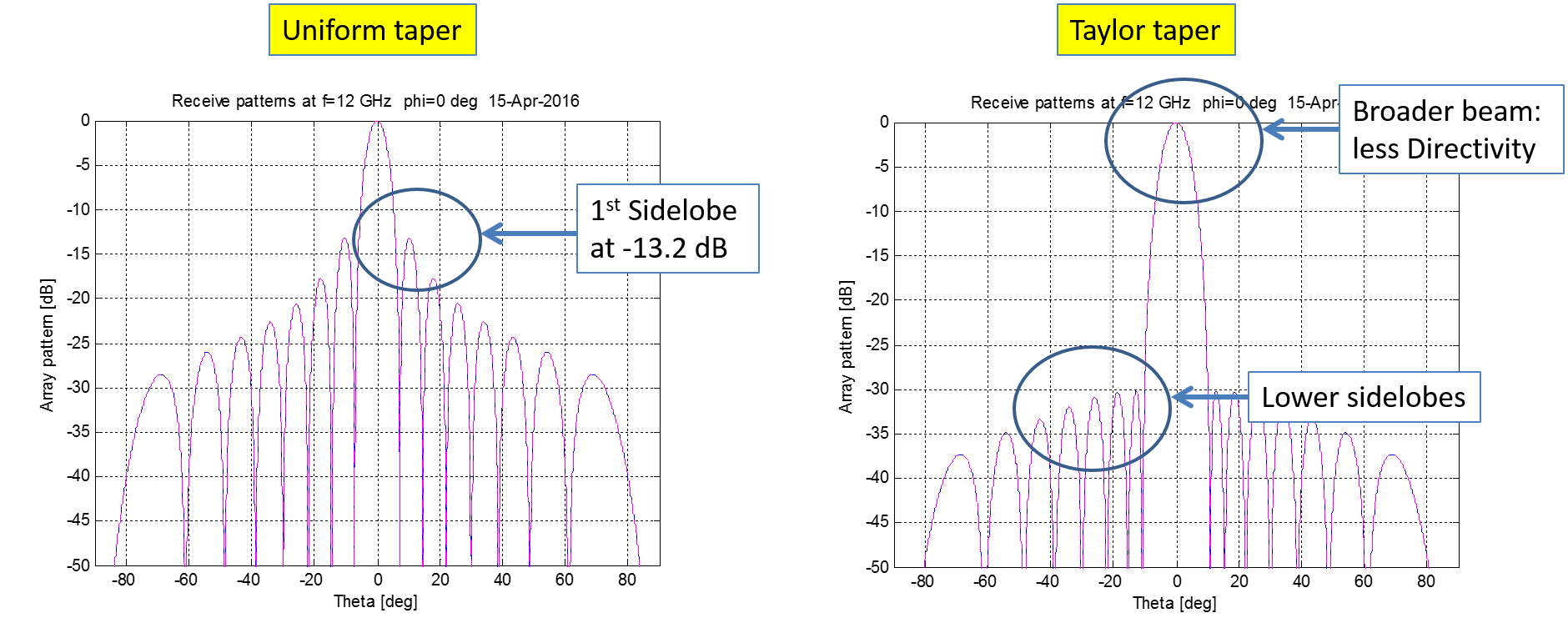,width=160mm}}

\caption{\it Normalized radiation pattern of a 16-element linear array with a Taylor taper with 30 dB peak sidelobe level. The main beam is directed towards broadside $\theta=0^0$.
The element distance $d_x=\lambda_0/2$. The corresponding radiation pattern of an array with uniform tapering is also shown. }
\label{fig:linartaylortaper}
\end{figure}

In section \ref{subsec:unitaper} we have shown that a linear array with uniform amplitude tapering, i.e. $|a_k|=1$
for all $k$, generates quite high sidelobes in the radiation pattern, corresponding to the behavior of the $sinc$
function. This results in a first sidelobe level of -13.2 dB w.r.t. the main beam power level. We can reduce the sidelobe level
of phased arrays considerably by using an amplitude weighting along the array elements.
The weighting function or amplitude taper needs to be chosen in such a way that contributions of the edge elements to the total array factor are reduced.
In other words, the illumination of the edges is reduced. One of the well-known taper functions is the cosine-power taper given by:
\begin{equation}
\displaystyle |a_k| = h + (1-h) \cos^m\left( \frac{\pi x_k}{(K-1)d_x}
\right), \ \ \ |x_k| \leq \frac{Kd_x}{2},
\end{equation}
in case of an array along the $x$-axis, where $x_k$ represents the $x$-coordinate
of array element $k$. Note that we have assumed in this case that the center of the array is located at $x=0$. The parameter $h$ allows us to use an offset (also known as {\it pedestal}).
When $h=1$, we again obtain a uniform taper. Commonly used taper functions are the cosine taper ($m=1$) and the cosinus-square taper ($m=2$).
The corresponding normalized radiation patterns are shown in Fig. \ref{fig:linarcostaper1} and Fig.
\ref{fig:linarcostaper2} in case of an array with $K=16$ elements and element spacing $d_x=\lambda_0/2$.
The cosine taper results in a peak sidelobe level of approx. -23 dB and the cosine-square taper results in a peak sidelobe level of -32 dB.
This is a significant improvement as compared to a uniform taper.
Other well-known taper functions are the Taylor taper and the Dolph-Tchebycheff taper
\cite{Hansen}. The Taylor tapering results in a monotonic decrease of the sidelobe level away from the main beam, whereas the Dolph-Tchebycheff taper creates a constant sidelobe level.
An example of the Taylor taper is shown in Fig. \ref{fig:linartaylortaper} for a 16-element linear array with a spacing of $d_x=\lambda_0/2$. The Taylor taper provides low sidelobes with a peak sidelobe level of 30 dB in this case. As a side effect, the beam width of the main beam increases, and as a result, the directivity will be lower as compared to an array with uniform tapering, see also next section.

\begin{figure}[hbt]
\vspace{-1cm}
\centerline{\psfig{figure=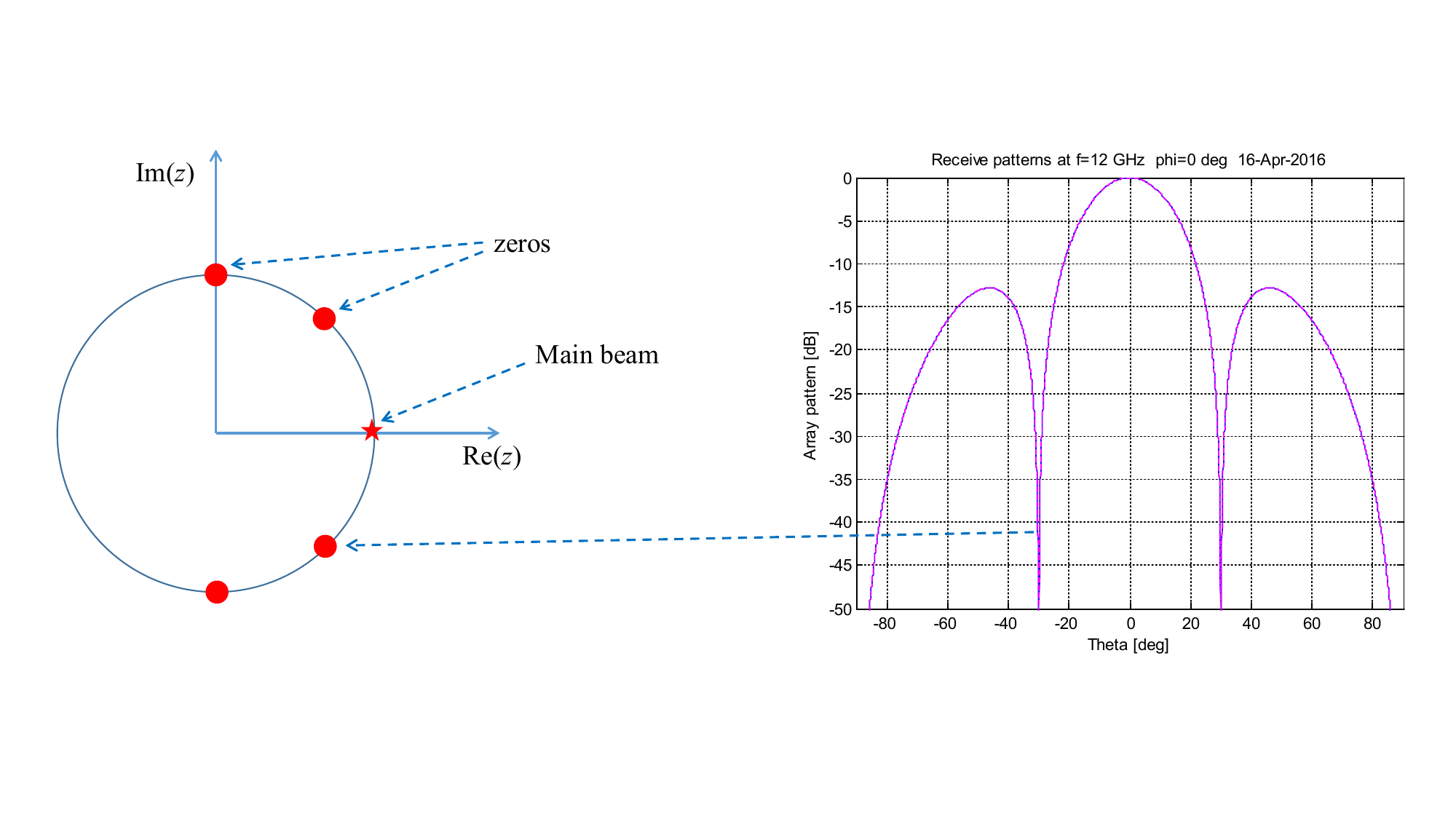,width=170mm}}
\vspace{-1.75cm}
\caption{\it Zeros of the Schelkunov polynomial plotted in the complex $z$-plane and relation to the nulls in the radiation pattern. Linear array with uniform illumination ($|a_k|=1$) and with $K=8$ elements spaced $\di d=\lambda_0/4$. The main beam is directed towards broadside ($\theta_0=0$).}
\label{fig:Synthesiscircle1}
\end{figure}

\begin{figure}[hbt]
\centerline{\psfig{figure=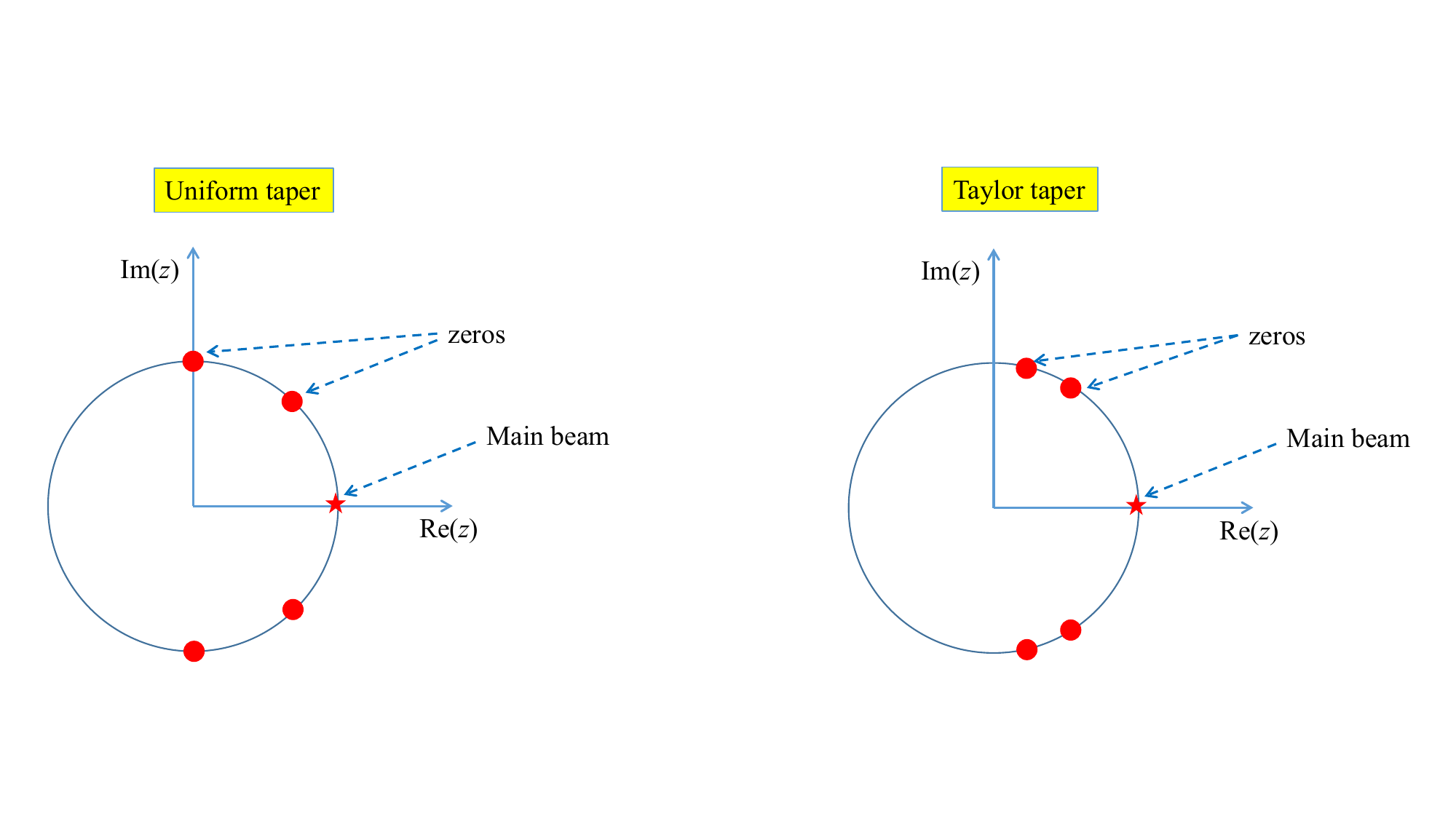,width=160mm}}

\caption{\it Zeros of the Schelkunov polynomial plotted in the complex $z$-plane for a uniform and Taylor taper. Linear array with $K=8$ elements with spacing of $\di d=\lambda_0/4$ and main beam directed towards broadside ($\theta_0=0$).}
\label{fig:Synthesiscircle2}
\end{figure}

\begin{figure}[hbt]
\centerline{\psfig{figure=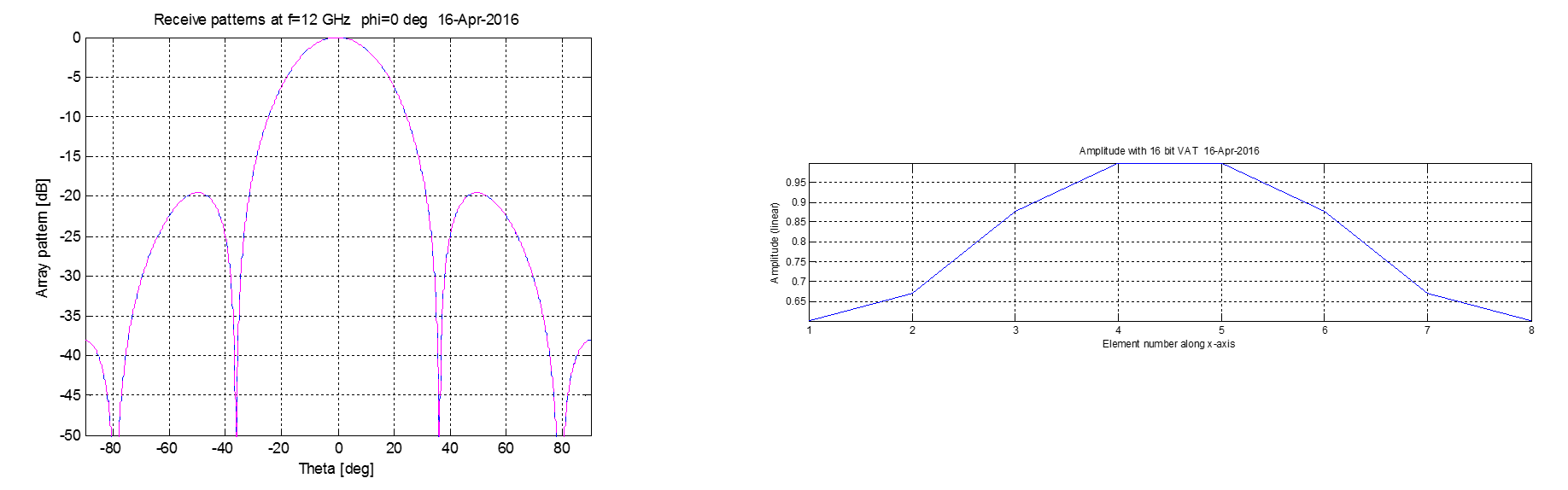,width=170mm}}

\caption{\it Radiation pattern (left) and amplitude taper along the array (right) of a linear array with $K=8$ elements with spacing of $\di d=\lambda_0/4$ and main beam directed towards broadside ($\theta_0=0$).}
\label{fig:Synthesiscircle3}
\end{figure}

One of the main benefits of phased arrays is that the pattern (main beam and sidelobes) can be controlled by using specific sets for the complex weight of each of the array elements.
Several pattern synthesis methods are described in literature that control the main beam and/or sidelobe level. An overview of these methods can be found in \cite{Mailloux}.
The synthesis method introduced by Schelkunov \cite{Schelkunov} provides some physical background in sidelobe control of pencil-beam patterns.
The array factor of a linear array (\ref{eq:Arfactorlin2}) can also be written as a polynomial of a new variable $z$, resulting in the so-called Schelkunov polynomial:
\begin{equation}
\begin{array}{lcl}
\displaystyle S(z) & = & \displaystyle a_1+a_2 z + a_3 z^2 + .... + a_K a^{K-1},
 \end{array} \label{eq:Schelkunov1}
\end{equation}
where $\di z=e^{\jmath k_0 d_x \sin\theta}$ and $\di a_k=|a_k|e^{-\jmath k_0 d_x \sin\theta_0}$.
Note that $|z|=1$, so located on a unit-circle in the complex $z$ plane.
The Schelkunov polynomial (\ref{eq:Schelkunov1}) can also be written in terms of a product of $K-1$ factors:
\begin{equation}
\begin{array}{lcl}
\displaystyle S(z) & = & \displaystyle a_K (z-z_1)(z-z_2) ..... (z-z_{K-1}),
 \end{array} \label{eq:Schelkunov2}
\end{equation}
where $z_1$, $z_2$, .... $z_{K-1}$ represent the zeros of the polynomial. A zero in the polynomial corresponds with a null-location in the radiation pattern of the linear array.
As an example, we will consider a linear array with $K=8$ elements with spacing of $\di d=\lambda_0/4$ and main beam directed towards broadside ($\theta_0=0$).
Let us first consider a uniform amplitude tapering. Fig. \ref{fig:Synthesiscircle1} shows the locations of the zeros in the Schelkunov polynomial plotted in the complex $z$-plane, together with the radiation pattern. The null at $\theta=-30^0$ in Fig. \ref{fig:Synthesiscircle1} corresponds with the following location in the complex $z$-plane:
\begin{equation}
\begin{array}{lcl}
\displaystyle z & = & \displaystyle e^{\jmath k_0 d_x \sin\theta} = e^{\jmath \frac{2 \pi}{\lambda_0} \frac{\lambda_0}{4} \sin\theta} \\
 & = & \di e^{\jmath \frac{\pi}{2} \sin\theta} = e^{\jmath \frac{\pi}{4}}
 \end{array} \label{eq:Schelkunov3}
\end{equation}
The separation between the nulls of a uniformly illuminated array in Fig. \ref{fig:Synthesiscircle1} is equidistant. If we want to create low sidelobes, we should try to put the nulls closer together and move the first nulls away from the main-lobe.
This is done when using a Taylor of Dolph-Chebychev tapering. This is illustrated in Fig. \ref{fig:Synthesiscircle2}, where the null location of a Taylor taper with -20 dB peak sidelobe is compared to the uniformly illuminated array. The corresponding radiation pattern and taper function along the array is provided in Fig. \ref{fig:Synthesiscircle3}. By comparing the location of the nulls with Fig. \ref{fig:Synthesiscircle3}, it can be observed that the nulls have indeed shifted away from the main beam and are now located closer together. In addition, the beam width of the main beam has broadened, which will result in a reduction of the directivity as we will see in the next section.
Low sidelobes can also be obtained by choosing a specific (often randomized) array grid. A special case is discussed in \cite{Smoldersseqrot2014}, where a sequential rotation technique is used to realize circularly-polarized arrays with low sidelobes.

\subsection{Directivity and taper efficiency}
The directivity of an array directly determines the range of a wireless link or radar system as stated by the radio- and radar equation (see chapter \ref{chap:fundpar}).
The directivity of a linear array of isotropic radiators can be determined by substituting the array factor (\ref{eq:Arfactorlin2}) into equation
(\ref{eq:richtfunctie}). In this way, we can obtain the directivity in the direction $\theta_0$:
\begin{equation}
\begin{array}{lcl}
\displaystyle D(\theta_0) & =  & \displaystyle
\frac{P(\theta_0)}{P_t / 4\pi} \\ & =  & \displaystyle \frac{4\pi
|S(\theta_0)|^2}{\displaystyle \int\limits_{V_{\infty}}
|S(\theta)|^2 d\Omega},
\end{array}
 \label{eq:richtfunctielinar1}
\end{equation}
where the integration is done over the entire visible space.
The integral can be solved by using a new coordinate $\xi$, where
$\xi$ is defined as the opening angle with the positive $y$-axis in the $y-z$-plane.
The differential solid angle is now given by $d\Omega=\cos{\theta}d\theta d\xi$.
In case of a linear array located along the $x$-axis, the array factor will not depend on the $\xi$ coordinate.
Let us assume that the main beam of the array is directed towards broadside, so $\theta_0=0^0$.
By using $u=\sin\theta$ and
$du=\cos\theta d\theta$, we can rewrite (\ref{eq:richtfunctielinar1}) in the following form:
\begin{equation}
\begin{array}{lcl}
\displaystyle D & =  & \displaystyle  \frac{\displaystyle 2 \left(
\sum_{k=1}^K |a_k| \right)^2}{\displaystyle \int\limits_{-1}^{1}
|S(u)|^2 du} \\ & =  & \displaystyle  \frac{\displaystyle 2 \left(
\sum_{k=1}^K |a_k| \right)^2}{\displaystyle \int\limits_{-1}^{1}
\sum_{k=1}^K \sum_{l=1}^K |a_k||a_l| e^{\jmath k_0 d_x u (k-l)}
du} \\
 & =  &
\displaystyle  \frac{\displaystyle 2 \left( \sum_{k=1}^K |a_k|
\right)^2}{\displaystyle \sum_{k=1}^K \sum_{l=1}^K |a_k||a_l|
\int\limits_{-1}^{1} e^{\jmath k_0 d_x u (k-l)} du} \\
 & =  &
\displaystyle  \frac{\displaystyle \left( \sum_{k=1}^K |a_k|
\right)^2}{\displaystyle \sum_{k=1}^K \sum_{l=1}^K |a_k||a_l|
sinc[k_0 d_x (k-l)]}. \\
\end{array}
 \label{eq:richtfunctielinar2}
\end{equation}
When $k_0 d_x = m\pi$ (which implies that $d_x=m\lambda_0/2$), we find that
$sinc[k_0 d_x (k-l)]$ can only be non zero when $k=l$. Therefore, the directivity can be written as:

\begin{equation}
\begin{array}{lcl}
\displaystyle D & =  & \displaystyle  \frac{\displaystyle \left(
\sum_{k=1}^K |a_k| \right)^2}{\displaystyle \sum_{k=1}^K |a_k|^2}.
\\
\end{array}
 \label{eq:richtfunctielinar3}
\end{equation}
As a result, we can observe that the directivity of a uniformly-illuminated linear array with element spacing $d_x=m \lambda_0/2$  is equal to
$D=K$. Consider for example, an array with $K=1000$ (isotropic)
elements. The corresponding directivity is in this case $D=10\log_{10}(1000)=30$ dB.

\begin{figure}[hbt]
\centerline{\psfig{figure=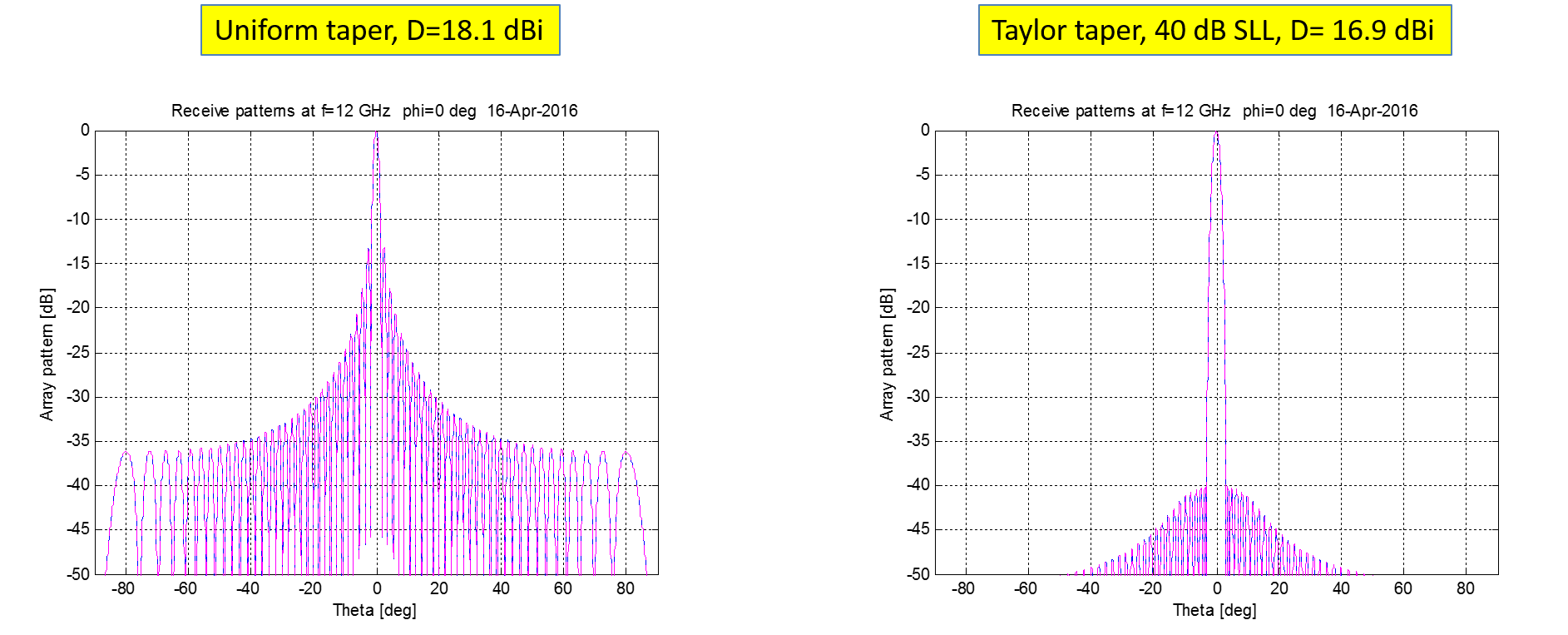,width=170mm}}

\caption{\it Effect of Taylor amplitude tapering on the radiation pattern of a linear array with $K=64$ elements with element spacing of $\di d=\lambda_0/4$ and main beam directed towards broadside ($\theta_0=0$). The taper efficiency $\eta_{app}=0.76$.}
\label{fig:Taylor40dBlineararray}
\end{figure}

When we apply an amplitude tapering with $|a_k| \leq 1$, the directivity will be reduced as compare to the uniform case according to:
\begin{equation}
\begin{array}{lcl}
\displaystyle D & =  & \displaystyle  K \eta_{tap},
\\
\end{array}
 \label{eq:richtfunctielinar4}
\end{equation}
where the taper efficiency $\eta_{tap}$ is given by
\begin{equation}
\begin{array}{lcl}
\displaystyle \eta_{tap} & =  & \displaystyle  \frac{\displaystyle \left(
\sum_{k=1}^K |a_k| \right)^2}{\displaystyle K \sum_{k=1}^K |a_k|^2}.
\\
\end{array}
 \label{eq:taperefficiency_linear}
\end{equation}
An example is shown in Fig. \ref{fig:Taylor40dBlineararray} for a linear array with $K=64$ elements and a spacing of $d_x=\lambda_0 /2$. The figure shows the radiation pattern in case of a Taylor taper with 40 dB peak sidelobe as compared to the uniform case. The Taylor taper provides low sidelobes, but also a directivity reduction of 1.2 dB ($\eta_{app}=0.76$). In other words, the effective antenna aperture has reduced with $24\%$ as compared to the uniformly-illuminated array.

\subsection{Beam broadening due to beam scanning}
The beam width of an antenna has been defined in chapter  \ref{chap:fundpar}.
In this section, we will again consider a uniformly illuminated linear array of isotropic radiators.
The array factor is given by expression (\ref{eq:Arfactorlin8}).
From this expression, we can observe that the array factor depends on the coordinate $u-u_0$, where $u_0$ is a fixed value corresponding to the direction of the main beam (scan angle).
Therefore, we can conclude that the array factor does not depend on the absolute value of $u$.
As a result, the 3 dB beam width ((i.e. the $u$
value for which the amplitude of the array factor is equal to $1/\sqrt{2}$ w.r.t. the maximum value) will also not depend on $u$.
To conclude, the beam width will be constant in the $u$-domain.
However, when we express the array factor in terms of $\theta$, the situation is different since the relation between $u$ and $\theta$ is given by:
\begin{equation}
u=\sin\theta,
\end{equation}
which implies that
\begin{equation}
\displaystyle d\theta=\frac{du}{\cos\theta}.
\end{equation}
As a result we can express a small difference $\Delta \theta$ around $\theta_0$ in the following way:
\begin{equation}
\displaystyle \Delta\theta \approx \frac{\Delta u}{\cos\theta_0}.
\end{equation}
Assume that $\Delta u$ indicates the difference in $u$ corresponding to the 3 dB beam width in the expression of the array factor (\ref{eq:Arfactorlin8}).
The corresponding beam width expressed in $\theta$ will then have a $\frac{1}{\cos\theta_0}$ dependence. As a result, we can write the 3 dB
beam width in the following form:
\begin{equation}
\displaystyle \theta_{HP} = \frac{\theta_0}{\cos\theta_0}.
\end{equation}
As expected, the beam of the array becomes broader when the main beam is scanned towards the horizon ($\theta_0= \pm 90^0$).

\section{Planar phased arrays of isotropic radiators}

\begin{figure}[hbt]
  \centerline{\psfig{figure=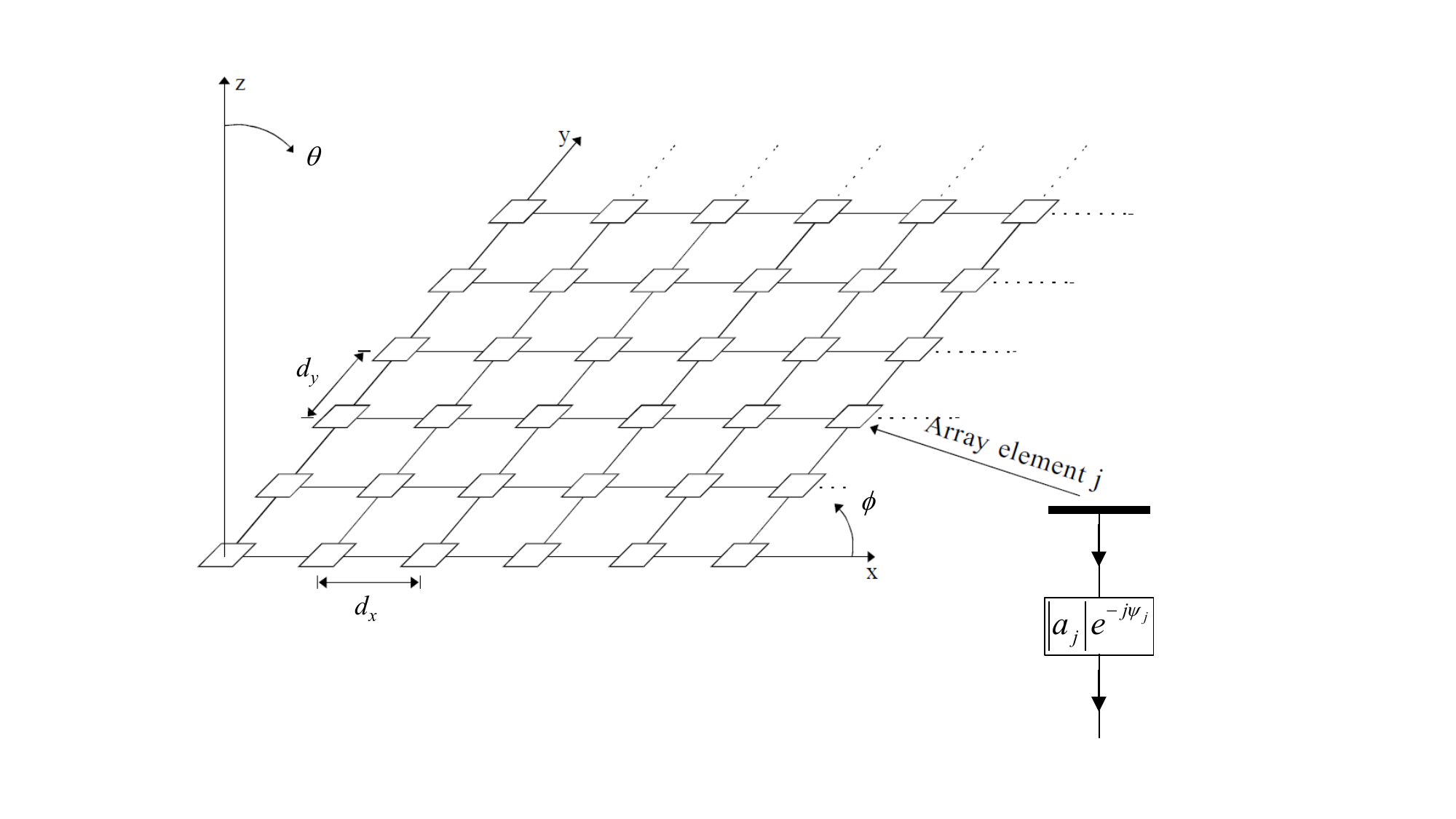,width=170mm}}
  \vspace{-0.5cm}
  \caption{\it Planar phased-array antenna consisting of
  $K \times L$ elements separated by a distance $d_x$ along the
  $x$-axis and $d_y$ along the $y$-axis. The element index is $j=(l-1)K+k$.
  The signals received by each antenna element are fed into an electronic receiver that takes care of the
  amplitude weighting $|a_{j}|$ and phase shifting $-\psi_{j}$ of the received signal.}
  \label{fig:planararray}
\end{figure}

The theory of linear arrays can be easily extended to planar arrays.
Fig. \ref{fig:planararray} shows the geometry of a two-dimensional planar array, in which the elements are
placed on a rectangular grid. Let us again first assume that all array elements are isotropic radiators.
The element distance along the $x$-axis and $y$-axis are given by $d_x$ and $d_y$, respectively.
$j$ denotes the element index according to $j=(l-1)K+k$. The position of the $j$-th element in the array is
denoted by $\vec{r}_j$ according to:
\begin{equation}
\begin{array}{lcl}
\vec{r}_j & = & x_j \vec{u}_x+y_j \vec{u}_y \\
 & = & (k-1)d_x \vec{u}_x + (l-1)d_y \vec{u}_y.
\end{array}
\end{equation}
We will again investigate the array in receive-mode first.
Now let us assume that a plane wave illuminates the array from the direction
$(\theta,\phi)$, denoted by the unit vector $\vec{u}_r$.
The array factor can now be obtained by summing all received contributions from the array elements according to:
\begin{equation}
\displaystyle S(\theta,\phi) = \sum_{k=1}^K \sum_{l=1}^L |a_j|
e^{\jmath [k_0 \vec{u}_r \cdot \vec{r}_j  -\psi_j]},
\label{eq:Arfactorplan1}
\end{equation}
where
\begin{equation}
\displaystyle a_j=|a_j|e^{-\jmath \psi_j} \label{eq:excoef}, \ \ \
\ \ j=(l-1)K+k,
\end{equation}
represents the excitation coefficient of element $j$ with phase $\psi_j$ and amplitude $|a_j|$.
The inner product $(\vec{r}_j \cdot
\vec{u}_r)$ can be found by
\begin{equation}
\begin{array}{lcl}
\displaystyle  \vec{u}_r \cdot \vec{r}_j & = & [(k-1)d_x \vec{u}_x
+ (l-1)d_y \vec{u}_y] \cdot [\sin\theta \cos\phi
\vec{u}_x+\sin\theta \sin\phi \vec{u}_y + \cos\theta \vec{u}_z] \\
& = & (k-1)d_x u + (l-1)d_y v.
\end{array}
\end{equation}
with $u$ and $v$ given by
\begin{equation}
\begin{array}{lcl}
u & = & \sin\theta \cos\phi, \\ v & = & \sin\theta \sin\phi.
\end{array}
\end{equation}
As a result, we can rewrite the array factor
(\ref{eq:Arfactorplan1}) in terms of
($u,v$)-coordinates according to
\begin{equation}
\displaystyle S(u,v) = \sum_{k=1}^K \sum_{l=1}^L |a_j| e^{\jmath
[k_0(k-1)d_x u+k_0(l-1)d_y v-\psi_j]}.
\label{eq:Arfactorplan2}
\end{equation}
When we want to direct the main beam of the array towards $(\theta_0,\phi_0)$, the phase shift $\psi_j$ at each array element should be chosen as:
\begin{equation}
\displaystyle \psi_j=k_0(k-1)d_x u_0 + k_0(l-1)d_y v_0,
\label{eq:phaseplanar}
\end{equation}
with $u_0=\sin\theta_0 \cos\phi_0$ and $v=\sin\theta_0 \sin\phi_0$.
Substituting (\ref{eq:phaseplanar}) in (\ref{eq:Arfactorplan2}), results in the expression for the array factor of a planar array with the main beam directed towards $(u_0,v_0)$:
\begin{equation}
\displaystyle S(u,v) = \sum_{k=1}^K \sum_{l=1}^L |a_j| e^{\jmath
k_0 [(k-1)d_x (u-u_0)+(l-1)d_y (v-v_0)]}. \label{eq:Arfactorplan3}
\end{equation}
Now consider a uniformly illuminated array with $|a_j|=1$, for all
$j$. The array factor can now be expressed in terms of a product of two array factors corresponding to two linear arrays with uniform illumination:
\begin{equation}
\begin{array}{lcl}
\displaystyle S(u,v) & = & \displaystyle \left( \sum_{k=1}^K
e^{\jmath k_0 (k-1)d_x (u-u_0)} \right) \left( \sum_{l=1}^L
e^{\jmath k_0 (l-1)d_y (v-v_0)} \right) \\ & = & \displaystyle
e^{\jmath (K-1) k_0 d_x (u-u_0)/2} e^{\jmath (L-1) k_0 d_y
(v-v_0)/2}  \\ &  & \displaystyle \ \ \ \times \left[ \frac{\sin(K
k_0 d_x (u-u_0)/2)}{\sin(k_0 d_x (u-u_0)/2)} \right] \left[
\frac{\sin(L k_0 d_y (v-v_0)/2)}{\sin(k_0 d_y (v-v_0)/2)} \right].
\end{array}
\label{eq:Arfactorplan4}
\end{equation}

\begin{figure}[hbt]
\centerline{\psfig{figure=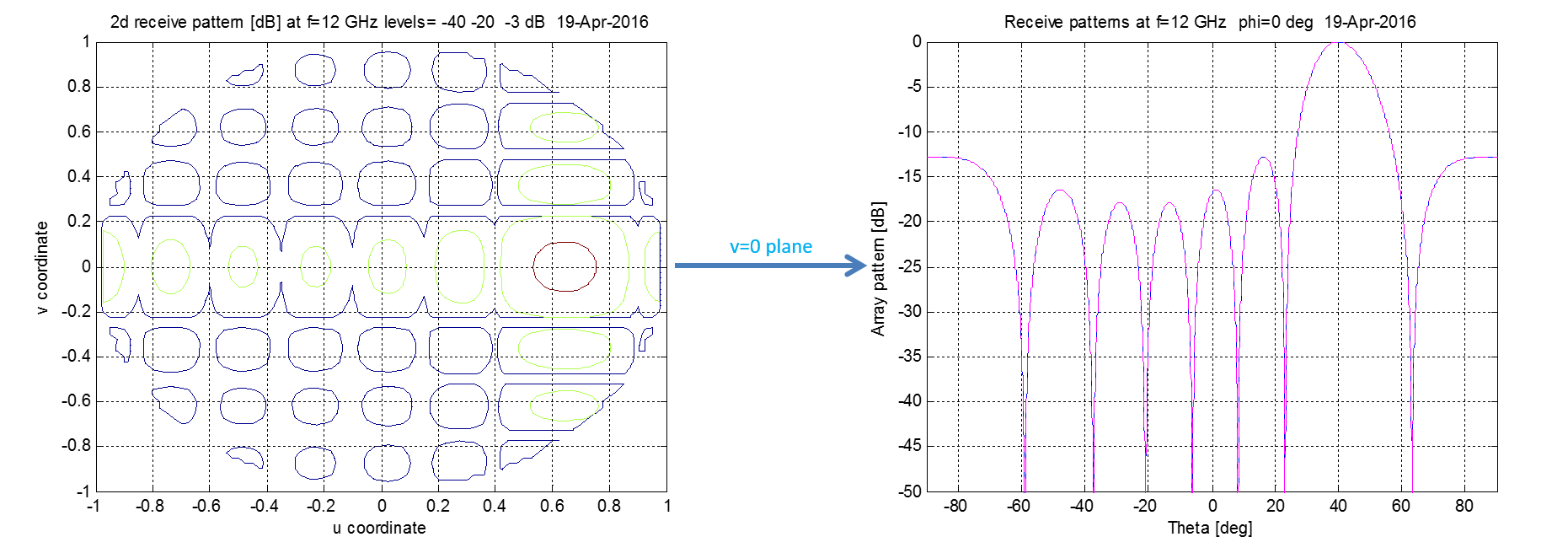,width=170mm}}

\caption{\it Normalized radiation pattern of a uniformly illuminated planar array of isotropic antennas with
$8\times 8=64$ elements. The main beam of the array is directed towards ($\theta_0=40^0,\phi_0=0^0$). The element spacing is
$d_x=d_y=\lambda_0/2$.} \label{fig:planarradpat1}
\end{figure}

Fig. \ref{fig:planarradpat1} shows the normalized radiation pattern of a planar array consisting of 64 isotropic antennas
($K=L=8$). The element spacing is equal to $\lambda_0/2$ in both directions. The main beam is scanned towards ($\theta_0=40^0, \phi_0=0^0$).
We can clearly observe the beam broadening of the main beam and of the first sidelobes.
The peak value of the first sidelobe is again -13.2 dB, corresponding to the $sinc$ function in expression (\ref{eq:Arfactorplan4}).
Lower sidelobe levels can be obtained by applying amplitude tapering, in a similar way as done for linear arrays.

\begin{figure}[hbt]
\centerline{\psfig{figure=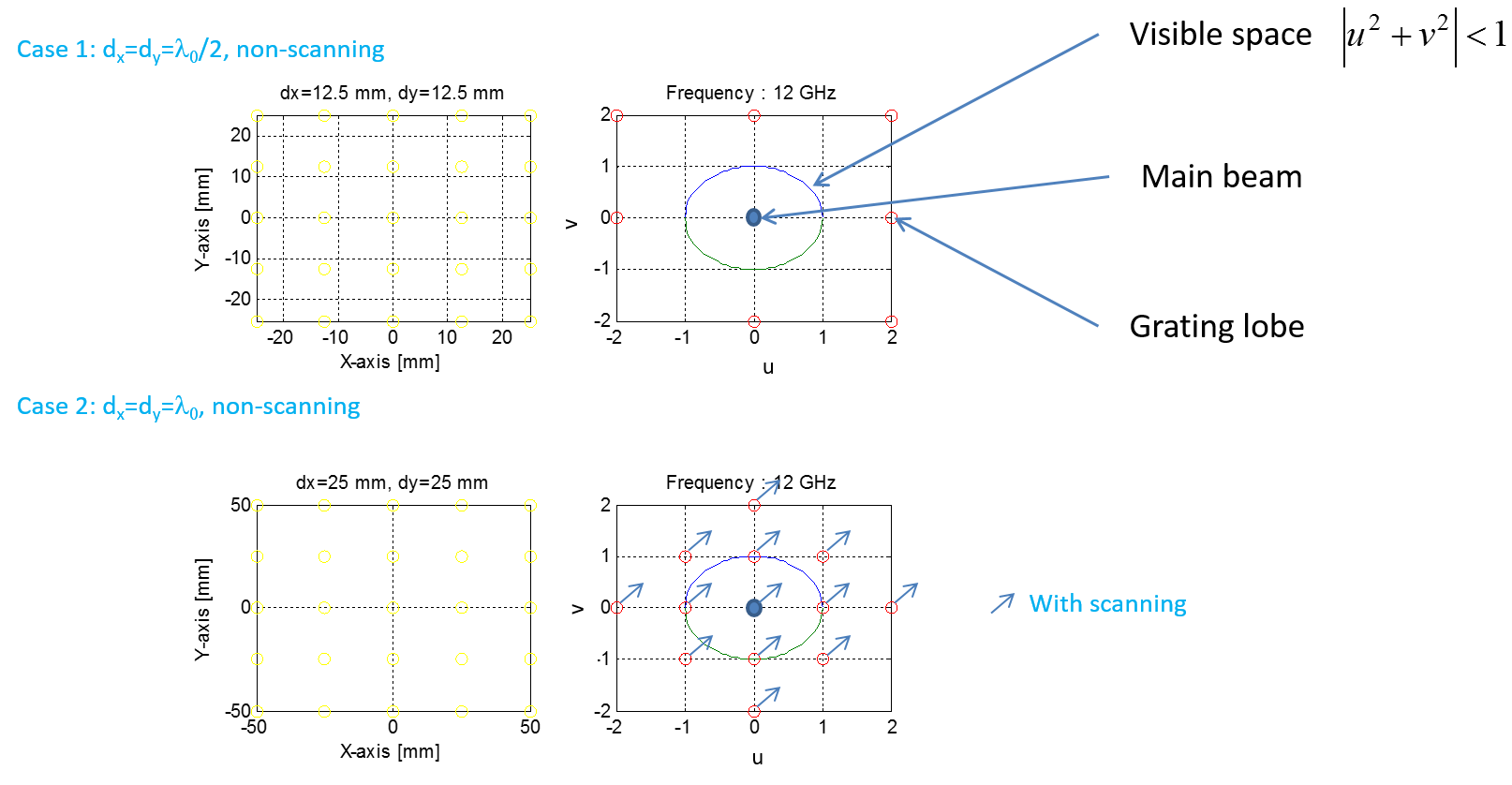,width=160mm}}
\caption{\it Grating lobe pattern in the $(u,v)$-plane of a
planar array placed on a rectangular grid with: {\it case 1}  $\displaystyle
\frac{d_x}{\lambda_0}= \frac{d_y}{\lambda_0}=0.5$ and {\it case 2} $\displaystyle
\frac{d_x}{\lambda_0} =\frac{d_y}{\lambda_0}=1$. The blue dot shows the main beam scanned towards broadside ($u_0=v_0=0$). The entire pattern will shift according to the direction of the arrows in case of scanning.} \label{fig:gratlobe1}
\end{figure}

In our analysis of linear arrays, we have already observed that multiple maxima (grating lobes) can occur in the array factor.
This happens when the distance between the array elements is too large.
Grating lobes in a planar array occur when the argument in the exponent of (\ref{eq:Arfactorplan3}) is a multiple of $2\pi$.
As a result, grating lobes will appear in the directions indicated by ($u_m,v_m$) when the following condition is fulfilled:
\begin{equation}
\displaystyle k_0 \left[ (k-1)d_x(u_m-u_0)+(l-1)d_y(v_m-v_0)
\right] = 2\pi m,
\end{equation}
for integer values of $m$, where $m=0$ corresponds to the main beam.
In case of a rectangular array grid, the grating lobes will occur along a rectangular grid in the $(u,v)$-plane given by:
\begin{equation}
\begin{array}{lcll}
\displaystyle u-u_0 & = & \displaystyle \frac{i}{d_x/\lambda_0} &
\ \ \ i=0,\pm 1, \pm 2, .... \\ \displaystyle v-v_0 & = &
\displaystyle \frac{j}{d_y/\lambda_0} & \ \ \ j=0,\pm 1, \pm 2,
....
\end{array}
\end{equation}
The rectangular grid of grating lobes in the $(u,v)$-plane is illustrated in Fig. \ref{fig:gratlobe1} in case of an array with $\displaystyle
\frac{d_x}{\lambda_0} =0.5$ and $\displaystyle
\frac{d_x}{\lambda_0} =1$ with the main beam directed towards broadside $u_0=v_0=0$.
We can clearly see that all grating lobes appear outside the unit circle. As a result, no grating lobes will appear in the visible space. In case of scanning, the whole pattern will shift according to the arrows as indicated in Fig. \ref{fig:gratlobe1}.
In case 2, the element spacing is larger than $\displaystyle \frac{\lambda_0}{2}$. As a result, grating lobes will appear in the visible space. In most applications grating lobes will deteriorate the performance of the antenna system.

\section{Arrays of non-ideal antenna elements}
\begin{figure}[hbt]
\centerline{\psfig{figure=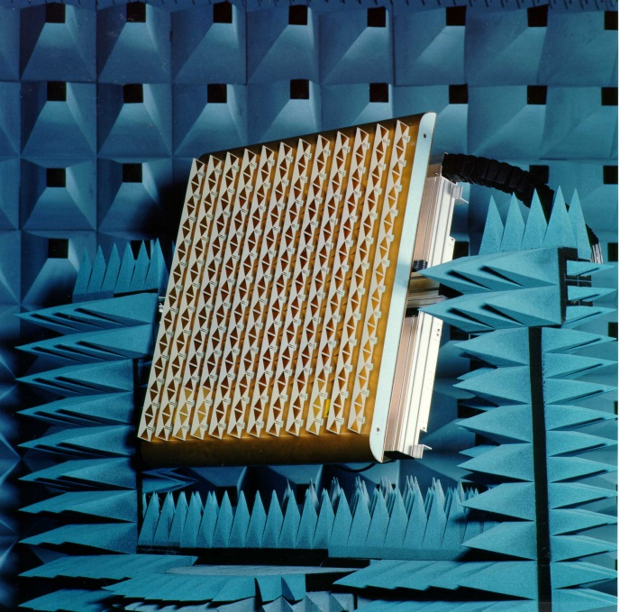,width=80mm}}

\caption{\it The one-square metre array (OSMA), a pathfinder for the square kilometer array (SKA) radio telescope \cite{Smolderscalibration}, \cite{SKA}. OSMA consists of 64 bow-tie elements with a bandwidth of approx. an octave from 1-2 GHz.}
\label{fig:OSMAphoto}
\end{figure}

Up to now, we have assumed that the individual array elements are ideal isotropic radiators with omni-directional radiation characteristics.
As already discussed before, isotropic radiators can physically not exist.
Therefore, we need to introduce physical antenna structures in our array theory.
However, most of the array features that we have already discussed in this chapter will not change significantly by using real antenna elements.
Several types of antennas can be used as a building block in a phased array system, such as dipole antennas (usually with a length of $\lambda_0/2$) and microstrip antennas. These individual antennas should have a broad beam width, else the scan range of the phased array will be very limited.
More details about the radiation characteristics of various types of antenna can be found in other chapters. An example of a realized array for radio-astronomy is shown in Fig \ref{fig:OSMAphoto}, where a 64-element planar array of bow-tie antennas is shown \cite{SKA}.

Now let us assume that the far-field pattern of an isolated array element is given by $\vec{f}(u,v)$, which can physically be interpreted as the electric field generated by an isolated element. The radiation pattern of an individual array element
$\vec{f}(u,v)$ is known as the {\it element factor} or
{\it element pattern}.
Let us for now assume that the radiation characteristics of all array elements are identical. We can then simply apply the superposition principle in order to find the far-field of a planar array of real antenna elements:
\begin{equation}
\displaystyle \vec{S}(u,v) = \sum_{k=1}^K \sum_{l=1}^L |a_j|
\vec{f}_j(u,v) e^{\jmath k_0 [(k-1)d_x (u-u_0)+(l-1)d_y
(v-v_0)]}. \label{eq:Arfactorelpat1}
\end{equation}
Since we have assumed here that all array elements are identical with
$\vec{f}_j=\vec{f}$, we can rewrite (\ref{eq:Arfactorelpat1}) in the following form:
\begin{equation}
\displaystyle \vec{S}(u,v) = \vec{f}(u,v) \sum_{k=1}^K
\sum_{l=1}^L |a_j| e^{\jmath k_0 [(k-1)d_x (u-u_0)+(l-1)d_y
(v-v_0)]}. \label{eq:Arfactorelpat2}
\end{equation}
The normalized radiation pattern can be determined from \ref{eq:Arfactorelpat2} by:
\begin{equation}
\displaystyle {F}(u,v) = \frac{|\vec{S}(u,v)|^2}{|\vec{S}(u_0,v_0)|^2}. \label{eq:Arfactorelpat3}
\end{equation}

\section{Mutual coupling}
In the previous sections we have assumed that mutual electromagnetic interaction between the individual array elements can be neglected.
However, in practise the array elements will be placed on an array grid with a typical element spacing of only $\lambda_0/2$.
Since most antenne elements only operate at resonance, the physical size of the antenna elements will not differ that much from the element spacing.
As a result, the individual antenna element will interact with each other, causing so-called {\it mutual coupling}.
Mutual coupling will affect the element pattern of each array element and will have an influence on the input impedance.
Since each antenna element will observe a (slightly) different electromagnetic environment, the element pattern and input impedance of each element will be different.
A rigorous electromagnectic analysis of mutual coupling in phased-arrays is outside the scope of this book. Numerical electromagnetic methods (e.g. MoM, see chapter \ref{chap:NumEM}) can be used to determine the mutual coupling. In addition, in \cite{Smolders} a method is provided to analyze mutual coupling in finite arrays of microstrip antennas.

\begin{figure}[hbt]
\centerline{\psfig{figure=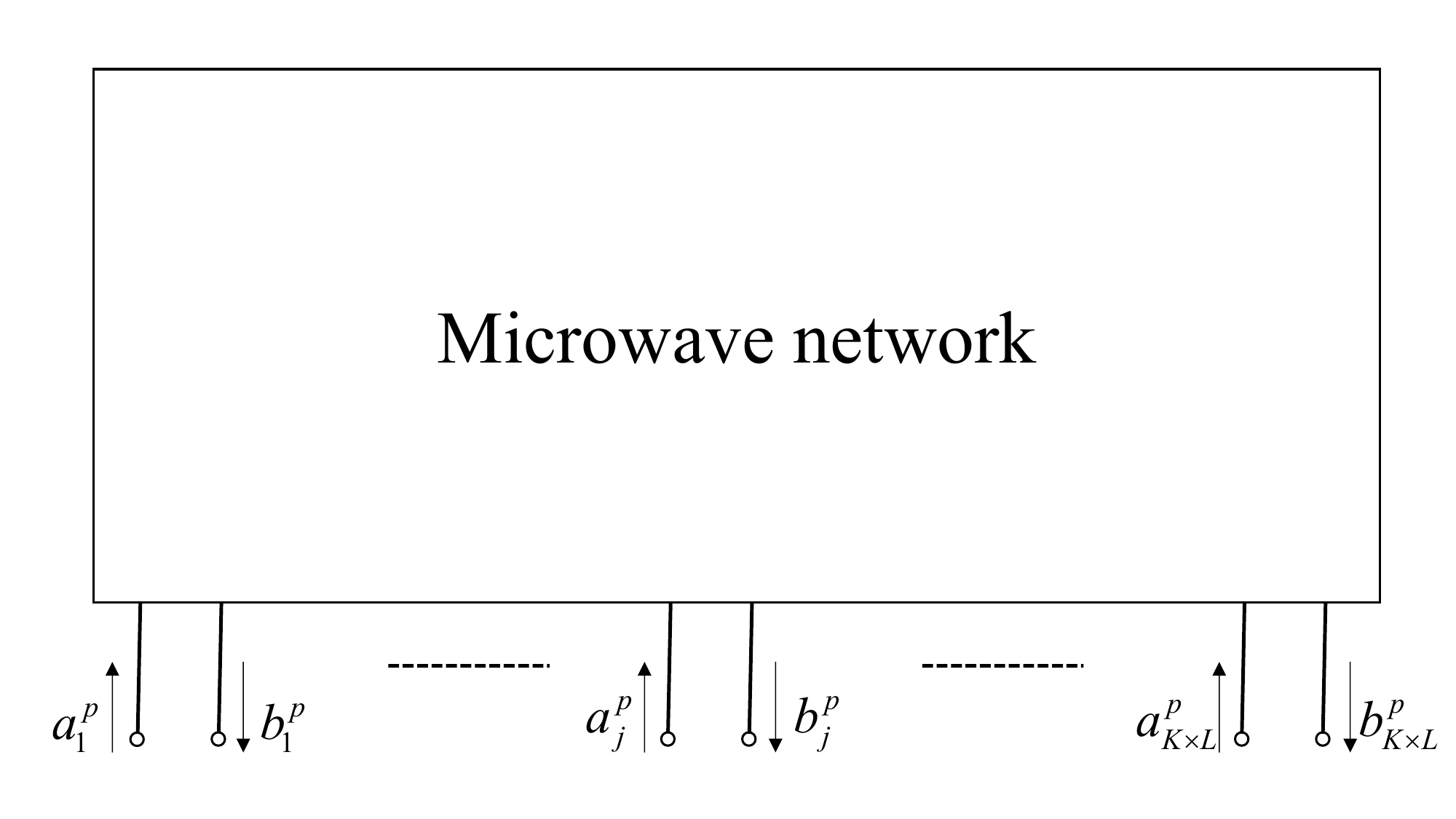,width=120mm}}
\vspace{-0.25cm}
\caption{\it Microwave network representation of a phased-array
antenna consisting of $K \times L$ antenna elements.}
\label{fig:micnetwork}
\end{figure}

In this book we will restrict ourselves to the introduction of the key performance parameters that describe the effect of mutual coupling.
We will use the concept of microwave networks (see chapter \ref{chap:Tlines}) in which we will use propagating waves along transmission lines instead of currents and voltages.
Let us consider the microwave network of Fig. \ref{fig:micnetwork}.
This {\it black-box} representation is an equivalent microwave network of a phased-array antenna consisting of  $K \times L$ elements
(see also Fig. \ref{fig:planararray}).
A forward wave will propagate with complex amplitude $a_j$ along the transmission line connected to the input port of each antenna element $j$.
In addition, a reflected wave with complex amplitude $b_j$ will propagate along the same transmission line in the opposite direction.
The complex amplitudes of both the forward and reflected waves can be directly measured with a vector network analyzer (VNA).
When we consider the array in transmit mode, the reflected signal, represented by $b_j$, is unwanted since it reduces the total signal that is radiated by array element $j$.
As an example to further explore the concept, we will consider a very small array with only two antenna elements, so $K=2$ and $L=1$.
The relation between the forward and reflected waves at the input of the network can be written in the following form:
\begin{equation}
\displaystyle \left[ \begin{array}{l} b_1 \\ b_2 \end{array}
\right] = \left[ \begin{array}{l} S_{11} \ \ S_{12} \\ S_{21} \ \
S_{22} \end{array} \right] \left[ \begin{array}{l} a_1 \\ a_2
\end{array} \right],
\end{equation}
where $[S]$ is the de scattering matrix.
Note that $S_{11}$ represents the reflection coefficient at port 1 when port 2 is terminated with a matched load impedance, also known as the reference impedance.
In practise the reference impedance will often by equal to $50 \Omega$.
Similarly, $S_{22}$ is the reflection coefficient at port 2 when port 1 is terminated with a matched load impedance.
The coefficient $S_{21}$ is the coupling coefficient between antenna
element 1 and 2. It is a result of the electromagnetic interaction between both antennas.
Since array antennas are passive, reciprocity applies. As a result $S_{21}=S_{12}$.
The coupling coefficient can be obtained by using
\begin{equation}
\displaystyle  S_{21} = \frac{b_2}{a_1} \left|_{a_2=0} \right. .
\end{equation}
In case the array is used in transmit mode, the reflected signal at port 1 will include two components according to
\begin{equation}
\displaystyle  R^{act}_1 = \frac{S_{11}a_1 + S_{12} a_2}{a_1}, \label{eq:activeR}
\end{equation}
where $a_1$ and $a_2$ represent the excitation coefficients which have already been introduced in (\ref{eq:excoef}).
Since $a_1$ and $a_2$ both depend on the scan angle of the main beam of the array $(\theta_0,\phi_0)$, the resulting reflection coefficient $R^{act}_1$ will also depend on the scan angle.
Therefore, $R^{act}_1$ is referred to as the {\it active reflection coefficient} of array element 1.
Similarly, $R^{act}_2$ is the active reflection coefficient of array element 2.
From (\ref{eq:activeR}), we can observe that a large coupling coefficient $S_{12}$ can significantly deteriorate $R^{act}_1$ and $R^{act}_2$
Note that in practise, the active reflection coefficients not only depend on the scan angle, but also on frequency.

For an array consisting of $K \times L$ elements, the active reflection coefficient of array element $j$ is given by
\begin{equation}
\displaystyle  R^{act}_j (u_0,v_0) = \frac{1}{a_j} \sum_{i=1}^{K \times L}
S_{ij} a_i (u_0,v_0), \label{eq:activeRKL}
\end{equation}
where the excitation coefficients $a_i$ with $i=1,2,...$
are given by (\ref{eq:excoef}).
Since part of the incident power will be reflected, a power loss equal to $L_j=1-|R^{act}_j|^2$ will occur. $L_j$ is the {\it scan loss} and can be interpreted as a reduction of the antenna gain of array element $j$. The element factor in \ref{eq:Arfactorelpat2} should then be replaced by:
\begin{equation}
\displaystyle \vec{f}_j(u,v) = \vec{f}(u,v) \sqrt{  1-{\left| R^{act}_j (u_0,v_0) \right|}^2 }.
\label{eq:elementpatmutual}
\end{equation}
We can also define the active input impedance of the $j$-th array element as:
\begin{equation}
\displaystyle  Z^{in,act}_j (u_0,v_0) = \frac{V_j}{I_j}=\sum_{i=1}^{K \times L}
Z_{ji} \frac{I_i}{I_j},
\label{eq:activeZin1}
\end{equation}
where $V_j$ and $I_j$ are the voltage and current at the input port of the $j$-th element and where $Z_{ji}$ is the mutual impedance between array element $j$ and $i$.
The relation between the active input impedance and the active reflection coefficient is now given by:
\begin{equation}
\displaystyle  R^{act}_j (u_0,v_0)= \frac{Z^{in,act}_j (u_0,v_0) - Z_0}{Z^{in,act}_j (u_0,v_0) + Z_0},
\label{eq:activeZin2}
\end{equation}
where $Z_0$ is the port impedance, usually $Z_0=50 \Omega$.

\begin{figure}[hbt]
\centerline{\psfig{figure=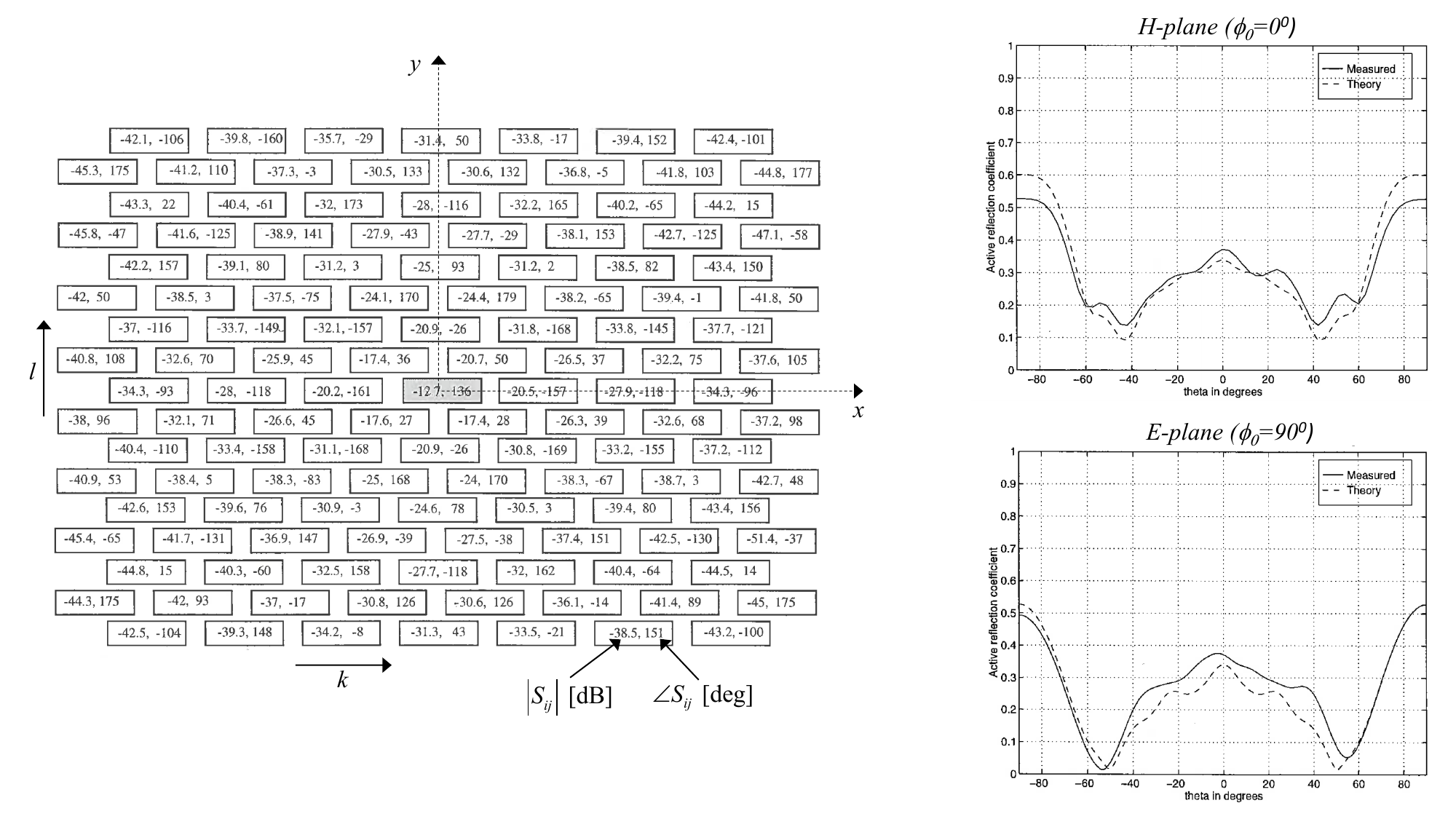,width=170mm}}

\caption{\it Measured coupling coefficients (left) of the center part of 127 array elements and the corresponding active reflection coefficient (right plots) in the H-plane ($\phi_0=0^0$) and E-plane ($\phi_0=90^0$). The coupling coefficients between the center element ($k=4$, $l=9$) and the other elements is shown. More details on the measured data can be found in \cite{Smolders1996}.}
\label{fig:aparmutualcoupling}
\end{figure}

\begin{figure}[hbt]
\vspace{-0.5cm}
\centerline{\psfig{figure=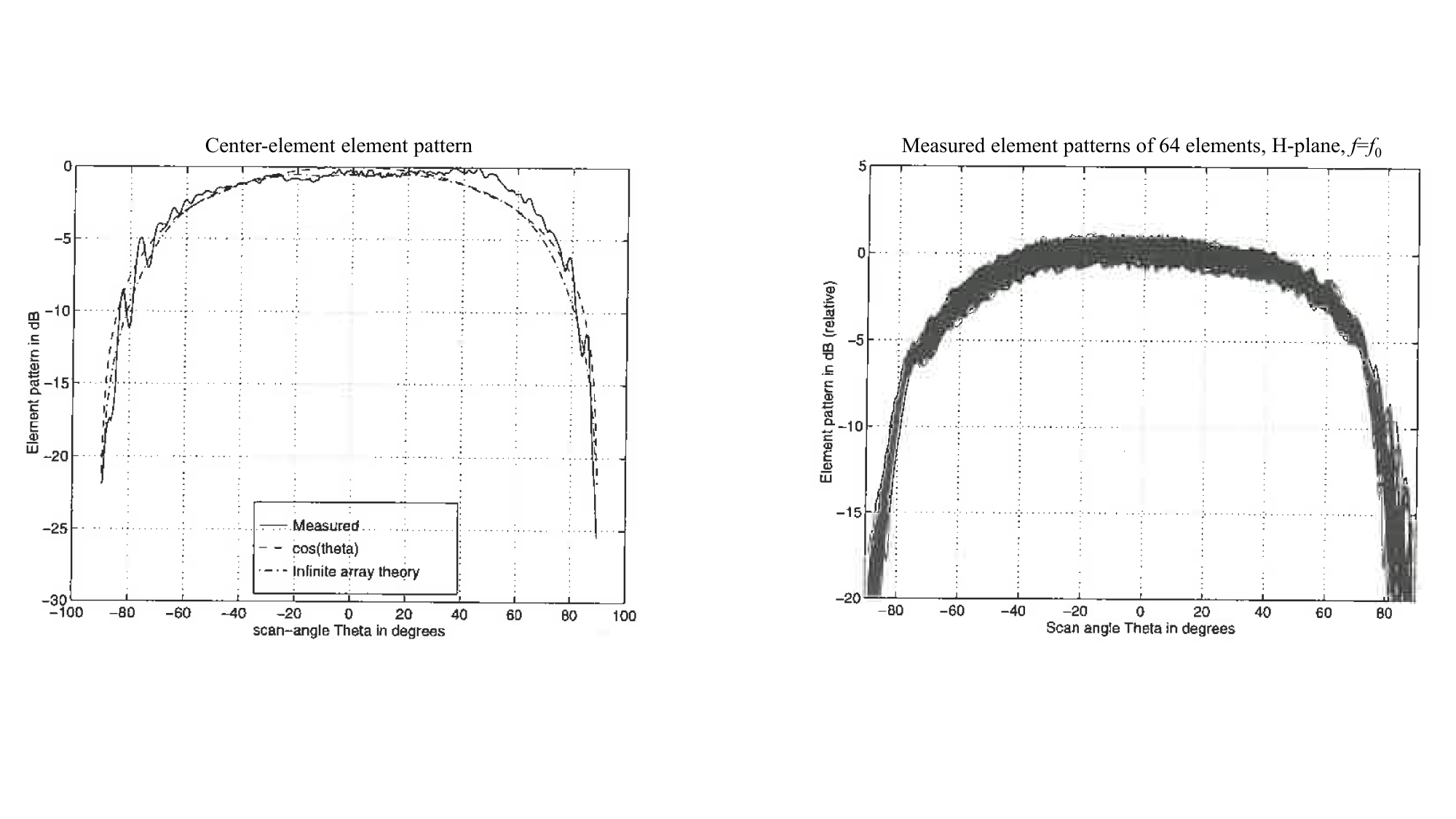,width=170mm}}
\vspace{-2cm}
\caption{\it Measured and simulated element pattern of the center element (left) and measured element patterns (right) of the center part of an array with 127 array elements in the H-plane ($\phi_0=0^0$). More details can be found in \cite{Smolders1996}.}
\label{fig:aparelementpattern}
\end{figure}

\begin{figure}[hbt]
\centerline{\psfig{figure=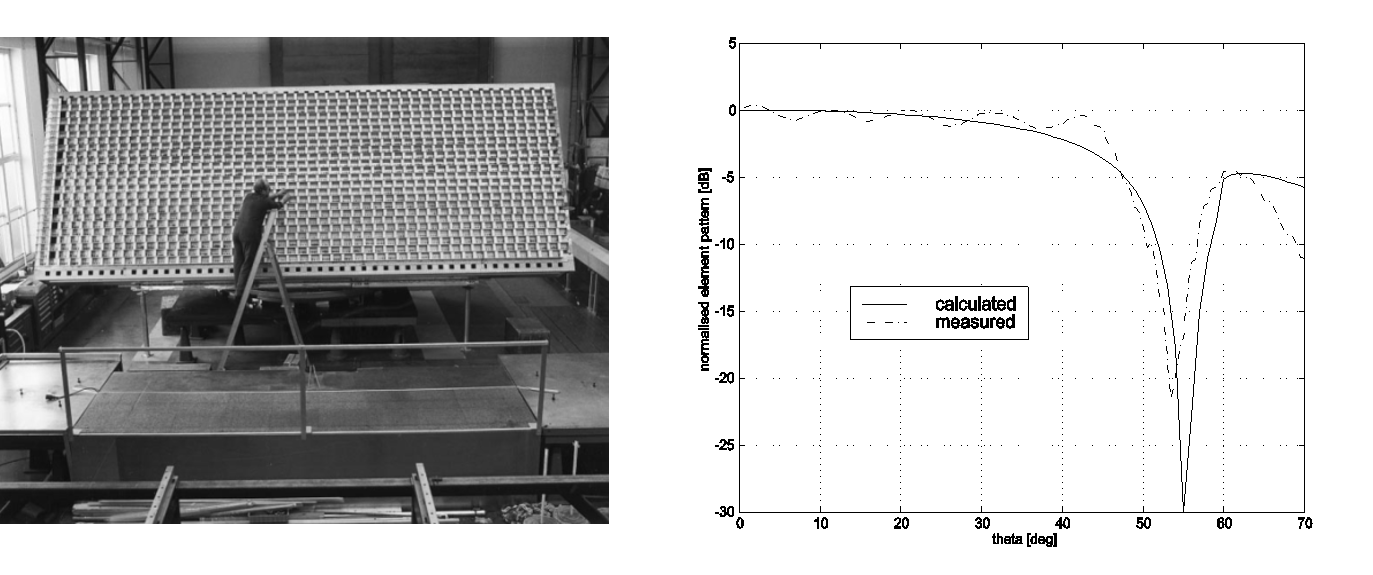,width=170mm}}

\caption{\it Photo and the measured element pattern of the center element of a 1024 phased-array of folded dipoles operating at L-band frequencies. More details can be found in \cite{Beurdensmolders}.}
\label{fig:smartL}
\end{figure}

The effect of mutual coupling on the performance of phased arrays can be best illustrated with some examples. Let us first consider a large X-band phased-array radar with 4096 open-ended waveguide radiators as presented in \cite{Smolders1996}. The measured coupling coefficients of a small sub-array of 127 elements near the center of the array and the associated active reflection coefficient are provided in Fig. \ref{fig:aparmutualcoupling}. It can be observed that the active reflection coefficient depends on the scan angle, but remains quite low up to a scan angle of $+/- 60^0$ in both planes. The measured element patterns in the H-plane ($\phi=0^0$) are provided in Fig. \ref{fig:aparelementpattern}. The element patterns fluctuate around a $\cos\theta$ pattern, resulting in about 3 dB of gain reduction at a scan angle of $\theta_0=55^0$.
Another example of the effect of mutual coupling is shown in Fig. \ref{fig:smartL}. Due to a periodic effect in this array of 1024 folded-dipoles, a so-called {\it blind scan angle} occurs at a scan angle of $\theta_0=60^0$. In this case, all power is reflected and as a result, the directivity will be very low. More details can be found in \cite{Beurdensmolders}.

\section{Beamforming}
\subsection{Beamforming concepts and architectures}

\subsection{Phase shifters (PHS) versus true time delay (TTD)}
\begin{figure}[hbt]
\centerline{\psfig{figure=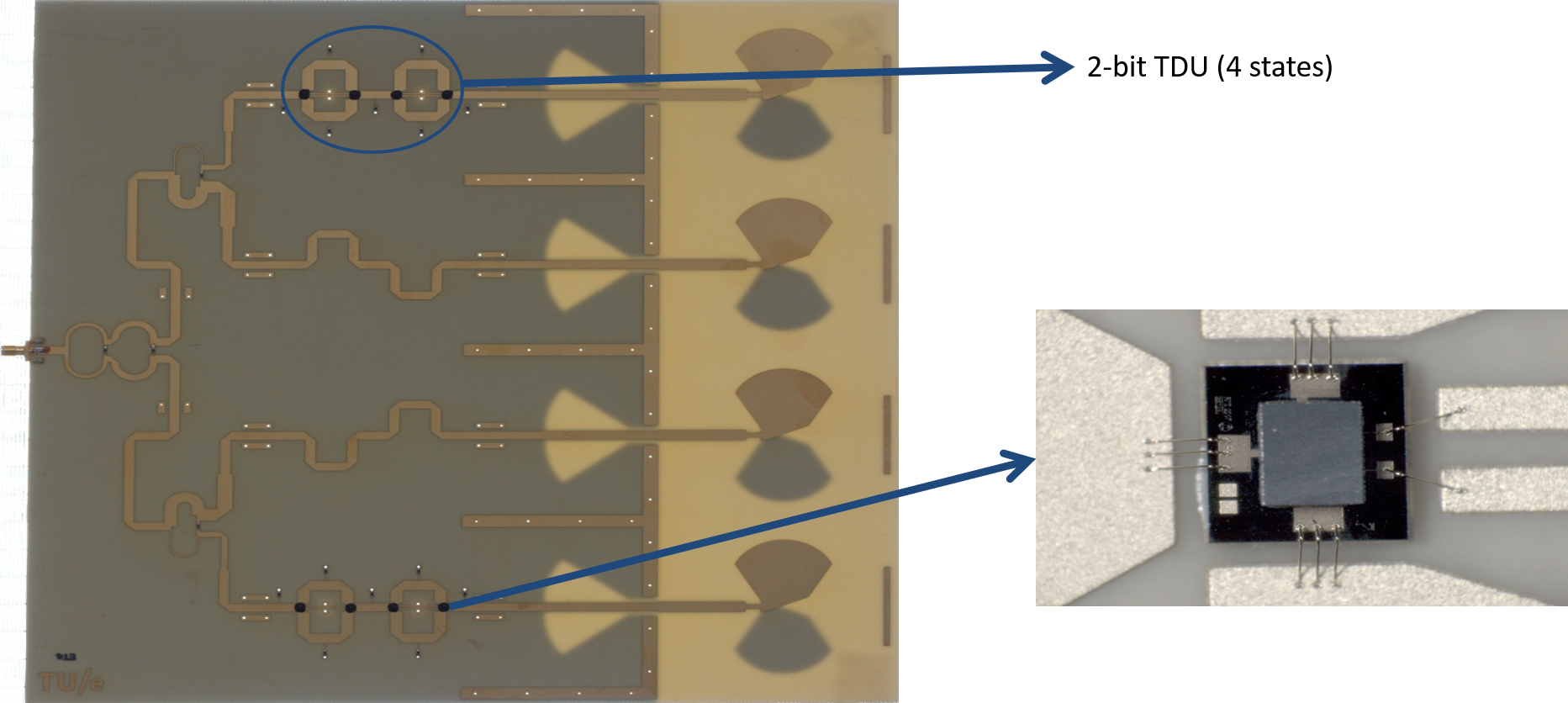,width=170mm}}

\caption{\it 4-element beamformer using TDU units including 4 dipole-like antenna elements. The switches in the TDU units are realized with MEMS technology \cite{MestromBeamformer}}
\label{fig:TDUunitphoto}
\end{figure}

\begin{figure}[hbt]
\centerline{\psfig{figure=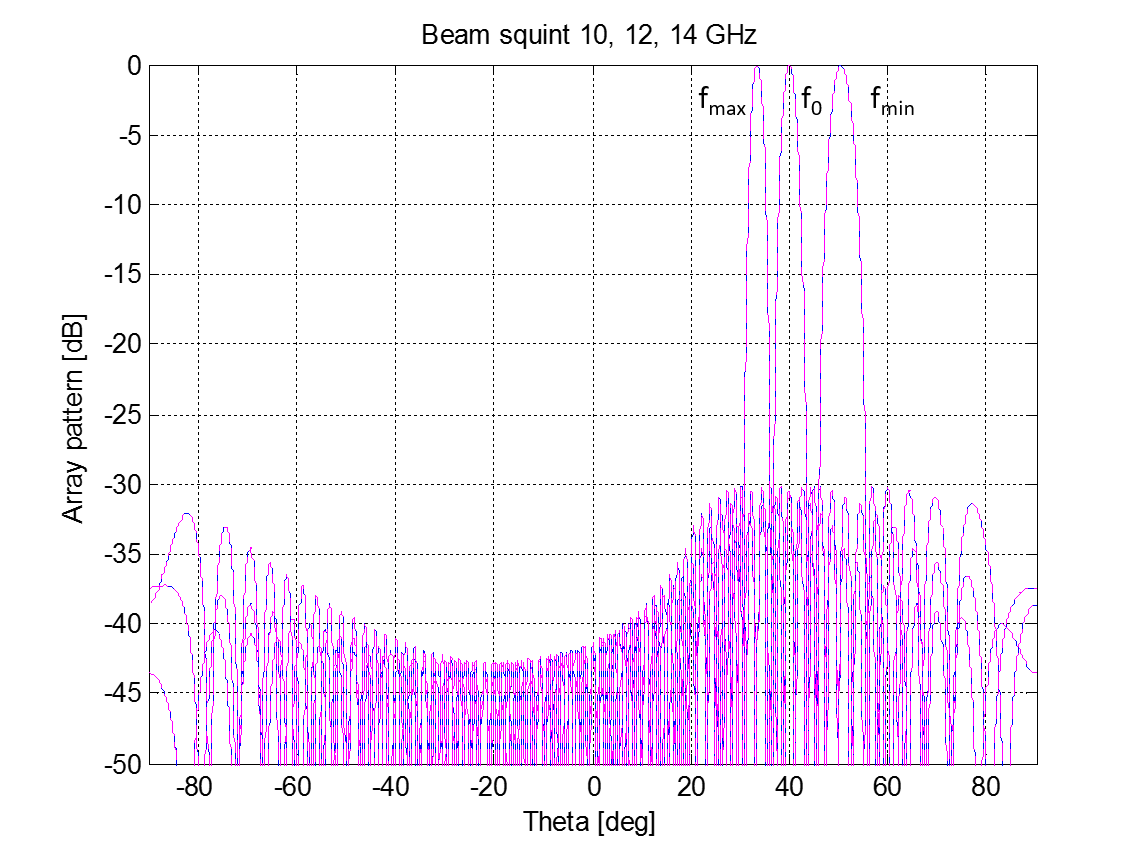,width=130mm}}

\caption{\it Beam squinting in a linear array of 64 elements. Phase-shifter settings are  optimized at the center frequency $f_0=12~$GHz. By using TDU units, the main beam would remain fixed at $\theta_0=40^0$ over the entire frequency range.}
\label{fig:beamsquint}
\end{figure}

Traditionally, phased-arrays use phase shifters (PHS) to steer the beam of the antenna. As an alternative true-time delays (TTD) can be used. In order to understand the differences and whether PHS or TTD should be used in an application, let us investigate the array factor in more detail.
The general form of the array factor is given by (\ref{eq:Arfactorlin2}):
\begin{equation}
\begin{array}{lcl}
\displaystyle S(u) & = &  \displaystyle \sum_{k=1}^K |a_k| e^{\jmath [k_0 (k-1) d_x
(u-u_0)]} = \sum_{k=1}^K |a_k| e^{\jmath \left[ \frac{2\pi f_0}{c} (k-1) d_x
u-\psi_k \right]},
 \end{array} \label{eq:PHSversusTDD1}
\end{equation}
where $f_0$ is the frequency of operation, $c$ is the speed of light and $\psi_k$ is the applied phase shift to array element $k$ at the frequency $f_0$.
In a PHS unit, the realized phase shift is constant versus frequency. In a TTD unit, the phase shift is realized with transmission lines of a specific length. As a consequence, the realized phase will depend linearly on frequency. An example of a 4-element beamformer using 2-bit TDU units is shown in Fig. \ref{fig:TDUunitphoto}.
When we look carefully at equation (\ref{eq:PHSversusTDD1}), we can observe that the required phase shift $\psi_k$ should have a linear dependency with frequency, since the term $\frac{2\pi f_0}{c} (k-1) d_x u$ also shows a linear behavior versus frequency. As a consequence, PHS units can only be applied in phased-array systems with a relative small operational frequency bandwidth. At least when a fixed scan angle is required.
The effect of so-called beam squinting is illustrated in Fig. \ref{fig:beamsquint}, where the radiation pattern of a 64-element linear array is shown at 10, 12 and 14 GHz. The center frequency of the array is $f_0=12~$GHz at which the optimal phase shift settings were chosen to scan the array towards $40^0$. The main beam is clearly squinting to another direction when the frequency of operation is changed according to:
\begin{equation}
\begin{array}{lcl}
\displaystyle \frac{f}{f_0} & = &  \displaystyle \frac{u_0}{u}, \\
\displaystyle u & = &  \displaystyle u_0 \frac{f_0}{f}, \\
 \end{array} \label{eq:PHSversusTDD2}
\end{equation}
where we assumed that $f_0$ is the design frequency of the array (in this case 12 GHz) and $f$ is the frequency of interest. Note that beam squinting can also be useful: it is a way to realize beam steering without the need of adjustable phase-shifters.

\subsection{The effect of phase and amplitude errors}
In this section we will investigate the effect of phase and amplitude errors on the radiation characteristics of phased arrays. Errors in phase and amplitude at the output of each array element will occur due to non-perfect electronics, mechanical tolerances and environmental changes, like a change in temperature. Let us first consider a linear array of isotropic radiators with uniform amplitude distribution ($|a_k|=1$). In addition, we will assume that the main beam of the array is directed towards broadside ($\theta_0=u_0=0$). This configuration is shown in Fig. \ref{fig:lineairarray}.
Without any errors the array factor is now given by (see also (\ref{eq:Arfactorlin2}))
\begin{equation}
\begin{array}{lcl}
\displaystyle S(u) & = & \displaystyle  \sum_{k=1}^K  e^{\jmath k_0 (k-1) d_x
u}.
 \end{array} \label{eq:Errors1}
\end{equation}
Let us now first determine the effect of phase errors. Lateron, we will generalize our analysis to include amplitude errors as well. Phase errors will affect the main beam (directivity, beam width) and the sidelobes.

As a first step we will investigate the main beam. Assume that the applied phase of each element has an error, say $\delta_k$ according to a Gaussian or uniform probability density function with mean value $0$ (since we are considering the non-scanning case here first). The array factor at $u=0$ can now be written as:
\begin{equation}
\begin{array}{lcl}
\displaystyle S_{m \delta} & = & \displaystyle  \sum_{k=1}^K  e^{\jmath k_0 (k-1) d_x
u-\jmath \delta_k} \left|_{u=0} \right.  =\sum_{k=1}^K  e^{-\jmath  \delta_k} = \sum_{k=1}^K  \left[ \cos\delta_k - \jmath \sin\delta_k \right].
 \end{array} \label{eq:Errors2}
\end{equation}
Assume that $\delta_k$ is small, so the sum over the $\sin\delta_k$ terms can be neglected, resulting in:
\begin{equation}
\begin{array}{lcl}
\displaystyle S_{m \delta} & = & \displaystyle  \sum_{k=1}^K   \cos\delta_k \approx
\sum_{k=1}^K \left[1-\frac{1}{2} \delta^2_k \right] = K- \frac{1}{2} \sum_{k=1}^K \delta^2_k ,
 \end{array} \label{eq:Errors3}
\end{equation}
where we have applied a Taylor expansion of $\cos\delta_k$. The main beam {\it average} complex power is given by:
\begin{equation}
\begin{array}{lcl}
\displaystyle \overline{\left|S_{m \delta}\right|^2} & = & \di \overline{ \left( K- \frac{1}{2} \sum_{k=1}^K \delta^2_k \right)^2} \\
& = & \di K^2 -K \overline{\sum_{k=1}^K \delta^2_k} + \frac{1}{4}  \overline{ \sum_{k,l}^K \delta^2_k \delta^2_l } ,
 \end{array} \label{eq:Errors4}
\end{equation}
Assume that the variance of $\delta_k$ of all elements is equal $\overline{\delta^2_k}=\overline{\delta^2}$. This implies that for large $K$ and small $\overline{\delta^2}$:
\begin{equation}
\begin{array}{lcl}
\displaystyle \overline{\delta^2_k \delta^2_l } & = & \di \overline{\delta^4} \ll 4K \overline{\delta^2}, \\
\di \overline{\left|S_{m \delta}\right|^2}& = & \di K^2 -K \overline{\sum_{k=1}^K \delta^2_k} \\
\di & = & \di K^2 \left( 1- \overline{\delta^2} \right),
 \end{array} \label{eq:Errors5}
\end{equation}
since the last term in (\ref{eq:Errors4}) can be neglected.

As a second step, we will determine the effect of phase errors on the sidelobes. Let us assume that we are at a null of the error-free radiation pattern. For the non-scanning case the array factor is now written as:
\begin{equation}
\begin{array}{lcl}
\displaystyle S_{s \delta} & = & \displaystyle  \sum_{k=1}^K e^{\jmath k_0 (k-1) d_x
u} e^{-\jmath \delta_k}  =\sum_{k=1}^K  e^{\jmath  \beta_k} \left[\cos\delta_k-\jmath\sin\delta_k \right] ,
 \end{array} \label{eq:Errors6}
\end{equation}
with $\beta_k=k_0(k-1)d_x u$. Since we are at a null of the error-free pattern, the $\cos\delta_k$ term vanishes:
\begin{equation}
\begin{array}{lcl}
\displaystyle S_{s \delta} & = & \displaystyle  = -\jmath \sum_{k=1}^K  e^{\jmath  \beta_k} \sin\delta_k \approx -\jmath \sum_{k=1}^K  e^{\jmath  \beta_k} \delta_k.
 \end{array} \label{eq:Errors7}
\end{equation}
We have now transformed the phase-error in an amplitude error occuring in the error-free array pattern. The average power at a null of a sidelobe now becomes:
\begin{equation}
\begin{array}{lcl}
\displaystyle \overline{\left|S_{s \delta} \right|^2} & = & \displaystyle \overline{ \left[ \sum_{k=1}^K  e^{\jmath  \beta_k} \delta_k \right] \left[ \sum_{l=1}^K  e^{-\jmath  \beta_l} \delta_l \right]} \\
& = & \displaystyle \sum_{k=1}^K \overline{\delta^2_k} + \overline{ \sum_{k \neq l} \delta_k \delta_l  e^{\jmath (\beta_k-\beta_l)}} \\
& = & \displaystyle K \overline{\delta^2} ,
 \end{array} \label{eq:Errors8}
\end{equation}
with again $\overline{\delta^2_k}=\overline{\delta^2}$ and we have assumed that $ \overline{\delta_k}$ and $ \overline{\delta_l}$ are uncorrelated.

The average normalized sidelobe power due to phase errors at a null of the error-free array pattern is now given by:
\begin{equation}
\begin{array}{lcl}
\displaystyle \overline{R^2_p} & = & \displaystyle \frac{\overline{\left|S_{s \delta} \right|^2}}{\overline{\left|S_{m \delta} \right|^2}} = \frac{K \overline{\delta^2}}{K^2 \left( 1- \overline{\delta^2} \right)} \approx \frac{\overline{\delta^2}}{K}.
 \end{array} \label{eq:Errors9}
\end{equation}
From (\ref{eq:Errors9}) we can conclude that large arrays can tolerate much larger phase errors for a given required sidelobe level. The is due to the fact that we have assumed the errors to be uncorrelated.

The directivity and associated antenna gain will be reduced in case of phase errors. Starting from the definition of the directivity (\ref{eq:richtfunctielinar1}), we can determine the directivity in case of errors $D_e$ of a linear array with uniform illumination with $u_0=\theta_0=0$ as:
\begin{equation}
\begin{array}{lcl}
\displaystyle D_e & =  & \displaystyle
\frac{P}{P_t / 4\pi} = \frac{\overline{\left|S_{m \delta} \right|^2}}{\overline{\left|S_{ideal} (u) \right|^2} + \overline{\left|S_{s \delta} (u) \right|^2}} \\
& =  & \displaystyle \frac{K^2\left(1-\overline{\delta^2} \right)}{K+K \overline{\delta^2}} =
K \frac{1-\overline{\delta^2}}{1+ \overline{\delta^2}} \\
& \approx  & \displaystyle K \frac{1}{1+ \overline{\delta^2}}.
\end{array}
 \label{eq:Errors10}
\end{equation}
In (\ref{eq:Errors10}), the average radiated power $\overline{\left|S_{ideal} (u) \right|^2}$ of an error-free linear array with uniform tapering ($|a_k|=1$) is equal to $K$.
The normalized directivity loss then becomes:
\begin{equation}
\begin{array}{lcl}
\displaystyle \frac{D_e}{D} & =  & \displaystyle
 \frac{1}{1+ \overline{\delta^2}},
\end{array}
 \label{eq:Errors11}
\end{equation}
since the directivity of an ideal linear array with uniform tapering is $D=K$.

Electronic phase shifting is commonly realized using discrete settings. Let us consider a phase shifter with $P$ bits. The total number of settings is $\di 2^P$. The incremental phase step [radians] is:
\begin{equation}
\begin{array}{lcl}
\displaystyle \Delta \psi_k & =  & \displaystyle
 \frac{2 \pi}{2^P}.
\end{array}
 \label{eq:Errors12}
\end{equation}
Assuming an {\it uniformly}-distributed random phase error in the range $[-\Delta \psi_k/2,\Delta \psi_k/2]$, we find that the variance is given by:
\begin{equation}
\begin{array}{lcl}
\displaystyle \overline{\delta^2} & =  & \displaystyle
 \frac{\left(\frac{2 \pi}{2^P}\right)^2}{12}=\frac{\pi^2}{3(2^{2P})}.
\end{array}
 \label{eq:Errors13}
\end{equation}

\begin{figure}[hbt]
\centerline{\psfig{figure=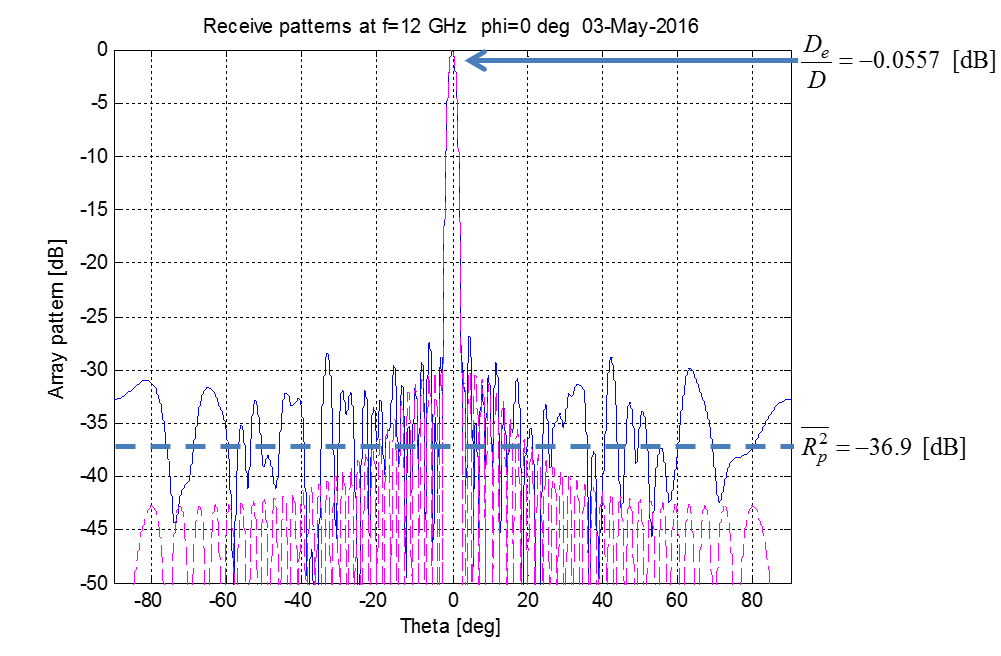,width=130mm}}

\caption{\it Effect of phase errors using 4-bit phase shifters on the radiation pattern of a linear array with uniform taper with $K=64$. The ideal pattern is designed to have a peak sidelobe level (SLL) of 30 dB using a Taylor taper with $n=8$.}
\label{fig:Errorpattern64}
\end{figure}

\vspace{0.5cm}
{\underline {\it Example: 4-bit phase shifter}}. \\
Let us consider a 4-bit phase shifter used in a linear array of $K=64$ elements with uniform tapering. The phase step is $22.5^0$ and the corresponding maximum phase error is now $\pm 11.25^0$. We now find that the variance is given by:
\begin{equation}
\begin{array}{lcl}
\displaystyle \overline{\delta^2} & =  & \displaystyle
 =\frac{\pi^2}{3(2^{2P})}=0.0129 ~~[rad^2],
\end{array}
 \label{eq:Errors14}
\end{equation}
which corresponds to an average phase error of $6.5^0$ (rms). The average sidelobe level at a null of the error-free pattern and the directivity loss and can be found by:
\begin{equation}
\begin{array}{lcl}
\displaystyle \overline{R^2_p} & = & \displaystyle  \frac{K \overline{\delta^2}}{K^2 \left( 1- \overline{\delta^2} \right)} = -36.9 ~~[\mbox{dB}], \\
\displaystyle \frac{D_e}{D} & =  & \displaystyle
 \frac{1}{1+ \overline{\delta^2}} = -0.0557 ~~[\mbox{dB}].
 \end{array} \label{eq:Errors14}
\end{equation}
The corresponding radiation pattern is shown in Fig. \ref{fig:Errorpattern64}. The ideal pattern has low sidelobes which is achieved by using a Taylor tapering with 30 dB peak sidelobe level.

\begin{figure}[hbt]
\centerline{\psfig{figure=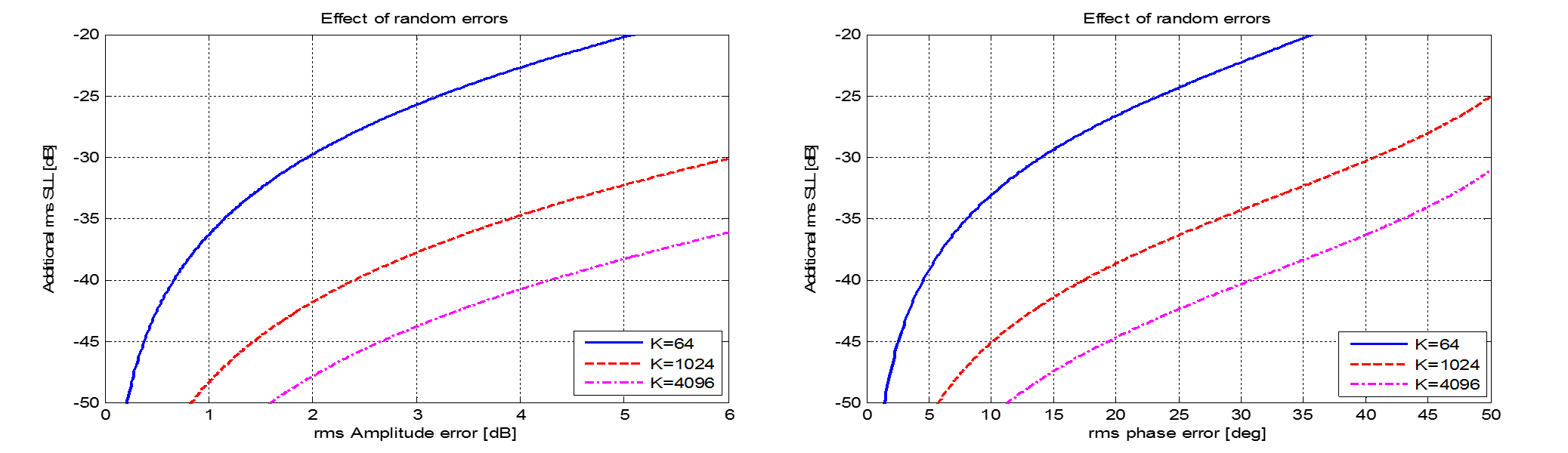,width=170mm}}

\caption{\it Design guidelines to determine the required accuracy of the phase shifter and variable gain amplifier used in a phased array with $K \times L$ antenna elements. Random errors are assumed.}
\label{fig:phaseamplitudeerrors}
\end{figure}

\vspace{0.5cm}
As a final step we can generalize to the case of a planar array with both amplitude and phase errors and where the main beam is scanned towards any direction. Let us consider a planar array with $K \times L$ elements with phase and amplitude error of element $j$ given by $\delta_j$ and $\Delta_j$, respectively. We again will assume that the variance of the phase and amplitude errors of all elements are identical and given by $\overline{\delta^2}$ and $\overline{\Delta^2}$, respectively.
It can be shown that the average sidelobe level at a null of the error-free pattern and the directivity loss are given by:
\begin{equation}
\begin{array}{lcl}
\displaystyle \overline{R^2_p} & = & \displaystyle  \frac{\overline{\delta^2}+\overline{\Delta^2} }{D_0}, \\
\displaystyle \frac{D_e}{D_0} & =  & \displaystyle
 \frac{1}{1+ \overline{\delta^2}+\overline{\Delta^2}},
 \end{array} \label{eq:Errors15}
\end{equation}
where $D_0$ is the directivity of the array.
Fig. \ref{fig:phaseamplitudeerrors} provides design guidelines to determine the required amplitude and phase accuracy of the electronic circuits used in a phased array. Three array sizes are compared with $K=64, ~1024, ~4096$ elements.

\subsection{Noise Figure and Noise Temperature}
\begin{figure}[hbt]
\centerline{\psfig{figure=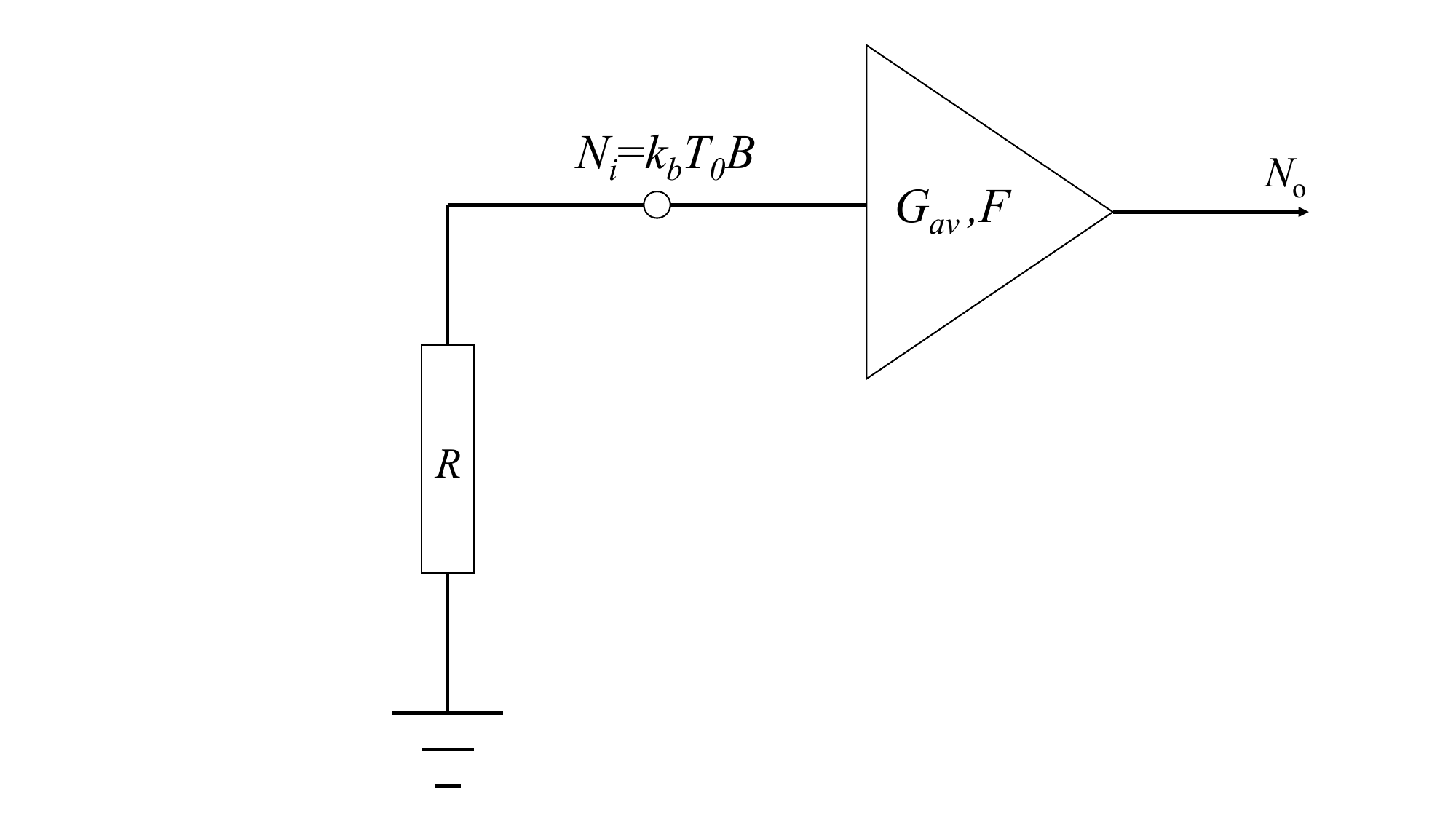,width=100mm}}

\caption{\it Noise in microwave receivers.}
\label{fig:noisemicrowave1}
\end{figure}

Let us first do a quick review of noise in microwave receivers by considering the configuration shown in Fig. \ref{fig:noisemicrowave1}, see also section \ref{sec:LNA}. A resistor is connected to the input of an amplifier. The maximum noise power $N$ delivered by a resistor $R$ operating at the environmental temperature $T_0$ is $N=k_b T_0 B$, where $k_b=1.38\times 10^{-23}~$J/K is Boltmann's constant and $B$ [Hz] is the frequency bandwidth of the system.
The amplifier is characterized by its available power gain $G_{av}$ and noise figure $F$, which is defined as the ratio of the signal to noise ratio at the input versus the signal to noise ratio at the output:
\begin{equation}
\begin{array}{lcl}
\displaystyle F & = & \displaystyle  \frac{\di \left( \frac{S_i}{N_i} \right)}{\di \left( \frac{S_o}{N_o} \right) }  =  \displaystyle  \frac{\di \left( \frac{S_i}{k_b T_0 B}\right) }{\di \left( \frac{G_{av} S_i}{N_o}\right) }  \\
& = & \displaystyle  \frac{N_o}{G_{av} k_b T_0 B} .
 \end{array} \label{eq:Noise1}
\end{equation}
Therefore, the additional noise $\Delta N^{amp}_i$ added by the amplifier seen at the input of a noise-free amplifier with available power gain $G_{av}$ is now:
\begin{equation}
\begin{array}{lcl}
\displaystyle \Delta N^{amp}_i & = & \displaystyle  \frac{N_o-G_{av} N_i}{G_{av}}=(F-1)k_b T_0 B .
 \end{array} \label{eq:Noise2}
\end{equation}
Now let us connect an antenna at the input of the receiver. We will assume that the noise generated by the antenna can be considered to be white noise, so it would not depend on the frequency within the bandwidth of interest. In this case, we can model the antenna noise as a thermal noise source, equivalent to the case of a resistor at the input of the receiver. The noise power $N_a$ from a single antenna element with antenna noise temperature $T_a$ is now equal to $N_a=k_b T_a B$. The antenna noise temperature can be determined by integrating the antenna gain function of the antenna over the sky-noise distribution:
\begin{equation}
\begin{array}{lcl}
\displaystyle T_a & = &  \displaystyle \frac{1}{4 \pi} \int\int T(\Omega) G(\Omega) d\Omega, \\
 \end{array} \label{eq:Noise3}
\end{equation}
where $\Omega$ is the solid angle with $d\Omega=\sin\theta d\theta d\phi$.
Note that $T_a$ is often much lower than the environmental temperature since we weigh the sky-noise temperature distribution $T(\Omega)$ with the gain $G(\Omega)$ of the antenna. For example in satellite TV operating in the Ku-band (11-14 GHz), a typical value is $T_a=100~$K.

\begin{figure}[hbt]
\centerline{\psfig{figure=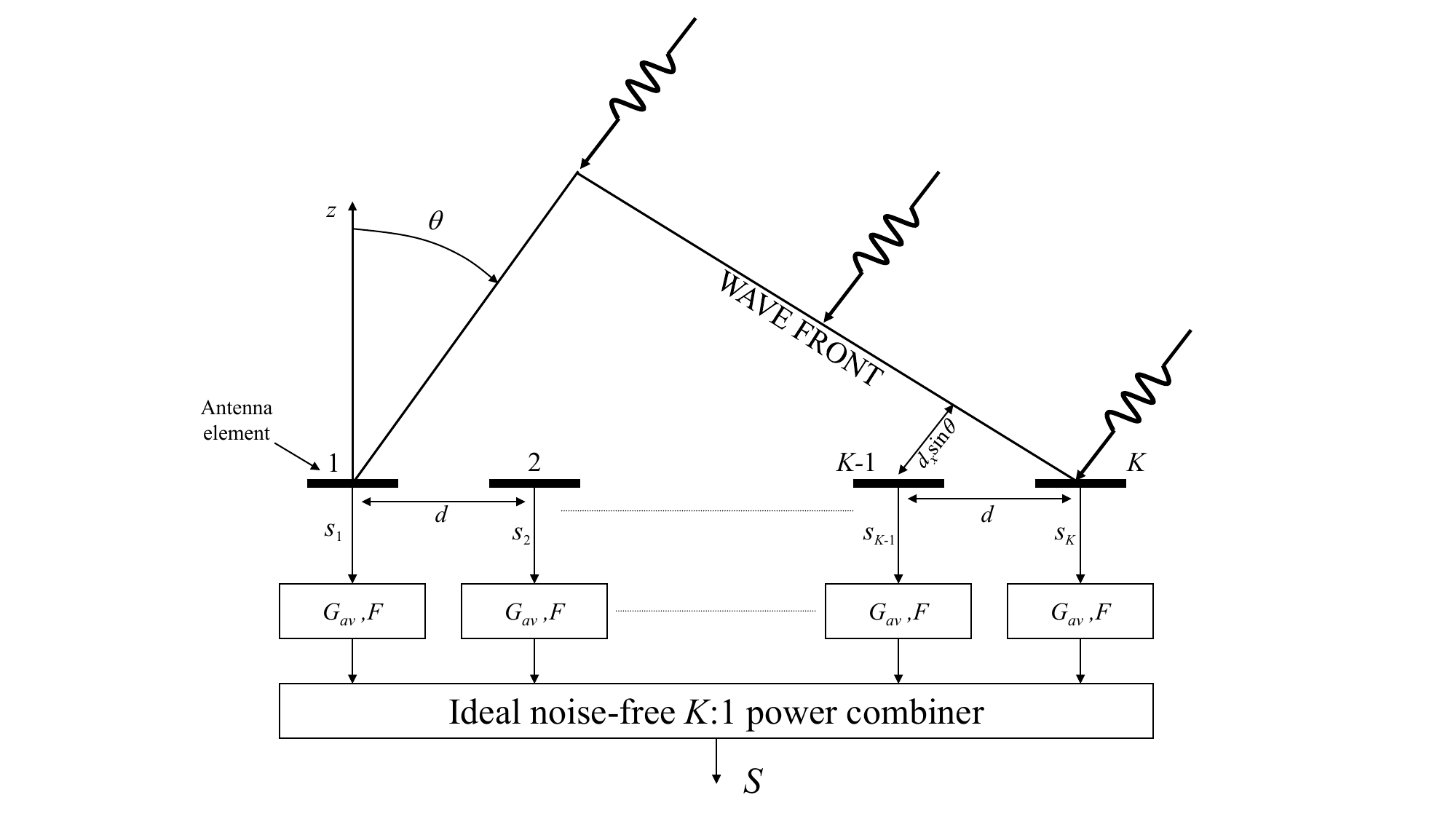,width=160mm}}

\caption{\it Array with K elements. Each antenna element is connected to a low-noise amplifier with available power gain $G_{av}$ and noise figure $F$.}
\label{fig:noiselineararray}
\end{figure}

\begin{figure}[hbt]
\vspace{-1cm}
\centerline{\psfig{figure=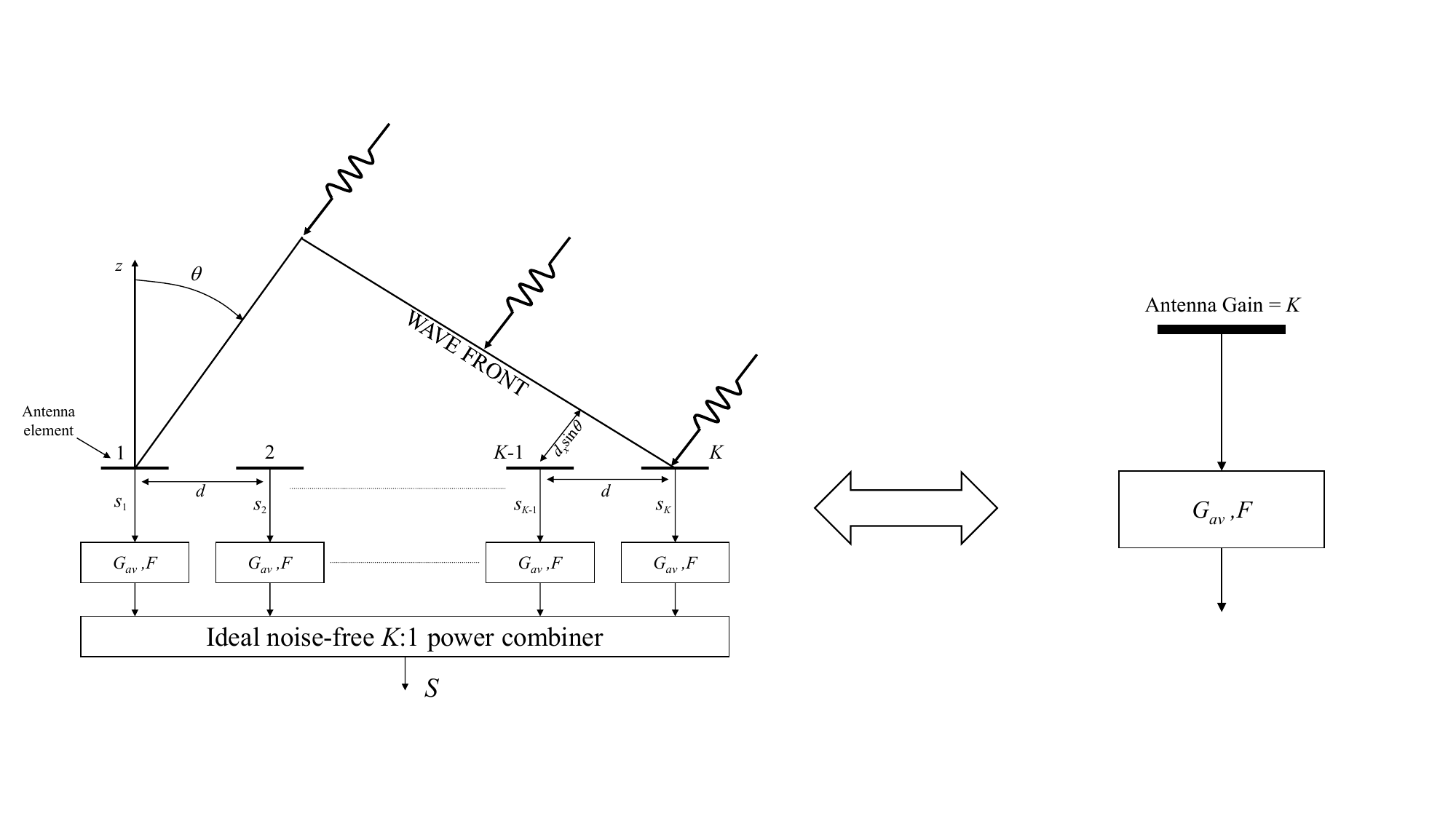,width=170mm}}
\vspace{-1.5cm}

\caption{\it Noise in a phased-array can be modelled by considering an equivalent single antenna with the same antenna gain and connected to a single low-noise amplifier with available power gain $G_{av}$ and noise figure $F$. In case of an ideal array with uniform taper the antenna gain is equal to $K$.}
\label{fig:noiseequivalent}
\end{figure}

Now consider a phased array with $K$ elements with uniform amplitude taper, as illustrated in Fig \ref{fig:noiselineararray}. Each antenna element is connected to a low-noise amplifier with available power gain $G_{av}$ and noise figure $F$. The noise power received by a single antenna element $k$ is $N_{ka}=k_b T_{ka} B$. Let us assume that $T_{ka}=T_a$.
The $K:1$ power combiner coherently combines the antenna noise power contributions from all elements, since (worst-case) the wanted signal and the antenna noise could be co-located at the same spherical coordinate $(u_0,v_0)$. The total output noise power due to the antenna noise is given by:
\begin{equation}
\begin{array}{lcl}
\displaystyle N^T_a & = &  \displaystyle KG_{av} N_{ka}= k_b T_a BKG_{av} .
 \end{array} \label{eq:Noise4}
\end{equation}
The noise from the electronics at the input of the power combiner is given by:
\begin{equation}
\begin{array}{lcl}
\displaystyle N_e & = &  \displaystyle  k_b T_0 B(F-1)G_{av}.
 \end{array} \label{eq:Noise5}
\end{equation}
The transmitted noise due to the electronics at the output of the power combiner now becomes:
\begin{equation}
\begin{array}{lcl}
\displaystyle N^o_e & = &  \displaystyle \frac{k_b T_0 B(F-1)G_{av}}{K}.
 \end{array} \label{eq:Noise6}
\end{equation}
And as a result, the total noise at the output of the combiner will become:
\begin{equation}
\begin{array}{lcl}
\displaystyle N^{To}_e & = &  \displaystyle KN^0_e =k_b T_0 B(F-1)G_{av}.
 \end{array} \label{eq:Noise7}
\end{equation}
The signal power at the output of the combiner is:
\begin{equation}
\begin{array}{lcl}
\displaystyle S_o & = &  \displaystyle K G_{av} S_a,
 \end{array} \label{eq:Noise8}
\end{equation}
where $S_a$ is the signal power received by a single array element. The signal-to-noise ratio at the output of the combiner is therefore given by:
\begin{equation}
\begin{array}{lcl}
\displaystyle \left( \frac{S}{N} \right)_{out} & = &  \displaystyle \frac{S_o}{N^T_a+N^{To}_e} =
\frac{KG_{av}S_a}{k_b T_aBKG_{av}+k_b T_0 B(F-1)G_{av}}\\
& = &  \displaystyle \frac{S_a}{k_b T_aB+k_b T_0 B(F-1)/K}.
 \end{array} \label{eq:Noise9}
\end{equation}
From (\ref{eq:Noise9}) we can conclude that the noise performance of a phased array is identical to the case of a single equivalent large antenna (e.g. a reflector antenna) with antenna gain equal to the antenna gain of the phased array. The equivalent antenna is connected to a low-noise amplifier with an available power gain $G_{av}$ and noise figure $F$ as illustrated in Fig. \ref{fig:noiseequivalent}.

\section{Sparse arrays}
\begin{figure}[hbt]
\centerline{\psfig{figure=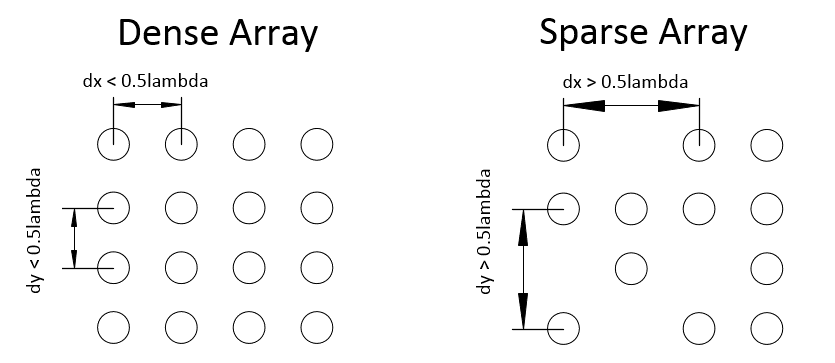,width=120mm}}

\caption{\it Dense regular versus sparse irregular arrays}
\label{fig:sparsearrays}
\end{figure}

\begin{figure}[hbt]
\centerline{\psfig{figure=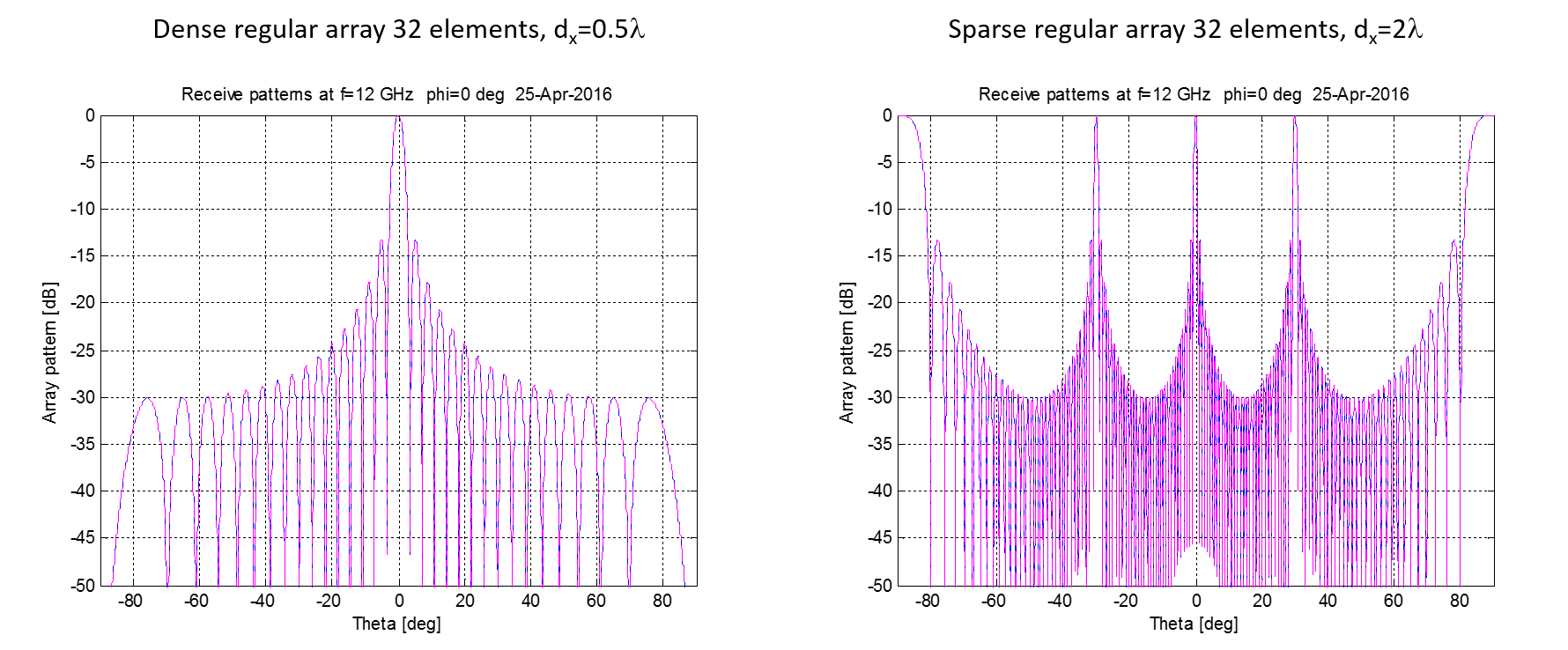,width=160mm}}

\caption{\it Radiation pattern of a dense regular array and sparse regular array. A linear array of 32 elements with uniform tapering is assumed.}
\label{fig:radpatsparseregular}
\end{figure}

\begin{figure}[hbt]
\centerline{\psfig{figure=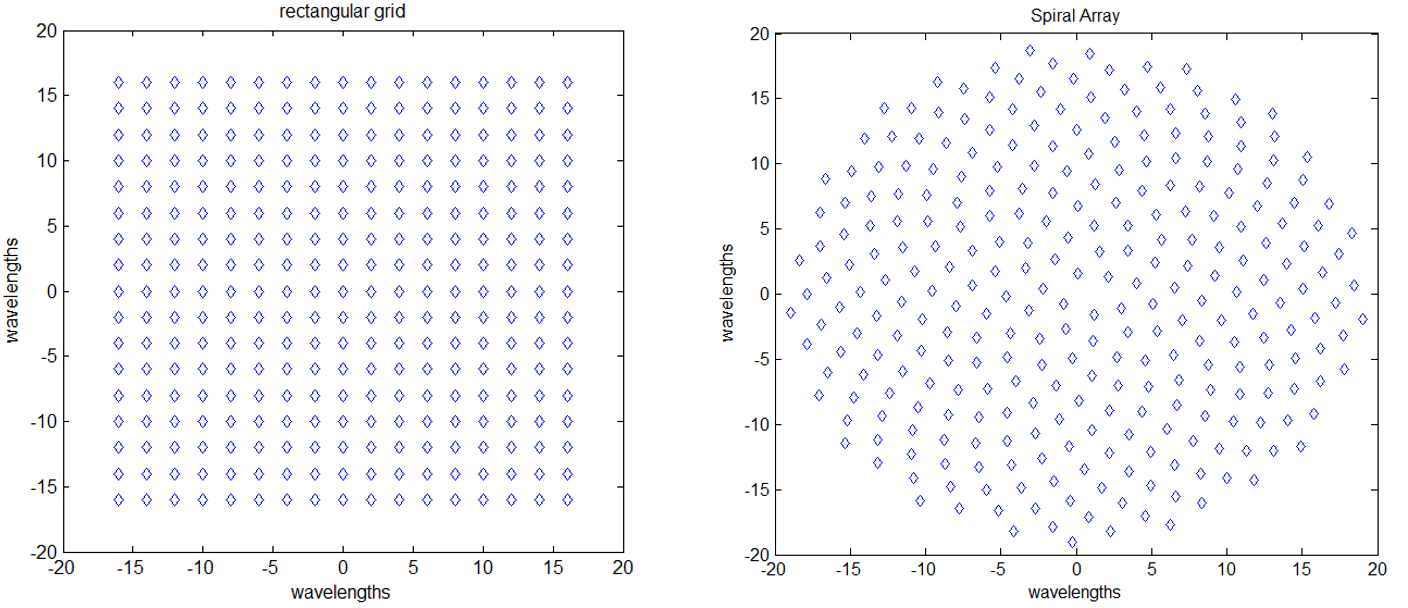,width=160mm}}

\caption{\it Sparse regular array (left) and sparse irregular array (sun-flower arrangement). Both arrays have 289 elements and have an average element spacing of about $2 \lambda_0$.}
\label{fig:sunflower}
\end{figure}

\begin{figure}[hbt]
\centerline{\psfig{figure=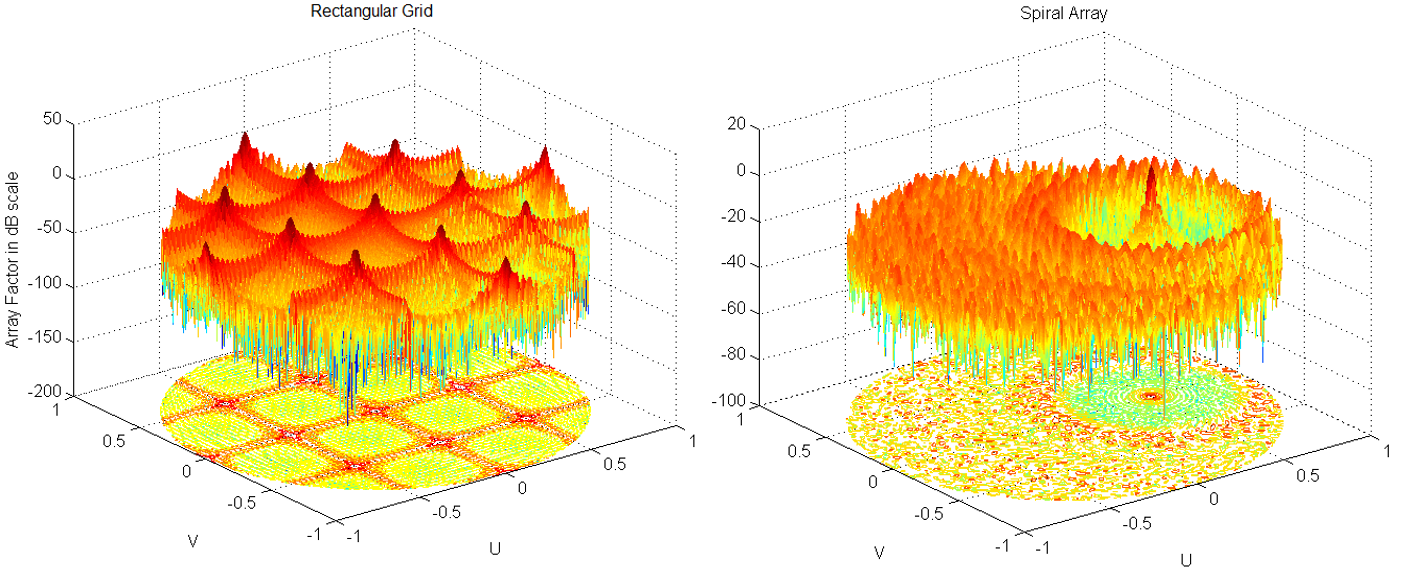,width=160mm}}

\caption{\it Radiation pattern in the $(u,v)$-plane of the sparse regular and sun-flower array. Main beam is scanned towards $(\theta_0,\phi_0)=(30^0,0^0)$. A uniform amplitude tapering is used. Courtesy Yijing Zhang 2015.}
\label{fig:radpatsunflower}
\end{figure}

Most application of phased-arrays do not tolerate grating lobes. In radar applications, grating lobes would create confusion about the position of the target and it would spread the radiated power into multiple directions. In telecommunication applications, grating lobes could allow strong interfering signals to saturate the receiver electronics. However, some applications could allow grating lobes to appear. An example is radio astronomy.
Sparse arrays are arrays of which the inter-element distance $(d_x,d_y)$  is much larger than $0.5 \lambda_0$. Fig. \ref{fig:sparsearrays} illustrates the basic concept of sparse arrays. Sparse array can have a regular grid or irregular grid.
Fig. \ref{fig:radpatsparseregular} shows the radiation pattern of a linear array with $K=32$ elements placed on a dense versus sparse regular grid. The sparse array clearly has a much narrower main beam. The price to pay is the appearance of several grating lobes.
A way to remove the unwanted grating lobes in sparse arrays is to use an irregular array grid. In this case the radiated power in the grating lobes is spread out over the entire visible space. As a result, the average sidelobe level will be higher and will depend on the amount of array elements.
An example of a very efficient irregular sparse array configuration is the {\it sun-flower} topology, as shown in Fig. \ref{fig:sunflower}. The average element distance in both the regular and sun-flower configuration in Fig. \ref{fig:sunflower} is about $2\lambda_0$.
Fig. \ref{fig:radpatsunflower} shows the corresponding radiation patterns of both arrays. Clearly, the sunflower array suppresses the grating lobes at the expense of a higher average sidelobe level.
When the element distribution in a sparse array is random or quasi-random, it can be shown that the average sidelobe level for large arrays with $K \times L$ elements, is approximated by:
\begin{equation}
\begin{array}{lcl}
\displaystyle SLL & = &  \displaystyle \frac{1}{KL}.
 \end{array} \label{eq:sparse1}
\end{equation}
So in case of a sparse-random array with 1000 elements, the average SLL will be -30 dB below the main beam gain.

\section{Calibration of phased-arrays}
\begin{figure}[hbt]
\centerline{\psfig{figure=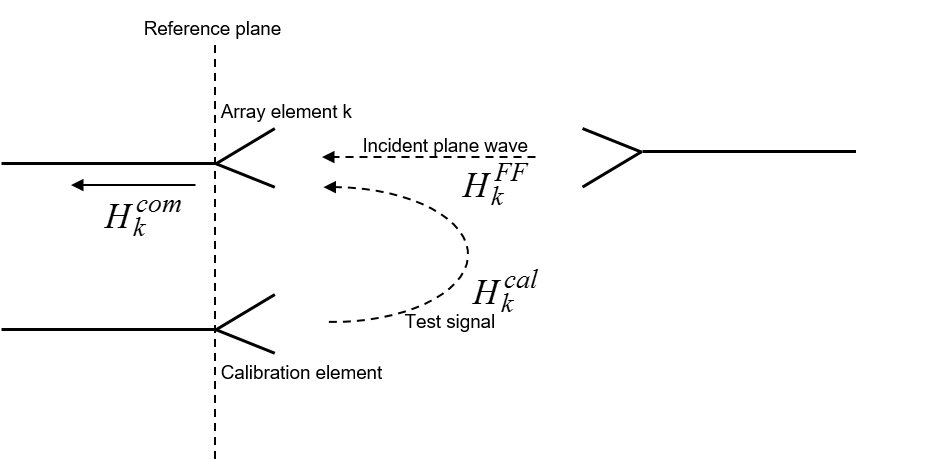,width=110mm}}

\caption{\it Calibration principle used in phased arrays. Only element $k$ is shown.}
\label{fig:calprinciple}
\end{figure}

Amplitude and phase errors will occur in phased-array systems. These errors can have a systematic time-invariant nature, but can also be caused by changes in the environmental conditions, e.g. temperature changes. Calibration must be used to correct for these errors. A schematic diagram of a calibration concept which is often used in modern phased arrays \cite{Smolderscalibration} is shown in Fig. \ref{fig:calprinciple}.
Two types of signals can be injected into each array element: an external sigal generated by a far-field source $H^{FF}_k$, or a local calibration signal $H^{cal}_k$ injected via a special calibration network within the array or by means of injection via mutual coupling with a special calibration element, as illustrated in Fig. \ref{fig:calprinciple}. The first step in the calibration procedure is to relate the received signals at the output of each array element from a test signal of one of the calibration elements to the received signals due to one or more incident plane waves from a far-field source. This part of the calibration procedure is called {\it off-line calibration}, since it needs to be done only once in a high-quality anechoic facility, e.g. by using a near-field scanner or compact antenna range. The element calibration coefficients are calculated and stored in a look-up table. With $a_k$ the received signal from the far-field source and $b_k$ the received signal from the calibration element, we obtain the following calibration coefficients in our look-up table:
\begin{equation}
\begin{array}{lcl}
\displaystyle c_k & = &  \displaystyle \frac{a_k}{b_k}=\frac{H^{FF}_k~H^{com}_k}{H^{cal}_k~H^{com}_k} = \frac{H^{FF}_k}{H^{cal}_k}.
 \end{array} \label{eq:cal1}
\end{equation}
The second step is to measure the amplitude and phase at the output of each array element due to a signal from one of the calibration elements. This part is called {\it on-line calibration}, since it can be done while the system is operation in the application environment.
The measured signal from the calibration element is now:
\begin{equation}
\begin{array}{lcl}
\displaystyle \tilde{b}_k & = &  \displaystyle H^{cal}_k~\tilde{H}^{com}_k.
 \end{array} \label{eq:cal2}
\end{equation}
By combining the results from both calibration steps, we can re-construct the signal transfer from a far-field source to array element $k$:
\begin{equation}
\begin{array}{lcl}
\displaystyle \tilde{a}_k & = &  \displaystyle H^{FF}_k~\tilde{H}^{com}_k = H^{cal}_k~\tilde{H}^{com}_k~\left(\frac{H^{FF}_k}{H^{cal}_k} \right)= \tilde{b}_k~c_k.
 \end{array} \label{eq:cal3}
\end{equation}
Phased arrays often consist of many elements, e.g. common values are 1024 and 4096 elements. Therefore, the on-line calibration could take a lot of time. Several techniques have been proposed in literature to overcome this problem. An example is the {\it multi-element phase togging (MEP) technique} \cite{Smolderscalibration}, in which the FFT principle is used to calibrate groups of array elements simultaneously.

\section{Focal-plane arrays}
\begin{figure}[hbt]
\centerline{\psfig{figure=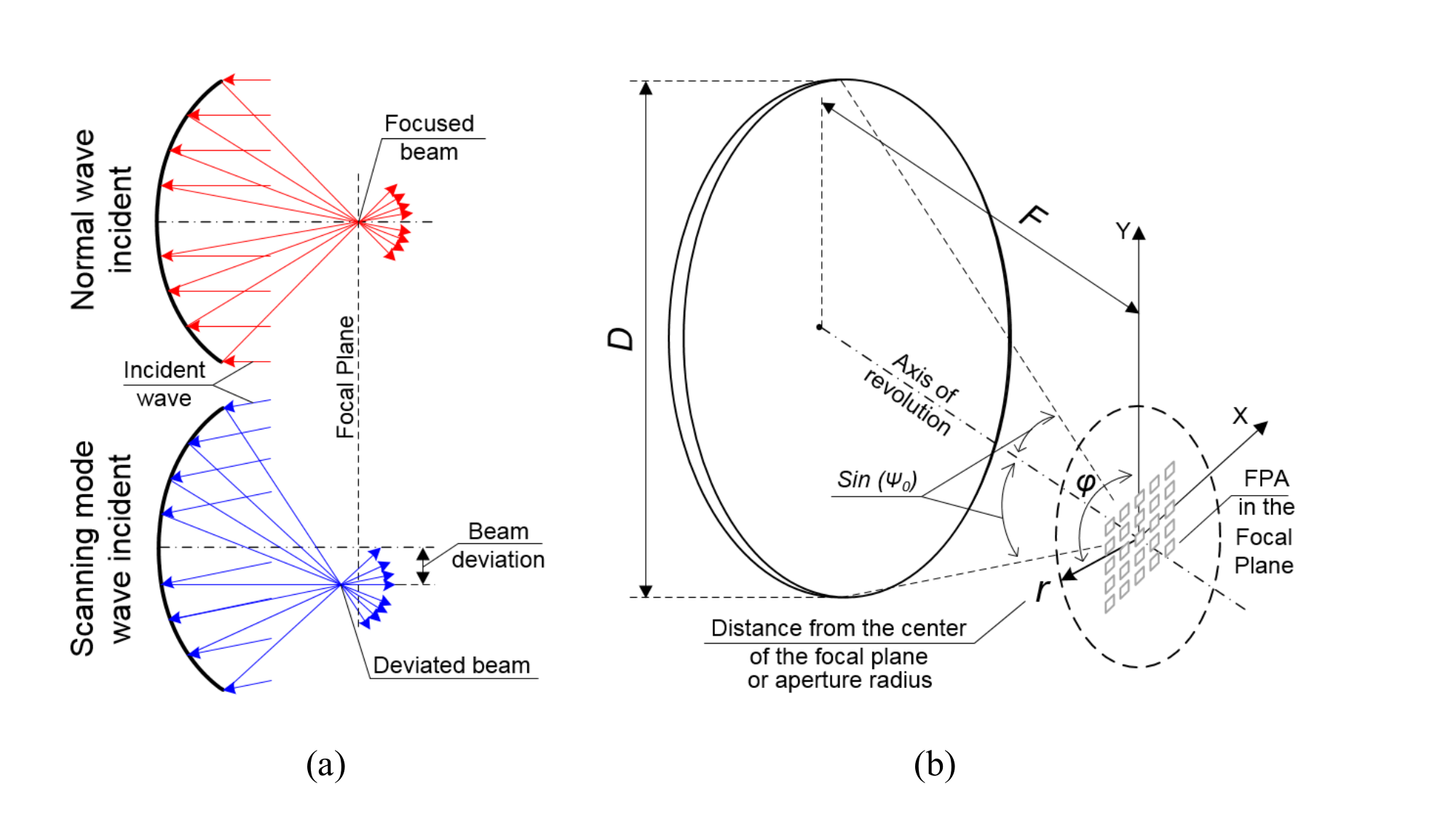,width=140mm}}
\vspace{-0.5cm}
\caption{\it Focal-plane array (FPA) concept in which a phased-array feed is placed in the focal-plane in front of a parabolic reflector antenna, from \cite{DubokTAP2017}. When an incident plane wave illuminates the reflector, the location of the beam focus moves along the phased-array feed with varying angle of incidence.}
\label{fig:FPA}
\end{figure}

\begin{figure}[hbt]
\centerline{\psfig{figure=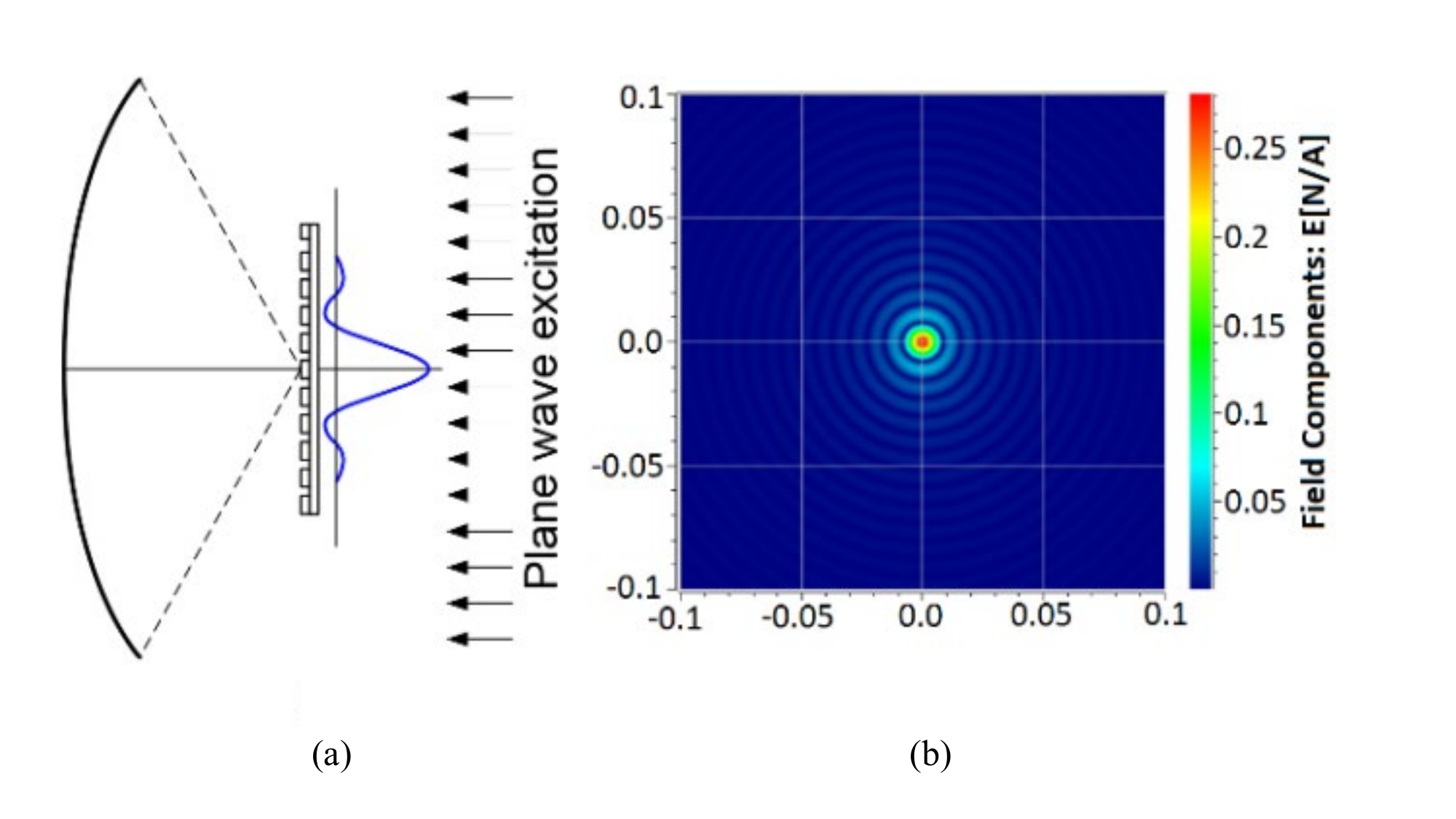,width=140mm}}
\vspace{-0.75cm}
\caption{\it (a) Focal-plane array illuminated by a plane wave (broadside incidence), (b) Field distribution in the focal plane according to (\ref{eq:Fieldfocalplane}) for a classical prime focus parabolic reflector with $F/D=0.6$ and $f=30$ GHz. Dimensions in focal plane in [m]. Figure from \cite{DubokTAP2017}.}
\label{fig:FPAfield}
\end{figure}

A major draw-back of phased-arrays is the large amount of individual antenna elements and associated electronic circuits that is required, leading to high cost and high power consumption. However, several applications do require a high antenna gain, but do not require a very wide scan-range, instead these applications only use a limited field-of-view (FoV). In this case, the hybrid concept of {\it focal-plane arrays (FPAs)} is very interesting. FPAs combine the benefits of phased-arrays and traditional reflector-based solutions by offering a high antenna gain, relative low costs and electronic beam scanning over a limited FoV. Therefore, FPAs have already become an interesting alternative for conventional horn-fed reflector antennas in a number of applications, e.g. in radio astronomy and in satellite communication. Moreover, emerging applications such as point-to-point wireless communications, 5G new-radio millimeter-wave (mm-wave) wireless and low-cost Ka-band multi-function radars could be areas in which FPAs can play a major role. Fig. \ref{fig:FPA} shows the concept of a basic FPA system. A {\it phased-array feed (PAF)} is placed in the focal-plane in front of a parabolic reflector antenna with diameter $D$ and focal length $F$. When an incident plane wave illuminates the reflector, the location of the beam focus moves along the phased-array feed depending on the scan-angle. Therefore, for each scan angle a limited number of array elements will be active at the same time. The antenna gain of a FPA is mainly determined by the size of the reflector and not by the number of array elements in the PAF. In this way, a high antenna gain can be realized with electronic beam-scanning capabilities.
For large $F/D$ ratio and small scan angles the electric field distribution in the focal plane of a prime-focus parabolic reflector antenna can be approximated by:
\begin{equation}
\begin{array}{lcl}
\displaystyle E(r) & = &  \displaystyle \frac{2 J_1(k_0 r \sin{\Psi_0})}{k_0 r \sin{\Psi_0}},
 \end{array} \label{eq:Fieldfocalplane}
\end{equation}
where $J_1$ is the Bessel function of the first kind with order 1, $\Psi_0=\pi/4$, the subtended angle of the reflector, $k_0$ is the free-space wavenumber and $r$ is the distance from the center of the focal plane (aperture radius), see Fig. \ref{fig:FPA}. The field distribution in the focal plane is illustrated in the example of Fig. \ref{fig:FPAfield} in case of broadside incidence of a plane wave. The rings in the field distribution are also known as {\it Airy rings}. We can clearly see that only a limited number of array elements will be active simultaneously for a particular angle of incidence (scan angle). In case of scanning, the pattern will shift along the focal plane.
The aperture efficiency $\eta$ as a function of the radius of the PAF feed $r_p$ of the FPA can be determined by integrating the electric field in the focal plane and normalize it to the total power $P_{tot}$ obtained by the reflector:
\begin{equation}
\begin{array}{lcl}
\displaystyle \eta (r_p) & = &  \displaystyle \frac{1}{P_{tot}} \int\limits_{0}^{2\pi} \int\limits_{0}^{r_p} |E(r)|^2 dr d\phi ,
 \end{array} \label{eq:FPAefficiency}
\end{equation}
More details on the scanning capabilities and limitations can be found in \cite{DubokTAP2017}, \cite{SmoldersFPA2019}.


%

%


\renewcommand{\theenumi}{\alph{enumi}}
\renewcommand{\labelenumi}{\theenumi) }
\renewcommand{\theequation}{\thechapter.\arabic{equation}}
\newcommand{\e}{\\[5mm]}
\newcommand{\vc}[1]{\underline{#1}}
\newcommand{\tu}[1]{\underline{\tilde{#1}}}
\newcommand{\isdef}{\stackrel{\rm def}{=}}
\newcounter{nr}

\def\bea{\begin{eqnarray*}}
\def\eea{\end{eqnarray*}}
\def\bd{\begin{displaymath}}
\def\ed{\end{displaymath}}
\def\nv#1{\vec{#1}}
\def\bd{\begin{displaymath}}
\def\ed{\end{displaymath}}
\def\be{\begin{eqnarray*}}
\def\ee{\end{eqnarray*}}
\def\oppS{{\cal S}}
\def\volV{{\cal V}}
\def\dl{{\rm d}}
\newcommand{\intsubscr}[1]{\displaystyle \int\limits_{\rput{0}{{}^{#1}}}}
\newcommand{\intdbl}[1]{\displaystyle \intsubscr{#1}\!\!\!\!\int}
\newcommand{\inttri}[1]{\displaystyle \int\!\!\!\!\intsubscr{#1}\!\!\!\!\int}
\newcommand{\intkr}[1]{\displaystyle \rlap{{\footnotesize $\bigcirc$}}{\intsubscr{#1}}}
\newcommand{\intopp}[1]{\displaystyle \rlap{{\footnotesize $\bigcirc$}}{\!\!\intdbl{#1}}}
\def\Re{\mbox{Re}\/}
\def\Im{\mbox{Im}\/}
\def\eps{\varepsilon}
\def\isdef{\hat{=}}
\def\R{\mbox{${\tt I \hspace{-5pt} R}$}}
\def\C{\mbox{${\rm l \hspace{-10pt} C}$}}
\def\hf{\mbox{$\frac{1}{2}$}}
\def\dhf{\mbox{$\frac{d}{2}$}}
\def\fracG#1#2{{\displaystyle \frac{#1}{#2}}}
\def\dg2{\mbox{$\frac{\delta y}{2}$}}
\def\rp{\nv{r} \, '}

\def\pard#1{{\frac{\partial}{\partial #1}}}

\appendix

\chapter{Vector calculus and coordinate systems}
\label{chap:AppendixA}

\section{Vector algebra}

\begin{eqnarray*}
  & & \vec{a}\cdot\vec{b} = \vec{b}\cdot\vec{a},\\
  & & \vec{a}\cdot(\vec{b}\times\vec{c})=\vec{b}\cdot(\vec{c}\times\vec{a})
  =\vec{c}\cdot(\vec{a}\times\vec{b}),\\
  & & \vec{a}\times\vec{b}=-(\vec{b}\times\vec{a}),\\
  & & \vec{a}\times(\vec{b}\times\vec{c})=(\vec{a}\cdot\vec{c})\vec{b}
  -(\vec{a}\cdot\vec{b})\vec{c}.\\
\end{eqnarray*}

\section{Vector operations in rectangular (Cartesian) coordinates}
\begin{figure}[hbt]
\centerline{\psfig{figure=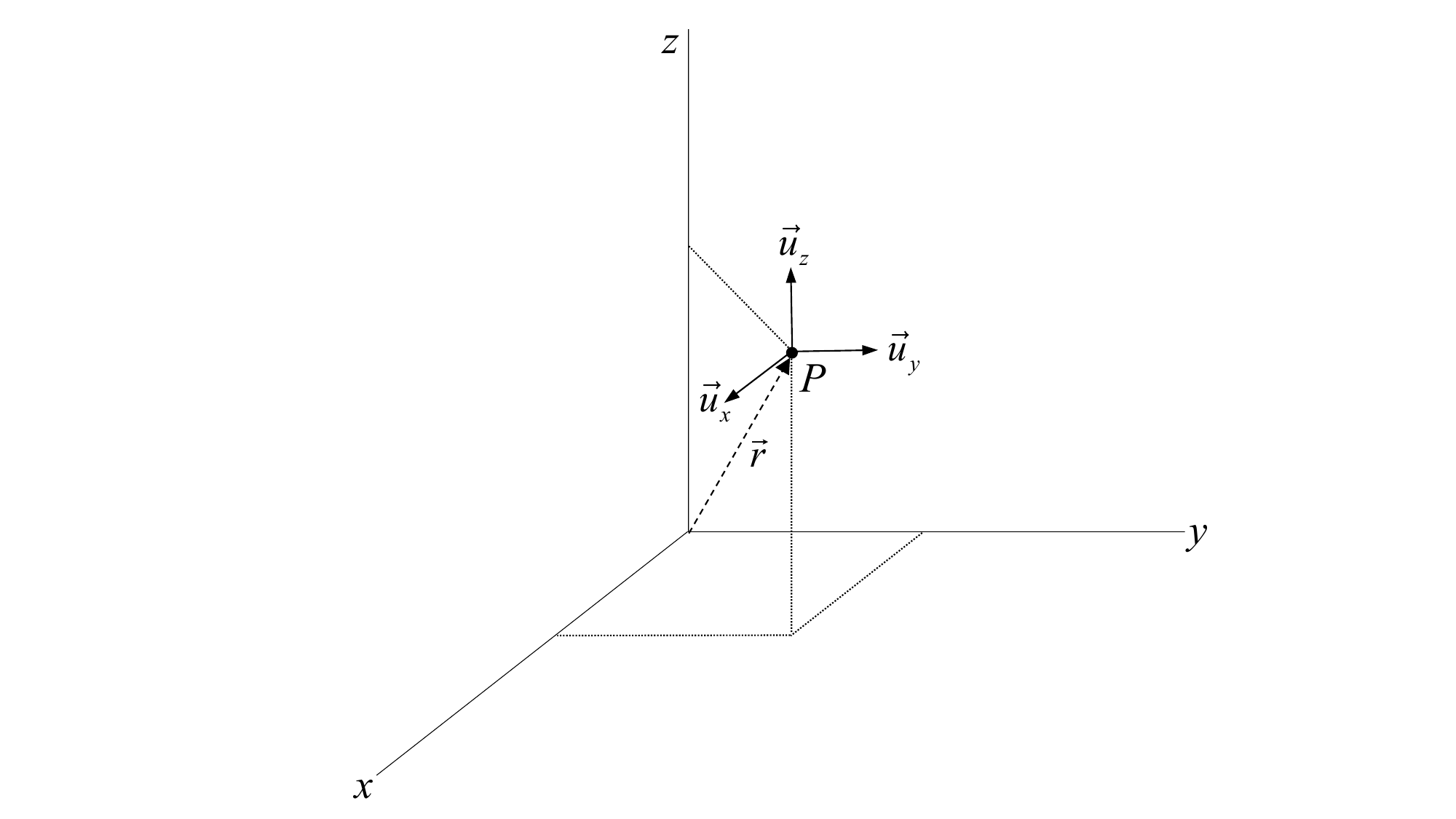,width=140mm}}
\caption{\it Rectangular coordinate system with $\vec{r}=x\vec{u}_x+y\vec{u}_y+z\vec{u}_z$.}
\end{figure}
\begin{eqnarray*}
   \vec{A} & = & \vec{u}_{x} A_x  +\vec{u}_{y} A_y +\vec{u}_{z} A_z, \\
   \nabla \Psi & = & \mbox{grad}\Psi= \vec{u}_{x} \frac{\partial\Psi}{\partial x}  +\vec{u}_{y} \frac{\partial\Psi}{\partial y} +\vec{u}_{z}\frac{\partial\Psi}{\partial z},\\
   \nabla \cdot \vec{A} & = & \mbox{div} \vec{A} = \frac{\partial A_x}{\partial x} + \frac{\partial A_y}{\partial y} +\frac{\partial A_z}{\partial z},\\
   \nabla \times \vec{A} & = & \mbox{curl} \vec{A}= \left|
  \begin{array}{ccc}
    \vec{u}_{x} & \vec{u}_{y} & \vec{u}_{z}\\
    \di \pard{x} & \di \pard{y} & \di \pard{z}\\
    A_{x} & A_{y} & A_{z}
  \end{array}
  \right|
  \begin{array}{c}
   \di  =\vec{u}_{x}(\frac{\partial A_z}{\partial y}- \frac{\partial A_y}{\partial z})+\vec{u}_{y}
    (\frac{\partial A_x}{\partial z} - \frac{\partial A_z}{\partial x})\\
    \di +\vec{u}_{z} (\frac{\partial A_y}{\partial x}-\frac{\partial A_x}{\partial y}),
  \end{array}\\
    \nabla^{2}\Psi & = & \Big( \frac{\partial^{2}}{\partial x^{2}} +  \frac{\partial^{2}}{\partial y^{2}} + \frac{\partial^{2}}{\partial z^{2}} \Big)\Psi, \\
  \nabla^{2}\vec{A} & = &  \vec{u}_{x} \nabla^{2}A_x+\vec{u}_{y} \nabla^{2}A_y+\vec{u}_{z} \nabla^{2}A_z.
\end{eqnarray*}

\section{Vector operations in cylindrical coordinates}

\begin{figure}[hbt]
\centerline{\psfig{figure=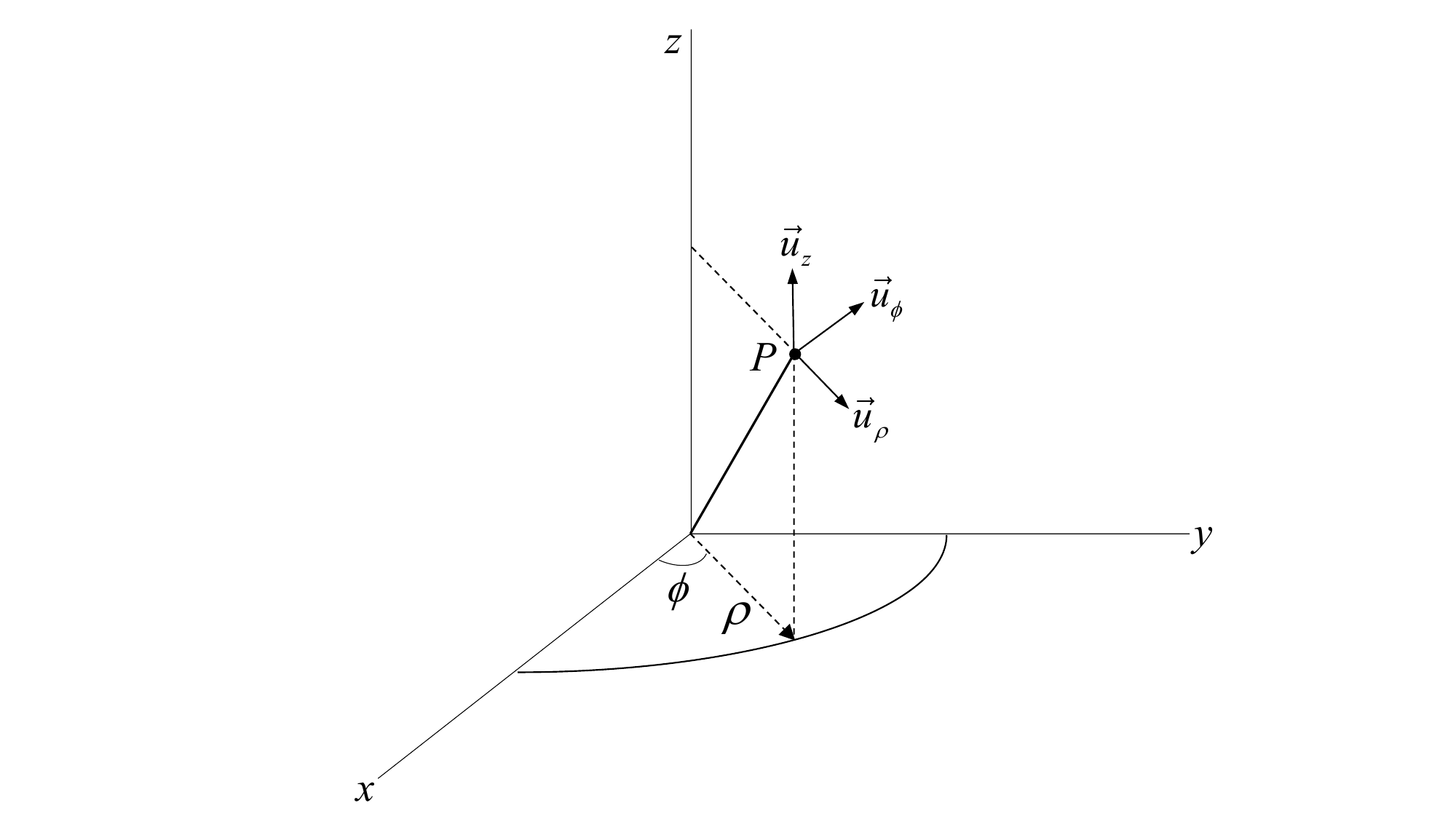,width=140mm}}
\caption{\it Cylindrical coordinate system with $x=\rho\cos\phi$, $y=\rho\sin\phi$, $z=z$.}
\end{figure}
\begin{eqnarray*}
  & & \nabla\Psi=\vec{u}_{\rho}\frac{\partial\Psi}{\partial \rho}\,+\,
  \vec{u}_{\phi}\frac{1}{\rho}\,\frac{\partial\Psi}{\partial\phi}\,+\,
  \vec{u}_{z}\frac{\partial\Psi}{\partial z},\\
  & & \nabla \cdot\vec{A}=\frac{1}{\rho}\,\frac{\partial}{\partial \rho}(rA_{\rho})+
  \frac{1}{\rho}\,\frac{\partial A_{\phi}}{\partial \phi}+
  \frac{\partial A_{z}}{\partial z},\\
  & & \nabla \times\vec{A}=\vec{u}_{\rho}\left[\frac{1}{\rho}\,
  \frac{\partial A_{z}}{\partial\phi}-\frac{\partial A_{\phi}}
  {\partial z}\right] \,+\, \vec{u}_{\phi}\left[\frac{\partial A_{\rho}}
  {\partial z}-\frac{\partial A_{z}}{\partial \rho}\right]\,+\,
  \vec{u}_{z}\frac{1}{\rho}\left[\frac{\partial}{\partial \rho}(\rho A_{\phi})
  -\frac{\partial A_{\rho}}{\partial\phi}\right],\\
  & & \nabla^{2}\Psi=\frac{1}{\rho}\, \frac{\partial}{\partial \rho}\left(
  \rho\frac{\partial\Psi}{\partial \rho}\right)+\frac{1}{\rho^{2}}
  \frac{\partial^{2}\Psi}{\partial\phi^{2}}+
  \frac{\partial^{2}\Psi}{\partial z^{2}}.\\
\end{eqnarray*}

\section{Vector operations in spherical coordinates}

\begin{figure}[hbt]
\centerline{\psfig{figure=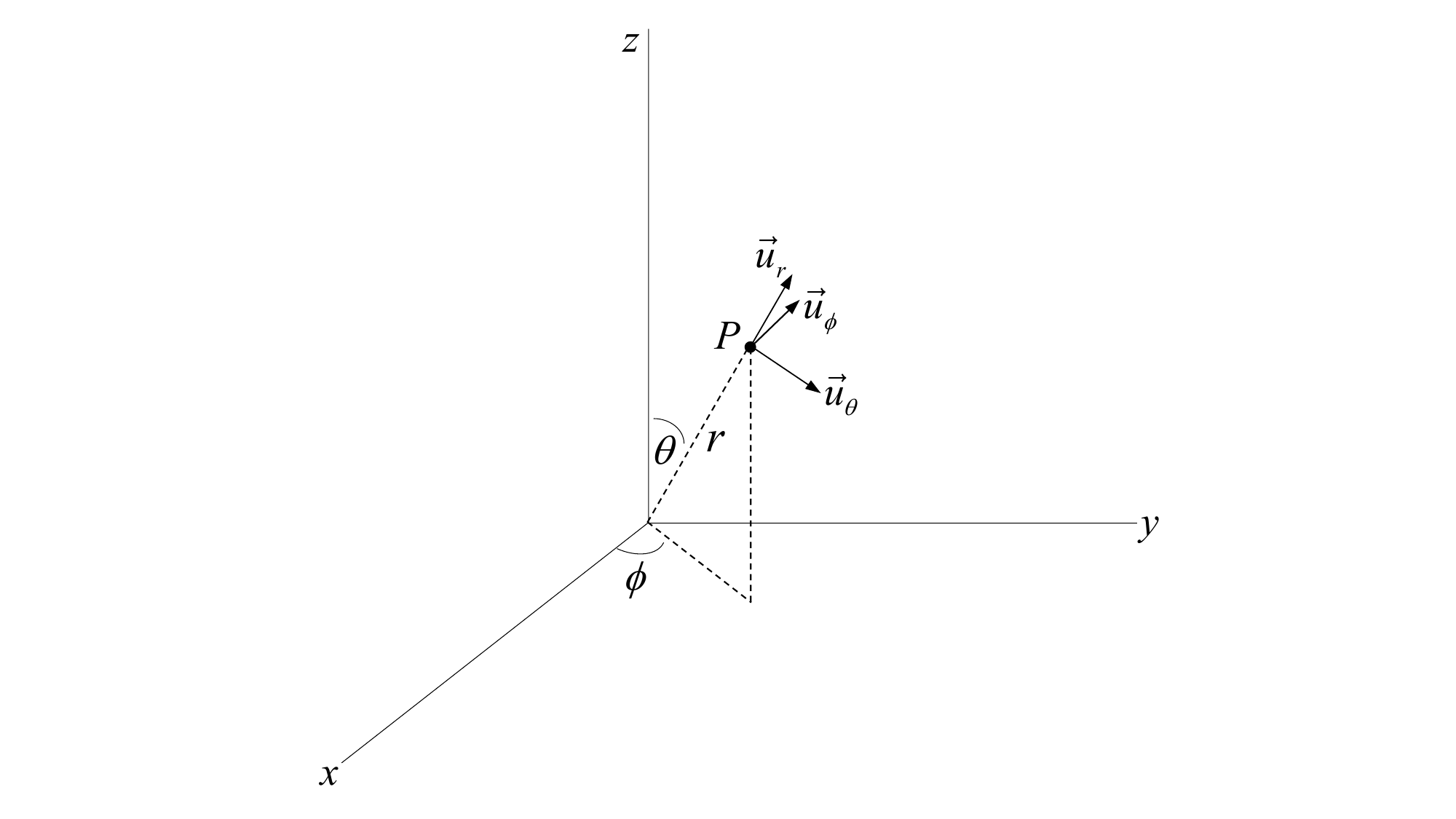,width=140mm}}
\caption{\it Spherical coordinate system with $x=r\sin\theta\cos\phi$, $y=r\sin\theta
  \sin\phi$, $z=r\cos\theta$ and $dS=r^2\sin\theta d\theta d\phi=r^2 d\Omega$, where $d\Omega=\sin\theta d\theta d\phi$ and $\Omega$ is the solid angle.}
\end{figure}
\begin{eqnarray*}
  & & \nabla\Psi=\vec{u}_{r}\frac{\partial\Psi}{\partial r}\,+\,
  \vec{u}_{\theta}\frac{1}{r}\,\frac{\partial\Psi}{\partial\theta}\,+\,
  \vec{u}_{\phi}\frac{1}{r\sin\theta}\,
  \frac{\partial\Psi}{\partial \phi},\\
  & & \nabla \cdot\vec{A}=\frac{1}{r^{2}}\,\frac{\partial}
  {\partial r}(r^{2}A_{r})+
  \frac{1}{r\sin\theta}\,\frac{\partial}{\partial \theta}
  (\sin\theta A_{\theta})+
  \frac{1}{r\sin\theta}\frac{\partial A_{\phi}}{\partial\phi},\\
  & & \nabla \times\vec{A}=\vec{u}_{r}\frac{1}{r\sin\theta}\left\{
  \frac{\partial}{\partial\theta}(\sin\theta A_{\phi})-
  \frac{\partial A_{\theta}}{\partial\phi}\right\}\,+\\
  & & \qquad\qquad\vec{u}_{\theta}\left[\frac{1}{r\sin\theta}\,
  \frac{\partial A_{r}}{\partial\phi}-\frac{1}{r}\,
  \frac{\partial}{\partial r}(r A_{\phi})\right]\,+\,
  \vec{u}_{\phi} \frac{1}{r} \,
  \left[\frac{\partial}{\partial r}(r A_{\theta})-
  \frac{\partial A_{r}}{\partial\theta}\right],\\
  & & \nabla^{2}\Psi=\frac{1}{r^{2}}\,\frac{\partial}{\partial r}\left(
  r^{2}\frac{\partial\Psi}{\partial r}\right)+\frac{1}{r^{2}\sin\theta}
  \,\frac{\partial}{\partial\theta}\left(\sin\theta \frac{\partial\Psi}
  {\partial\theta}\right)+
  \frac{1}{r^{2}\sin^{2}\theta}\frac{\partial^{2}\Psi}{\partial\phi^{2}}.\\
\end{eqnarray*}

\section{Commonly used vector products}

\begin{eqnarray*}
  & & \displaystyle \vec{u}_r \times \vec{u}_{\theta} = \vec{u}_{\phi},\\
  & & \displaystyle \vec{u}_r \times \vec{u}_{\phi} = - \vec{u}_{\theta},\\
  & & \displaystyle \vec{u}_{\theta} \times \vec{u}_{\phi} = \vec{u}_{r}.
\end{eqnarray*}
\begin{eqnarray*}
  & & \displaystyle \vec{u}_r \times \vec{u}_{x} =   \vec{u}_{\phi} \cos{\theta} \cos{\phi} + \vec{u}_{\theta} \sin{\phi} ,\\
  & & \displaystyle \vec{u}_r \times \vec{u}_{y} =   \vec{u}_{\phi} \cos{\theta} \sin{\phi} - \vec{u}_{\theta} \cos{\phi} ,\\
  & & \displaystyle \vec{u}_r \times \vec{u}_{z} = - \vec{u}_{\phi} \sin{\theta}.
  \end{eqnarray*}
\begin{eqnarray*}
 & & \displaystyle \vec{u}_r \times \vec{u}_r \times \vec{u}_x =   \vec{u}_{\phi} \sin{\phi} - \vec{u}_{\theta} \cos{\theta} \cos{\phi},\\
 & & \displaystyle \vec{u}_r \times \vec{u}_r \times \vec{u}_y = - \vec{u}_{\phi} \cos{\phi} - \vec{u}_{\theta} \cos{\theta} \sin{\phi},\\
 & & \displaystyle \vec{u}_r \times \vec{u}_r \times \vec{u}_z =   \vec{u}_{\theta} \sin{\theta}.
\end{eqnarray*}
\begin{eqnarray*}
 & & \di\vec{u}_r \cdot \vec{r}_0 = x_0 \sin{\theta}\cos{\phi} + y_0\sin{\theta}\sin{\phi} = x_0 u + y_0 v, ~\mbox{when $\vec{r}_0=(x_0,y_0,0)$} \\
 & & u=\sin{\theta}\cos{\phi}, ~v=\sin{\theta}\sin{\phi}.
\end{eqnarray*}
\begin{eqnarray*}
 & & \di\vec{u}_x = \vec{u}_r \sin{\theta} \cos{\phi} + \vec{u}_{\theta} \cos{\theta} \cos{\phi} - \vec{u}_{\phi} \sin{\phi}, \\
 & & \di\vec{u}_y = \vec{u}_r \sin{\theta} \sin{\phi} + \vec{u}_{\theta} \cos{\theta} \sin{\phi} + \vec{u}_{\phi} \cos{\phi}, \\
 & & \di\vec{u}_z = \vec{u}_r \cos{\theta} - \vec{u}_{\theta} \sin{\theta}, \\
 & & \di\vec{u}_r = \vec{u}_x \sin{\theta} \cos{\phi} + \vec{u}_{y} \sin{\theta} \sin{\phi} + \vec{u}_{z} \cos{\theta}, \\
 & & \di\vec{u}_{\theta} = \vec{u}_x \cos{\theta} \cos{\phi} + \vec{u}_{y} \cos{\theta} \sin{\phi} - \vec{u}_{z} \sin{\theta}, \\
 & & \di\vec{u}_{\phi} =   - \vec{u}_x \sin{\phi} + \vec{u}_{y} \cos{\phi}.
\end{eqnarray*}

\section{Vector identities}

\bea
  \nabla(\Phi +\Psi) & = & \nabla\Phi + \nabla\Psi,\\
  \nabla \cdot(\vec{A}+\vec{B}) & = & \nabla \cdot\vec{A}+\nabla \cdot\vec{B},\\
  \nabla \times(\vec{A}+\vec{B}) & = & \nabla \times\vec{A}+\nabla \times\vec{B}. \\
\eea
\bea
  \nabla(\Phi\Psi) & = & \Phi\,\nabla\Psi + \Psi\nabla\Phi,\\
  \nabla \cdot(\Phi\vec{A}) & = & \Phi\,\nabla \cdot\vec{A} +
  \vec{A}\cdot\nabla\Phi,\\
  \nabla \times(\Phi\vec{A}) & = & \Phi\,\nabla \times\vec{A} -
  \vec{A}\times\nabla\Phi.
\eea
\bea
  \nabla(\vec{A}\cdot\vec{B}) & = & (\vec{B}\cdot\nabla) \vec{A}+
  (\vec{A}\cdot\nabla) \vec{B}+\vec{B}\times(\nabla \times\vec{A})+
  \vec{A}\times(\nabla \times\vec{B}),\\
  \nabla \cdot(\vec{A}\times\vec{B}) & = & \vec{B}\cdot(\nabla \times\vec{A})-
  \vec{A}\cdot(\nabla \times\vec{B}),\\
  \nabla \times(\vec{A}\times\vec{B}) & = & (\vec{B}\cdot\nabla)\vec{A}-
  (\vec{A}\cdot\nabla)\vec{B}-\vec{B}\,(\nabla \cdot\vec{A})+\vec{A}\,(\nabla \cdot\vec{B}).
\eea
\bea
  \nabla^{2}\vec{A} & = & \nabla \nabla \cdot \vec{A}-\nabla \times\nabla \times \vec{A},\\
  \nabla \cdot \nabla \Psi & = & \nabla^2 \Psi, \\
  \nabla \cdot \nabla \times \vec{A} & = & 0,\\
  \nabla \times \nabla \Phi & = & \vec{0}.
\eea

\end{document}